\documentclass[12pt,twoside]{book}

\usepackage{graphicx}
\usepackage{amsmath,amssymb}
\usepackage{ascmac}
\usepackage{color}
\usepackage{fancybox}
\usepackage{bm}
\usepackage{bbm}
\usepackage{mathrsfs}
\usepackage{cite}
\usepackage[bf,compact,topmarks,pagestyles]{titlesec}
\usepackage[bf]{titlesec}
\usepackage{titletoc}
\usepackage{setspace}
\usepackage{fancyhdr}

  \contentsmargin{2.55em}
  \dottedcontents{part}[1.5em]{\addvspace{20pt}\bf}{2.3em}{1pc}
  \dottedcontents{section}[1.5em]{\addvspace{5pt}}{2.3em}{1pc}
  \dottedcontents{subsection}[3.8em]{\addvspace{5pt}}{3.2em}{1pc}

 \titleformat{\part}[block]
  {\filcenter\huge
   \addtolength{\titlewidth}{2pc}%
   \addvspace{90pt}%
\bf\fontsize{25pt}{0}
}
  {Part \thepart}{1em}{\\\vspace{5mm}}
   \titlespacing{\part}
  {1pc}{*2}{*2}[1pc]

\makeatletter

\@addtoreset{equation}{chapter}

\makeatother
\begin{document}
\frontmatter
\begin{center}
{\Large Doctor Thesis}
\end{center}
\vspace{10mm}
\begin{center}
{\bf\LARGE
Studies on Extended Higgs Sectors\\\vspace{3mm}
as a Probe of\\\vspace{5mm}
New Physics Beyond the Standard Model}\\
\vspace{10mm}
{\Large Kei Yagyu}\\
\vspace{10mm}
{\large \it
Department of Physics, University of Toyama,\\
3190 Gofuku, Toyama 930-8555, Japan}\\
\end{center}
\begin{center}
{\large March 2012}
\end{center}
\begin{figure}[h]
\begin{center}
\includegraphics[width=40mm]{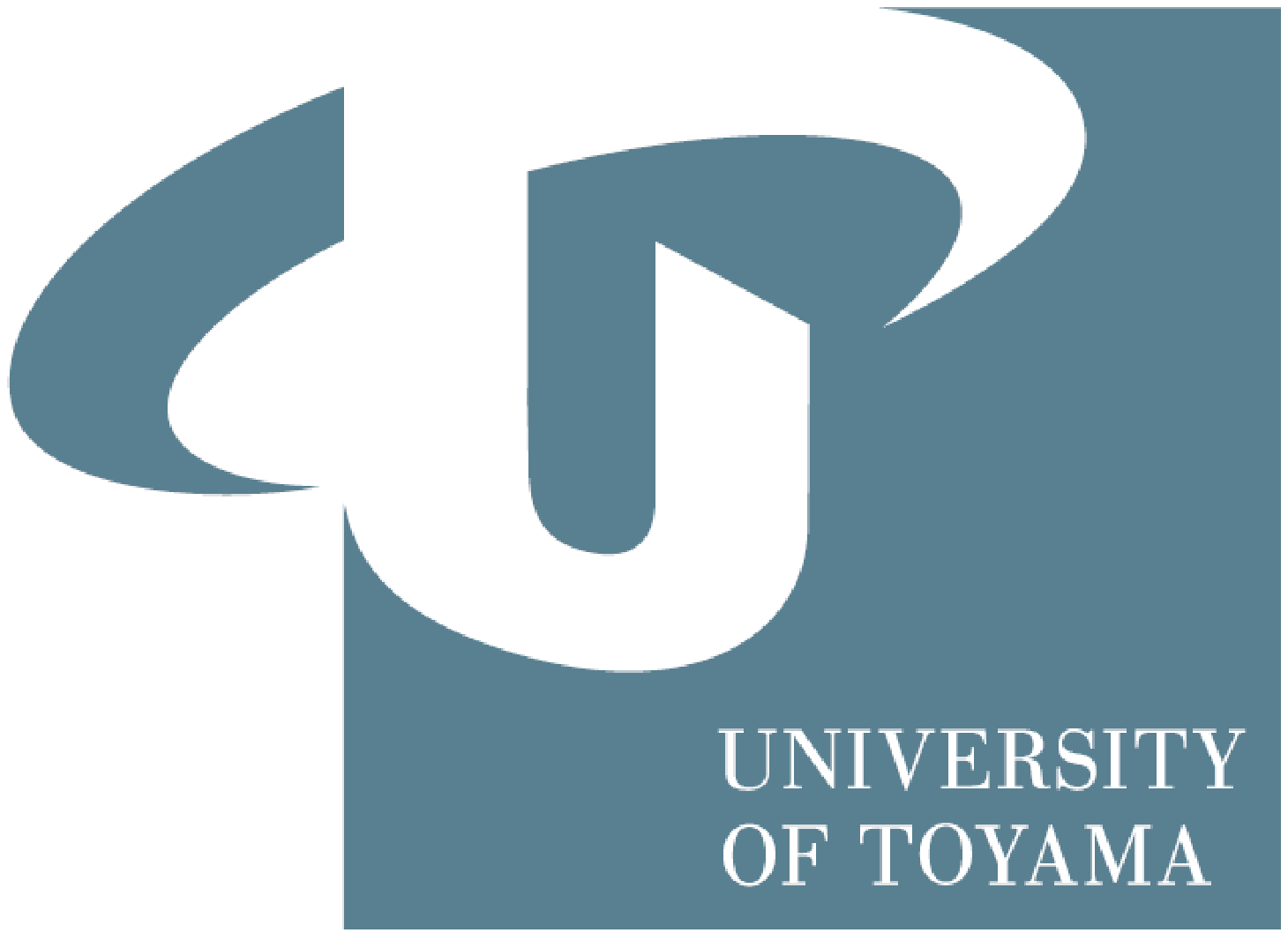}\hspace{50mm}
\includegraphics[width=30mm]{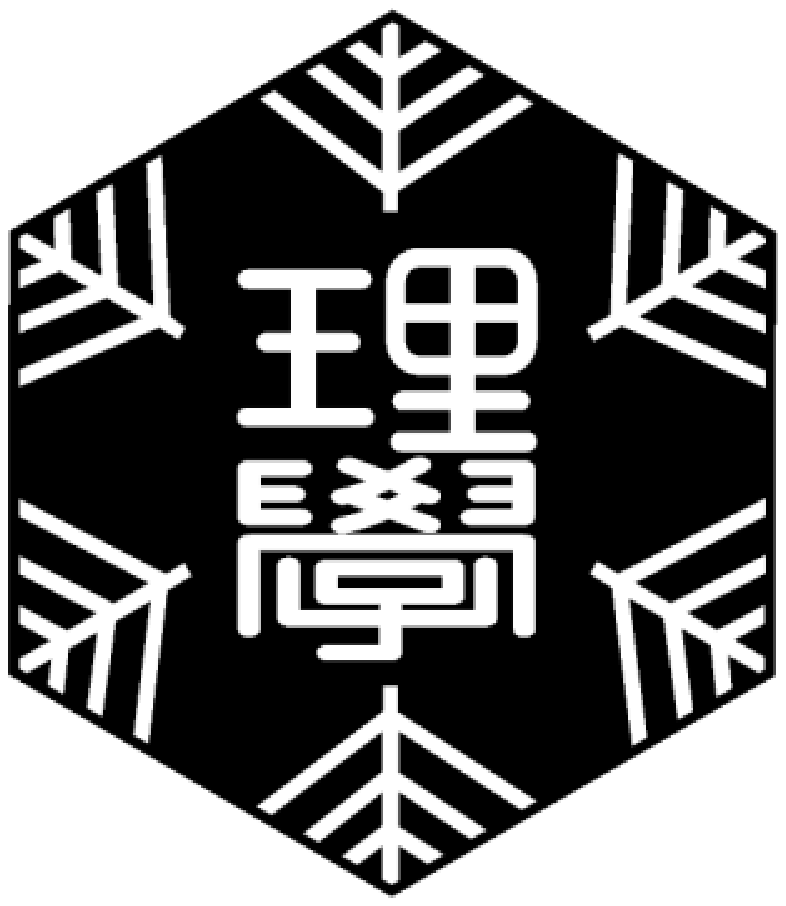}
\end{center}
\end{figure}

\newpage
\begin{center}
{\Large ACKNOWLEDGMENTS}
\end{center}
\begin{spacing}{1.5}
{\large
I would like to express my sincere gratitude to my supervisor, 
Prof.~Shinya Kanemura for providing me much instruction in particle physics and its philosophy. 
I am very grateful to Prof.~Mayumi Aoki, Prof.~Tetsuo Shindou, Prof.~Stefano Moretti, Prof.~Rui Santos, 
Dr.~Koji Tsumura, Dr.~Renato Guedes, Mr.~Kazuya Yanase and Mr.~Yuki Mukai 
for the fruitful collaborations. 
I am also grateful to the other faculty members of the laboratory, Prof.~Takeshi Kurimoto and 
Prof.~Mitsuru Kakizaki for useful discussions. 
I would like to thank Prof.~Yasuhiro Okada, Prof.~Fusakazu Matsushima and Prof.~Yoshiki Moriwaki for careful
reading of this thesis. 
I would also like to thank all the members in my laboratory, 
Dr.~Makoto Nakamura, 
Mr.~Takehiro Nabeshima, 
Mr.~Hiroyuki Taniguchi, 
Mr.~Tsutomu Kaburagi, 
Mr.~Hiroyasu Nakahori, 
Ms.~Mariko Kikuchi, 
Mr.~Naoki Machida and 
Mr.~Toshinori Matsui
for their hospitality.  }
\end{spacing}
\newpage
\begin{center}
{\Large Abstract}
\end{center}
It is the most important issue in particle physics to understand the mechanism of electroweak symmetry breaking. 
In the standard picture for elementary particles based on the quantum field theory, 
the vacuum expectation value (VEV) of the Higgs scalar field triggers the spontaneous
breakdown of the electroweak gauge symmetry. 
Although the Standard Model (SM) for elementary particles 
has been successful in describing high-energy phenomena at colliders, 
the Higgs boson has not been discovered yet. 
Currently, Higgs boson searches are underway at the CERN Large Hadron Collider (LHC). 
The Higgs boson is strongly expected to be discovered in near future 
as long as the SM is effectively correct. 
The discovery of the Higgs boson does not mean the end of particle physics, because 
there remain problems which cannot be explained within the framework of the SM.  
First of all, it is well known that the Higgs scalar boson causes so-called the hierarchy problem, in which 
the quadratic ultraviolet divergence appears in the renormalization of the Higgs boson mass. 
The renormalization of such quadratic divergences leads to a huge unnatural fine tuning. 
Second, neutrino oscillation has been established by experiments. 
Tiny masses of neutrinos are necessary to explain such phenomena, while neutrinos are massless in the SM. 
Therefore, we need to extend the SM so as to have tiny but nonzero neutrino masses. 
Third, there is no candidate for dark matter in the particle content in the SM, 
although the existence of dark matter has confined at the experiment. 
Finally, it has been clarified that the baryon asymmetry of the Universe cannot be realized in the SM. 
To solve these problems, new physics models beyond the SM have been proposed. 
In such new models, the Higgs sector is often extended from that of the SM, 
where it takes the minimal form with only one isospin doublet Higgs field. 
The structure of a Higgs sector strongly depends on the property of the corresponding new physics model at
the high energy scale. 
Therefore, 
by comparing predicted observables with future experimental data such as the Higgs boson mass, the width, decay 
branching ratios and so on, we can determine new paradigm for physics beyond the SM.
In this thesis, we discuss theoretical properties of various Higgs sectors, and we analyze
constraints from current experimental data,  
and then we study collider signatures in each Higgs sector from this view point. 

\vspace{2mm}
In Part~I, we focus on the phenomenology of various extended Higgs sectors. 

First, as a simple but important extended Higgs sector, 
we discuss the two Higgs doublet model (THDM). 
The THDM appears in several classes of new physics models such as the CP-violation, 
the minimal supersymmetric standard model (MSSM), radiative seesaw models and so on.   
In the THDM, 
the softly-broken discrete $Z_2$ symmetry is often imposed to avoid the flavor changing neutral current at the tree level.  
Under the $Z_2$ symmetry, there are four types of the Yukawa interaction (type-I, type-II, type-X and type-Y). 
The type of Yukawa interaction can 
be related to each new physics scenarios. 
For example, the type-II THDM is predicted in the Higgs sector of the MSSM, 
while so-called the type-X THDM appears in some TeV scale 
models which can explain neutrino masses, dark matter and baryon symmetry of the Universe.  
We investigate the phenomenological differences among these THDMs at the LHC and the International Linear Collider (ILC) 
to discriminate new physics models. 
We find that in the type-II THDM, additional Higgs bosons such as the CP-odd Higgs boson and the heavier CP-even Higgs boson   
can dominantly decay into $b\bar{b}$, while in the type-X THDM, those mainly decay into $\tau^+\tau^-$.  
By using this difference the Higgs sector in the MSSM and the type-X THDM may be able to be distinguished at colliders. 

Second, we consider the Higgs model with the $Y=1$ triplet Higgs field so-called the Higgs triplet model (HTM) 
which is motivated by generating tiny neutrino masses. 
In this model, the rho parameter $\rho$ can deviate from unity at the tree level, since the custodial 
symmetry is broken in the kinetic term of Higgs fields. 
Thus, the electroweak parameters are described by the four input parameters 
such as 
$\alpha_{\text{em}}$, $G_F$, $m_Z$ and $\sin\theta_W$ instead of three parameters 
$\alpha_{\text{em}}$, $G_F$ and $m_Z$ with the relation of $\cos\theta_W=m_W/m_Z$ in models with $\rho=1$ at the tree level. 
We calculate the one-loop correction to the W boson mass as well as the rho parameter in order to 
clarify the possible mass spectrum of extra scalar bosons under the constraint from the electroweak precision data. 
We find that the hierarchical mass spectrum among the scalar bosons mainly originated from the triplet field 
is favored by the data, especially  
in the case where doubly-charged scalar bosons are the lightest of all them.  
We then discuss the phenomenology of the HTM in light of the case with 
the mass splitting among the triplet-like scalar bosons. 
We outline that all the masses of these scalar bosons would be 
reconstructed by using the transverse mass distribution and the invariant mass distribution at the LHC. 
We also calculate the deviation of the decay rate of the Higgs boson decay into two photons in the HTM from that in the SM. 

Third, we study properties of charged Higgs bosons $H^\pm$, especially 
focusing on the $H^\pm W^\mp Z$ vertex. 
This vertex strongly depends on the structure of Higgs sectors depending on the breakdown of a 
custodial symmetry of the model, so that we can constrain Higgs sectors with custodial symmetry breaking. 
We study the possibility of measuring the $H^\pm W^\mp Z$ vertex 
via single charged Higgs boson production associated with the 
$W^\pm$ boson at the ILC by using the 
recoil method. 
We find that the $H^\pm W^\mp Z$ vertex would be testable with the similar accuracy to the rho parameter 
at the ILC. 

Finally, we discuss various supersymmetric (SUSY) Higgs sectors, where 
extra chiral superfields are added to the MSSM, 
which are motivated to solve several physics problems. 
For example, 
the next-to-MSSM in which a neutral singlet superfield is added to the MSSM is motivated to solve the $\mu$-problem, 
models with extra triplet superfields can be used to explain neutrino masses and so on. 
In particular, we focus on decoupling properties of extended SUSY Higgs sectors. 
When masses of new particles are heavy then effects of new physics to the low energy observables decouple which 
is well known as the Appelquest's theorem.  
However, if masses of new particles are mainly determined by the VEV of the Higgs field 
then this theorem does not hold. 
In such a case, nondecoupling effects appear in the low energy observables. 
We investigate such effects in various SUSY Higgs sectors with additional interaction terms from the 
F-term contribution in the Higgs potential at the tree level. 
We also discuss extended SUSY Higgs sectors without such F-term contributions. 
As a concrete example, we consider the model with four Higgs doublet fields. 
In this model, 
if there are mixings between the MSSM-like doublet fields and
the extra doublet fields due to a large soft-breaking B-term,
nonvanishing effects can appear in the MSSM observables such as 
the masses of the CP-even scalar bosons, those of charged Higgs bosons and the mixing angle 
between the CP-even scalar states. 

\vspace{2mm}
In Part~II, we discuss new physics models at the TeV scale, 
where neutrino masses, dark matter and/or 
baryon asymmetry of the Universe can be explained. 

First, 
we discuss a model proposed by M. Aoki, S. Kanemura and O. Seto (2009), 
in which neutrino oscillation, dark matter, and baryon asymmetry of the Universe
would be simultaneously explained by the TeV scale physics without fine tuning. 
The Higgs sector of this model is composed of the two Higgs doublet fields 
with the singlet neutral and charged scalar fields where singlet fields are odd under an unbroken $Z_2$ symmetry. 
We discuss not only the constraints on the parameter space 
from the current experimental data 
but also theoretical bounds from the triviality and the vacuum stability. 
We find that the model can be consistent up to the scale above 10 TeV  
in the parameter region where the neutrino data, 
the lepton flavor violation data,  
the thermal relic abundance of dark matter 
as well as the requirement from the strongly first order phase transition 
are satisfied. 

Second, we investigate a SUSY extension of the radiative seesaw model 
proposed by A.~Zee (1986) and K.~S.~Babu (1988) independently, in which 
neutrino masses are induced at the two-loop level by addition to charged singlet fields. 
One of the problem in the original Zee-Babu model is absence of the dark matter candidate. 
By the SUSY extension of the model, 
the lightest superpartner particle can be a dark matter candidate. 
We show that the neutrino data can be reproduced with satisfying current data from lepton 
flavour violation even in the scenario where not all the superpartner particles are heavy.
In this model, in addition to the doubly-charged isospin singlet scalar bosons, 
their SUSY partner fermions appear.  
We also discuss the outline of phenomenology for these particles at
the LHC.

Finally, we consider models which contain the isospin doublet scalar fields with $Y = 3/2$. 
Such a doublet field $\Phi_{3/2}$ is composed of a doubly charged scalar boson as well 
as a singly charged one. 
We discuss a simple model with $\Phi_{3/2}$, and we study its collider 
phenomenology at the LHC. 
We then consider a new model for radiatively generating neutrino masses 
with a dark matter candidate, in which $\Phi_{3/2}$ and an extra $Y = 1/2$ doublet as well as vector-like 
singlet fermions carry the odd quantum number for an unbroken discrete $Z_2$ symmetry. 
\tableofcontents
\mainmatter
\chapter{Introduction}
\section{Overview}
Although the Standard Model (SM) for particle physics has been successful for over three decades, 
the Higgs sector, which is introduced for the spontaneous breakdown of the electroweak gauge symmetry, remains unknown. 
In the SM, weak gauge bosons obtain their masses through the Higgs mechanism; i.e., 
the Nambu-Goldstone (NG) bosons are absorbed into the longitudinal components of $W$ and $Z$ bosons.  
At the same time, the masses for quarks and leptons are generated via the Yukawa interaction.  
Therefore, the Higgs boson is the origin of the mass for the elementary particles. 

Exploration of the Higgs boson is the most important issue in current high energy physics. 
The upper bound for the Higgs boson mass can be obtained by taking into account the  
tree level unitarity for elastic scattering processes of the 
longitudinal component of the gauge bosons, e.g.,  
$W_L^+W_L^-\to W_L^+W_L^-$~\cite{LQT}.
If we assume the validity of the perturbative calculation, then
the Higgs boson mass should be lower than around 1 TeV in the SM. 
If there is no Higgs boson or if the Higgs boson is heavier than 1 TeV, 
the $WW$ scattering amplitude should be grown as a function of the 
squired center of mass energy. 
As a result, the unitarity is broken at around the 1 TeV. 
Therefore, the Higgs boson does not exist, 
there must appear some new phenomena beyond the SM in the $WW$ scattering process. 

The CERN Large Hadron Collider (LHC) has built in order to survey the essence of 
electroweak symmetry breaking. 
At the LHC, the most important production process for the Higgs boson is the gluon fusion process
($gg\to H$)~\cite{gluon_fusion}. 
There are the other important production processes; 
the vector-boson fusion process ($qq'\to qq' H$)~\cite{VBF}, 
the vector-boson associated production process ($q\bar{q}\to WH/ZH$)~\cite{WZ_associate} and 
the top-quark pair associated production process ($q\bar{q}/gg\to t\bar{t} H$)~\cite{Htt_production}. 
From the recent Higgs boson searches at the LHC, 
the mass of the Higgs boson in the SM has already constrained to be between 115 GeV and 127 GeV or 
to be higher than 600 GeV at the 95\% confidence level~\cite{LHC-Higgs}. 
By the combination with the electroweak precise measurement at the LEP, 
we may expect that a light Higgs boson exists as long as the Higgs boson interactions are of 
SM-like and that it will be discovered in near future.

When the Higgs boson is discovered at the LHC, is physics of the elementary particle completed?  
The answer is ``No". 
There are several problems which cannot be explained within the SM. 
\begin{description}
\item[1.)]
As a purely theoretical issue, the SM has the gauge hierarchy problem~\cite{hierarchyproblem}. 
The quantum corrections to the Higgs boson mass depend on the quadratic power of the cutoff scale $\Lambda$, 
so that if $\Lambda$ is assumed to be at an ultra-high energy such as the grand unification scale, 
it is required a huge fine tuning in renormalization of the Higgs boson mass. 
\item[2.)] 
There are phenomena which cannot be explained within the SM such as the
neutrino oscillation~\cite{neutrino-oscillation}, 
existence of the dark matter~\cite{DM} and baryon asymmetry of the Universe~\cite{BAU}. 
\end{description}
Many new physics models have been proposed which are motivated to solve these problems. 
Regarding the hierarchy problem, 
it comes from the Higgs sector in the SM, where 
the elementary Higgs scalar boson is just introduced to break the electroweak symmetry. 
The mass terms for the gauge bosons and the fermions in the SM are forbidden by the gauge symmetry and 
the chiral symmetry, respectively, 
so that the radiative corrections to these particles are proportional to 
the logarithmic divergence at most. 
On the other hand, since the mass terms for the scalar bosons 
cannot be forbidden by the symmetry in the SM, 
the quadratic divergence appears in the radiative corrections to the Higgs boson mass. 

Therefore, new physics models which are motivated by solving the hierarchy problem 
have been proposed by considering 
what is the essence of the electroweak symmetry breaking. 
For example, in supersymmetric (SUSY) models, although the existence of the elementary Higgs scalar boson is admitted, 
the quadratic divergence due to a particle in the loop is cancelled by that due to a superpartner particles. 
Dynamical symmetry breaking~\cite{DSB} is the other idea to solve the hierarchy problem. 
In such models, the elementary Higgs scalar boson is not introduced, but 
the composite states of fermions play a role of the Higgs scalar boson, so that 
the quadratic divergence does not appear because of the chiral symmetry. 
The ideas of the little Higgs mechanism~\cite{little_higgs_mechanism} 
and the gauge Higgs unification~\cite{GHU1, GHU2} have also been proposed to solve the hierarchy problem. 
These new physics models often predict extended Higgs sectors as the low energy effective theory. 

Apart from the hierarchy problem, 
there are scenarios which can explain the phenomena beyond the SM listed as 2.) 
by the extension of the Higgs sector. 
The type II seesaw model~\cite{typeII_seesaw}, 
where a Higgs triplet field with $Y=1$ is added to the SM, can generate tiny neutrino masses at the tree level. 
Radiative seesaw models~\cite{zee,zee-2loop,babu-2loop,Krauss:2002px,Ma:2006km,aks_prl} 
can also explain tiny neutrino masses at the loop level,  
where additional scalar bosons, e.g., charged scalar bosons are running in the loop.  
Imposing an unbroken discrete symmetry such as $Z_2$ symmetry to a part of the Higgs sector, 
dark matter candidates can be obtained. 
The inert doublet model~\cite{dark_higgs} is one of the examples. 
Furthermore, baryon asymmetry 
can be generated at the electroweak phase transition 
by the nondecoupling property~\cite{nondec} and additional 
CP violating phases in the extended Higgs sector~\cite{ewbg-thdm,ewbg-thdm2}.

In both model explaining the problem 1) and that explaining the problems 2), 
the Higgs sector is often extended. 
The structure of a Higgs sector strongly depends on the property of the corresponding new physics model at the high energy scale, 
so that experimental reconstruction of the Higgs sector is extremely important to determine 
new paradigm for physics beyond the SM. 
In this thesis, we discuss the phenomenology of extended Higgs sectors from this view point. 

The Higgs sector in the SM takes a minimal form, which is composed of only one isospin doublet Higgs field. 
However, there is no strong reason for taking into account the minimal Higgs sector.  
Since the Higgs sector has not been confirmed yet, there are various possibilities for extended Higgs sectors.  
For example, 
we can consider extended Higgs sectors with extra $SU(2)$ singlet, doublet and triplet fields 
adding to the minimal Higgs sector. 
The important point is how we can constrain these possibilities. 
There are two important experimental observables to constrain the structure of the Higgs sector:
the electroweak rho parameter and the flavor changing neutral current (FCNC). 

The experimental value of the rho parameter $\rho_{\text{exp}}$ is quite close to unity; 
$\rho_{\exp}=1.0008^{+0.0017}_{-0.0007}$ \cite{PDG}.  
This fact suggests that a global 
$SU(2)$ symmetry (custodial symmetry) plays an important 
role in the Higgs sector. 
The theoretical prediction of the rho parameter is determined by 
the number of scalar fields as well as 
their representations under the isospin $SU(2)_L$ and 
the hypercharge $U(1)_Y$.  
In the Higgs model which contains complex scalar fields with the isospin $T_i$ and
the hypercharge $Y_i$ as well as real ($Y=0$) scalar fields 
with the isospin $T_i'$, the rho parameter is given at the tree level by 
\begin{align}
\rho_{\textrm{tree}}&=\frac{\sum_i\left[|v_i|^2(T_i(T_i+1)-Y_i^2)+u_i^2T_i'(T_i'+1)\right]}{2\sum_i|v_i|^2Y_i^2},\label{rho_tree}
\end{align}
where $v_i$ ($u_i$) represents the vacuum expectation value (VEV) of the complex (real) scalar field \cite{HHG}. 
In models with only scalar doublet fields (and singlets), $\rho_{\text{tree}}=1$ is predicted because of the custodial symmetry 
in the kinetic term of the Higgs sector. 
On the other hand, addition of 
Higgs fields with the isospin larger than one half can 
shift the rho parameter from unity at the tree level, whose 
deviation is proportional to VEVs of these exotic scalar 
fields such as triplet fields. 
The rho parameter, therefore, has been used to constrain a class of Higgs models. 

In the SM, FCNC 
phenomena are suppressed due to the electromagnetic gauge symmetry and 
the Glashow-Iliopoulos-Maiani mechanism \cite{GIM}, 
so that the experimental bounds 
on the FCNC processes, which are mediated by the neutral gauge bosons,  
are satisfied. 
At the same time, neutral Higgs boson mediating FCNC processes are absent at the tree level in the SM. 
Since the matrix for the Yukawa interaction and that for the fermion masses are given by 
the same Yukawa matrix,  
those two matrices can be diagonalized simultaneously. 
In models with more 
than one Higgs doublet, this is not true in general, because 
two or more Yukawa matrices for each fermion cannot be 
simultaneously diagonalized. It is well known that, to 
avoid such Higgs-boson-associated FCNC interactions, 
each fermion should couple to only one of the Higgs 
doublets. This can be realized in a natural way by imposing 
a discrete $Z_2$ symmetry \cite{GW}. 

In addition to these two basic experimental constraints, 
by focusing on the physics of charged Higgs bosons $H^\pm$, which  
are contained in most of the extended Higgs models, 
we can obtain the other constraint to the Higgs sector. 
In particular, the $H^\pm W^\mp Z$ vertex can be a useful probe 
of the extended Higgs sector~\cite{Grifols-Mendez,HWZ,HWZ-Kanemura,logan}.  
The magnitude of the vertex depends on the violation of the custodial symmetry such as the 
rho parameter. 
In general, this can be independent of the rho parameter. 
If a charged Higgs boson is from a doublet field, 
the $H^\pm W^\mp Z$ vertex vanishes at the tree level. 
The vertex is then one-loop induced 
and its magnitude is proportional to the violation of the global symmetry 
in the sector of particles in the loop. 
On the other hand, 
in models with exotic representations such as triplet scalar fields 
this vertex appears at the tree level. 
Therefore, the determination of the $H^\pm W^\mp Z$ vertex 
can be a complementary tool to the rho parameter in testing the 
{\it exoticness} of the Higgs sector. 

We should discuss extended Higgs sectors based on physics motivations which can satisfy these experimental data. 
In Part~I, as concrete examples, 
we discuss the phenomenology of extended Higgs sectors such as 
the two Higgs doublet model (THDM), the Higgs triplet model (HTM) and Higgs sectors in the supersymmetry. 
In Part~II, we discuss applications of extended Higgs sectors to models which can explain 
tiny neutrino masses, dark matter and/or baryon asymmetry of the Universe at the TeV scale.  
\section{The Two Higgs Doublet Model}
The THDM is an important example as an extended Higgs model, 
since that appears in the Higgs sector of the minimal supersymmetric standard model (MSSM)~\cite{MSSM}. 
Radiative seesaw models, e.g., the Zee model~\cite{zee} contain two Higgs doublets. 
In addition, additional CP-violating phases appear in the THDM, 
which is required by the scenario of electroweak baryogenesis.  
Recently, in Ref.~\cite{Sher_rept2}, 
theoretical and phenomenological aspects in the THDM have been reviewed, 
where extensive references to the original literature are included.
In the THDM, neutral scalar boson mediating FCNC processes appear at the tree level. 
To avoid such a dangerous FCNC process, 
a softly-broken discrete $Z_2$ symmetry is often introduced~\cite{GW}. 
Under the $Z_2$ symmetry, there are four types of the Yukawa interaction 
depending on its charge assignment for quarks and leptons~\cite{Barger,Grossman}. 
We call these types of Yukawa interaction as type-I, type-II, type-X, and type-Y~\cite{typeX}. 
We discuss the phenomenological difference among these types of Yukawa interaction. 

The type in the Yukawa interaction can 
be related to each new physics scenarios. For example, the 
Higgs sector of the MSSM is the THDM with a SUSY relation 
among the parameters of the Higgs sector, whose 
Yukawa interaction is of type-II, in which only a Higgs 
doublet couples to up-type quarks and the other couples to 
down-type quarks and charged leptons. 
On the other hand, a model has been proposed in Ref.~\cite{aks_prl}, where 
neutrino masses, dark matter and baryogenesis would be explained simultaneously at the TeV scale. 
In this model the Higgs sector contains the two Higgs doublets, and its Yukawa interaction corresponds 
to the type-X THDM, in which only a Higgs 
doublet couples to quarks and the other couples to leptons~\cite{typeX,typeX2}. 
Therefore, in order to select the true model from various 
new physics candidates that predict THDMs, it is important to experimentally 
determine the type of Yukawa interaction. 

We first study the total widths and 
the decay branching ratios of the extra Higgs bosons such as the 
CP-odd (CP-even) Higgs boson $A$ ($H$) and the charged Higgs bosons $H^\pm$ in the four types of Yukawa interaction.  
We then summarize constraints on the 
mass of the charged Higgs bosons from current experimental bounds, especially from the 
$B$-meson decay data such as $B\to X_s\gamma$~\cite{bsgNLO} and $B\to \tau\nu$~\cite{Btaunu_AR,Btaunu_MS,Btaunu_Du}
depending on the type of Yukawa interaction.
The $B\to X_s\gamma$ results give a severe  
lower bound, $m_{H^+}>295$ GeV, at the next-to-next-to leading 
order in the (non-SUSY) type-II THDM and the type-Y THDM~\cite{bsgNNLO1, bsgNNLO2}, while 
in the type-I and the type-X THDM,  
the mass of $\mathcal{O}$(100) GeV is allowed in the wide regions of the parameter space. 
We finally discuss the possibility of discriminating 
the types of Yukawa interaction at the LHC and also at the international Linear Collider (ILC). 
We discuss the signal of extra neutral and 
charged Higgs bosons at the LHC, which may be useful to 
distinguish the type of Yukawa interaction. 
Recently, a detailed simulation study for the type-X THDM has been performed for multi-$\tau$ signatures at the LHC~\cite{Yokoya}. 
In that paper, assuming the integrated luminosity of 100 fb$^{-1}$, the excess can be seen in various three-
and four-lepton channels via the processes 
$q\bar{q}\to Z^*\to HA$ and $q\bar{q}'\to W^{\pm*}\to HH^\pm$ ($AH^\pm$).  

\section{The Higgs Triplet Model}
The HTM is also a well-motivated extended Higgs model, where the $Y=1$ Higgs triplet field is added to the SM,  
since tiny masses of neutrinos can be generated at the tree level. 
Assuming that the triplet scalar field carries two units of lepton number, 
the lepton number conservation is violated in a trilinear interaction terms 
among the Higgs doublet field and the Higgs triplet field. 
Majorana masses for neutrinos are then generated 
through the Yukawa interaction of the lepton doublet and the triplet scalar field. 
When we take the lepton number violating coupling to be eV scale, 
the masses of the component fields of the triplet can be taken to be of the TeV scale 
or less. 
In such a case, the model can be tested by directly detecting the triplet-like scalar bosons, such as 
the doubly-charged ($H^{\pm\pm}$), the singly-charged ($H^\pm$), the neutral CP-even ($H$) 
and the neutral CP-odd ($A$) scalar bosons. 
In Ref.~\cite{small_mu}, a simple suppression mechanism for $\mu$ has been proposed, where 
lepton number violating coupling is induced at the one-loop level. 

In addition to the appearance of these scalar bosons, 
a striking prediction of the HTM is the relationships among the masses of the 
triplet-like scalar bosons, i.e.,  
$m_{H^{++}}^2-m_{H^{+}}^2\simeq m_{H^{+}}^2-m_A^2$ and  $m_A^2\simeq m_H^2$, 
where $m_{H^{++}}$, $m_{H^{+}}$, $m_A$ and $m_H$ are the masses of $H^{\pm\pm}$, $H^\pm$, $A$ and $H$, respectively. 
The squired mass difference is determined by the VEV of the doublet scalar field ($\simeq 246$ GeV), 
and a scalar self-coupling constant. 
As such a mass difference is not forbidden by the symmetry of the model,  
we may be able to distinguish the model from the others which contain charged 
Higgs bosons by testing these mass relations. 

It is important that how the mass differences among the triplet-like scalar bosons 
are constrained by experiments. 
We study radiative corrections to the electroweak observables in the HTM to constrain the mass difference~\cite{ky_rho}. 
In the model with $\rho \neq 1$ at the tree level like the HTM, 
apart from the models with $\rho=1$ at the tree level, 
a new input parameter has to be introduced 
in addition to the usual three input parameters such as: $\alpha_{\text{em}}$, $G_F$ and $m_Z$.   
In Ref.~\cite{blank_hollik}, the on-shell renormalization scheme is constructed in the Higgs model with the $Y=0$ triplet field, in which 
four input electroweak parameters: $\alpha_{\text{em}}$, $G_F$, $m_Z$ and $\sin^2\theta_W$ are chosen to 
describe all the other electroweak observables. 
The radiative corrections to the electroweak observables have been calculated 
in the $Y=0$ triplet model~\cite{blank_hollik,real-triplet,Chen:2005jx} and in the left-right symmetric model~\cite{Chen:2005jx}.  

In our analysis, we first define the on-shell renormalization scheme for the electroweak sector of 
the HTM by using the method in Ref.~\cite{blank_hollik}. We then 
calculate radiative corrections to the electroweak observables such as $m_W$ and $\rho$ as 
a function of the four input parameters: $\alpha_{\text{em}}$, $G_F$, $m_Z$ and $\sin^2\theta_W$. 
We examine the preferable values of the mass spectrum 
of the triplet-like scalar bosons and the VEV of the triplet field under the constraint from the electroweak precision data. 
We find that the hierarchical mass spectrum with large mass splitting is favored especially for the case of 
$m_A$ $(\simeq m_H)>m_{H^+}>m_{H^{++}}$. 
On the contrary, the inverted hierarchical case with $m_{H^{++}}>m_{H^+}>m_A$ $(\simeq m_H)$ is 
relatively disfavored.  

We then discuss the phenomenology of the HTM at the LHC, especially focusing on the case with the mass difference. 
Phenomenology with the mass difference among the triplet-like scalar bosons~\cite{4-lepton,Perez:2008ha,Akeroyd-Sugiyama,Melfo:2011nx,Chakrabarti:1998qy,aky_triplet} 
is drastically different from that in the case without the mass difference~
\cite{HTM_delm0,HTM_delm0_2,HTM_delm0_3,Kadastik:2007yd,4-lepton,Perez:2008ha,Akeroyd-Sugiyama,Melfo:2011nx,Chakrabarti:1998qy,aky_triplet,Chiang:2012dk}. 

In the case without the mass difference, 
$H^{++}$ decays into the same sign dilepton $\ell^+ \ell^+$ 
or the diboson $W^+W^+$, depending on the size of $v_\Delta$, $m_{H^{++}}$ and also the detail of neutrino masses, 
where $v_\Delta$ is the VEV of the triplet field. 
On the other hand, in the case with the mass difference, 
there are two cases, where $H^{++}$ is the heaviest or the lightest of all the triplet-like scalar bosons. 
In the case where $H^{++}$ is the lightest, while $H^+$ can decay into $H^{++}W^{-(*)}$~\cite{Akeroyd-Sugiyama} 
the decay pattern of $H^{++}$ is the same as in the case without the mass difference.   
On the contrary, in the case  where $H^{++}$ is the heaviest, 
the cascade decay of $H^{++}$ dominates; i.e., $H^{++}\to H^+W^{+(*)}\to H W^{+(*)}W^{+(*)}$ ($A W^{+(*)}W^{+(*)}$) 
as long as $v_\Delta$ is neither too small nor too large
\footnote{Recently the importance of this cascade decay has been mentioned in Refs.~\cite{Akeroyd-Sugiyama,Melfo:2011nx}.}. 

In this thesis, we focus on the phenomenology of the HTM in the case with $m_{H^{++}}>m_{H^{+}}>m_H$ ($m_A$)~\cite{aky_triplet}
\footnote{As mentioned above, although this scenario is disfavored by the electroweak precision data, 
we can still consider the scenario by the extension of the minimal HTM.}. 
In this case, the limit of the mass of $H^{++}$ from the recent results at the LHC 
cannot be applied, so that the triplet-like scalar bosons with the mass of $\mathcal{O}(100)$ GeV are still allowed.  
We outline that all the masses of the triplet-like scalar bosons may be able to be reconstructed 
by measuring the Jacobian peak~\cite{Jacobian} in the transverse mass distribution as well as 
the invariant mass distributions of the systems which 
are generated via the decays of the triplet-like scalar bosons.

\section{Supersymmetric Higgs sectors}
We also study extended Higgs sectors in the supersymmetry. 
In general, SUSY Higgs sectors are written by a simple formula; i.e., those are constructed from 
F-term, D-term and soft-breaking terms. 
In the MSSM, the interaction terms in the Higgs potential are given only by D-term contributions. 
This fact predicts the mass $m_h$ of the lightest CP-even Higgs boson $h$ less than that of the Z boson at the tree level. 
At the one-loop level the trilinear top-Yukawa term in the superpotential gives a significant 
F-term contribution to $m_h$~\cite{mh-MSSM,mh-MSSM1,dabelstein}, 
by which $m_h$ can be above the lower bound from the direct search results at the LEP experiment. 
Predictions to $m_h$ can be different drastically in the extended SUSY standard models, 
where additional chiral superfields are added to the MSSM. 
These SUSY models can be classified into two models; i.e., 
those with interaction terms from the tree level F-term contributions 
and those without such F-term contributions. 
In the former models, 
we discuss 
effects of F-term contributions to $m_h$ and the triple $h$ coupling $hhh$ 
at the one-loop level when $h$ looks the SM Higgs boson~\cite{ksy}. 
We find that in the model with an extra neutral singlet superfield or extra triplet superfields, 
possible allowed regions of $m_h$ can be much larger than that in the MSSM 
if we allow the appearance of a strong coupling constant at the TeV scale. 
The deviation of $hhh$ from the SM prediction can be significant in some models; 
e.g., a model with four doublets and charged singlet superfields.  
In extended SUSY standard models without interaction terms from F-term contributions 
in the Higgs potential at the tree level, 
there are no such significant effects to $m_h$ and $hhh$ coupling. 
However, even without such F-term contributions, 
large deviations can be seen in the MSSM observables 
due to the mixing among the MSSM-like Higgs bosons and the extra Higgs bosons. 
As a simplest example, we discuss 
the model, where two extra doublet superfields are added to the MSSM~\cite{aksy_4hdm}. 
\section{Extended Higgs sectors and the phenomena beyond the SM}

In Part II, we discuss TeV scale models including extended Higgs sectors 
which can explain the phenomena beyond the SM; i.e., 
neutrino masses, dark matter and/or baryon asymmetry of the Universe.  

First, we study a model proposed in Ref.~\cite{aks_prl}, in which 
neutrino oscillation, dark matter, and
the baryon asymmetry of the Universe can be simultaneously explained by the TeV-scale physics without fine tuning. 
We investigate theoretical constraints from the vacuum stability and the triviality 
in the model~\cite{aky_rge}. 
We calculate the scale dependence of the coupling constants from 
the renormalization group equations at the one-loop level in the model.  
As the result, we find that the model can be consistent up to above 10 TeV, 
where the model would be expected to be replaced by a more fundamental model. 
We also confirm that the model can explain the constraints from the current experimental data, 
for example $\mu\to e\gamma$, $\mu\to eee$, the WMAP data and the neutrino oscillation data.

Next, we consider the SUSY extension of the 
Zee-Babu model~\cite{zee-2loop,babu-2loop} which can generate neutrino masses at the two-loop level 
by adding charged singlet fields~\cite{aksy_szb}. 
In the original (non-SUSY) Zee-Babu model, 
there remain the hierarchy problem as well as the dark matter problem. 
By introducing supersymmetry, the quadratic divergence in the one-loop correction
to the Higgs boson mass can be cancelled. 
In addition, due to the R-parity, the stability of the
lightest super partner particle such as the neutralino are guaranteed, which may be identified as a
candidate of dark matter.
In this model, doubly-charged iso-singlet scalar bosons and those SUSY partner fermions appear. 
We discuss the collider signature of these particles, and we
show that the distinctive signature may be measured in the invariant mass distribution 
of the system which is produced via the decay of the doubly-charged scalar bosons and fermions. 

Finally, we consider the isospin doublet field $\Phi_{3/2}$ with $Y=3/2$~\cite{aky_32}, in which 
doubly-charged scalar states as well as singly-charged ones are contained. 
Although it has been well studied the phenomenology of doubly-charged scalar bosons from the 
isospin singlet ($Y=2$) and the triplet ($Y=1$), 
the phenomenology of doubly-charged scalar bosons from $\Phi_{3/2}$ has hardly been
studied. 
First, we discuss the phenomenology of the simple model with $\Phi_{3/2}$ at the LHC. 
We find that the transverse mass distribution may be useful to measure the masses of scalar bosons 
including the doubly-charged scalar bosons in the model.  
We then show that $\Phi_{3/2}$ can be applied to the model, where neutrino masses 
can explain at the one-loop level and there are the dark matter candidates. 

\section{Organization of this thesis}
Part~I is composed of three sections: chapter~2, chapter~3 and chapter~4. 
In chapter~2, we give a brief review of the SM Higgs sector. 
Phenomenology of extended Higgs sectors are discussed in chapter~3, where 
we consider the THDM and the HTM. 
In chapter~4, we discuss the SUSY Higgs sectors and their decoupling properties.
Succeeding five sections from chapter~5 to chapter~8 are included in Part~II. 
In chapter~5, we outline the radiative seesaw models. 
In chapter~6, after we introduce the model proposed in Ref.~\cite{aks_prl}, 
we then discuss the 
theoretical constraints from the vacuum stability and the triviality, and also study 
the constraints from the current experimental data. 
In chapter~7, we discuss the SUSY extension of the Zee-Babu model and its phenomenology at the LHC. 
In chapter~8, we discuss the model with isospin doublet with $Y=3/2$ field and its phenomenology at the LHC. 
We also discuss that such a $Y=3/2$ doublet field can apply to the radiative seesaw model. 
The conclusion of this thesis is given in chapter~9. 

\part{Phenomenology of Higgs sectors}

\chapter{The Standard Model Higgs sector}
In this chapter, we review the Higgs sector in the SM. 
In the SM, the Higgs sector takes the minimal form, where there is only one $SU(2)_L$ doublet Higgs scalar field $\Phi$. 
After the spontaneous breakdown of the $SU(2)_L\times U(1)_Y$ gauge symmetry, 
weak gauge bosons, fermions and the Higgs boson obtain their masses in the kinetic term of the Higgs doublet field, 
the Yukawa Lagrangian and 
the Higgs potential, respectively. 
In Table~\ref{pc_SM}, the charge assignments for the SM particles 
under the $SU(3)_c\times SU(2)_L\times U(1)_Y$ gauge symmetry 
are listed. 
Throughout this thesis, 
the relationship among the hypercharge $Y$, the isospin $I_3$ and the electromagnetic charge $Q$ is defined by $Q=I_3+Y$. 
In Table~\ref{pc_SM}, $Q_L^i$ ($L_L^i$) is the $i$th generation left-handed quark (lepton) doublet, 
and $e_R^i$ and $u_R^i$ ($d_R^i$) are the right-handed charged lepton and the up-quark (down-quark) singlet, respectively. 
We first discuss how all the masses of the SM particles are generated. 
Second, the bounds of the Higgs boson mass are discussed. 
Third, we calculate the decay rates of the Higgs boson.  
\begin{table}[t]
\begin{center}
{\renewcommand\arraystretch{1}
\begin{tabular}{|c||ccc|}\hline
&$SU(3)_c$&$SU(2)_L$&$U(1)_Y$ \\\hline\hline
$Q_L^i$&\textbf{3}&\textbf{2}&$+\frac{1}{6}$\\\hline
$u_R^i$&$\textbf{3}$&\textbf{1}&$+\frac{2}{3}$ \\\hline
$d_R^i$&$\textbf{3}$&\textbf{1}&$-\frac{1}{3}$ \\\hline
$L_L^i$&\textbf{1}&\textbf{2}&$-\frac{1}{2}$ \\\hline
$e_R^i$&\textbf{1}&\textbf{1}&-1 \\\hline
$\Phi$&\textbf{1}&\textbf{2}&$+\frac{1}{2}$ \\\hline
\end{tabular}}
\caption{Particle content and its charge assignment under the $SU(3)_c\times SU(2)_L\times U(1)_Y$ gauge symmetry in the SM.}
\label{pc_SM}
\end{center}
\end{table}
\section{The masses of particles}
\subsection{The masses for the Higgs boson and the gauge bosons}
The Higgs Lagrangian is composed of the kinetic term and the Higgs potential as
\begin{align}
\mathcal{L}_{\text{Higgs}}&=|D^\mu\Phi|^2-V_{\text{SM}},
\quad V_{\text{SM}}=-\mu^2(\Phi^\dagger\Phi)+\lambda(\Phi^\dagger\Phi)^2,\label{pot_sm}
\end{align}
where the Higgs doublet field $\Phi$ can be parameterized as
\begin{align}
\Phi=\left[
\begin{array}{c}
w^+\\
\frac{1}{\sqrt{2}}(h+v+iz)
\end{array}\right],
\end{align}
with $w^\pm$ and $z$ are the NG bosons which are absorbed by the longitudinal components of $W^\pm$ boson and $Z$ boson, 
respectively, and $v\simeq \text{246 GeV}$ is the VEV of $\Phi$. 
The covariant derivative for $\Phi$ is given as 
\begin{align}
D_\mu \Phi =\left(\partial_\mu -i\frac{g}{2}\tau^a W_\mu^a -i\frac{g'}{2}B_\mu\right)\Phi, 
\end{align}
where $W_\mu^a$ $(a=1,2,3)$ is the $SU(2)_L$ gauge boson, and $B_\mu$ is the $U(1)_Y$ gauge boson. 
By imposing the vacuum condition: 
\begin{align}
\left.\frac{\partial V_{\text{SM}}}{\partial h}\right|_0=0, 
\end{align}
we obtain 
\begin{align}
\mu^2=v^2{\lambda}. \label{vc_sm}
\end{align}
Plugging Eq.~(\ref{vc_sm}) into the Higgs potential in Eq.~(\ref{pot_sm}), 
the mass $m_h$ of the physical neutral Higgs boson $h$ is obtained as
\begin{align}
m_h^2=2\lambda v^2.
\end{align}
The mass terms of the weak gauge bosons are derived in the kinetic term $|D_\mu\Phi|^2$ as 
\begin{align}
|D_\mu\Phi|^2&=\frac{g^2v^2}{8}\left[(W_\mu^1)^2+(W_\mu^2)^2\right]+\frac{v^2}{8}(W_\mu^3,B_\mu)
\left(
\begin{array}{cc}
g^2 & -gg'\\
-gg' & g^{'2}
\end{array}\right)\left(
\begin{array}{c}
W^{3\mu}\\
B^\mu
\end{array}\right)+\cdots.
\end{align}
The charged gauge bosons $W^\pm$ and the neutral gauge bosons are obtained by 
\begin{align}
W_\mu^\pm =\frac{1}{\sqrt{2}}(W_\mu^1\mp i W_\mu^2),\quad 
\left(
\begin{array}{c}
Z_\mu \\
A_\mu
\end{array}\right)=
\left(
\begin{array}{cc}
\cos\theta_W & -\sin\theta_W \\
\sin\theta_W & \cos\theta_W
\end{array}\right)
\left(
\begin{array}{c}
W_\mu^3 \\
B_\mu
\end{array}\right), 
\end{align}
where $\theta_W$ is the weak mixing angle with
\begin{align}
\cos\theta_W = \frac{g}{\sqrt{g^2+g^{'2}}},\quad\sin\theta_W = \frac{g'}{\sqrt{g^2+g^{'2}}}. 
\end{align}
The masses of $W^\pm$ and $Z$ are then 
\begin{align}
m_W^2 = \frac{g^2v^2}{4},\quad m_Z^2=\frac{g^2+g^{'2}}{4}v^2,
\end{align}
and the mass of the photon $A_\mu$ remains zero. 
\subsection{The fermion masses}
The Yukawa Lagrangian is given by
\begin{align}
\mathcal{L}_{\text{SM}}^Y&=
-\left[\bar{Q}_L^iY_d^{ij}\Phi d_R^j+\bar{Q}_L^iY_u^{ij}\tilde{\Phi}u_R^j+\bar{L}_L^iY_e^{ij}\Phi e_R^j+\text{h.c.}\right],
\end{align}
where, $\tilde{\Phi}=i\tau_2 \Phi^*$ and $Y_{u,d,e}$ are the 3$\times$3 complex matrices. 
By taking the following base transformations: 
\begin{align}
&u_R^i\to V_u^{ij}u_R^{j},\quad d_R^i\to V_d^{ij}d_R^{j},\quad e_R^i\to V_e^{ij}e_R^{j},\notag\\
&Q_L^i=\left(
\begin{array}{c}
u_L^i\\
d_L^i
\end{array}\right)\to \left(\begin{array}{c}
U_u^{ij}u_L^{j}\\
U_d^{ij}d_L^{j}
\end{array}\right)=
U_u^{ij}\left(\begin{array}{c}
u_L^{j}\\
V_{\text{CKM}}d_L^{j}
\end{array}\right),\notag\\
&L_L^i=\left(
\begin{array}{c}
\nu_L^i\\
e_L^i
\end{array}\right)\to U_\ell^{ij}\left(\begin{array}{c}
\nu_L^{j}\\
e_L^{j}
\end{array}\right), 
\end{align}
where $V_{\text{CKM}}=U_u^\dagger U_d$ is the Cabibbo-Kobayashi-Maskawa matrix~\cite{CKM}, 
the matrices $Y_{u,d,e}$ can be diagonalized as 
\begin{align}
\mathcal{L}_{\text{SM}}^Y&\to -\left[\bar{Q}_L (U_d^\dagger Y_dV_d) \Phi d_R+\bar{Q}_L(U_u^\dagger Y_uV_u)\tilde{\Phi}u_R
+\bar{L}_L (U_\ell^\dagger Y_eV_e)\Phi e_R+\text{h.c.}\right]\notag\\
&=-\left[\bar{Q}_L Y_d^{\text{diag}} \Phi d_R+\bar{Q}_LY_u^{\text{diag}} \tilde{\Phi}u_R
+\bar{L}_L Y_e^{\text{diag}} \Phi e_R+\text{h.c.}\right].  \label{yukawa_sm}
\end{align}
The diagonalized matrices $Y_{u,d,e}^{\text{diag}}$ can be expressed as 
\begin{align}
Y_d^{\text{diag}} =
\left(
\begin{array}{ccc}
y_d & 0 & 0\\
0 & y_s & 0\\
0 & 0 & y_b
\end{array}\right) ,\quad 
Y_u^{\text{diag}} =
\left(
\begin{array}{ccc}
y_u & 0 & 0\\
0 & y_c & 0\\
0 & 0 & y_t
\end{array}\right),\quad
Y_e^{\text{diag}} =
\left(
\begin{array}{ccc}
y_e & 0 & 0\\
0 & y_\mu & 0\\
0 & 0 & y_\tau
\end{array}\right). 
\end{align}
Replacing $\Phi$ in Eq.~(\ref{yukawa_sm}) by the VEV of the Higgs doublet field $\langle\Phi\rangle=(0,v/\sqrt{2})^T$, the 
masses of the fermions are 
\begin{align}
\mathcal{L}_{\text{SM}}^Y &\to -\left[\bar{Q}_LY_d^{\text{diag}} \langle\Phi\rangle d_R+\bar{Q}_L Y_u^{\text{diag}}\langle\tilde{\Phi}\rangle u_R+\bar{L}_L Y_e^{\text{diag}} \langle\Phi\rangle e_R+\text{h.c.}\right]\notag\\
&=-\frac{v}{\sqrt{2}}\bar{d}_L Y_d^{\text{diag}} d_R-\frac{v}{\sqrt{2}}\bar{u}_LY_u^{\text{diag}} u_R
-\frac{v}{\sqrt{2}}\bar{e}_LY_e^{\text{diag}} e_R+\text{h.c.}.
\end{align}
Therefore, the fermion masses are written as 
\begin{align}
m_f=\frac{y_fv}{\sqrt{2}}.
\end{align}
We note that the coupling constants between the Higgs boson and the fermions $h\bar{f}f$ are given by $Y_f^{\text{diag}}$, 
so that there is no interaction which causes scalar mediating 
FCNC processes at the tree level in the SM. 
In addition, the tree level neutral gauge boson mediating FCNC processes are also forbidden by the GIM mechanism~\cite{GIM}. 
\section{Bounds for the Higgs boson mass}
In the SM, the only unknown parameter is the Higgs boson mass $m_h$ or the Higgs self-coupling constant $\lambda$. 
Here, we discuss the constraints for the Higgs boson mass in the theoretical point of view. 
\subsection{Perturbative unitarity bound}

The upper bound for the Higgs boson mass can be obtained by assuming the unitarity of the S-matrix. 
This approach has been first proposed by Lee, Quigg and Thacker~\cite{LQT}.
From the optical theorem, the total cross section $\sigma_{\text{tot}}$ can be written by the imaginary part of 
the scattering amplitude with the scattering angle $\theta=0$ as 
\begin{align}
\sigma_{\text{tot}}=\frac{1}{s}\textrm{Im}[\mathcal{M}(\theta=0)]\label{optical}, 
\end{align}
where $s$ is the squired center of mass energy. 
Since the main contribution of $\sigma_{\text{tot}}$ comes from 2 body $\to$ 2 body process, 
we can write 
\begin{align}
\sigma_{\text{tot}}\gtrsim \frac{1}{s}\int d\cos\theta \frac{|\mathcal{M}|^2}{32\pi}. \label{2b2b}
\end{align}
On the other hand, the amplitude $\mathcal{M}$ can be expanded in terms of the $J$th partial wave amplitude $a_J$ as 
\begin{align}
\mathcal{M}=16\pi\sum_{J=0}^{\infty}(2J+1)P_J(\cos\theta)a_J\label{partial}.
\end{align}
By the combination of Eqs.~(\ref{optical}), (\ref{2b2b}) and Eq.~(\ref{partial}), we obtain 
\begin{align}
&\text{Im}(a_J)\gtrsim |a_J|^2\notag\\
&\leftrightarrow \text{Re}(a_J)^2+\left[\text{Im}(a_J)-\frac{1}{2}\right]^2\gtrsim\left(\frac{1}{2}\right)^2. \label{circle}
\end{align}
This inequality suggests that $a_J$ has to be on the circle with the radius $1/2$ and the center of coordinate (0,1/2) 
in the complex plane. 
Therefore, we can require that Re($a_J$) is satisfied 
\begin{align}
|\textrm{Re}(a_J)|<1/2,
\label{ll}
\end{align}
at the tree level. 
We apply this constraint to the $W_L^+W_L^-\to W_L^+W_L^-$ process, where $W_L$ is the longitudinal component of the $W$ boson. 
In Appendix~A, detailed calculations for this process are given. 
In the high energy limit, 0th partial wave amplitude can be calculated as 
\begin{align}
a_0
&\sim -\frac{G_Fm_h^2}{4\sqrt{2}\pi}. 
\end{align}
From Eq.~(\ref{ll}), we can obtain the upper bound of the mass of the Higgs boson: 
\begin{align}
m_h^2<\frac{2\pi\sqrt{2}}{G_F}\sim (871\textrm{GeV})^2. \label{lqt1}
\end{align}
The stronger constraint can be obtained by including the other scattering processes; 
\begin{align}
\frac{1}{\sqrt{2}}Z_LZ_L,\quad\frac{1}{\sqrt{2}}hh,\quad hZ_L. 
\label{4}
\end{align}
In the high energy limit, the scattering process for the longitudinal component of 
the massive gauge bosons can be replaced by that for corresponding NG boson modes which 
is known as the equivalence theorem~\cite{equivalence_theorem}. 
The 0th partial wave amplitude can be written as the 4$\times$4 matrix in the basis of 
($w^+w^-$, $\frac{1}{\sqrt{2}}zz$, $\frac{1}{\sqrt{2}}hh$, $hz$): 
\begin{align}a_0=
\frac{-G_Fm_h^2}{4\pi\sqrt{2}}\left(\begin{array}{cccc}
1 & \frac{1}{\sqrt{8}}&\frac{1}{\sqrt{8}}&0\\
\frac{1}{\sqrt{8}}&\frac{3}{4}&\frac{1}{4}&0\\
\frac{1}{\sqrt{8}}&\frac{1}{4}&\frac{3}{4}&0\\
0&0&0&\frac{1}{2}
\end{array}\right). 
\end{align}
By diagonalizing this matrix, we obtain the eigenvalues of ($3/2,1/2,1/2,1/2$) in the unit of 
$\frac{-G_Fm_h^2}{4\pi\sqrt{2}}$. 
By imposing the condition Eq.~(\ref{ll}) to each eigenvalue, we obtain 
\begin{align}
m_h^2 < (710\text{ GeV})^2. 
\end{align}
\subsection{Triviality and vacuum stability bounds}

In the perturbative unitarity bound which is discussed just above, 
the upper bound for the Higgs boson mass can be obtained by only assuming the unitarity of the S-matrix. 
As the other approach to constrain the Higgs boson mass, 
there are the triviality and the vacuum stability bounds which are assumed the cutoff scale 
$\Lambda$ of the theory~\cite{RGE_SM,Inoue,Komatsu,Lindner,Grzadkowski}. 
First, recall the renormalization group equation (RGE) of the Higgs self-coupling $\lambda$ at the one-loop level: 
\begin{align}
\frac{d\lambda}{d\log Q}&\simeq\frac{1}{16\pi^2}\left[24\lambda^2+12y_t^2\lambda-6y_t^4\right], \label{rge_sm1}
\end{align}
where $Q$ is an arbitrary scale. Full set of the RGEs in the SM is listed in Appendix~B. 
From this equation, large initial values of $\lambda$ compared with that of the top-Yukawa coupling $y_t$ 
suggest that $\lambda$ is getting large value as $Q$ is increasing. 
Thus, the value of $\lambda$ becomes too large to rely on the perturbative calculation or 
close to the infinity which is known as the Landau pole~\cite{Landau_pole}. 
On the other hand, 
when initial values of $\lambda$ is small compared with that of $y_t$, 
due to the contribution from $-6y_t^4$ term in Eq.~(\ref{rge_sm1}), $\lambda$ is getting small value as $Q$ is increasing. 
This suggests that the value of $\lambda$ becomes negative in some scale $Q$, so that the vacuum 
stability is broken. 
Consequently, if we require that the model appears neither the Landau pole nor the vacuum instability 
up to the cutoff scale $\Lambda$, 
then we can obtain the constraint for the initial value of $\lambda$ as a function of $\Lambda$.  
Since the Higgs boson mass is determined by $2\lambda v^2$, the constraint for $\lambda$ can be translated into that for the Higgs boson mass. 
In Ref.~\cite{Hambye:1996wb}, the upper bound for the Higgs boson mass has been obtained by using 
the two-loop level RGE with one-loop matching conditions.  
For $\Lambda=10^{19}$ GeV, the upper bound is $m_h<180\pm 4\pm5$ GeV, 
the first error including the theoretical uncertainty and the second error 
reflecting the top-quark mass uncertainty due to $m_t=175\pm6$ GeV. 

\section{Decay of the Higgs boson}
\begin{figure}[t]
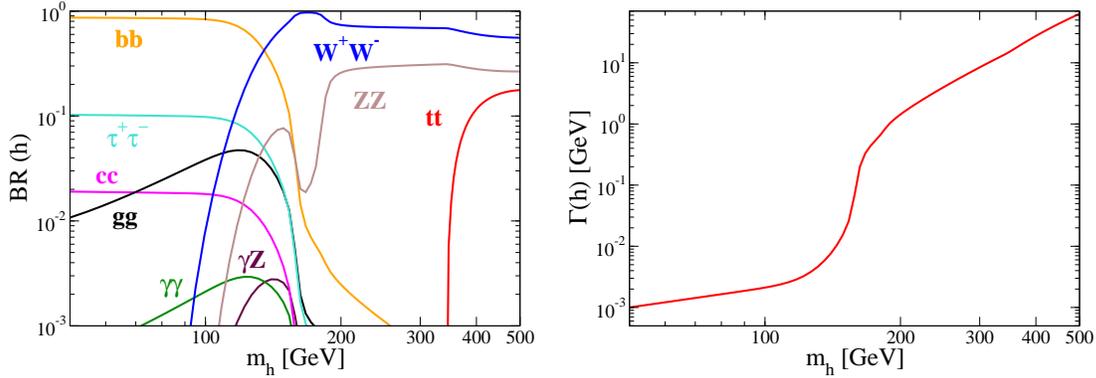

\begin{center}
\includegraphics[width=70mm]{BR_SMHiggs.eps}\hspace{3mm}
\includegraphics[width=70mm]{Gam_SMHiggs.eps}
\caption{(Left) The decay branching ratio of the SM Higgs boson. (Right) The decay width of the SM Higgs boson.}
\label{SMHiggs_decay}
\end{center}
\end{figure}

The decay rates of the Higgs boson only depends on the mass of the Higgs boson in the SM. 
There are various decay modes of the Higgs boson; 
(1) fermion pair decays ($h\to f\bar{f}$), 
(2) gauge boson pair decays ($h\to W^+W^-$ and $h\to ZZ$), 
(3) loop-induced decays ($h\to \gamma\gamma$, $h\to \gamma Z$ and $h\to gg$) and 
(4) three body decays ($h\to WW^* \to Wf\bar{f}'$ and $h\to ZZ^* \to Zf\bar{f}$). 
The formulae for these decay rates are listed in Appendix~C. 
The decay branching ratio and the decay width of the Higgs boson are shown in Fig.~\ref{SMHiggs_decay}. 

In this figure, the masses of the top quark, bottom quark and charm quark are taken to be 171.2 GeV, 3.0 GeV and 0.44 GeV. 
The decay branching ratio is drastically different in the region of 
$m_h\lesssim 135$ GeV from that in the region of $m_h> 135$ GeV. 
In the region of $m_h\lesssim 135$ GeV, 
the Higgs boson mainly decays into $b\bar{b}$, while in the region of $m_h> 135$ GeV, 
the gauge boson pair decay modes can be dominant.  
The decay width is grown rapidy when $m_h$ is greater than 130-140 GeV, 
because of the partial decay width of $h\to f\bar{f}$ modes is proportional to $m_h$, 
while that of the $h\to VV$ ($V=W,Z$) modes is proportional to $m_h^3$. 


\chapter{Extended Higgs sectors}
\section{The Two Higgs Doublet Model}
\subsection{Model}
The THDM contains two isospin doublet scalar fields $\Phi_1$ and $\Phi_2$ with $Y=1/2$. 
The most general Higgs potential is given by
\begin{align}
V&=m_1^2|\Phi_1|^2+m_2^2|\Phi_2|^2-(m_3^2\Phi_1^\dagger \Phi_2 +\text{h.c.})\notag\\
& +\frac{1}{2}\lambda_1|\Phi_1|^4+\frac{1}{2}\lambda_2|\Phi_2|^4+\lambda_3|\Phi_1|^2|\Phi_2|^2+\lambda_4|\Phi_1^\dagger\Phi_2|^2\notag\\
& +\frac{1}{2}[\lambda_5(\Phi_1^\dagger\Phi_2)^2+\lambda_6|\Phi_1|^2\Phi_1^\dagger \Phi_2+\lambda_7|\Phi_2|^2\Phi_1^\dagger\Phi_2+\text{h.c.}], \label{pot_thdm1}
\end{align}
where $m_1$, $m_2$ and $\lambda_1$-$\lambda_4$ are real while 
$m_3$ and $\lambda_5$-$\lambda_7$ are complex in general. 
The Higgs doublets can be parameterized as 
\begin{align}
\Phi_i=\left[\begin{array}{c}
w_i^+\\
\frac{1}{\sqrt{2}}(v_i+h_i+iz_i)
\end{array}\right],\hspace{3mm}(i=1,2).
\end{align}
The most general Yukawa Lagrangian is given by 
\begin{align}
&\mathcal{L}_{\text{THDM}}^Y=\notag\\
&-\left[\bar{Q}_L(Y_{d1}\Phi_1 + Y_{d2}\Phi_2)d_R
+\bar{Q}_L(Y_{u1}\tilde{\Phi}_1+Y_{u2}\tilde{\Phi}_2)u_R
+\bar{L}_L(Y_{e1}\Phi_1+Y_{e2}\Phi_2) e_R
+\text{h.c.}\right], \label{yukawa_thdm}
\end{align}
where $Y_{f1}$ and $Y_{f2}$, ($f=u,d,e$) are the 3$\times$3 complex matrices and $\tilde{\Phi}_i=i\tau_2\Phi_i^*$, ($i=1,2$). 
In the THDM, neutral scalar bosons mediated FCNC processes appear at the tree level in general. 
To understand the FCNC problem, 
we introduce so-called the Georgi basis or the Higgs basis~\cite{Georgi_Base} as  
\begin{align}
\left(\begin{array}{c}
\Phi_1\\
\Phi_2
\end{array}\right)=R(\beta)\left(\begin{array}{c}
\Phi\\
\Psi
\end{array}\right)
,\quad 
R(\theta)\equiv \left(\begin{array}{cc}
\cos\beta & -\sin\beta\\
\sin\beta & \cos\beta
\end{array}\right), \label{gb11}
\end{align}
where 
\begin{align}
\Phi=\left[
\begin{array}{c}
w^+\\
\frac{1}{\sqrt{2}}(v+h_1'+iz)
\end{array}\right],\quad
\Psi=\left[
\begin{array}{c}
H^+\\
\frac{1}{\sqrt{2}}(h_2'+iA)
\end{array}\right], \label{gb22}
\end{align}
with $v\simeq$ 246 GeV and $\tan\beta=v_2/v_1$. 
In Eq.~(\ref{gb22}), 
$w^\pm$ and $z$ are the NG bosons which are absorbed into the longitudinal component of 
$W^\pm$ and $Z$, respectively. 
The scalar boson states in the Georgi basis are related to the original ones through the following relations: 
\begin{align}
\left(\begin{array}{c}
w_1^\pm\\
w_2^\pm
\end{array}\right)=R(\beta)
\left(\begin{array}{c}
w^\pm\\
H^\pm
\end{array}\right),\quad 
\left(\begin{array}{c}
z_1\\
z_2
\end{array}\right)
=R(\beta)\left(\begin{array}{c}
z\\
A
\end{array}\right),\quad
\left(\begin{array}{c}
h_1\\
h_2
\end{array}\right)=R(\beta)
\left(\begin{array}{c}
h_1'\\
h_2'
\end{array}\right). \label{gb33}
\end{align}
In the CP-conserving case, $H^\pm$ are the charged scalar bosons, $A$ is the CP-odd scalar boson 
and $h_1'$ and $h_2'$ are the CP-even scalar states whose 
mass matrix is non-diagonal at this stage. 
In this basis, the Yukawa Lagrangian can be rewritten as 
\begin{align}
&\mathcal{L}_{\text{THDM}}^Y=\notag\\
&-\left[\bar{Q}_L\left(\frac{\sqrt{2}}{v}M_d\Phi + Y_d\Psi\right)d_R
+\bar{Q}_L\left(\frac{\sqrt{2}}{v}M_u\tilde{\Phi}+Y_u\tilde{\Psi}\right)u_R
+\bar{L}_L\left(\frac{\sqrt{2}}{v}M_e\Phi+Y_e\Psi\right) e_R
+\text{h.c.}\right], 
\end{align}
where $M_f$ is the non-diagonal 3$\times$3 fermion mass matrix, while $Y_f$ is the 
complex 3$\times$3 matrix which is non-diagonal in general. 
These matrices can be written in terms of the original Yukawa matrices $Y_{f1}$ and $Y_{f2}$ as
\begin{subequations}
\begin{align}
\frac{\sqrt{2}}{v}M_f&=Y_{f1}\cos\beta+Y_{f2}\sin\beta,\\
Y_f&=-Y_{f1}\sin\beta+Y_{f2}\cos\beta. 
\end{align}
\label{yyy}
\end{subequations}
In the mass eigenstates of fermions, the matrix $Y_f$ is rotated by the unitary matrices which 
diagonalize the mass matrix $M_f$. 
However, the rotated matrix $Y_f'$ is also non-diagonal in general, since 
there is no reason for $Y_f$ to be proportional to $M_f$. 
Therefore, through the non-diagonal elements of $Y_f$, 
neutral scalar boson mediating FCNC processes can appear at the tree level. 

There are several ways to avoid such a FCNC problem; 
\begin{description}
\item[(1)] Introducing a texture to the Yukawa matrices~\cite{texture}. 
\item[(2)] Assuming that the masses of the neutral scalar bosons which are mediated the FCNC processes are sufficiently large. 
\item[(3)] Assuming that the two Yukawa matrices $Y_{f1}$ and $Y_{f2}$ are proportional to each other~\cite{pich}. 
\item[(4)] Imposing an additional symmetry such as the $Z_2$ symmetry~\cite{GW}
to forbid the one of the Yukawa matrices $Y_{f1}$ and $Y_{f2}$. 
\end{description}
A model with the prescription of (1) listed in the above 
is one of a realization for a phenomenologically viable model which deduces 
so-called the type-III THDM~\cite{type3}, where tree-level FCNC processes appear. 
In Ref.~\cite{texture}, by the assumption that
non-diagonal Yukawa couplings are proportional to the geometric mean of the
two fermion masses, $g_{ij}\propto m_im_j$, the FCNC interactions can be suppressed. 

The way of (2) is the obvious possibility and not so phenomenologically interesting. 
In this case, the THDM is just reduced to the SM. 

The THDM with the assumption of (3), 
which is so-called the Yukawa alignment, has been discussed in Ref.~\cite{pich}. 
Following~\cite{pich}, we assume that the one of the two Yukawa matrices is proportional to the other one: 
\begin{align}
Y_{f2}=\zeta_f Y_{f1}, 
\end{align} 
where $\zeta_f$ is the complex constant. 
In this case, Eq.~(\ref{yyy}) can be rewritten as 
\begin{subequations}
\begin{align}
\frac{\sqrt{2}}{v}M_f&=Y_{f1}(\cos\beta+\zeta_f\sin\beta),\\
Y_f&=Y_{f1}(-\sin\beta+\zeta_f\cos\beta), 
\end{align}
\end{subequations}
so that the matrix $Y_f$ is given by the same matrix for the fermion masses as 
\begin{align}
Y_f=\frac{\zeta_f-\tan\beta}{1+\zeta_f\tan\beta}\times \frac{\sqrt{2}}{v}M_f. 
\end{align}
Therefore, the matrices $Y_f$ and $M_f$ can be diagonalized simultaneously. 

In this section, we study the THDM with a softly-broken discrete $Z_2$ symmetry 
which is corresponding to the prescription of (4) as the simplest but the natural way\footnote{ 
In Ref.~\cite{omura}, the THDM with an additional local $U(1)$ symmetry instead of the $Z_2$ symmetry 
has been discussed.}.  
Hereafter, we discuss the $Z_2$ invariant THDM. 
Under the $Z_2$ symmetry, we suppose that the Higgs doublets are translated into  
$\Phi_1\to +\Phi_1$ and $\Phi_2\to -\Phi_2$. 
The $Z_2$ invariant Higgs potential can be written as 
\begin{align}
V&=m_1^2|\Phi_1|^2+m_2^2|\Phi_2|^2-m_3^2(\Phi_1^\dagger \Phi_2 +\text{h.c.})\notag\\
& +\frac{1}{2}\lambda_1|\Phi_1|^4+\frac{1}{2}\lambda_2|\Phi_2|^4+\lambda_3|\Phi_1|^2|\Phi_2|^2+\lambda_4|\Phi_1^\dagger\Phi_2|^2
+\frac{1}{2}\lambda_5[(\Phi_1^\dagger\Phi_2)^2+\text{h.c.}], \label{pot_thdm2}
\end{align}
where the terms of $\lambda_6$ and $\lambda_7$ in Eq.~(\ref{pot_thdm1}) are forbidden by the $Z_2$ symmetry.  
In the $Z_2$ invariant Higgs potential, there are six real parameters and two complex parameters. 
In Eq.~(\ref{pot_thdm2}), we assume the CP-conserving Higgs potential; i.e., the imaginary parts of $m_3$ and $\lambda_5$ are neglected. 
From the vacuum condition: 
\begin{align}
m_1^2&=m_3^2\tan\beta-\frac{v^2}{2}(\lambda_1\cos^2\beta+\bar{\lambda}\sin^2\beta),\\
m_2^2&=m_3^2\cot\beta-\frac{v^2}{2}(\lambda_1\sin^2\beta+\bar{\lambda}\cos^2\beta), 
\end{align}
we can eliminate $m_1^2$ and $m_2^2$ in the Higgs potential,  
where $\bar{\lambda}=\lambda_3+\lambda_4+\lambda_5$.  
The masses of $H^\pm$ and $A$ can be calculated as 
\begin{align}
m_{H^\pm}^2=M^2-\frac{v^2}{2}(\lambda_4+\lambda_5),\quad m_A^2&=M^2-v^2\lambda_5, \label{mass1}
\end{align}
where $M$ is the soft breaking $Z_2$ symmetry parameter:
\begin{align}
M^2=\frac{m_3^2}{\sin\beta\cos\beta}. \label{bigm}
\end{align}
The mass matrix for the neutral CP-even scalar states is 
\begin{align}
V_{\text{THDM}}^{\text{CP-even}}&=\frac{1}{2}(h_1',h_2')
\left(\begin{array}{cc}
M_{11}^2 & M_{12}^2\\
M_{12}^2 &M_{22}^2
\end{array}\right)
\left(\begin{array}{c}
h_1' \\
h_2'
\end{array}\right), 
\end{align}
where matrix elements are 
\begin{subequations}
\begin{align}
M_1^2&=v^2(\lambda_1\cos^4\beta+\lambda_2\sin^4\beta)+\frac{v^2}{2}\bar{\lambda}\sin^22\beta,\\
M_2^2&=M^2+v^2\sin^2\beta\cos^2\beta(\lambda_1+\lambda_2-2\bar{\lambda}),\\
M_{12}^2&=\frac{v^2}{2}\sin2\beta(-\lambda_1\cos^2\beta+\lambda_2\sin^2\beta)+\frac{v^2}{2}\sin2\beta\cos2\beta\bar{\lambda}.
\end{align}
\label{mateven}
\end{subequations}
We here introduce the mixing angle $\alpha$ 
to diagonalize the mass matrix for the CP-even scalar states as:
\begin{align}
\left(\begin{array}{c}
h_1' \\
h_2'
\end{array}\right)=R(\alpha-\beta)
\left(\begin{array}{c}
H\\
h
\end{array}\right). 
\end{align}
The mass eigenvalues are 
\begin{align}
m_{H,h}^2&=\frac{1}{2}\left[M_1^2+M_2^2\pm\sqrt{(M_1^2+M_2^2)^2+4M_{12}^2}\right].
\end{align}
The mixing angle $\alpha-\beta$ is expressed in terms of the mass matrix elements in Eq.~(\ref{mateven})
\begin{align}
\tan 2(\alpha-\beta)=\frac{2M_{12}^2}{M_1^2-M_2^2}.
\end{align} 
The original eight parameters in the Higgs potential: $\lambda_1-\lambda_5$ and $m_1^2-m_3^2$ are 
described by the four physical scalar boson masses: $m_{H^\pm}$, $m_A$, $m_H$, $m_h$, 
two mixing angles $\alpha$ and $\beta$, the VEV $v$ and the soft-breaking scale of the $Z_2$ symmetry $M$. 
It is useful to rewrite the quartic couplings $\lambda_1-\lambda_5$ to the physical parameters as
\begin{subequations}
\begin{align}
\lambda_1
&=\frac{1}{v^2\cos^2\beta}\left[-\sin^2\beta M^2+\cos^2\alpha m_H^2+\sin^2\alpha m_h^2\right],\\
\lambda_2
&=\frac{1}{v^2\sin^2\beta}\left[-\cos^2\beta M^2+\sin^2\alpha m_H^2+\cos^2\alpha m_h^2\right],\\
\lambda_3&=-\frac{M^2}{v^2}+\frac{2m_{H^\pm}^2}{v^2}+\frac{1}{v^2}\frac{\sin 2\alpha}{\sin2\beta}(m_H^2-m_h^2),\\
\lambda_4&=\frac{1}{v^2}(M^2+m_A^2-2m_{H^\pm}^2),\\
\lambda_5&=\frac{1}{v^2}(M^2-m_A^2).
\end{align}
\end{subequations}

The most general Yukawa interaction under the $Z_2$ symmetry 
can be written as 
\begin{align}
{\mathcal L}^Y_\text{THDM} =
&-y_u{\overline Q}_L\widetilde{\Phi}_uu_R^{}
-y_d{\overline Q}_L\Phi_dd_R^{}
-y_e{\overline L}_L\Phi_e e_R^{}+\text{h.c.},
\end{align}
where $\Phi_f$ ($f=u,d$ or $e$) is either $\Phi_1$ or $\Phi_2$. 
There are four independent $Z_2$ charge assignments 
on quarks and charged leptons, as summarized in TABLE~\ref{Tab:type}~\cite{Barger,Grossman}. 
In the type-I THDM, all quarks and charged leptons obtain their masses
from the VEV of $\Phi_2$.
In the type-II THDM, masses of up-type quarks are generated by the VEV
of $\Phi_2$, while those of down-type quarks and charged leptons
are acquired by that of $\Phi_1$. The Higgs sector of the
MSSM is a special THDM whose Yukawa interaction is of type-II. 
The type-X Yukawa interaction (all quarks couple to $\Phi_2$
while charged leptons couple to $\Phi_1$) 
is used in the model in Ref.~\cite{aks_prl}.
The remaining one is referred to as the type-Y THDM. 
\begin{table}[t]
\begin{center}
\begin{tabular}{|c||c|c|c|c|c|c|}
\hline & $\Phi_1$ & $\Phi_2$ & $u_R^{}$ & $d_R^{}$ & $\ell_R^{}$ &
 $Q_L$, $L_L$ \\  \hline
Type-I  & $+$ & $-$ & $-$ & $-$ & $-$ & $+$ \\
Type-II & $+$ & $-$ & $-$ & $+$ & $+$ & $+$ \\
Type-X  & $+$ & $-$ & $-$ & $-$ & $+$ & $+$ \\
Type-Y  & $+$ & $-$ & $-$ & $+$ & $-$ & $+$ \\
\hline
\end{tabular}
\end{center}
\caption{Variation in charge assignments of the $Z_2$ symmetry~\cite{typeX}.} \label{Tab:type}
\end{table}

The Yukawa interactions are expressed in terms of mass eigenstates
of the Higgs bosons as
\begin{align}
{\mathcal L}^Y_\text{THDM} =
&-\sum_{f=u,d,e} \left( \frac{m_f}{v}\xi_h^f{\overline
f}fh+\frac{m_f}{v}\xi_H^f{\overline
f}fH+i\frac{m_f}{v}\xi_A^f{\overline f}\gamma_5fA\right)\notag\\
&-\left[\frac{\sqrt2V_{ud}}{v}\overline{u}
\left(m_u\xi_A^u\text{P}_L+m_d\xi_A^d\text{P}_R\right)d\,H^+
+\frac{\sqrt2m_\ell\xi_A^\ell}{v}\overline{\nu_L^{}}e_R^{}H^+
+\text{h.c.}\right],\label{yukawa_thdm}
\end{align}
where $P_{L/R}$ are projection operators for left-/right-handed fermions,
and the factors $\xi^f_\varphi$ are listed in TABLE~\ref{yukawa_tab}.

For the successful electroweak symmetry breaking,
a combination of quartic coupling constants should satisfy
the condition of vacuum stability~\cite{VS_thdm,VS_thdm2,Kasai}.
We also take into account bounds from perturbative unitarity 
to restrict parameters in the Higgs potential~\cite{PU_thdm,PU_thdm2}.
The top and bottom Yukawa coupling constants cannot be taken too large.
The requirement $|Y_{t,b}|^2<\pi$ at the tree level can provide a milder
constraint $0.4\lesssim\tan\beta\lesssim 91$, where $|Y_t|=(\sqrt{2}/v)
m_t \cot\beta$ and $|Y_b|=(\sqrt{2}/v) m_b \tan\beta$.
\begin{table}[t]
\begin{center}
{\renewcommand\arraystretch{1.5}
\begin{tabular}{|c||ccccccccc|}\hline
&$\xi_h^u$ &$\xi_h^d$&$\xi_h^d$&$\xi_H^u$&$\xi_H^d$&$\xi_H^\ell$&$\xi_A^u$&$\xi_A^d$&$\xi_A^\ell$ \\\hline\hline
Type-I &$\frac{\cos\alpha}{\sin\beta}$&$\frac{\cos\alpha}{\sin\beta}$&$\frac{\cos\alpha}{\sin\beta}$&$\frac{\sin\alpha}{\sin\beta}$&$\frac{\sin\alpha}{\sin\beta}$&$\frac{\sin\alpha}{\sin\beta}$&$-\cot\beta$&$\cot\beta$&$\cot\beta$\\\hline
Type-II &$\frac{\cos\alpha}{\sin\beta}$&$-\frac{\sin\alpha}{\cos\beta}$&$-\frac{\sin\alpha}{\cos\beta}$&$\frac{\sin\alpha}{\sin\beta}$&$\frac{\cos\alpha}{\cos\beta}$&$\frac{\cos\alpha}{\cos\beta}$&$-\cot\beta$&$-\tan\beta$&$-\tan\beta$\\\hline
Type-X &$\frac{\cos\alpha}{\sin\beta}$&$\frac{\cos\alpha}{\sin\beta}$&$-\frac{\sin\alpha}{\cos\beta}$&$\frac{\sin\alpha}{\sin\beta}$&$\frac{\sin\alpha}{\sin\beta}$&$\frac{\cos\alpha}{\cos\beta}$&$-\cot\beta$&$\cot\beta$&$-\tan\beta$\\\hline
Type-Y &$\frac{\cos\alpha}{\sin\beta}$&$-\frac{\sin\alpha}{\cos\beta}$&$\frac{\cos\alpha}{\sin\beta}$&$\frac{\sin\alpha}{\sin\beta}$&$\frac{\cos\alpha}{\cos\beta}$&$\frac{\sin\alpha}{\sin\beta}$&$-\cot\beta$&$-\tan\beta$&$\cot\beta$\\\hline
\end{tabular}}
\caption{The mixing factors in Yukawa interactions in Eq.~(\ref{yukawa_thdm})~\cite{typeX}.}
\label{yukawa_tab}
\end{center}
\end{table}

\subsection{Decay}
Here, we discuss the difference in decays of the Higgs bosons 
for the types of Yukawa interactions in the THDM. 
We calculate the decay rates of the Higgs bosons 
and evaluate the total widths and the branching ratios. 
In particular, we show the result with $\sin(\beta-\alpha)=1$~\cite{SMlike_thdm,KOSY}, 
where $h$ is the SM-like Higgs 
boson while the VEV of $H$ is zero. 
The decay pattern of $h$ is almost the same as that of the SM 
Higgs boson with the same mass at the leading order except 
for the loop-induced channels when $\sin(\beta-\alpha)=1$. 
In this case, $H$ does not decay into the gauge 
boson pair at tree level, so it mainly decays into fermion 
pairs\footnote{In the case with a more complicated mass spectrum a heavy Higgs boson 
can decay into the states which contain lighter Higgs 
bosons~\cite{Mukai}.}. 
We note that $A$ and $H^\pm$ do not decay into the gauge boson pair at the 
tree level for all 
values of $\sin(\beta-\alpha)$. 

The decay patterns are therefore completely 
different among the different types of Yukawa interactions~\cite{Barger,Grossman}. 
For the decays of $H$ and $A$, we take into account the decay channels of 
$q\bar q$, $\ell^+\ell^-$, ($WW^{(\ast)}$, $ZZ^{(\ast)}$) at the tree level, 
and $gg$, $\gamma\gamma$, $Z\gamma$ at the leading order, where $q$ represents 
$s$, $c$, $b$ (and $t$), and $\ell$ represents $\mu$ and $\tau$. 
Running masses for $b$, $c$, and $s$ quarks are fixed as  
$\overline{m_b}=3.0$ GeV, $\overline{m_c}=0.81$ GeV and 
$\overline{m_s}=0.077$ GeV, respectively. 
For the decay of the charged Higgs boson, the modes into 
$tb$, $cb$, $cs$, $\tau\nu$, and $\mu\nu$ 
are taken 
into account as long as they are kinematically allowed. 
%

\begin{figure}
\begin{center}
\begin{minipage}{0.275\hsize}
\vspace{0.8ex}
\includegraphics[width=4.5cm,angle=-90]{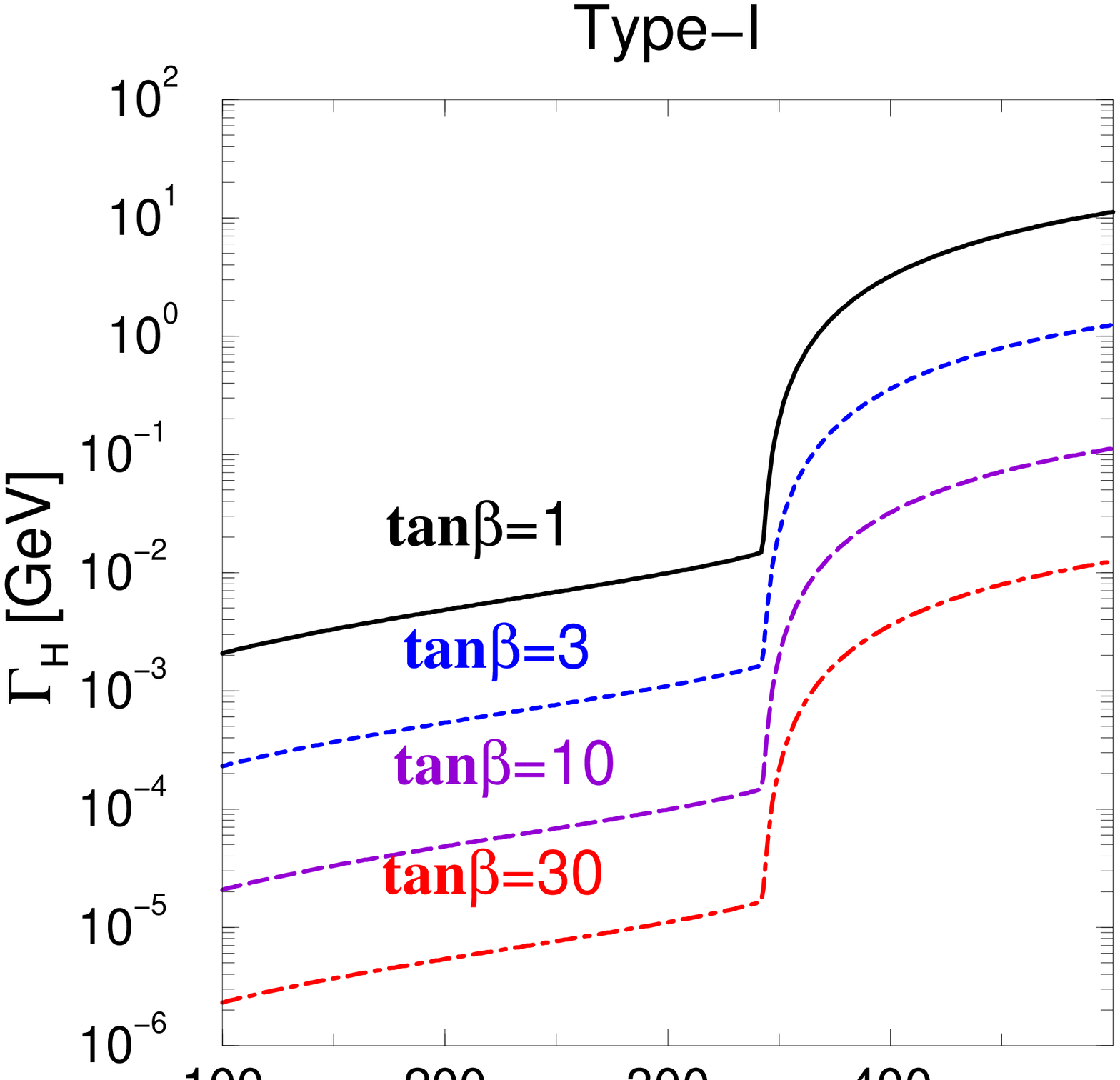}
\end{minipage}
\begin{minipage}{0.23\hsize}
\includegraphics[width=4.4cm,angle=-90]{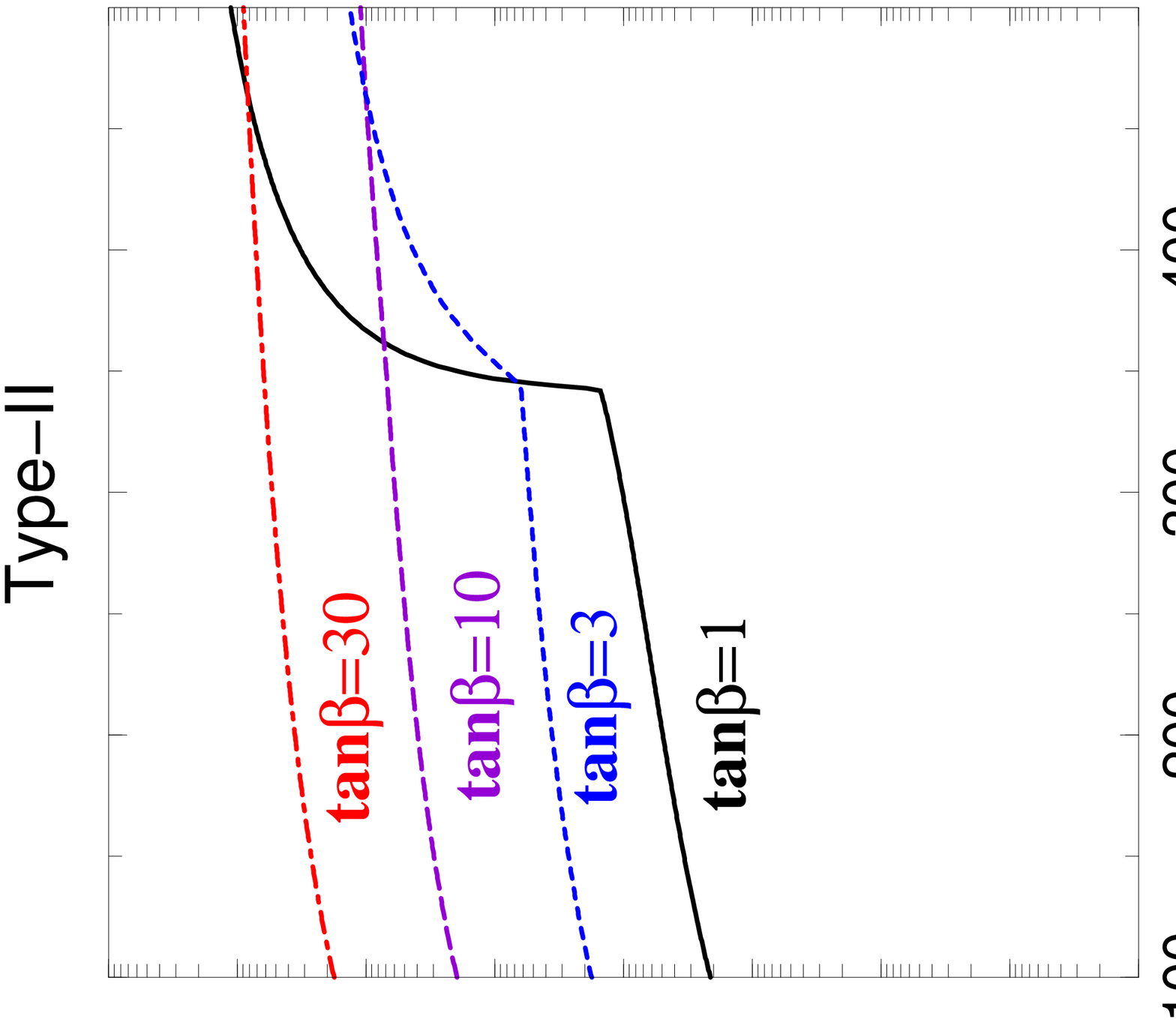}
\end{minipage}
\begin{minipage}{0.23\hsize}
\includegraphics[width=4.4cm,angle=-90]{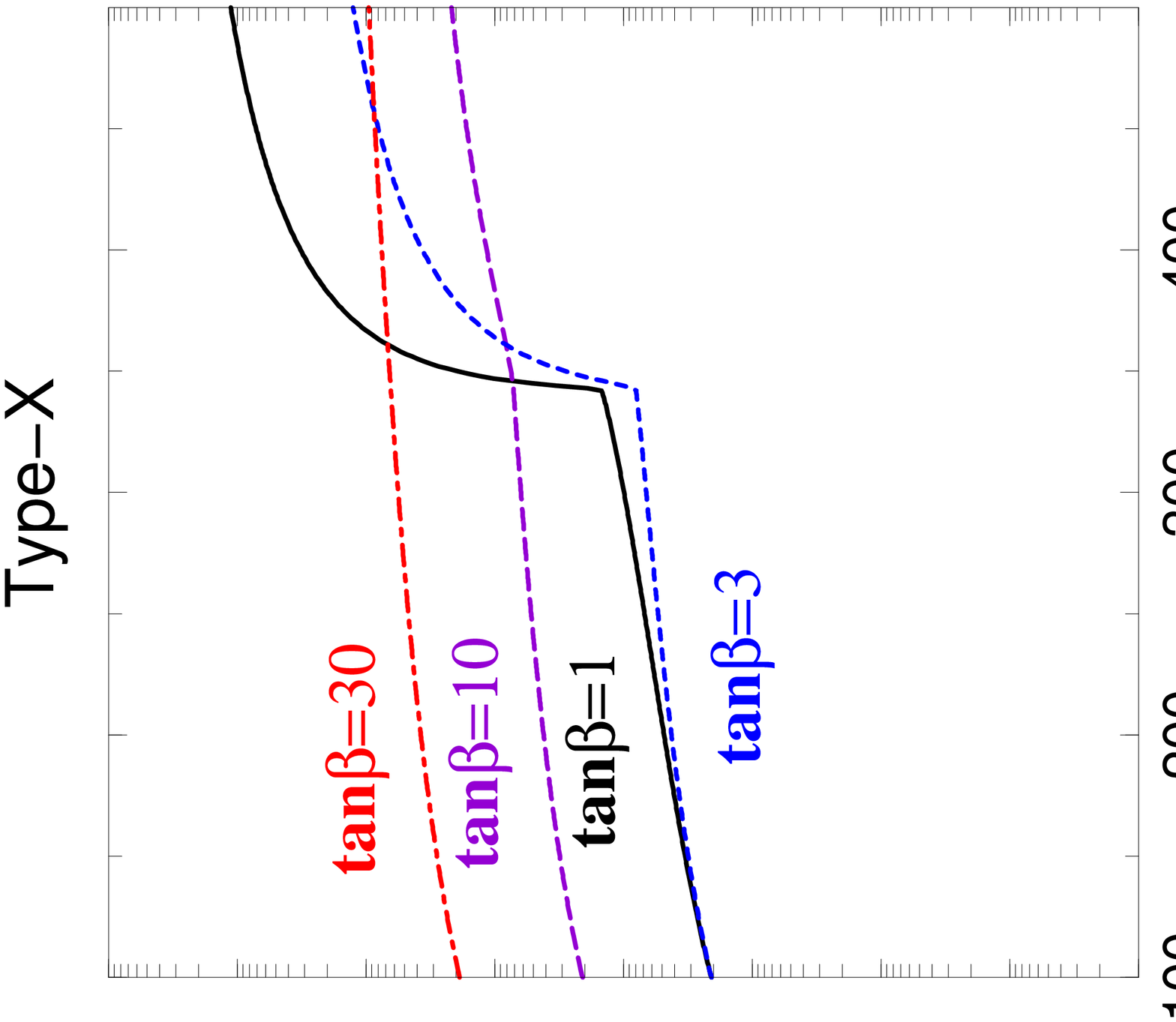}
\end{minipage}
\begin{minipage}{0.23\hsize}
\includegraphics[width=4.4cm,angle=-90]{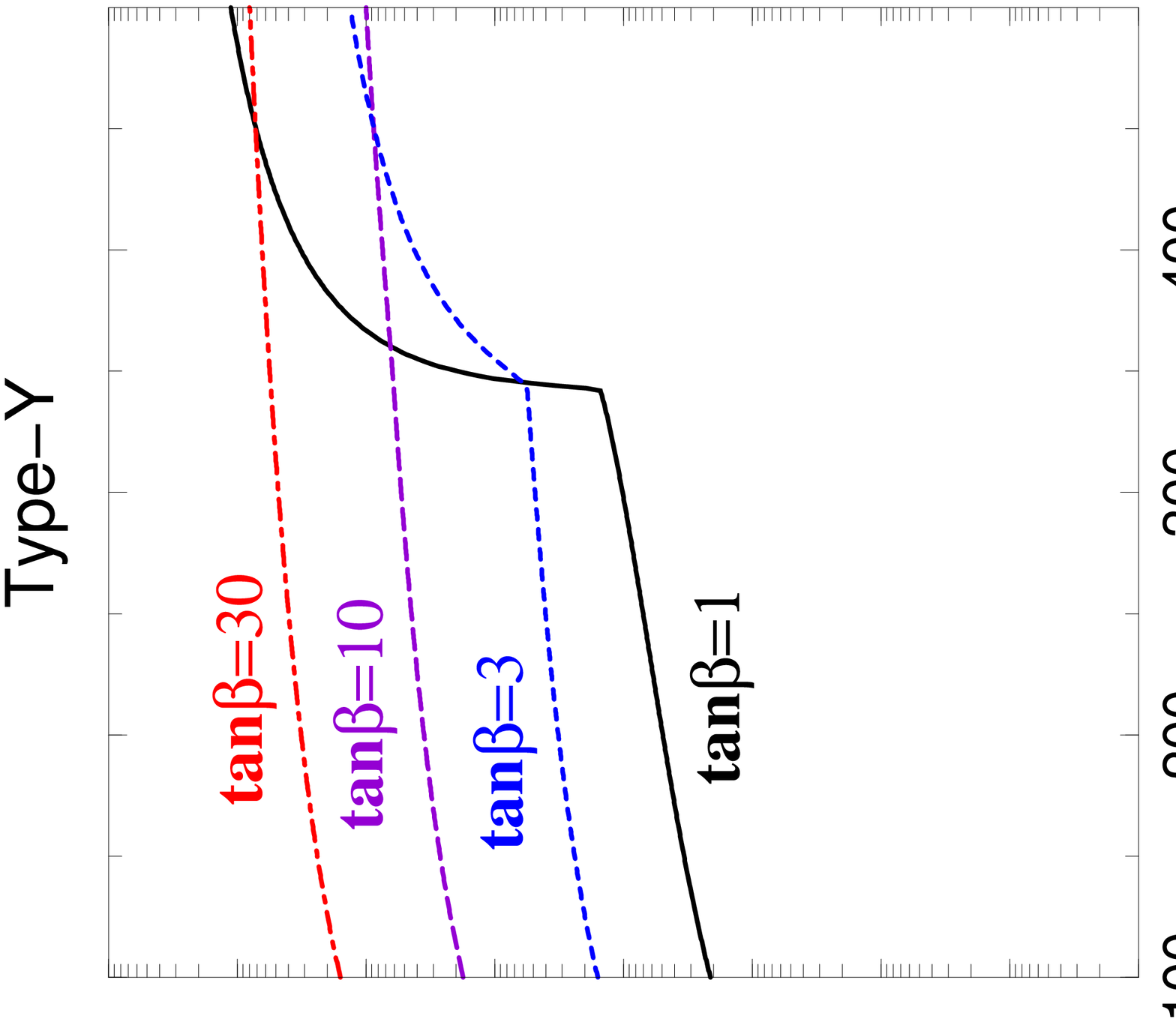}
\end{minipage}
\end{center}
\begin{center}
\begin{minipage}{0.275\hsize}
\vspace{0.8ex}
\includegraphics[width=4.5cm,angle=-90]{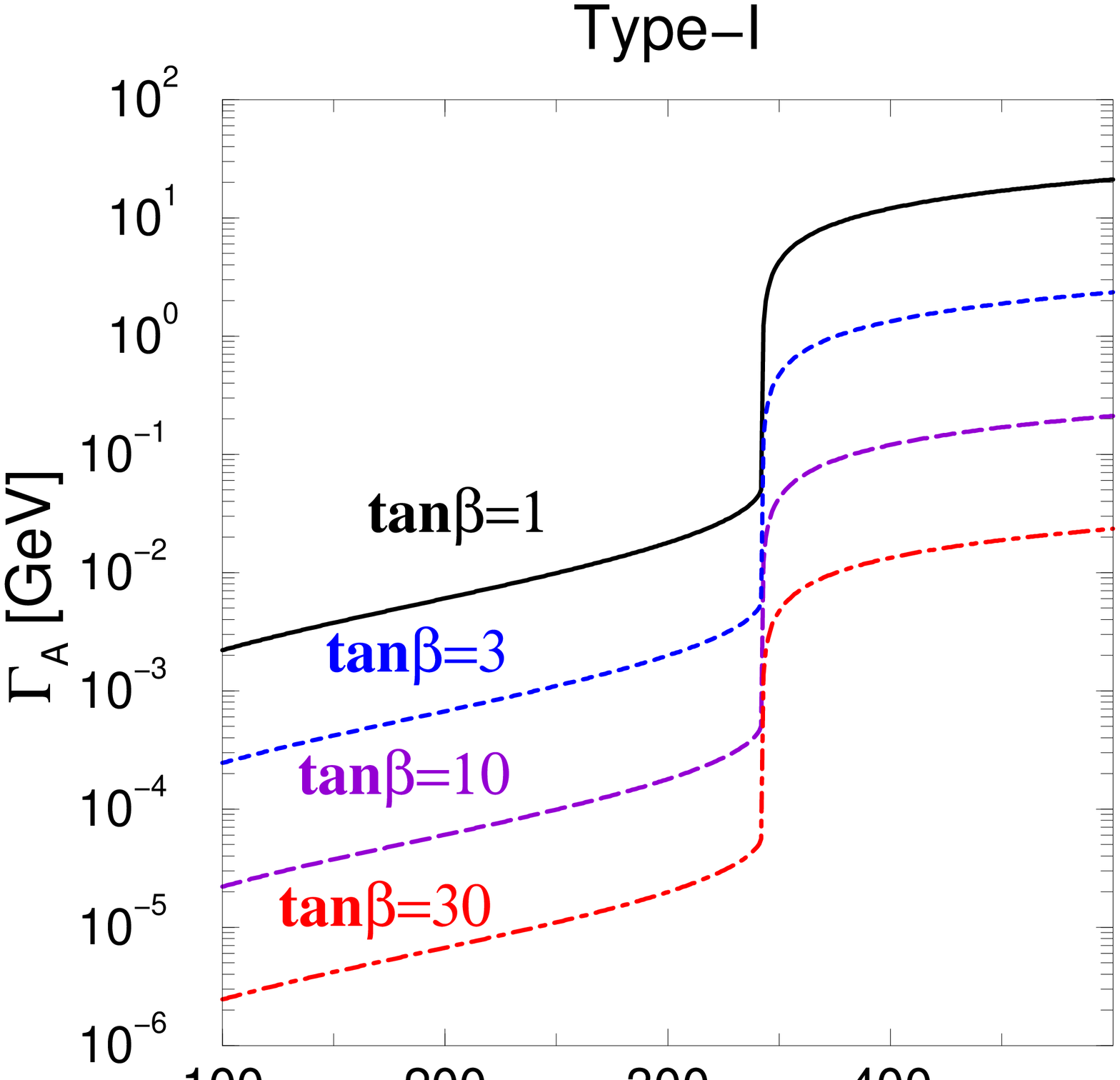}
\end{minipage}
\begin{minipage}{0.23\hsize}
\includegraphics[width=4.4cm,angle=-90]{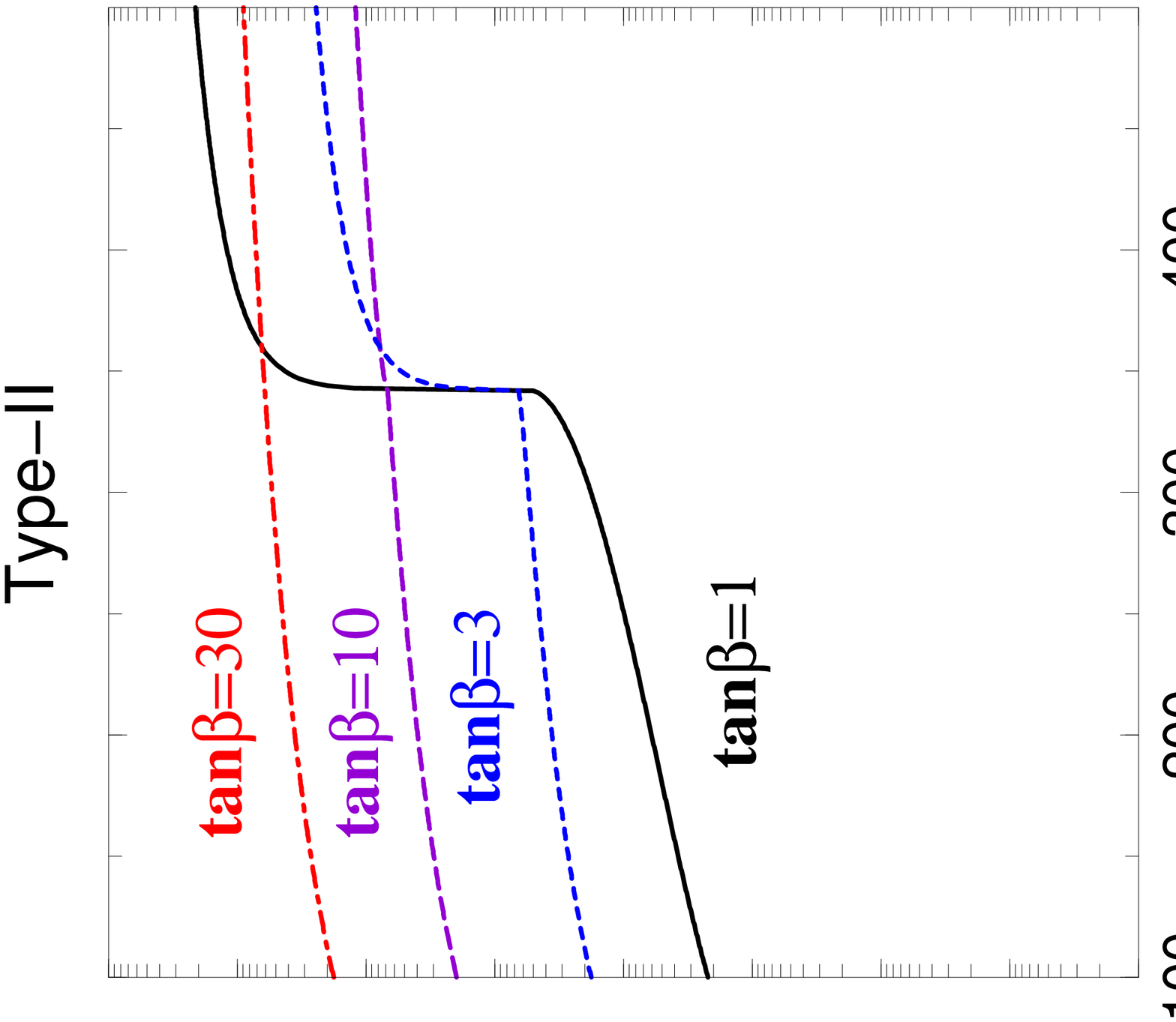}
\end{minipage}
\begin{minipage}{0.23\hsize}
\includegraphics[width=4.4cm,angle=-90]{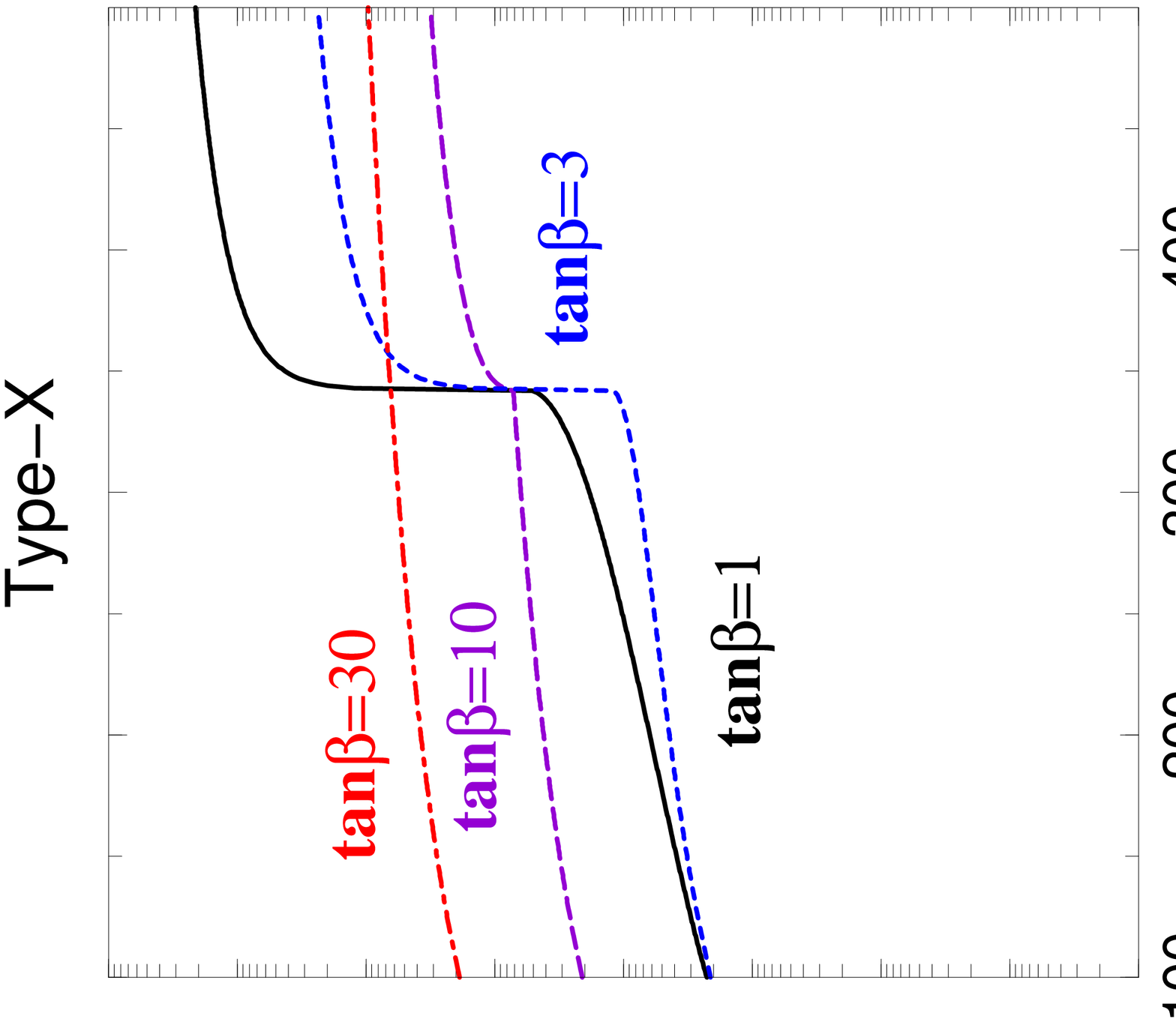}
\end{minipage}
\begin{minipage}{0.23\hsize}
\includegraphics[width=4.4cm,angle=-90]{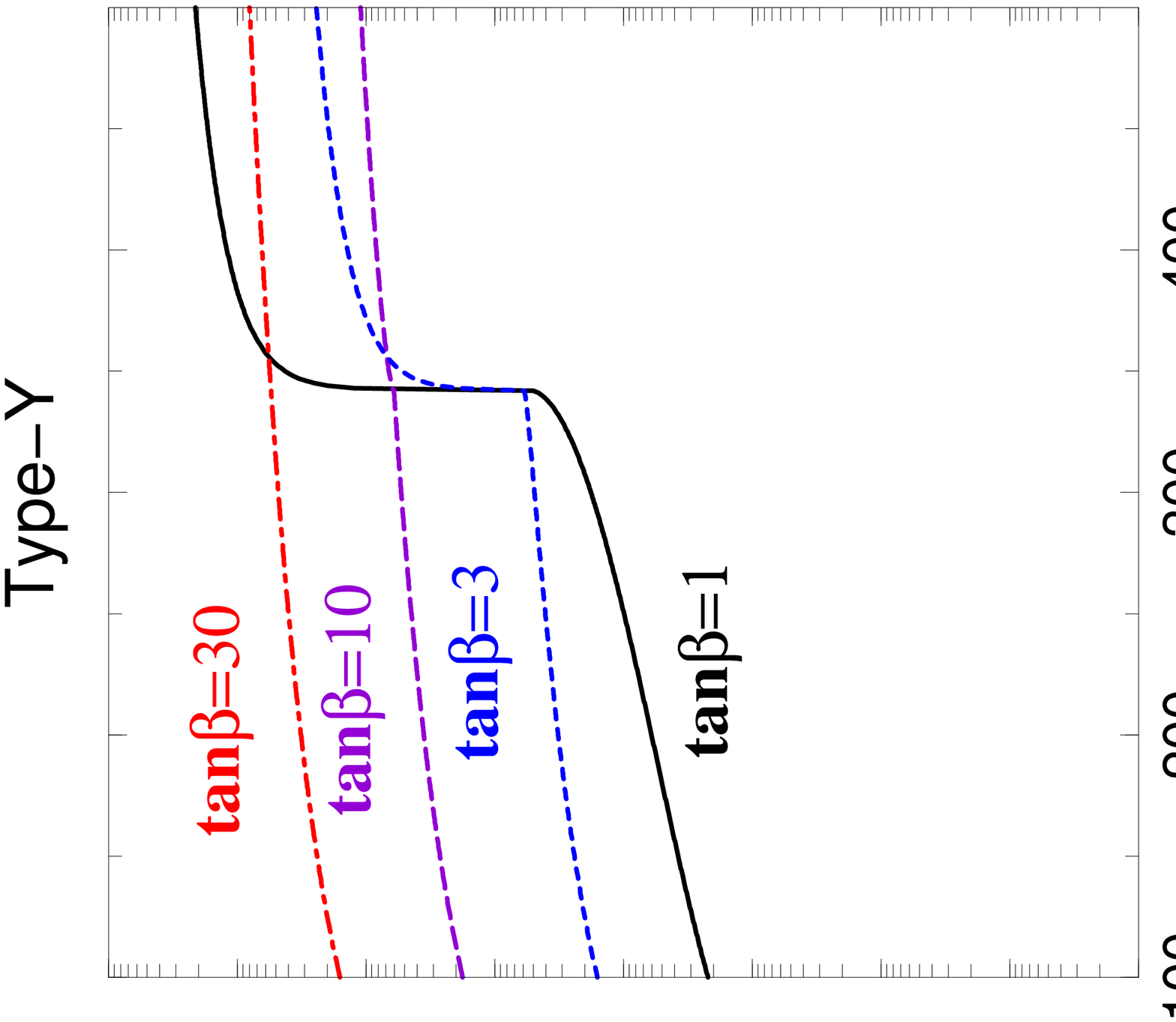}
\end{minipage}
\end{center}
\begin{center}
\begin{minipage}{0.275\hsize}
\vspace{0.8ex}
\includegraphics[width=4.5cm,angle=-90]{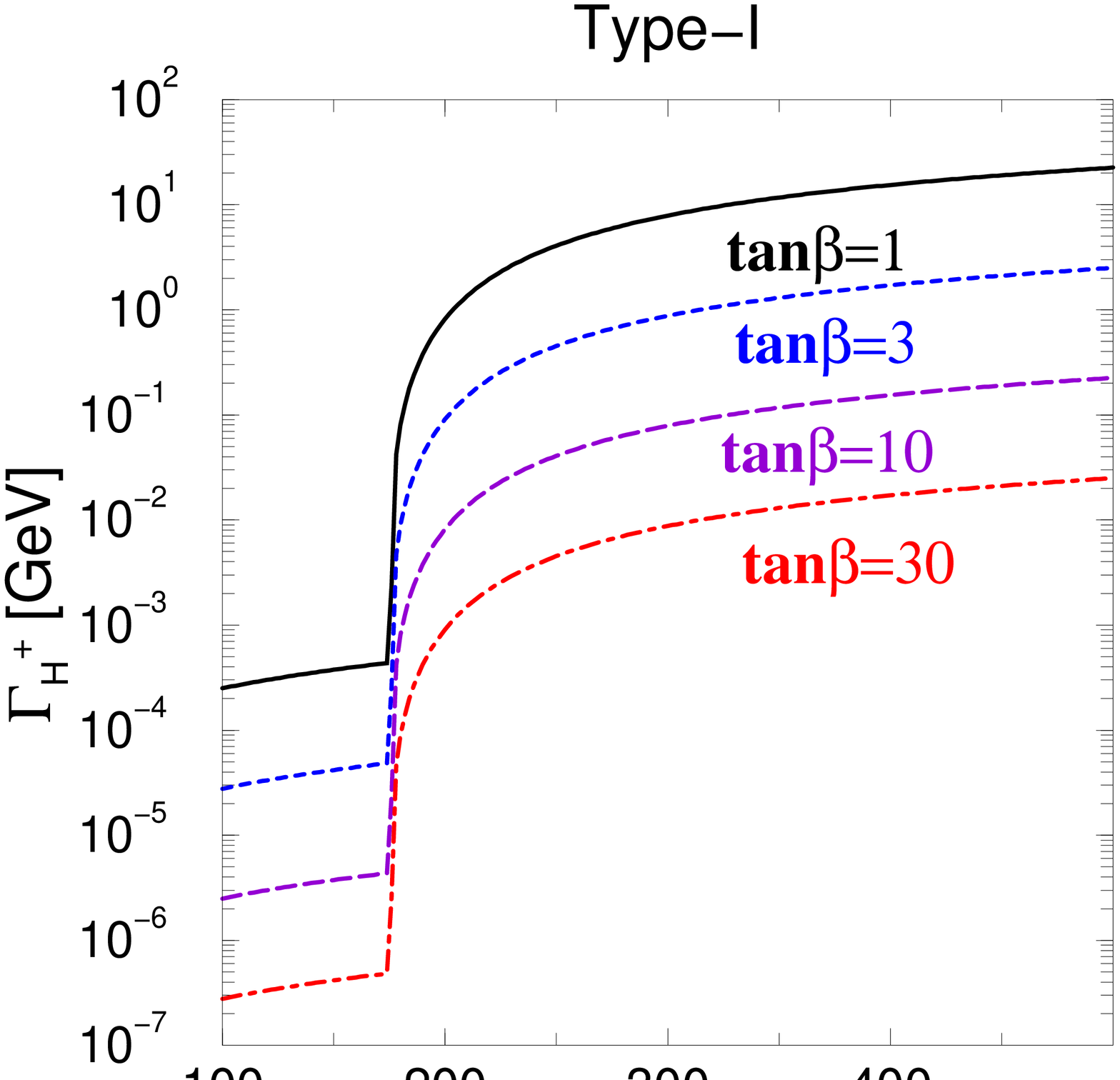}
\end{minipage}
\begin{minipage}{0.23\hsize}
\includegraphics[width=4.4cm,angle=-90]{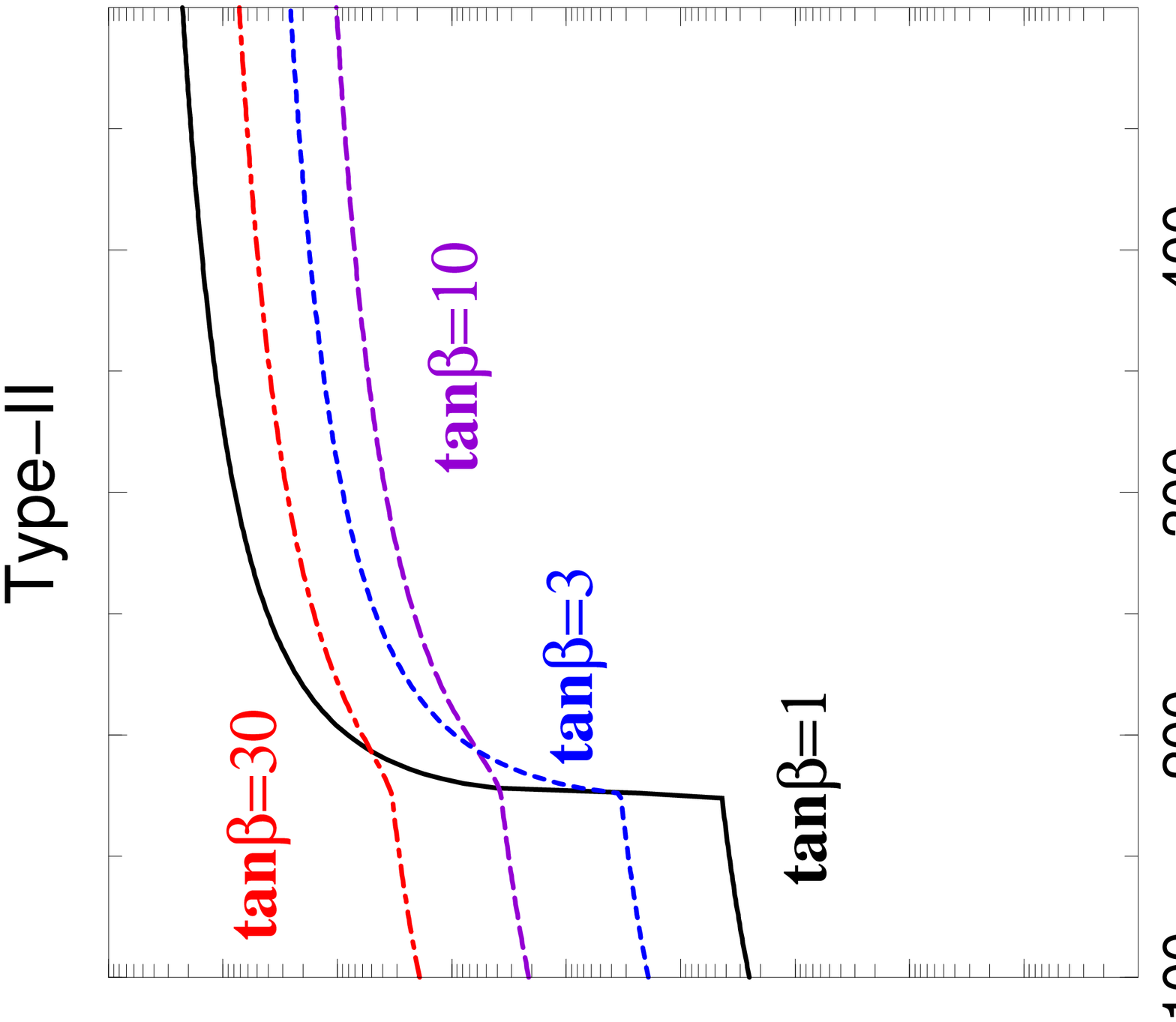}
\end{minipage}
\begin{minipage}{0.23\hsize}
\includegraphics[width=4.4cm,angle=-90]{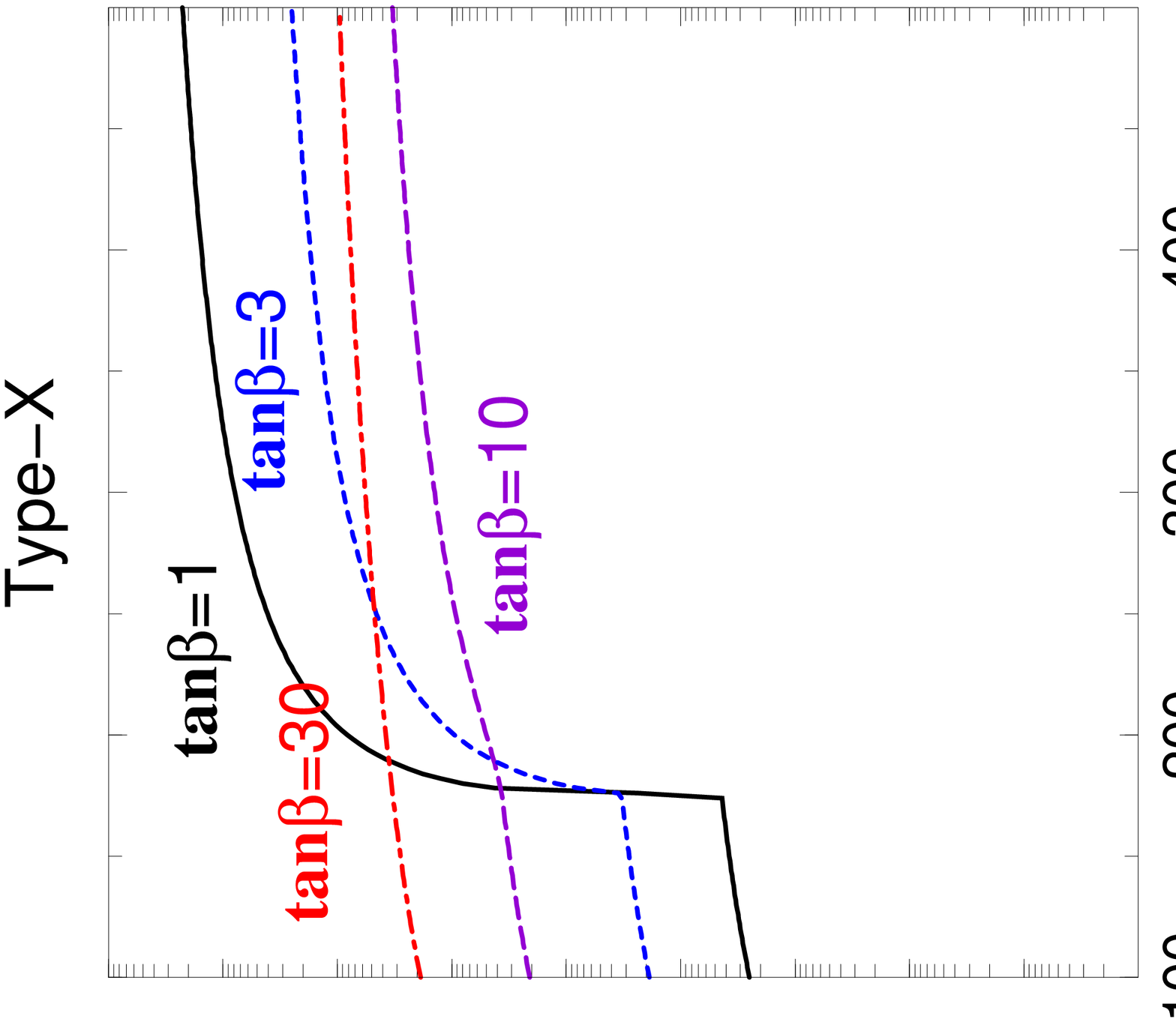}
\end{minipage}
\begin{minipage}{0.23\hsize}
\includegraphics[width=4.4cm,angle=-90]{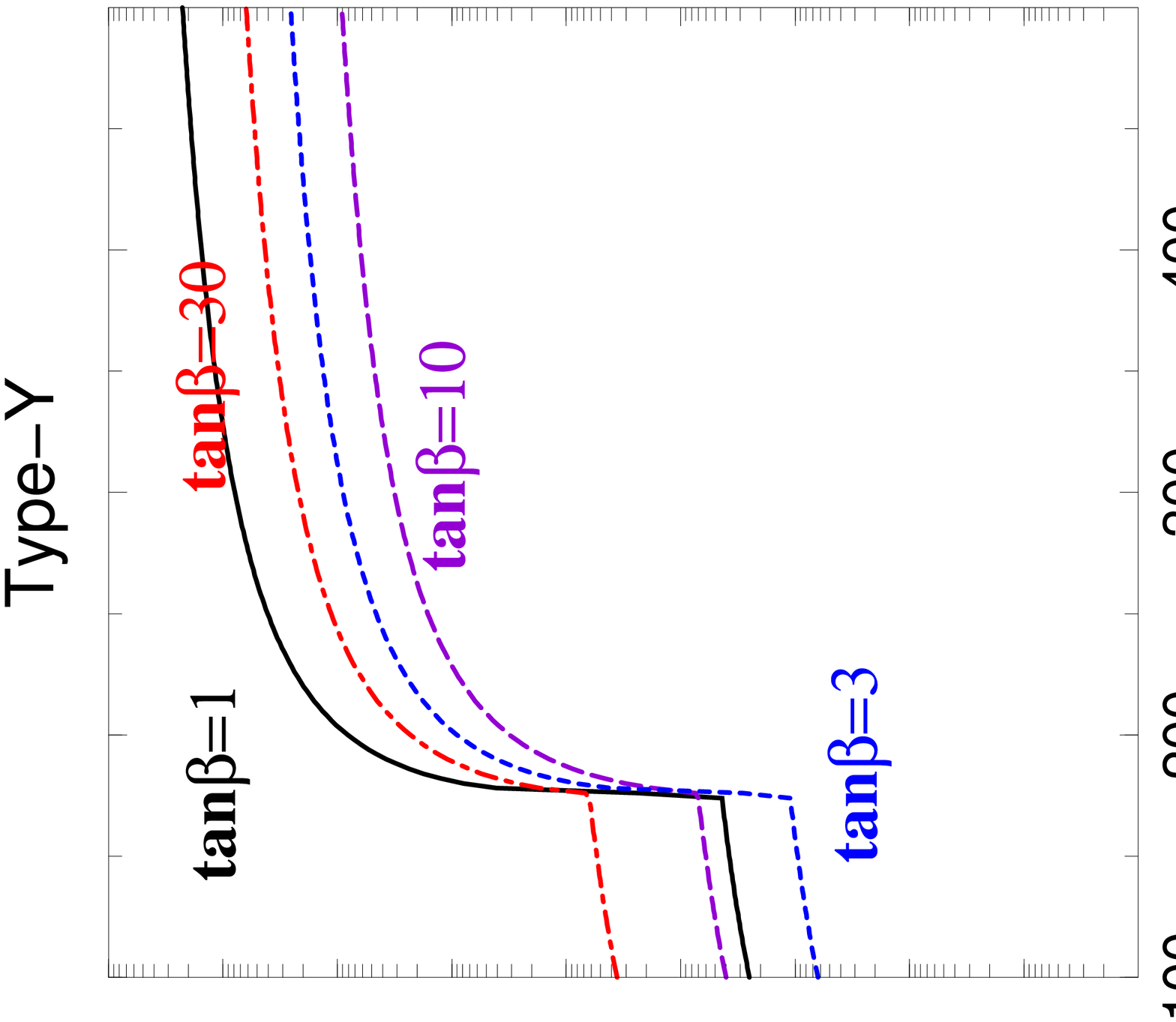}
\end{minipage}
\end{center}
\caption{Total decay widths of $H$, $A$ and $H^\pm$ in the four
 different  types of THDM as a function of the decaying scalar boson mass
for several values of $\tan\beta$ under the assumption $m_\Phi^{}=m_H^{}=m_A^{}=m_{H^\pm}^{}$. 
The SM-like limit $\sin(\beta-\alpha) =1$ is
taken, where $h$ is the SM-like Higgs boson~\cite{typeX}.}
\label{FIG:width_mass}
\end{figure}

In FIG.~\ref{FIG:width_mass}, the total widths of $H$, $A$ and $H^\pm$ 
are shown as a function of the mass of decaying Higgs bosons for 
several values of $\tan\beta$ in the four different types of Yukawa 
interactions. We assume $\sin(\beta-\alpha)=1$ and 
$m_\Phi^{}=m_H^{}=m_A^{}=m_{H^\pm}^{}$. 
The widths strongly depend on the types of Yukawa interactions for 
each $\tan\beta$ value before and after the 
threshold of the $t\bar t$ ($tb$) decay mode opens. 

\begin{figure}
\begin{center}
\begin{minipage}{0.285\hsize}
\includegraphics[width=4.4cm,angle=-90]{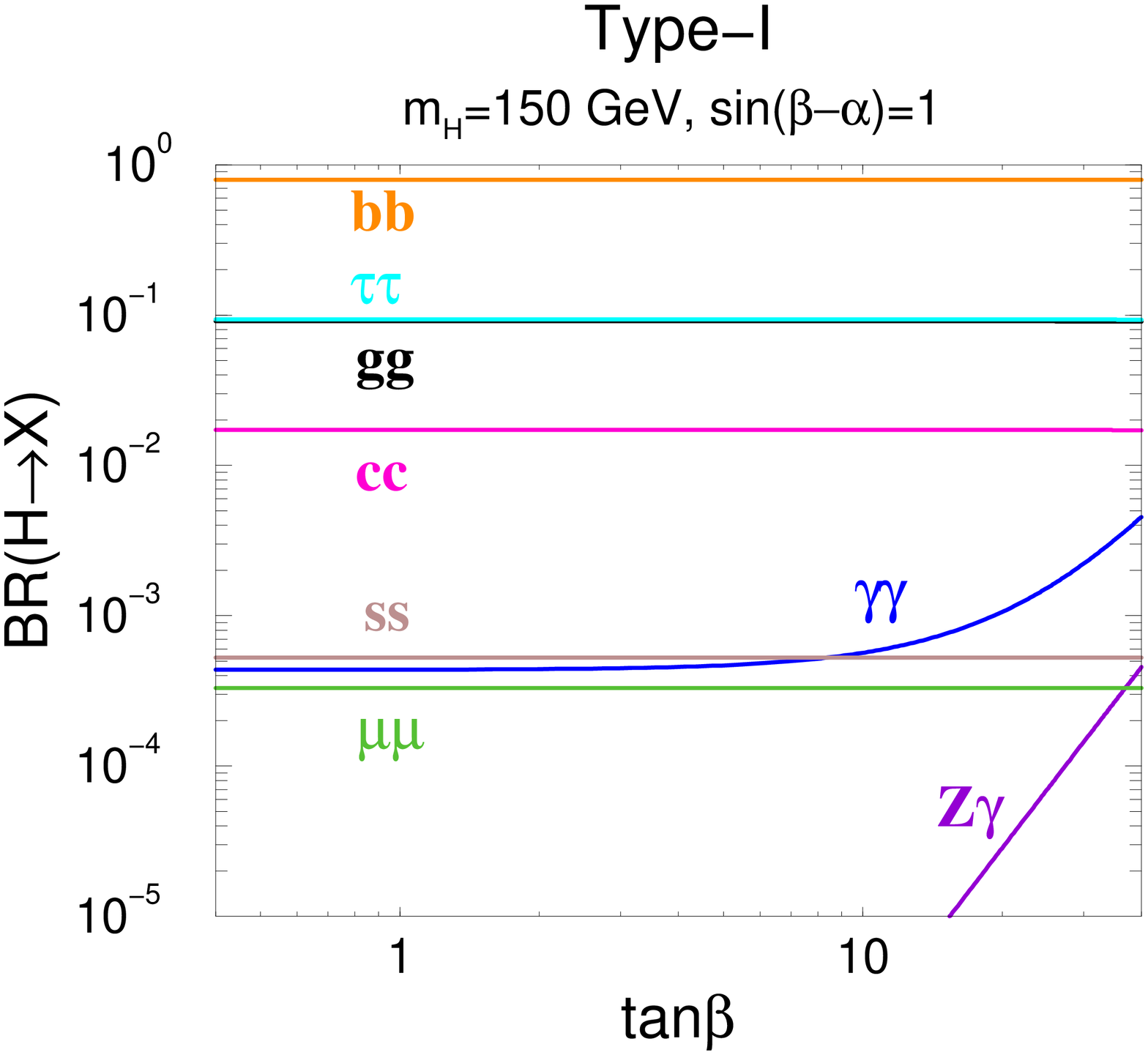}
\end{minipage}
\begin{minipage}{0.23\hsize}
\includegraphics[width=4.4cm,angle=-90]{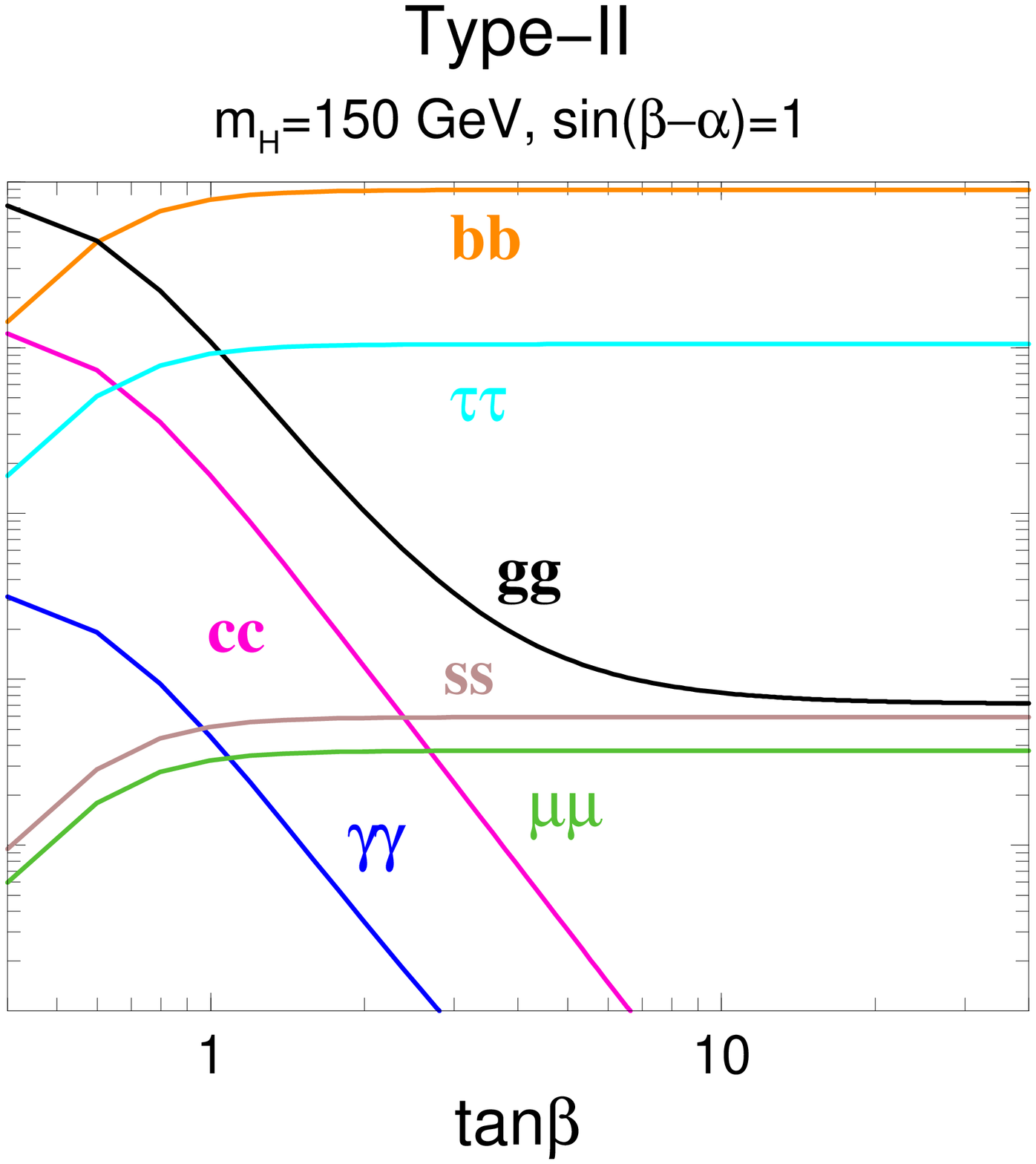}
\end{minipage}
\begin{minipage}{0.23\hsize}
\includegraphics[width=4.4cm,angle=-90]{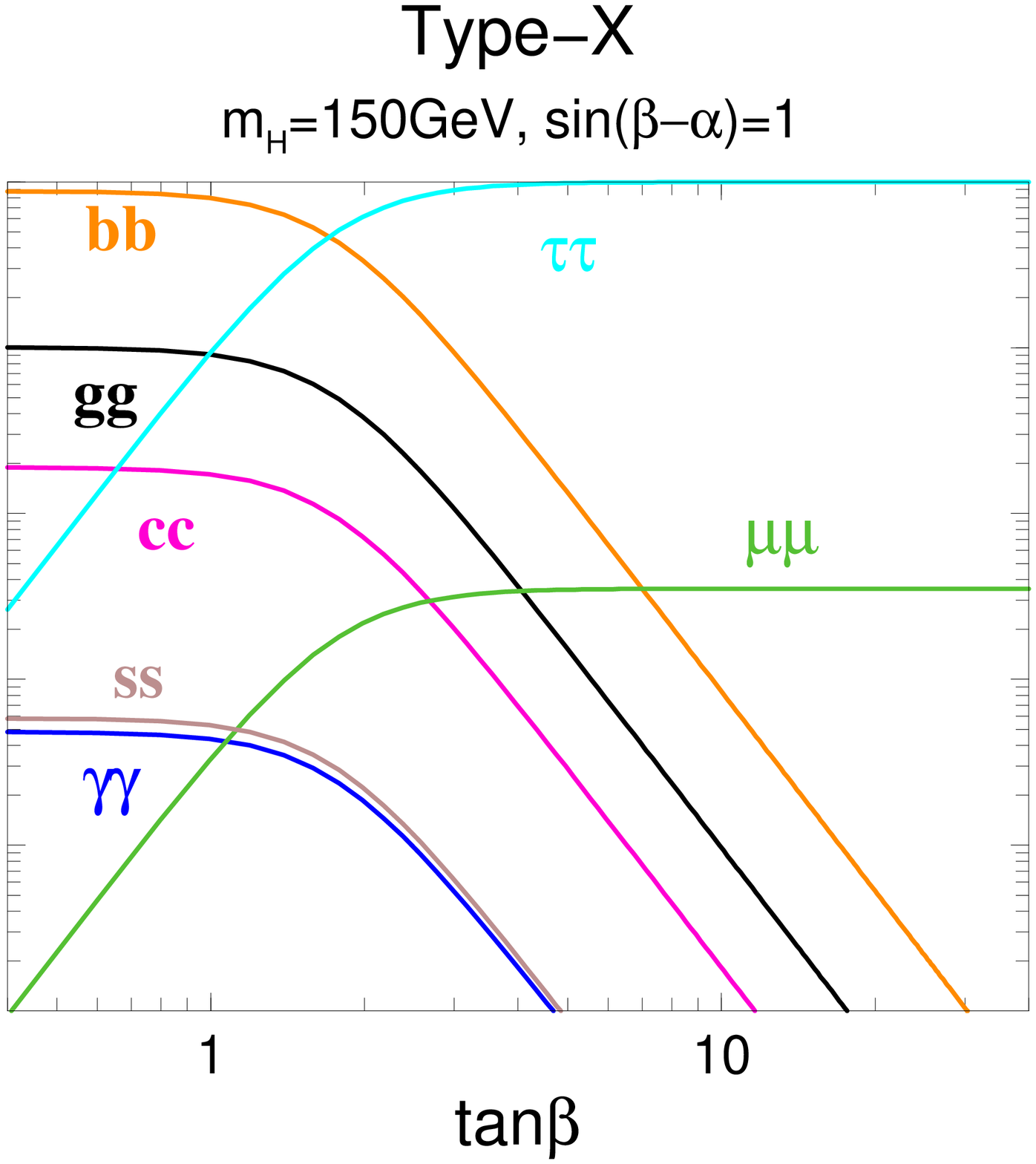}
\end{minipage}
\begin{minipage}{0.23\hsize}
\includegraphics[width=4.4cm,angle=-90]{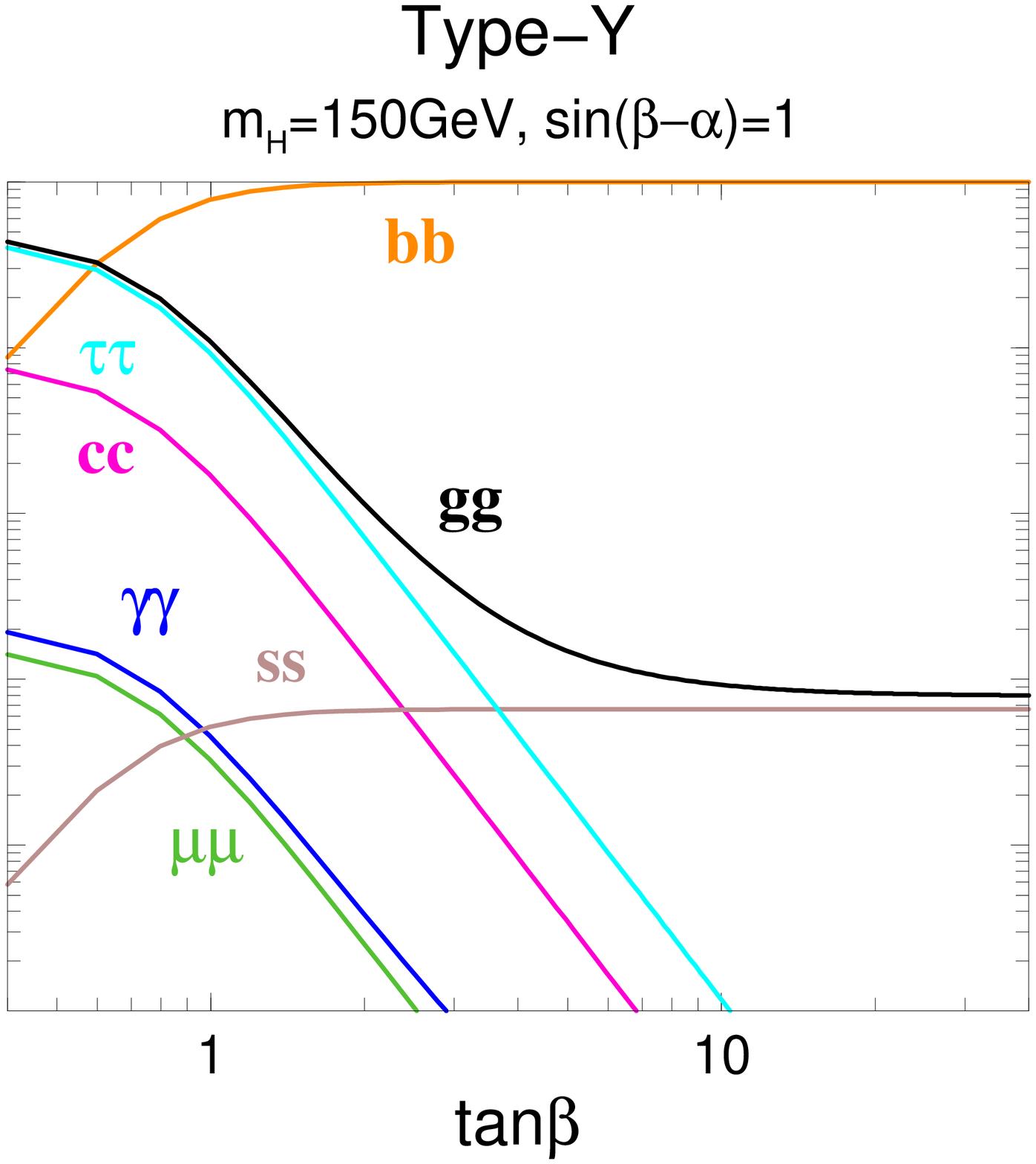}
\end{minipage}
\begin{minipage}{0.285\hsize}
\includegraphics[width=4.4cm,angle=-90]{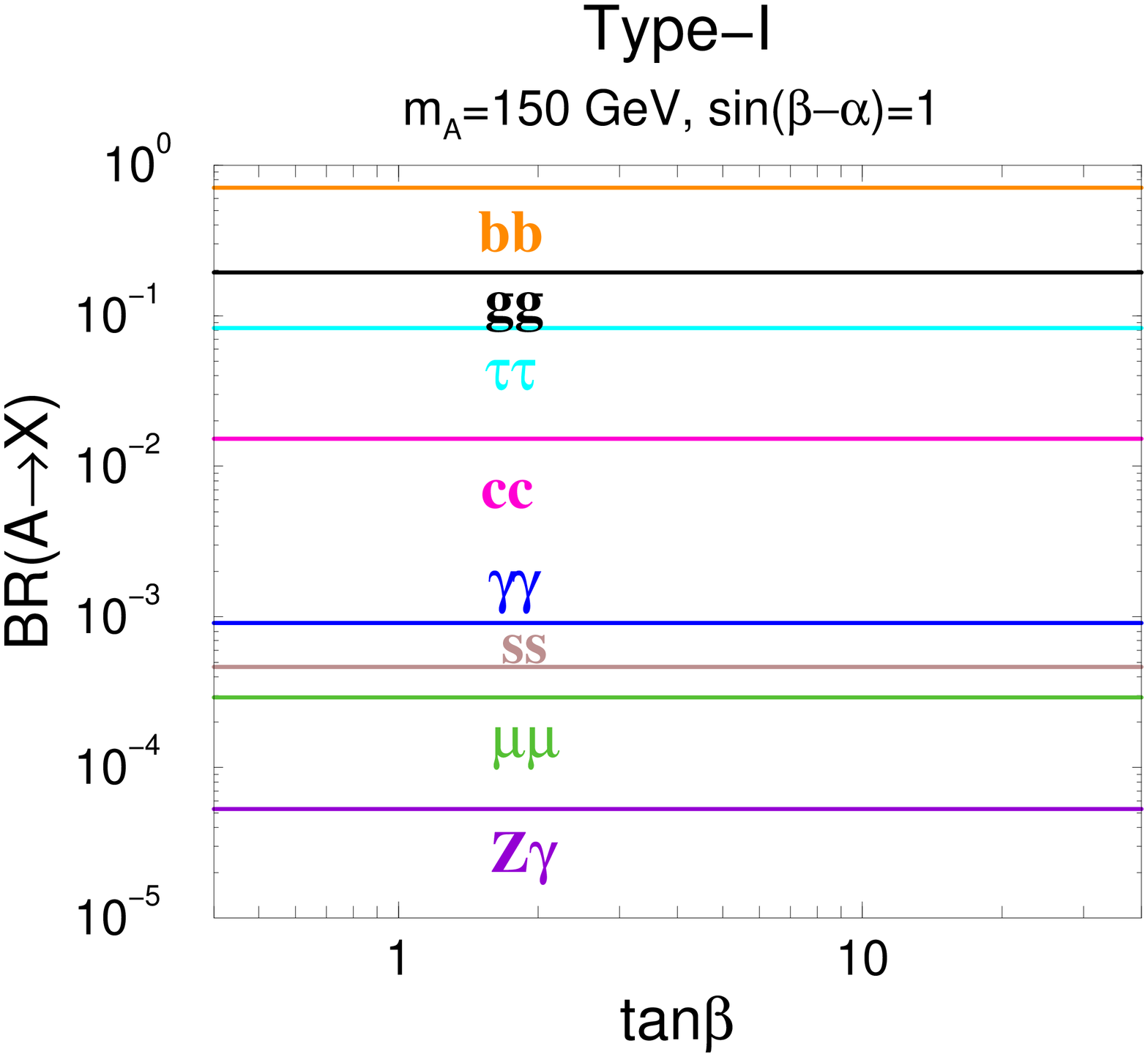}
\end{minipage}
\begin{minipage}{0.23\hsize}
\includegraphics[width=4.4cm,angle=-90]{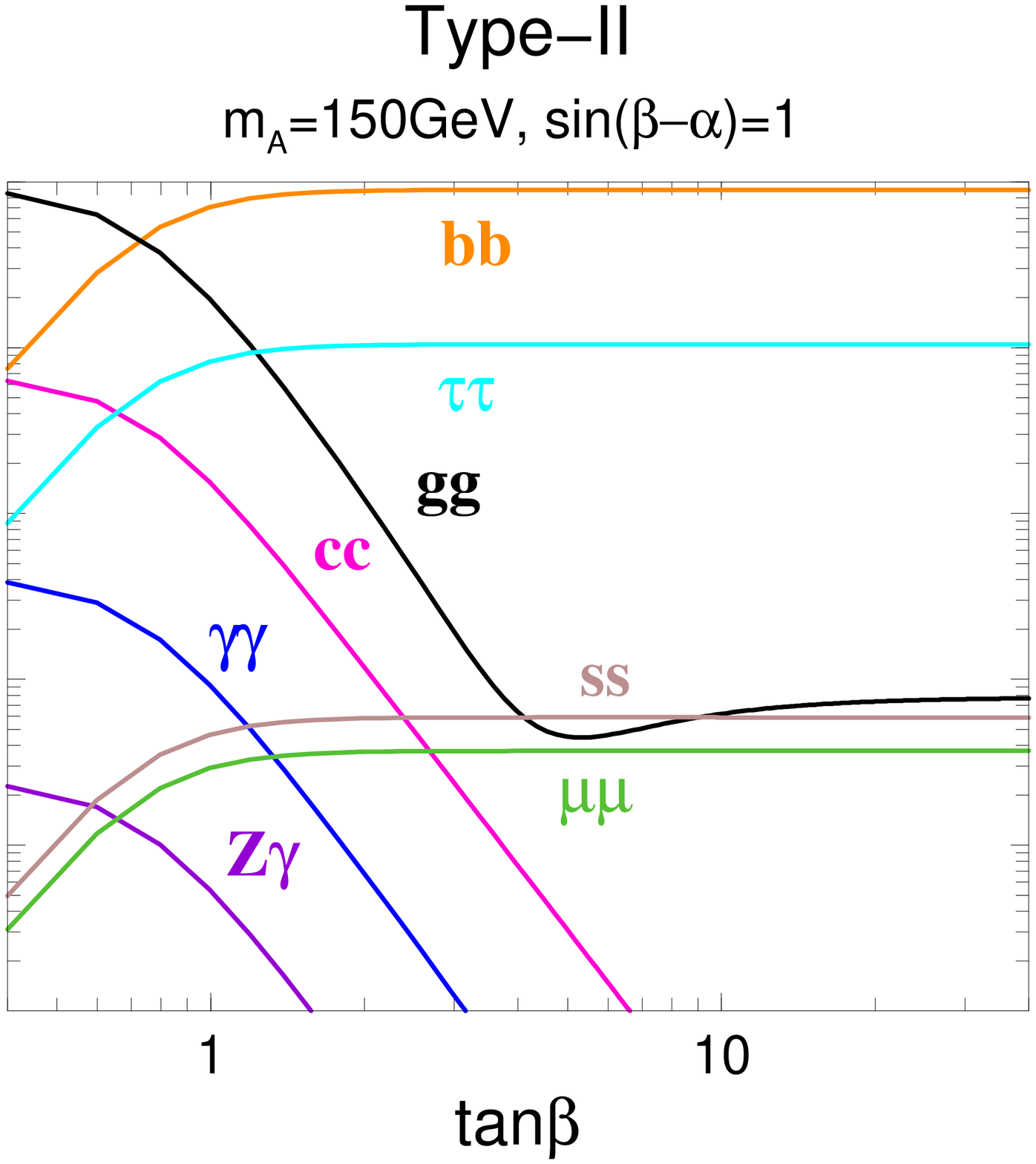}
\end{minipage}
\begin{minipage}{0.23\hsize}
\includegraphics[width=4.4cm,angle=-90]{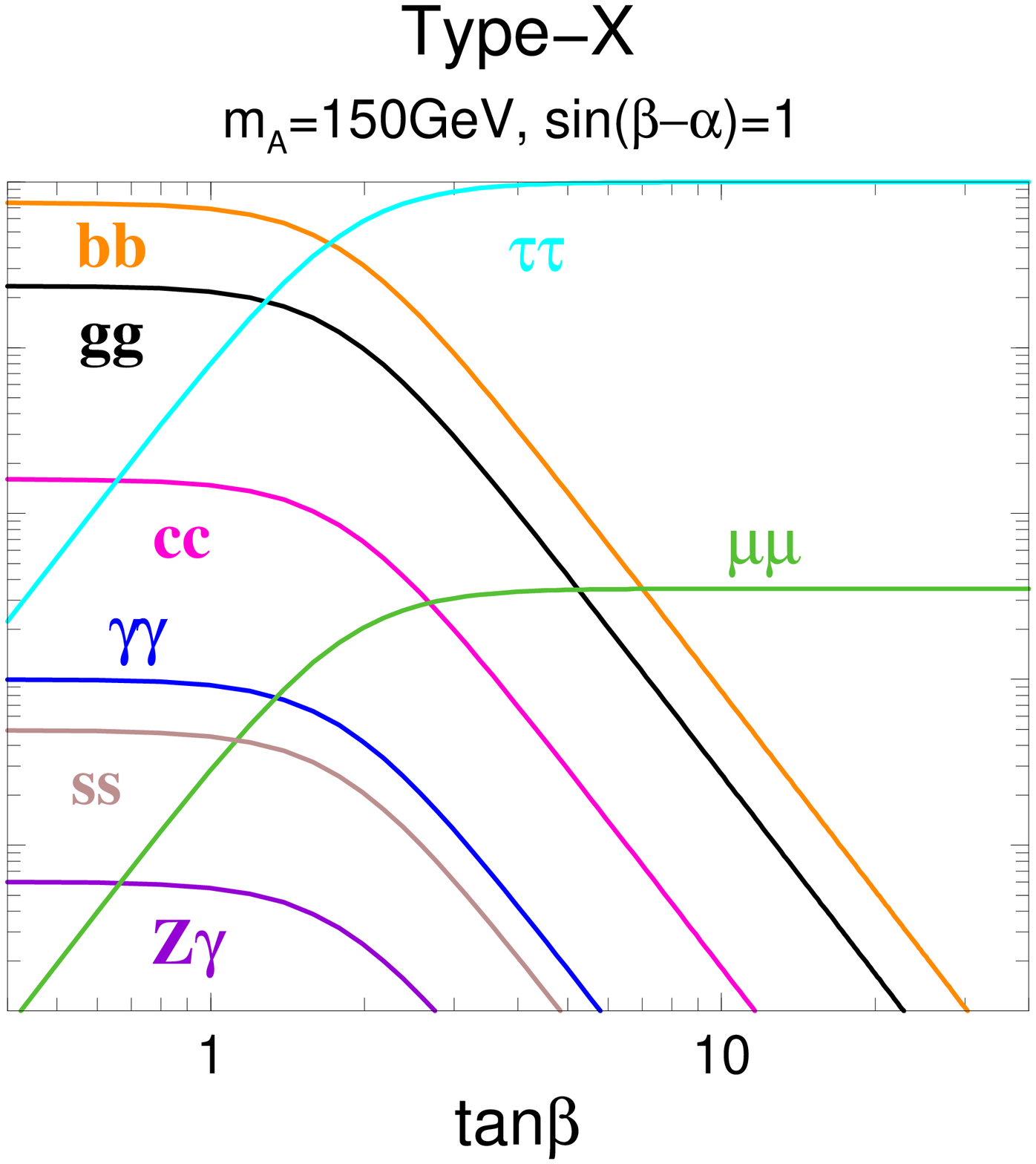}
\end{minipage}
\begin{minipage}{0.23\hsize}
\includegraphics[width=4.4cm,angle=-90]{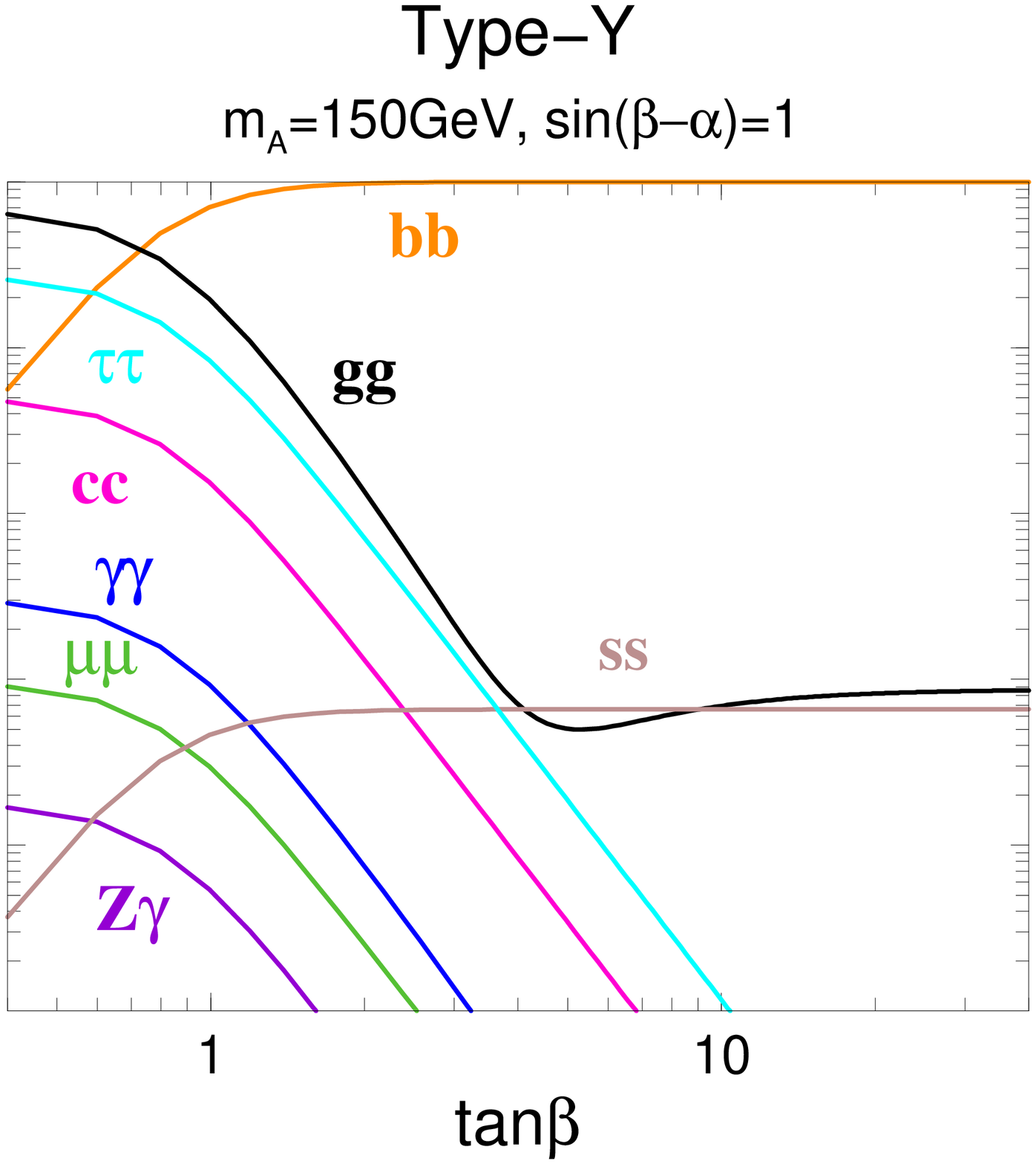}
\end{minipage}
\begin{minipage}{0.285\hsize}
\includegraphics[width=4.4cm,angle=-90]{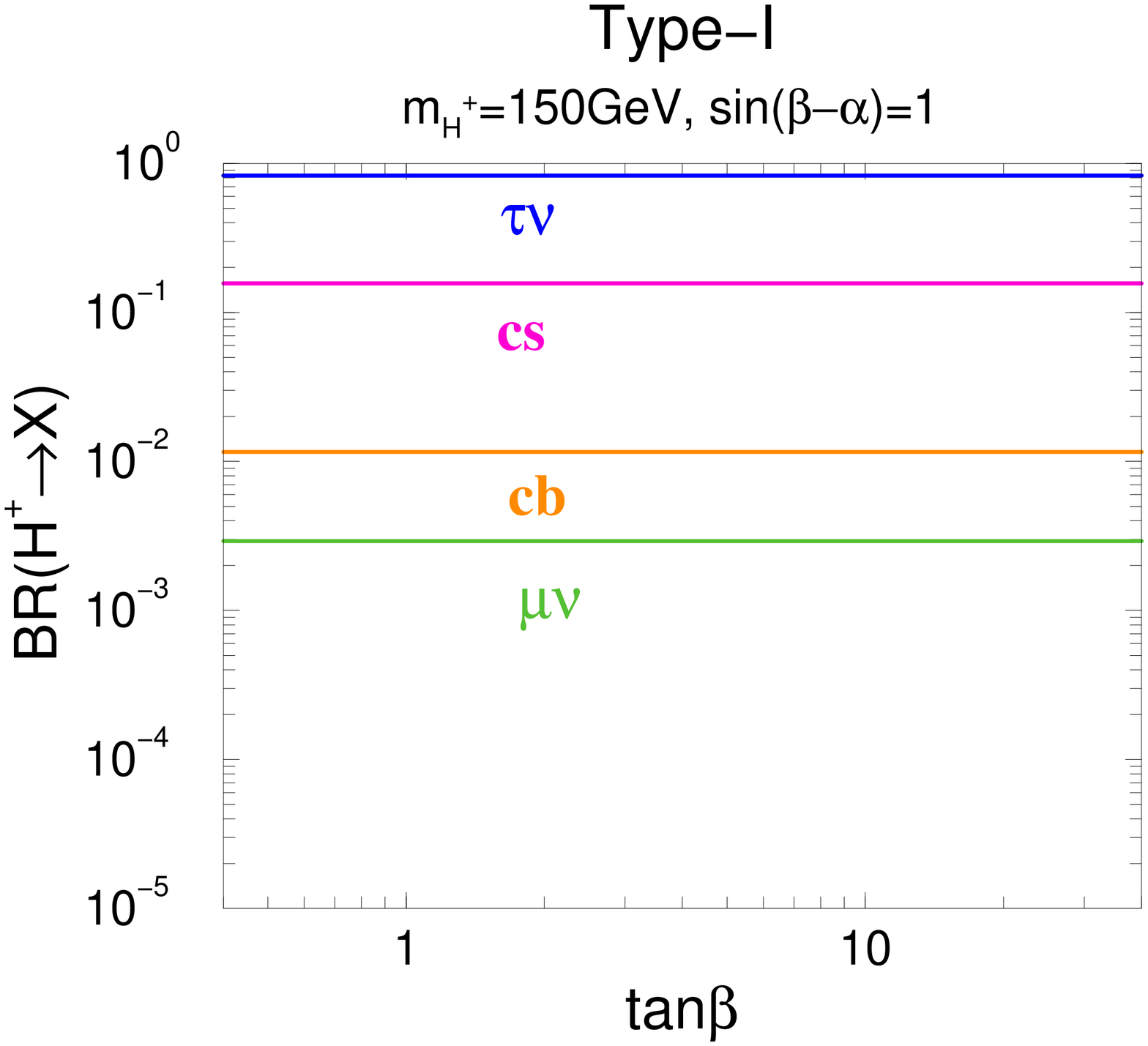}
\end{minipage}
\begin{minipage}{0.23\hsize}
\includegraphics[width=4.4cm,angle=-90]{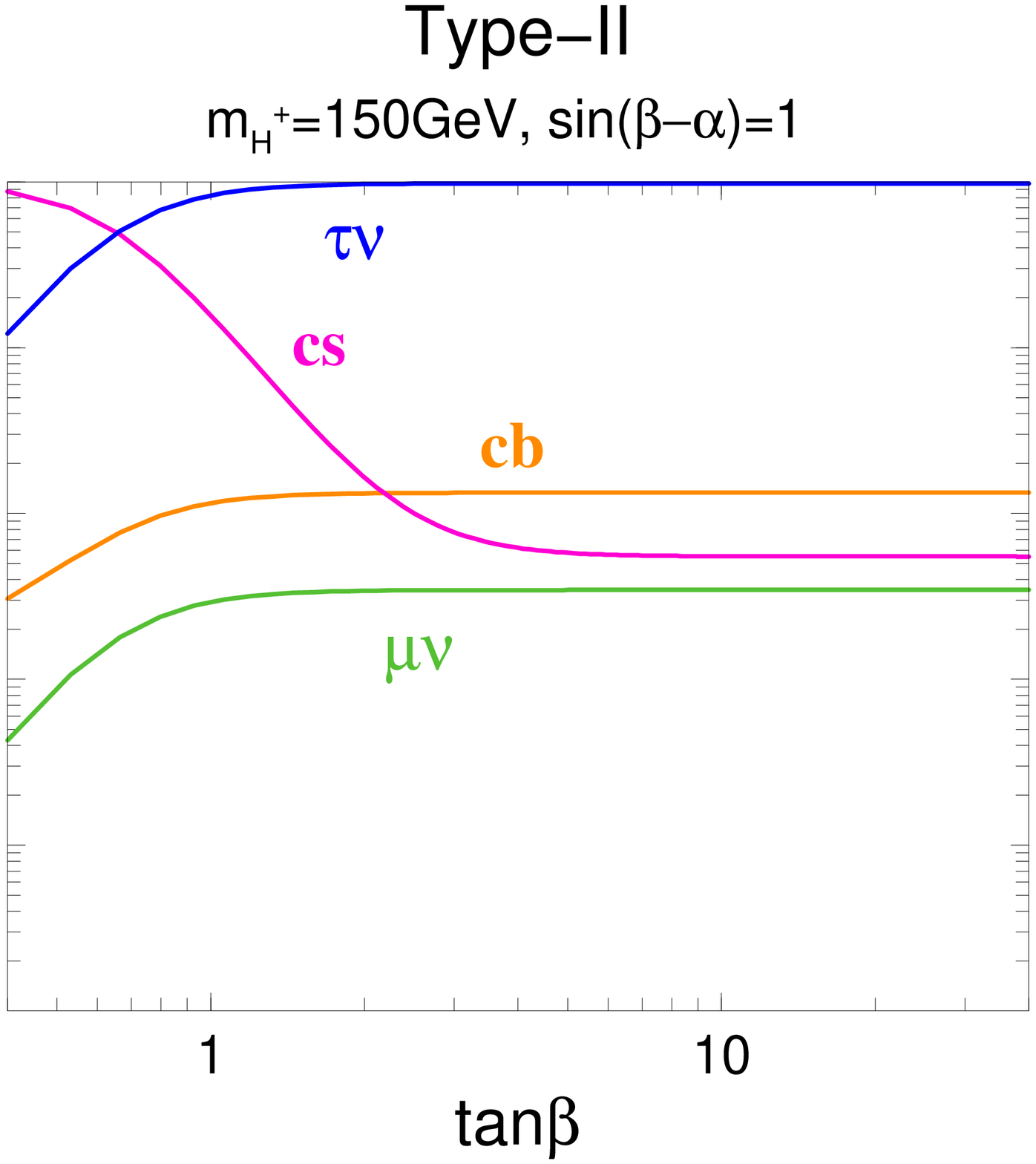}
\end{minipage}
\begin{minipage}{0.23\hsize}
\includegraphics[width=4.4cm,angle=-90]{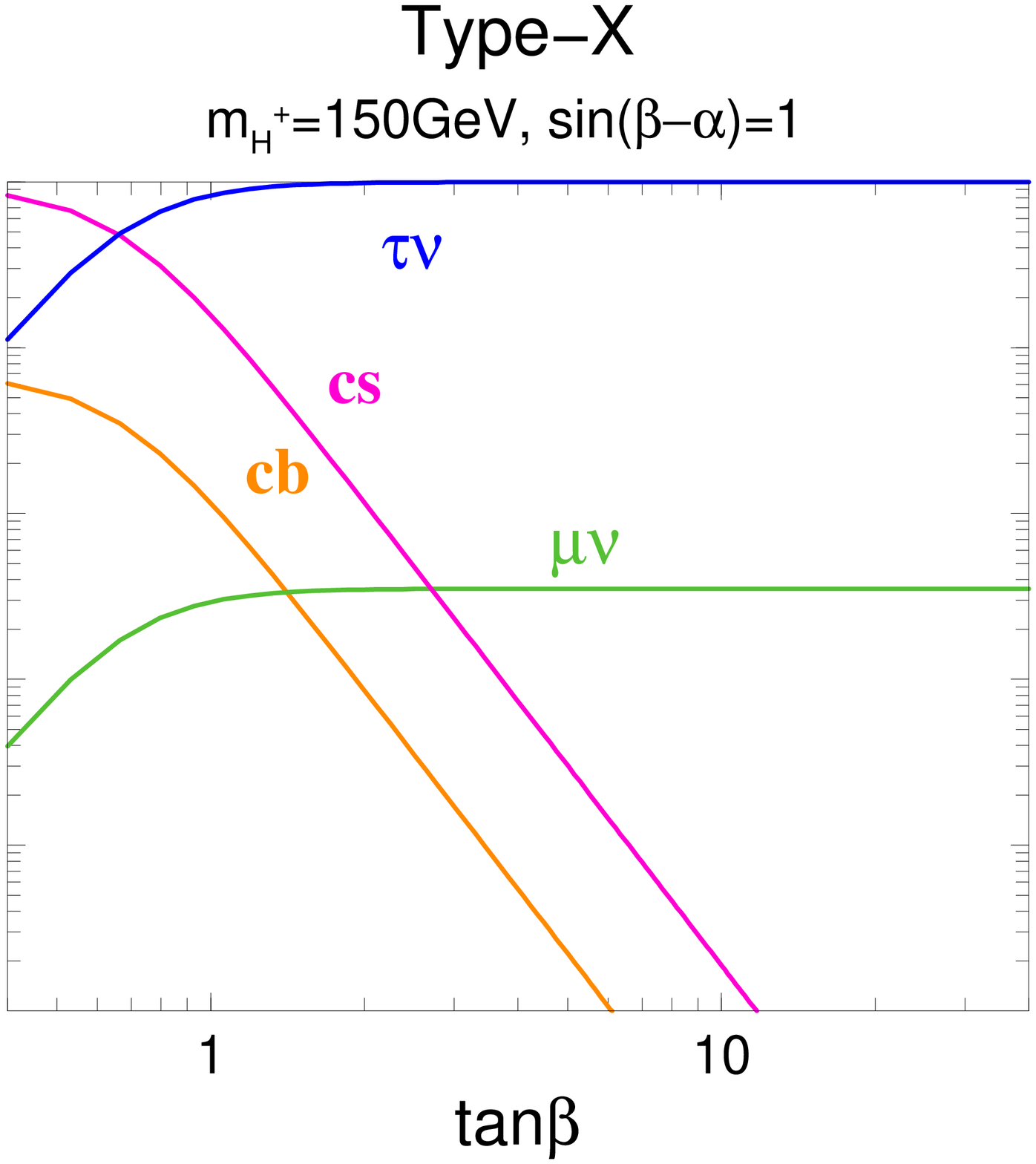}
\end{minipage}
\begin{minipage}{0.23\hsize}
\includegraphics[width=4.4cm,angle=-90]{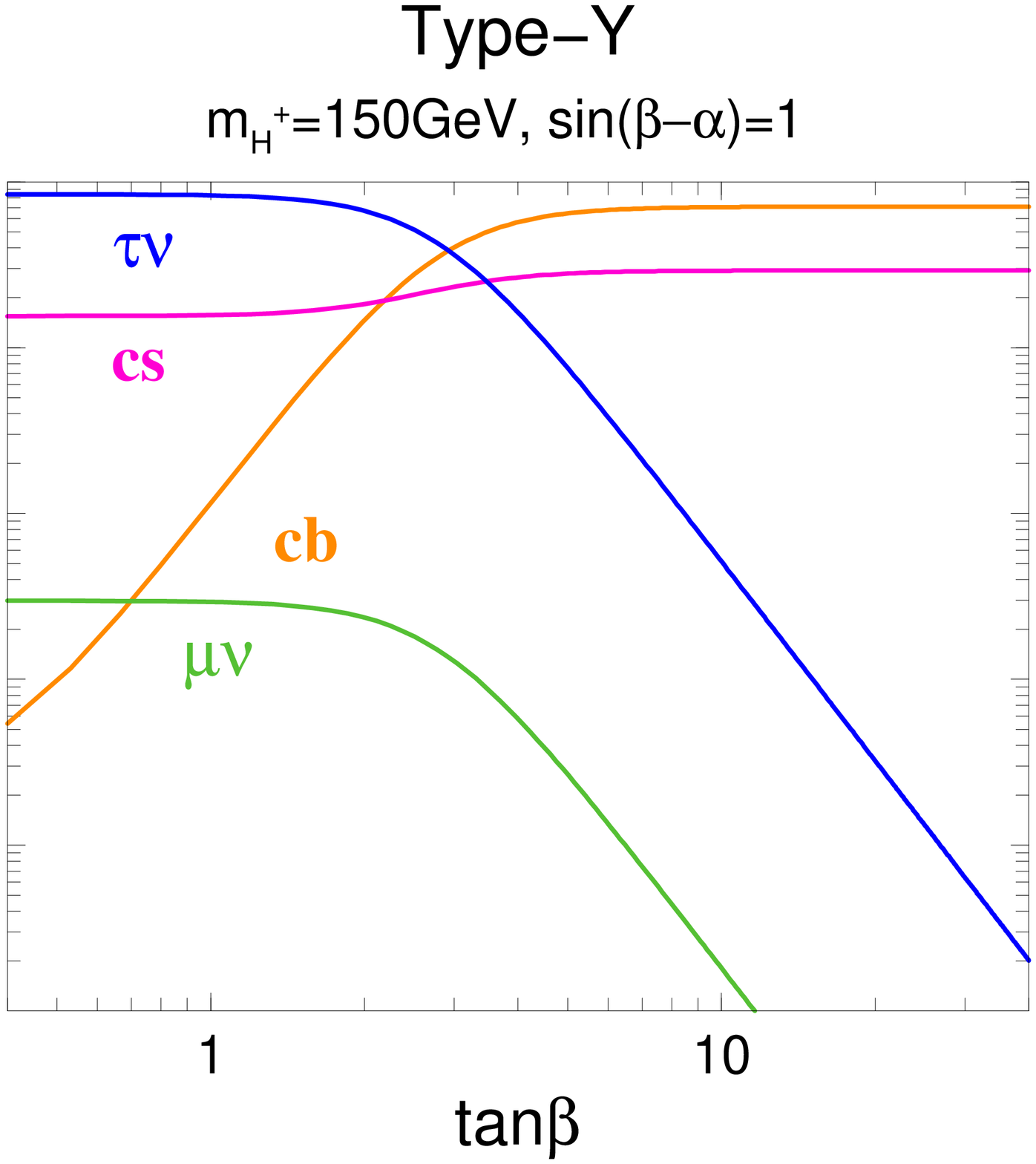}
\end{minipage}
\caption{Decay branching ratios of $H$, $A$ and $H^\pm$
in the four different types of THDM as a function of $\tan\beta$
for $m_H^{}=m_A^{}=m_{H^\pm}^{}=150$ GeV.
The SM-like limit $\sin(\beta-\alpha) =1$ is taken, where $h$
is the SM-like Higgs boson~\cite{typeX}.}
\label{FIG:br_150}
\end{center}
\end{figure}

In FIG.~\ref{FIG:br_150}, the decay branching ratios of $H$, $A$ and 
$H^\pm$ are shown for $m_\Phi^{}=150$ GeV and $\sin(\beta-\alpha)=1$
as a function of $\tan\beta$. 
In the type-I THDM, the decay of $H$ into a gauge boson pair 
$\gamma\gamma$ or $Z\gamma$ can increase for large $\tan\beta$ values, 
because all the other fermionic decays (including the $gg$ mode) 
are suppressed but the charged scalar loop contribution to $\gamma\gamma$ 
and $Z\gamma$ decay modes is not always suppressed 
for large $\tan\beta$. 
Such an enhancement of the bosonic decay modes cannot be seen 
in the decay of $A$ since there is no $AH^+H^-$ coupling. 
In the type-X THDM, the main decay mode of $H$ and $A$ is 
$\tau^+\tau^-$ for $\tan\beta \gtrsim 2$, and the leptonic decays 
$\tau^+\tau^-$ and $\mu^+\mu^-$ become almost $99\%$ and $0.35\%$ for 
$\tan\beta \gtrsim 10$, 
 while the $b\bar b$ (or $gg$) mode is always the main decay mode 
 in all other cases. 
In the type-Y THDM, the leptonic decay modes of $H$ and $A$ are rapidly 
suppressed for large $\tan\beta$ values, and only the branching ratios 
of $b\bar b$ and $gg$ modes are sizable for $\tan\beta \gtrsim 10$. 
In charged Higgs boson decays with $m_{H^\pm}^{}=150$ GeV, 
the decay into $\tau\nu$ is dominant in the type-I THDM, the type-II THDM and 
the type-X THDMs for $\tan\beta \gtrsim 1$, 
while hadronic decay modes can also be dominant in the type-Y THDM.


%
%
\subsection{Constraints from the current experimental data on THDMs}

One of the direct signal of the THDM is the discovery of extra Higgs bosons,
which have been searched at the LEP and Tevatron~\cite{LEP,Tevatron}.
Indirect contributions of Higgs bosons to precisely measurable observables
can also be used to constrain Higgs boson parameters.
In this section, we summarize these experimental bounds.

A direct mass bound is given from the LEP direct search results as
$m_{H^0}^{}\gtrsim92.8$ GeV for CP-even Higgs bosons and
$m_A^{}\gtrsim93.4$ GeV for CP-odd Higgs bosons in supersymmetric models.
The bound for charged Higgs boson has also been set as $m_{H^\pm}^{}\gtrsim79.3$ GeV~\cite{LEP}.

In the THDM, one-loop contributions of scalar loop diagrams
to the rho parameter are expressed as~\cite{HHG}
\begin{align}
\delta\rho_\text{THDM}^{}&= \frac{\sqrt{2}G_F}{16\pi^2}
\Big\{F_5(m_A,m_{H^\pm})-\cos^2(\alpha-\beta)\left[F_5(m_h^2,m_A^2)-F_5(m_h,m_{H^\pm})\right]\notag\\
&-\sin^2(\alpha-\beta)\left[F_5(m_H,m_A)-F_5(m_H,m_{H^\pm})\right]\Big\},
\end{align}
where $F_5(x,y)=\frac{1}{2}(m_1^2+m_2^2)-\frac{m_1^2m_2^2}{m_1^2-m_2^2}\ln\frac{m_1^2}{m_2^2}$ with
$F_5(m,m)=0$\footnote{
There are other (relatively smaller in most of the parameter space)
contributions to the rho parameter in the THDM, i.e.,
those from the diagrams where the gauge boson (as well as HG 
boson) and the Higgs boson are running together in the
loop\cite{HHG}. We have included these effect in our numerical analysis.}.
These quadratic mass contributions can cancel out when
Higgs boson masses satisfy the following relation: (i) $m_A^{}\simeq m_{H^\pm}^{}$,
(ii) $m_H^{}\simeq m_{H^\pm}^{}$ with $\sin(\beta-\alpha)\simeq1$,
and (iii) $m_h^{}\simeq m_{H^\pm}^{}$ with $\sin(\beta-\alpha)\simeq0$.
These relations correspond to the custodial symmetry
invariance~\cite{Csym,Csym2}.
This constraint is independent of the type of Yukawa interaction.


\begin{figure}[tb]
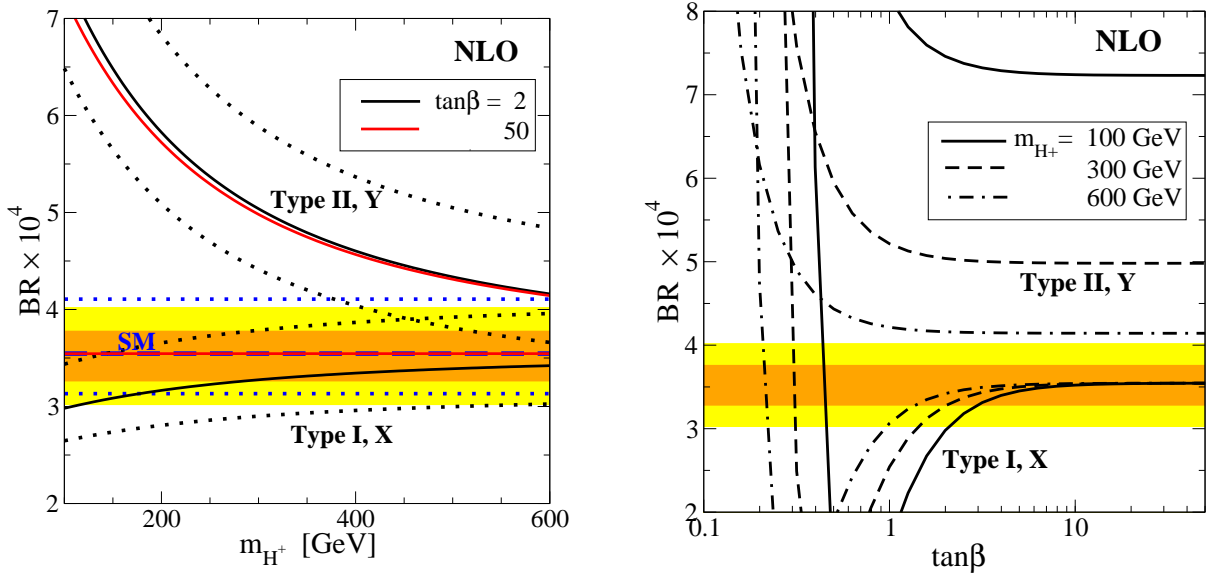

\begin{minipage}{0.49\hsize}
\includegraphics[width=7.5cm,angle=0]{BR_mh.eps}
\end{minipage}
\begin{minipage}{0.49\hsize}
\includegraphics[width=7.5cm,angle=0]{BR_tanb.eps}
\end{minipage}
\caption{Predictions of the decay branching ratio for $b\to s\gamma$
are shown at the NLO approximation as a function of $m_{H^\pm}^{}$ and $\tan\beta$.
The dark (light) shaded band represents $1\sigma$ $(2\sigma)$ allowed region
of current experimental data. In the left panel, solid (dashed) curves denote
the prediction for $\tan\beta=2$ $(50)$ in various THDMs. In the right panel, solid,
dashed and dot-dashed curves are those for $m_{H^\pm}^{}=100, 300$ and $600$ GeV,
respectively~\cite{typeX}.}
\label{FIG:bsg}
\end{figure}

It has been known that the charged Higgs boson mass in the type-II THDM
is stringently constrained by the precision measurements of the
radiative decay of $b\to s\gamma$ at Belle~\cite{Ref:BELLE} and
BABAR~\cite{Ref:BABAR} as well as CLEO~\cite{Ref:CLEO}.
The process $b\to s\gamma$ receives contributions from the $W$ boson loop
and the charged Higgs boson loop in the THDM.
A notable point is that these two contributions
always work constructively in the type-II (type-Y) THDM,
while this is not the case in the type-I (type-X) THDM~\cite{Barger}.
In FIG.~\ref{FIG:bsg}, we show the branching ratio of $B\to X_s\gamma$ for
each type of THDM as a function of $m_{H^\pm}^{}$
(left-panel) and $\tan\beta$ (right-panel), which are evaluated at the next-to-leading 
order (NLO) following the formulas in Ref.~\cite{bsgNLO}.
The SM prediction at the NLO is also shown for comparison. 
The theoretical uncertainty is about $15\%$\footnote{In Ref.~\cite{bsgNLO}, 
the theoretical uncertainty is smaller because the value for the error
in $m_c^\text{pole}/m_b^\text{pole}$ is taken to be $7\%$, which gives main 
uncertainty in the branching ratio.} 
in the branching ratio ( as indicated by dotted curves in FIG.~\ref{FIG:bsg}), 
which mainly comes from the pole mass of charm quark $m_c^\text{pole}=1.65\pm 0.18$ 
GeV ~\cite{PDG}.
The experimental bounds of the branching ratio are also indicated, where
the current world average value is given by 
${\mathcal B}(B\to X_s\gamma)=(3.52\pm 0.23\pm0.09)\times 10^{-4}$~\cite{Ref:HFAG}.
It is seen in FIG.~\ref{FIG:bsg} that the branching ratio in
the type-I (type-X) THDM lies within the 2 $\sigma$ experimental error 
in all the regions of $m_{H^{\pm}}$ indicated for $\tan\beta\gtrsim 2$, 
while that in the type-II (type-Y) THDM is far from the value indicated by 
the data for a light charged Higgs boson region $(m_{H^\pm}^{}\lesssim 200$ 
GeV$)$.
In the right figure, a cancellation occurs in the type-I (type-X) THDM
since there are destructive interferences between the $W$ boson and
the $H^\pm$ contributions.
It is emphasized that the charged Higgs boson could be light
in the type-I (type-X) THDM under the constraint from $B\to X_s\gamma$ results.
We note that in the MSSM the chargino contribution can compensate
the charged Higgs boson contribution~\cite{Ref:bsgMSSM}.
This cancellation weakens the limit on $m_{H^\pm}^{}$ from $b\to s\gamma$ 
in the type-II THDM,
and allows a light charged Higgs boson
as in the type-I (type-X) THDM.

We give some comments on the NNLO analysis, 
although it is basically out of the scope of this paper.
At the NNLO, the branching ratio for $b\to s\gamma$ 
has been evaluated in the SM in
Ref.~\cite{bsgNNLO1,bsgNNLO2}.
The predicted value at the NNLO is less than that at the NLO approximation 
in a wide range of renormalization scale. 
In Ref.~\cite{bsgNNLO1}, the SM branching ratio is $(3.15\pm 0.23)\times
10^{-4}$, and the lower bound of the $m_{H^\pm}^{}$, after adding the
NLO charged Higgs contribution, is estimated as 
$m_{H^\pm}^{} \gtrsim 295$ GeV ($95\%$ CL) in the type-II (type-Y)
THDM~\cite{bsgNNLO1}\footnote{In Ref.~\cite{bsgNNLO2} 
the NNLO branching ratio in the SM is calculated as $(2.98\pm 0.26)\times
10^{-4}$, and the mass bound is a little bit relaxed.}.
On the other hand, in the type-I (type-X) THDM, although the branching ratio 
becomes smaller as compared to the NLO evaluation, no serious bound on 
$m_{H^\pm}^{}$ can be found for $\tan\beta \gtrsim 2$. 
Therefore, charged Higgs boson mass is not expected to be strongly constrained in 
the type-I (type-X) THDM even at the NNLO, and our main conclusion that 
the type-I (type-X) THDM 
is favored for $m_{H^\pm}^{}\lesssim 200$ GeV based on the NLO
analysis should not be changed.

\begin{figure}[tb]
\begin{minipage}{0.49\hsize}
\includegraphics[width=7.5cm]{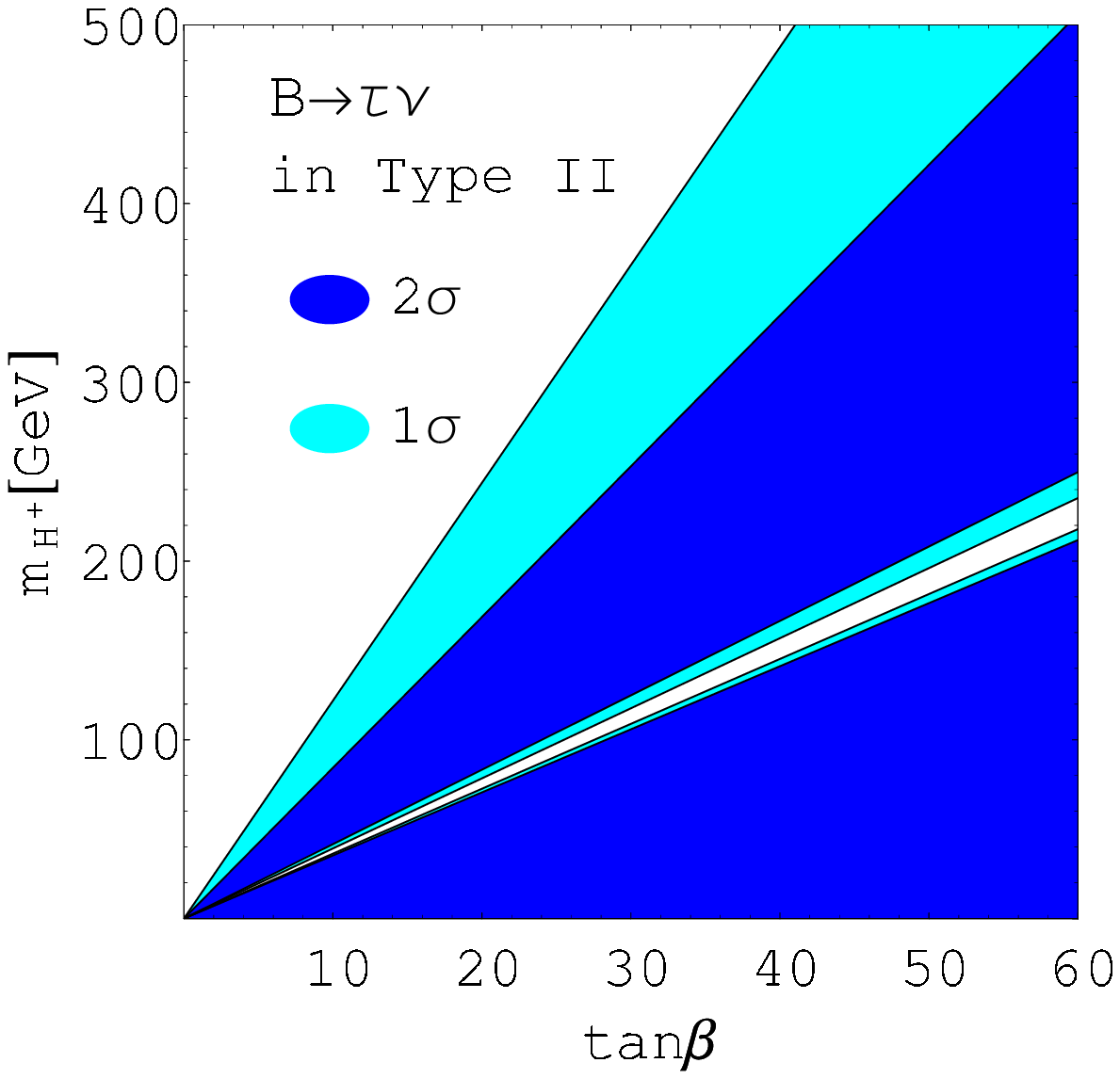}
\end{minipage}
\begin{minipage}{0.49\hsize}
\includegraphics[width=7.5cm]{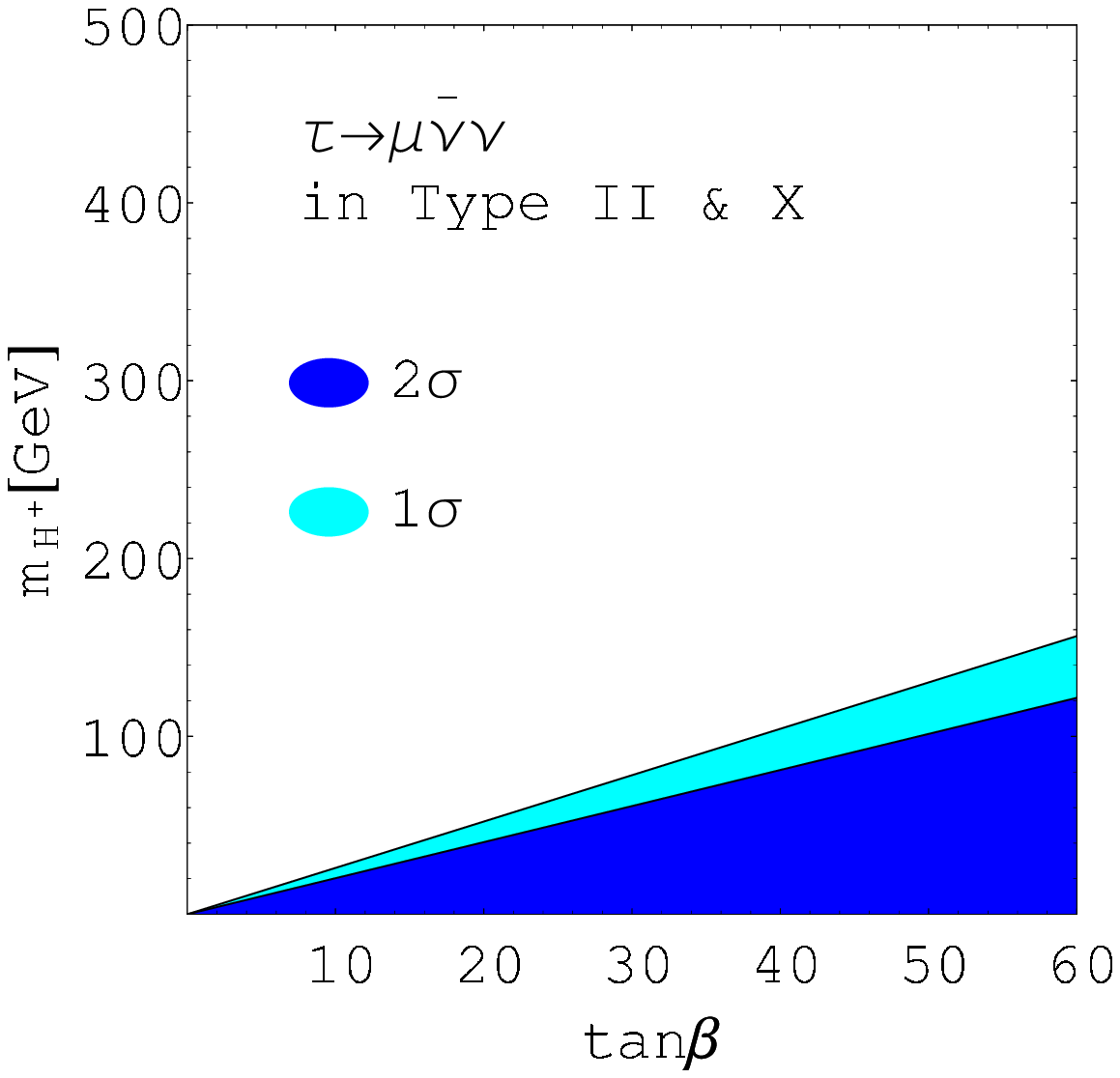}
\end{minipage}
\caption{Bounds from $B\to\tau\nu$ (left panel) and tau leptonic decay (right panel) on $m_{H^\pm}^{}$ as a function of $\tan\beta$ are shown. The dark (light) shaded region
corresponds to the $2\sigma$ $(1\sigma)$ exclusion of these experimental results.
In the type-II THDM the wide parameter space is constrained by $B\to\tau\nu$,
while only the tau leptonic decays are important for the type-X THDM~\cite{typeX}.}
\label{FIG:mH+tanb}
\end{figure}

The decay $B\to\tau\nu$ has been discussed in the type-II THDM~\cite{Btaunu_MS,Btaunu_Du}.
The data for ${\mathcal B}(B^+\to\tau^+\nu_\tau)=(1.4\pm0.4)\times 10^{-4}$
are obtained at the $B$ factories~\cite{PDG,Ref:BtaunuExp}.
The decay branching ratio can be written as~\cite{Btaunu_MS,Ref:IP}
\begin{align}
\frac{{\mathcal B}(B^+\to\tau^+\nu_\tau)_\text{THDM}}{{\mathcal B}(B^+\to\tau^+\nu_\tau)_\text{SM}}
\simeq\left(1-\frac{m_B^2}{m_{H^\pm}^2}\xi_A^d\xi_A^\ell\right)^2.
\end{align}
In FIG.~\ref{FIG:mH+tanb}, the allowed region from the $B\to\tau\nu$ results
is shown in the type-II THDM. The dark (light) shaded region denotes
the $2\sigma$ $(1\sigma)$ exclusion from the $B\to\tau\nu$ measurements.
The process is important only in the type-II THDM with large $\tan\beta$ values.
The other types of Yukawa interactions do not receive constraints form this process.

The rate for the leptonic decay of the tau lepton $\tau\to\mu\overline{\nu}\nu$
can be deviated from the SM value by the presence of a light charged Higgs boson~\cite{Ref:TauMuNuNu}.
The partial decay rate is approximately expressed as~\cite{Btaunu_AR,Btaunu_MS}
\begin{align}
\frac{\Gamma_{\tau\to\mu\overline{\nu}\nu}^\text{THDM}}{\Gamma_{\tau\to\mu\overline{\nu}\nu}^\text{SM}}
&\simeq 1-\frac{2m_\mu^2}{m_{H^\pm}^2}{\xi_A^\ell}^2 \kappa\left(\frac{m_\mu^2}{m_\tau^2}\right)+\frac{m_\mu^2m_\tau^2}{4m_{H^\pm}^4}{\xi_A^\ell}^4,
\end{align}
where $\kappa(x)=g(x)/f(x)$ is defined by
$f(x)=1-8x-12x^2\ln x+8x^3-x^4,$ and $g(x)=1+9x-9x^2-x^3+6x(1+x)\ln x$.
In the type-II (type-X) THDM, the leptonic Yukawa interaction
can be enhanced in the large $\tan\beta$ region. Hence, both the models are weakly
constrained by tau decay data, as in FIG.~\ref{FIG:mH+tanb}.

The precision measurement for the muon anomalous magnetic moment
can give mass bound on the Higgs boson in the SM~\cite{Ref:g-2MuSM}.
This constraint can be applied for more
general interaction, including THDMs~\cite{Ref:g-2MuTHDM}.
At the one-loop level, the contribution is given by
\begin{align}
\delta a_\mu^{1-\text{loop}}
&\simeq \frac{G_Fm_\mu^4}{4\pi^2\sqrt2}
\left[\sum_{\phi^0=h,H}\frac{{\xi_{\phi^0}^\ell}^2}{m_{\phi^0}^2}
\left(\ln\frac{m_{\phi^0}^2}{m_\mu^2}-\frac76\right)
+\frac{{\xi_A^\ell}^2}{m_A^2}
\left(-\ln\frac{m_A^2}{m_\mu^2}+\frac{11}6\right)
-\frac{{\xi_A^\ell}^2}{6m_{H^\pm}^2}\right].
\end{align}
This process is also purely leptonic and only gives milder bounds
on the Higgs boson masses for very large $\tan\beta$ values
in the type-II (type-X) THDM. It gives no effective bound on the type-I (type-Y) THDM. 
It is also known that the two-loop (Barr-Zee type) diagram  
can significantly affect $a_\mu$~\cite{Ref:BZ,Ref:King}. 
The contribution can be large because of the enhancement factors 
of $m_f^2/m_\mu^2$ and also of the mixing factors $\xi_\phi^f$ as
~\cite{Ref:King}
\begin{align}
\delta a_\mu^\text{BZ}
&\simeq \frac{N_c^fQ_f^2G_F\alpha m_\mu^2}{4\pi^3v^2}
\left[-\sum_{\phi^0=h,H} 
\xi_{\phi^0}^\ell\xi_{\phi^0}^ff\left(\frac{m_f^2}{m_{\phi^0}^2}\right) 
+\xi_A^\ell\xi_A^fg\left(\frac{m_f^2}{m_A^2}\right) \right],
\end{align}
where 
\begin{align}
f(z)&=\frac{z}2\int_0^1dx\frac{1-2x(1-x)}{x(1-x)-z}\ln\frac{x(1-x)}{z},\\
g(z)&=\frac{z}2\int_0^1dx\frac1{x(1-x)-z}\ln\frac{x(1-x)}{z}.
\end{align}
The contribution from this kind of diagram is only important 
for large $\tan\beta$ values with smaller Higgs boson masses in the type-II THDM. 
For the other types of THDM, it would give a much less effective bound 
on the parameter space.

\subsection{Collider signals in the Type-X THDM
at the LHC and the ILC
}

We discuss the collider phenomenology of the models at the LHC and the ILC.
There have already been many studies on the production and decays
of the Higgs bosons in the type-II THDM, especially in the context of
the MSSM, while the phenomenology of the other types of THDMs has not
yet been studied sufficiently.
Recently, the type-X THDM has been introduced in the model to explain
phenomena such as neutrino masses, dark matter, and baryogenesis at the
TeV scale~\cite{aks_prl}.
We therefore concentrate on the collider signals in the type-X THDM, and
discuss how we can distinguish the model from the type-II THDM (the
MSSM), mainly in scenarios with a light charged Higgs boson
($100$ GeV $\lesssim m_{H^\pm}^{} \lesssim 300$ GeV).
(Such a light charged Higgs boson is severely constrained by the $b\to s
\gamma$ result in the non-supersymmetric type-II THDM and the type-Y THDM,
while it can be allowed in the MSSM and the type-X (type-I) THDM.)
As we are interested in the differences in the types of the Yukawa
interactions,
we focus here on the case of $m_{H^\pm}^{} \simeq m_A^{} \simeq m_H^{}$ with
$\sin(\beta-\alpha) \simeq 1$ for definiteness.
\\\\
{\bf Charged Higgs boson searches at the LHC}
\\\\
\begin{figure}[t]
\begin{center}
\includegraphics[width=70mm]{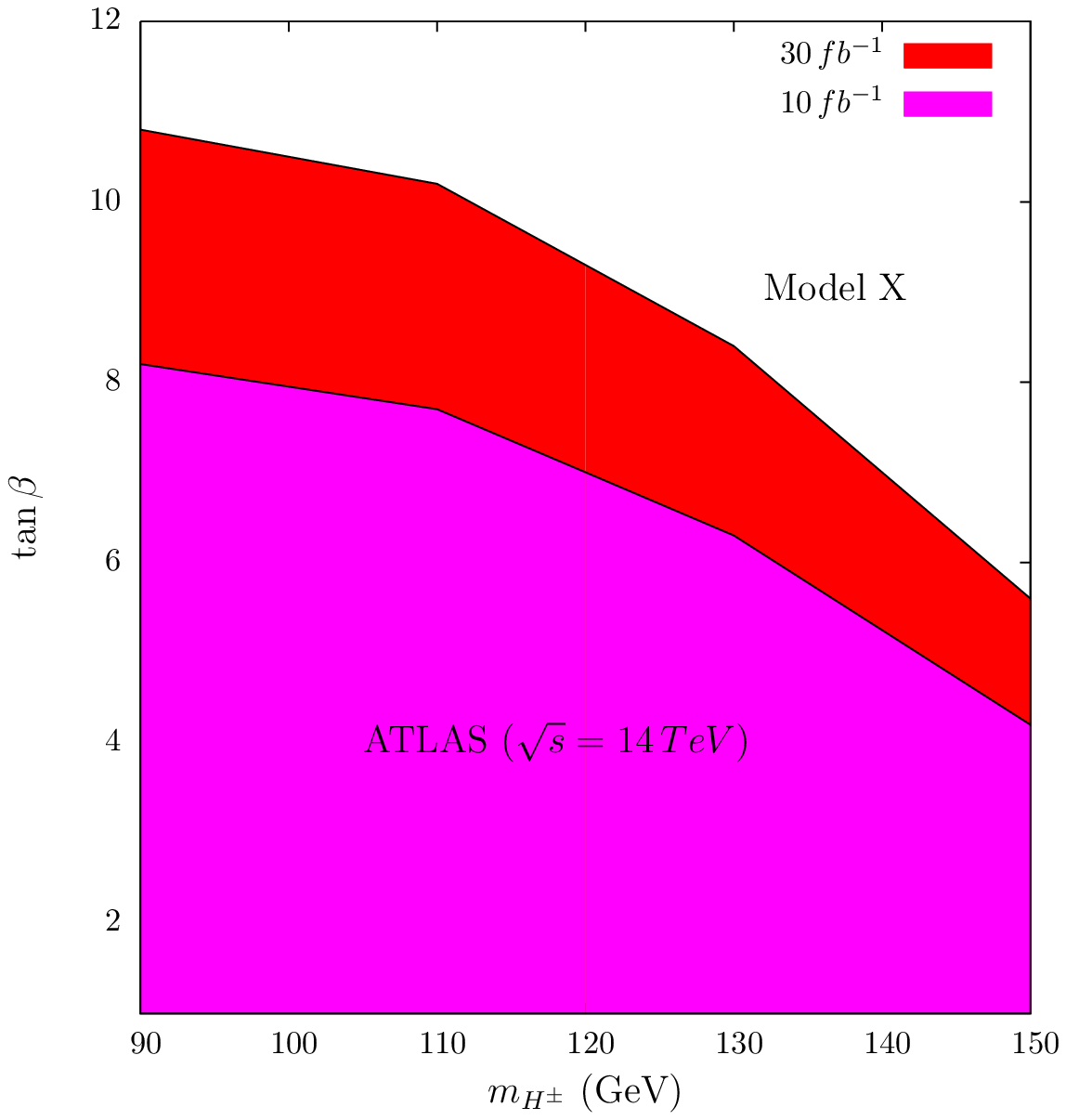}\hspace{5mm}
\includegraphics[width=70mm]{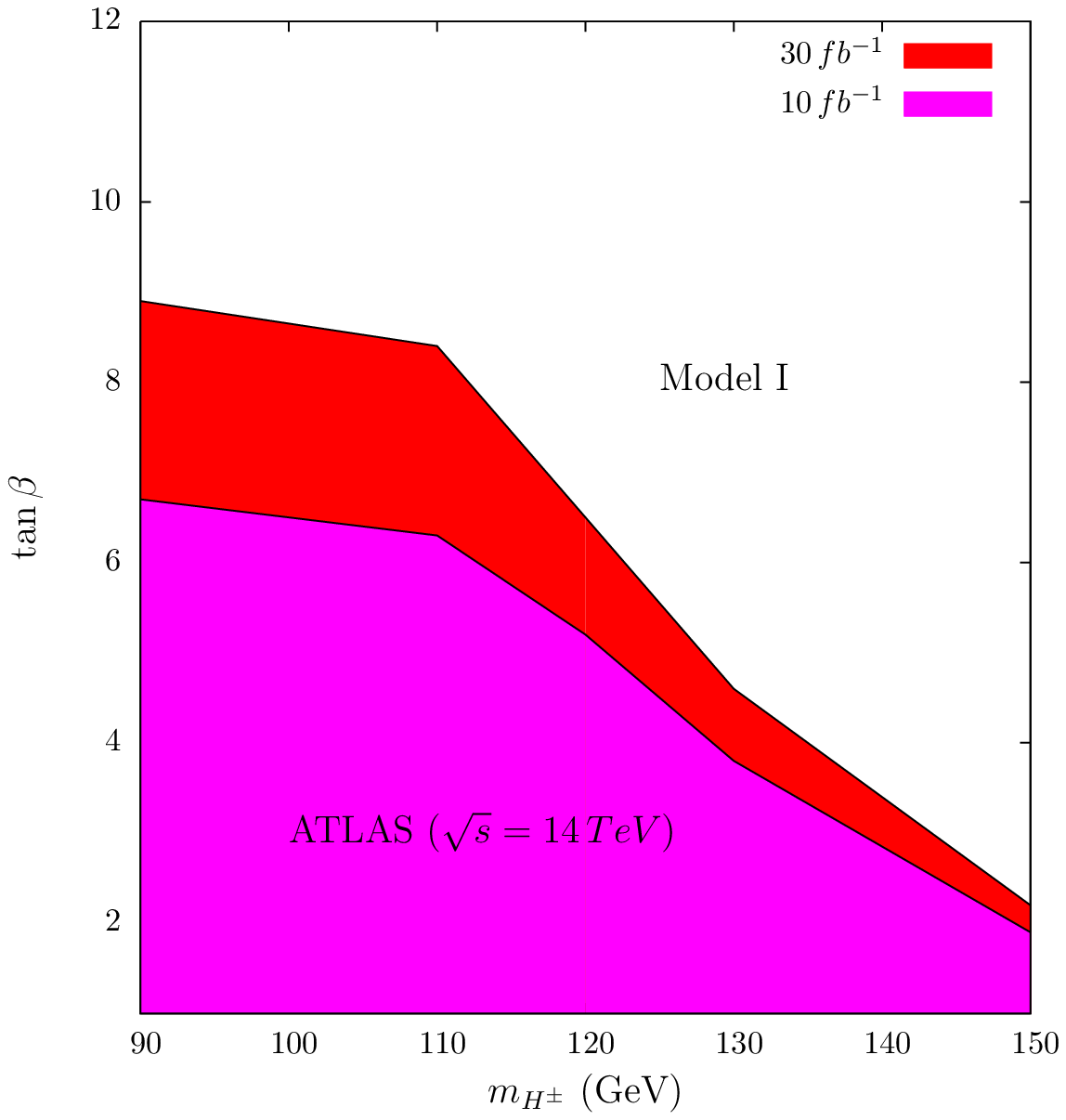}
\caption{Region of the parameter space excluded by the ATLAS collaboration for 
$\sqrt{s} = 14$ TeV in the type-X THDM (left) and the type-I THDM (right) using the channel 
$pp \to t \bar{t} \to H^{\pm} b W^{\pm} \bar{b} \to \tau \nu b \bar{b} q \bar{q}$~\cite{RUI}. The plot
was done using the data in~\cite{ATLAS}.}
\end{center}
\label{fig:ATLAS}
\end{figure}
A light charged Higgs boson with $m_{H^\pm}^{} \lesssim m_t -m_b$ can be
produced in the decay of top quarks at the LHC.
The discovery potential for the charged Higgs boson via
the $t\overline{t}$ production has been studied in the MSSM~\cite{mH+TeVatron}.
Assuming an integrated luminosity of $30$ fb${}^{-1}$, the expected signal
significance of the event $t\overline{t}\to H^\pm W^\mp b\overline{b}\to
\ell\nu\tau\nu_\tau b\overline{b}$ is greater than $5 \sigma$ for
$m_{H^\pm}^{}\lesssim 130$ GeV for $\tan\beta\lesssim2$
and $\tan\beta\gtrsim 20$~\cite{mH+TeVatron}.
The same analysis can also be applied for the type-X THDM as well as the type-I THDM, in which
a similar number of $H^\pm$ can be produced when $\tan\beta\sim{\cal O}(1)$.
The main decay mode ($\tau\nu$) is common in the type-II THDM, the type-X THDM and the type-X THDM
, except for very low $\tan\beta$ values.
In Fig.~\ref{fig:ATLAS}, 
we present the region of the parameter space that can be excluded at 
95\% CL in the type-X THDM (left panel) and the type-I THDM (right panel) at the LHC~\cite{RUI} after collecting 10 and 30 
fb$^{-1}$ of integrated luminosity using the results from the ATLAS collaboration~\cite{ATLAS}.

For $m_{H^\pm}^{}\gtrsim m_t$, charged Higgs bosons can be produced in
$q\bar q/gg\to t\overline{b}H^-$, $gb\to tH^-$~\cite{Ref:tbH,Ref:ttH},
$gg$ $(q\bar q)\to H^+H^-$~\cite{Ref:H+H-,Ref:GGH2}
and $gg$ $(b\bar b)\to H^\pm W^\mp$~\cite{Ref:WH}.
These processes, except for the $H^+H^-$ production, 
are via the Yukawa coupling of $t \bar b H^-$, so
that the cross sections are significant for $\tan\beta\sim{\mathcal O}(1)$
or $\tan\beta \gtrsim 10$--$20$ in the type-II THDM
and only for $\tan\beta\sim{\mathcal O}(1)$ in the type-X THDM.
The type of Yukawa interaction in the THDM
can then be discriminated by measuring the difference in decay
branching ratios of $H^\pm$.
In the type-II THDM $H^\pm$ mainly decay into $tb$, while
$\tau\nu$ is dominant for $\tan\beta \gtrsim 10$ in the type-X THDM.
\\\\
{\bf Neutral Higgs boson $(A$ and $H)$ production at the LHC}
\\\\
At the LHC, the type of the Yukawa interaction may be determined
in the search for neutral Higgs bosons through the direct production
via gluon fusion $gg\to A/H$~\cite{Ref:GGH2,Ref:GGH},
vector boson fusion $V^\ast V^\ast \to H$, $V=W,Z$~\cite{Ref:VBF1,Ref:VBF2},
and also via associated production $pp \to b\bar b A$
$(b\bar b H)$~\cite{Ref:bbH,Ref:QQH}.
The production process $pp \to t\bar t A$ $(t\bar t H)$~\cite{Ref:ttH,Ref:QQH,Ref:ttH2}
can also be useful for $\tan\beta\sim 1$.
We discuss the possibility of discriminating the type of the Yukawa
interaction by using the production and decay processes of $A$ and $H$
for $\sin(\beta-\alpha)=1$.
Additional neutral Higgs bosons $A$ and $H$ are directly produced
by the gluon fusion mechanism at the one-loop level.
When $\sin(\beta-\alpha)\simeq 1$, the production rate can be significant
due to the top quark loop contribution for $\tan\beta \sim 1$
and, in the MSSM (the type-II THDM), also
for large $\tan\beta$ via the bottom quark loop contributions~\cite{Ref:GGH2}.
Notice that there is no rate of $V^\ast V^\ast \to A$ because there is no 
$VVA$ coupling, and that the production of $H$ from the vector boson
fusion $V^\ast V^\ast \to H$ is relatively unimportant when
$\sin(\beta-\alpha)\simeq 1$.
The associate production process $pp\to b\overline{b}A$
$(b\overline{b}H)$ can be significant for large
$\tan\beta$ values in the MSSM (the type-II THDM)~\cite{Ref:bbH}.

In the MSSM (the type-II THDM),
the produced $A$ and $H$ in these processes decay mainly into $b\bar b$
when $\sin(\beta-\alpha) \simeq 1$,
which would be challenging to detect because of huge QCD backgrounds.
Instead, the decays into a lepton pair $\tau^+\tau^-$ ($\mu^+\mu^-$)
would be useful for searches of $A$ (and $H$).
However, the decay branching ratios of $A \to
\tau^+\tau^-$ ($\mu^+\mu^-$) are less than $0.1$ ($0.0004$).
A simulation study~\cite{Ref:atlasTDR} shows that the Higgs boson search via the
associate production $b\bar b A$ ($b\bar b H$) is better than
that via the direct production from gluon fusion to see  both
$\tau^+\tau^-$ and $\mu^+\mu^-$ modes, especially in the large $\tan\beta$ area.
The largest background is the Drell-Yan process from
$\gamma^\ast/Z^\ast \to \tau^+\tau^-$ (and $\mu^+\mu^-$).
The other ones, such as $t\bar t$, $b\bar b$ and $W+jet$, also contribute to
the backgrounds.
The rate of the $\tau^+\tau^-$ process from the signal is much larger than
that of the $\mu^+\mu^-$ one.  However, the resolution for tau leptons is much
broader than that for muons, so that for relatively small $m_A$ ($m_H$)
the $\mu^+\mu^-$ mode can be more useful than the $\tau^+\tau^-$
mode~\cite{Ref:HanMuMu}.

In the type-X THDM, signals from the associate production $pp \to b\bar bA$
are very difficult to detect. The production cross section
is at most $150$ fb for $m_A^{}=150$ GeV at $\tan\beta=1$~\cite{Ref:atlasTDR}, 
where the branching ratio
$A\to \tau^+\tau^-$ and $A\to \mu^+\mu^-$ are small,
and the produced signals are less for larger values of $\tan\beta$.
On the other hand, the direct production from $gg\to A/H$ can be
used to see the signal.
The cross sections are significant for $\tan\beta \sim 1$, and
they decrease for larger values of $\tan\beta$ by a factor of $1/\tan^2\beta$.
However, the branching ratios of $A/H \to \tau^+\tau^-$
dominate over those of $A/H\to b\bar b$ for $\tan\beta
\gtrsim 2$ and become almost $100\%$ for $\tan\beta \gtrsim 4$ (see FIG.~\ref{FIG:br_150}).
Therefore, large significances can be expected around $\tan\beta \sim 2$
in the type-X THDM.
\begin{figure}[tb]
\begin{minipage}{0.49\hsize}
\includegraphics[width=7.5cm,angle=-90]{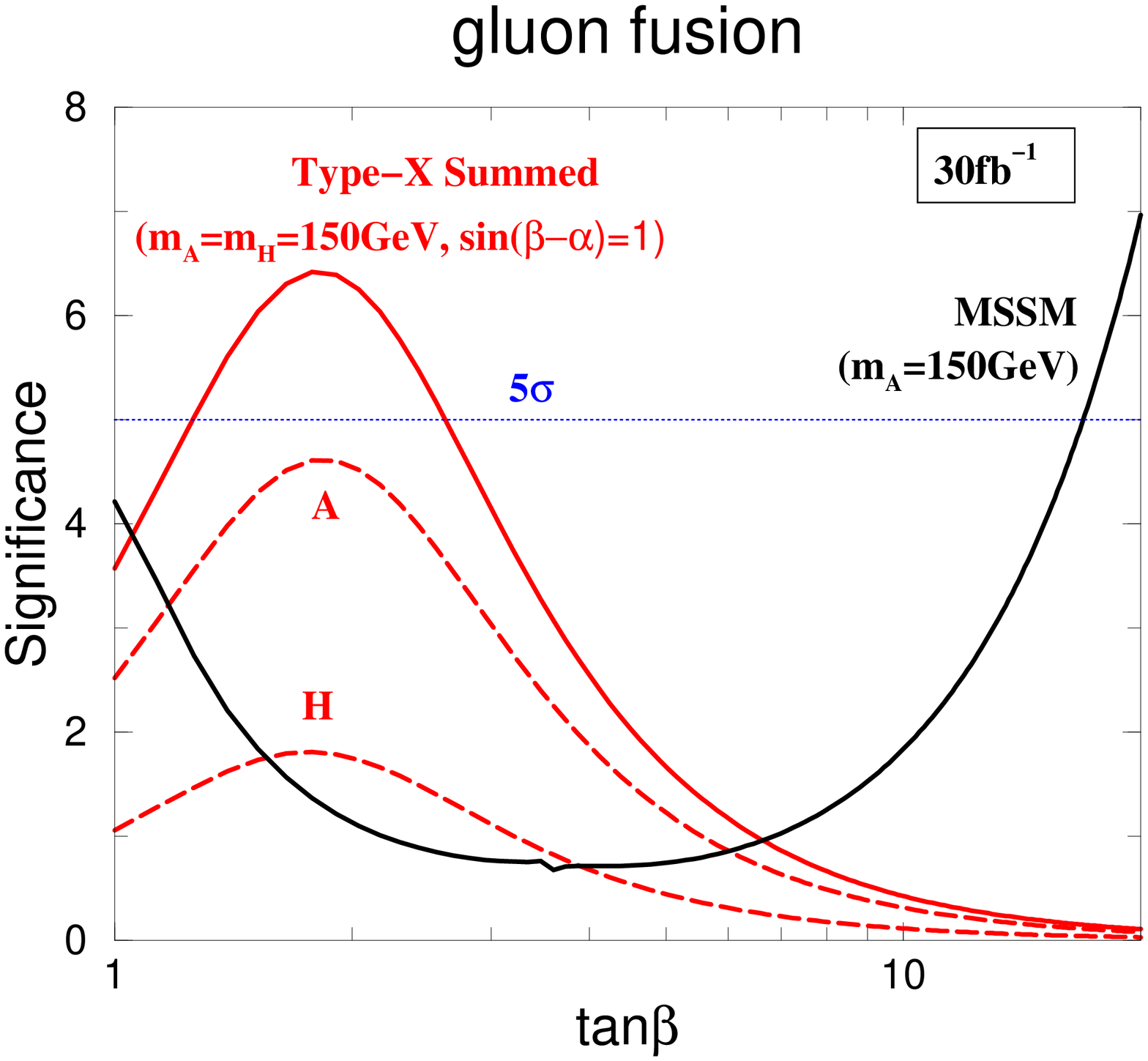}
\end{minipage}
\begin{minipage}{0.49\hsize}
\includegraphics[width=7.5cm,angle=-90]{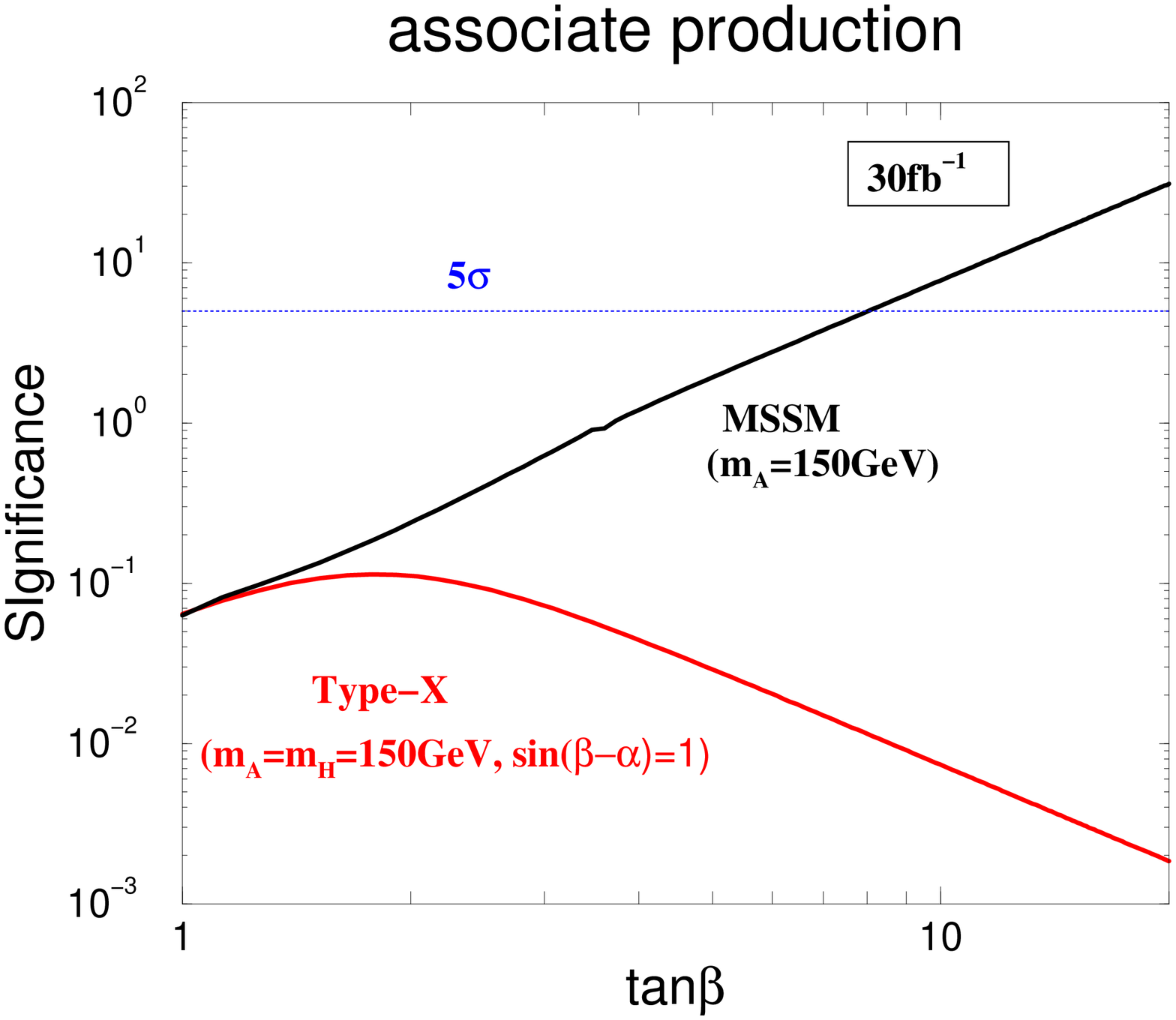}
\end{minipage}
\caption{Signal significance ($S/\sqrt{B}$) for gluon fusion $gg\to A/H$ (left panel)
and the associated production $pp\to b\bar b A$ $(b\bar b H)$ (right panel)
with the $\tau^+\tau^-$ final state in the type-X THDM and the MSSM.
In both figures, the dashed and the solid curves represents
the expected values of signal significance for $A/H$ and summed over $A$ and $H$.
The red (thin) curves denote the significance in the type-X THDM, while
the black (thick) solid curves denote that in the MSSM~\cite{typeX}.}
\label{FIG:ggA}
\end{figure}

In FIG.~\ref{FIG:ggA}, we show the expected signal significance of the
direct production $pp (gg)\to A/H \to \tau^+\tau^-$ in the type-X THDM and
the MSSM at the LHC, assuming an integrated luminosity of $30$ fb$^{-1}$.
The mass of the CP-odd Higgs boson $m_A^{}$ is taken to be $150$ GeV
in both models, while $m_H^{}$ is $150$ GeV for the results
of the type-X THDM and that in the MSSM is deduced using the MSSM mass relation.
For the detailed analysis of background simulation, we
employed the one shown in the ATLAS TDR~\cite{Ref:atlasTDR}.
The basic cuts, such as the high $p_T^{}$ cut and the standard $A/H\to\tau^+\tau^-$ reconstruction, are assumed~\cite{Ref:atlasTDR}.
We can see that, for the search of the direct production, 
the signal significance in the type-X THDM can be larger than
that in the MSSM for $\tan\beta \lesssim 5$.
In particular, the signal in the type-X THDM can be expected to be
detectable $(S/\sqrt{B}>5)$ when $\tan\beta \sim 2$
for the luminosity of $30$ fb$^{-1}$.
For smaller values of $m_A$ ($m_H$), the production cross section becomes large
so that the signal rate is more significant, but the separation
from the Drell-Yan background becomes more difficult
because the resolution of the tau lepton is broad. 
Therefore, the significance becomes worse for $m_A^{} (m_H^{}) \lesssim 130$ GeV.

\begin{figure}[tb]
\begin{center}
\includegraphics[width=7.5cm,angle=-90]{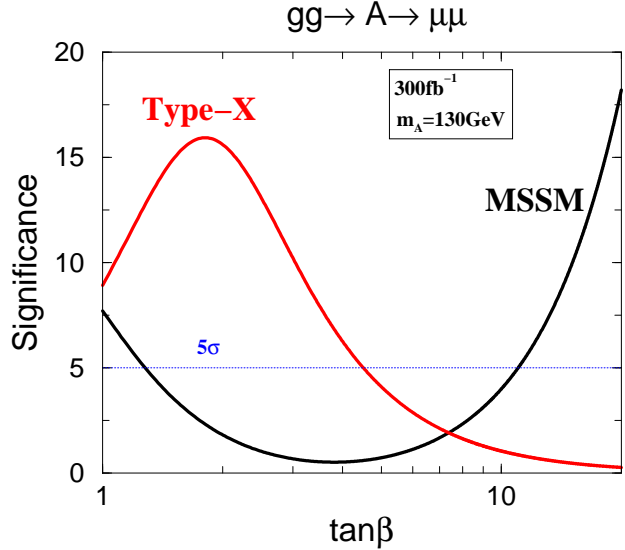}
\end{center}
\caption{Signal significance ($S/\sqrt{B}$) for the $\mu^+\mu^-$ final state 
from gluon fusion $gg\to A$ in the type-X THDM and the MSSM.
The red (thin) curves denote significance in the type-X THDM, while
the black (thick) solid curves denote that in the MSSM~\cite{typeX}.}
\label{FIG:mumu}
\end{figure}

When $A$ and $H$ are lighter than $130$ GeV, the $\mu^+\mu^-$ mode can be more
useful than the $\tau^+\tau^-$ mode. The resolution of muons is much better
than that of tau leptons, so that the invariant mass cut is very effective
in reducing the background from $\gamma^\ast/Z^\ast \to \mu^+\mu^-$.
The feasibility of the process $gg\to A/H \to \mu^+\mu^-$
has been studied in the SM and the MSSM in Refs.~\cite{Ref:atlasTDR,Ref:HanMuMu}.
We evaluate the signal significances of $gg\to A/H \to \mu^+\mu^-$  in
the type-X THDM by using the result in Ref.~\cite{Ref:HanMuMu}.
In TABLE~\ref{Tab:signif_mumu},
we list the results of the significance in the SM and the
type-X THDM. According to Tao Han's paper, the
basic kinematic cuts of $p_T^{}>20$ GeV, $|\eta|<2.5$ and
the invariant mass cuts as
$m_{A/H}^{}-2.24$ GeV $ < M_{\mu\mu}^{} < m_{A/H}^{}+2.24$ GeV
are used~\footnote{This choice for the invariant mass cut is rather severe; i.e., 
it requires the precise determination of $m_A^{}$ and $m_H^{}$. 
If the range of the invariant mass cut is taken to be double, 
roughly speaking, background events also become double. 
This would suppress the signal significance in TABLE \ref{Tab:signif_mumu} by 
a factor of $\sim 1/\sqrt2$.}. 
The integrated luminosity is assumed to be $300$ fb$^{-1}$.
For the results in the type-X THDM, $\tan\beta=2$ and
 $\sin(\beta-\alpha)=1$ are taken. 
The results show that the significance can be substantial for
$m_A^{}\gtrsim 115$ GeV when $\tan\beta=2$.
For smaller masses of the extra Higgs bosons, the cross section for the
signal processes can be larger, but the invariant mass cut becomes
less effective in the reduction of the Drell-Yan background
because of the smaller mass difference between $m_Z^{}$ and
$m_{A}^{}$; hence, the signal significance becomes worse.
The $\tan\beta$ dependence in the signal significance for the muon final states 
is also shown in FIG.~\ref{FIG:mumu}. The shape of the curves is similar to that 
for the tau lepton final state in FIG~\ref{FIG:ggA}. 
\begin{table}[t]
\begin{center}
\begin{tabular}{|c||c|c|c|c|c|}
\hline  $m_\phi$ [GeV] & SM ($H_\text{SM}$) & MSSM($A$) & Type-X ($H$) & Type-X ($A$) & Type-X(Sum) ($m_A^{} \simeq m_H^{}$) \\  \hline
115  & 2.41 & 1.31 & 4.31 & 12.0 & 16.3 \\
120  & 2.51 & 1.49 & 4.89 & 13.4 & 18.3 \\
130  & 2.25 & 1.81 & 5.78 & 15.6 & 21.4 \\
140  & 1.61 & 2.11 & 6.60 & 17.5 & 24.1 \\
\hline
\end{tabular}
\end{center}
\caption{Expected signal significances for
$gg\to \phi \to \mu^+\mu^- (\phi=H_\text{SM}, H, A)$
 in the SM, the MSSM, and the type-X THDM.
For the results in the MSSM $\tan\beta=2$ is taken, 
and for that in the type-X THDM $\tan\beta=2$  
and $\sin(\beta-\alpha)=1$ are taken.
 The integrated luminosity is assumed to be $300$ fb$^{-1}$~\cite{typeX}.
 }\label{Tab:signif_mumu}
\end{table}

In summary, we would be able to distinguish the type-X THDM from the MSSM by
measuring the leptonic decays of the additional Higgs bosons produced via
the direct search processes $gg \to A/H \to \tau^+\tau^-$ ($\mu^+\mu^-$) and
the associated processes $pp\to b\bar b A$ $(b\bar b H)$.
First, if a light scalar boson is found via $gg \to h \to \gamma\gamma$ or
$W^+W^- \to h \to \gamma\gamma$ ($\tau^+\tau^-$) and if
the event number is consistent with the prediction in the SM,
then we know that the scalar boson is of the SM or at least
SM-like: in the THDM framework this means $\sin(\beta-\alpha)\simeq 1$,
assuming that it is the lightest one.
Second, under the situation above, 
when the associated production $pp\to b\bar b A$ $(b\bar b H)$
is detected at a different invariant mass than the mass of the
SM-like one and no $gg\to A/H \to \tau^+\tau^-$ ($\mu^+\mu^-$) is found
at that point, we would be able to identify the MSSM Higgs sector
(or the type-II THDM) with high $\tan\beta$ values.
On the other hand, the type-X THDM with low $\tan\beta$ would be
identified by finding the signal from the gluon fusion process without
that from the $b\bar b \tau^+\tau^-$.
If signals from the direct production processes are found 
but the number is not sufficient, then the value of $\tan\beta$
would be around $6$--$10$ ($m_t \cot\beta \sim m_b \tan\beta$), 
where the rates in the MSSM and the type-X THDM are similar. In this case, it
would be difficult to distinguish these models from the above
processes. 
As we discussed in the next subsection, Higgs pair production processes
$pp\to AH^\pm, HH^\pm$, and $AH$ can be useful to measure
the Yukawa interaction through branching fractions, because these production
mechanism do not depend on $\tan\beta$ in such a situation.

\begin{figure}[tb]
\begin{center}
\includegraphics[width=8cm]{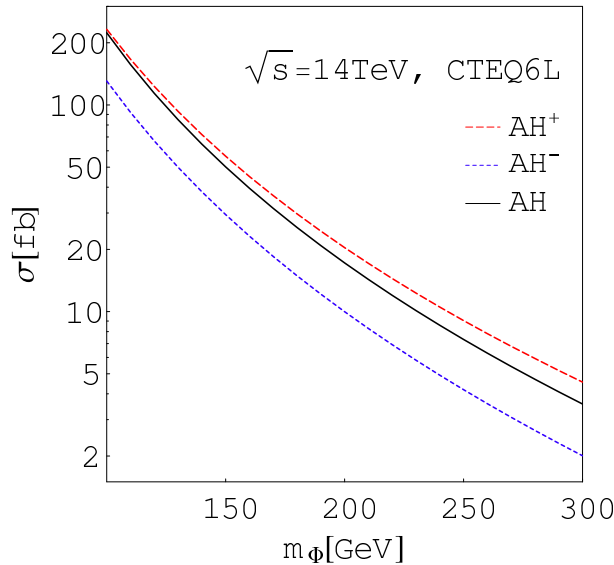}
\caption{Production cross sections of $pp\to AH^\pm$ and $AH$ are
shown at the leading order as a function of scalar boson masses in the THDM
where $m_\Phi=m_H^{}=m_A^{}=m_{H^\pm}^{}$ are chosen.
The long-dashed, dashed and solid curves denote the pair production
of $AH^+$, $AH^-$ and $AH$, respectively. The cross section of
$pp\to HH^\pm$ is the same as those of $pp\to AH^\pm$ when $\sin(\beta-\alpha)=1$~\cite{typeX}. }
\label{FIG:AH}
\end{center}
\end{figure}

We have concentrated on $\sin(\beta-\alpha)\simeq 1$ in this analysis because  
the parameter is motivated in Ref.~\cite{aks_prl}. 
Here we comment on the situation without the condition $\sin(\beta-\alpha)=1$. 
If $\sin(\beta-\alpha)$ is not close to unity, our conclusion can be modified. 
The production cross sections of $gg\to A/H$ and $pp\to b{\bar b}A(b{\bar b}H)$ 
can be enhanced in the type-X THDM for $\tan\beta\gtrsim 1$ since the 
factor $(\sin\alpha/\sin\beta)$ of quark-Higgs couplings can be large in a 
specific region of the parameter space. The signal of 
the CP odd Higgs boson $A$ can then be significant. On the other hand, 
the CP even Higgs boson $H$ can decay into $WW^*$ when $\sin(\beta-\alpha)$ 
is not unity. This would reduce leptonic branching fractions. 
The signal can be enhanced only for large $\tan\beta$ regions 
because leptonic decays are significant only for such a parameter space.
We also note that $H$ can be produced significantly 
by the vector boson fusion mechanism in a mixing case. 
\\\\
{\bf Pair production of extra Higgs bosons at the LHC}
\\\\
The types of the Yukawa interactions can be studied using the process of
$q\bar q' \to {W^\pm}^\ast \to AH^\pm$ ($HH^\pm$)
~\cite{Ref:GGH2,Ref:KY,Akeroyd:2003jp,Ref:CKY,Ref:TOBE} and
$q\bar q \to Z^\ast \to AH$~\cite{Ref:GGH2},
unless the extra Higgs bosons $H$, $A$ and/or $H^\pm$ are too
heavy~\footnote{When the mixing between $h$ and $H$ is large, the $hH^\pm$
production can also be important~\cite{Ref:TOBE}.}.
Hadronic cross sections for these processes are shown at the leading
order in
FIG.~\ref{FIG:AH} as a function of the mass of the produced scalar boson
$\Phi$, where $m_\Phi^{}=m_H^{}=m_A^{}=m_{H^\pm}^{}$.
Expected rates of $AH^\pm$(sum) and $AH$ are about $143$ fb 
and $85$ fb for $m_{H^\pm}^{} = 130$ GeV and about $85$ fb
and $50$ fb for $m_{H^\pm}^{} = 150$ GeV, respectively.
The NLO QCD corrections are expected to enhance these rates 
typically by about $20\%$~\cite{Ref:KY,Ref:CKY}.
The production rates are common in all the types of THDMs, because
the cross sections are determined only by the masses of the produced scalar bosons.
Therefore, they are very sensitive to the difference in the decay branching
ratios of the produced Higgs bosons.
In the MSSM, the $b\bar b \tau^\pm \nu$ ($b\bar b\tau^+\tau^-$)
events can be
the main signal of the $AH^\pm$ and $HH^\pm$ ($AH$) processes,
while in the type-X THDM ($\tan\beta \gtrsim 2$),
$\tau^+\tau^-\tau^\pm\nu$ ($\tau^+\tau^-\tau^+\tau^-$) would be the main signal events.

In the MSSM, the parton level 
background analysis for the $AH^\pm$ ($HH^\pm$) production process has been
performed in Ref.~\cite{Ref:CKY,Ref:TOBE} by using the $b\overline{b}\tau\nu$ 
final state.
The largest background comes from $q\overline{q'}\to Wg^*\to
Wb\overline{b}$, which can be reduced by basic kinematic cuts and
the invariant mass cut of $b\bar b$, as well as by 
the kinematic cut to select hard hadrons from the parent $\tau^\pm$ from $H^\pm$.
It has been shown that a sufficient signal significance can be
obtained for smaller masses of Higgs bosons~\cite{Ref:CKY}.

In the type-X THDM, the produced $AH^\pm$ ($HH^\pm$) and $AH$ pairs
can be studied via the leptonic decays.
Hence these channels can be useful to discriminate the type-X THDM from
the MSSM.
Assuming an integrated luminosity of $300$ fb$^{-1}$,
$8.6 \times 10^4$ and $5.1 \times 10^4$ of the
signal events are produced from both $AH^\pm$ and $HH^\pm$ production
for $m_\Phi^{}=130$ GeV and $150$ GeV, respectively, where
$m_\Phi^{} = m_H^{} = m_A^{} = m_{H^\pm}^{}$.
$A$ and $H$ ($H^\pm$) decay into $\tau^+\tau^-$ ($\tau\nu$)
by more than $95$\% and $95$\% ($99$\%) for $\tan\beta\gtrsim 4$, respectively.
The purely leptonic signal would have an advantage in the signal to
background ratio because the background from the intermediate state
$q\overline{q'}\to Wg^*$ would be negligible.
For $\tan\beta=7$, the produced $AH^\pm$ and $HH^\pm$ pairs almost all 
(more than $99\%$)
go to $\tau^+\tau^-\tau\nu$ final states.
The numbers of signal and background are summarized in TABLE~\ref{Tab:AH+}. 
The signal to background ratio for $\tau^+\tau^-\tau\nu$ final state 
is not so small ${\mathcal O}(0.1$--$1)$,  
before cuts~\footnote{The $\gamma W^\pm$ production may give a much larger 
cross section for background events. It may also be reduced considerably 
by kinematic cuts.}.
The backgrounds are expected to be reduced by using
high-$p_T^{}$ cuts, hard hadrons from the parent tau leptons from $H^\pm$,
and invariant mass cuts for $\tau^+\tau^-$ from $A$ and $H$, 
in addition to the basic cuts. 
However, the signal significance strongly depend on the rate of
miss identification of hadrons as tau leptons, so that
a realistic simulation is necessary.

\begin{table}[t]
\begin{center}
\begin{tabular}{|c||c|c|c|c|c|}
\hline & $AH^\pm, HH^\pm (m_\Phi^{}=130 \text{ GeV})$ & $AH^\pm, HH^\pm(m_\Phi^{}=150 \text{ GeV})$ & $ZW^\pm$\\
\hline $\tau^+\tau^-\tau\nu$ & $8.4 \times 10^4$ & $5.0 \times 10^4$ & 
$3.2\times 10^{4}$\\
\hline $\mu^+\mu^-\tau\nu$ & $3.0\times 10^2$ & $1.8\times 10^2$ & $3.1\times 10^{4}$\\
\hline
\end{tabular}
\end{center}
\caption{Events for the $\tau^+\tau^-\tau\nu$ 
and $\mu^+\mu^-\tau\nu$ final states 
from the Higgs boson pair production and $ZW^\pm$ background~\cite{typeX}. 
The signal events are summed over $AH^\pm$ and $HH^\pm$. 
The integrated luminosity is taken to be $300$ fb$^{-1}$. 
Values for the decay branching ratios are taken to be 
${\mathcal B}(A/H\to \tau^+\tau^-)=0.99$, 
${\mathcal B}(A/H\to \mu^+\mu^-)=0.0035$, and 
${\mathcal B}(H^\pm\to \tau\nu)=0.99$, 
which correspond to the values for $\tan\beta\gtrsim 7$.   
The cross section of $pp\to ZW^\pm$ is evaluated 
as $\sigma_{ZW}=27.7$ pb by CalcHEP~\cite{CalcHEP}. 
}\label{Tab:AH+}
\end{table}

We also would be able to use the $\mu^+\mu^-\tau^+\nu$
events to identify the $AH^+$ and $HH^+$ production in the
type-X THDM, by using the much better resolution of $\mu^+\mu^-$
in performing the invariant mass cuts.
For $300$ fb$^{-1}$, the $AH^+$ and $HH^+$ process
can produce about
${\mathcal O}(100)$ 
of $\mu^+\mu^-\tau^+ \nu$ events for $m_A^{}=m_H^{}=130$ GeV.
The number of background events 
is about $3.1 \times 10^5$ of $\mu^+\mu^-\tau^+ \nu$ from $ZW^\pm$ production. 
Signals and background for $\mu^+\mu^-\tau^+\nu$ events are 
also summarized in TABLE~\ref{Tab:AH+}. 
The background can be expected to be reduced by imposing
a selection of the events around the invariant mass of $m_A^{} \sim
M_{\mu\mu}$ and the high $p_T^{}$ cuts. Hard hadrons from the
decay of $\tau$'s from $H^+$ can also be used to reduce the
background. In the MSSM, much smaller signals are expected,
so that this process can be a useful probe of the type-X THDM.

In a similar way, we may use $AH$ production~\cite{Ref:GGH2}
to identify the type-X THDM.
For the $\tau^+\tau^-\tau^+\tau^-$ decay mode, the signal is evaluated approximately as
$2.5\times 10^4$ events, assuming $300$ fb$^{-1}$ for $m_A^{}=m_H^{}=130$ GeV
and $\tan\beta=7$.
The main background may come from the $q\bar q\to ZZ$ process.
We also consider the $\mu^+\mu^-\tau^+\tau^-$ decay mode.
The number of signal event is ${\mathcal O}(100)$
for an integrated luminosity $300$ fb$^{-1}$. 
The numbers of signal and background event are listed in 
TABLE~\ref{Tab:AH}. 
It would be valuable to use the detailed background simulation.
\begin{table}[t]
\begin{center}
\begin{tabular}{|c||c|c|c|c|c|}
\hline & $AH (m_\Phi^{}=130 \text{ GeV})$ & $AH (m_\Phi^{}=150 \text{ GeV})$ & $ZZ$\\
\hline $\tau^+\tau^-\tau^+\tau^-$ & $2.5\times 10^5$ & $1.5\times 10^5$ & 
$3.6\times 10^{3}$\\
\hline $\tau^+\tau^-\mu^+\mu^-$ & $1.8\times 10^2$ & $1.0\times 10^2$ & 
$7.1\times 10^{3}$\\
\hline
\end{tabular}
\end{center}
\caption{Events for the $\tau^+\tau^-\tau^+\tau^-$ 
and $\tau^+\tau^-\mu^+\mu^-$ final states 
from the Higgs boson pair production and $ZZ$ background~\cite{typeX}. 
The integrated luminosity is taken to be $300$ fb$^{-1}$. 
The cross section of $pp\to ZZ$ is evaluated 
as $\sigma_{ZZ}=10.5$ pb by CalcHEP~\cite{CalcHEP}. 
}\label{Tab:AH}
\end{table}
\\\\
{\bf Pair production of extra Higgs bosons at the ILC}
\\\\
At the ILC, we would be able to test the types of the Yukawa interactions
via the pair production of the additional Higgs bosons
$e^+e^- \to AH$~\cite{Ref:GGH2,Ref:eeAH}.
\begin{figure}[t]
\begin{center}
\includegraphics[width=7.5cm,angle=-90]{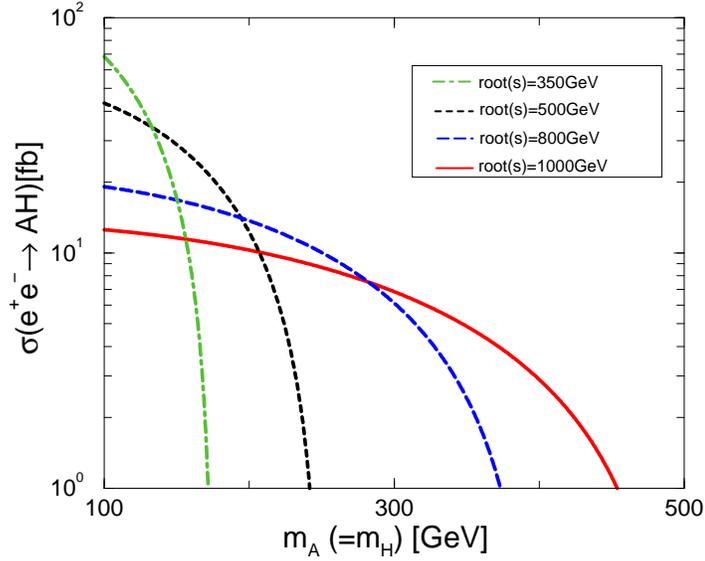}
\caption{The production cross section of $e^+e^- \to AH$ is shown as a
function of the Higgs boson mass.
The dot-dashed, dashed, long-dashed, and solid curves correspond to 
$\sqrt{s}=350, 500, 800$, and $1000$ GeV, respectively~\cite{typeX}.}
\label{FIG:eeAH}
\end{center}
\end{figure}
In Fig.~\ref{FIG:eeAH}, the production cross section is shown for
$\sqrt{s}=350, 500$, $800$, and $1000$ GeV as a function of $m_A^{}$
assuming $m_{A}^{}=m_H^{}$ in the THDM. The production rate
is determined only by $m_{A}^{}$, and $m_H^{}$ at the leading order,
and  is independent of $\tan\beta$.
(In the MSSM, it depends indirectly on $\tan\beta$
via the mass spectrum.)
The signal of the type-X THDM can be identified by searching
for the events of $\tau^+\tau^-\tau^+\tau^-$ and
$\mu^+\mu^-\tau^+\tau^-$.
When $\sqrt{s}=500$ GeV, assuming an integrated luminosity of
$500$ fb$^{-1}$, the event number is estimated
as $1.6\times 10^4$ $(1.8\times 10^2)$
in the type-X (type-II) THDM for
$\tau^+\tau^-\tau^+\tau^-$, and $1.1\times 10^2$ $(0.6)$
for $\mu^+\mu^-\tau^+\tau^-$
assuming  $m_H^{}=m_A^{}=m_{H^\pm}^{}=130$ GeV, $\sin(\beta-\alpha)=1$
and $\tan\beta=10$. This number does not change much for $\tan\beta\gtrsim 3$.
The main background comes from the $Z$ pair production, whose rate is
about $4\times 10^2$ fb. The event numbers from the background are then
$2.3\times 10^2$ for $\tau^+\tau^-\tau^+\tau^-$ and $4.6\times 10^2$
for $\mu^+\mu^-\tau^+\tau^-$.
Therefore, the signal should be easily detected in the type-X THDM, by
which we would be able to distinguish the type-X from the type-II (the
MSSM).

The detailed measurement of the masses of additional
Higgs bosons and Yukawa coupling constants will make it possible to
determine the scenario of physics beyond the SM through the Higgs physics.

\clearpage
\section{The Higgs Triplet Model}

In this section, we discuss the extended Higgs sector which contains the 
Higgs triplet fields. 
The Higgs triplet fields are introduced in various new physics models such as the left-right symmetric model~\cite{LRmodels}, 
the type II seesaw model~\cite{typeIII_seesaw} and so on. 
Here, we focus on the Higgs Triplet Model (HTM) which is motivated by the type II seesaw model, 
where neutrino masses can be generated at the tree-level. 
In the HTM, the Higgs triplet field $\Delta$ with $Y=1$ is added to the SM. 

First, we define the Lagrangian of the HTM, and we discuss tree level formulae such as 
masses for physical scalar bosons, gauge bosons and neutrinos as well as the rho parameter. 
Second, the on-shell renormalization scheme is introduced for the electroweak precision observables. 
We then calculate the predictions for $m_W$ and $\rho$ at the one-loop level.  
Third, we discuss the phenomenology of the HTM; i.e., 
we calculate the decay branching ratios of scalar bosons which are originated from the triplet Higgs field, 
and we show how masses of these scalar bosons can be reconstructed at the LHC. 
Finally, we evaluate the deviation from the SM prediction of the decay rate of $h\to \gamma\gamma$. 

\subsection{Tree level formulae}
The scalar sector of the HTM is composed of the isospin doublet field $\Phi$ with 
hypercharge $Y=1/2$ and the triplet field $\Delta$ with $Y=1$. 
The relevant terms in the Lagrangian are given by 
\begin{align}
\mathcal{L}_{\text{HTM}}=\mathcal{L}_{\text{kin}}+\mathcal{L}_{Y}-V(\Phi,\Delta), 
\end{align}
where $\mathcal{L}_{\text{kin}}$, $\mathcal{L}_{Y}$ and $V(\Phi,\Delta)$ are 
the kinetic term, the Yukawa interaction and the scalar potential, respectively. 
The kinetic term of the Higgs fields is given by 
\begin{align}
\mathcal{L}_{\text{kin}}&=(D_\mu \Phi)^\dagger (D^\mu \Phi)+\text{Tr}[(D_\mu \Delta)^\dagger (D^\mu \Delta)], 
\end{align}
where the covariant derivatives are defined as
\begin{align}
D_\mu \Phi=\left(\partial_\mu+i\frac{g}{2}\tau^aW_\mu^a+i\frac{g'}{2}B_\mu\right)\Phi, \quad
D_\mu \Delta=\partial_\mu \Delta+i\frac{g}{2}[\tau^aW_\mu^a,\Delta]+ig'B_\mu\Delta. 
\end{align}
The Yukawa interaction for neutrinos is given by 
\begin{align}
\mathcal{L}_Y&=h_{ij}\overline{L_L^{ic}}i\tau_2\Delta L_L^j+\text{h.c.}, \label{nu_yukawa}
\end{align}
where $h_{ij}$ is the $3\times 3$ complex symmetric Yukawa matrix. 
Notice that the triplet field $\Delta$ carries the lepton number of 2.  
The most general form of the Higgs potential under the gauge symmetry is given by 
\begin{align}
V(\Phi,\Delta)&=m^2\Phi^\dagger\Phi+M^2\text{Tr}(\Delta^\dagger\Delta)+\left[\mu \Phi^Ti\tau_2\Delta^\dagger \Phi+\text{h.c.}\right]\notag\\
&+\lambda_1(\Phi^\dagger\Phi)^2+\lambda_2\left[\text{Tr}(\Delta^\dagger\Delta)\right]^2+\lambda_3\text{Tr}[(\Delta^\dagger\Delta)^2]
+\lambda_4(\Phi^\dagger\Phi)\text{Tr}(\Delta^\dagger\Delta)+\lambda_5\Phi^\dagger\Delta\Delta^\dagger\Phi, \label{pot_htm}
\end{align}
where $m$ and $M$ are the dimension full real parameters, $\mu$ is the dimension full complex parameter 
which violates the lepton number, and 
$\lambda_1$-$\lambda_5$ are the coupling constants which are real. 
We here take $\mu$ to be real. 
The scalar fields $\Phi$ and $\Delta$ can be parameterized as
\begin{align}
\Phi=\left[
\begin{array}{c}
\varphi^+\\
\frac{1}{\sqrt{2}}(\varphi+v_\Phi+i\chi)
\end{array}\right],\quad \Delta =
\left[
\begin{array}{cc}
\frac{\Delta^+}{\sqrt{2}} & \Delta^{++}\\
\Delta^0 & -\frac{\Delta^+}{\sqrt{2}} 
\end{array}\right]\text{ with } \Delta^0=\frac{1}{\sqrt{2}}(\delta+v_\Delta+i\eta),
\end{align}
where $v_\Phi$ and $v_\Delta$ 
are the VEVs of the doublet Higgs field and the triplet Higgs field, respectively which satisfy 
$v^2\equiv v_\Phi^2+2v_\Delta^2\simeq$ (246 GeV)$^2$. 

From the stationary condition at the vacuum $(v_\Phi, v_\Delta)$, we obtain 
\begin{align}
m^2&=\frac{1}{2}\left[-2v_\Phi^2\lambda_1-v_\Delta^2(\lambda_4+\lambda_5)+2\sqrt{2}\mu v_\Delta\right],\\
M^2&=M_\Delta^2-\frac{1}{2}\left[2v_\Delta^2(\lambda_2+\lambda_3)+v_\Phi^2(\lambda_4+\lambda_5)\right], \label{vc}
\text{ with } M_\Delta^2\equiv \frac{v_\Phi^2\mu}{\sqrt{2}v_\Delta}, 
\end{align}
and we can eliminate $m^2$ and $M^2$. 
The mass matrices for the scalar bosons can be diagonalized by rotating the 
scalar fields as 
\begin{align}
\left(
\begin{array}{c}
\varphi^\pm\\
\Delta^\pm
\end{array}\right)&=
\left(
\begin{array}{cc}
\cos \beta_\pm & -\sin\beta_\pm \\
\sin\beta_\pm   & \cos\beta_\pm
\end{array}
\right)
\left(
\begin{array}{c}
w^\pm\\
H^\pm
\end{array}\right),\quad 
\left(
\begin{array}{c}
\chi\\
\eta
\end{array}\right)=
\left(
\begin{array}{cc}
\cos \beta_0 & -\sin\beta_0 \\
\sin\beta_0   & \cos\beta_0
\end{array}
\right)
\left(
\begin{array}{c}
z\\
A
\end{array}\right),\notag\\
\left(
\begin{array}{c}
\varphi\\
\delta
\end{array}\right)&=
\left(
\begin{array}{cc}
\cos \alpha & -\sin\alpha \\
\sin\alpha   & \cos\alpha
\end{array}
\right)
\left(
\begin{array}{c}
h\\
H
\end{array}\right),
\end{align}
with the mixing angles
\begin{align}
\tan\beta_\pm=\frac{\sqrt{2}v_\Delta}{v_\Phi},\quad \tan\beta_0 = \frac{2v_\Delta}{v_\Phi}, \quad
\tan2\alpha &=\frac{v_\Delta}{v_\Phi}\frac{2v_\Phi^2(\lambda_4+\lambda_5)-4M_\Delta^2}{2v_\Phi^2\lambda_1-M_\Delta^2-v_\Delta^2(\lambda_2+\lambda_3)}, \label{tan2a}
\end{align}
In addition to the three NG bosons $w^\pm$ and $z$ which are absorbed by the longitudinal components 
of the $W$ boson and the $Z$ boson, 
there are seven physical mass eigenstates $H^{\pm\pm}$, $H^\pm$, $A$, $H$ and $h$. 
The masses of these physical states are expressed as 
\begin{align}
m_{H^{++}}^2&=M_\Delta^2-v_\Delta^2\lambda_3-\frac{\lambda_5}{2}v_\Phi^2,\label{mhpp}\\
m_{H^+}^2&= \left(M_\Delta^2-\frac{\lambda_5}{4}v_\Phi^2\right)\left(1+\frac{2v_\Delta^2}{v_\Phi^2}\right),\label{mhp}\\
m_A^2 &=M_\Delta^2\left(1+\frac{4v_\Delta^2}{v_\Phi^2}\right), \label{mA}\\
m_H^2&=\mathcal{M}_{11}^2\sin^2\alpha+\mathcal{M}_{22}^2\cos^2\alpha+\mathcal{M}_{12}^2\sin2\alpha,\label{mH}\\
m_h^2&=\mathcal{M}_{11}^2\cos^2\alpha+\mathcal{M}_{22}^2\sin^2\alpha-\mathcal{M}_{12}^2\sin2\alpha,
\end{align}
where $\mathcal{M}_{11}^2$, $\mathcal{M}_{22}^2$ and $\mathcal{M}_{12}^2$ are 
the elements of the mass matrix $\mathcal{M}_{ij}^2$ for the CP-even scalar states in the $(\varphi,\delta)$ basis which are 
given by
\begin{align}
\mathcal{M}_{11}^2&=2v_\Phi^2\lambda_1,\\
\mathcal{M}_{22}^2&=M_\Delta^2+2v_\Delta^2(\lambda_2+\lambda_3),\\
\mathcal{M}_{12}^2&=-\frac{2v_\Delta}{v_\Phi}M_\Delta^2+v_\Phi v_\Delta(\lambda_4+\lambda_5).
\end{align}
The six parameters $\mu$ and $\lambda_1$-$\lambda_5$ in the Higgs potential in Eq.~(\ref{pot_htm}) 
can be written in terms of the physical scalar masses, the mixing angle $\alpha$ and VEVs $v_\Phi$ and $v_\Delta$ as
\begin{align}
\mu&=\frac{\sqrt{2}v_\Delta^2}{v_\Phi^2}M_\Delta^2 =\frac{\sqrt{2}v_\Delta}{v_\Phi^2+4v_\Delta^2}m_A^2,\\
\lambda_1 & = \frac{1}{2v_\Phi^2}(m_h^2\cos^2\alpha+m_H^2\sin^2\alpha),\\
\lambda_2 & = \frac{1}{2v_\Delta^2}\left[2m_{H^{++}}^2+v_\Phi^2\left(\frac{m_A^2}{v_\Phi^2+4v_\Delta^2}-\frac{4m_{H^+}^2}{v_\Phi^2+2v_\Delta^2}\right)
+m_H^2\cos^2\alpha+m_h^2\sin^2\alpha\right],\\
\lambda_3 & = \frac{v_\Phi^2}{v_\Delta^2}\left(\frac{2m_{H^+}^2}{v_\Phi^2+2v_\Delta^2}-\frac{m_{H^{++}}^2}{v_\Phi^2}-\frac{m_A^2}{v_\Phi^2+4v_\Delta^2}\right),\\
\lambda_4 & = \frac{4m_{H^+}^2}{v_\Phi^2+2v_\Delta^2}-\frac{2m_A^2}{v_\Phi^2+4v_\Delta^2}+\frac{m_h^2-m_H^2}{2v_\Phi v_\Delta}\sin2\alpha,\\
\lambda_5 & = 4\left(\frac{m_A^2}{v_\Phi^2+4v_\Delta^2}-\frac{m_{H^+}^2}{v_\Phi^2+2v_\Delta^2}\right).
\end{align}
Bounds from the tree-level unitarity and the vacuum stability have been
studied in Ref.~\cite{type2_unitarity}.
The masses of the W boson and the Z boson are obtained at the tree level as 
\begin{align}
m_W^2 = \frac{g^2}{4}(v_\Phi^2+2v_\Delta^2),\quad m_Z^2 =\frac{g^2}{4\cos^2\theta_W}(v_\Phi^2+4v_\Delta^2).
\end{align}
The electroweak rho parameter can deviate from unity at the tree level; 
\begin{align}
\rho \equiv \frac{m_W^2}{m_Z^2\cos^2\theta_W}=\frac{1+\frac{2v_\Delta^2}{v_\Phi^2}}{1+\frac{4v_\Delta^2}{v_\Phi^2}}. \label{rho_triplet}
\end{align}
As the experimental value of the rho parameter is near unity, 
$v_\Delta^2/v_\Phi^2$ is required to be much smaller than unity at the tree level. 
In this case, the state $h$ behaves mostly as the SM Higgs boson, while 
the other states are almost originated from the components of the triplet field. 
Then, from Eqs.~(\ref{mhpp}), (\ref{mhp}), (\ref{mA}) and (\ref{mH}), 
there are interesting relations among the masses; 
\begin{align}
&m_{H^{++}}^2-m_{H^{+}}^2\simeq m_{H^+}^2-m_A^2\left(\simeq-\frac{\lambda_5}{4}v_\Phi^2\equiv \xi\right), \label{mass1}\\
&m_H^2\simeq m_A^2\left(\simeq M_\Delta^2\right). \label{mass2}
\end{align}
These characteristic mass relations would be used as a probe of the Higgs potential in the HTM~\cite{aky_triplet}. 
If the masses of the triplet-like Higgs bosons are hierarchical, 
there are two patterns of the mass hierarchy among the triplet-like scalar bosons; i.e., 
when $\lambda_5$ is positive (negative), the mass hierarchy is $m_{\phi^0}>m_{H^+}>m_{H^{++}}$ ($m_{H^{++}}>m_{H^+}>m_{\phi^0}$), 
where $m_{\phi^0}=m_A$ or $m_H$. 
We here define the mass difference between $H^{\pm\pm}$ and $H^\pm$ as 
\begin{equation}
\Delta m \equiv m_{H^{++}}-m_{H^+}. 
\end{equation}

In the HTM, tiny Majorana masses of neutrinos are generated by the Yukawa interaction with the VEV of the triplet field, which  
is proportional to the lepton number violating coupling constant $\mu$ as
\begin{align}
(m_\nu)_{ij}=\sqrt{2}h_{ij} v_\Delta=h_{ij}\frac{\mu v_\Phi^2}{M_\Delta^2}. \label{mn}
\end{align}
If $\mu\ll M_\Delta$ the smallness of the neutrino masses are explained by the so-called type II seesaw mechanism~\cite{typeII_seesaw}. 
In this section, we assume that the lepton number violating parameter $\mu$ is sufficiently smaller than the electroweak scale 
so that the mass scale of the triplet-like field is $\mathcal{O}$(100-1000) GeV. 

We here give some comments on radiative corrections to the relation 
$m_{H^{++}}^2-m_{H^+}^2 \simeq m_{H^+}^2 - m_{\phi^0}^2$. 
This relation can be changed when we take into account radiative corrections to the Higgs potential.  
At the one-loop level, the relation in mass differences can be rewritten as
\begin{align}
  \frac{m_{H^{++}}^2-m_{H^{+}}^2}{m_{H^+}^2-m_{\phi^0}^2} \simeq 1 + \delta_{\phi^0},\quad (\phi^0=H\text{ or }A),
\end{align}
where $\delta_{\phi^0}$ is the deviation from the tree level prediction due to radiative
corrections, which is given as a function of the masses and mixing angles. 
In principle, we may test the HTM with this kind of the corrected mass relation
instead of the tree level formula by measuring the masses of the bosons.
A detailed study of
radiative corrections to the Higgs potential in the HTM is an important and interesting issue which
will be performed in elsewhere~\cite{akky}.
\subsection{One-loop corrections to electroweak parameters}

Here, we calculate one-loop corrected electroweak observables   
in the on-shell scheme which was at first proposed by Blank and Hollik~\cite{blank_hollik} in the model with 
a triplet Higgs field with $Y=0$. 
In the SM, and in all the models with $\rho=1$ at the tree level, 
the kinetic term of the Higgs field contains three parameters $g$, $g'$ and $v$. 
All the electroweak parameters are determined by giving a set of three input parameters which 
are well known; i.e., for example, $\alpha_{\text{em}}$, $G_F$ and $m_Z$~\cite{hollik_sm,Aoki:1982ed}. 
On the other hand, in models with $\rho\neq 1$ at the tree level like the HTM, 
an additional input parameter is necessary to describe electroweak parameters. 
Therefore, in addition to the three input parameters $\alpha_{\text{em}}$, $G_F$ and $m_Z$, 
we take the weak angle $\sin^2\theta_W$ as the fourth input parameter in our calculation as in Ref.~\cite{blank_hollik}. 
The experimental values of these input parameters are given by~\cite{PDG}
\begin{align}
&\alpha_{\text{em}}^{-1}(m_Z) = 128.903(15),\quad G_F= 1.16637(1)\times 10^{-5} \text{ GeV}^{-2},\notag\\
&m_Z=91.1876(21)\text{ GeV},\quad \hat{s}_W^2(m_Z)= 0.23146(12),  \label{data}
\end{align}
where $\hat{s}_W^2(m_Z)$ is defined as the ratio of the coefficients of the vector part and the axial vector part in the 
$Z\bar{e}e$ vertex;
\begin{align}
1-4\hat{s}_W^2(m_Z)=\frac{\text{Re}(v_e)}{\text{Re}(a_e)}, 
\end{align}
here $v_e$ and $a_e$ are defined in Eq.~(\ref{ren_zee}). 
Tree level formulae for the other electroweak parameters are given in terms of the four input parameters:
\begin{align}
g^2&=\frac{4\pi\alpha_{\text{em}}}{\hat{s}_W^2},\label{gsq}\\
m_W^2&=\frac{\pi\alpha_{\text{em}}}{\sqrt{2}G_F\hat{s}_W^2},\label{mwsq}\\
v^2&=\frac{1}{\sqrt{2}G_F},\label{v} \\
v_\Delta^2 &= \frac{\hat{s}_W^2\hat{c}_W^2}{2\pi\alpha_{\text{em}}}m_Z^2-\frac{\sqrt{2}}{4G_F}\label{vdel}, 
\end{align}
where $\hat{s}_W^2=\hat{s}_W^2(m_Z)$ and $\hat{c}_W^2=1-\hat{s}_W^2$
\footnote{In the limit of $v_\Delta\to 0$, we obtain the following relation
\begin{align}
\frac{\hat{s}_W^2\hat{c}_W^2}{\pi\alpha_{\text{em}}}m_Z^2=\frac{1}{\sqrt{2}G_F}.
\end{align}
By using this relation, $m_W$ can be expressed by 
$m_W^2 =m_Z^2\hat{c}^2_W$. 
This relation can be found in models with $\rho=1$ at the tree level such as the SM. }.

The deviation from the relation in Eq.~(\ref{mwsq}) due to radiative corrections can be parameterized as 
\begin{align}
G_F = \frac{\pi \alpha_{\text{em}}}{\sqrt{2}m_W^2\hat{s}_W^2}(1+\Delta r), \label{gf}
\end{align}
where $\Delta r$ is 
\begin{align}
\Delta r = -\frac{\delta G_F}{G_F}+\frac{\delta\alpha_{\text{em}}}{\alpha_{\text{em}}}-\frac{\delta \hat{s}_W^2}{\hat{s}_W^2}\label{delr1}
-\frac{\delta m_W^2}{m_W^2}.
\end{align}
The counter terms are obtained by imposing the renormalization conditions given in Ref.~\cite{hollik_sm} as
\begin{align}
\frac{\delta G_F}{G_F} & = -\frac{\Pi_T^{WW}(0)}{m_W^2}-\delta_{VB},\label{ct1}\\
\frac{\delta\alpha_{\text{em}}}{\alpha_{\text{em}}}&=
\frac{d}{dp^2}\Pi_T^{\gamma\gamma}(p^2)\Big|_{p^2=0}+\frac{2\hat{s}_W}{\hat{c}_W}\frac{\Pi_T^{\gamma Z}(0)}{m_Z^2},\label{ct2}\\
\frac{\delta m_W^2}{m_W^2} &=\frac{\Pi_T^{WW}(m_W^2)}{m_W^2}\label{ct3}, 
\end{align}
and in Ref.~\cite{blank_hollik} as 
\begin{align}
\frac{\delta \hat{s}_W^2}{\hat{s}_W^2} = \frac{\hat{c}_W}{\hat{s}_W}\frac{\Pi_T^{\gamma Z}(m_Z^2)}{m_Z^2}-\delta_{V}', 
\end{align}
where $\delta_{VB}$ and $\delta_{V}'$ are the vertex and the box diagram corrections
to $G_F$ and the radiative corrections to the $Z\bar{e}e$ vertex, respectively.  
These are calculated as \cite{blank_hollik,hhkm} 
\begin{align}
\delta_{VB} & = \frac{\alpha_{\text{em}}}{4\pi \hat{s}_W^2}\left[6+\frac{10-10\hat{s}_W^2-3(R/\hat{c}_W^2)(1-2\hat{s}_W^2)}{2(1-R)}\ln R\right],\quad R\equiv\frac{m_W^2}{m_Z^2},\\
\delta_{V}' & = \frac{v_e}{2\hat{s}_W^2}\left[\frac{\Gamma_V^{Z\bar{e}e}(m_Z^2)}{v_e}-\frac{\Gamma_A^{Z\bar{e}e}(m_Z^2)}{a_e}\right], 
\end{align}
where $\Gamma_{V,A}^{Z\bar{e}e}$ are defined through the renormalized $Z\bar{e}e$ vertex as \cite{blank_hollik}
\begin{align}
\hat{\Gamma}_\mu^{Z\bar{e}e}(m_Z^2)&=i\frac{g}{2\hat{c}_W}\left[(v_e+a_e\gamma_5)\gamma_\mu+\gamma_\mu\Gamma_V^{Z\bar{e}e}(m_Z^2)
+\gamma_\mu\gamma_5\Gamma_A^{Z\bar{e}e}(m_Z^2)\right],\\
v_e&=-\frac{1}{2}+2\hat{s}_W^2,\quad a_e = -\frac{1}{2}. \label{ren_zee}
\end{align}

In Eqs.~(\ref{ct1})-(\ref{ct3}), $\Pi_T^{XY}(p^2)$ ($X,Y=W,Z,\gamma$) are the 1PI diagrams for the gauge boson self-energies. 
We show the list of all the analytic expressions for the gauge boson self-energies in the HTM with the $Y=1$ triplet field in Appendix~E. 
The radiative correction $\Delta r$ can be obtained by 
\begin{align}
\Delta r &= \frac{\Pi_T^{WW}(0)-\Pi_T^{WW}(m_W^2)}{m_W^2}+\frac{d}{dp^2}\Pi_T^{\gamma\gamma}(p^2)\Big|_{p^2=0}
+\frac{2\hat{s}_W}{\hat{c}_W}\frac{\Pi_T^{\gamma Z}(0)}{m_Z^2}-\frac{\hat{c}_W}{\hat{s}_W}
\frac{\Pi_T^{\gamma Z}(m_Z^2)}{m_Z^2}\notag\\
&+\delta_{VB}+\delta_{V}'. 
\end{align}
From Eqs.~(\ref{rho_triplet}) and (\ref{gf}), 
the renormalized W boson mass as well as the renormalized rho parameter are given by
\begin{align}
m_W^2&= \frac{\pi \alpha_{\text{em}}}{\sqrt{2}G_F\hat{s}_W^2}(1+\Delta r),\\
\rho &=  \frac{\pi \alpha_{\text{em}}}{\sqrt{2}G_Fm_Z^2\hat{s}_W^2\hat{c}_W^2}(1+\Delta r). 
\end{align}
Therefore, with four input parameters ($\alpha_{\text{em}}$, $G_F$, $m_Z$ and $\hat{s}_W^2$), 
$\Delta r$ determines both the one-loop corrected mass of the W boson and the rho parameter in the HTM. 

\begin{figure}[t]
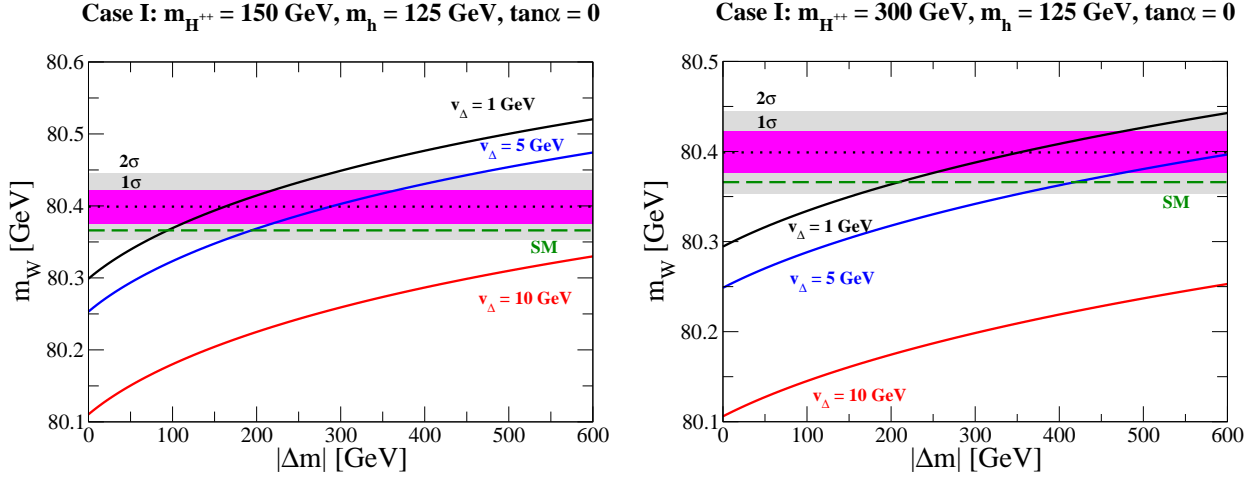

\begin{center}
\includegraphics[width=80mm]{mw_IH_ml150_new.eps}\hspace{3mm}
\includegraphics[width=80mm]{mw_IH_ml300_new.eps}
\end{center}
\caption{The one-loop corrected values of $m_W$ as a function of the absolute value of $\Delta m$ 
for each fixed value of $v_\Delta$ in Case I ($m_{\phi^0} > m_{H^{+}}>m_{H^{++}}$). 
We take $m_h=125$ GeV, $m_t=173$ GeV and $\tan\alpha=0$ in the both figures. 
The pink (gray) shaded region represents the 1$\sigma$ (2$\sigma$) error for the experimental data of $m_W^{\text{exp}}$ 
(=80.399 $\pm$ 0.23 GeV~\cite{PDG}). 
In the left (right) figure, we take $m_{H^{++}}=$ 150 GeV (300 GeV). 
The dashed line shows the SM prediction of $m_W$ at the one-loop level with the SM Higgs boson mass to be 125 GeV~\cite{ky_rho}. }
\label{mw1}
\end{figure}

In the following, we show numerical results for the radiative corrections to $m_W^2$ as well as $\rho$ 
in the HTM. 
The radiative correction depends on the mass spectrum and the mixing angle 
in the Higgs potential; i.e., $m_{H^{++}}$, $m_{H^{+}}$, $m_A$, $m_H$, $m_h$ and $\tan\alpha$.  
Although these six parameters are all free parameters, 
under the requirement of $v_\Delta^2\ll v^2$ the approximated formulae given in Eqs.~(\ref{mass1}) and (\ref{mass2})
tell us to pick up the following three parameters such as the mass of the SM-like Higgs boson $m_h$, 
the mass of the lightest triplet-like scalar boson $m_A$ (or $m_{H^{++}}$) and 
the mass difference $\Delta m$ between $H^{\pm\pm}$ and $H^\pm$. 
In the following analysis, we consider the scenarios with mass splitting for the triplet-like Higgs bosons; 
namely, for Case I ($m_{\phi^0} >m_{H^+}>m_{H^{++}}$) and Case II ($m_{H^{++}}>m_{H^+}>m_{\phi^0}$). 
We take pole masses of the top quark $m_t=173$ GeV and  
the bottom quark $m_b=4.7$ GeV and $\alpha_s(m_Z)=0.118$~\cite{PDG}. 
We take into account the leading order QCD correction in the calculation of the one-loop corrected $m_W$ as well as the rho parameter.

In Fig.~\ref{mw1}, 
the renormalized value of $m_W$ is shown as a function of $|\Delta m|$ for several values of $v_\Delta$ 
in Case I with the data $m_W^{\text{exp}}=80.399\pm 0.023$ GeV~\cite{PDG}. 
The mass of the SM-like Higgs boson $h$ is taken as $m_h=125$ GeV, and the mixing angle $\tan\alpha$ is set on zero. 
The mass of the lightest triplet-like Higgs boson $m_{H^{++}}$ is taken to be 150 GeV (left figure) 
and 300 GeV (right figure). 
It is seen that the predicted value of $m_W$ for the degenerate mass case ($|\Delta m|=0$) is 
outside the region within the $2\sigma$ error. 
The prediction satisfies the data when $|\Delta m|$ has a non-zero value. 
When $m_{H^{++}}=150$ GeV, the favored value of $|\Delta m|$ by the data within the $2\sigma$ error 
is $80$ GeV$\lesssim |\Delta m|\lesssim 280$ GeV ($190$ GeV$\lesssim |\Delta m|\lesssim430$ GeV) for 
$v_\Delta=1$ GeV (5 GeV). 
The preferred value of $|\Delta m|$ for smaller values of $v_\Delta$ than 1 GeV is similar to that for $v_\Delta=1$ GeV. 
When $m_{H^{++}}=300$ GeV, the allowed values of $|\Delta m|$ are larger than the case of $m_{H^{++}}=150$ GeV for 
the same value of $v_\Delta$. 
Smaller mass splitting which satisfies the data corresponds to the smaller value of $v_\Delta$ while 
largest value of $|\Delta m|$ ($\sim$ 500-600 GeV), which comes from perturbative unitarity, corresponds to 
$v_\Delta\sim \mathcal{O}(10)$ GeV. 
We note that the result is almost unchanged even if we vary $\tan\alpha$ in the region of $0<\tan\alpha<1$ as long as 
$H^{\pm\pm}$ is not too heavy.  

\begin{figure}[t]
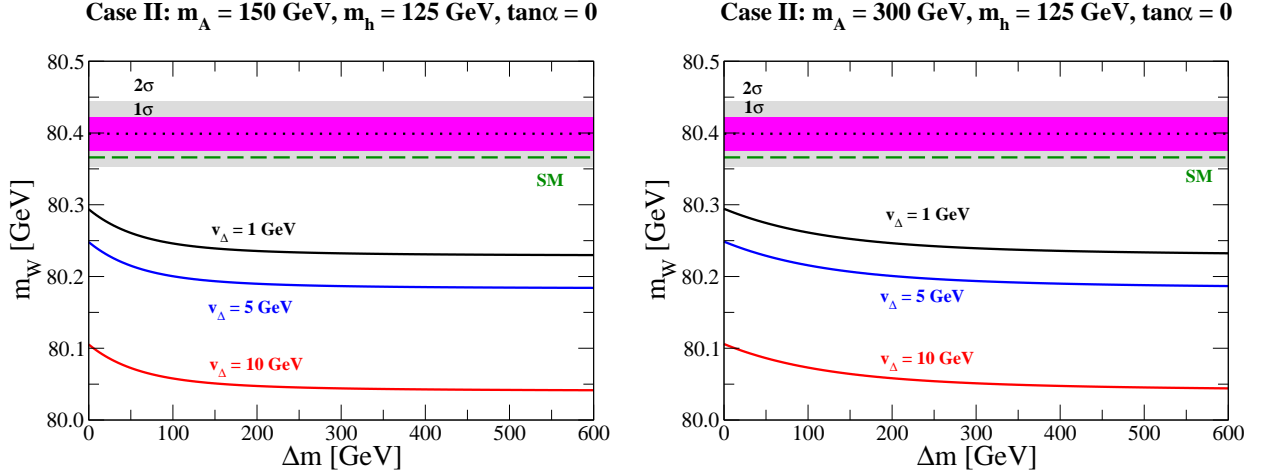

\begin{center}
\includegraphics[width=80mm]{mw_NH_ml150_new.eps}\hspace{3mm}
\includegraphics[width=80mm]{mw_NH_ml300_new.eps}
\end{center}
\caption{The one-loop corrected values of $m_W$ as a function of $\Delta m$ 
for each fixed value of $v_\Delta$ in Case II ($m_{H^{++}}>m_{H^+}>m_{\phi^0}$). 
We take $m_h=125$ GeV, $m_t=173$ GeV and $\tan\alpha=0$ in the both figures. 
The pink (gray) shaded region represents the 1$\sigma$ (2$\sigma$) error for the experimental data of $m_W^{\text{exp}}$ 
(=80.399 $\pm$ 0.23 GeV~\cite{PDG}). 
In the left (right) figure, we take $m_{A}=$ 150 GeV (300 GeV). 
The dashed line shows the SM prediction of $m_W$ at the one-loop level with the SM Higgs boson mass to be 125 GeV~\cite{ky_rho}. }
\label{mw2}
\end{figure}

\begin{figure}[t]
\begin{center}
\includegraphics[width=70mm]{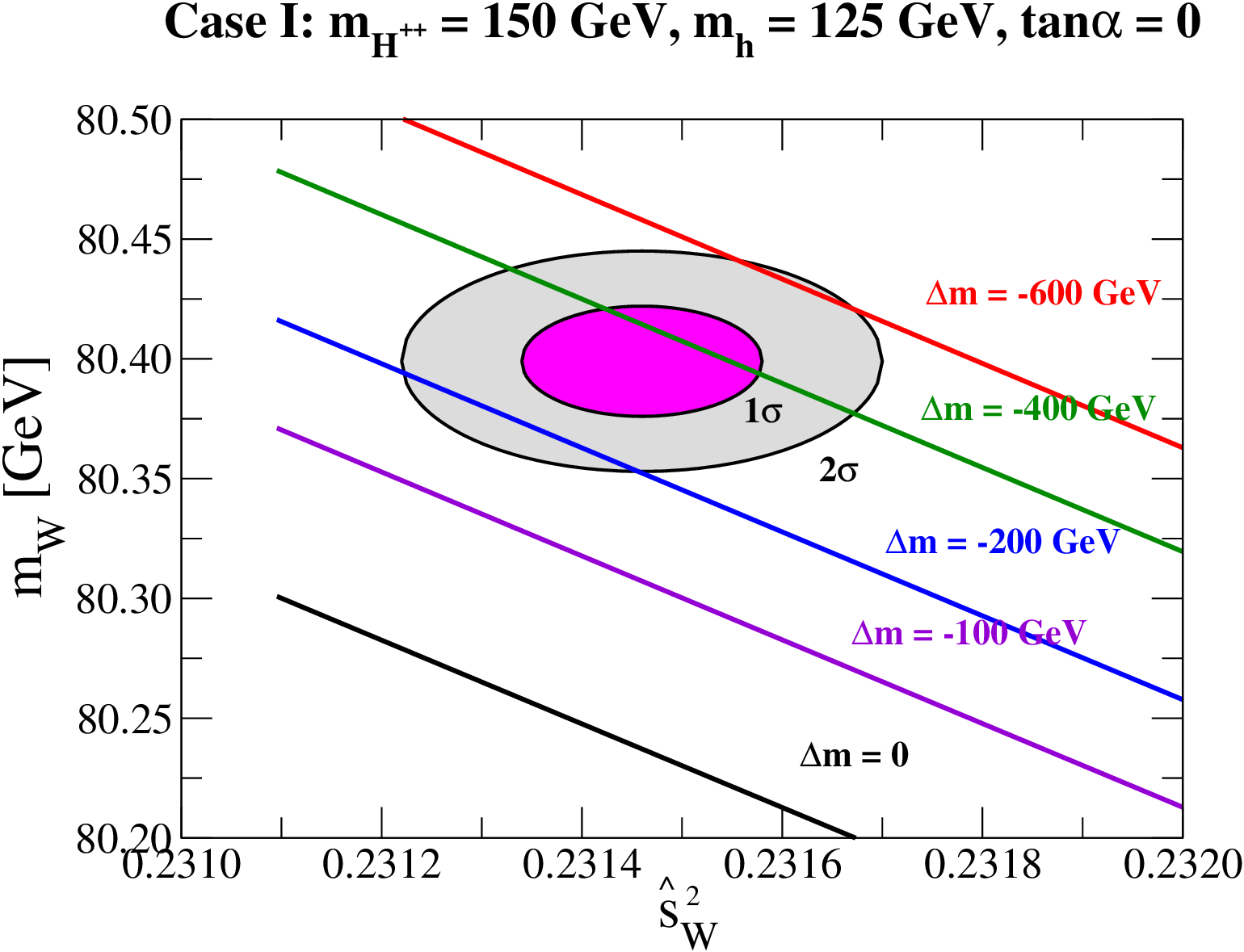}\hspace{3mm}
\includegraphics[width=70mm]{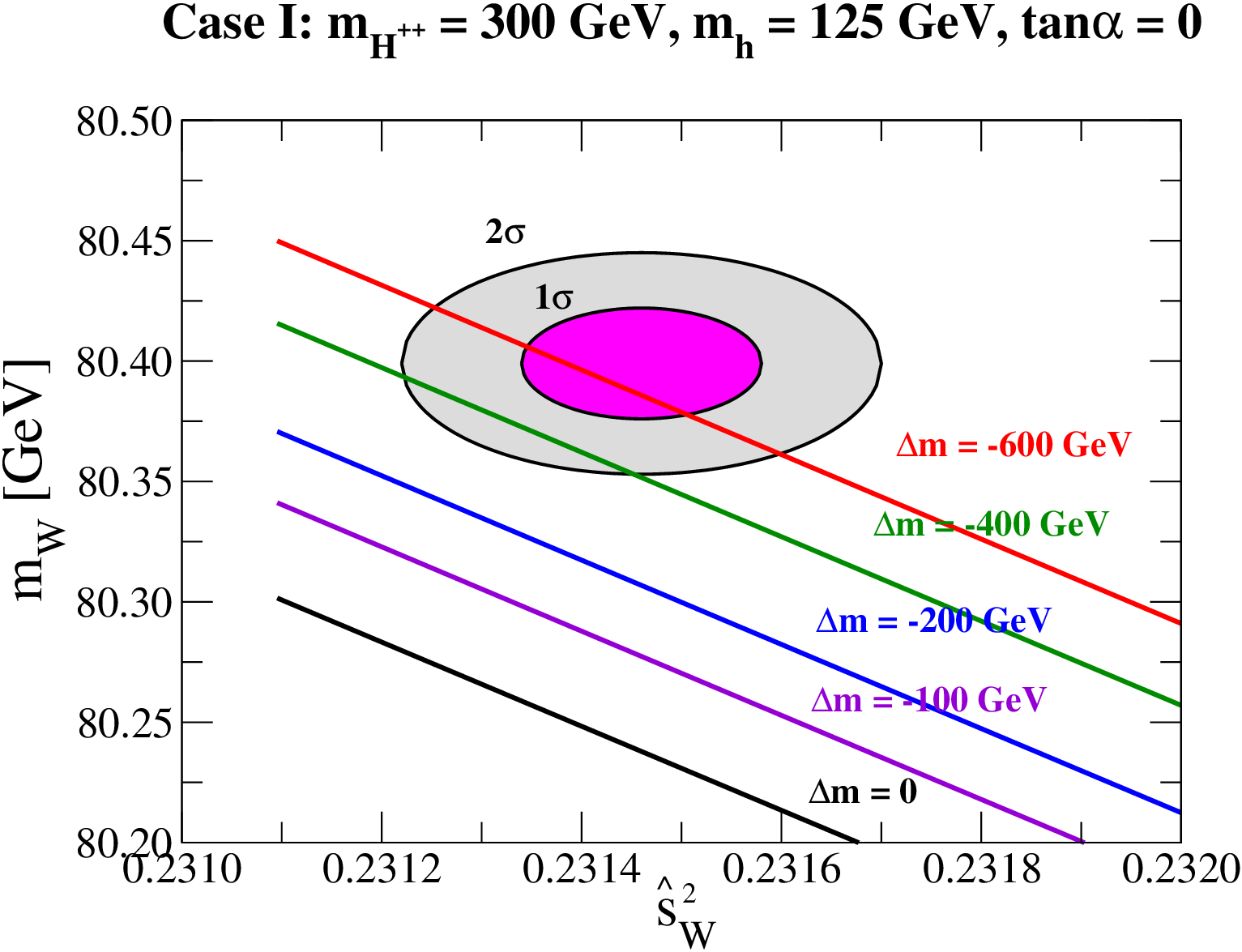}\\\vspace{6mm}
\includegraphics[width=70mm]{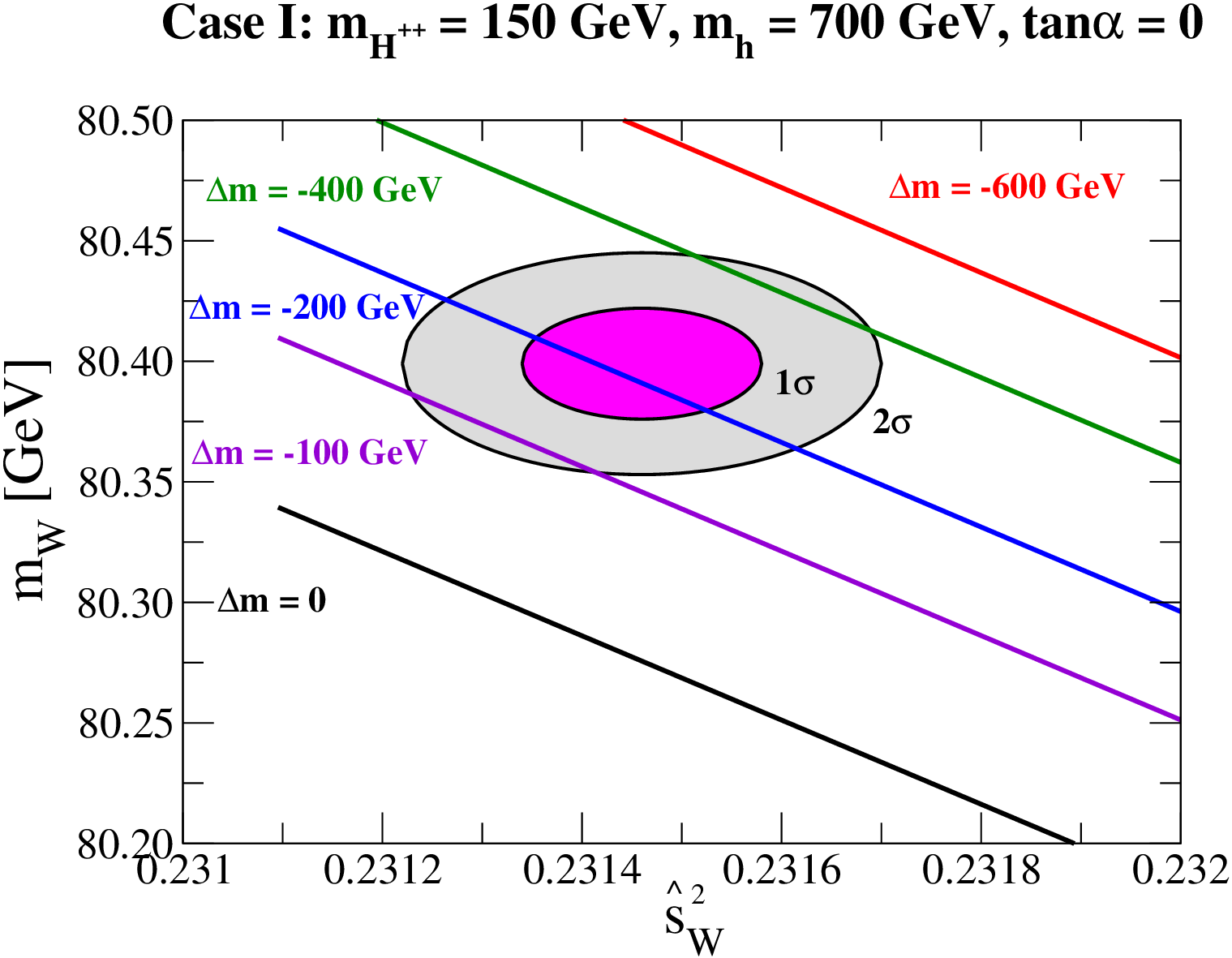}\hspace{3mm}
\includegraphics[width=70mm]{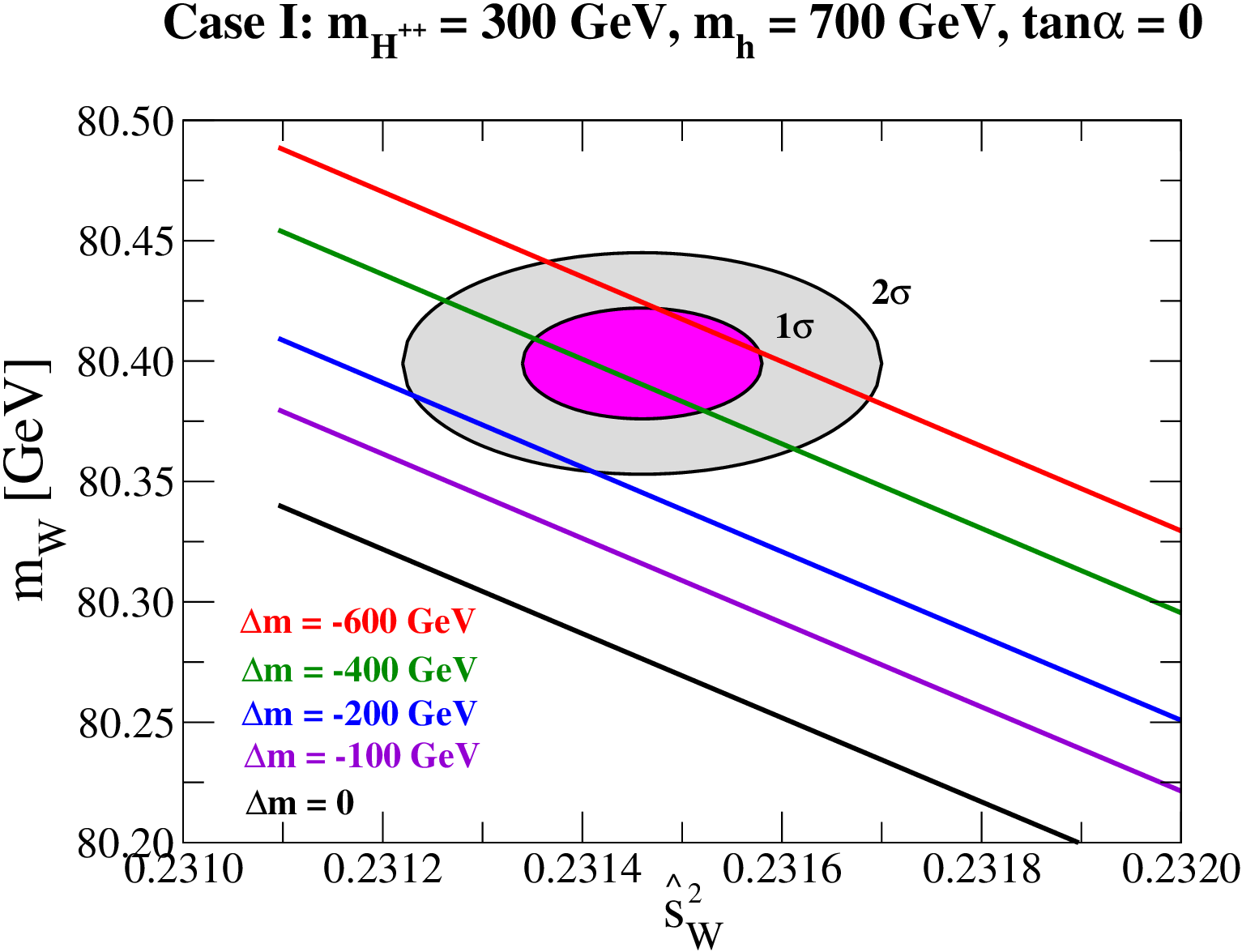}
\end{center}
\caption{The one-loop corrected values of $m_W$ as a function of $\hat{s}_W^2$ in Case I ($m_{\phi^0} > m_{H^{+}}>m_{H^{++}}$). 
We take $m_t=173$ GeV and $\tan\alpha=0$ in all the figures. 
The pink (gray) shaded region represents the 1$\sigma$ (2$\sigma$) error for the experimental data of 
$m_W^{\text{exp}}$ (=80.399 $\pm$ 0.23 GeV~\cite{PDG}) and $\hat{s}_W^{2\text{ exp}}$ (=0.23146 $\pm$ 0.00012 GeV~\cite{PDG}). 
In the left (right) two figures, we take $m_h=$ 125 GeV (700 GeV). 
The mass of the lightest triplet-like Higgs boson is taken to be 150 GeV and 300 GeV~\cite{ky_rho}. }
\label{mw3}
\end{figure}
In Fig.~\ref{mw2}, 
the renormalized value of $m_W$ is shown as a function of $\Delta m$ for several values of $v_\Delta$ 
in Case II with the data $m_W^{\text{exp}}=80.399\pm 0.023$ GeV~\cite{PDG}. 
The mass of the SM-like Higgs boson $h$ is taken as $m_h=125$ GeV, and the mixing angle $\tan\alpha$ is set on zero. 
The mass of the lightest triplet-like Higgs boson $m_A$ is taken to be 150 GeV (left figure) 
and 300 GeV (right figure). 
It is found that Case II is strongly constrained by the electroweak precision data for entire range of $\Delta m$. 
The situation is worse for larger values of $\Delta m$ and also for larger values of $v_\Delta$. 
Therefore, Case I can be more consistent with the electroweak precision data than the degenerate mass case and also Case II. 
We note that the result is almost unchanged even if we vary $\tan\alpha$ in the region of $0<\tan\alpha<1$ as long as 
$A$ is not too heavy.

\begin{figure}[t]
\begin{center}
\includegraphics[width=70mm]{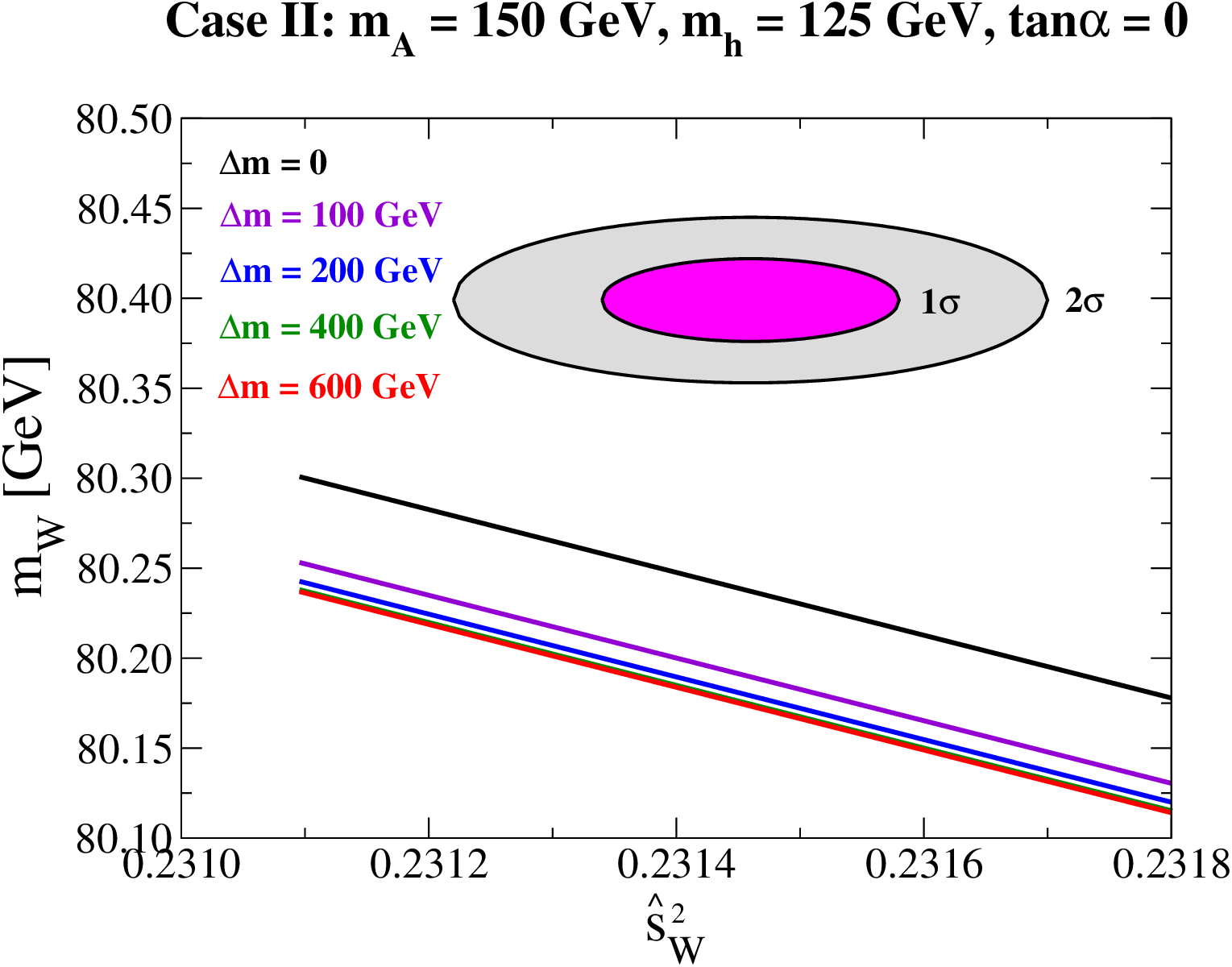}\hspace{3mm}
\includegraphics[width=70mm]{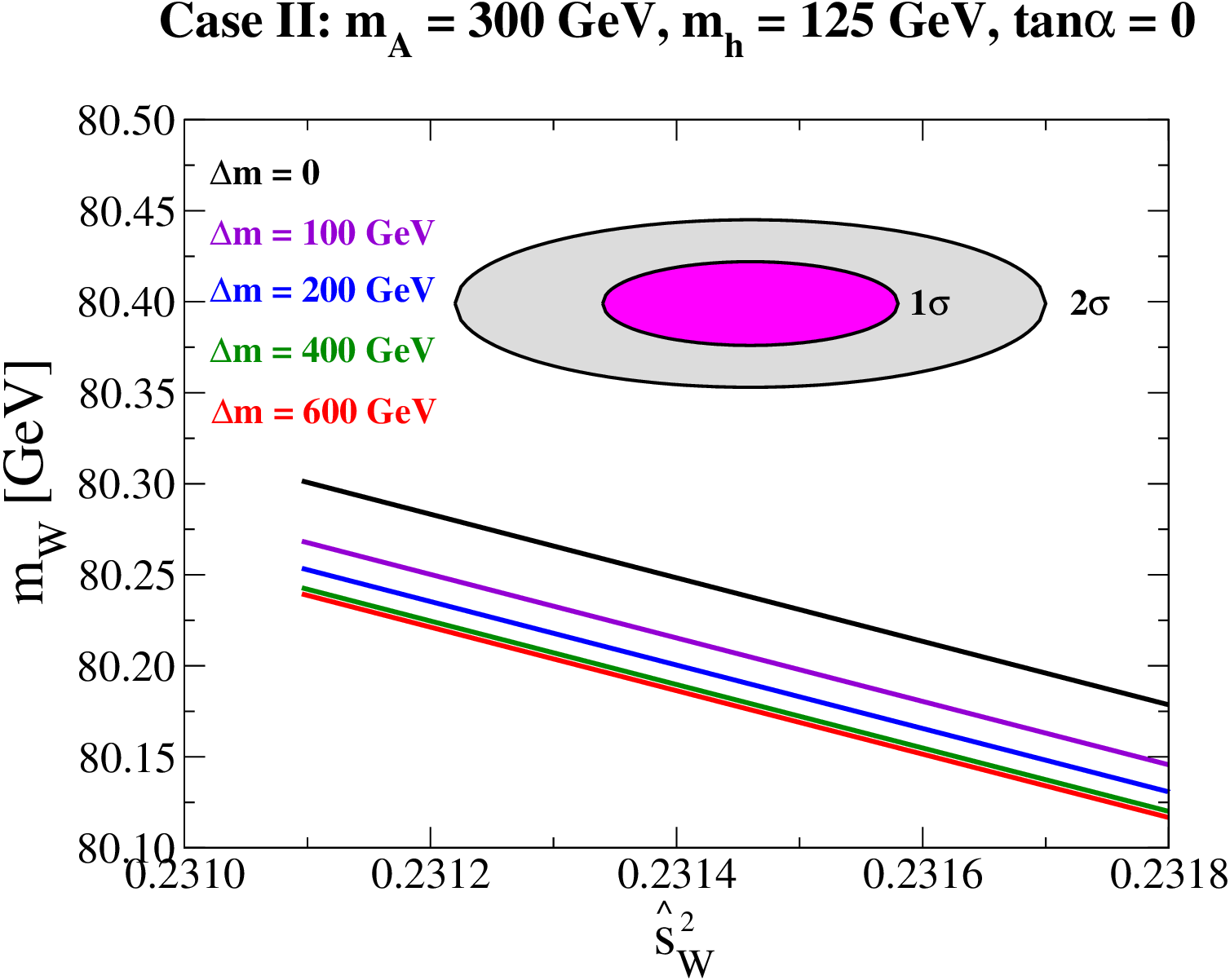}\\\vspace{6mm}
\includegraphics[width=70mm]{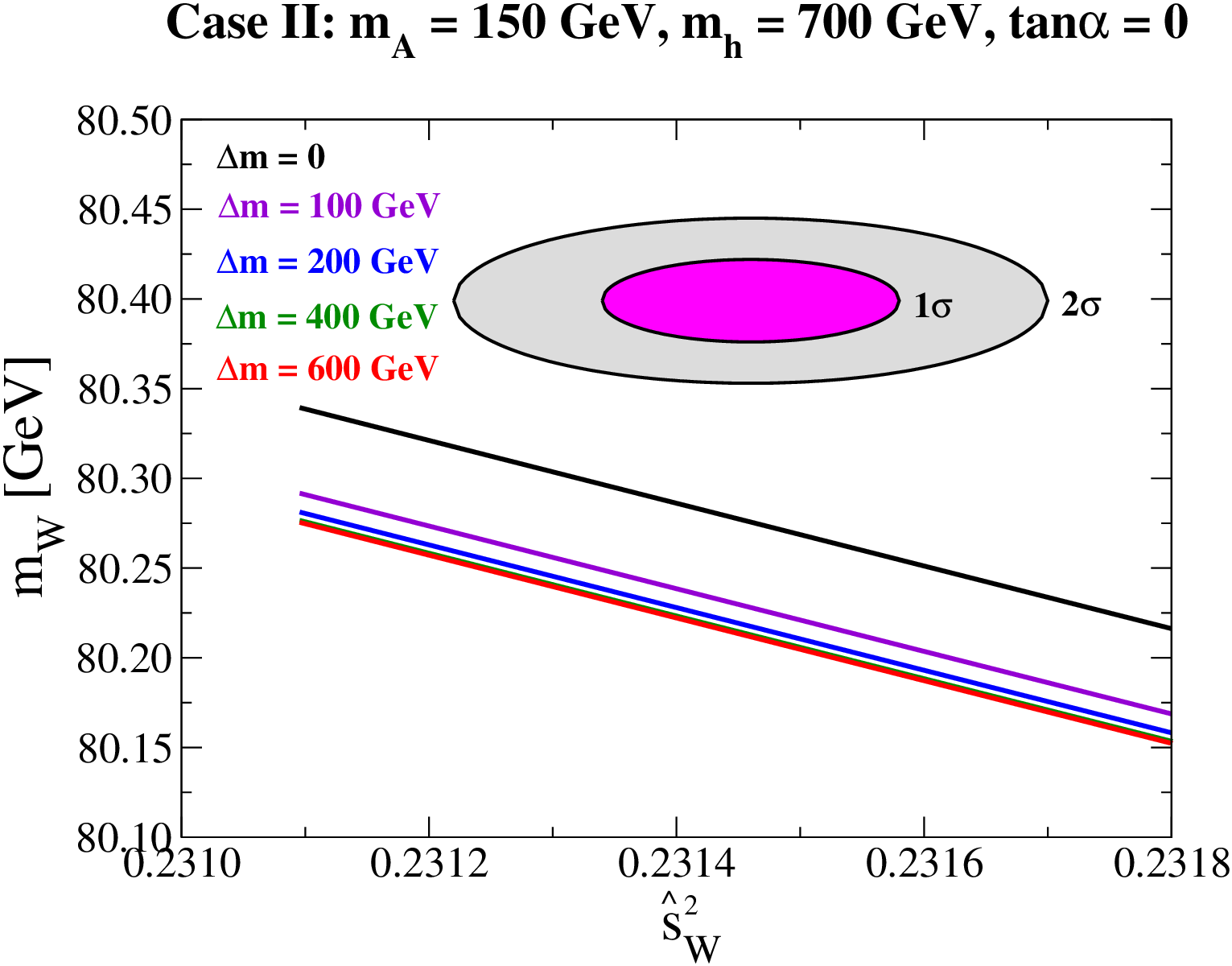}\hspace{3mm}
\includegraphics[width=70mm]{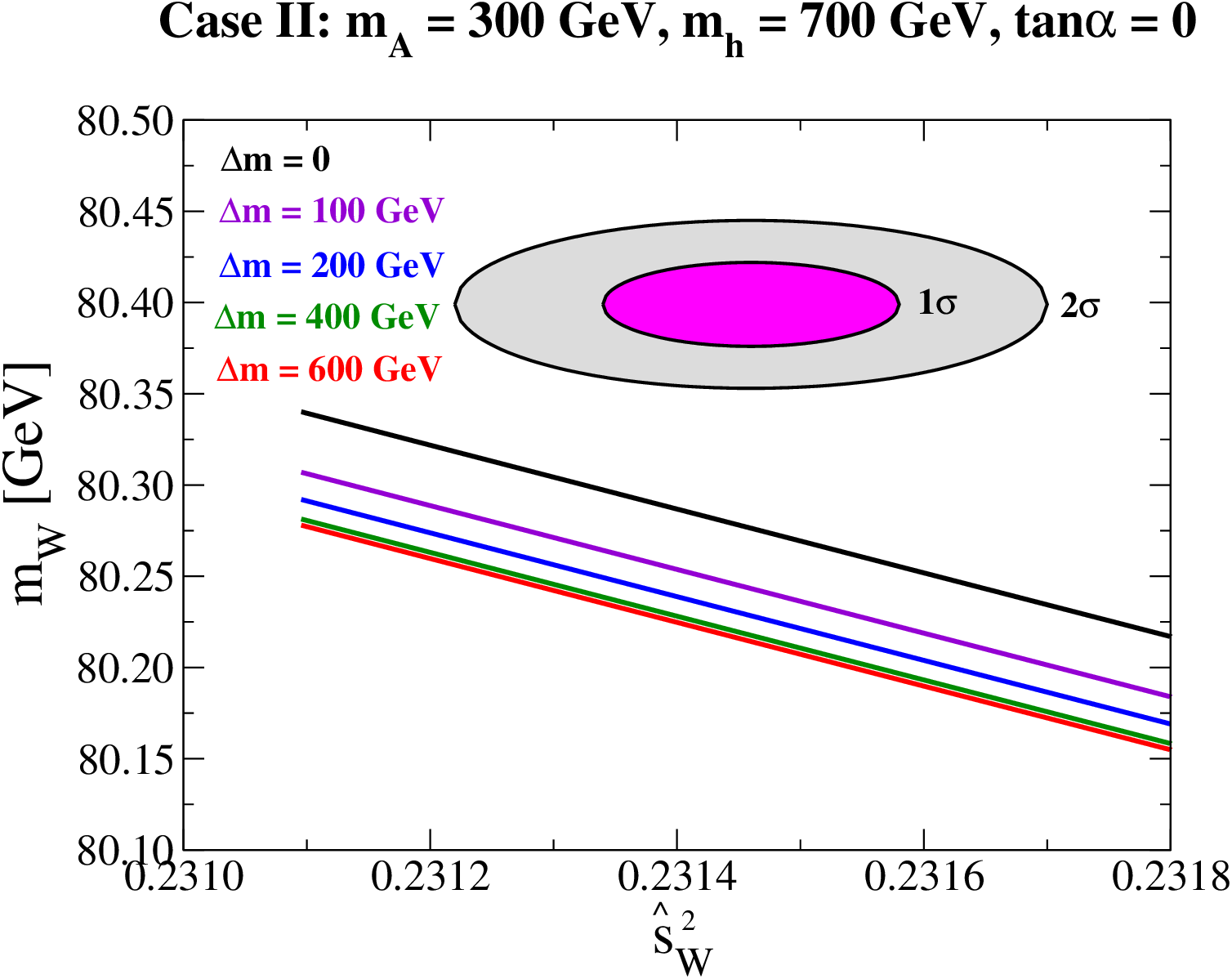}
\end{center}
\caption{The one-loop corrected values of $m_W$ as a function of $\hat{s}_W^2$ in Case II ($m_{H^{++}}>m_{H^+}>m_{\phi^0}$). 
We take $m_t=173$ GeV and $\tan\alpha=0$ in all the figures. 
The pink (gray) shaded region represents the 1$\sigma$ (2$\sigma$) error for the experimental data of 
$m_W^{\text{exp}}$ (=80.399 $\pm$ 0.23 GeV~\cite{PDG}) and $\hat{s}_W^{2\text{ exp}}$ (=0.23146 $\pm$ 0.00012 GeV~\cite{PDG}). 
In the left (right) two figures, we take $m_h=$ 125 GeV (700 GeV). 
The mass of the lightest triplet-like Higgs boson is taken to be 150 GeV and 300 GeV~\cite{ky_rho}. }
\label{mw4}
\end{figure}
In Fig.~\ref{mw3}, 
we show the renormalized values for $m_W$ for each value of $\Delta m$ as a function of the input parameter $\hat{s}_W^2$ in Case I.  
The values of $(m_{H^{++}},m_h)$ are taken to be (150 GeV,125 GeV), (300 GeV,125 GeV), (150 GeV,700 GeV) and (300 GeV,700 GeV) 
in the figures located at the upper left, the upper right, the lower left and the lower right panels, respectively. 
In all figures, the mixing angle $\tan\alpha$ is set to be zero. 
Regions indicated by the data of $m_W$ and $\hat{s}_W^2$ within the $1\sigma$ error and the $2\sigma$ error are 
also shown in each figure for the fixed value of $m_t$ (=173 GeV). 
When $m_h=125$ GeV (upper figures), the predicted values of $m_W$ for $\Delta m =0$ are 
far from the allowed region by the data. 
For $m_{H^{++}}=150$ GeV and $m_h=125$ GeV (upper left), 
the prediction is consistent with the data within the $2\sigma$ error when about
$160$ GeV$\lesssim |\Delta m| \lesssim 600$ GeV is taken. 
On the other hand, for $m_{H^{++}}=300$ GeV and $m_h=125$ GeV (upper right), 
smaller values are predicted for $m_W$ as compared to the case with $m_{H^{++}}=150$ GeV for 
non-zero value of $|\Delta m|$. 
They approach to the predicted values of $m_W$ with $|\Delta m|=0$ in the large mass limit for $H^{\pm\pm}$. 
It is consistent with the data when we take $\Delta m\gtrsim 400$ GeV in this case. 
When $m_h=700$ GeV (lower figures), the predicted values of $m_W$ for $\Delta m =0$ are 
far from the allowed region by the data but closer than the case of $m_h=125$ GeV. 
For $m_{H^{++}}=150$ GeV and $m_h=700$ GeV (lower left), 
the prediction is consistent with the data within the $2\sigma$ error when about 
$100$ GeV$\lesssim |\Delta m| \lesssim 400$ GeV is taken. 
On the other hand, for $m_{H^{++}}=300$ GeV and $m_h=700$ GeV (lower right), 
it is consistent with the data when we take $\Delta m\gtrsim 200$ GeV in this case. 
The edge of each curve at $\hat{s}_W^2\simeq 0.2311$ corresponds to $v_\Delta=0$. 
We note that the result is almost unchanged even if we vary $\tan\alpha$ in the region of $0<\tan\alpha<1$ as long as 
$H^{\pm\pm}$ is not too heavy.

In Fig.~\ref{mw4}, 
we show the renormalized values for $m_W$ for each value of $\Delta m$ as a function of the input parameter $\hat{s}_W^2$ in Case II.  
The values of $(m_A,m_h)$ are taken to be (150 GeV,125 GeV), (300 GeV,125 GeV), (150 GeV,700 GeV) and (300 GeV,700 GeV) 
for the figures located at the upper left, the upper right, the lower left and the lower right, respectively. 
In all figures, the mixing angle $\alpha$ is set to be zero. 
Regions indicated by the data within the $1\sigma$ error and the $2\sigma$ error are 
also shown in each figure for the fixed value of $m_t$ (=173 GeV). 
For $m_{A}=150$ GeV and $m_h=125$ GeV (upper left), 
the predicted values for $m_W$ with non-zero $\Delta m$ ($>0$) are 
smaller than that with $\Delta m=0$. 
The situation is unchanged for the other choices of 
($m_A$,$m_h$)$=$(300 GeV,125 GeV), (150 GeV,700 GeV) and (300 GeV,700 GeV). 
Therefore, the hierarchical scenario with non-zero $\Delta m$ is 
highly constrained by the combined data of $m_W$ and $\hat{s}_W^2$. 
The edge of each curve at $\hat{s}_W^2\simeq 0.2311$ corresponds to $v_\Delta=0$. 
We note that the result is almost unchanged even if we vary $\tan\alpha$ in the region of $0<\tan\alpha<1$ as long as 
$A$ is not too heavy. 

\begin{figure}[t]
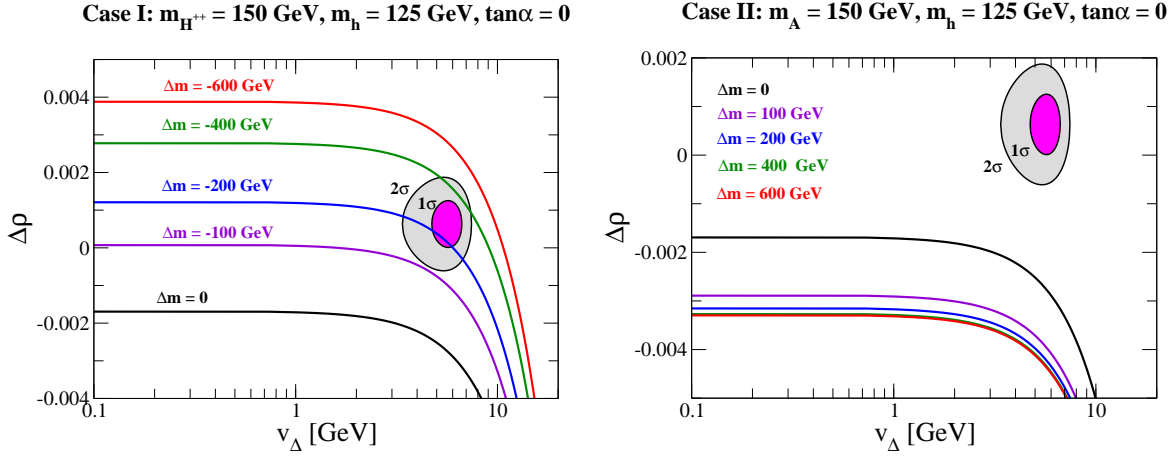

\begin{center}
\includegraphics[width=75mm]{rho_IH_new.eps}\hspace{3mm}
\includegraphics[width=75mm]{rho_NH_new.eps}
\end{center}
\caption{The deviation of the one-loop corrected values of the rho parameter from 
those of the SM one-loop prediction 
as a function of $v_\Delta$. 
We take $m_h=125$ GeV, $m_t=173$ GeV and $\tan\alpha=0$ in the both figures. 
The pink (gray) shaded region represents the 1$\sigma$ (2$\sigma$) error for the experimental data of 
$\Delta \rho^{\text{exp}}$ (=0.000632$\pm$0.000621) which is derived from the data of the $T$ parameter (=0.07$\pm$0.08~\cite{PDG}). 
In the left figure, the mass hierarchy of the triplet-like Higgs bosons is taken to be Case I 
($m_{\phi^0} > m_{H^{+}}>m_{H^{++}}$), and $m_{\phi^0}$ is fixed to be 150 GeV. 
In the right figure, the mass hierarchy of the triplet-like Higgs bosons is taken to be Case II 
($m_{H^{++}}>m_{H^+}>m_{\phi^0}$), and $m_{H^{++}}$ is fixed to be 150 GeV~\cite{ky_rho}.}
\label{rho}
\end{figure}

\begin{figure}[t]
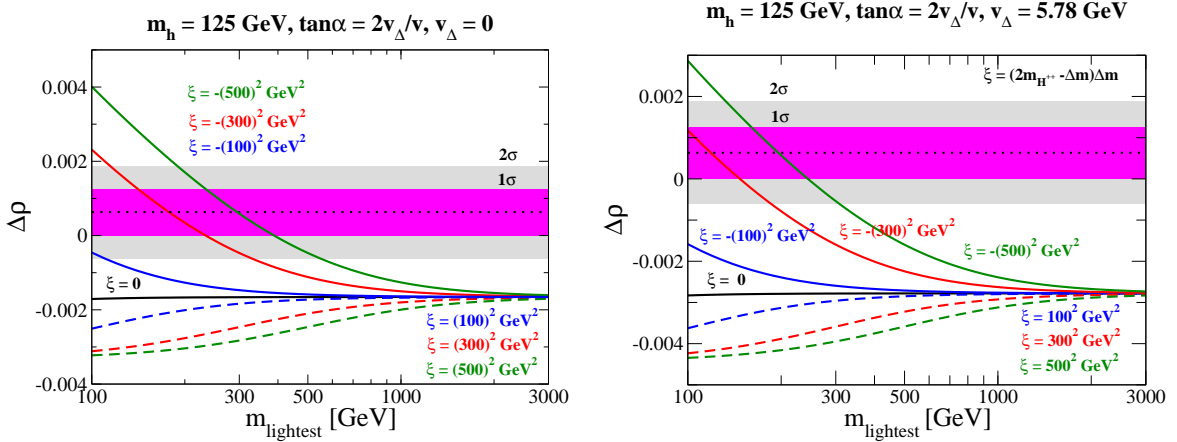

\begin{center}
\includegraphics[width=75mm]{rho_as_vd0.eps}\hspace{3mm}
\includegraphics[width=75mm]{rho_as_new.eps}

\end{center}
\caption{The deviation of the one-loop corrected values of the rho parameter from 
those of the SM one-loop prediction 
as a function of the mass of the lightest triplet-like Higgs boson $m_{\text{lightest}}$ 
for each fixed value of $\xi$ ($\equiv m_{H^{++}}^2-m_{H^+}^2$). 
We take $m_t=173$ GeV, $m_h=125$ GeV and $\tan\alpha=2v_\Delta/v$ in the both figures. 
In the left (right) figure, $v_\Delta$ is taken to be 0 (5.78 GeV). 
The pink (gray) shaded region represents the 1$\sigma$ (2$\sigma$) error for the experimental data of 
$\Delta \rho^{\text{exp}}$ (=0.000632$\pm$0.000621) which is derived from the data of the 
$T$ parameter (=0.07$\pm$0.08~\cite{PDG})~\cite{ky_rho}. 
}
\label{rho_as}
\end{figure}

In Fig.~\ref{rho}, the deviation in the one-loop corrected rho parameter in the HTM
from that of the SM one-loop prediction ($\Delta \rho\equiv \rho-\rho_{\text{SM}}(m_h^{\rm ref})$)
is shown as a function of $v_\Delta$, 
where $v_\Delta$ is defined in Eq.~(\ref{data}). 
In order to describe the allowed region of $\Delta \rho$, 
we employ the data for the electroweak $T$ parameter~\cite{Peskin-Takeuchi} of 
$T=0.07\pm 0.08$~\cite{PDG}, in which $T=0$ is chosen for the reference value of the SM Higgs boson mass $m_h^{\rm ref}$ 
to be 117 GeV and $m_t=173$ GeV. 
We then obtain $\Delta\rho^{\text{exp}}=0.000632\pm 0.000621$, where $m_h^{\rm ref}=125$ GeV is chosen by taking into account 
the recent direct search results at the LHC~\cite{LHC-Higgs}.  
In the left figure, the results in Case I are shown, while in the right figure 
those in Case II are shown for several values of $\Delta m$.  
The mass of the SM-like Higgs boson is taken to be $m_h=125$ GeV, and the mixing angle $\tan\alpha$ is set to be zero. 
In Case I (left figure), the predicted values of $\Delta \rho$ for $\Delta m=0$ are 
outside the region within the $2\sigma$ error under the data 
$\Delta\rho^{\text{exp}}$ and $\hat{s}_W^2$ given in Eq.~(\ref{data}) with $m_t=173$ GeV. 
But the effect of non-zero $|\Delta m|$ makes $\Delta\rho$ larger. 
The allowed value for $v_\Delta$ within the $2\sigma$ error is 
about 3.5 GeV $\lesssim v_\Delta\lesssim 8$ GeV for about 100 GeV $\lesssim |\Delta m|\lesssim 440$ GeV.  
Notice that, as shown in Fig.~\ref{mw3}, 
the favored value of $|\Delta m|$ from the data of $m_W$ and $\hat{s}_W^2$ is about 
$200$ GeV $\lesssim|\Delta m|\lesssim 600$ GeV in Case I.  
Therefore,  
we may conclude that the combined data indicate the favored value of $v_\Delta$ to be 3.5-8 GeV in Case I 
with $m_{H^{++}}=150$ GeV. 
Next, the result in Case II is shown in the right figure, where 
the effect of $\Delta m$ ($>0$) gives the negative contribution to $\Delta \rho$. 
However, it can be seen that 
Case II is already highly constrained by the data of $m_W$ and $\hat{s}_W^2$ 
with $m_A=150$ GeV.

We here give a comment on the decoupling property of the heavy triplet-like Higgs bosons in the electroweak observables. 
In Fig.~\ref{rho_as}, we show $\Delta\rho$ as a function of the lightest of all the triplet-like Higgs bosons for each value $\xi$ 
($\equiv m_{H^{++}}^2-m_{H^+}^2$). 
We again take $m_h=125$ GeV and $m_t=173$ GeV. 
In the left figure, $v_\Delta$ is fixed to be 0 
while in the right figure, $v_\Delta$ is fixed to be the central value indicated by the data ($v_\Delta=5.78$ GeV). 
In this figure, $\tan\alpha$ is chosen to be $2v_\Delta/v$, 
which is the asymptotic value in the limit of $m_{\text{lightest}}\to\infty$. 
It can be seen that the one-loop contribution of these particles decouples in the large mass limit even in the case with $v_\Delta=0$. 
The asymptotic value in this limit is determined in the renormalization scheme 
with the four input parameters $\alpha_{\text{em}}$, $G_F$, $m_Z$ and $\hat{s}_W^2$ without 
the tree level relation of $m_W^2=\hat{c}_W^2 m_Z^2$. 
Therefore, it is not surprising that the asymptotic value in the HTM does not coincide with the SM value 
($\Delta\rho=0$ in this figure) with the three input parameters with $m_W^2=\hat{c}_W^2 m_Z^2$. 
In the large mass limit and $v_\Delta\to 0$, the one-loop corrected rho parameter can be expressed as 
\begin{align}
\rho_{\text{HTM}}^{\text{as}}\xrightarrow[m_{\text{lightest}}\to \infty]{}
1+\frac{g^2N_c}{16\pi^2}\Big[&\frac{2}{3}\ln m_t-\frac{1}{9}+\frac{4}{9}\hat{s}_W^2Q_t^2-4I_tQ_t\left(\frac{1}{3}\ln m_t+\frac{1}{18}\right)\notag\\
&+\frac{8}{3}\hat{s}_W^2Q_b^2\ln m_b-\frac{1}{9}(-4\hat{s}_W^2Q_b^2+2I_bQ_b)(-5+6\ln m_Z)\Big]\notag\\
&+0.0027+\delta_{VB}+\delta_{V}'\notag\\
&\simeq 1.00834, \label{rho_HTM_as}
\end{align}
where 0.0027 is the contribution from the SM bosonic loop. 
On the other hand, the one-loop corrected rho parameter in the SM can be expressed as 
\begin{align}
\rho_{\text{SM}}\simeq 1+\frac{s_W^2}{c_W^2-s_W^2}\left[\frac{c_W^2}{s_W^2}\frac{N_c\sqrt{2}}{16\pi^2}G_Fm_t^2-\delta_{VB}\right]
\simeq 1.0102. \label{rho_SM_top}
\end{align}
From Eqs.~(\ref{rho_HTM_as}) and (\ref{rho_SM_top}), it can be seen that 
the rho parameter in the HTM can deviate from that in the SM even in the large triplet-like scalar mass and $v_\Delta\to 0$ limit, 
and the value of $\rho_{\text{SM}}-\rho_{\text{HTM}}^{\text{as}}$is approximately 0.00186. 
In the SM and all the models with $\rho=1$ at the tree level, 
$\delta \rho$ ($\equiv \rho-1$) measures the violation of the custodial $SU(2)$ symmetry~\cite{Csym,Csym2}. 
Such effects 
appear as the quadratic power-like contributions of the mass difference between particles 
in the $SU(2)$ multiplet; i.e., 
$\delta\rho\simeq$ $(m_u-m_d)^2/v^2$ for $m_u\simeq m_d$ via the chiral fermion loop diagram~\cite{fermion_loop}, 
and $\delta\rho\simeq$ $(m_{H^+}-m_A)^2/v^2$ for $m_{H^+}\simeq m_A$ via the additional scalar boson loop diagram~\cite{
Toussaint,Bertolini,Peskin_Wells,Osland,Taniguchi} 
in the general two Higgs doublet model.
On the other hand, in the models with $\rho\neq 1$ at the tree level like the HTM, 
such quadratic power-like mass contributions are absorbed by 
renormalization of the new independent input parameter $\delta \hat{s}_W^2$.  
Consequently, only a logarithmic dependence on the masses of the particles in the 
loop diagram remain. 
In other words, in these models the rho parameter is no more the parameter which measures the violation 
of the $SU(2)$ custodial symmetry in the sector of particles in the loop. 
This has already been known in the calculation of the model with the $Y=0$ triplet field~\cite{blank_hollik,Chen:2005jx}.

\subsection{Decay of the triplet-like scalar bosons}

In the previous section, we have discussed the constraint from the electroweak 
precision data such as $m_W$ and the rho parameter. 
We have concluded that the hierarchical mass scenario for Case II as well as the degenerated mass scenario are 
highly constrained by the data. 
On the other hand, 
the hierarchical mass scenario for Case I is allowed by the data in the case where 
the mass of $H^{\pm\pm}$ is of $\mathcal{O}(100$-$200)$ GeV with $\Delta m$ to be several hundred GeV and $v_\Delta$ of several GeV. 
Although the degenerated mass scenario and Case II are highly constrained by the electroweak precision data\footnote{
In the minimal Higgs triplet model 
in which a doublet Higgs field with $Y=1/2$ and a triplet Higgs field with $Y=1$ are contained, 
these two cases are highly constrained. 
However, this constraint would be relaxed when we consider extended Higgs models such as 
the HTM with inert doublet scalar fields or inert triplet scalar fields.}, 
we discuss the decay branching ratios of the triplet-like scalar bosons 
in Case I, Case II and also the degenerate mass scenario.

\begin{figure}[t]
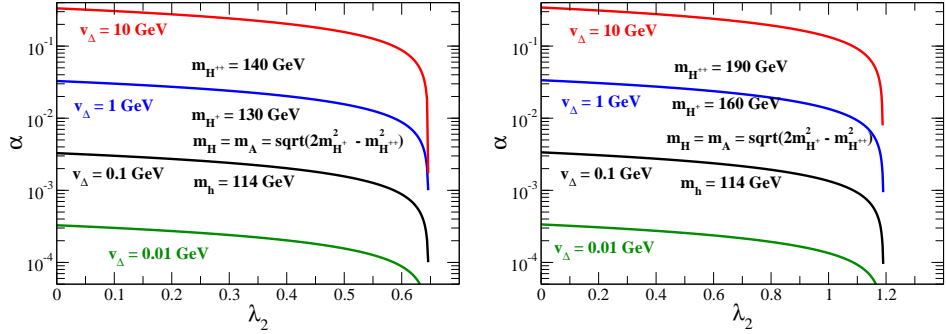

\begin{center}
\includegraphics[width=60mm]{alpha_lam2.eps}\hspace{3mm}
\includegraphics[width=60mm]{alpha_lam2_2.eps}
\caption{The mixing angle $\alpha$ as a function of $\lambda_2$ for each fixed value of $v_\Delta$ in the case of $m_h=114$ GeV. 
In the left (right) figure, we take $m_{H^{++}}=140$ GeV and $\Delta m=10$ GeV ($m_{H^{++}}=190$ GeV and $\Delta m=30$ GeV). }
\end{center}
\label{alpha_lam2}
\end{figure}

\begin{figure}[t]
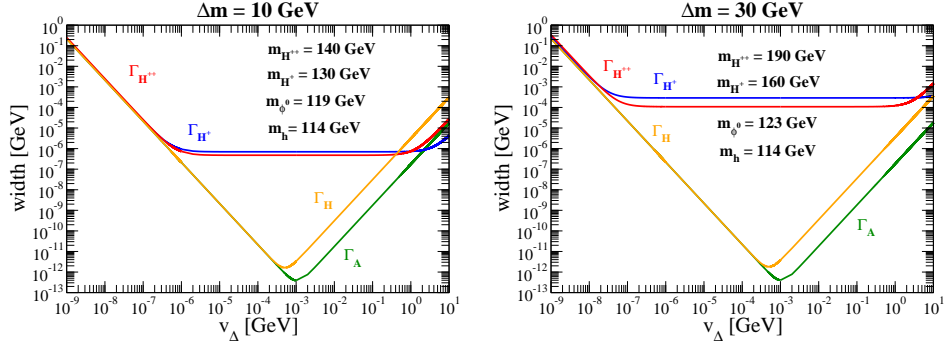

\begin{center}
\includegraphics[width=60mm]{width_140_input.eps}\hspace{3mm}
\includegraphics[width=60mm]{width_190_input.eps}
\caption{Decay width of $H^{++}$, $H^+$, $H$ and $A$ as a function of $v_\Delta$. 
We take $m_{H^{++}}=140$ GeV (190 GeV), $\Delta m=$10 GeV (30~GeV) 
in the left (right) figure. In both the figures, $m_h$ is fixed to be 114 GeV~\cite{aky_triplet}. 
}
\label{decay-rate}
\end{center}
\end{figure}

\begin{figure}[!t]
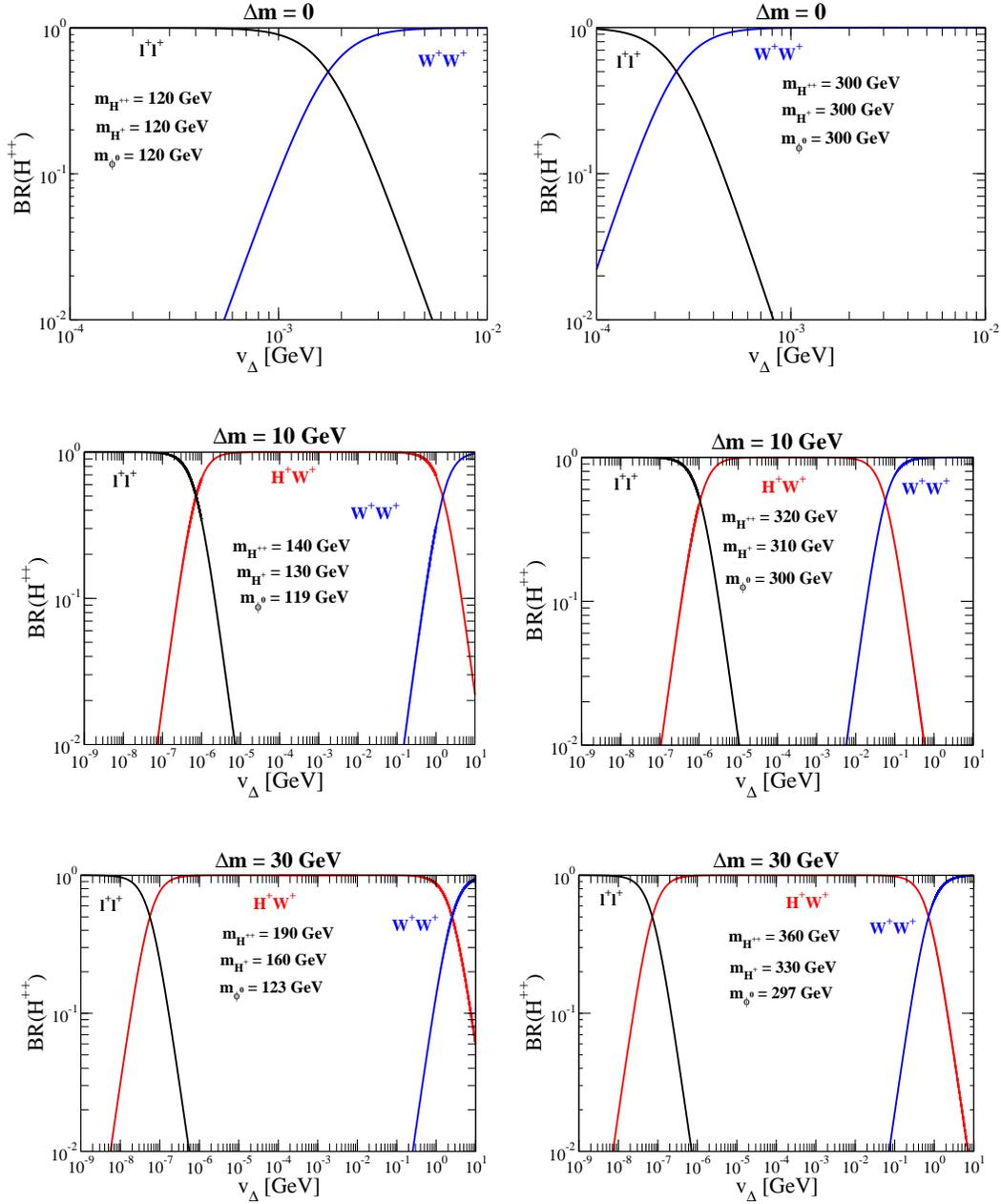

\begin{center}
\includegraphics[width=69mm]{br_Hpp_dm0_120.eps}\hspace{3mm}
\includegraphics[width=65mm]{br_Hpp_dm0_300.eps}\\
\vspace{7mm}
\includegraphics[width=65mm]{br_Hpp_dm10_140.eps}\hspace{3mm}
\includegraphics[width=65mm]{br_Hpp_dm10_320.eps}\\\vspace{7mm}
\includegraphics[width=65mm]{br_Hpp_dm30_190.eps}\hspace{3mm}
\includegraphics[width=65mm]{br_Hpp_dm30_360.eps}
\caption{Decay branching ratio of $H^{++}$ as a function of $v_\Delta$. 
In the upper left (right) figure, $m_{H^{++}}$ is 
fixed to be 120 GeV (300 GeV), and $\Delta m$ is taken to be zero. 
In the middle left (right) figure, $m_{H^{++}}$ is 
fixed to be 140 GeV (320 GeV), and $\Delta m$ is taken to be 10 GeV. 
In the bottom left (right) figure, 
$m_{H^{++}}$ is 
fixed to be 190 GeV (360 GeV), and $\Delta m$ is taken to be 30 GeV~\cite{aky_triplet}.}
\label{fig1}
\end{center}
\end{figure}

The decay modes of the triplet-like scalar bosons can be classified into three modes: 
(i) decay via the Yukawa coupling defined in Eq.~(\ref{nu_yukawa}), 
(ii) that via $v_\Delta$ and 
(iii) that via the gauge coupling. 
The magnitude of the Yukawa coupling constant and $v_\Delta$ are related from the neutrino mass as in Eq.~(\ref{mn}). 
The main decay modes of $H^{++}$ and $H^+$ depend on the size of $v_\Delta$ and $\xi$. 
The decay mode (iii) particularly is important in the case of $\xi \neq 0$.  
Typically, in this case, the heaviest triplet-like scalar boson decays into the second heaviest one associated with the $W$ boson.  
The formulae of the decay rates of $H^{\pm\pm}$, $H^\pm$, $H$ and $A$ are listed in Appendix~A. 
Here, the leptonic decay modes through the Yukawa coupling $h_{ij}$ 
are summed over all flavors 
and each element of $h_{ij}$ is taken to be 0.1 eV/($\sqrt{2}v_\Delta$). 
In the calculation of the decay rates for the triplet-like scalar bosons, 
we use the relations in Eqs.~(\ref{mass1}) and (\ref{mass2}), 
so that five mass parameters: $m_{H^{++}}$, $m_{H^+}$, $m_A$, $m_H$ and $m_h$ can be described by 
three parameters: $m_{H^{++}}$, $m_h$ and $\Delta m$ or $m_A$, $m_h$ and $\Delta m$. 
Furthermore, we here take $\alpha$ to not be an independent parameter but dependent parameter 
which is determined by $v_\Delta$, $m_{H^{++}}$ ($m_A$), $m_h$, $\Delta m$ and $\lambda_2$. 
In Fig.~(\ref{alpha_lam2}), 
the mixing angle $\alpha$ is shown as a function of $\lambda_2$ for each fixed 
value of $v_\Delta$ in the case of $m_h=114$ GeV. 
In the left figure, 
we take $m_{H^{++}}=140$ GeV and $\Delta m=10$ GeV, while 
in the right figure, we take $m_{H^{++}}=190$ GeV and $\Delta m=30$ GeV. 
In the both figure, we can see that the mixing angle $\alpha$ is not so sensitive to $\lambda_2$. 
In the following analysis, we take $\lambda_2$ to be zero.

In FIG.~\ref{decay-rate}, the decay width for the triplet-like scalar bosons is shown 
in the case of $\Delta m =10$ GeV and $\Delta m =30$ GeV. 
Since there is a decay mode through the gauge coupling 
the minimum value of the decay widths of $H^{++}$ and $H^+$ are $\mathcal{O}(10^{-6})$ GeV 
for $\Delta m=10$ GeV and $\mathcal{O}(10^{-4})$ GeV for $\Delta m=30$ GeV. 
On the other hand, the decay widths of $H$ and $A$ become minimum at $v_\Delta\simeq 10^{-4}-10^{-3}$ GeV 
with the magnitude of $\mathcal{O}(10^{-13}-10^{-12})$ GeV.  
This result is consistent with Ref.~\cite{Perez:2008ha}. 

We consider the decay branching ratio of $H^{++}$. 
In the case with $\Delta m =0$ and 
$m_{H^{++}}$=140 GeV, $H^{++}$ decays into $\ell^+\ell^+$ with $v_\Delta \lesssim 10^{-3}$ GeV 
or $W^+W^+$ with $v_\Delta \gtrsim 10^{-3}$ GeV. 
The value of $v_\Delta$ where the main decay mode changes from $H^{++}\to \ell^+\ell^+$ to $H^{++}\to W^+W^+$ is shifted 
at $v_\Delta \simeq 10^{-4}$ GeV when $m_{H^{++}}=300$ GeV. 
In the case of $\Delta m$ =10 GeV, 
$H^{++}$ decays into $H^+W^{+*}$ in the region of $10^{-6}\text{ GeV} \lesssim v_\Delta \lesssim 1\text{ GeV}$ 
($10^{-6}\text{ GeV} \lesssim v_\Delta \lesssim 0.1\text{ GeV}$) for $m_{H^{++}}$=140 GeV (320 GeV). 
Similarly, in the case of $\Delta m$ =30 GeV,  
$H^{++}$ decays into $H^+W^{+*}$ in the region of $10^{-7}\text{ GeV} \lesssim v_\Delta \lesssim 1\text{ GeV}$ 
for $m_{H^{++}}$=190 GeV and 360 GeV. 
In FIG.~\ref{fig1}, the decay branching ratio of $H^{++}$ is shown as 
a function of $v_\Delta$. 

\begin{figure}[!t]
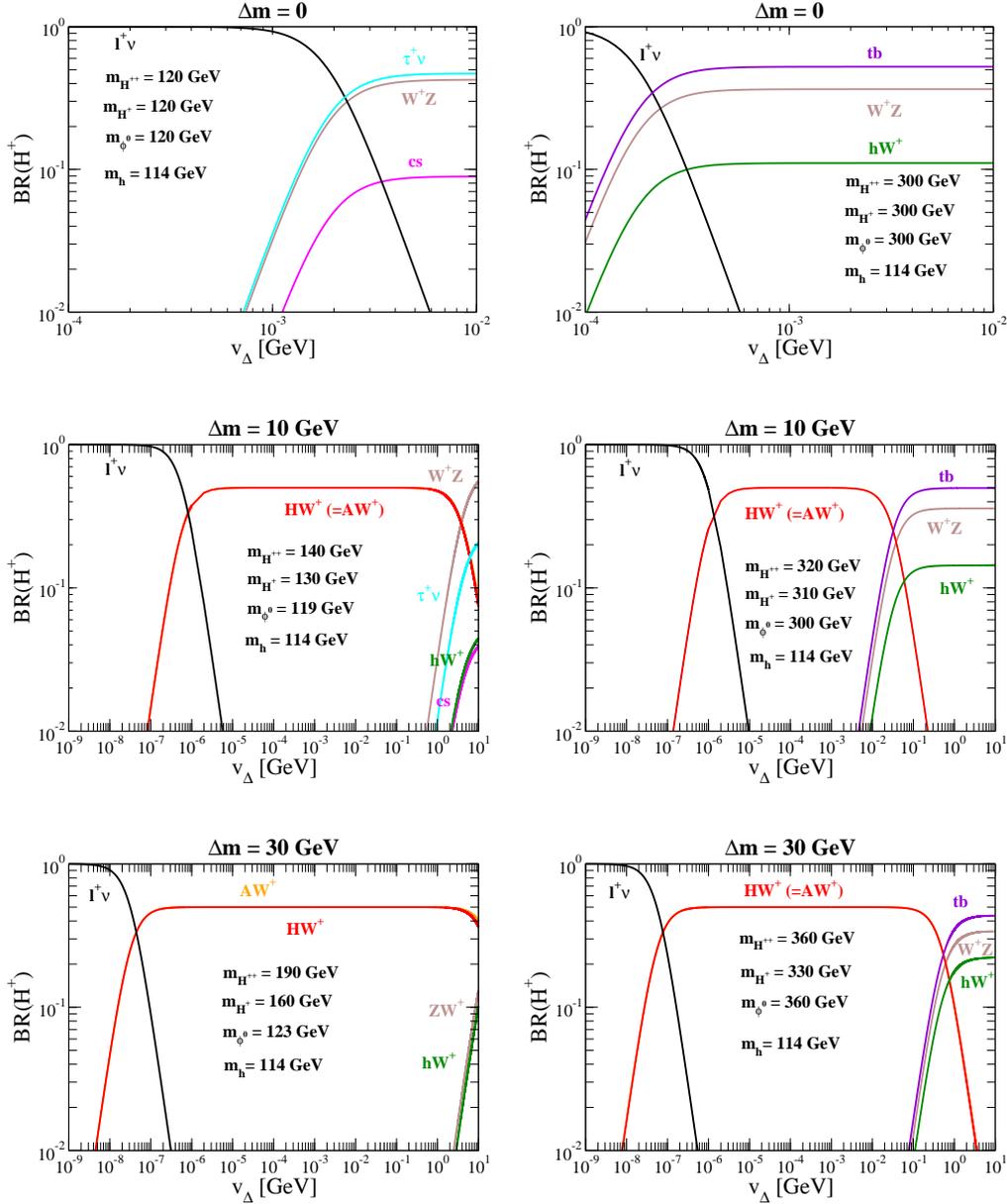

\begin{center}
\includegraphics[width=65mm]{br_ch_120_input.eps}\hspace{3mm}
\includegraphics[width=65mm]{br_ch_300_input.eps}\\
\vspace{7mm}
\includegraphics[width=65mm]{br_ch_140_input.eps}\hspace{3mm}
\includegraphics[width=65mm]{br_ch_320_input_new.eps}\\
\vspace{7mm}
\includegraphics[width=65mm]{br_ch_190_input.eps}\hspace{3mm}
\includegraphics[width=65mm]{br_ch_360_input.eps}
\caption{Decay branching ratio of $H^+$ as a function of $v_\Delta$. 
In all the figures, $m_h$ is taken to be 114 GeV. 
In the upper left (right) figure, $m_{H^+}$ is 
fixed to be 120 GeV (300 GeV), and $\Delta m$ is taken to be zero. 
In the middle left (right) figure, $m_{H^+}$ is 
fixed to be 130 GeV (310 GeV), and $\Delta m$ is taken to be 10 GeV. 
In the bottom left (right) figure, $m_{H^+}$ is 
fixed to be 160 GeV (330 GeV), and $\Delta m$ is taken to be 30 GeV~\cite{aky_triplet}.}
\label{fig2}
\end{center}
\end{figure}

The decay branching ratio of $H^+$ is shown in FIG.~\ref{fig2}. 
In the case of $\Delta m=0$, $H^+$ decays into $\ell^+\nu$ with $v_\Delta < 10^{-4}-10^{-3}$ GeV.  
When $v_\Delta > 10^{-4}-10^{-3}$ GeV, 
$H^+$ decays into $\tau^+\nu$, $W^+ Z$ and $c\bar{s}$ for $m_{H^+}=120$ GeV, while 
$H^+$ decays into $t\bar{b}$, $W^+ Z$ and $hW^+$ for $m_{H^+}=300$ GeV. 
In the case of $\Delta m$ =10 GeV, similarly to the decay of $H^{++}$, $H^+$ decays into $\phi^0 W^{+*} $ 
in the region of $10^{-6}\text{ GeV} \lesssim v_\Delta \lesssim 1\text{ GeV}$ 
($10^{-6}\text{ GeV} \lesssim v_\Delta \lesssim 10^{-2}\text{ GeV}$ )
for $m_{H^+}=130$ GeV (310 GeV). 
In the case of $\Delta m$ =30 GeV, $H^+$ decays into $\phi^0 W^{+*} $ 
in the region of $10^{-7}\text{ GeV} \lesssim v_\Delta \lesssim 10\text{ GeV}$ 
($10^{-7}\text{ GeV} \lesssim v_\Delta \lesssim 10^{-1}\text{ GeV}$) for $m_{H^+}=160$ GeV (330 GeV).

\begin{figure}[!t]
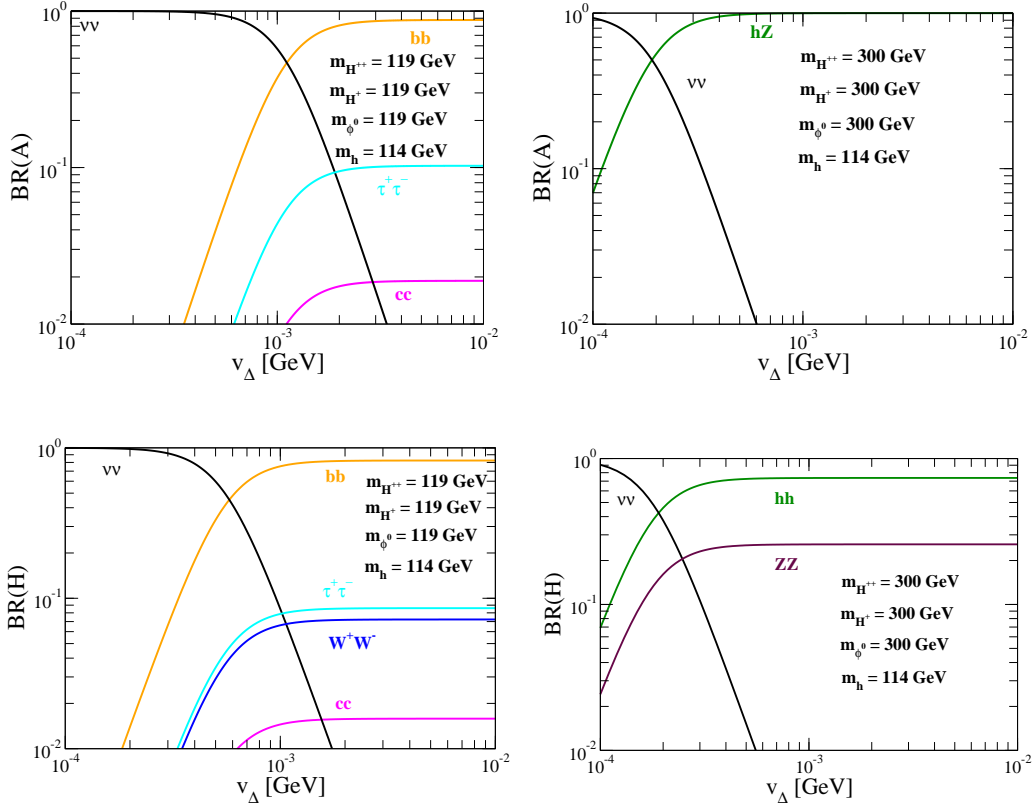

\begin{center}
\includegraphics[width=65mm]{br_A_120_input.eps}\hspace{3mm}
\includegraphics[width=66mm]{br_A_300_input.eps}\\
\vspace{7mm}
\includegraphics[width=67mm]{br_bh_119_input.eps}\hspace{3mm}
\includegraphics[width=65mm]{br_bh_300_input.eps}\\
\caption{Decay branching ratios of $A$ and $H$ as a function of $v_\Delta$. 
In all the figures, $m_h$ is taken to be 114 GeV. 
In the upper left (right) figure, the branching ratio of $A$ is shown in the case of $m_A=119$ GeV (300 GeV). 
In the lower left (right) figure, the branching ratio of $H$ is shown in the case of $m_H=119$ GeV (300 GeV)~\cite{aky_triplet}. }
\label{fig3}
\end{center}
\end{figure}

The decay branching ratios of $H$ and $A$ are shown in FIG.~\ref{fig3}. 
Both $H$ and $A$ decay into neutrinos in the region of $v_\Delta < 10^{-4}-10^{-3}$ GeV. 
When $v_\Delta> 10^{-4}-10^{-3}$ GeV, 
both $H$ and $A$ decay into $b\bar{b}$ with $m_{\phi^0} = 119$ GeV 
while $H$ ($A$) decay into $hh$ and $ZZ$ ($hZ$) with $m_{\phi^0} = 300$ GeV. 

Finally, we comment on the case of $\xi <0$. 
In this case, $H$ and $A$ can decay into $H^\pm W^{\mp(*)} $ depending on the magnitude of 
$\xi$ and $v_\Delta$. 
At the same time, 
$H^+$ can decay into $H^{++} W^{-(*)}$. 
The decay of $H^{++}$ is the 
same as in the case without the mass difference. 

\subsection{Mass determination of the triplet-like scalar bosons at the LHC}

We discuss how the HTM with $\xi > 0$ can be tested at the LHC.  
At the LHC, the triplet-like scalar bosons $H^{\pm\pm}$, $H^\pm$, $H$ and $A$ 
are mainly produced through 
the Drell-Yan processes, for instance, $pp\to H^{++}H^{--}$, $pp\to H^+H^-$, 
$pp\to H^{\pm\pm}H^\mp$ and $pp\to H^\pm \phi^0$ and $pp\to HA$. 
In particular, latter three processes are important when we consider the case of $\xi >0$. 
The cross sections for the latter three production processes are shown in FIG.~\ref{cs}. 

We comment on vector boson fusion production processes. 
There are two types of the vector boson fusion processes. 
First one is the process via $VV\Delta$ vertices, where $V=Z$ or $W^\pm$. 
The cross section of this process is small, since the $VV\Delta$ vertex is proportional to $v_\Delta$
\footnote{The magnitude of $v_\Delta$ may be determined indirectly via $B_{ee}/B_{WW}$ or 
$\Gamma_{ee}$ and $0\nu\beta\beta$ where $H^{++}\to \ell^+\ell^+$, $W^+W^+$ are dominant \cite{Kadastik:2007yd}. 
On the other hand, it could be 
directly measured via $qq\to q'q W^{\pm*}Z^*\to q'q H^\pm$
at the LHC~\cite{Asakawa:2005gv} and via $e^+e^-\to Z^* \to H^\pm W^\mp$ at the ILC~\cite{Cheung:1994rp,Yanase}.}. 
The other one is the process via the gauge coupling constant. 
In particular, 
$qq\to q'q'H^{++}\phi^0$ is the unique process 
whose difference of the electric charge between produced scalar bosons is two.  
This production cross section is 0.51 fb (0.13 fb) at $\sqrt{s}=14$ TeV ($\sqrt{s}=7$ TeV ) assuming 
mass parameters Set~1 which is given just below.

\begin{figure}[!t]
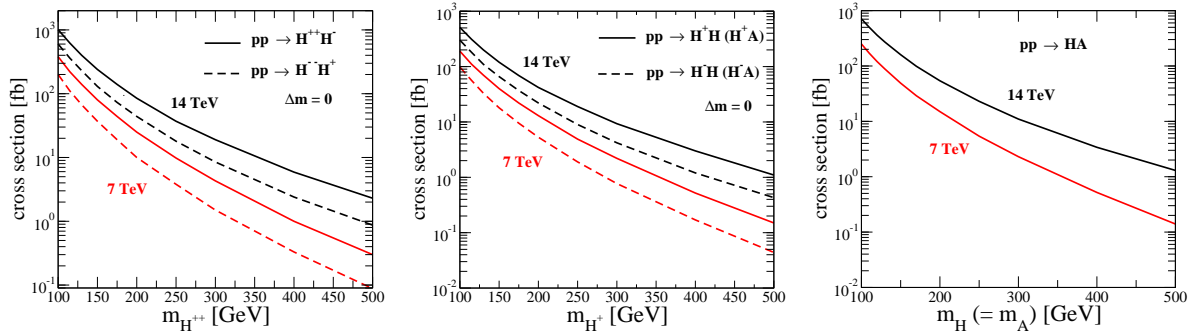

\begin{center}
\includegraphics[width=50mm]{cs_HppHm_dm0.eps}\hspace{2mm}
\includegraphics[width=50mm]{cs_HpA_dm0.eps}\hspace{2mm}
\includegraphics[width=50mm]{cs_HA_new.eps}
\caption{Production cross sections for the triplet-like scalar bosons in the Drell-Yan process~\cite{aky_triplet}. }
\label{cs}
\end{center}
\end{figure}

\begin{table}[!t]
\begin{center}
{\renewcommand\arraystretch{1.1}
\begin{tabular}{|l||c|c|c|}\hline
Process &$\Delta m=0$ &$\Delta m=$10 GeV & $\Delta m=$30 GeV \\\hline\hline
$pp\to  H^{++}H^-$&310 fb (110 fb)&350 fb (120 fb)&140 fb (43 fb)\\\hline
$pp\to  H^+H$     &150 fb (53 fb) &230 fb (81 fb) &150 fb (50 fb)\\\hline
$pp\to  HA  $     &200 fb (65 fb) 
&370 fb (130 fb) 
&330 fb (110 fb)\\\hline
\end{tabular}}
\caption{Production cross sections for the triplet-like scalar bosons 
in the case of $\Delta m =0$ with $m_{H^{++}}=140$ GeV, 
those of the case for Set~1 and Set~2. 
The numbers without (with)
the bracket are the production cross sections at $\sqrt{s}=14$ TeV ($\sqrt{s}=7$ TeV)~\cite{aky_triplet}.}
\label{cs1}
\vspace{3mm}
{\renewcommand\arraystretch{1.1}
\begin{tabular}{|c||c|c|c|c|}\hline
Scenario &Decay of $H^{++}$&Decay of $H^+$&Decay of $H$&Decay of $A$ \\\hline\hline
\hspace{-5.5mm}(1a) ($v_\Delta = 5$ GeV)&$W^+W^{+*}$ [0.93]&$W^{+*}Z$ [0.37], $\tau^+\nu$ [0.14] &$b\bar{b}$ [0.82]&$b\bar{b}$ [0.89]\\\hline
(1b) ($v_\Delta = 10^{-2}$ GeV)&$H^+W^{+*}$ [1.0]&$AW^{+*}$ [0.5], $HW^{+*}$ [0.5]&$b\bar{b}$ [0.82]&$b\bar{b}$ [0.89]\\\hline
(1c) ($v_\Delta = 10^{-5}$ GeV)&$H^+W^{+*}$ [1.0]&$AW^{+*}$ [0.5], $HW^{+*}$ [0.5]&$\nu\nu$ [1.0]&$\nu\nu$ [1.0]\\\hline
(1d) ($v_\Delta = 10^{-8}$ GeV)&$\ell^+\ell^+$ [1.0]&$\ell^+\nu$ [1.0]&$\nu\nu$ [1.0]&$\nu\nu$ [1.0] \\\hline
\end{tabular}}
\caption{The main decay mode of the triplet-like scalar bosons in Scenario (1a) to Scenario (1d). 
The masses of the triplet-like scalar bosons are taken to be as for Set~1. 
The number in ( ) represents the sample value of $v_\Delta$ corresponding to the scenario. 
The number in [ ] represents the value of the decay branching ratio corresponding to the 
value of $v_\Delta$ displayed in ( ) in the same row. 
Here, $\ell\ell$ mode and $\ell\nu$ mode are summed over all flavors~\cite{aky_triplet}. }
\label{tdecay1}
\vspace{3mm}
{\renewcommand\arraystretch{1.1}
\begin{tabular}{|l||c|c|c|c|}\hline
Scenario &Decay of $H^{++}$&Decay of $H^+$&Decay of $H$&Decay of $A$ \\\hline\hline
(2a) [$v_\Delta = 5$ GeV]&$W^+W^{+*}$ [0.76]&$AW^{+*}$ [0.47] $HW^{+*}$ [0.46] &$b\bar{b}$ [0.78]&$b\bar{b}$ [0.89]\\\hline
(2b) [$v_\Delta = 10^{-2}$ GeV]&$H^+W^{+*}$ [1.0]&$AW^{+*}$ [0.5] $HW^{+*}$ [0.5]&$b\bar{b}$ [0.78]&$b\bar{b}$ [0.89]\\\hline
(2c) [$v_\Delta = 10^{-5}$ GeV]&$H^+W^{+*}$ [1.0]&$AW^{+*}$ [0.5] $HW^{+*}$ [0.5]&$\nu\nu$ [1.0]&$\nu\nu$ [1.0]\\\hline
(2d) [$v_\Delta = 10^{-8}$ GeV]&$\ell^+\ell^+$ [0.97]&$\ell^+\nu$ [0.91]&$\nu\nu$ [1.0]&$\nu\nu$ [1.0] \\\hline
\end{tabular}}
\caption{
The main decay mode of the triplet-like scalar bosons in Scenario (2a) to Scenario (2d). 
The masses of the triplet-like scalar bosons are taken to be as for Set~2. 
The number in ( ) represents the sample value of $v_\Delta$ corresponding to the scenario. 
The number in [ ] represents the value of the decay branching ratio corresponding to the 
value of $v_\Delta$ displayed in ( ) in the same row. 
Here, $\ell\ell$ mode and $\ell\nu$ mode are summed over all flavors~\cite{aky_triplet}. }
\label{tdecay2}
\end{center}
\end{table}

We consider the following two sets for mass parameters: 
\begin{align}
\text{(Set~1)}\quad
m_{H^{++}}=140 \text{ GeV},\quad m_{H^+}=130 \text{ GeV},\quad m_H=m_A=119 \text{ GeV},\quad m_h = 114 \text{ GeV},\notag\\
\text{(Set~2)}\quad
m_{H^{++}}=190 \text{ GeV},\quad m_{H^+}=160 \text{ GeV},\quad m_H=m_A=123 \text{ GeV},\quad m_h = 114 \text{ GeV},\notag
\end{align}
which correspond to the cases with $\xi$=(52 GeV)$^2$ and $\xi$=(102 GeV)$^2$, respectively. 
In the following numerical analysis, $\lambda_2=0$ is taken.  
In these parameter sets, the production cross sections for the triplet-like scalar bosons are listed in TABLE~\ref{cs1}.  
We can classify scenarios by the following four regions of $v_\Delta$ for Set~1: 
\begin{align}
\begin{array}{cl}
\text{Scenario (1a)} &\quad v_\Delta \gtrsim 1 \text{ GeV} ,\\
\text{Scenario (1b)} &\quad 10^{-3} \text{ GeV} \lesssim v_\Delta \lesssim 1 \text{ GeV}, \\
\text{Scenario (1c)} &\quad 10^{-6} \text{ GeV} \lesssim v_\Delta \lesssim 10^{-3} \text{ GeV}, \\
\text{Scenario (1d)} &\quad v_\Delta \lesssim 10^{-6} \text{ GeV}. 
\end{array}\notag
\end{align}
We can also classify scenarios by the following four regions of $v_\Delta$ for Set~2: 
\begin{align}
\begin{array}{cl}
\text{Scenario (2a)} &\quad v_\Delta \gtrsim 1 \text{ GeV} ,\\
\text{Scenario (2b)} &\quad 10^{-4} \text{ GeV} \lesssim v_\Delta \lesssim 1 \text{ GeV}, \\
\text{Scenario (2c)} &\quad 10^{-7} \text{ GeV} \lesssim v_\Delta \lesssim 10^{-4} \text{ GeV}, \\
\text{Scenario (2d)} &\quad v_\Delta \lesssim 10^{-7} \text{ GeV}. 
\end{array}\notag
\end{align}

\begin{table}[!t]
{\renewcommand\arraystretch{1.3}
\begin{tabular}{|c||c|c|c|}\hline
& $m_{H^{++}}$ & $m_{H^+}$ & $m_H/m_A$\\\hline\hline
(1a) &{\small$pp\to H^{++}H^-\to$}

&{\small$pp\to H^{+}H\to (\ell^+jjE_T\hspace{-4mm}/\hspace{3mm})(j_bj_b)$ } &
{\small$pp\to HA\to (j_bj_b)(j_bj_b)$ }\\
&{\small$(\ell^+\ell^+E_T\hspace{-4mm}/\hspace{3mm})(jjjj)$} & {\small[11 fb] (3.8 fb)} & {\small [270 fb] (95 fb)}
\\
& {\small [2.8 fb] (0.95 fb)} 
&{\small$pp\to H^{+}H\to (\ell^+E_T\hspace{-4mm}/\hspace{3mm})(j_bj_b)$ }  &
{\small$pp\to H^{+}H\to (\ell^+E_T\hspace{-4mm}/\hspace{3mm})(j_bj_b)$ }\\
& 
& {\small[9.3 fb] (3.3 fb)} & {\small[9.3 fb] (3.3 fb)}
\\\hline
(1b) &{\small$pp\to H^{++}H^-\to$}&{\small $pp\to H^+H\to (\ell^+j_bj_bE_T\hspace{-4mm}/\hspace{3mm})(j_bj_b)$} &{\small$pp\to HA\to (j_bj_b)(j_bj_b)$} \\
&{$\small(\ell^+\ell^+j_bj_bE_T\hspace{-4mm}/\hspace{3mm})(jjj_bj_b)$} & {\small[36 fb] (13 fb)} & {\small[270 fb] (95 fb)}\\
& {\small [8.4 fb] (2.9 fb)}
& &{\small$pp\to H^+H\to (\ell^+j_bj_bE_T\hspace{-4mm}/\hspace{3mm})(j_bj_b)$} \\
& 
&  & {\small[36 fb] (13 fb)}\\\hline
(1c) &\multicolumn{3}{c|}{Challenging}\\\hline
(1d) &\multicolumn{3}{c|}{Excluded}\\\hline
\end{tabular}}
\caption{
The processes which can be used to reconstruct the masses of the triplet-like scalar bosons 
are summarized. 
The numbers in [ ] and ( ) represent the cross section for the final state of the process at $\sqrt{s}=14$ TeV and $\sqrt{s}=7$ TeV, 
respectively, for Set~1.   
The values of the decay branching ratios of the triplet-like scalar bosons are listed in TABLE~\ref{tdecay1}. 
In this table, the $b$-tagging efficiency is assumed to be 100 \%~\cite{aky_triplet}. }
\label{t3}
\vspace{3mm}
{\renewcommand\arraystretch{1.3}
\begin{tabular}{|c||c|c|c|}\hline
& $m_{H^{++}}$ & $m_{H^+}$ & $m_H/m_A$\\\hline\hline
(2a)&
{\small$pp\to H^{++}H^-\to$ }
&{\small$pp\to H^{+}H\to (\ell^+j_bj_bE_T\hspace{-4mm}/\hspace{3mm})(j_bj_b)$ } &
{\small$pp\to HA\to (j_bj_b)(j_bj_b)$ }\\
&{\small$(\ell^+\ell^+E_T\hspace{-4mm}/\hspace{3mm})(jjj_bj_b)$} &{\small[21 fb] (6.9 fb)} &{\small [230 fb] (76 fb)}\\
&{\small [2.7 fb] (0.84 fb)} & & 
\\\hline
(2b)&{\small$pp\to H^{++}H^-\to$}&{\small $pp\to H^+H\to (\ell^+j_bj_bE_T\hspace{-4mm}/\hspace{3mm})(j_bj_b)$} &{\small$pp\to HA\to (j_bj_b)(j_bj_b)$} \\
&{\small$(\ell^+\ell^+j_bj_bE_T\hspace{-4mm}/\hspace{3mm})(jjj_bj_b)$} &{\small[22 fb] (7.2 fb)}&{\small[230 fb] (76 fb)}\\
&{\small [3.2 fb] (0.99 fb)} &  & \\\hline
(2c)&\multicolumn{3}{c|}{Challenging}\\\hline
(2d)&\multicolumn{3}{c|}{Excluded}\\\hline
\end{tabular}}
\caption{
The processes which can be used to reconstruct the masses of the triplet-like scalar bosons 
are summarized. 
The numbers in [ ] and ( ) represent the cross section for the final state of the process at $\sqrt{s}=14$ TeV and $\sqrt{s}=7$ TeV, 
respectively, for Set~2.   
The values of the decay branching ratios of the triplet-like scalar bosons are listed in TABLE~\ref{tdecay2}. 
In this table, the $b$-tagging efficiency is assumed to be 100 \%~\cite{aky_triplet}.  }
\label{t4}
\end{table}

\begin{figure}[!t]
\vspace{8mm}
\begin{center}
\includegraphics[width=65mm]{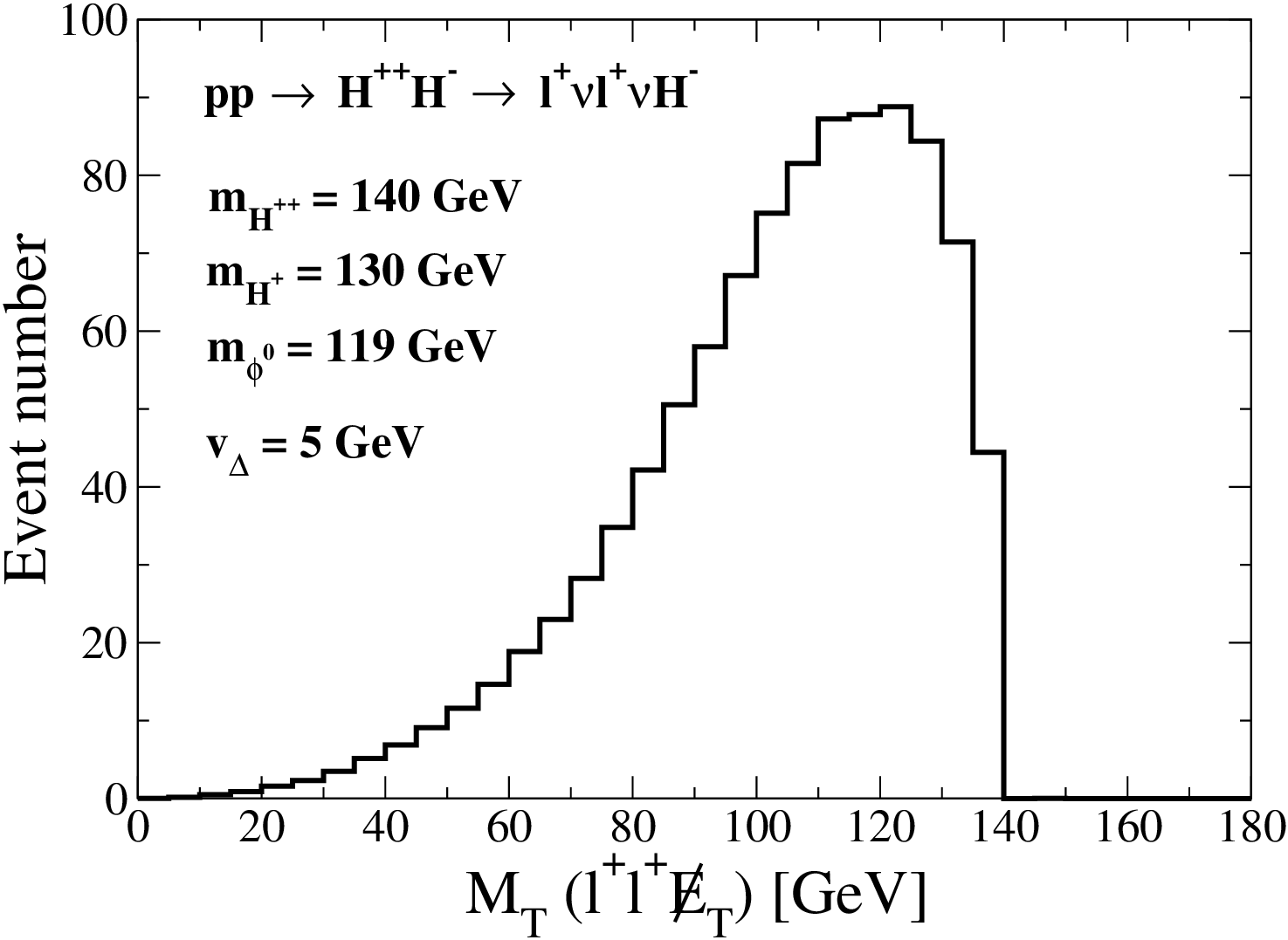}\hspace{3mm}
\includegraphics[width=65mm]{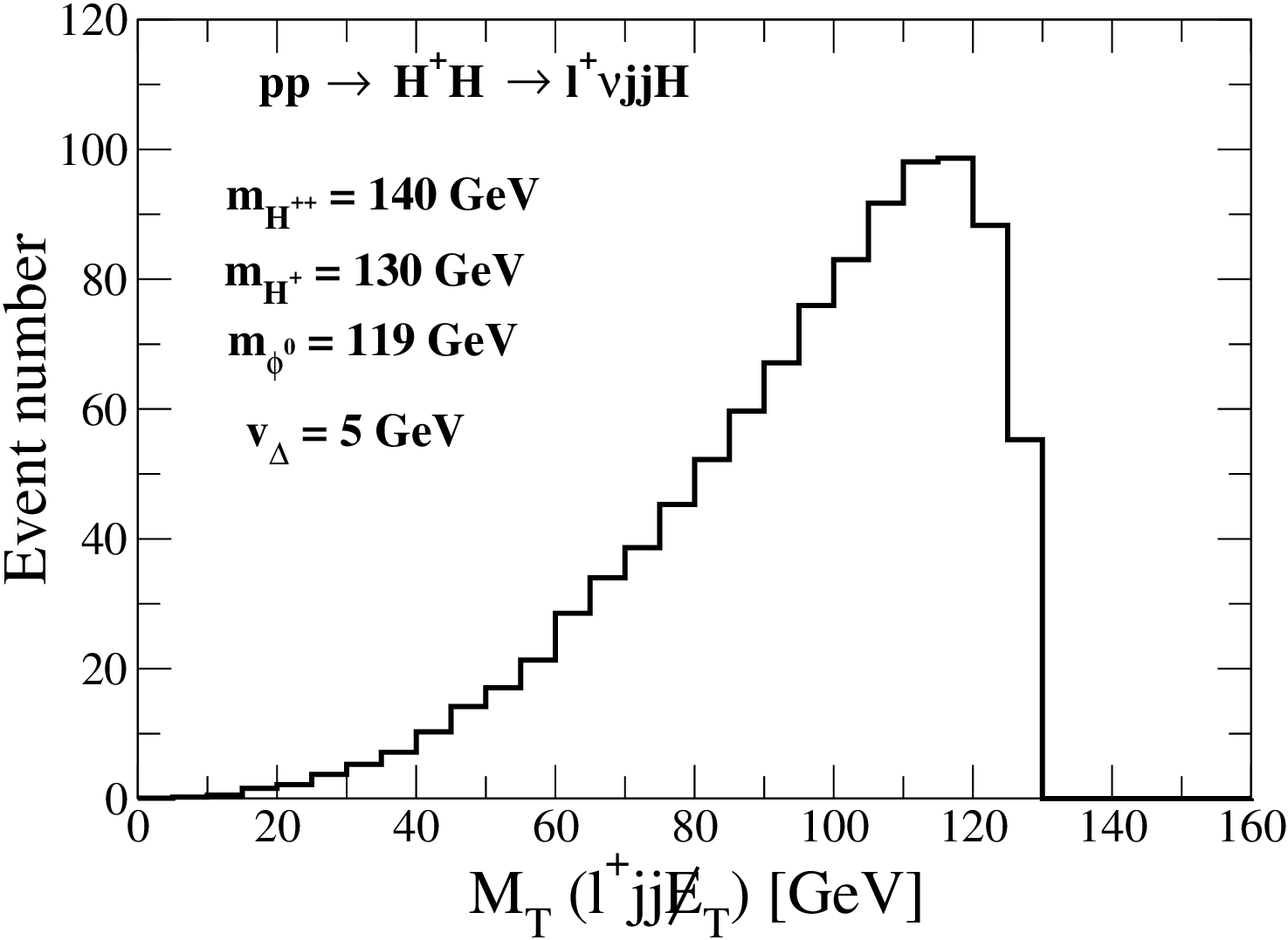}\\\vspace{7mm}
\includegraphics[width=65mm]{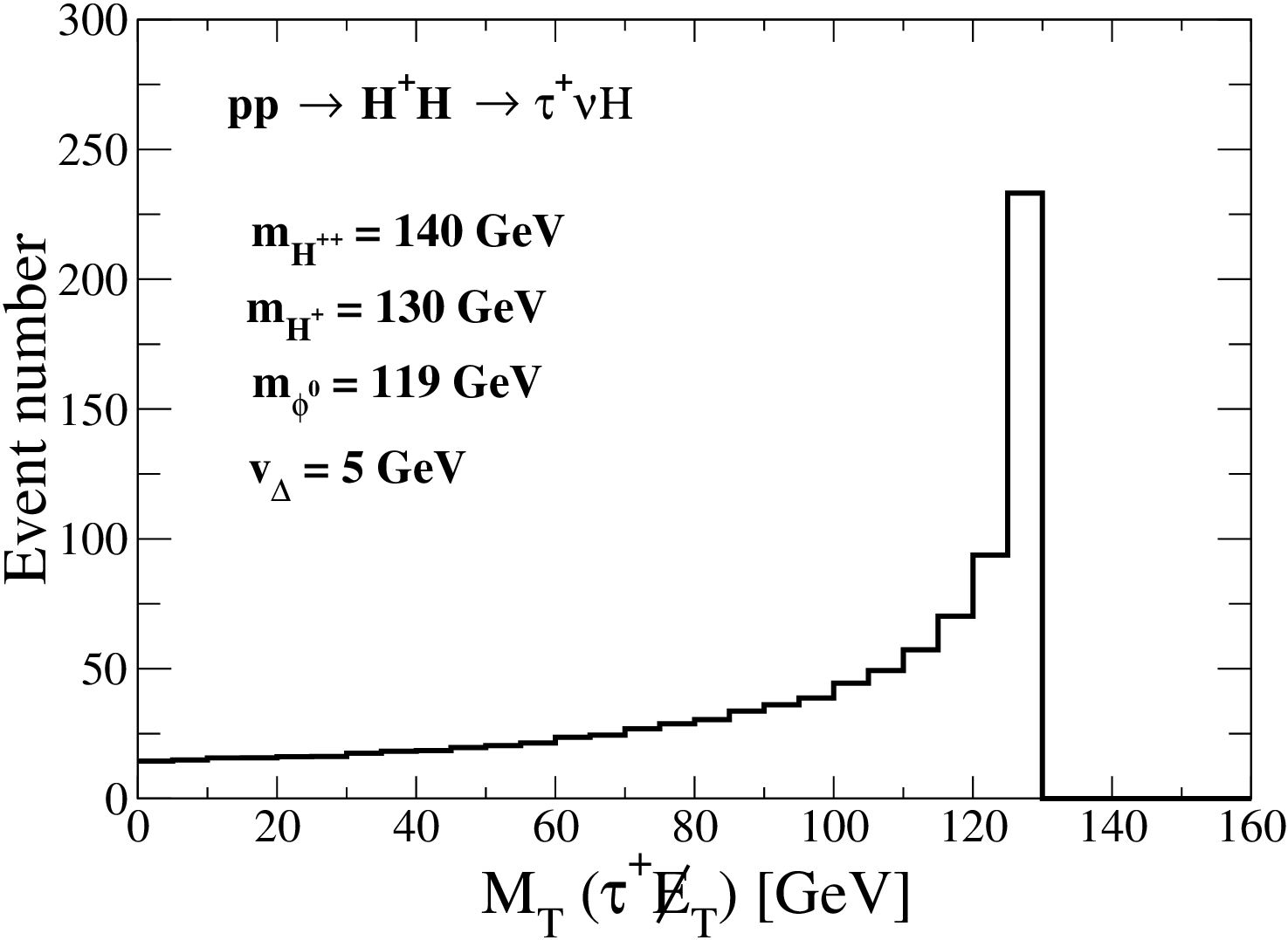}\hspace{3mm}
\includegraphics[width=65mm]{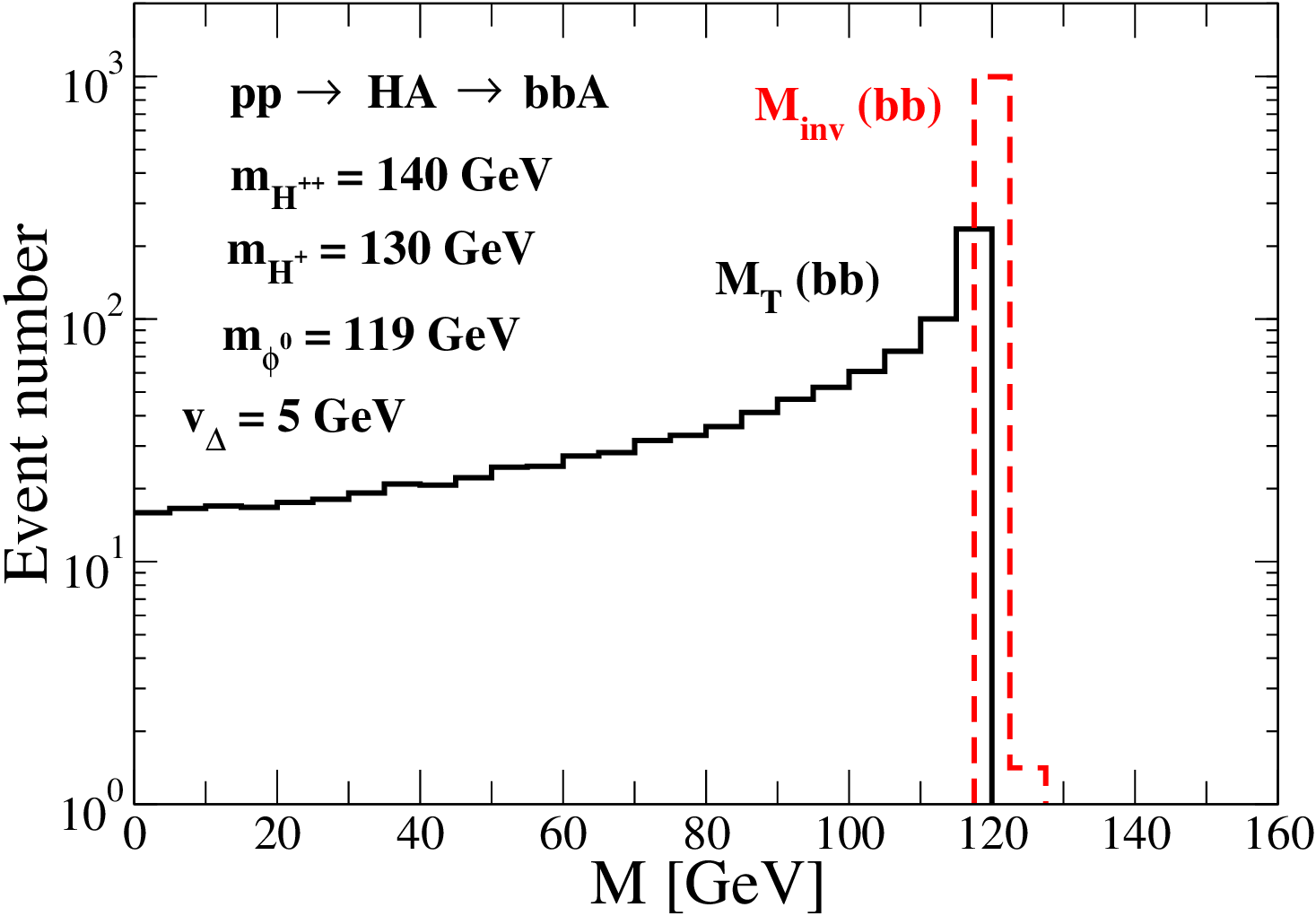}
\caption{The transvers mass distributions for each system in Scenario (1a). 
The total event number is assumed to be 1000. 
In the bottom-right figure, 
the horizontal axis $M$ represents the 
transverse mass distribution for the $b\bar{b}$ system $M_T(bb)$ (solid) or the
invariant mass distribution for the $b\bar{b}$ system $M_{\text{inv}}(bb)$ (dashed)~\cite{aky_triplet}. }
\label{mt2}
\vspace{15mm}
\includegraphics[width=50mm]{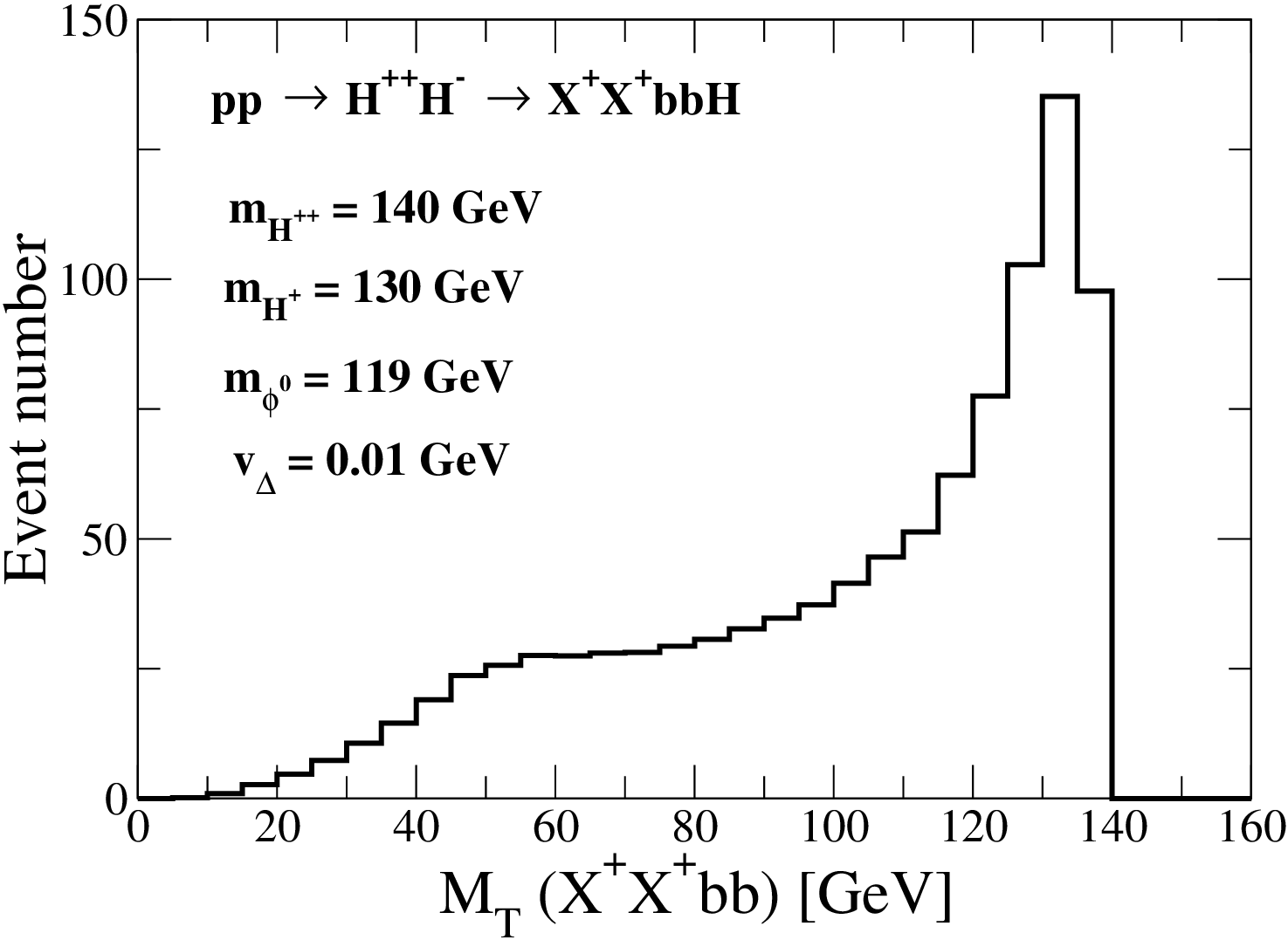}\hspace{3mm}
\includegraphics[width=50mm]{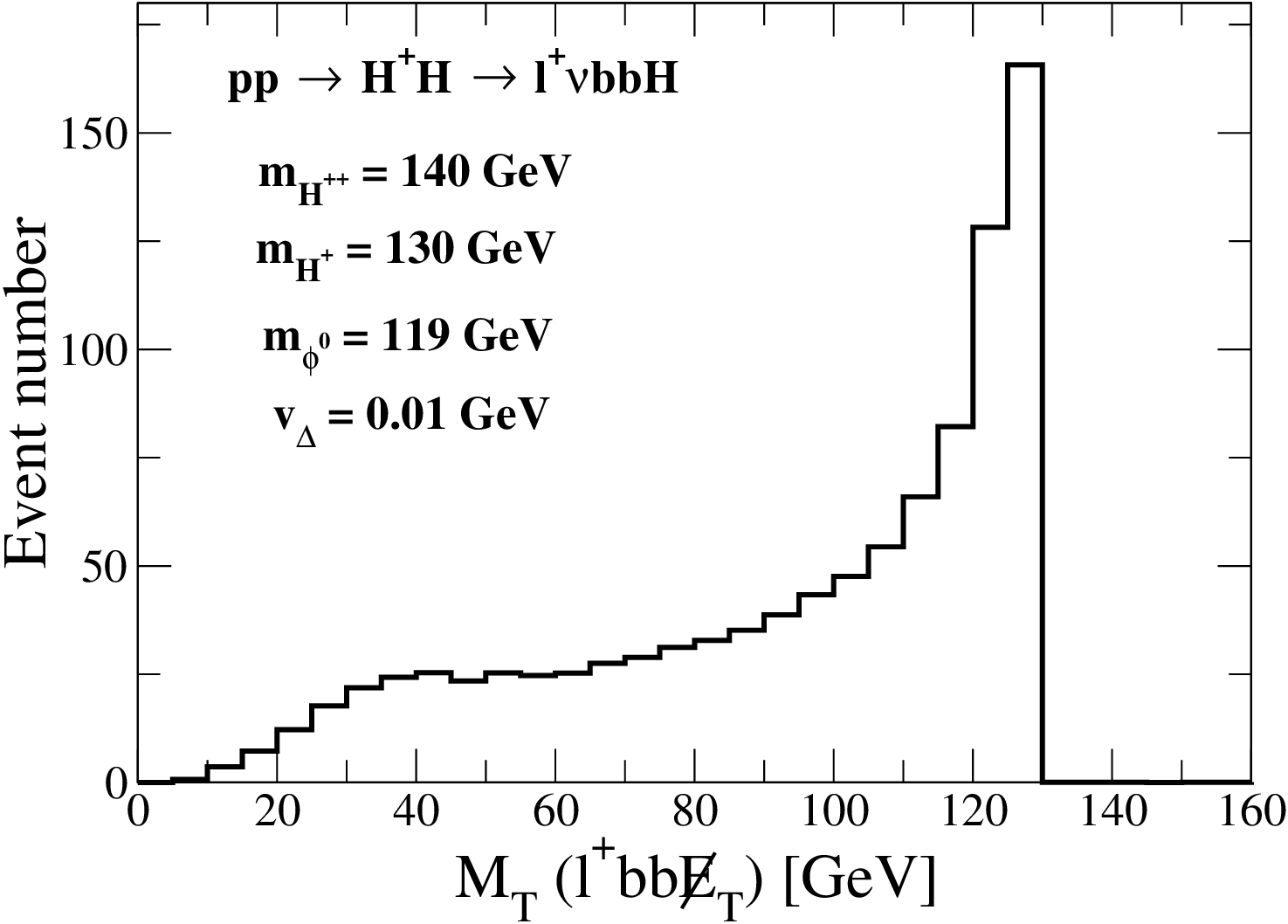}\hspace{3mm}
\includegraphics[width=50mm]{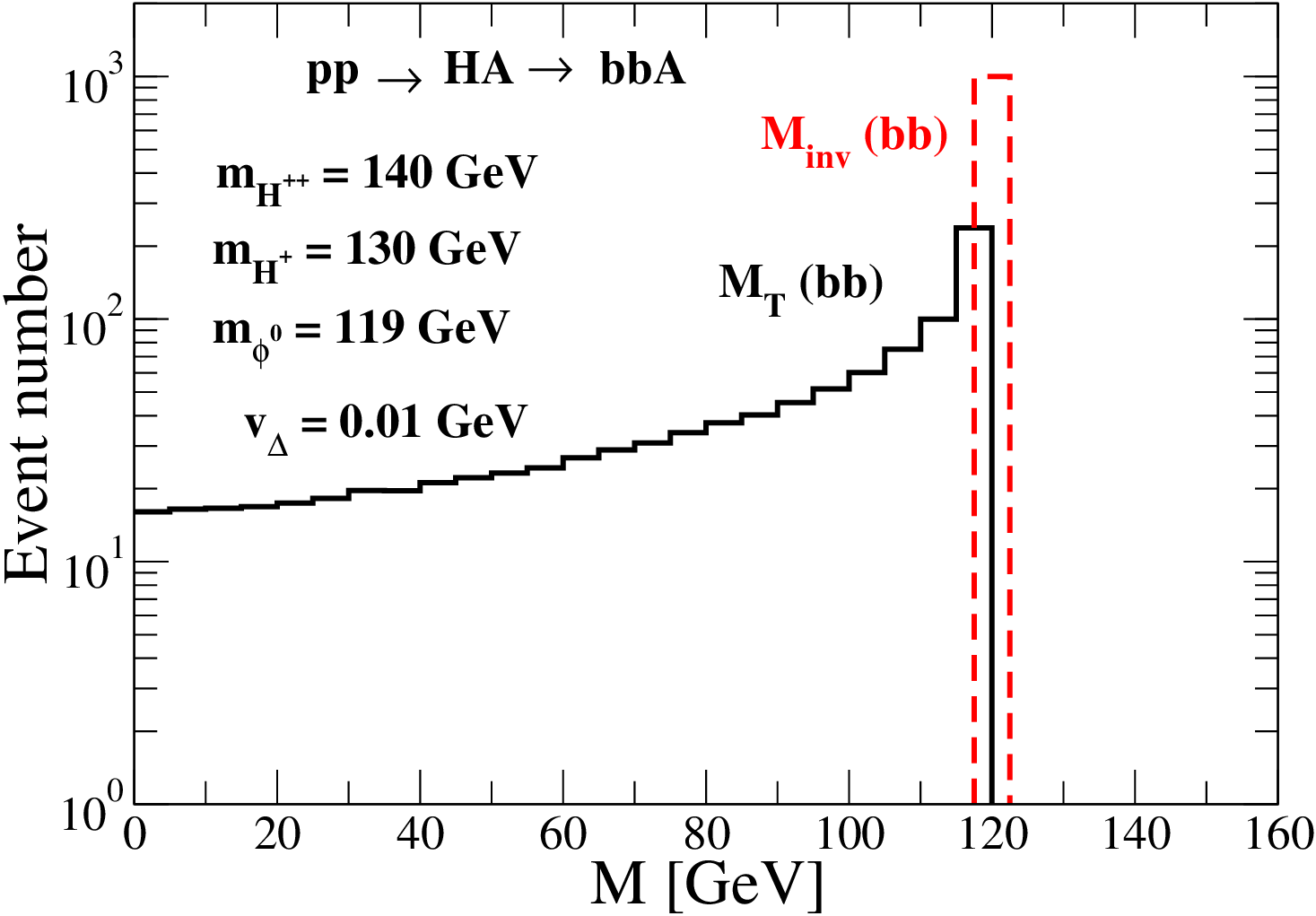}
\caption{The transvers mass distributions for each system in Scenario (1b). 
The total event number is assumed to be 1000. 
In the right figure, 
the horizontal axis $M$ represents the 
transverse mass distribution for the $b\bar{b}$ system $M_T(bb)$ (solid) or the 
invariant mass distribution for the $b\bar{b}$ system $M_{\text{inv}}(bb)$ (dashed)~\cite{aky_triplet}. }
\label{mt1}
\end{center}
\end{figure}

In each scenario, main decay modes of the triplet-like scalar bosons are listed in TABLE~\ref{tdecay1} 
and TABLE~\ref{tdecay2}. 
We here analyse the signal for Set~1 which may be used to reconstruct the masses of the triplet-like scalar bosons. 
The signal distributions discussed below are calculated by using CalcHEP~\cite{CalcHEP}. 

\begin{description}
\item[Scenario (1a)];\\    
We can measure $m_{H^{++}}$ by observing 
the endpoint in the transverse mass distribution of the $\ell^+\ell^+ E_T\hspace{-4mm}/\hspace{3mm}$ system 
in the process $pp\to H^{++}H^-\to (W^{+*}W^+)(W^{-*}Z)\to (\ell^+\ell^+ E_T\hspace{-4mm}/\hspace{3mm})(jjjj)$, (FIG.~\ref{mt2} upper left). 
At the same time, we can also determine $m_{H^+}$ by measuring the 
endpoint in the transverse mass distribution of the 
$\ell^+ jjE_T\hspace{-4mm}/\hspace{3mm}$ system or the $\ell^+ E_T\hspace{-4mm}/\hspace{3mm}$ system 
in the process $pp\to H^{+}\phi^0\to (W^{+*}Z)(b\bar{b})\to (\ell^+ jjE_T\hspace{-4mm}/\hspace{3mm})(j_bj_b)$ or 
$pp\to H^{+}\phi^0\to (\tau^+\nu)(b\bar{b})\to (\ell^+E_T\hspace{-4mm}/\hspace{3mm})(j_bj_b)$, 
(FIG.~\ref{mt2} upper right and lower left). 
In addition, $m_{\phi^0}$ can be determined by using 
the invariant mass distribution 
or by observing the endpoint in the transverse mass distribution 
of the $b\bar{b}$ system in 
the process $pp\to HA\to (b\bar{b})(b\bar{b})\to (j_bj_b)(j_bj_b)$, (FIG.~\ref{mt2} lower right).
\item[Scenario (1b)];\\
We can determine $m_{H^{++}}$ by measuring 
the endpoint in the transverse mass distribution of the $\ell^+\ell^+j_bj_bE_T\hspace{-4mm}/\hspace{3mm}$ system 
in the process $pp\to H^{++}H^-\to (W^{+*}W^{+*}b\bar{b})(W^{-*}b\bar{b})\to (\ell^+\ell^+j_bj_bE_T\hspace{-4mm}/\hspace{3mm})(jjj_bj_b)$, 
(FIG.~\ref{mt1} left). 
Analysing the transverse mass distribution for the $\ell^+\ell^+j_bj_bE_T\hspace{-4mm}/\hspace{3mm}$ system,  
we treat that a lepton pair $\ell^+\nu$ from $W^{+*}$ as one massless fermion as represented $X^+$ in FIG.~\ref{mt1}. 
This procedure is justified since the angle between $\ell^+$ and $\nu$ is distributed almost around $0^\circ$. 
We can also determine $m_{H^+}$ 
by measuring the endpoint in the transverse mass distribution of the $\ell^+ j_bj_bE_T\hspace{-4mm}/\hspace{3mm}$ system 
in the process $pp\to H^{+}\phi^0\to (W^{+*}b\bar{b})(b\bar{b})\to (\ell^+ j_bj_bE_T\hspace{-4mm}/\hspace{3mm})(j_bj_b)$, (FIG.~\ref{mt1} center). 
In addition, 
$m_{\phi^0}$ can be reconstructed 
by measuring the invariant mass distribution of the $b\bar{b}$ system 
and by observing the endpoint of the transverse mass distribution of the $b\bar{b}$ system 
in the process $pp\to HA\to (b\bar{b})(b\bar{b})\to (j_bj_b)(j_bj_b)$ (FIG.~\ref{mt1} right). 
\item[Scenario (1c)];\\
The final state of the decay of the triplet-like scalar bosons always include neutrinos, 
so that the reconstruction of 
the masses of the triplet-like scalar bosons would be challenging.
\item[Scenario (1d)];\\
This scenario is already excluded from the direct search results at the LHC for the processes of 
$pp\to H^{++}H^{--}(H^{\pm\pm}H^\mp)\to \ell^+\ell^+\ell^-\ell^-(\ell^\pm\ell^\pm\ell^\mp\nu)$. 
\end{description}

In TABLE~\ref{t3}, processes which can use the reconstruction of the masses of the triplet-like scalar bosons 
are summarized in each scenario. 
The cross sections for the final states of each process are also listed. 
In the case of Set~2, the masses of the triplet like scalar bosons may be able to reconstruct in the similar way to the case 
of Set~1. 
Thus, we show only the signal cross sections for the final states for Set~2 in TABLE~\ref{t4}.

\subsection{The two photon decay of the SM-like Higgs boson}

We discuss the radiative effect of triplet-like Higgs bosons on 
the decay rate of $h\to \gamma\gamma$ in the HTM under the constraint from the electroweak precision data. 
The $h\gamma\gamma$ vertex is generated at the one-loop level,  
so that the significant one-loop contributions of additional charged particles can appear. 
In the HTM, there are doubly- and singly-charged Higgs bosons which would give substantial 
one-loop contributions to the decay rate of $h\to \gamma\gamma$. 
In Ref.~\cite{Arhrib_hgg}, this decay process have been discussed in the HTM under the constraint from 
perturbative unitarity and vacuum stability. 
We here analyze the decay rate taking into account our new results of the radiative corrections to the 
electroweak parameters.  

\begin{figure}[!t]
\begin{center}
\includegraphics[width=65mm]{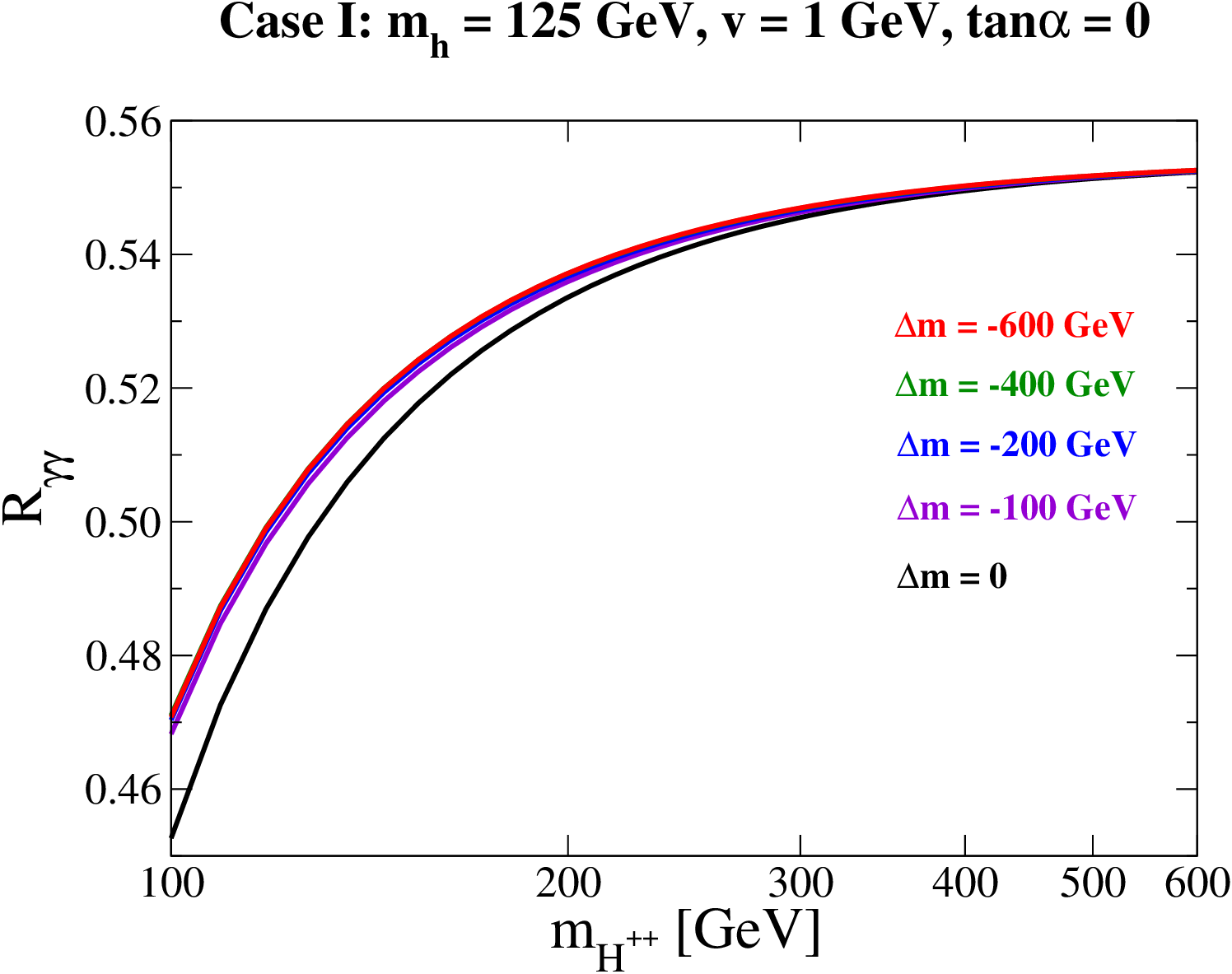}\hspace{3mm}
\includegraphics[width=65mm]{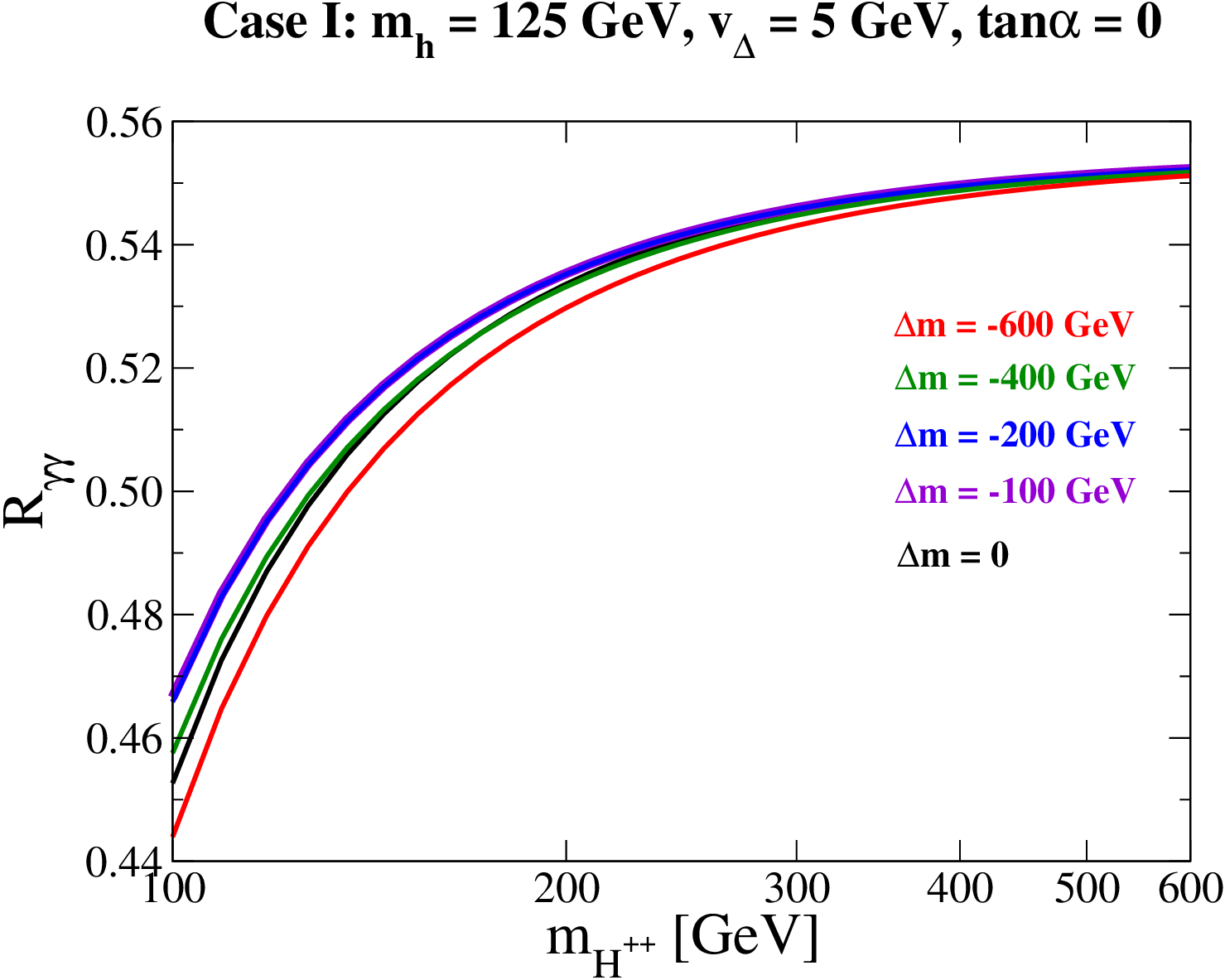}
\caption{The ratio of the decay rate for $h\to \gamma\gamma$ in the HTM to that in the SM as a function of $m_{H^{++}}$ 
for each fixed value of $\Delta m$ ($<0$) in Case I $(m_{\phi^0}>m_{H^+}>m_{H^{++}})$. 
In the both figures, we take $m_t=173$ GeV, $m_h=125$ GeV and $\tan\alpha=0$. 
In the left (right) figure, we take $v_\Delta=1$ GeV (5 GeV)~\cite{ky_rho}.}
\label{hgg1}
\vspace{5mm}
\includegraphics[width=65mm]{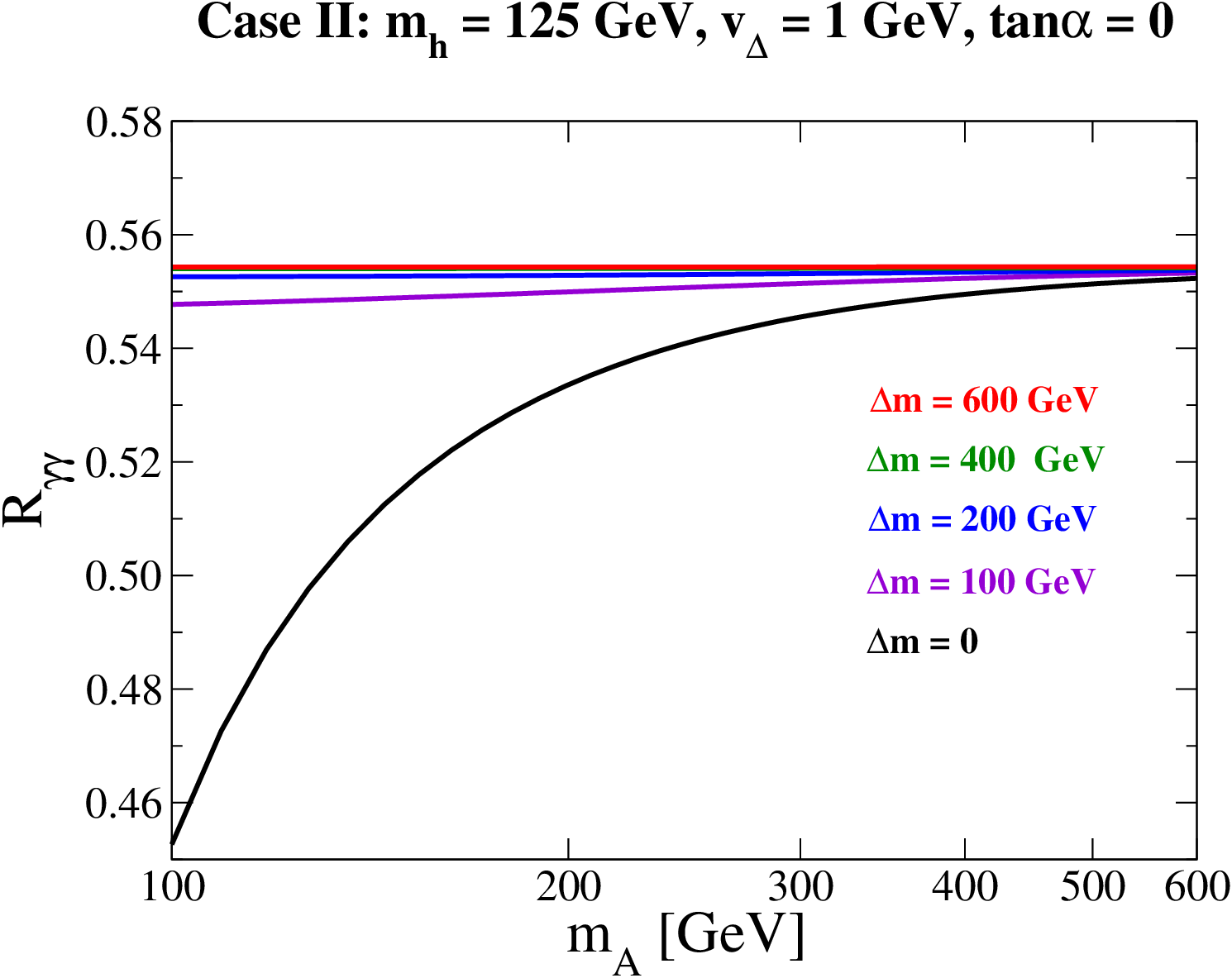}\hspace{3mm}
\includegraphics[width=65mm]{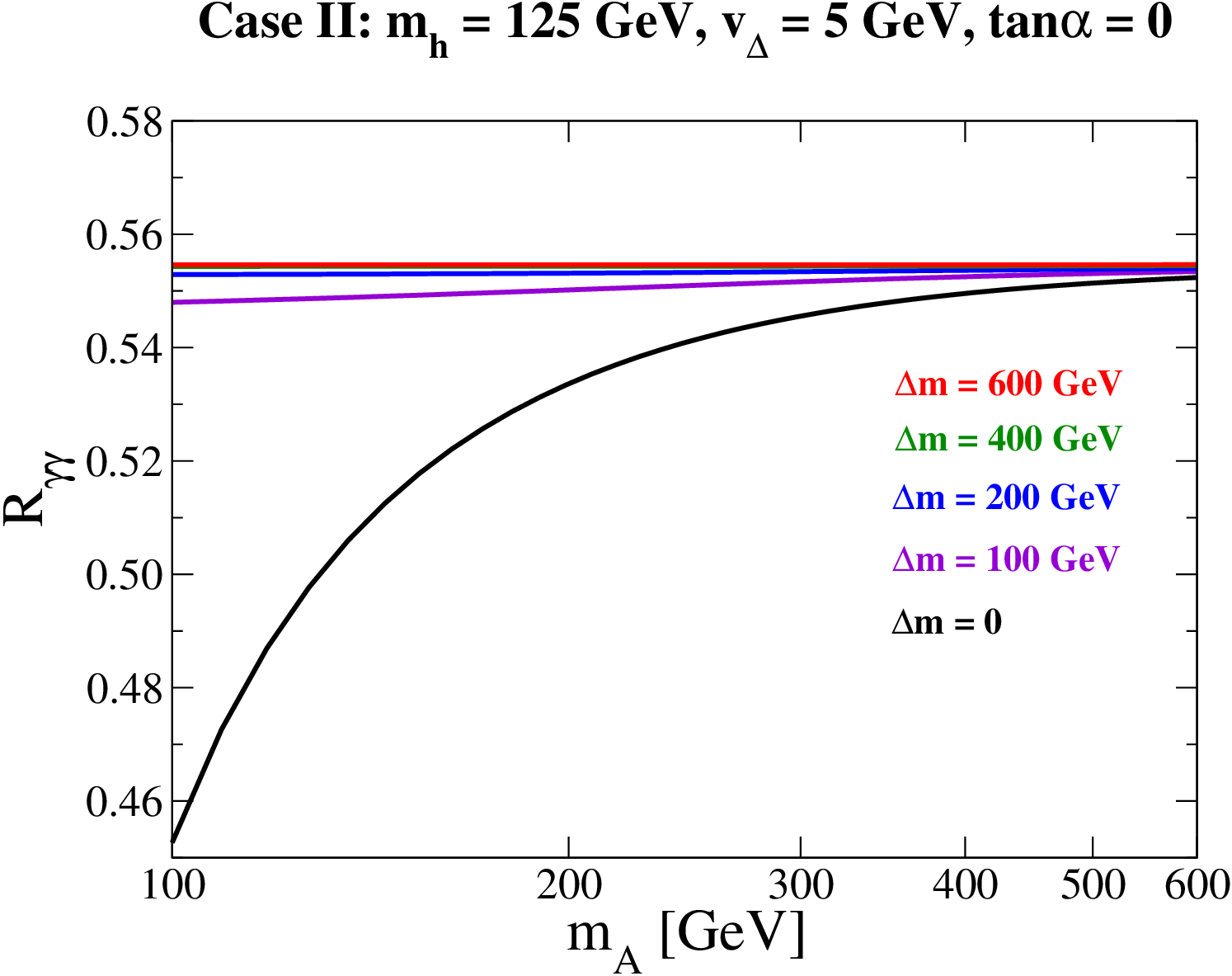}
\end{center}
\caption{The ratio of the decay rate for $h\to \gamma\gamma$ in the HTM to that in the SM as a function of $m_A$ 
for each fixed value of $\Delta m$ ($>0$) in Case II $(m_{H^{++}}>m_{H^+}>m_{\phi^0})$. 
In the both figures, we take $m_t=173$ GeV, $m_h=125$ GeV and $\tan\alpha=0$. 
In the left (right) figure, we take $v_\Delta=1$ GeV (5 GeV)~\cite{ky_rho}.}
\label{hgg2}
\end{figure}

The decay rate of $\phi\to \gamma\gamma$ is calculated at the one-loop level by~\cite{hgg}
\begin{align}
\Gamma(\phi\to \gamma\gamma)&=\frac{G_F\alpha_{\text{em}}^2m_\phi^3}{128\sqrt{2}\pi^3}
\Bigg|-2\sum_fN_f^cQ_f^2\tau_f[1+(1-\tau_f)f(\tau_f)]+2+3\tau_W+3\tau_W(2-\tau_W)f(\tau_W)\notag\\
&+Q_{H^{++}}^2\frac{2vc_{hH^{++}H^{--}}}{m_\phi^2}[1-\tau_{H^{++}}f(\tau_{H^{++}})]
+Q_{H^{+}}^2\frac{2vc_{hH^{+}H^{-}}}{m_\phi^2}[1-\tau_{H^{+}}f(\tau_{H^{+}})]\Bigg|^2, \label{hgg}
\end{align}
where the function $f(x)$ is given by 
\begin{align}
f(x)=\left\{
\begin{array}{c}
[\arcsin(1/\sqrt{x})]^2, \quad \text{if }x\geq 1,\\
-\frac{1}{4}[\ln \frac{1+\sqrt{1-x}}{1-\sqrt{1-x}}-i\pi]^2, \quad \text{if }x< 1
\end{array}\right.. 
\end{align}
In Eq.~({\ref{hgg}}),  $Q_\varphi$ is the electric charge of the field $\varphi$, 
$N_f^c$ is the color factor and $\tau_\varphi= 4m_\varphi^2/m_\phi^2$.  
In the HTM, the coupling constants $c_{h H^+H^-}$ and $c_{h H^{++}H^{--}}$ are given by 
\begin{align}
c_{h H^+H^-}&=\frac{1}{v_\Delta}\left[
m_{H^+}^2\left(\sqrt{2}s_{\beta_\pm}c_{\beta_\pm}c_\alpha+2s_{\beta_\pm}^2s_\alpha\right)
-m_A^2s_\alpha\left(c_{\beta_0}^2+\frac{s_{\beta_0}^2}{2}\right)
+m_h^2\left(\frac{s_{\beta_\pm}^3c_\alpha}{\sqrt{2}c_{\beta_\pm}}+c_{\beta_\pm}^2s_\alpha\right)\right], \label{hHpHm}\\
c_{h H^{++}H^{--}}&=\frac{1}{v_\Delta}\left[
2m_{H^{++}}^2s_\alpha+m_h^2s_\alpha
-2m_{H^+}^2\left(2c_{\beta_\pm}^2s_\alpha-\sqrt{2}s_{\beta_\pm}c_{\beta_\pm}c_\alpha\right)
-m_A^2\left(s_{\beta_0}c_{\beta_0}c_\alpha-c_{\beta_0}^2s_\alpha\right)
\right].\label{hHppHmm}
\end{align}

In the case with $\alpha\simeq 0$ and $v_\Delta\simeq 0$, 
coupling constants in Eqs.~(\ref{hHpHm}) and (\ref{hHppHmm}) can be written as the simple form: 
\begin{align}
c_{h H^+H^-}&\simeq\frac{2m_{H^+}^2}{v},\\
c_{h H^{++}H^{--}}&\simeq\frac{2m_{H^{++}}^2}{v}.
\end{align}

\begin{figure}[!t]
\begin{center}
\includegraphics[width=65mm]{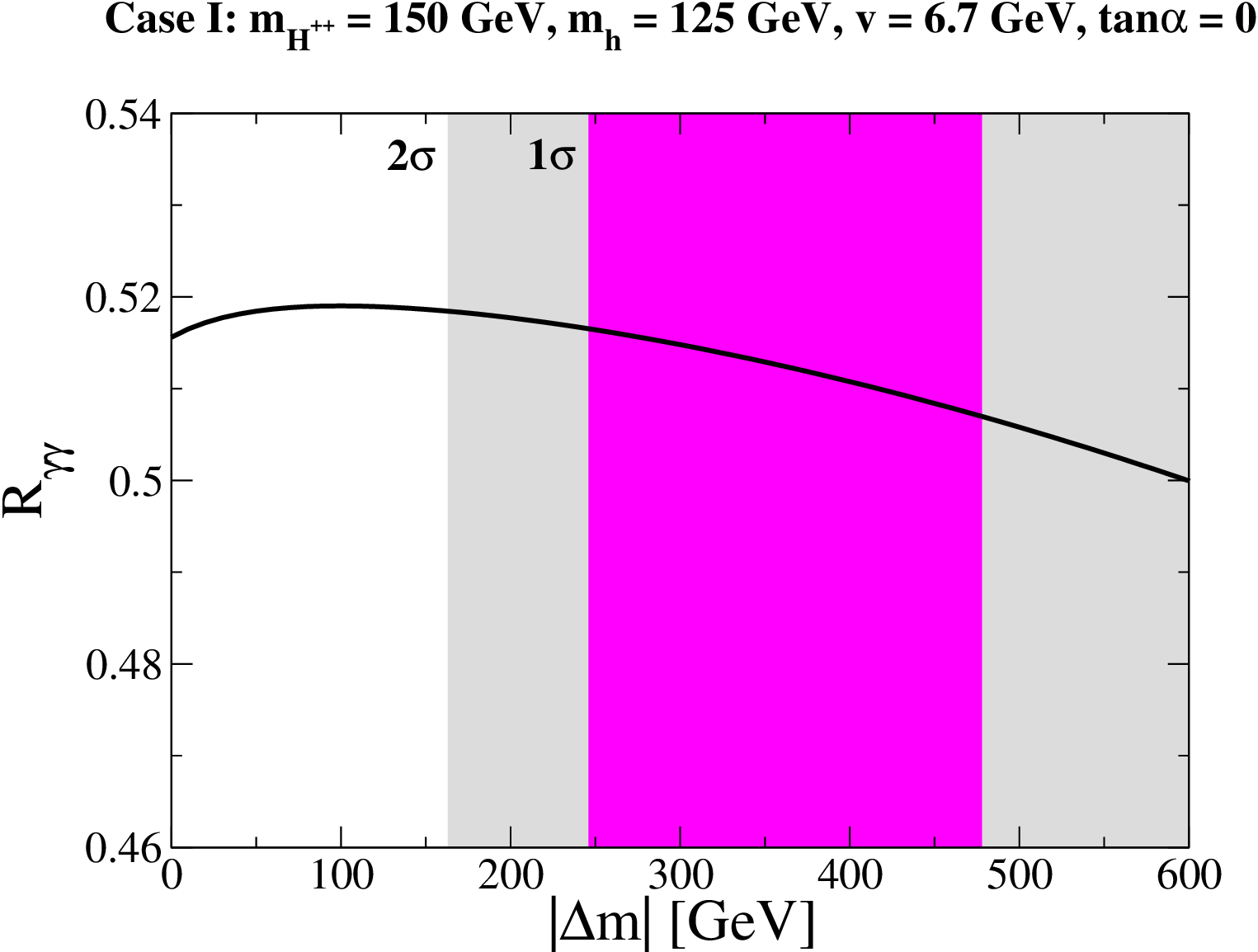}\hspace{3mm}
\includegraphics[width=65mm]{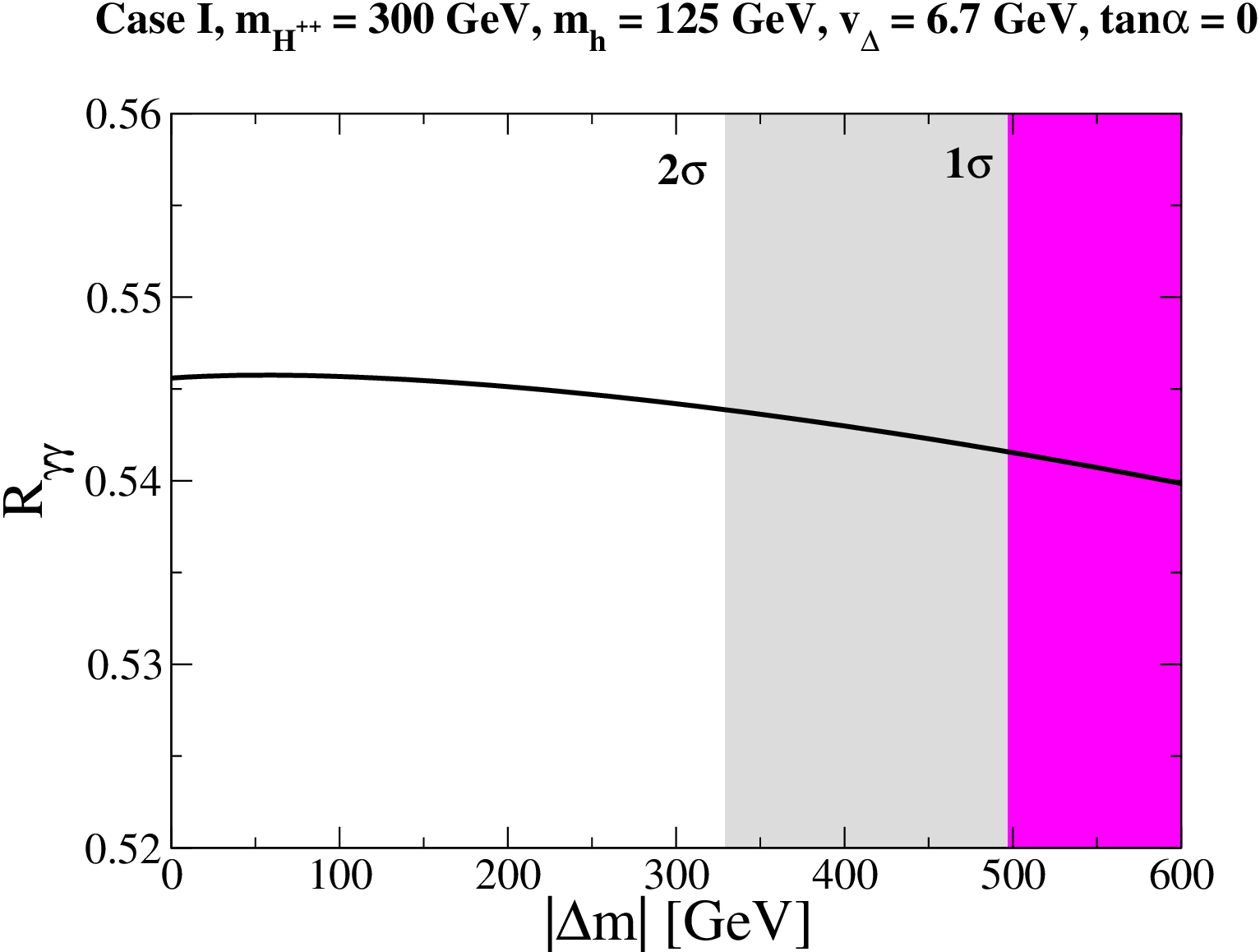}\hspace{3mm}
\caption{The ratio of the decay rate for $h\to \gamma\gamma$ in the HTM to that in the SM as a function of 
the absolute value of $\Delta m$ in Case I $(m_{\phi^0}>m_{H^+}>m_{H^{++}}$). 
We take $m_t=173$ GeV, $m_h=125$ GeV, $v_\Delta=6.7$ GeV and $\tan\alpha=0$ in all the figures. 
In the left (right) figure, we take $m_{H^{++}}=150$ GeV (300 GeV). 
The pink (gray) shaded region represents the 1$\sigma$ (2$\sigma$) allowed region of $\Delta m$ under the constraint from 
the data for $m_W^{\text{exp}}$ and $\hat{s}_W^{2\text{ exp}}$~\cite{ky_rho}.
}
\label{hgg3}
\vspace{5mm}
\includegraphics[width=65mm]{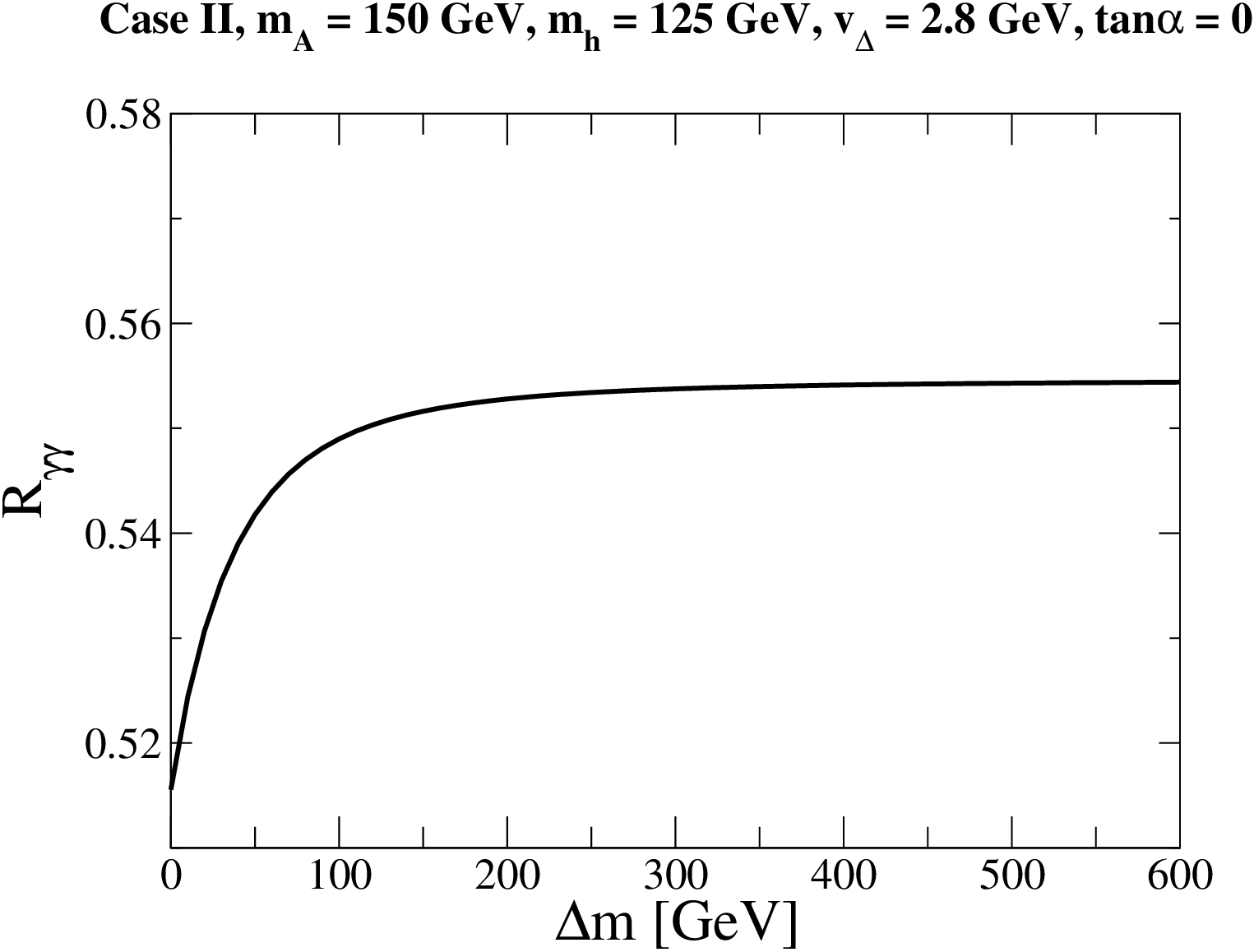}\hspace{3mm}
\includegraphics[width=65mm]{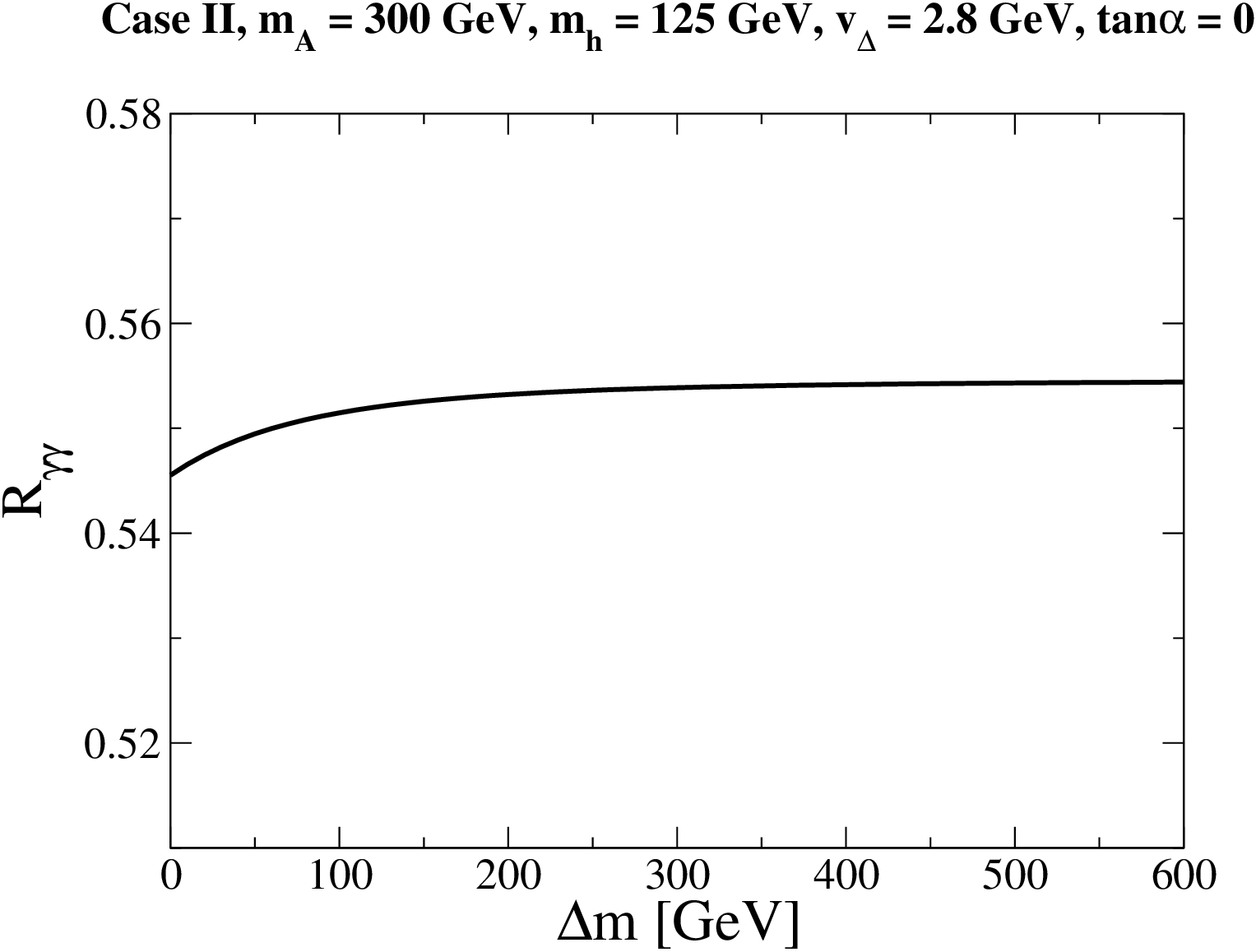}\hspace{3mm}
\end{center}
\caption{The ratio of the decay rate for $h\to \gamma\gamma$ in the HTM to that in the SM as a function of 
$\Delta m$ in Case II $(m_{H^{++}}>m_{H^+}>m_{\phi^0})$. 
We take $m_t=173$ GeV, $m_h=125$ GeV, $v_\Delta=2.8$ GeV and $\tan\alpha=0$ in all the figures. 
In the left (right) figure, we take $m_{A}=150$ GeV (300 GeV). 
There is no consistent region whth the data for $m_W^{\text{exp}}$ and $\hat{s}_W^{2\text{ exp}}$~\cite{ky_rho}. }
\label{hgg4}
\end{figure}

In Fig.~\ref{hgg1}, the ratio of the decay rates 
$R_{\gamma\gamma}\equiv \Gamma(h\to \gamma\gamma)_{\text{HTM}}/\Gamma(\phi_{\text{SM}}\to \gamma\gamma)_{\text{SM}}$ 
is shown as a function of $m_{H^{++}}$ for each value of $\Delta m$ at $m_h(=m_{\phi_{\text{SM}}})=125$ GeV 
and $\tan\alpha=0$ in Case I ($m_{\phi^0}>m_{H^+}>m_{H^{++}}$). 
For the left figure, $v_\Delta$ is taken to be 1 GeV, while 
it is taken to be 5 GeV for the right figure. 
In the both figures, $R_{\gamma\gamma}<1$ because 
the one-loop contributions of the singly-charged Higgs boson and the doubly-charged Higgs boson 
to $\Gamma(\phi\to \gamma\gamma)$ have the same sign which is destructive to the contribution of the SM loop diagrams. 
The magnitude of the deviation from the SM can be significant, which amounts to larger than 40\%. 
For $v_\Delta=1$ GeV the deviation is smaller when larger $\Delta m$ is taken.   
The deviation becomes smaller and insensitive to $\Delta m$ in the large mass region for $H^{\pm\pm}$.

One might think that the deviation would approach to zero in the large mass limit for $H^{\pm\pm}$. 
This can be true in a generic case. 
However, such decoupling is not applied to the present case. 
Since the coupling constants $c_{hH^{+}H^{-}}$ and $c_{hH^{++}H^{--}}$ are both proportional to 
the mass squired of triplet-like Higgs bosons, 
the large mass limit with a fixed value of $\Delta m$ with $\alpha=0$ can only be realized by taking 
these coupling constants to be infinity. 
It is known that in such a case, Appelquist's decoupling theorem~\cite{decoupling_theorem} does not hold, 
and the one-loop contributions of $H^\pm$ and $H^{\pm\pm}$ remain 
in the large mass limit as non-decoupling effects. 
We note that,   
in this case with $\alpha=0$, we have the relation $m_A^2\simeq M_\Delta^2=(\lambda_4+\lambda_5)v_\Phi^2/2$ from Eq.~(\ref{tan2a}), 
so that all the masses of triplet-like Higgs bosons cannot be taken to be  larger than 
TeV scales because of the perturbative unitarity constraint. 
On the contrary, 
if $\alpha=0$ is relaxed, 
$m_A$ is a free parameter, which satisfies 
$m_A^2\simeq M_\Delta^2=(\mu/v_\Delta)v_\Phi^2/\sqrt{2}$ from Eq.~(\ref{vc}),  
and it can be taken to be much larger than the electroweak scale when   
$\mu/v_\Delta\gg 1$ is assumed. 
Then, the correction due to the triplet field is suppressed by a factor of 
$v^2/m_A^2$. 
Namely, the decoupling theorem holds in this case.

In Fig.~\ref{hgg2}, $R_{\gamma\gamma}$ is shown as a function of $m_A$ 
for each value of $\Delta m$ at $m_h(=m_{\phi_{\text{SM}}})=125$ GeV 
and $\alpha=0$ 
in Case II ($m_{H^{++}}>m_{H^+}>m_{\phi^0}$). 
It is seen that as compared to Case I $R_{\gamma\gamma}$ 
is sensitive to the choice of $\Delta m$. 
Similarly to Case I, the deviation from the SM value is negative. 
However, smaller deviation is obtained for larger $\Delta m$ for the both cases with 
$v_\Delta=1$ GeV and $v_\Delta=5$ GeV in the region of relatively lower values of $m_A$.

In Fig.~\ref{hgg3}, we show the results of $R_{\gamma\gamma}$ as a function of $|\Delta m|$ 
in Case I with indicating the allowed regions of each confidence level under the 
electroweak precision data. 
The mass of $H^{\pm\pm}$ is taken to be 150 GeV (left) and 300 GeV (right). 
In all the figures, we take $m_h=125$ GeV, $\tan\alpha=0$ and $v_\Delta=6.7$ GeV. 
The magnitude of the ratio $R_{\gamma\gamma}$ strongly depends on $m_{H^{++}}$. 
Therefore, we may be able to obtain the indirect information of the mass spectrum of the 
triplet-like Higgs bosons by measuring the decay rate of $h\to \gamma\gamma$.

Finally, in Fig.~\ref{hgg4}, $R_{\gamma\gamma}$ is shown as a function of $\Delta m$ 
in Case II with indicating the allowed regions of each confidence level under the 
electroweak precision data. 
The mass of $A$ is taken to be 150 GeV (left) and 300 GeV (right). 
In all the figures, we take $m_h=125$ GeV, $\tan\alpha=0$ and $v_\Delta=2.8$ GeV. 
As compared to the case shown in Fig.~\ref{hgg3}, 
the mass dependence on $m_A$ is small among the three values of $m_A$. 
As we already discussed, Case II is not preferred by the electroweak precision data, 
and only the region with larger deviation than $2\sigma$ appears in each figure.

\clearpage
\section{Testing Higgs models via the $H^\pm W^\mp Z$ vertex}
A common feature in the extended Higgs models 
is the appearance of physical charged scalar components. 
Most of  the extended Higgs models contain 
singly charged Higgs bosons $H^\pm$ such as the THDM, the HTM, etc., which are discussed in the previous sections. 
Hence, we may be able to discriminate each Higgs model through 
the physics of charged Higgs bosons. 
In particular, the $H^\pm W^\mp Z$ vertex can be a useful probe 
of the extended Higgs sector~\cite{Grifols-Mendez,HWZ,HWZ-Kanemura,logan}.  
Assuming that there are several physical charged scalar states 
$H_{\alpha}^\pm$ ($\alpha \geq 2$) and the NG modes 
$H_1^\pm$,  The vertex parameter $\xi_\alpha$ in 
$\mathcal{L}=
igm_W \xi_\alpha H_\alpha^+ W^- Z+\textrm{h.c.}$ 
is calculated at the tree level as~\cite{Grifols-Mendez}
\begin{align}
\sum_{\alpha \geq 2} |\xi_\alpha|^2 
&=\frac{1}{\cos^2\theta_W}\left[\frac{2g^2}{m_W^2}
\Big\{\sum_i[T_i(T_i+1)-Y_i^2]|v_i|^2Y_i^2\Big\}-\frac{1}
{\rho_{\textrm{tree}}^2}\right], 
\end{align}
where $\rho_{\rm tree}$ is given in Eq.~(\ref{rho_tree}). 
A non-zero value of $\xi_\alpha$ appears at the tree level 
only when $H_\alpha^\pm$ comes from an exotic representation such as 
triplets. 
Similarly to the case of the rho parameter, 
the vertex is related to the custodial symmetry. 
In general, this can be independent of the rho parameter. 
If a charged Higgs boson $H^\pm_\alpha$ is from a doublet field, 
$\xi_\alpha$ vanishes at the tree level. 
The vertex is then one-loop induced 
and its magnitude is proportional to the violation of the global symmetry 
in the sector of particles in the loop. 
Therefore, the determination of the $H^\pm W^\mp Z$ vertex 
can be a complementary tool to the rho parameter in testing the 
{\it exoticness} of the Higgs sector. 

In this section, we discuss how accurately the $H^\pm W^\mp Z$ vertex 
can be determined at the collider experiments. 
At the LHC, the vertex would be determined by using the single $H^\pm$ 
production from the $WZ$ fusion\cite{Asakawa:2005gv}. 
The results are strongly model dependent, 
and the vertex may not be measured unless the $H^\pm$ is light enough 
and $|\xi_\alpha|^2$ is greater than $10^{-2}$. 
If kinematically allowed, the $H^\pm W^\mp Z$ vertex may also 
be measured via the 
decay process of $H^\pm \to W^\pm Z$~\cite{HWZ,HWZ-Kanemura}. 

We here focus on the process $e^+e^- \to W^\pm H^\mp$ 
at the ILC~\cite{Godbole-Mukhopadhyaya-Nowakowski,Cheung:1994rp,eeHW1,eeHW2,Kanemura_Moretti_Odagiri}. 
At the ILC, the neutral Higgs boson is produced via 
the Higgs strahlung process $e^+e^-\to ZH$~\cite{ZH}. 
The mass of the Higgs boson can be determined 
in a model independent way by using the so-called recoil method~\cite{recoil}, 
where the information of the Higgs boson can be extracted 
by measuring the leptonic decay products of the recoiled $Z$ boson. 
In this section, we employ this method to test the $H^\pm W^\mp Z$ vertex 
via $e^+e^- \to W^\pm H^\mp$. 
We analyze the signal and backgrounds at the parton level by using 
CalcHEP~\cite{CalcHEP}. 
We take into account the beam polarization and the expected 
resolution for the two-jet system.    
We find that assuming that $H^\pm$ decays into lepton pairs, 
the $H^\pm W^\mp Z$ vertex can be explored accurately by measuring the 
signal of the two-jet with one charged lepton with missing momentum.
For relatively light charged Higgs bosons,   
the signal significance with the value of $|\xi_\alpha|^2 \sim O(10^{-3})$ 
can be as large as two  after appropriate kinematic cuts 
for the collision energy $\sqrt{s} = 300$ GeV and the 
integrated luminosity 1 ab$^{-1}$, 
even when the initial state radiation (ISR) is taken into account.  

\subsection{The $H^\pm W^\mp Z$ vertex}

\begin{figure}[t]
\begin{center}
\includegraphics[width=90mm]{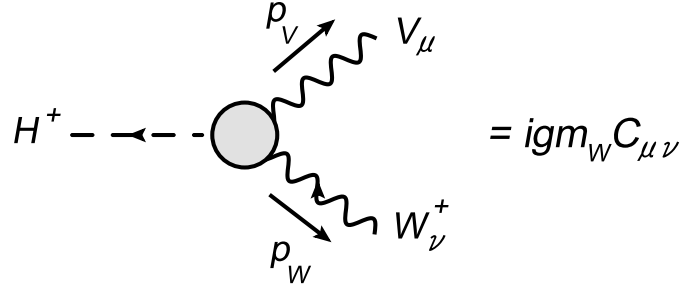}
\caption{The $H^\pm W^\mp V$ vertex ($V=Z$ or $\gamma$)~\cite{Yanase}.}
\label{hwvf}
\end{center}
\end{figure}
The $H^\pm W^\mp V$ vertex ($V=Z$ or $\gamma$) 
is defined in FIG. \ref{hwvf}, where   
$C^{\mu\nu}$ is expressed in terms of the form factors 
$F_{HWV}$, $G_{HWV}$ and $H_{HWV}$ as 
\begin{align}
C^{\mu\nu}=F_{HWV}g^{\mu\nu}+G_{HWV}\frac{p_W^\mu p_V^\nu}{m_W^2}
       +iH_{HWV}\frac{p_{W\rho} p_{V\sigma}}{m_W^2}\epsilon^{\mu\nu\rho\sigma},
\label{HWV}
\end{align}
with $\epsilon_{\mu\nu\rho\sigma}$ being the anti-symmetric tensor 
with $\epsilon_{0123}=+1$, and $p_V^\mu$ and $p_W^\mu$ being the 
outgoing momenta of $V$ and $W$ bosons, respectively. 
Among the form factors, $F_{HW\gamma}=0$ is derived at the tree level
 due to the $U(1)_{\rm em}$ gauge invariance 
in any extended Higgs models. 
These form factors $F_{HWV}$, $G_{HWV}$ and $H_{HWV}$ 
are respectively related to the coefficients 
$f_{HWV}$, $g_{HWV}$ and $h_{HWV}$ in the effective 
Lagrangian~\cite{HWZ,HWZ-Kanemura}; 
\begin{align}
\mathcal{L}_{\text{eff}}&=gm_Wf_{HWV}H^\pm W_\mu^\mp V^\mu 
                          +g_{HWV}H^\pm F_V^{\mu\nu} F_{W\mu\nu} 
+(ih_{HWV} \epsilon_{\mu\nu\rho\sigma}H^\pm F_V^{\mu\nu} F_W^{\rho\sigma} +\text{H.c.}),
\end{align}
where $F_V^{\mu\nu}$, and $F_W^{\mu\nu}$ are the field strengths. 
We note that $f_{HWZ}$ is the coefficient of the dimension 
three operator, while  
the $g_{HWV}$ and $h_{HWV}$ are those of the dimension five operator,  
so that only $f_{HWZ}$ may appear at the tree level. 
Therefore, the dominant contribution to the $H^\pm W^\mp V$ vertex 
is expected to be from $F_{HWZ}$.  

In the Higgs model with only doublet scalar fields (plus singlets) 
all the form factors including $F_{HWZ}$ vanish 
at the tree level~\cite{Grifols-Mendez}, because of the 
custodial invariance in the kinetic term. 
The form factors  $F_{HWV}$, $G_{HWV}$ and $H_{HWV}$ ($V=\gamma$ and $Z$) 
are generally induced at the loop level. 
In particular, the leading one-loop contribution to $F_{HWZ}$ are 
proportional to the violation of 
the custodial symmetry in the sector of the particle in the loop.
For example, in THDM, the custodial 
symmetry is largely broken via the $t$-$b$ loop contribution as well 
as via the Higgs sector with the mass difference between 
the CP-odd Higgs boson ($A^0$) and the charged Higgs boson 
$H^\pm$~\cite{HWZ-Kanemura}.   
The one-loop induced form factors are theoretically constrained 
from above by perturbative unitarity \cite{PU_thdm, PU_thdm2}. 
In such a case, the effect of the custodial symmetry violation also 
can deviate the rho parameter from unity at the one loop level. 
However, when the lightest of CP-even neutral Higgs bosons 
is approximately regarded as  the SM-like Higgs boson, 
the rho parameter can be unity even with a large mass splitting between 
$A^0$ and $H^\pm$  
when the masses of the heavier CP-even neutral Higgs 
boson $H^0$ and $H^\pm$ are common~\cite{Csym}. 
This means that the appearance of the 
$H^\pm W^\mp Z$ vertex and the deviation from unity in the rho parameter 
are not necessarily correlated at the one-loop level, 
and they can be independent quantities, in principle.

The simplest models in which the $H^\pm W^\mp Z$ vertex appears 
at the tree level are those with triplet scalar fields.
In the model with an isospin doublet field ($Y=1/2$) 
and either an real triplet field $\eta$ ($Y=0$) 
or an additional complex triplet field $\Delta$ ($Y=1$),  
concrete expressions for the tree-level formulae for  
$|F_{HWZ}|^2$ and that of $\rho_{\textrm{tree}}$ are 
shown in TABLE~\ref{f_models}, 
where $v$, $v_\eta$ and $v_\Delta$  are respectively 
VEVs of the doublet scalar field and 
the additional triplet scalar field $\eta$ and $\Delta$. 
These triplet scalar fields also contribute to 
the rho parameter at the tree level, so that their VEVs 
are constrained by the current rho parameter data,  
$\rho_{\exp}=1.0008^{+0.0017}_{-0.0007}$; i.e., 
$v_\eta \lesssim 6$ GeV for the real triplet field $\eta$, 
and $v_\Delta \lesssim 8$ GeV for the complex triplet $\Delta$ 
(95 \% CL). 
We note that in order to obtain the similar accuracy to the rho 
parameter data by measuring the $H^\pm W^\mp Z$ vertex, 
the vertex  has to be measured with the detectability to    
$|F_{HWZ}|^2 \sim {\cal O}(10^{-3})$.

\begin{table}[t] 
\begin{center}
{\renewcommand\arraystretch{1.4}
\begin{tabular}{|c||c|c|c|}\hline
Model & SM with $\eta$  ($Y=0$)& SM with $\Delta$ ($Y=1$) 
& the GM model \\\hline\hline
$|F_{HWZ}|^2=$&$\frac{4v^2 v^{2}_\eta}{\cos^2\theta_W(v^2+4v^{2}_\eta)^2}$
&$\frac{2v^2 v^{2}_\Delta}{\cos^2\theta_W(v^2+2v^{2}_\Delta)^2 }$& 
$\frac{4 v_\Delta^2}{\cos^2\theta_W(v^2+4v_\Delta^2)}$ 
\\\hline 
$\rho_{\text{tree}}=$&$1+\frac{4v^{2}_\eta}{v^2}$
&$\frac{1+2\frac{v^{2}_\Delta}{v^2}}{1+4\frac{v^{2}_\Delta}{v^2}}$&$1$ \\\hline
\end{tabular}}
\caption{The tree-level expression 
for $F_{HWZ}$ and rho parameter at the tree level~\cite{Yanase} 
in the model with a real triplet field, that with a complex triplet field 
and the Georgi-Machacek (GM) model~\cite{Georgi-Machacek}.}
\label{f_models}
\end{center}
\end{table}

Finally, we mention about the model with a real triplet field $\eta$ 
and a complex triplet field $\Delta$ in addition to the SM, 
which is proposed by Georgi-Machacek and 
Chanowiz-Golden~\cite{Georgi-Machacek, Gunion-Vega-Wudka,Godbole-Mukhopadhyaya-Nowakowski, AK_GM}.
In this model, an alignment of the VEVs for $\eta$ and $\Delta$ 
are introduced ($v_\eta= v_\Delta/\sqrt{2}$), 
by which the Higgs potential is invariant under   
the custodial $SU(2)$ symmetry at the tree level. 
Physical scalar states in this model can be classified 
using the transformation property against 
the custodial symmetry; i.e., 
the five-plet, the three-plet and the singlet. 
Only the charged Higgs boson from the five-plet state has 
the non-zero value of $F_{HWZ}$ at the tree level. 
Its value is proportional to the VEV $v_\Delta$ for the triplet scalar 
fields. 
However, the value of $v_\Delta$ is not strongly constrained 
by the rho parameter data, because 
the tree level contribution to the rho parameter 
is zero due to the custodial symmetry: 
see TABLE~\ref{f_models}. 
Consequently, the magnitude of $|F_{HWZ}|^2$ 
can be of order one.

\subsection{The $e^+e^-\to H^\pm W^\mp$ process}

\begin{figure}[t]
\begin{center}
\includegraphics[width=70mm]{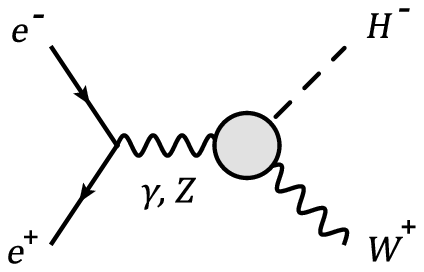}
\caption{The $e^+e^- \to H^-W^+$ process~\cite{Yanase}.}
\label{eehwa}
\end{center}
\end{figure}

The process $e^+ e^-\to H^-   W^+$~\cite{eeHW1,eeHW2,Kanemura_Moretti_Odagiri} is depicted in FIG.~\ref{eehwa}. 
This process is directly related to the $H^\pm W^\mp Z$ vertex. 
The helicity amplitudes are calculated by 
\begin{align}
\mathcal{M}(\tau,\lambda)
&=\sum_{V=Z,\gamma}igm_WC_V\frac{1}{s-m_V^2}j_\mu(\tau) C^{\mu\nu}
\epsilon_\nu(\lambda),
\end{align}
where $\sqrt{s}$ is the center-of-mass energy, 
$j_\mu(\tau)$ is the electron current, and 
$\epsilon_\nu(\lambda)$ is the polarization vector of the $W^+$ 
boson~\cite{eeHW1}. 
The helicities of the electron and the $W^+$ boson 
can be $\tau=\pm 1$ and $\lambda=0,\pm 1$, respectively.  
The coefficient $C_V$ is given by
\begin{align}
C_V=\left\{
\begin{array}{cc}
eQ_e,&\hspace{4mm}\text{for }V=\gamma,\\
\frac{g}{\cos\theta_W}(T_e^3-\sin^2\theta_WQ_e),
&\hspace{4mm}\text{for }V=Z,\label{c}
\end{array}
\right.
\end{align}
with $Q_e=-1$, $T_e^3=-1/2$ $(0)$ for $\tau=-1$ ($+1$). 
The squared amplitude is evaluated as
\begin{align}
&|\mathcal{M}(\tau)|^2\equiv
\sum_{\lambda=0,\pm}|\mathcal{M}(\tau,\lambda)|^2\notag\\
&=g^2\left|C_\gamma\frac{F_{HW\gamma}}{s}+
C_Z\frac{F_{HWZ}}{s-m_Z^2}\right|^2\left[\frac{\sin^2\theta}{4}
(s+m_W^2-m_{H^\pm}^2)^2+sm_W^2(\cos^2\theta+1)\right],\label{amp2}
\end{align}
where $\theta$ is the angle between the momentum of $H^\pm$ and the 
beam axis, $m_{H^\pm}$ is the mass of $H^\pm$ 
and the form factors $G_{HWV}$ and $H_{HWV}$ are taken to be zero. 
The helicity specified cross sections are written 
in terms of the squared amplitude in Eq. (\ref{amp2}), 
\begin{align}
\sigma(s; \tau)=
\frac{1}{32\pi s}\beta\left(\frac{m_{H^\pm}^2}{s},\frac{m_W^2}{s}\right)
\int_{-1}^{1}d\cos\theta|\mathcal{M}(\tau)|^2,
\end{align}
where $\sigma(s; +1)=\sigma(e^+_Le^-_R\to H^-W^+)$ and 
$\sigma(s; -1)=\sigma(e^+_Re^-_L\to H^-W^+)$, and 
\begin{align}
\beta\left(x,y\right)=\sqrt{1+x^2+y^2-2xy-2x-2y}.
\end{align}
The helicity averaged cross section is given by 
$\sigma(e^+e^-\rightarrow H^-W^+)=(\sigma(s, +1)+\sigma(s, -1))/4$.

\begin{figure}[t]
\begin{center}
\includegraphics[width=54mm]{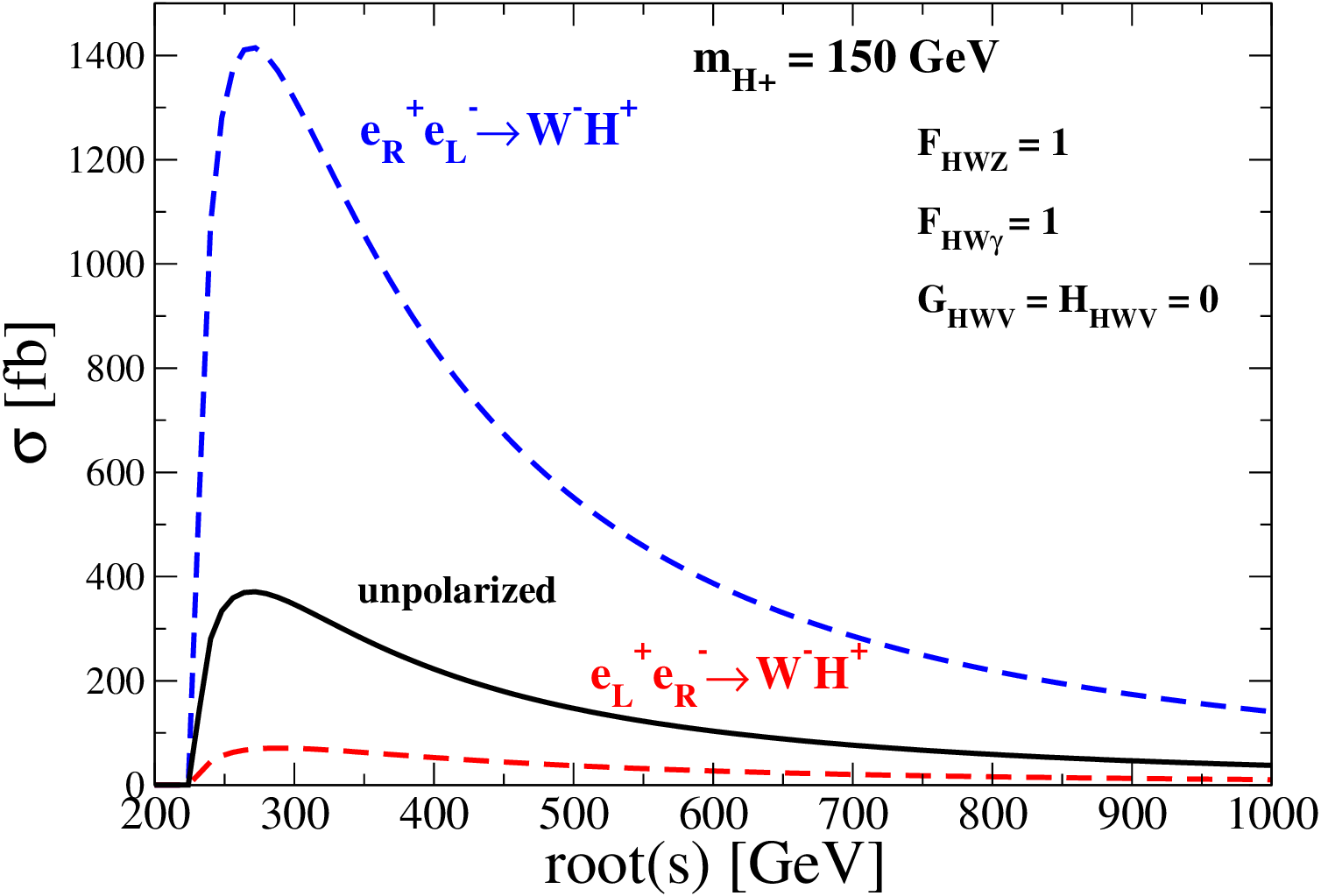}
\includegraphics[width=54mm]{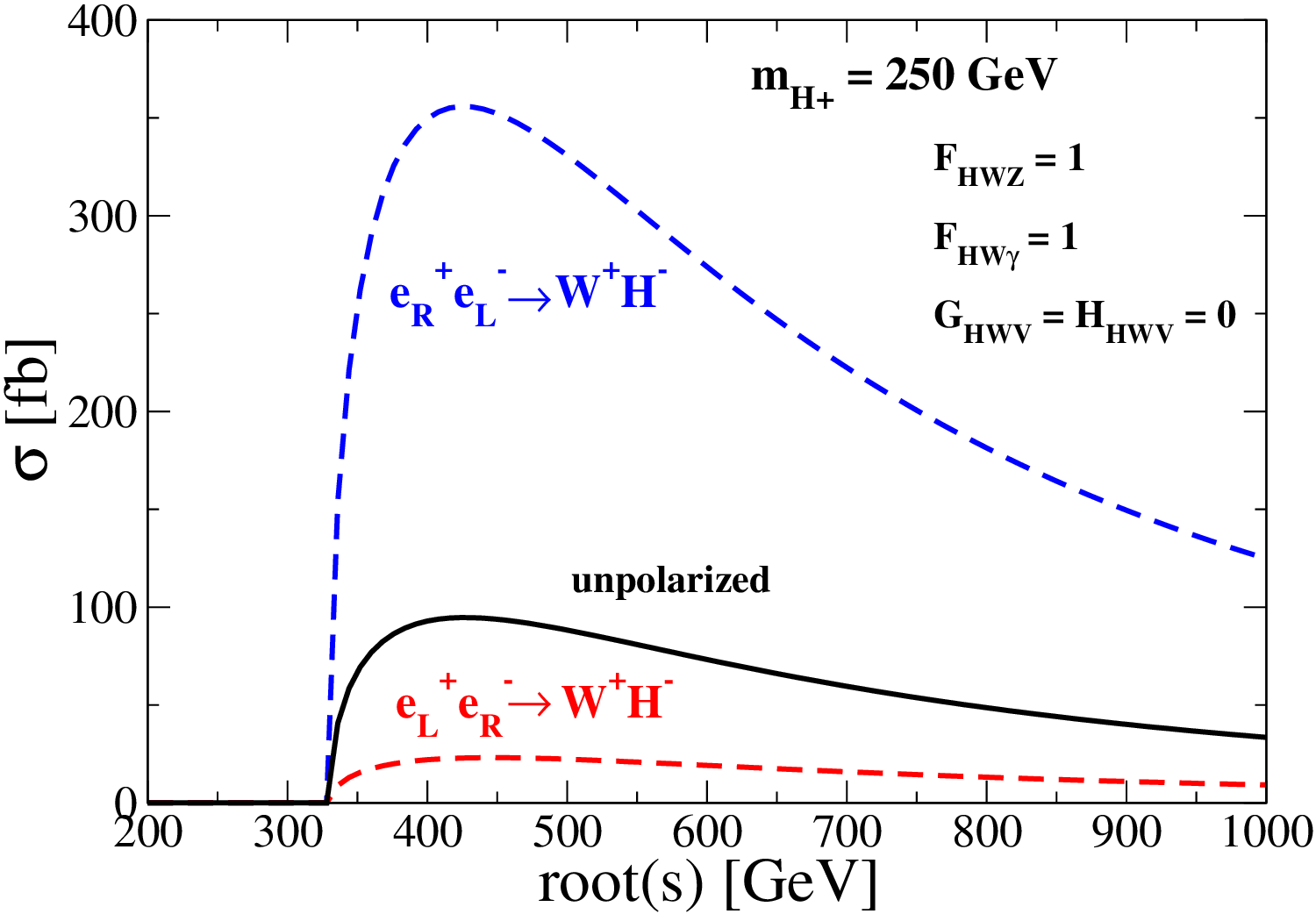}
\includegraphics[width=54mm]{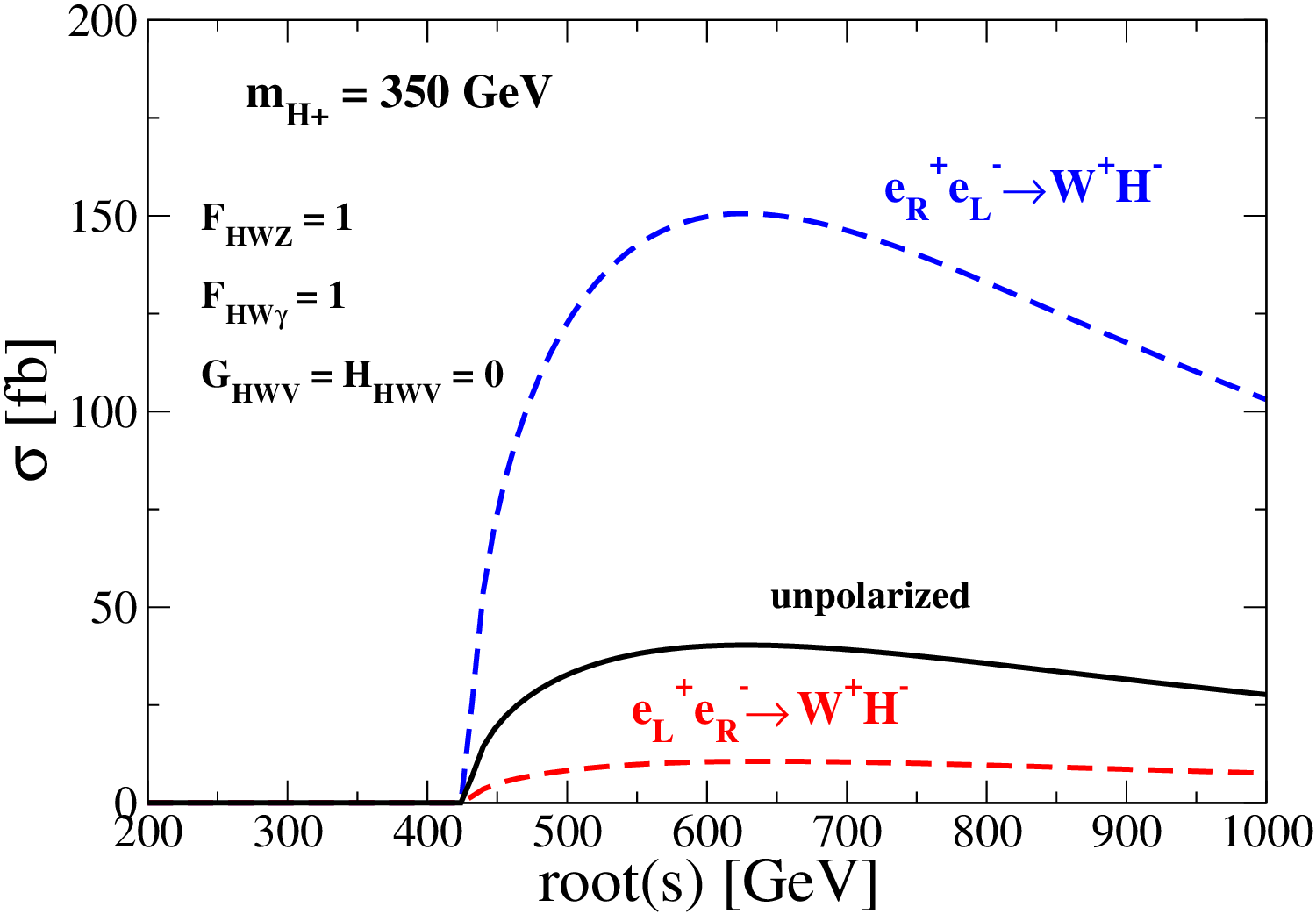}
\caption{The total cross section as a function of $\sqrt{s}$ in the case 
of $F_{HWZ}=F_{HW\gamma}=1$~\cite{Yanase}.} \label{mch150_fz1_0}
\end{center}
\end{figure}

\begin{figure}[t]
\begin{center}
\includegraphics[width=54mm]{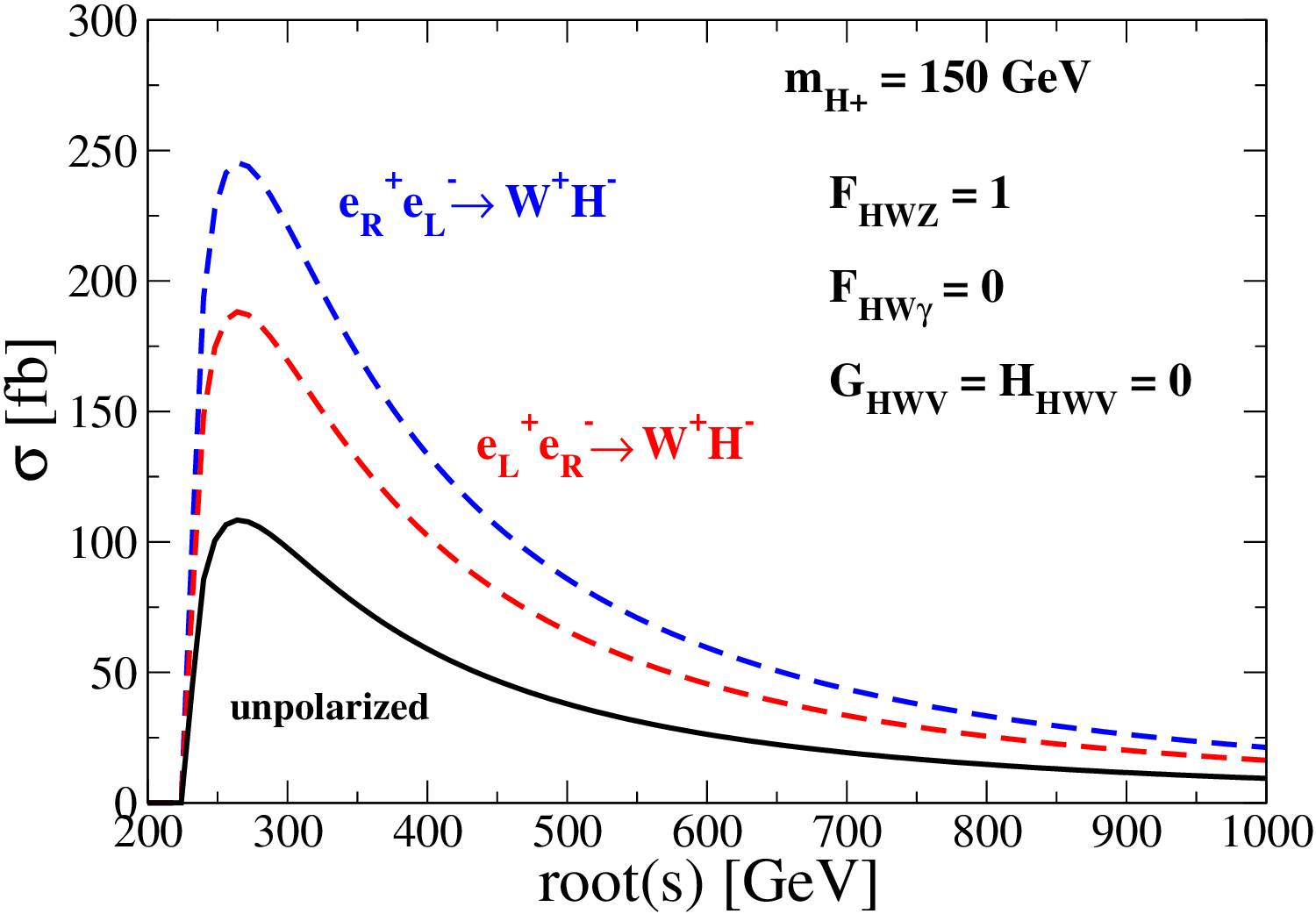}
\includegraphics[width=54mm]{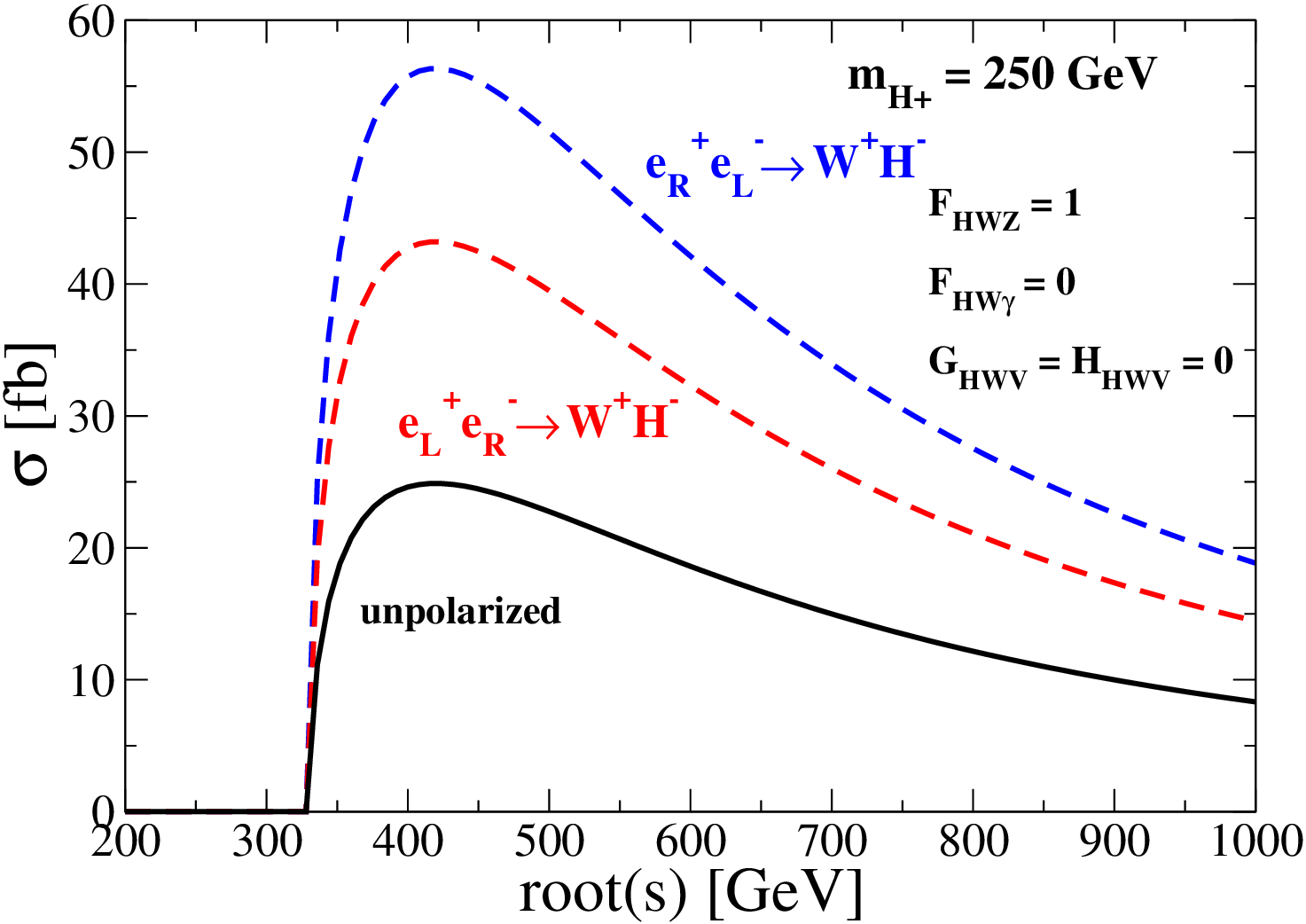}
\includegraphics[width=54mm]{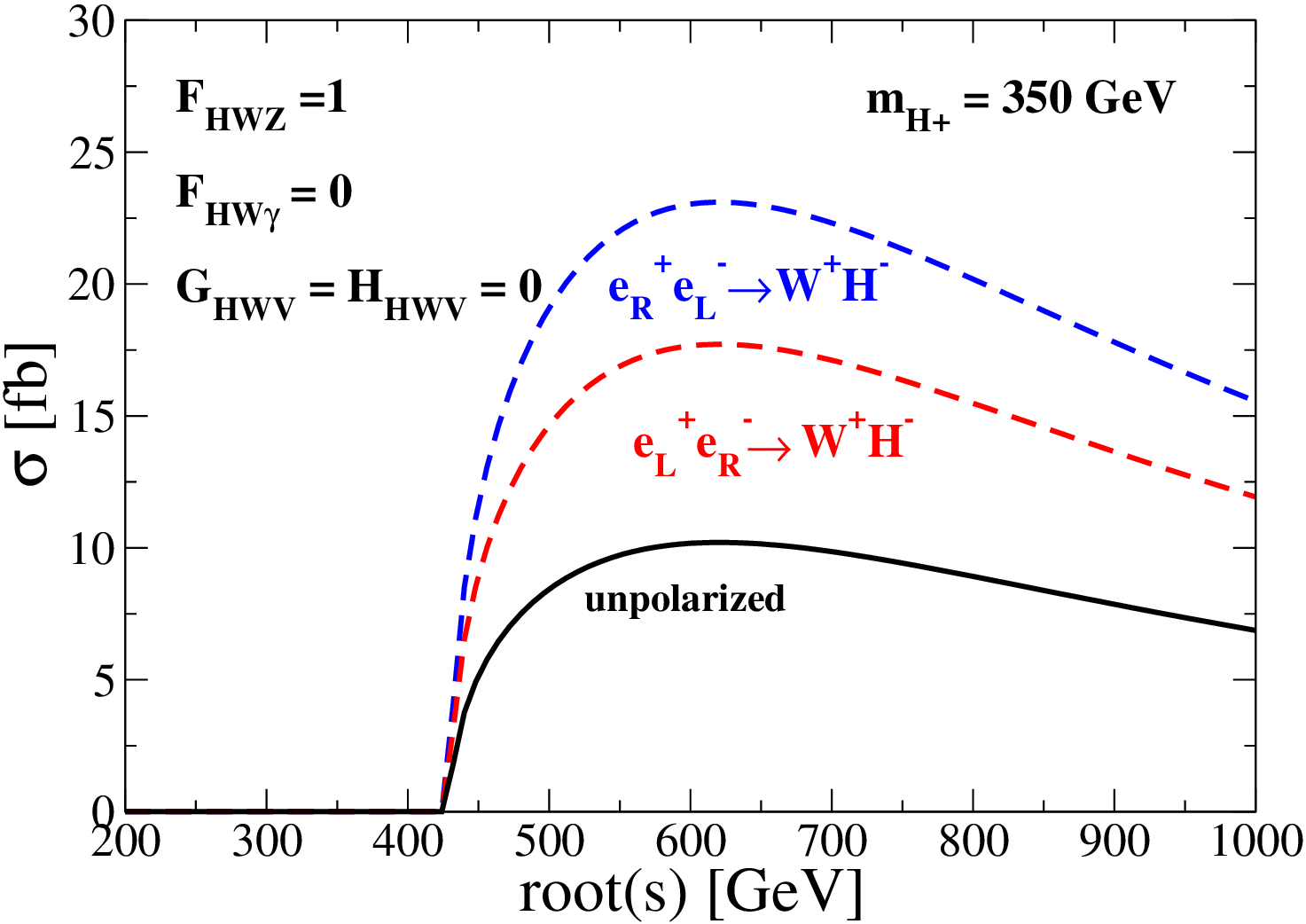}
\caption{The total cross section as a function of $\sqrt{s}$ 
in the case of $F_{HWZ}=1, F_{HW\gamma}=0$~\cite{Yanase}.}
\label{mch150_fz1_02}
\end{center}
\end{figure}

In FIGs. \ref{mch150_fz1_0} and \ref{mch150_fz1_02}, we show that the 
$\sqrt{s}$ dependence
 of the helicity dependent and the helicity averaged cross sections. 
Notice that the behavior of these cross sections drastically changes 
depending on the initial electron helicity 
in the case of $F_{HWZ}\simeq F_{HW\gamma}$. 
On the contrary, there is no such a difference 
in the case of $F_{HWZ}\gg F_{HW\gamma}$. 
As mentioned before, $F_{HW\gamma}$ is zero at the tree level in any models 
because of the $U(1)_{\textrm{em}}$ gauge invariance. 
The relation of $F_{HWZ}\gg F_{HW\gamma}$ or $F_{HWZ}\simeq F_{HW\gamma}$ 
can be tested by using the initial electron helicities.  



\subsection{Recoil method and the assumption for the ILC performance}

We investigate the possibility of measuring 
the $H^\pm W^\mp Z$ vertex by using a recoil method at the ILC. 
It has been known that this method is a useful tool 
for measuring the mass of the SM-like Higgs boson $H_{\rm SM}$  
without assuming the decay branching fraction of the Higgs 
boson~\cite{recoil}. 
In the Higgs-strahlung process $e^+e^- \to ZH_{\text{SM}} $~\cite{ZH}, 
the Higgs boson mass can be obtained as the recoil mass $m_{\text{recoil}}$ 
from two leptons produced from the $Z$ boson, whose 
energy is $E_{\ell\ell}$, and the invariant mass is $M_{\ell\ell}$. 
They satisfy the relation, 
\begin{align}
m_{\text{recoil}}^2(\ell\ell)=s-2\sqrt{s}E_{\ell\ell}+M_{\ell\ell}^2.
\end{align}
The information of the Higgs boson mass can be extracted by 
measuring $E_{\ell\ell}$ and $m_{\ell\ell}$ 
in a model independent way. 

\begin{figure}[t]
\begin{center}
\includegraphics[width=70mm]{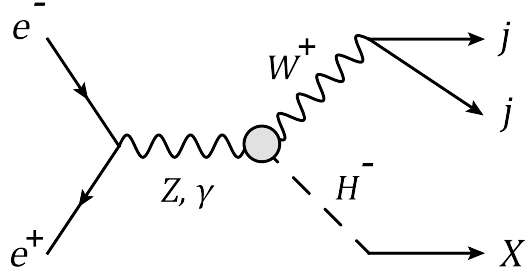}
\caption{The signal process~\cite{Yanase}.}
\label{ee_jjX}
\end{center}
\end{figure}

In this section, we apply this method to $e^+e^-\to W^\pm H^\mp $ in order to  
measure the $H^\pm W^\mp Z$ vertex. 
In order to identify the process, 
we consider the hadronic decays $W\to jj$ 
instead of the leptonic decay of the produced $W$ boson, 
and obtain information of the $H^\pm W^\mp Z$ vertex by using the 
recoil of the two-jet system. 
The recoiled mass of $H^\pm$ is given in terms of the 
two-jet energy $E_{jj}$ and 
the two-jet invariant mass $M_{jj}$ as
\begin{align}
m_{\text{recoil}}^2(jj)=s-2\sqrt{s}E_{jj}+M_{jj}^2.\label{recoil2}
\end{align}
This process is shown in FIG.~\ref{ee_jjX}.
It is clear that the detector performance for 
the resolution of two jets is crucial in such an analysis. 
In particular, the jets from the $W$ boson in the signal process 
has to be precisely measured in order to be separated with those from the 
$Z$ boson in the background process. 
At the ILC, the resolution for the two jet system with the energy $E$ 
in the unit of GeV is expected to be $\sigma_E = 0.3 \sqrt{E}$ GeV, 
by which the background 
from $Z\to jj$ can be considerably reduced.  
We here adopt the similar value for $\sigma_E$ ($\sim 3$ GeV) 
in our later analysis. 

At the ILC, the polarized electron and positron beams can be used, by 
which the background from the $W$ boson pair production process can be 
reduced. We here use the following beams polarized as  
\begin{align}
\frac{N_{e_R^-}-N_{e_L^-}}{N_{e_L^-}+N_{e_R^-}} =0.8,\quad 
\frac{N_{e_L^+}-N_{e_R^+}}{N_{e_L^+}+N_{e_R^+}} =0.5,\label{porl}
\end{align}
which are expected to be attained at the ILC~\cite{ILCTDR2007}, 
where 
$N_{e_{R,L}^-}$ and $N_{e_{R,L}^+}$ are numbers of right- (left-) handed electron and 
positron in the beam flux per unit time. 
The total cross sections for the  signal and the backgrounds can be evaluated 
from the helicity specified cross sections as  
\begin{align}
\sigma_{\text{tot}} (e^+e^-\to X)=&x_-x_+\sigma(e_L^+e_R^- \to X)+(1-x_-)(1-x_+) 
\sigma(e_R^+e_L^- \to X)\notag\\
&+x_-(1-x_+)\sigma(e_R^+e_R^- \to X)+x_+(1-x_-)\sigma(e_L^+e_L^- \to X),
\end{align}
where $x_- = N_{e_R^-}/(N_{e_L^-}+N_{e_R^-})$ and 
      $x_+ = N_{e_L^+}/(N_{e_L^+}+N_{e_R^+})$.  

The high-energy electron and positron beams lose their incident energies 
by the ISR. 
In our analysis, we also take into account such effect, and see how the 
results without the ISR are changed by including the effect of the ISR. 

\subsection{Signal and Backgrounds}

The size of the signal cross section is determined by 
the center of mass energy $\sqrt{s}$, the mass $m_{H^\pm}^{}$ and 
the form factors $F_{HWZ}$ and $F_{HW\gamma}$. 
In the following analysis, we consider the case of 
$(F_{HWZ}, F_{HW\gamma}) \equiv (\xi,0)$. 
This approximately corresponds to most of the cases we are interested, such as 
the triplet models.   
In order to examine the possibility of constraining $|\xi|^2$, 
we here assume that the mass of the charged Higgs boson is already 
known with some accuracy 
by measuring the other processes at the LHC or at the ILC. 
Then $|\xi|^2$ is the only free parameter in the production cross section.  

\begin{figure}[t]
\begin{center}
\includegraphics[width=120mm]{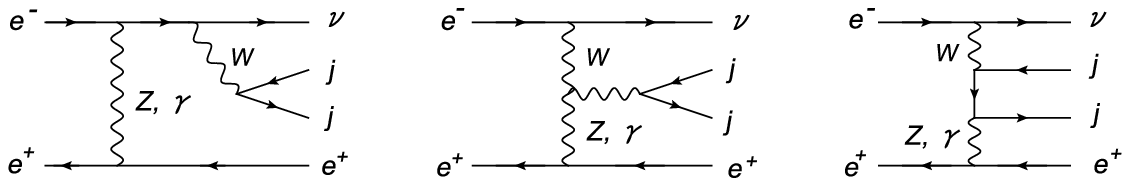}\\
\vspace{5mm}
\includegraphics[width=80mm]{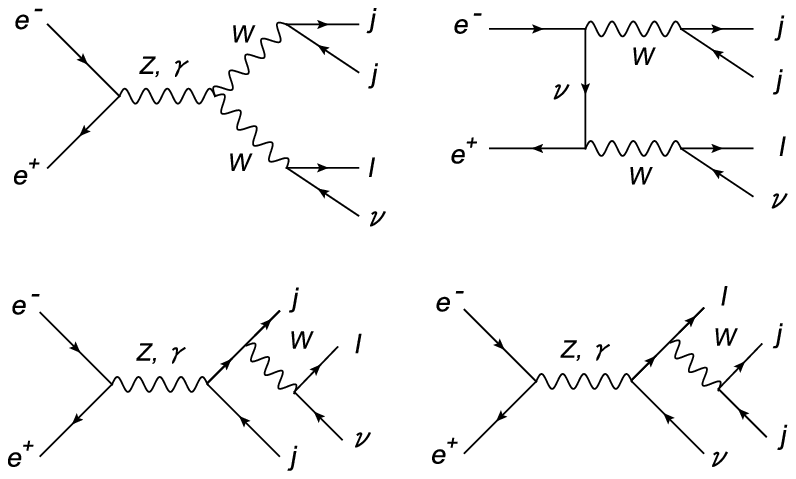}
\caption{The $e^+e^-\to \ell\nu jj$ backgrounds~\cite{Yanase}.}
\label{enjj_BG}
\end{center}
\end{figure}

\begin{figure}[t]
\begin{center}
\includegraphics[width=80mm]{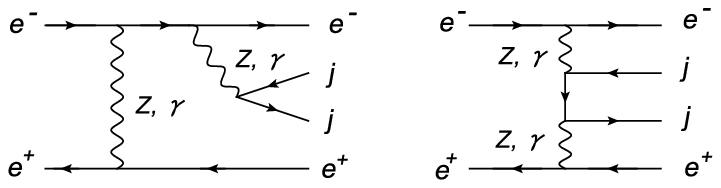}\\
\vspace{5mm}
\includegraphics[width=120mm]{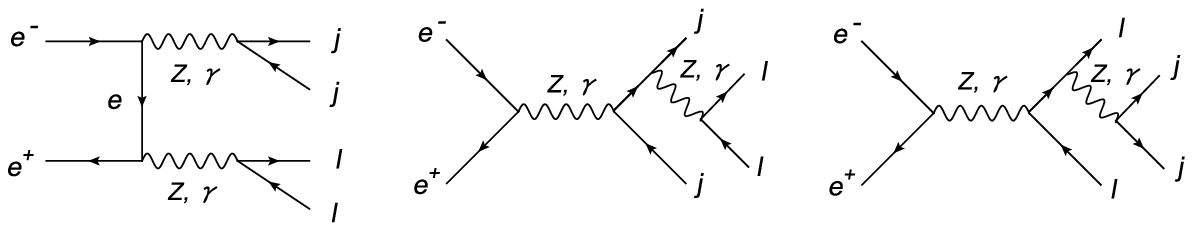}
\caption{The $e^+e^-\to \ell\ell jj$ backgrounds~\cite{Yanase}.}
\label{eejj_BG}
\end{center}
\end{figure}

In order to perform the signal and background analysis, 
we here assume that the decay of the produced charged Higgs boson is 
lepton specific; i.e., $H^\pm \to \ell \nu$ where $\ell$ is either 
$e$, $\mu$ or $\tau$.  
The final state of the signal is then $e^+e^- \to H^\pm W^\mp \to \ell \nu jj$. 
We first consider $m_{H^\pm} < m_W+m_Z$ to avoid the complexness 
with the possible decay mode of $H^\pm \to W^\pm Z$, whose 
branching ratio strongly depends on the model. 
The main backgrounds come from the $W$ boson pair production process 
$e^+e^- \to W^+W^-$ and the single $W$ production processes 
in FIG.~\ref{enjj_BG}. 
For the $e^\pm \nu jj$ final state, additional processes 
shown in FIG.~\ref{enjj_BG} (upper figures) 
can also be a significant background. 
In addition, we take into account the processes with the final 
state of $\ell\ell jj$ shown in FIG.~\ref{eejj_BG}. They can be 
backgrounds if one of the outgoing leptons escapes from 
the detection at the detector. We here assume that 
the efficiency for lepton identification is 90 \%. 

\begin{figure}[t]
\begin{center}
\includegraphics[width=70mm]{distr_tjj.eps}\hspace{5mm}
\includegraphics[width=70mm]{distr_ejj.eps}\\
\vspace{10mm}
\includegraphics[width=70mm]{distr_clep.eps}\hspace{5mm}
\includegraphics[width=70mm]{distr_mll.eps}
\end{center}
\caption{Distributions of the signal 
for $m_{H^\pm}=110$, 130, 150 and 170 GeV as well as the backgrounds 
after the invariant mass $M_{jj}$ cut in Eq.~(\ref{mjj}) without the ISR   
as a function of the transverse momentum 
$p_T^{jj}$ (upper left), the energy of the $jj$ system (upper right), 
the angle $\theta_{\rm lep}$ of a charged lepton with the beam 
axis (lower left), and 
the invariant mass $M_{\ell\nu}$ of the charged lepton and the missing momentum 
in the final state (lower right).  
$|\xi|^2$ is taken to be 1~\cite{Yanase}.}
\label{distr1}
\vspace{1cm}
\begin{center}
\includegraphics[width=70mm]{distr_recoil.eps}
\end{center}
\caption{Distributions of the signal for $m_{H^\pm}=110$, 130, 150 and 170 GeV
 as well as the backgrounds after the cut in Eq.~(\ref{mjj}) 
without the ISR as a function of the recoil mass $m_{\rm recoil}$~\cite{Yanase}.}
\label{distr_recoil}
\end{figure}

We impose the basic cuts for all events such as 
\begin{align}
10^\circ < A_j < 170^\circ,\quad 5^\circ < A_{jj} < 175^\circ,
\quad 10\; \textrm{GeV}< E_{jj}, \label{basic}
\end{align}
where $A_j$ is the angle between a jet and the beam axis, 
$A_{jj}$ is the angle between the two jets 
and $E_{jj}$ is the energy of the two jets. 
In the numerical evaluation, we use CalcHEP~\cite{CalcHEP}.

After the basic cuts, the event numbers of both the signal and the 
backgrounds are listed in TABLE~\ref{result1} for the case without ISR, and 
in TABLE~\ref{result2} for that with ISR, 
where the center of mass energy is set $\sqrt{s}=300$ GeV, the mass of the charged Higgs boson $m_{H^\pm}$ is 150 GeV,  
and the parameter $|\xi|^2$ for the $H^\pm W^\mp Z$ vertex is set to be $10^{-3}$. 
For both the cases signal over background 
ratios are less than $10^{-4}$ before 
imposing the other kinematic cuts than the basic cuts in Eq.~(\ref{basic}). 
In the following we first discuss the case without the ISR, 
then present the results for that with the ISR.

In order to improve the signal over background 
ratio, we impose additional kinematic cuts.
The two jets come from the $W$ boson for the signal, so that 
the invariant mass cut is useful to reduce the backgrounds where 
a parent of the two jets is not the $W$ boson. 
We here impose the condition; 
\begin{align}
m_W-n\sigma_E < M_{jj} < m_W+n\sigma_E,  \label{mjj}   
\end{align}
where $\sigma_E$ represents the resolution of the detector 
which we assume 3 GeV, and $n$ is taken to be 2 here.

In FIG.~\ref{distr1}, the differential cross sections of the signal and the 
backgrounds are shown for the events after the $M_{jj}$ cut in Eq.~(\ref{mjj}) 
as a function of the transverse momentum $p_T^{jj}$, 
the energy of the $jj$ system, the angle $\theta_{\rm lep}$ of a charged 
lepton with the beam axis, and 
the invariant mass $M_{\ell\nu}$ of the charged lepton and the missing momentum 
in the final state.  
For the signal, the results are shown 
for $|\xi|^2=1$ with the mass of the charged Higgs boson to be 
110, 130, 150 and 170 GeV. 
The $E_{jj}$ distribution shown in FIG.~\ref{distr1} (upper-right) can be 
translated into the distribution as a function of $m_{\rm recoil}$ by using the 
relation in Eq.~(\ref{recoil2}), which is shown in FIG.~\ref{distr_recoil}. 
The signal events form the peak at $m_{\rm recoil} \sim m_{H^\pm}$. 

In the following, we discuss the case with $m_{H^\pm}=150$ GeV\footnote{
Notice that the additional cuts given in Eqs.~(\ref{cut1})-(\ref{cut4}) 
are optimized for the case with $m_{H^\pm}=150$ GeV. 
From FIG.~\ref{distr1}, we can find and impose 
such additional cuts optimized for each value of $m_{H^\pm}$.
}. 
According to FIG.~\ref{distr1}, we impose the following 
four kinematic cuts sequentially: 
\begin{align}
  75\text{ GeV} <p_T^{jj}<100\text{ GeV},   \label{cut1} 
\end{align}
and 
\begin{align}
  115\text{ GeV} <E_{jj}<125\text{ GeV}   \label{cut2} 
\end{align}
for the $jj$ system in the final state.  
In TABLE~\ref{result1}, the resulting values for the cross sections 
for the signal and backgrounds are shown in each step of the cuts.
The backgrounds can be reduced in a considerable extent. 
For $|\xi|^2=10^{-3}$, the signal significance reaches 
to ${\cal O}(1)$ assuming the integrated luminosity of 
1 ab$^{-1}$.  

Until now, we have imposed the cuts on the $jj$ system, 
and no information from the $\ell \nu$ system has been used.  
Here, in order to further improve the signal significance,   
we impose new cuts related to the $\ell \nu$ system in order, which 
are determined from FIG.~\ref{distr1}; 
\begin{align}
|\cos\theta_{\text{lep}}| < 0.75, \label{cut3} 
\end{align}
and 
\begin{align}
144\text{ GeV} <M_{\ell\nu}<156\text{ GeV}. \label{cut4} 
\end{align}
As shown in TABLE~\ref{result1}, for $|\xi|^2=10^{-3}$  
the signal significance after these cuts can reach 
to $S/\sqrt{B} \simeq 2.5$ and 
the signal over background ratio can be $S/B \sim 10$ \%, assuming 
the integrated luminosity of $1$ ab$^{-1}$.

\begin{figure}[t]
\begin{center}
\includegraphics[width=70mm]{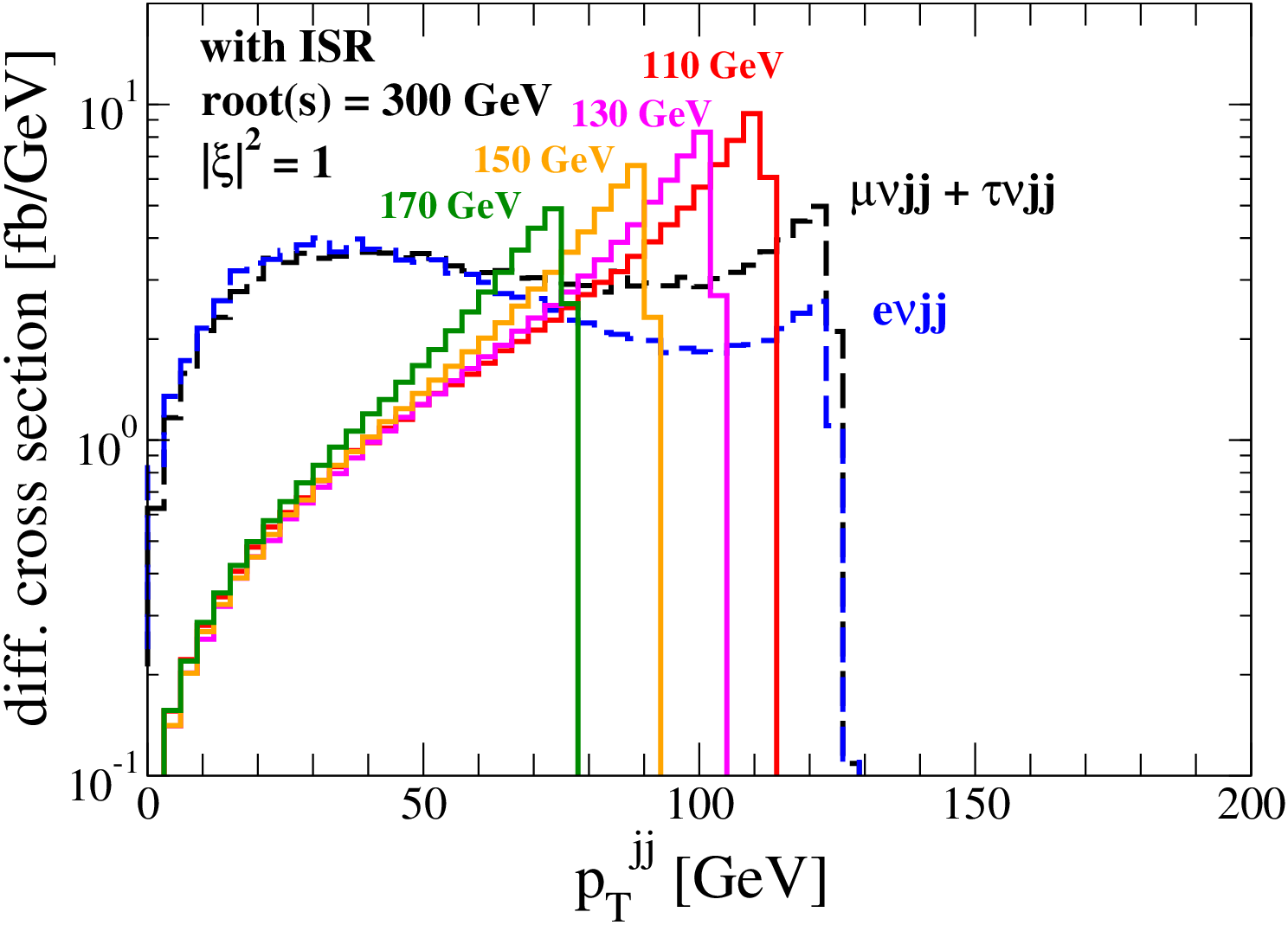}\hspace{5mm}
\includegraphics[width=70mm]{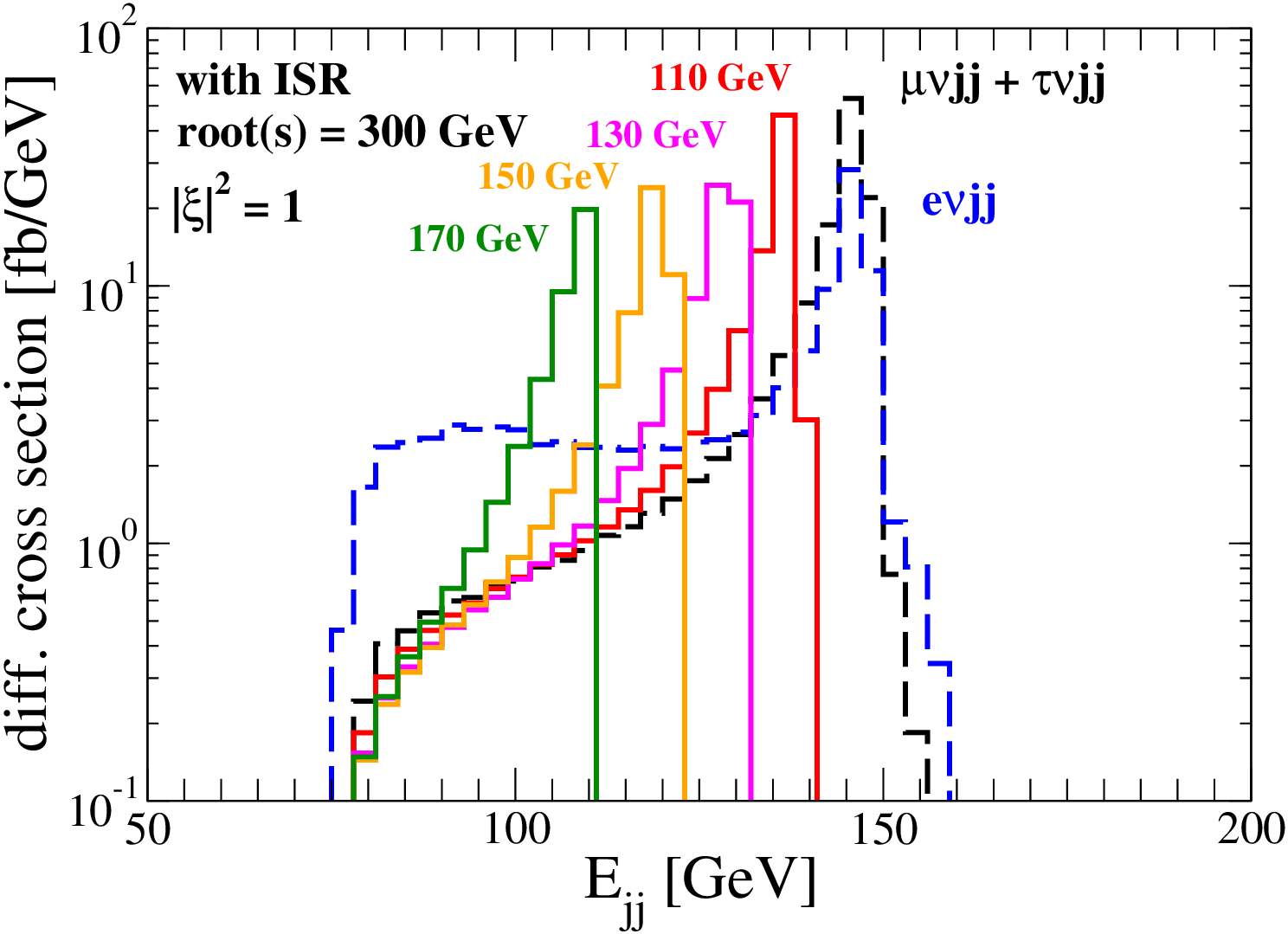}\\
\vspace{10mm}
\includegraphics[width=70mm]{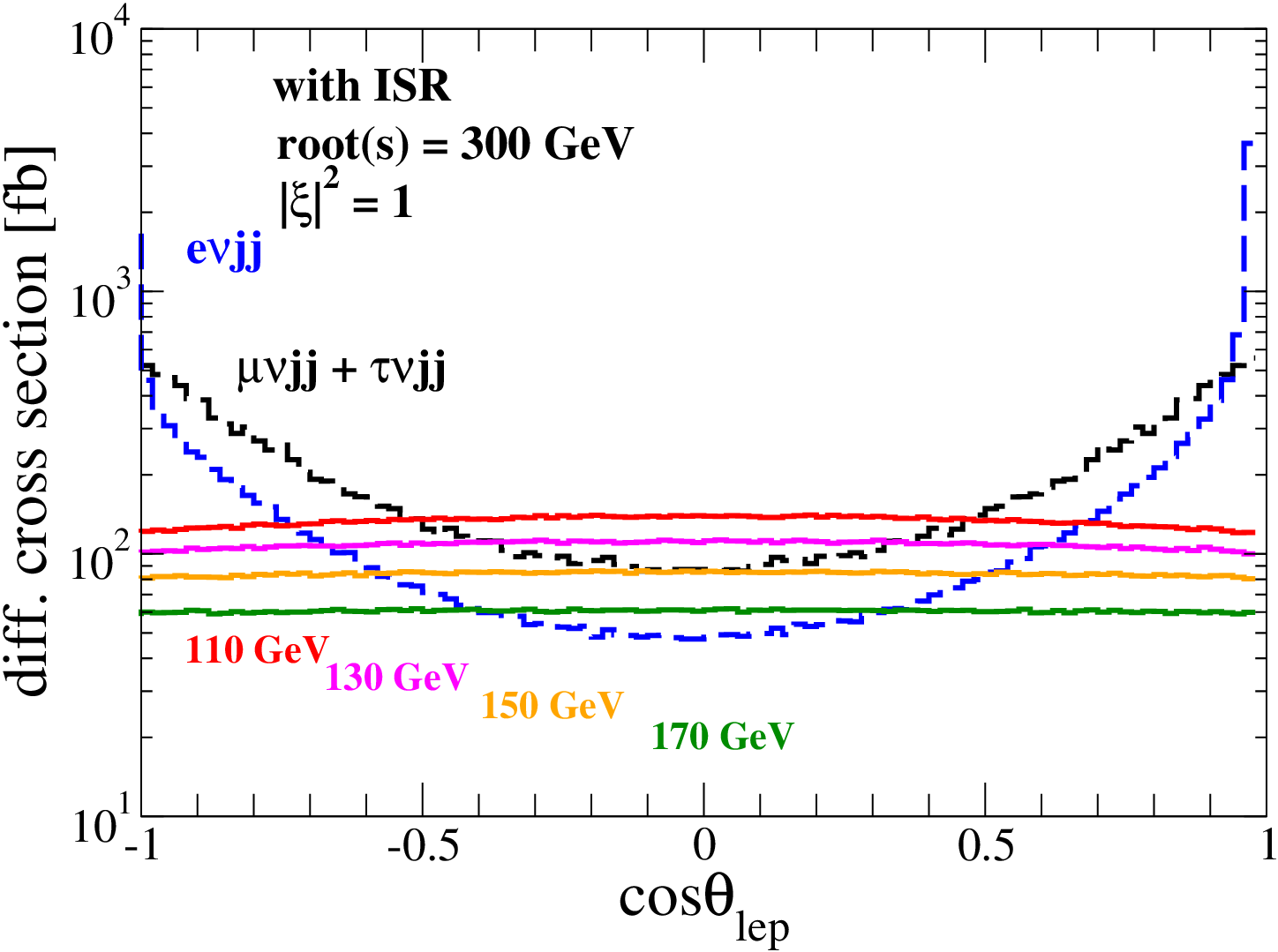}\hspace{5mm}
\includegraphics[width=70mm]{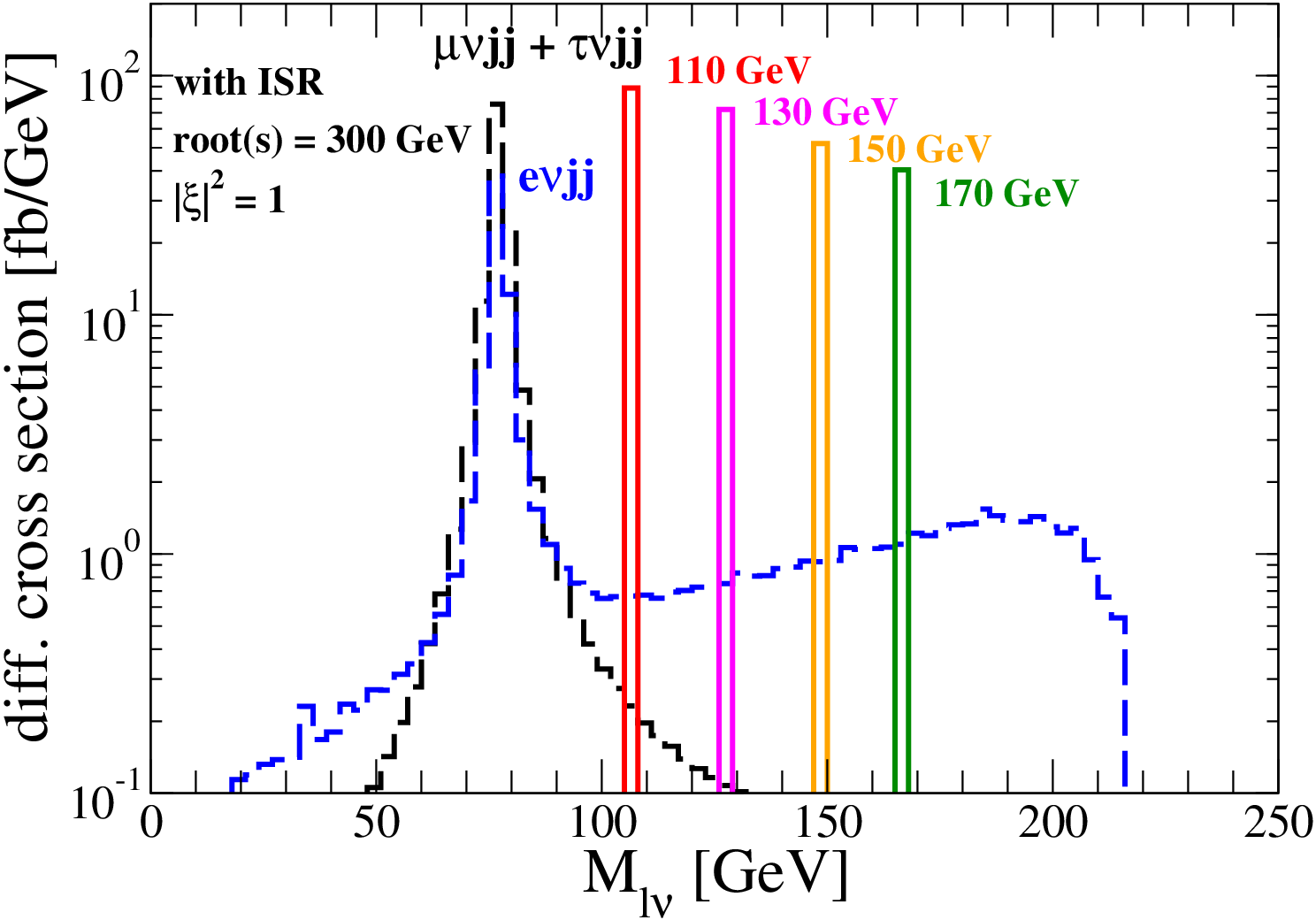}
\end{center}
\caption{Distributions of the signal for $m_{H^\pm}=110$, 
130, 150 and 170 GeV as well as the backgrounds 
after the invariant mass $M_{jj}$ cut in Eq.~(\ref{mjj}) with the ISR   
as a function of the transverse momentum 
$p_T^{jj}$ (upper left), the energy of the $jj$ system (upper right), 
the angle $\theta_{\rm lep}$ of a charged lepton with the beam 
axis (lower left), and 
the invariant mass $M_{\ell\nu}$ of the charged lepton and the missing momentum 
in the final state (lower right).  
$|\xi|^2$ is taken to be 1~\cite{Yanase}.}
\label{distr2}
\vspace{10mm}
\begin{center}
\includegraphics[width=70mm]{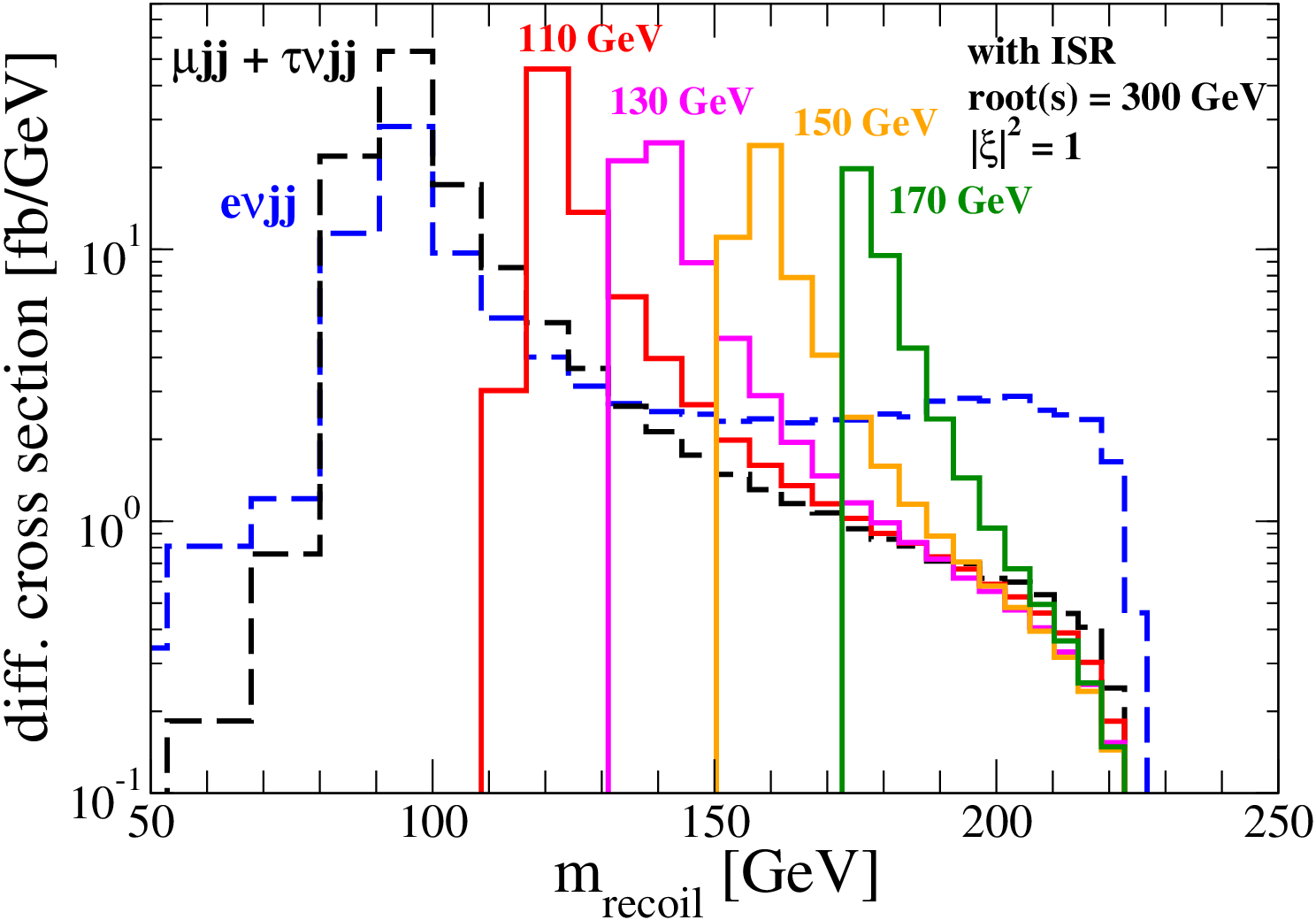}
\end{center}
\caption{Distributions of the signal for $m_{H^\pm}=110$, 130, 150 and 170 GeV
 as well as the backgrounds after the cut in Eq.~(\ref{mjj}) 
with the ISR as a function of the recoil mass $m_{\rm recoil}$~\cite{Yanase}.}
\label{distr_recoil_isr}
\end{figure}
Next let us see how this results can be changed by including the ISR. 
The beam parameters at $\sqrt{s}=500$~GeV are given 
in Ref.~\cite{ILCTDR2007}, such as the bunch $x+y$ size, the bunch length 
and the number of particles per a bunch. 
We here use the default values defined in CalcHEP~\cite{CalcHEP}; i.e., 
the bunch $x+y$ size $=560$ nm, bunch length $=400$ 
$\mu$m, and the number of particles/bunch $=2 \times 10^{10}$ at 
$\sqrt{s}=300$~GeV\footnote{
We have confirmed that the results are almost unchanged 
even when we use the values given in Ref.~\cite{ILCTDR2007}.}.

In FIG.~\ref{distr2}, the similar distributions to those in 
FIG.~\ref{distr1} but with the ISR are given 
for the signal and the backgrounds after 
the invariant mass $M_{jj}$ cut in Eq.~(\ref{mjj}).  
The biggest change can be seen in the $E_{jj}$ distribution. 
The background events originally located at the point just below 150 GeV 
in the case without the ISR,  
which corresponds to the $W$ boson mass, 
tend to move in the lower $E_{jj}$ regions, 
so that the signal over background ratio becomes worse.  
The recoil mass distribution is shown in FIG.~\ref{distr_recoil_isr}. 

Consequently, the signal significance after all the cuts is smeared 
from $2.5$ to $2.0$, while the signal over background 
ratio is changed from $8.7\times 10^{-2}$ to $7.5 \times 10^{-2}$. 
Cross sections of the signal and the backgrounds with the ISR are 
listed in TABLE~\ref{result2} 
with the values of $S/\sqrt{B}$ and $S/B$ for each 
stage of kinematic cuts. 
We stress that even taking the ISR into account, 
the $H^\pm W^\mp Z$ vertex 
with $|\xi|^2 > 10^{-3}$ can be  excluded with 95\%~CL.

\begin{table}[t]
\begin{center}
{\renewcommand\arraystretch{1.3}
\begin{tabular}{|l||c|c|c|c||c|c|}\hline
 &Basic & $M_{jj}$ \hspace{5mm}  &  $p_T^{jj}$\hspace{5mm}  & $E_{jj}$ \hspace{5mm} & $\cos\theta_{\text{lep}}$ & $M_{\ell \nu}$\hspace{5mm}  \\\hline\hline
$e_R^+e_L^-\to \ell^\pm \nu jj$ & 7.2$\times 10^{-3}$ &6.4$\times 10^{-3}$ &4.4$\times 10^{-3}$  &4.4$\times 10^{-3}$  &3.3$\times 10^{-3}$ &3.3$\times 10^{-3}$\\\hline
$e_L^+e_R^-\to \ell^\pm \nu jj$ & 1.4$\times 10^{-1}$&1.3$\times 10^{-1}$ &8.5$\times 10^{-2}$  &8.5$\times 10^{-2}$ &6.7$\times 10^{-2}$ &6.7$\times 10^{-2}$ \\\hline\hline
Total signal & 1.5$\times 10^{-1}$ &1.4$\times 10^{-1}$&8.9$\times 10^{-2}$  &8.9$\times 10^{-2}$ &7.0$\times 10^{-2}$&7.0$\times 10^{-2}$\\\hline\hline
$e_R^+e_L^-\to \mu^\pm \nu jj+\tau^\pm \nu jj$ &340 &300  &53  &2.9$\times 10^{-1}$ &2.2$\times 10^{-1}$&1.3$\times 10^{-1}$\\\hline
$e_L^+e_R^-\to \mu^\pm \nu jj+\tau^\pm \nu jj$ &80 &71 &13  &2.8$\times 10^{-1}$ &2.1$\times 10^{-1}$&1.1$\times 10^{-1}$\\\hline
$e_R^+e_L^-\to e^\pm \nu jj$ &220 &190 &31 &1.6 &6.4$\times 10^{-1}$&3.4$\times 10^{-1}$\\\hline
$e_L^+e_R^-\to e^\pm \nu jj$ &40 &36 &6.4 &1.4$\times 10^{-1}$ &1.1$\times 10^{-1}$&5.7$\times 10^{-2}$\\\hline
$e_R^+e_R^-\to e_R^- \bar{\nu} jj$ &100 &92 &11 &3.8 &2.2$\times 10^{-1}$&1.2$\times 10^{-1}$\\\hline
$e_L^+e_L^-\to e_L^+ \nu jj$ &40 &31 &4.3 &1.3 &7.2$\times 10^{-2}$&4.1$\times 10^{-2}$\\\hline\hline
Total $\ell\nu jj$ background&820 &720 &120 &7.4 &1.5 &8.0$\times 10^{-1}$\\\hline\hline
$e_R^+e_L^-\to \mu^+ \mu^- jj+\tau^+ \tau^- jj$ &1.2 &3.7$\times 10^{-2}$ &5.5$\times 10^{-3}$ &1.1$\times 10^{-4}$ &9.4$\times 10^{-5}$ &5.0$\times 10^{-5}$\\\hline
$e_L^+e_R^-\to \mu^+ \mu^- jj+\tau^+ \tau^- jj$ &19 &1.0  &1.4$\times 10^{-1}$ &3.0$\times 10^{-3}$ &2.5$\times 10^{-3}$ &1.4$\times 10^{-3}$\\\hline
$e_R^+e_L^-\to e^+ e^- jj$ &8.4 &9.0$\times 10^{-2}$ &4.6$\times 10^{-3}$ &5.8$\times 10^{-4}$ &2.6$\times 10^{-4}$ &1.3$\times 10^{-4}$ \\\hline
$e_L^+e_R^-\to e^+ e^- jj$ &220 &2.4 &1.2$\times 10^{-1}$ &1.5$\times 10^{-2}$ &6.7$\times 10^{-3}$ &3.4$\times 10^{-3}$ \\\hline
$e_R^+e_R^-\to e^+ e^- jj$ &59 &7.2$\times 10^{-1}$ &2.4$\times 10^{-2}$ &4.5$\times 10^{-3}$ &2.0$\times 10^{-3}$ &1.0$\times 10^{-3}$ \\\hline
$e_L^+e_L^-\to e^+ e^- jj$ &19 &1.0  &8.0$\times 10^{-3}$ &1.4$\times 10^{-3}$ &6.7$\times 10^{-4}$ &3.7$\times 10^{-4}$ \\\hline\hline
Total $\ell\ell jj$ background &330 &5.2 &3.0$\times 10^{-1}$ &2.5$\times 10^{-2}$  &1.2$\times 10^{-2}$ &6.4$\times 10^{-3}$\\\hline\hline
$S/\sqrt{B}$ (assuming 1 ab$^{-1}$) &1.4$\times 10^{-1}$&1.6$\times 10^{-1}$&2.6$\times 10^{-1}$&1.0&1.8&2.5\\\hline
$S/B$ (assuming 1 ab$^{-1}$) &1.3$\times 10^{-4}$&1.9$\times 10^{-4}$&7.4$\times 10^{-4}$&1.2$\times 10^{-2}$&4.6$\times 10^{-2}$&8.7$\times 10^{-2}$\\\hline
\end{tabular}}
\caption{The results without ISR. 
The cross sections of both the signal and the backgrounds are 
shown for $\sqrt{s}=300$ GeV in the unit of fb.  
For the signal, $m_{H^\pm}$ is 150 GeV and $|\xi|^2$ 
is taken to be 10$^{-3}$. 
For the $\ell\ell jj$ processes, the misidentity rate of one of 
the leptons is assumed to be 0.1. 
The signal significance $S/\sqrt{B}$ and the ratio $S/B$ are evaluated for 
the integrated luminosity to be 1 ab$^{-1}$~\cite{Yanase}.}
\label{result1}
\end{center}
\end{table}

\begin{table}[t]
\begin{center}
{\renewcommand\arraystretch{1.3}
\begin{tabular}{|l||c|c|c|c||c|c|}\hline
 &Basic & $M_{jj}$ \hspace{5mm}  &  $p_T^{jj}$\hspace{5mm}  & $E_{jj}$ \hspace{5mm} & $\cos\theta_{\text{lep}}$ & $M_{\ell \nu}$\hspace{5mm}  \\\hline\hline
$e_R^+e_L^-\to \ell^\pm \nu jj$ &6.8$\times 10^{-3}$ &6.0$\times 10^{-3}$ &3.3$\times 10^{-3}$ &3.1$\times 10^{-3}$ &2.4$\times 10^{-3}$ &2.4$\times 10^{-3}$\\\hline
$e_L^+e_R^-\to \ell^\pm \nu jj$ &1.3$\times 10^{-1}$ &1.2$\times 10^{-1}$ &6.6$\times 10^{-2}$ &6.3$\times 10^{-2}$ &5.0$\times 10^{-2}$ &4.9$\times 10^{-2}$\\\hline\hline
Total signal &1.4$\times 10^{-1}$ &1.3$\times 10^{-1}$ &6.9$\times 10^{-2}$ &6.6$\times 10^{-2}$ &5.2$\times 10^{-2}$ &5.1$\times 10^{-2}$ \\\hline\hline
$e_R^+e_L^-\to \mu^\pm \nu jj+\tau^\pm \nu jj$ &350 &310 &55 &2.9 &2.2 &1.1$\times 10^{-1}$\\\hline
$e_L^+e_R^-\to \mu^\pm \nu jj+\tau^\pm \nu jj$ &84 &76 &17 &1.8 &1.4 &9.7$\times 10^{-2}$\\\hline
$e_R^+e_L^-\to e^\pm \nu jj$ &210 &190 &32 &2.8 &1.6 &2.8$\times 10^{-1}$ \\\hline
$e_L^+e_R^-\to e^\pm \nu jj$ &42 &38 &8.5 &9.0$\times 10^{-1}$ &7.0$\times 10^{-1}$ &4.9$\times 10^{-2}$\\\hline
$e_R^+e_R^-\to e_R^- \bar{\nu} jj$ &92 &81 &10 &3.2 &2.2$\times 10^{-1}$ &1.0$\times 10^{-1}$\\\hline
$e_L^+e_L^-\to e_L^+ \nu jj$ &32 &29 &3.7 &1.1 &7.8$\times 10^{-2}$ &3.4$\times 10^{-2}$\\\hline\hline
Total $\ell\nu jj$ background &810 &720 &130 &13 &6.2 &6.7$\times 10^{-1}$ \\\hline\hline
$e_R^+e_L^-\to \mu^+ \mu^- jj+\tau^+ \tau^- jj$ &1.2 &4.2$\times 10^{-2}$ &5.9$\times 10^{-3}$ &3.7$\times 10^{-4}$ &3.1$\times 10^{-4}$&4.6$\times 10^{-5}$ \\\hline
$e_L^+e_R^-\to \mu^+ \mu^- jj+\tau^+ \tau^- jj$ &22 &1.2 &1.5$\times 10^{-1}$ &9.9$\times 10^{-3}$ &8.3$\times 10^{-3}$ &1.2$\times 10^{-3}$ \\\hline
$e_R^+e_L^-\to e^+ e^- jj$ &9.6 &9.2$\times 10^{-2}$ &4.1$\times 10^{-3}$ &6.3$\times 10^{-4}$ &3.2$\times 10^{-4}$ &1.0$\times 10^{-4}$\\\hline
$e_L^+e_R^-\to e^+ e^- jj$ &230 &2.4 &1.0$\times 10^{-1}$ &1.7$\times 10^{-2}$ &9.2$\times 10^{-3}$ &2.9$\times 10^{-3}$ \\\hline
$e_R^+e_R^-\to e^+ e^- jj$ &70 &6.4$\times 10^{-1}$ &2.3$\times 10^{-2}$ &4.2$\times 10^{-3}$ &2.1$\times 10^{-3}$ &9.1$\times 10^{-4}$\\\hline
$e_L^+e_L^-\to e^+ e^- jj$ &24 &2.2$\times 10^{-1}$ &7.4$\times 10^{-3}$ &1.4$\times 10^{-3}$ &6.3$\times 10^{-4}$ &3.1$\times 10^{-4}$\\\hline\hline
Total $\ell\ell jj$ background &360 &4.6 &2.9$\times 10^{-1}$ &3.4$\times 10^{-2}$ &2.1$\times 10^{-2}$ &5.5$\times 10^{-3}$\\\hline\hline
$S/\sqrt{B}$ (assuming 1 ab$^{-1}$)&1.3$\times 10^{-1}$&1.5$\times 10^{-1}$&1.9$\times 10^{-1}$&5.8$\times 10^{-1}$&6.6$\times 10^{-1}$&2.0\\\hline
$S/B$ (assuming 1 ab$^{-1}$)&1.2$\times 10^{-4}$&1.8$\times 10^{-4}$&5.3$\times 10^{-4}$&5.1$\times 10^{-3}$&8.4$\times 10^{-3}$&7.5$\times 10^{-2}$\\\hline
\end{tabular}}
\caption{The results with the ISR. 
The cross sections of both the signal and the backgrounds are shown for $\sqrt{s}=300$ GeV in the unit of fb.  
For the signal, $m_{H^\pm}$ is 150 GeV and  
$|\xi|^2$ is taken to be 10$^{-3}$. 
For the $\ell\ell jj$ processes, the misidentity rate of one 
of the leptons is assumed to be 0.1. 
The signal significance $S/\sqrt{B}$ and the ratio $S/B$ are evaluated for 
the integrated luminosity to be 1 ab$^{-1}$~\cite{Yanase}.}
\label{result2}
\end{center}
\end{table}

\clearpage

\chapter{Decoupling property of SUSY Higgs sectors}
In this chapter, we focus on SUSY Higgs sectors. 
Supersymmetry is expected to be a good candidate of 
new physics.  It can solve the hierarchy problem by the consequence of
the nonrenormalization theorem~\cite{nonreno-theorem}.
The stabilized Higgs boson mass makes it possible to directly connect
the electroweak scale with very high scales such as the Planck scale or 
that of grand unification.
SUSY extensions of the SM
with the R parity also provide dark matter candidates~\cite{SUSY-DM}. 
In addition, various mechanisms of generating tiny neutrino
masses~\cite{typeI_seesaw,typeII_seesaw,typeIII_seesaw,zee,zee-2loop,babu-2loop,Ma:2006km,Krauss:2002px,aks_prl} 
as well as those of baryogenesis~\cite{B-AD,B-Lepto,ewbg-thdm}
may also be compatible to SUSY models. 

The MSSM is a SUSY extension of the SM with the minimal number of particle content. 
In the MSSM, two Higgs doublets are introduced because of the anomaly cancellation. 
The most striking phenomenological prediction of the model 
is that on the mass ($m_h$) of the lightest CP-even Higgs boson $h$.
It can be calculated to be less than the mass of the Z boson 
at the tree level. Such an upper bound on $m_h$   
comes from the fact that the interaction terms in the 
Higgs potential are given only by D-term contributions
which are determined by the gauge coupling constants. 
At the one-loop level the trilinear top-Yukawa term in the
superpotential gives a significant F-term contribution to
$m_h$~\cite{mh-MSSM,mh-MSSM1,dabelstein}, by which $m_h$ can be above
the lower bound from the direct search results at the CERN LEP
experiment~\cite{LEP}.
The calculation has been improved with higher order
corrections~\cite{mh-MSSM2-rge,mh-MSSM2,mh-MSSM3}.
Apart from $m_h$, the masses of $H$, $H^\pm$ and
the mixing angle $\alpha$ are a function of only two
input parameters at the tree level; i.e., $m_A$ and $\tan\beta$,
where $m_A$ is the mass of CP-odd Higgs boson $A$, $H$ is the heavier
CP-even Higgs boson, $H^\pm$ are the charged Higgs bosons,
$\tan\beta$ is the ratio of VEVs of
the two Higgs bosons and $\alpha$ is the angle which diagonalizes the CP-even scalar states.
In particular, there is a simple tree-level relationship among the masses of 
$H^\pm$, $A$ and the W boson $W^\pm$ as $m_{H^\pm}^2=m_A^2+m_W^2$,
where $m_{H^\pm}$ and $m_W$ are respectively the masses of $H^\pm$ and $W^\pm$.
These characteristic predictions can be used
to confirm the MSSM.

However, these characteristic features which are seen in the MSSM Higgs sector 
can be changed when we consider extended SUSY standard models which are motivated to 
solve various physics problem. 
For example, 
the model with a neutral gauge singlet field to 
the MSSM, which is known as the next-to-MSSM (NMSSM)~\cite{NMSSM,mh-NMSSM,mh-NMSSM2}, 
solves the $\mu$ problem \cite{mu_probrem}.
Models with additional charged singlet 
fields can be used for radiative neutrino mass generation \cite{zee-2loop,babu-2loop}. 
Those with additional doublet fields may be required for dark doublet 
models \cite{dark_higgs}, and the model with triplets may be motivated for 
the SUSY left-right model \cite{LRmodels} or those with so-called the type-II 
seesaw mechanism \cite{typeII_seesaw,Rossi}. 

These extended SUSY Higgs sectors with additional chiral superfields 
can be classified to the two groups: i.e., 
1.) models with additional F-term contributions to the interaction terms 
in the Higgs potential such as the NMSSM, 
a model with triplet superfields added to the MSSM and so on, 
2.) those without such F-term contributions, e.g., the model with four Higgs doublet superfields (4HDM). 
In the models classified to 1.),  
additional F-term contributions to the mass of the lightest 
CP-even Higgs boson $m_h$ and the triple $h$ coupling $hhh$ can be significant even when $h$ looks the SM Higgs boson.
In the models classified to 2.),  even without F-term contributions in the Higgs potential, 
large deviations can be seen in the MSSM observables 
due to the mixing among the MSSM-like Higgs bosons and the extra Higgs bosons at the tree level.

In this chapter, we first discuss the MSSM Higgs sector. 
Next, we discuss models with additional $F$-term contributions to interaction terms in the Higgs potential. 
Finally, we discuss 
the 4HDM as a simplest example for a model without such F-term contributions. 
\section{The Higgs sector of the Minimal Supersymmetric Standard Model}
\begin{table}[t]
\begin{center}
{\renewcommand\arraystretch{1.2}
\begin{tabular}{|c||cc|ccc|}\hline
&Spin 0 & Spin 1/2&$SU(3)_c$&$SU(2)_L$&$U(1)_Y$ \\\hline\hline
$\hat{Q}_i$&$\tilde{Q}_{Li}$&$Q_{Li}$&\textbf{3}&\textbf{2}&$+\frac{1}{6}$\\\hline
$\hat{U}_i^c$&$\tilde{u}_{Ri}^*$&$u_{Ri}^c$&$\bar{\textbf{3}}$&\textbf{1}&$-\frac{2}{3}$ \\\hline
$\hat{D}_i^c$&$\tilde{d}_{Ri}^*$&$d_{Ri}^c$&$\bar{\textbf{3}}$&\textbf{1}&$+\frac{1}{3}$ \\\hline
$\hat{L}_i$&$\tilde{L}_{Li}$&$L_{Li}$&\textbf{1}&\textbf{2}&$-\frac{1}{2}$ \\\hline
$\hat{E}_i^c$&$\tilde{e}_{Ri}^*$&$e_{Ri}^c$&\textbf{1}&\textbf{1}&+1 \\\hline
$\hat{H}_d$&$H_d$&$\tilde{H}_d$&\textbf{1}&\textbf{2}&$-\frac{1}{2}$ \\\hline
$\hat{H}_u$&$H_u$&$\tilde{H}_u$&\textbf{1}&\textbf{2}&$+\frac{1}{2}$ \\\hline
\end{tabular}}
\caption{Charge assignment for the chiral superfield in the MSSM under the $SU(3)_c\times SU(2)_L\times U(1)$ gauge symmetry.}
\label{chiralsuper-mssm}
\end{center}
\end{table}
\begin{table}[t]
\begin{center}
{\renewcommand\arraystretch{1.2}
\begin{tabular}{|c||cc|ccc|}\hline
&Spin 1/2 & Spin 1&$SU(3)_c$&$SU(2)_L$&$U(1)_Y$ \\\hline\hline
$\hat{G}$&$\tilde{G}$&$G_\mu$&\textbf{8}&\textbf{1}&0\\\hline
$\hat{W}$&$\tilde{W}$&$W_\mu$&\textbf{1}&\textbf{3}&0 \\\hline
$\hat{B}$&$\tilde{B}$&$B_\mu$&\textbf{1}&\textbf{1}&0 \\\hline
\end{tabular}}
\caption{Charge assignment for the vector superfield in the MSSM under the $SU(3)_c\times SU(2)_L\times U(1)$ gauge symmetry.}
\label{vectorsuper-mssm}
\end{center}
\end{table}
The charge assignment for the chiral superfields which are denoted as the simbol with the hat in the MSSM are listed in Table~\ref{chiralsuper-mssm}.
In the MSSM, two Higgs doublets are introduced because of the anomary cancellation. In addition, 
there is another reason to introduce the two Higgs doublets. 
We cannot use the 
hermitian conjugate of chiral superfields in the superpotential, so that 
the Higgs doublet with $Y=1/2$ and that with $Y=-1/2$ are necessary to give masses for the up-type quarks (and also the charged leptons) 
and the up-type quarks, respectively.  

The superpotential of the MSSM is given as 
\begin{align}
W_{\text{MSSM}}&=-(Y_u)^{ij}\hat{U}_i^c(\hat{H}_u\cdot \hat{Q}_j)+(Y_d)^{ij}\hat{D}_i^c(\hat{H}_d\cdot \hat{Q}_j)
+(Y_e)^{ij}\hat{E}_i^c(\hat{H}_d \cdot\hat{L}_j)+\mu (\hat{H}_u\cdot \hat{H}_d). 
\end{align}
The soft-SUSY breaking terms are 
\begin{align}
\mathcal{L}_{\text{MSSM}}^{\text{soft}}
&=-\frac{1}{2}(M_1\tilde{B}\tilde{B}+M_2\tilde{W}\tilde{W}+M_3\tilde{G}\tilde{G})\notag\\
&-\Big[(M_{\tilde{Q}}^2)_{ij}\tilde{Q}_{Li}^\dagger \tilde{Q}_{Lj}+(M_{\tilde{u}}^2)_{ij}\tilde{u}_{Ri}^*\tilde{u}_{Rj}
+(M_{\tilde{d}}^2)_{ij}\tilde{d}_{Ri}^*\tilde{d}_{Rj}+(M_{\tilde{L}}^2)_{ij}\tilde{L}_{Li}^\dagger\tilde{L}_{Lj}
+(M_{\tilde{e}}^2)_{ij}\tilde{e}_{Ri}^*\tilde{e}_{Rj}\notag\\
&\hspace{5mm}+(M_-^2)H_d^\dagger H_d+(M_+^2)H_u^\dagger H_u\Big]-\Big(B\mu H_d\cdot H_u+\text{h.c.}\Big)\notag\\
&-\Big[(Y_u)_{ij}(A_u)_{ij}\tilde{u}_{Ri}^*H_u\cdot \tilde{Q}_{Lj}+(Y_d)_{ij}(A_d)_{ij}\tilde{d}_{Ri}^*H_d\cdot\tilde{Q}_{Lj}
+(Y_e)_{ij}(A_e)_{ij}\tilde{e}_{Ri}^*H_d\cdot\tilde{L}_{Lj}+\text{h.c.}\Big], \label{Lsoft_mssm}
\end{align}
In SUSY models, the Higgs potential can be calculated as  
\begin{align}
V_H=|D|^2+|F|^2-\mathcal{L}_{\text{soft}}, \label{eq:lag}
\end{align}
where $\mathcal{L}_{\text{soft}}^{}$ is the soft SUSY breaking Lagrangian, 
$|D|^2$ is the D-term, and $|F|^2$ is the F-term. 
The D-term and the F-term are expressed by the given superpotential $W$ as
\begin{align}
|D|^2&=\frac{1}{2}(g_a)^2
(\varphi^*_{i}T^a_{ij}\varphi_{j})^2,\quad|F|^2=\left|\frac{\partial W}{\partial \varphi_i}\right|^2, \label{Fterm}
\end{align}
where $\varphi_j$ represent 
scalar component fields of chiral superfields in the model.
In the MSSM, the Higgs potential can be written as 
\begin{align}
V_{\text{MSSM}}=&
m_1^2 H_d^\dagger H_d+m_2^2 H_u^\dagger H_u+(B\mu H_d\cdot H_u+\text{h.c.})\notag\\
&+\frac{g^2}{8}(H_d^\dagger \tau^a H_d+H_u^\dagger \tau^a H_u)^2+\frac{g^{\prime 2}}{8}(H_d^\dagger H_d-H_u^\dagger H_u)^2\notag\\
&\hspace{-5mm}=m_1^2 H_d^\dagger H_d+m_2^2 H_u^\dagger H_u+(B\mu H_d\cdot H_u+\text{h.c.})\notag\\
&+\frac{g^2+g^{\prime 2}}{8}(H_d^\dagger H_d-H_u^\dagger H_u)^2-\frac{g^2}{2}(H_d^\dagger H_u)(H_u^\dagger H_d), 
\end{align}
where $m_1^2$ and $m_2^2$ are $|\mu|^2+(M_-^2)$ and $|\mu|^2+(M_+^2)$, respectively. 
The Higgs doublets $H_d$ and $H_u$ can be parameterized as
\begin{align}
H_d=\left[
\begin{array}{c}
\frac{1}{\sqrt{2}}(h_d+v_d-iz_d)\\
-w_d^-
\end{array}\right],\quad 
H_u=\left[
\begin{array}{c}
w_u^+\\
\frac{1}{\sqrt{2}}(h_u+v_u+iz_u)
\end{array}\right],
\end{align}
where $v_d= v\cos\beta$ and $v_u=v\sin\beta$ are the VEV of the Higgs doublet with $v_d^2+v_u^2=v^2\simeq $ (246 GeV)$^2$. 
By the vacuum condition $m_1$ and $m_2$ can be eliminated as: 
\begin{subequations}
\begin{align}
\left. \frac{\partial V_{\text{MSSM}}}{\partial h_d}\right|_0 = 
m_1^2\cos\beta +\frac{1}{2}m_Z^2\cos\beta\cos2\beta+B\mu\sin\beta=0,\\
\left. \frac{\partial V_{\text{MSSM}}}{\partial h_u}\right|_0 =
m_2^2\sin\beta -\frac{1}{2}m_Z^2\sin\beta\cos2\beta +B\mu\cos\beta=0. 
\end{align}
\label{mssm-vc}
\end{subequations}
The two-point terms in the Higgs potential can be calculated as 
\begin{align}
V_{\text{MSSM}}^{\text{mass}}&=(w_d^+,w_u^+)\left(
\begin{array}{cc}
m_W^2\sin^2\beta-B\mu\tan\beta & B\mu-\frac{m_W^2}{2}\sin2\beta\\
B\mu-\frac{m_W^2}{2}\sin2\beta & m_W^2\cos^2\beta-B\mu\cot\beta 
\end{array}\right)
\left(
\begin{array}{c}
w_d^-\\
w_u^-
\end{array}\right)\notag\\
&+\frac{1}{2}(z_1,z_2)\left(
\begin{array}{cc}
-B\mu\tan\beta & B\mu\\
B\mu & -B\mu\cot\beta 
\end{array}\right)
\left(
\begin{array}{c}
z_d\\
z_u
\end{array}\right)\notag\\
&+\frac{1}{2}(h_d,h_u)\left(
\begin{array}{cc}
m_Z^2\cos^2\beta-B\mu\tan\beta & B\mu-\frac{m_Z^2}{2}\sin2\beta\\
B\mu-\frac{m_Z^2}{2}\sin2\beta & m_Z^2\sin^2\beta-B\mu\cot\beta 
\end{array}\right)
\left(
\begin{array}{c}
h_d\\
h_u
\end{array}\right).
\end{align}
The mass eigenstates are obtained by the mixing angles $\beta$ and $\alpha$:
\begin{align}
&\left(
\begin{array}{c}
z_d\\
z_u
\end{array}\right)=
\left(
\begin{array}{cc}
\cos\beta & -\sin\beta\\
\sin\beta & \cos\beta
\end{array}\right)\left(
\begin{array}{c}
z\\
A
\end{array}\right),\quad
\left(
\begin{array}{c}
w_d^\pm\\
w_u^\pm
\end{array}\right)=
\left(
\begin{array}{cc}
\cos\beta & -\sin\beta\\
\sin\beta & \cos\beta
\end{array}\right)\left(
\begin{array}{c}
w^\pm\\
H^\pm
\end{array}\right),\notag\\
&\left(
\begin{array}{c}
h_d\\
h_u
\end{array}\right)=
\left(
\begin{array}{cc}
\cos\alpha & -\sin\alpha\\
\sin\alpha & \cos\alpha
\end{array}\right)\left(
\begin{array}{c}
H\\
h
\end{array}\right),\label{rotba}
\end{align}
where $w^\pm$ and $z$ are the NG bosons which are absorbed by the longitudinal components of $W^\pm$ and $Z$. 
All the other mass eigenstates are the physical scalar bosons, 
those are the pair of the singly-charged scalar bosons $H^\pm$, neutral CP-even scalar bosons $H$ and $h$  
and a neutral CP-odd scalar boson $A$. 
Originally, the number of the parameter is five in the Higgs potential ($v_u$, $v_d$, $m_1$, $m_2$ and $B\mu$) at the tree level.  
Three of the five parameters are determined by $v$ and the vacuum conditions Eq.~(\ref{mssm-vc}). 
Thus, remaining parameters are $\tan\beta$ and $B\mu$. 
The $B\mu$ parameter can  be rewritten as the mass of $A$: 
\begin{align}
m_A^2&=-\frac{B\mu}{\sin\beta\cos\beta}. 
\end{align}
The masses of the other physical scalar bosons are expressed in terms of $\beta$ and $m_A$: 
\begin{align}
m_{H^+}^2&=m_W^2+m_A^2,\\
m_{h,H}^2&=\frac{1}{2}\left[m_A^2+m_Z^2\mp\sqrt{(m_A^2+m_Z^2)^2-4m_Z^2m_A^2\cos^22\beta}\right], \label{mh-mssm}
\end{align}
and the mixing angle $\alpha$ is 
\begin{align}
\tan2\alpha &= \tan2\beta\frac{m_A^2+m_Z^2}{m_A^2-m_Z^2}.
\end{align}
From Eq.~(\ref{mh-mssm}), the mass of the lightest Higgs boson $h$ can be expressed in the large $m_A$ limit:
\begin{align}
m_h \xrightarrow[m_A \gg m_Z]{} m_Z\cos2\beta \leq m_Z. 
\end{align}
Thus, at the tree level, $m_h$ cannot exceed the LEP bound. 
This upper limit for $m_h$ can be changed by considering the one-loop correction. 
In this section, we take into account the effects of the one-loop level correction by the effective 
potential method.
The effective potential is given as 
\begin{align}
V^{\text{eff}} &= -\frac{\mu_0^2}{2}\varphi^2+\frac{\lambda_0}{4}\varphi^4
+\sum_{f}\frac{(-1)^{s_f}N_c^fN_s^f}{64\pi^2}m_f^4(\varphi)\left[\ln\frac{m_f^2(\varphi)}{Q^2}-\frac{3}{2} \right], \label{ep}
\end{align}  
where $\mu_0^2$ 
and $\lambda_0$ are the bare squared mass and the coupling constant, 
$\varphi=v+\langle h \rangle$, $N_c^f$ is the color number, $s_f$
($N_s^f$ ) is the spin (degree of freedom) of the field $f$ in the
loop, $m_f(\varphi)$ is the field dependent mass of $f$, and $Q$ is an 
arbitrary scale. 
When the extra Higgs 
scalars are heavy enough, only the lightest Higgs boson $h$ stays at
the EW scale, and behaves as the SM-like one. The effective potential
in Eq.~(\ref{ep}) can then be applied with a good approximation. 
The vacuum, the mass $m_h$ and the $hhh$ coupling constant $\lambda_{hhh}$ are
determined at the one-loop order by the conditions; 
\begin{align}
\left.\frac{\partial V^{\text{eff}}}{\partial \varphi}\right|_{\varphi = v}=0,\quad 
\left.\frac{\partial^2 V^{\text{eff}}}{\partial \varphi^2}\right|_{\varphi = v}=m_h^2,\quad 
\left.\frac{\partial^3 V^{\text{eff}}}{\partial\varphi^3}\right|_{\varphi = v}=\lambda_{hhh}. \label{hhh_ren}
\end{align}
Here we consider the top quark and its scalar partner (stop) effects at the one-loop level in the MSSM. 
The effective potential is
\begin{align}
V_{\text{MSSM}}^{\text{eff(1)}}=
\frac{3}{32\pi^2}\left[m_{\tilde{t}_1}^4\left(\log\frac{m_{\tilde{t}_1}^2}{Q^2}-\frac{3}{2}\right)+m_{\tilde{t}_2}^4\left(\log\frac{m_{\tilde{t}_2}^2}{Q^2}-\frac{3}{2}\right)-2m_t^4\left(\log\frac{m_t^2}{Q^2}-\frac{3}{2}\right)\right], 
\end{align}
where $m_t=\frac{y_t}{\sqrt{2}}\varphi\sin\beta$ and $m_{\tilde{t}_{1,2}}$ are the masses of the top quark and the 
stops $\tilde{t}_1$ and $\tilde{t}_2$, respectively. 
The stop masses are obtained by diagonalizing the mass matrix: 
\begin{align}
\mathcal{M}^2_{\text{stop}}=
\left[
\begin{array}{cc}
m_t^2+M_{\tilde{t}_L}^2+m_Z^2\left(\frac{1}{2}-\frac{1}{6}\sin^2\theta_W\right)\cos2\beta &m_tX_t\\
m_tX_t &  m_t^2+M_{\tilde{t}_R}^2+\frac{2}{3}m_Z^2\cos2\beta
\end{array}
\right], 
\end{align}
where $(M^2_{\tilde{t}_L})=(M_{\tilde{Q}}^2)_{33}$, $M_{\tilde{t}_L}^2=(M_{\tilde{u}}^2)_{33}$, and 
$X_t=A_t-\mu\cot\beta $ with $A_t=(A_u)_{33}/y_t$ and  
\begin{align}
m_{\tilde{t}_{1,2}}^2=\frac{1}{2}\left[\text{tr}\mathcal{M}^2_{\text{stop}}\mp \sqrt{(\text{tr}\mathcal{M}^2_{\text{stop}})^2
-4\text{det}\mathcal{M}^2_{\text{stop}}}\right]. 
\end{align}
The mass of the lightest Higgs boson $h$ can be calculated at the one-loop level:
\begin{align}
m_h^2(\text{MSSM}) \simeq m_Z^2\cos^22\beta +\frac{3m_t^4}{4\pi^2v^2}\ln\frac{m_{\tilde{t}_1}^2m_{\tilde{t}_2}^2}{m_t^4}+
\frac{3m_t^2X_t^2\sin^2\beta}{4\pi^2(m_{\tilde{t}_2}^2-m_{\tilde{t}_1}^2)}\ln\frac{m_{\tilde{t}_2}^2}{m_{\tilde{t}_1}^2}. 
\label{mh-MSSM-loop}
\end{align}
The one-loop renormalized triple $h$ coupling $\lambda_{hhh}$ is also calculated according to Eq.~(\ref{hhh_ren}) as
\begin{align}
\lambda_{hhh}^{\text{MSSM}}\simeq \left[\frac{3m_h^2(\text{MSSM})}{v}\right]\left[
1-\frac{m_t^4}{\pi^2v^2m_h^2(\text{MSSM})}
+\frac{m_t^6(m_{\tilde{t}_1}^2+m_{\tilde{t}_2}^2)}{2\pi^2v^2m_{\tilde{t}_1}^2m_{\tilde{t}_2}^2m_h^2(\text{MSSM})}\right].
\end{align}
For later convenience, we here define $\bar{\lambda}_{hhh}^{\text{Model}}$ as 
\begin{align}
\bar{\lambda}_{hhh}^{\text{Model}}\equiv\left[\frac{3m_h^2(\text{Model})}{v}\right]\left[
1-\frac{m_t^4}{\pi^2v^2m_h^2(\text{Model})}
+\frac{m_t^6(m_{\tilde{t}_1}^2+m_{\tilde{t}_2}^2)}{2\pi^2v^2m_{\tilde{t}_1}^2m_{\tilde{t}_2}^2m_h^2(\text{Model})}\right]. 
\end{align}
\section{Nondecoupling effects in supersymmetric Higgs sectors}
In general, new physics can be tested not only by direct searches
but also by indirect searches. The indirect searches are performed
by precise experiments to find effects of a heavy new physics particle
on the observables which are well predicted in the low energy
theory such as the SM. Such new particle effects on the low
energy observables usually decouple in the large mass limit after
the renormalization calculation is completed.
This is known as the decoupling theorem proposed by Appelquist
and Carazzone~\cite{decoupling_theorem}. 
It is also known that the decoupling theorem does not hold when the new
particles receive their masses from the VEV of the Higgs boson.
In fact, there is a class of the new physics models where nondecoupling effects of
heavy particles can appear on the low energy observables.
For example, chiral fermions such as
quarks and charged leptons cannot have the mass term because of the
chiral symmetry, so that their masses are generated after the chiral
symmetry is spontaneously broken by the VEV.
Therefore, the effect of a heavy chiral fermion does not decouple, and
it appears as powerlike or logarithmic contributions of the mass in the
predictions for the low energy observables. 
Another example is the additional scalar fields in extended Higgs
sectors. 
To see this type of nondecoupling effects, 
we here discuss the quantum effect on the $hhh$ coupling in the non-SUSY 
THDM, where $h$ is regarded as the SM-like Higgs boson. 
The $hhh$ coupling can receive large nondecoupling effects 
from the loop contribution of extra Higgs bosons, when their 
masses are generated mainly by EWSB~\cite{KOSY,nondec}.
When $h$ is the SM like Higgs boson, physical masses
of the extra scalar bosons
are expressed by 
\begin{equation}
m_{\Phi_i}^2=M^2+\frac{\lambda_i^{} v^2}{2}\;,
\label{m2M2lamv2}
\end{equation}
where 
$\Phi_i^{}$ represents $H$, $H^{\pm}$ or $A$, and 
$M$ is the invariant mass scale which is defined in Eq.~(\ref{bigm}), 
and $\lambda_i^{}$ is a coupling for $\Phi_i^{\dagger} \Phi_i^{} h h$.
The one-loop contribution to the renormalized $hhh$ coupling is calculated 
as~\cite{KOSY,nondec}
\begin{equation}
\frac{\lambda_{hhh}^{\text{THDM}}}{\lambda_{hhh}^{\text{SM}}}
\simeq 1+
\frac{1}{12\pi^2m_h^2v^2}
\left\{
m_{H^0}^4
\left(1-\frac{M^2}{m_{H^0}^2}\right)^3
+
m_{A^0}^4
\left(1-\frac{M^2}{m_{A^0}^2}\right)^3
+
2m_{H^{\pm}}^4
\left(1-\frac{M^2}{m_{H^{\pm}}^2}\right)^3
\right\}
\;.
\end{equation}
One finds that for $M^2\gg \lambda_i^{}v^2$ it becomes
\begin{equation}
\frac{\lambda_{hhh}^{\text{THDM}}}{\lambda_{hhh}^{\text{SM}}}
\simeq 
1+
\frac{v^2}{96\pi^2m_h^2}
\left(
\lambda_{H^0}^{3}
+\lambda_{A^0}^{3}
+2\lambda_{H^{\pm}}^{3}
\right)\left(\frac{v^2}{M^2}\right)
\;,
\end{equation}
which vanishes in the large $M$ limit according to 
the decoupling theorem\cite{decoupling_theorem}.
On the contrary, when the physical scalar masses 
are mainly determined by the $\lambda_i^{} v^2$ term,
the loop contribution to the $hhh$ coupling does not decouple, and 
the quartic powerlike contributions of $m_{\Phi_i^{}}^{}$ remain;
\begin{equation}
\frac{\lambda_{hhh}^{\text{THDM}}}{\lambda_{hhh}^{\text{SM}}}
\simeq 1+
\frac{1}{12\pi^2m_h^2v^2}
\left(
m_{H^0}^4
+
m_{A^0}^4
+
2m_{H^{\pm}}^4
\right)
\;.
\end{equation}
Consequently, a significant quantum effect can be realized for the $hhh$ 
coupling when $m_{\Phi_i}^2 > m_h^2$.
The size of the correction from the SM value can be of 100\% for $m_h^{}=120$ GeV,
$M\simeq 0$, and $m_{H^0}^{}\simeq m_{A^0}\simeq m_{H^{\pm}}\simeq 400$ GeV under the constraint from 
perturbative unitarity\cite{PU_thdm,PU_thdm2}.
Such a large non-decoupling effect on the $hhh$ coupling 
is known to be related to the strongly first 
order electroweak phase transition~\cite{ewbg-thdm2}  
which is required for the electroweak baryogenesis\cite{ewbg-thdm}.

In addition to $hhh$ coupling, indirect effects of nondecoupling particles 
such as additional scalar bosons and chiral fermions 
appear in the low energy observables at the tree level, or at loop levels such as the electroweak $S$,
$T$ and $U$ parameters~\cite{Peskin-Takeuchi} and $\gamma\gamma h$ vertex~\cite{hgg}. 

Let us consider the effect of the
heavy particles in  
SUSY standard models. 
In general, a  SUSY Higgs potential is
composed of the D-term, the F-term
and the soft-breaking term given in Eqs.~(\ref{eq:lag}) and (\ref{Fterm}). 
Quartic coupling constants in the
potential can come from both the
D-term and the F-term.  
As we have discussed in the previous section, 
in the MSSM, because of the multi-doublet structure
only D-terms contribute to them, 
which are given by gauge coupling constants.  
Consequently, the mass of the lightest 
CP-even Higgs boson is determined by the gauge
coupling constants and the VEVs at the tree level, which is less than $m_Z$.
A substantial F-term contribution through the top Yukawa interaction enters into the Higgs potential
at the one-loop level via the superpotential. 
This contribution is proportional to $m_t^4$ as expressed in Eq.~(\ref{mh-MSSM-loop}). 
This one-loop correction shows a nondecoupling property in the large
mass limit of stops. Consequently $m_h$ can be
above the LEP bound at least when one of the stops is heavy
enough.
There are also contributions from the $\mu$ parameter
and the soft-breaking $A_t$ ($A_b$) parameter when there is the left-right
mixing in the stop (sbottom) sector (see the last term in Eq.~(\ref{mh-MSSM-loop})). 
Their one-loop effects can also be
nondecoupling and then can be significant to some extent 
when they are taken to be as large as the scale of the
soft-SUSY-breaking mass $m_{\rm SUSY}$.
In the NMSSM and the MSSM with triplets, $m_h$ can be 
significantly enhanced by the F-term
contribution. 
Notice that these F-term contributions should vanish in the
SUSY limit due to the nonrenormalization theorem.    
These F-term contributions also affect the prediction
on the other SM observables such as the triple Higgs boson coupling constants
$\lambda_{hhh}$ similarly to the case of non-SUSY extended Higgs models.
We discuss this class of SUSY models in section 4.3. 

On the other hand, a typical example for extended SUSY Higgs sectors
without interactions from the tree-level F-term is that 
with only multi-doublet structures, such as the 4HDM.
In this class of models, if there is no mixing between
the light two doublet fields and the additional ones, 
the effects of the extra fields on the MSSM observables become
suppressed due to the decoupling theorem
when the extra doublet fields are heavy, and the model
behaves like the MSSM. 
However, nonvanishing effects can appear through the mixing between
the light two doublet fields and the additional ones via the soft breaking B-term. 
These effects appear at the tree level, so that
they would give substantial modifications in the predictions
in the MSSM for the low energy observables.
We stress that these nonvanishing effects due to the B-term mixing 
are not the nondecoupling effects which appear in the large mass limit
for the new particles when $\lambda v^2 \gtrsim M^2$ as a
consequence of violation of the decoupling theorem.
In this sense, we call the nonvanishing B-term mixing effect
as the {\it quasi-nondecoupling effect}.
Notice that the quasi-nondecoupling effect only appears
in the predicted values in the MSSM.
It gives modifications in the MSSM predictions  
such as the masses of $h$, $H$ and $H^\pm$
and the mixing angle $\alpha$
as well as coupling constants for the
MSSM-like Higgs bosons. Such an effect, however, does disappear
in the predictable SM coupling constants
of $h\gamma\gamma$, $hWW$, $hZZ$ and $hhh$ 
in the SM-like limit ($m_A \to \infty$)
according to the decoupling theorem.

Therefore, we would like to address the question of how the extra
doublet fields in the extended SUSY model can affect the
observables which appear in the MSSM, such as  
the mass $m_{\phi}$ ($\phi$ represent $h$, $H$ and $H^\pm$),   
the mixing angle $\alpha$ and the vertex $F_{\phi'VV}$ ($V=W^\pm$ and 
$Z$; $\phi'=h$ and $H$). 
Deviations from the renormalized MSSM observable parameters may
be expressed as 
\begin{align}
 m_{\phi} &\simeq m_{\phi}^{\rm
 MSSM} \left(1+
 \delta_{\phi} \right) \;, \\
 \sin^2(\beta-\alpha_{\rm eff}) &\simeq
 [\sin^2(\beta-\alpha)]^{\rm MSSM}
 \left(1 + \delta_s \right) \;, \label{eq:delta_s}\\
 F_{\phi' VV} &\simeq F_{\phi' VV}^{\rm
 MSSM} \left(1 + \delta_{\phi' VV} \right) \;, 
\end{align}
where $\delta_\phi$, $\delta_s$ or $\delta_{\phi'VV}$ 
represent the effect of the extra heavy scalar
fields on each observable in the extended SUSY models.

In section 4.4, we study 
$\delta_h$, $\delta_H$, $\delta_{H^\pm}$ and $\delta_s$ in the 4HDM.
The MSSM predictions are evaluated  at the one-loop level using
the on-shell renormalization scheme in Ref.~\cite{dabelstein}.

\section{SUSY Higgs sectors with nondecoupling effects}
In this section, we consider various extension of SUSY Higgs sectors. 
One way of the extension of the MSSM may be adding new chiral 
superfields such as isospin singlets (neutral $\hat{S}$, singly charged 
$\hat{\Omega}_{L,R}$ or doubly charged $\hat{K}_{L,R}$), doublets ($\hat{H}_u^\prime$ 
and $\hat{H}_d^\prime$), or triplets ($\hat{\xi}$ 
with the hypercharge $Y=0$ or $\hat{\Delta}_{L,R}$ with $Y=\pm1$), whose properties 
are defined in Table~\ref{pc_esusy}. 
For anomaly cancellation, 
charged superfields are introduced in pair in each model. 
As we are interested in the variation in the
Higgs sector, these new fields are supposed to be colour singlet. 
Although there can be further possibilities such as introduction of 
new vector superfields which contain gauge fields for extra gauge
symmetries, models with extra dimensions, those with the R-parity 
violation, etc., we here do not discuss them. 

Although models in Table~\ref{pc_esusy} can be imposed 
additional exact or softly-broken discrete symmetries for various reasons, 
we here do not specify them as they do not affect 
our discussions. 
In this section, we discuss the three SUSY models in addition to the MSSM: 
the MSSM with extra $\hat{S}$ which is so-called the next-to MSSM (NMSSM), 
that with pair of extra triplets $\hat{\Delta}_{L,R}$ (TMSSM) and 
that with pair of extra doublet fields $\hat{H}_u^\prime$ and $\hat{H}_d^\prime$ and pair of charged singlet fields 
$\hat{\Omega}_{L,R}$ (4D$\Omega$). 
\begin{table}[t]
\begin{center}
{\renewcommand\arraystretch{1.3}
\begin{tabular}{|c||ccccc||cc||ccc|}\hline
& $\hat{S}$ & $\hat{\Omega}_R^c$ & $\hat{\Omega}_L$ & $\hat{K}_R^c$ 
& $\hat{K}_L$ & $\hat{H}_u^\prime$ & $\hat{H}_d^\prime$ & $\hat{\xi}$ &  $\hat{\Delta}_R^c$ & $\hat{\Delta}_L$\\\hline\hline
$SU(2)_L$ &\textbf{1}&\textbf{1}&\textbf{1}&\textbf{1}&\textbf{1}&\textbf{2}&\textbf{2}&\textbf{3}&\textbf{3}&\textbf{3} \\\hline
$U(1)_Y$ &$0$&$+1$&$-1$&$+2$&$-2$&$+1/2$&$-1/2$&$0$&$+1$&$-1$\\\hline
\end{tabular}}
\caption{Properties of the additional chiral superfields}
\label{pc_esusy}
\end{center}
\end{table}
\subsection{The next-to-MSSM}
The superpotential in the NMSSM is 
\begin{align}
W_{\text{NMSSM}}=W_{\text{MSSM}}+\lambda_{HHS} \hat{S}\hat{H}_u\cdot \hat{H}_d+\frac{\kappa}{3} \hat{S}^3+\frac{\mu_S}{2}\hat{S}^2. 
\end{align}
The soft-breaking terms are 
\begin{align}
\mathcal{L}_{\text{NMSSM}}^{\text{soft}}=\mathcal{L}_{\text{MSSM}}^{\text{soft}}
-M_S^2|S|^2-\left(A_\lambda H_u\cdot H_dS+\frac{A_\kappa}{3}S^3+B_S\mu_SS^2+\text{h.c.}\right). 
\end{align}
The Higgs potential is 
\begin{align}
V_{\text{NMSSM}}&=
m_1^2 H_d^\dagger H_d+m_2^2 H_u^\dagger H_u+m_S^2|S|^2+(B\mu H_d\cdot H_u+B_S\mu_SS^2+\text{h.c.})\notag\\
&+\frac{g^2+g^{\prime 2}}{8}(H_d^\dagger H_d-H_u^\dagger H_u)^2-\frac{g^2}{2}(H_d^\dagger H_u)(H_u^\dagger H_d)\notag\\
&+|\lambda_{HHS}|^2\left[|H_u\cdot H_d|^2+|S|^2(|H_u|^2+|H_d|^2)\right]+|\kappa|^2S^2|S|^2\notag\\
&+(\lambda_{HHS} \mu_S^* S^*H_u\cdot H_d+\kappa\mu_S^*S^2S^*+\lambda_{HHS}^*\kappa S^2H_u\cdot H_d
+A_\lambda H_u\cdot H_dS+\frac{A_\kappa}{3}S^3+\text{h.c.}).
\end{align}
The singlet scalar boson $S$ can be parameterized as 
\begin{align}
S=\frac{1}{\sqrt{2}}(S_\varphi+iS_\chi). 
\end{align}
Here, we consider the case where the singlet scalar boson does not obtain the VEV. 
In this case, the mixing among the scalar bosons from the doublet Higgs field and those from 
singlet field can be neglected. 
The one-loop level $m_h^2$ and the triple $h$ coupling in the NMSSM can be calculated as
\begin{align}
m_h^2(\text{NMSSM}) &\simeq m_h^2(\text{MSSM})+\frac{v^2}{2}|\lambda_{HHS}|^2\sin^22\beta+\frac{|\lambda_{HHS}|^4 v^2}{32\pi^2}
\ln\frac{m_{S_\varphi}^2m_{S_\chi}^2}{m_{\tilde{S}}^4},\label{mh_nmssm}\\
\lambda_{hhh}^{\text{NMSSM}}&\simeq\bar{\lambda}_{hhh}^{\text{NMSSM}}
+\left[\frac{3m_h^2(\text{NMSSM})}{v}\right]\frac{|\lambda_{HHS}|^6v^4}{96\pi^2m_h^2(\text{NMSSM})}
\left(\frac{1}{m_{S_\varphi}^2}+\frac{1}{m_{S_\chi}^2}\right),
\end{align}
where $m_{S_\varphi}^2$, $m_{S_\chi}^2$ and $m_{\tilde{S}}$ are the masses of $S_\varphi$, $S_\chi$ and $\tilde{S}$, respectively. 
The masses of $m_{S_\varphi}^2$ and $m_{S_\chi}^2$ can be expressed as
\begin{align}
m_{S_{\varphi,\chi}}^2 =\mathscr{M}_{S_{\varphi,\chi}}^2+\frac{|\lambda_{HHS}|^2}{2}v^2,
\end{align}
where $\mathscr{M}_{S_{\varphi,\chi}}^2$ is the invariant mass parameters. 
\subsection{Model with extra triplet superfields}
The superpotential in the TMSSM is 
\begin{align}
W_{\text{TMSSM}}=W_{\text{MSSM}}+\frac{h_\Delta^{ij}}{2}\hat{L}_i\cdot \hat{\Delta}_R^c \hat{L}_j +\frac{\lambda_{HH\Delta_L}}{2} \hat{H}_d\cdot \hat{\Delta}_R^c\hat{H}_d
+\frac{\lambda_{HH\Delta_R}}{2} \hat{H}_u\cdot \hat{\Delta}_L\hat{H}_u+\mu_\Delta \text{tr}(\hat{\Delta}_R^c \hat{\Delta}_L). 
\end{align}
The soft-breaking terms are 
\begin{align}
\mathcal{L}_{\text{TMSSM}}^{\text{soft}}&=\mathcal{L}_{\text{MSSM}}^{\text{soft}}
-M_{\Delta_R}^2\text{tr}(\Delta_R^\dagger\Delta_R)
-M_{\Delta_L}^2\text{tr}(\Delta_L^\dagger\Delta_L)\notag\\
&-\left[\frac{(A_\Delta)_{ij}}{2} \tilde{L}_i\cdot \Delta_R^*\tilde{L}_j
+\frac{A_1}{2} H_d \cdot \Delta_R^* H_d+\frac{A_2}{2} H_u \cdot \Delta_L H_u+B_\Delta\mu_\Delta\text{tr}(\Delta_R^*\Delta_L)+\text{h.c.}\right].
\end{align}
The Higgs potential is 
\begin{align}
&V_{\text{TMSSM}}=m_1^2 H_d^\dagger H_d+m_2^2 H_u^\dagger 
+m_{\Delta_R}^2\text{tr}(\Delta_R^\dagger\Delta_R)+m_{\Delta_L}^2\text{tr}(\Delta_L^\dagger\Delta_L)\notag\\
&+(B\mu H_d\cdot H_u+B_\Delta\mu_\Delta\text{tr}(\Delta_R^*\Delta_L)+\text{h.c.})\notag\\
&+\frac{g^2}{8}\left[H_d^\dagger \tau^a H_d+H_u^\dagger \tau^a H_u+\text{tr}(\Delta_R^T\tau^a \Delta_R^*)+\text{tr}(\Delta_L^\dagger\tau^a \Delta_L)\right]^2\notag\\
&+\frac{g^{\prime 2}}{8}\left[-H_d^\dagger H_d+H_u^\dagger H_u+2\text{tr}(\Delta_R^T\Delta_R^*)
-2\text{tr}(\Delta_L^\dagger\Delta_L)\right]^2\notag\\
&+\left(\lambda_{HH\Delta_L}\mu H_u^\dagger \Delta_R^* H_d+\lambda_{HH\Delta_R}\mu H_d^\dagger \Delta_L H_u
+\lambda_{HH\Delta_L}\mu_\Delta H_d\cdot \Delta_L^\dagger H_d+\lambda_{HH\Delta_R} \mu_\Delta H_u\cdot \Delta_R^T H_u +\text{h.c.}
\right)\notag\\
&+|\lambda_{HH\Delta_L}|^2(H_d\cdot \Delta_R^*)(H_d\cdot \Delta_R^*)^\dagger 
+|\lambda_{HH\Delta_R}|^2(H_u\cdot \Delta_L)(H_u\cdot \Delta_L)^\dagger\notag\\
&+\frac{|\lambda_{HH\Delta_L}|^2}{4}(H_d^\dagger H_d)^2+\frac{|\lambda_{HH\Delta_R}|^2}{4}(H_u^\dagger H_u)^2.
\end{align}
The scalar bosons from the triplet fields $\Delta_R^*$ and $\Delta_L$ can be parametrized as 
\begin{align}
\Delta_R^*=
\left(
\begin{array}{cc}
\frac{1}{\sqrt{2}}\Delta_R^+ & \Delta_R^{++}\\
\frac{1}{\sqrt{2}}(v_L+\Delta_{L\varphi}+i\Delta_{L\chi} ) & -\frac{1}{\sqrt{2}}\Delta_R^+
\end{array}\right),\quad \Delta_L=
\left(
\begin{array}{cc}
\frac{1}{\sqrt{2}}\Delta_L^- & \frac{1}{\sqrt{2}}(v_L+\Delta_{L\varphi}+i\Delta_{L\chi} ) \\
\Delta_L^{--}& -\frac{1}{\sqrt{2}}\Delta_L^+
\end{array}\right).
\end{align}
The one-loop level $m_h^2$ and the triple $h$ coupling in the NMSSM can be calculated as
\begin{align}
m_h^2(\text{TMSSM}) &\simeq m_h^2(\text{MSSM})
+\frac{v^2}{2}(|\lambda_{HH\Delta_L}|^2\cos^4\beta+|\lambda_{HH\Delta_R}|^2\sin^4\beta)\notag\\
&+\frac{|\lambda_{HH\Delta_L}|^4 v^2\cos^4\beta}{32\pi^2}
\left(\ln\frac{m_{\Delta_{R\varphi}}^2m_{\Delta_{R\chi}}^2}{m_{\tilde{\Delta}_R^0}^4}+2\ln\frac{m_{\Delta_R^+}^2}{m_{\tilde{\Delta}_R^+}^2}\right)\notag\\
&+\frac{|\lambda_{HH\Delta_R}|^4 v^2\sin^4\beta}{32\pi^2}
\left(\ln\frac{m_{\Delta_{L\varphi}}^2m_{\Delta_{L\chi}}^2}{m_{\tilde{\Delta}_L^0}^4}+2\ln\frac{m_{\Delta_L^+}^2}{m_{\tilde{\Delta}_L^+}^2}\right)
,\\
\lambda_{hhh}^{\text{TMSSM}}&\simeq\bar{\lambda}_{hhh}^{\text{TMSSM}}
+\left[\frac{3m_h^2(\text{TMSSM})}{v}\right]\frac{|\lambda_{HH\Delta_L}|^6v^4\cos^6\beta}{96\pi^2m_h^2(\text{TMSSM})}
\left(\frac{1}{m_{\Delta_{R\varphi}}^2}+\frac{1}{m_{\Delta_{R\chi}}^2}+\frac{2}{m_{\Delta_R^+}^2}\right)\notag\\
&+\left[\frac{3m_h^2(\text{TMSSM})}{v}\right]\frac{|\lambda_{HH\Delta_R}|^6v^4\sin^6\beta}{96\pi^2m_h^2(\text{TMSSM})}
\left(\frac{1}{m_{\Delta_{L\varphi}}^2}+\frac{1}{m_{\Delta_{L\chi}}^2}+\frac{2}{m_{\Delta_L^+}^2}\right),
\end{align}
where $m_{\Delta_{R\varphi}}^2$, $m_{\Delta_{R\chi}}^2$, 
$m_{\Delta_R^+}^2$, $m_{\tilde{\Delta}_R^0}$ and $m_{\tilde{\Delta}_R^+}$ are the masses of 
$\Delta_{R\varphi}$, $\Delta_{R\chi}$, $\Delta_R^+$, $\tilde{\Delta}_R^0$ and $\tilde{\Delta}_R^+$, respectively. 
In the above definition, the parameters which are denoted as $L$ instead of $R$ are masses of the corresponding fields which are 
denoted as $L$ instead of $R$. The scalar bosons masses can be expressed as 
\begin{align}
m_{\Delta_{R\varphi,R\chi}}^2 &=\mathscr{M}_{\Delta_{R\varphi,R\chi}}^2+\frac{|\lambda_{HH\Delta_L}|^2}{2}v^2\cos^2\beta,\\
m_{\Delta_{L\varphi,L\chi}}^2 &=\mathscr{M}_{\Delta_{L\varphi,L\chi}}^2+\frac{|\lambda_{HH\Delta_R}|^2}{2}v^2\sin^2\beta,\\
m_{\Delta_R^+}^2 &=\mathscr{M}_{\Delta_R^+}^2+\frac{|\lambda_{HH\Delta_L}|^2}{2}v^2\cos^2\beta,\\
m_{\Delta_L^+}^2 &=\mathscr{M}_{\Delta_L^+}^2+\frac{|\lambda_{HH\Delta_R}|^2}{2}v^2\sin^2\beta,
\end{align}
where $\mathscr{M}_{\Delta_{R\varphi,R\chi}}^2$, $\mathscr{M}_{\Delta_{L\varphi,L\chi}}^2$, 
$\mathscr{M}_{\Delta_R^+}^2$ and $\mathscr{M}_{\Delta_L^+}^2$ are the invariant mass parameters. 
\subsection{Model with four Higgs doublets and charged singlet superfields}
The 4D$\Omega$ contains the four Higgs doublets, so that 
in general, FCNC processes can appear at the tree level. 
We here impose the softly-broken $Z_2$ symmetry to avoid such processes. 
We assign that $\hat{H}_u'$, $\hat{H}_d'$, $\hat{\Omega}_R^c$ and $\hat{\Omega}_L$ are odd, 
while the other fields are even under this $Z_2$ symmetry \cite{kss}. 
The superpotential in the 4D$\Omega$ is 
\begin{align}
W_{\text{4D$\Omega$}}=W_{\text{MSSM}}
+\lambda_{HH\Omega_R}\hat{H}_d\cdot \hat{H}_d'\hat{\Omega}_R^c
+\lambda_{HH\Omega_L}\hat{H}_u\cdot \hat{H}_u'\hat{\Omega}_L+
\mu' \hat{H}_u'\cdot \hat{H}_d'+\mu_\Omega \hat{\Omega}_R^c\hat{\Omega}_L. 
\end{align}
The soft-breaking terms are 
\begin{align}
\mathcal{L}_{\text{4D$\Omega$}}^{\text{soft}}&=\mathcal{L}_{\text{MSSM}}^{\text{soft}}
-M_{H_d'}^2H_d^{\prime\dagger} H_d'+M_{H_u'}^2H_u^{\prime\dagger} H_u'
-M_{\Omega_R}^2(\omega_R^*\omega_R)
-M_{\Omega_L}^2(\omega_L^*\omega_L)\notag\\
&-\left[(A_f)_{ij} \omega_R^*\tilde{L}_i\cdot \tilde{L}_j
+A_1 \omega_R^*H_d \cdot  H_d'+A_2 \omega_L H_u \cdot H_u+B'\mu'H_u'\cdot H_d'+B_\Omega\mu_\Omega\omega_R^*\omega_L +\text{h.c.}\right].
\end{align}
The Higgs potential is 
\begin{align}
&V_{\text{4D$\Omega$}}=
m_1^2 H_d^\dagger H_d+m_2^2 H_u^\dagger H_u +m_3^2 H_d^{\prime\dagger} H_d'+m_4^2 H_u^{\prime\dagger}H_u'
+m_{\omega_R}^2\omega_R^*\omega_R+m_{\omega_L}^2\omega_L^*\omega_L\notag\\
&+(B\mu H_u\cdot H_d+B'\mu' H_u' \cdot H_d'+B_\Omega \mu_\Omega \omega_R^*\omega_L+\text{h.c.})\notag\\
&+\frac{g^2+g^{\prime 2}}{8}
(H_u^\dagger H_u +H_d^\dagger H_d-H_u^{\prime\dagger} H_u'-H_d^{\prime\dagger} H_d')^2\notag\\
&+\frac{g^2}{2}\Big[(H_d^\dagger H_u)(H_u^\dagger H_d)+(H_d^\dagger H_u')(H_u^{\prime\dagger} H_d)+
(H_d^{\prime\dagger} H_u)(H_u^\dagger H_d')+(H_d^{\prime\dagger} H_u')(H_u^{\prime\dagger} H_d')\notag\\
&+(H_d^\dagger H_d')(H_d^{\prime\dagger} H_d)-(H_d^\dagger H_d)(H_d^{\prime\dagger} H_d')
+(H_u^\dagger H_u')(H_u^{\prime\dagger} H_u)-(H_u^\dagger H_u)(H_u^{\prime\dagger} H_u')
\Big]\notag\\
&+\frac{g^{\prime 2}}{2}(\omega_R^*\omega_R-\omega_L^*\omega_L)^2+
\frac{g^{\prime 2}}{2}(H_u^\dagger H_u+H_u^{\prime\dagger} H_u'-H_d^\dagger H_d-H_d^{\prime\dagger} H_d')(\omega_R^*\omega_R-\omega_L^*\omega_L)\notag\\
&+|\lambda_{HH\Omega_R}|^2\left[(H_d^\dagger H_d +H_d^{\prime\dagger} H_d' )\omega_R^*\omega_R+(H_d\cdot H_d')^*(H_d\cdot H_d')\right]\notag\\
&+|\lambda_{HH\Omega_L}|^2\left[(H_u^\dagger H_u +H_u^{\prime\dagger} H_u' )\omega_L^*\omega_L+(H_u\cdot H_u')^*(H_u\cdot H_u')\right]\notag\\
&+[A_1 \omega_R^*H_d \cdot  H_d'+A_2 \omega_L H_u \cdot H_u'
+\mu^*\lambda_{HH\Omega_R}H_u^\dagger H_d' \omega_R^*+\mu \lambda_{HH\Omega_L}^* H_u^{\prime \dagger} H_d\omega_L^*
+\mu^{\prime *}\lambda_{HH\Omega_R}H_u^{\prime \dagger}H_d\omega_R^*\notag\\
&+\mu' \lambda_{HH\Omega_L}^*H_u^\dagger H_d' \omega_L^*+\mu_\Omega^*\lambda_{HH\Omega_R}H_d\cdot H_d'\omega_L^*
+\mu_\Omega\lambda_{HH\Omega_L}^*H_u^\dagger\cdot H_u^{\prime \dagger}\omega_R^*+\text{h.c.}]. 
\end{align}
The one-loop level $m_h^2$ and the triple $h$ coupling in the NMSSM can be calculated as
\begin{align}
&m_h^2(\text{4D$\Omega$}) \simeq m_h^2(\text{MSSM})
+\frac{|\lambda_{HH\Omega_R}|^4 v^2\cos^4\beta}{16\pi^2}
\ln\frac{m_{\omega_R}^2}{m_{\tilde{\omega}}^2}
+\frac{|\lambda_{HH\Omega_L}|^4 v^2\sin^4\beta}{16\pi^2}
\ln\frac{m_{\omega_L}^2}{m_{\tilde{\omega}}^2}
,\\
&\lambda_{hhh}^{\text{4D$\Omega$}}\simeq\bar{\lambda}_{hhh}^{\text{4D$\Omega$}}
+\left[\frac{3m_h^2(\text{4D$\Omega$})}{v}\right]
\frac{|\lambda_{HH\Omega_R}|^6v^4m_{\omega_R}^2\cos^6\beta}{48\pi^2m_h^2(\text{4D$\Omega$})}
+\left[\frac{3m_h^2(\text{4D$\Omega$})}{v}\right]\frac{|\lambda_{HH\Omega_L}|^6
v^4m_{\omega_L}^2\sin^6\beta}{48\pi^2m_h^2(\text{4D$\Omega$})},
\end{align}
where $m_{\omega_R}^2$, $m_{\omega_L}^2$, and $m_{\tilde{\omega}}$ are the masses of 
$\omega_R$, $\omega_L$ and $\tilde{\omega}$, respectively. 
The scalar bosons masses can be expressed as 
\begin{align}
m_{\omega_R}^2&=\mathscr{M}^2_{\omega_R}+\frac{|\lambda_{HH\Omega_R}|^2}{2}v^2\cos^2\beta,
m_{\omega_L}^2&=\mathscr{M}^2_{\omega_L}+\frac{|\lambda_{HH\Omega_L}|^2}{2}v^2\sin^2\beta,
\end{align}
where $\mathscr{M}_{\omega_R}$ and $\mathscr{M}_{\omega_L}$ 
are the invariant mass parameters.

\subsection{Possible allowed regions of $m_h$ and $\lambda_{hhh}$ in various SUSY Higgs models}
Here, we evaluate numerical calculation for $m_h$ and the deviation of $\lambda_{hhh}$ from the SM prediction in each 
SUSY Higgs model at the one-loop level.  

\begin{figure}[t]
\begin{center}
\includegraphics[width=80mm]{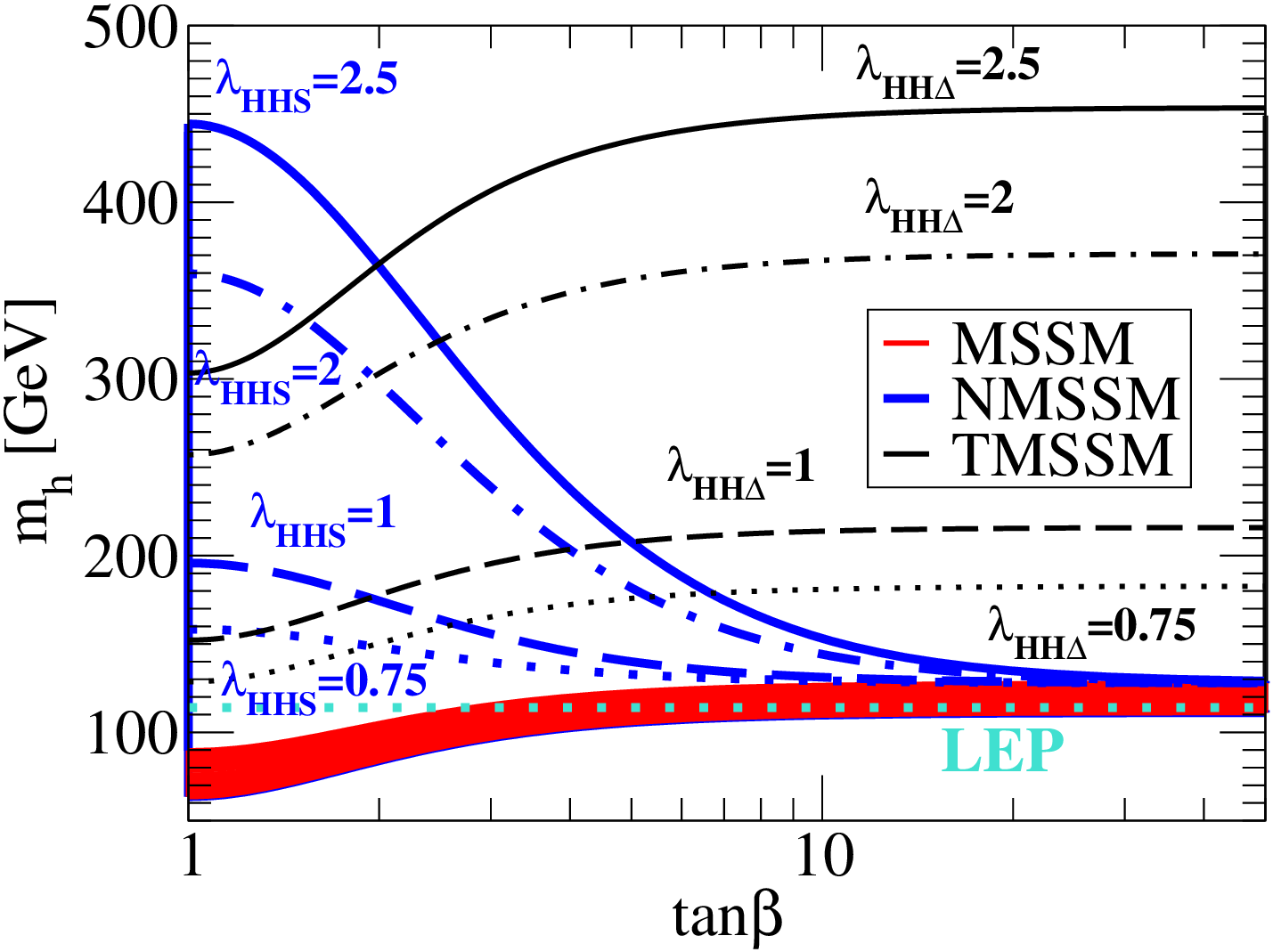}
\end{center}
\caption{	 
The upper bounds on $m_h$ as a function of $\tan\beta$ 
for fixed values of $\lambda_{HHS}$ and $\lambda_{HH\Delta_L}^{}=\lambda_{HH\Delta_R}^{}\equiv \lambda_{HH\Delta}$
in the NMSSM and the TMSSM, respectively. 
The red-filled region indicates the possible allowed region in the MSSM~\cite{ksy}.}
\label{ksy1}
\vspace{10mm}
\begin{center}
\includegraphics[width=65mm]{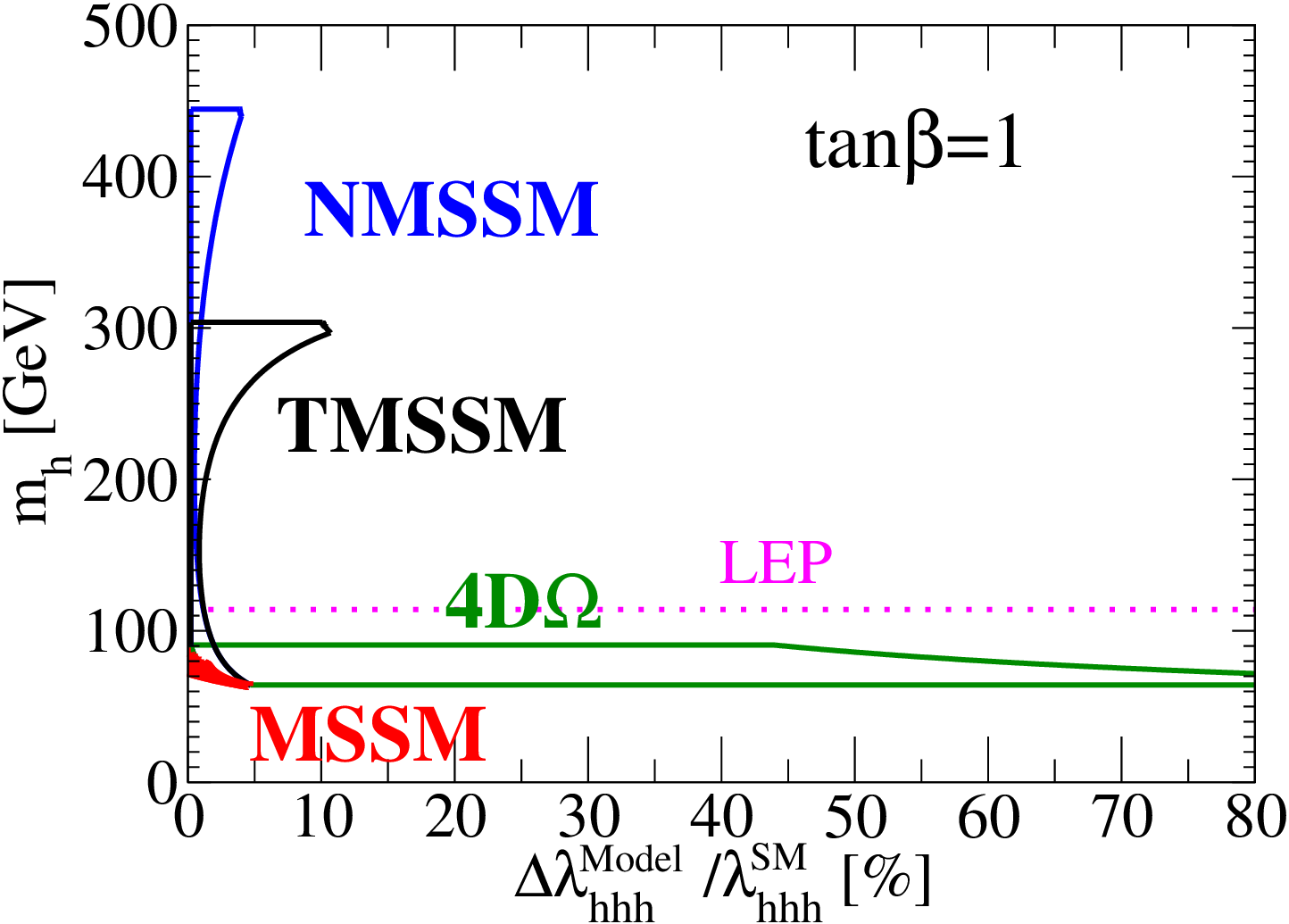}\hspace{3mm}
\includegraphics[width=65mm]{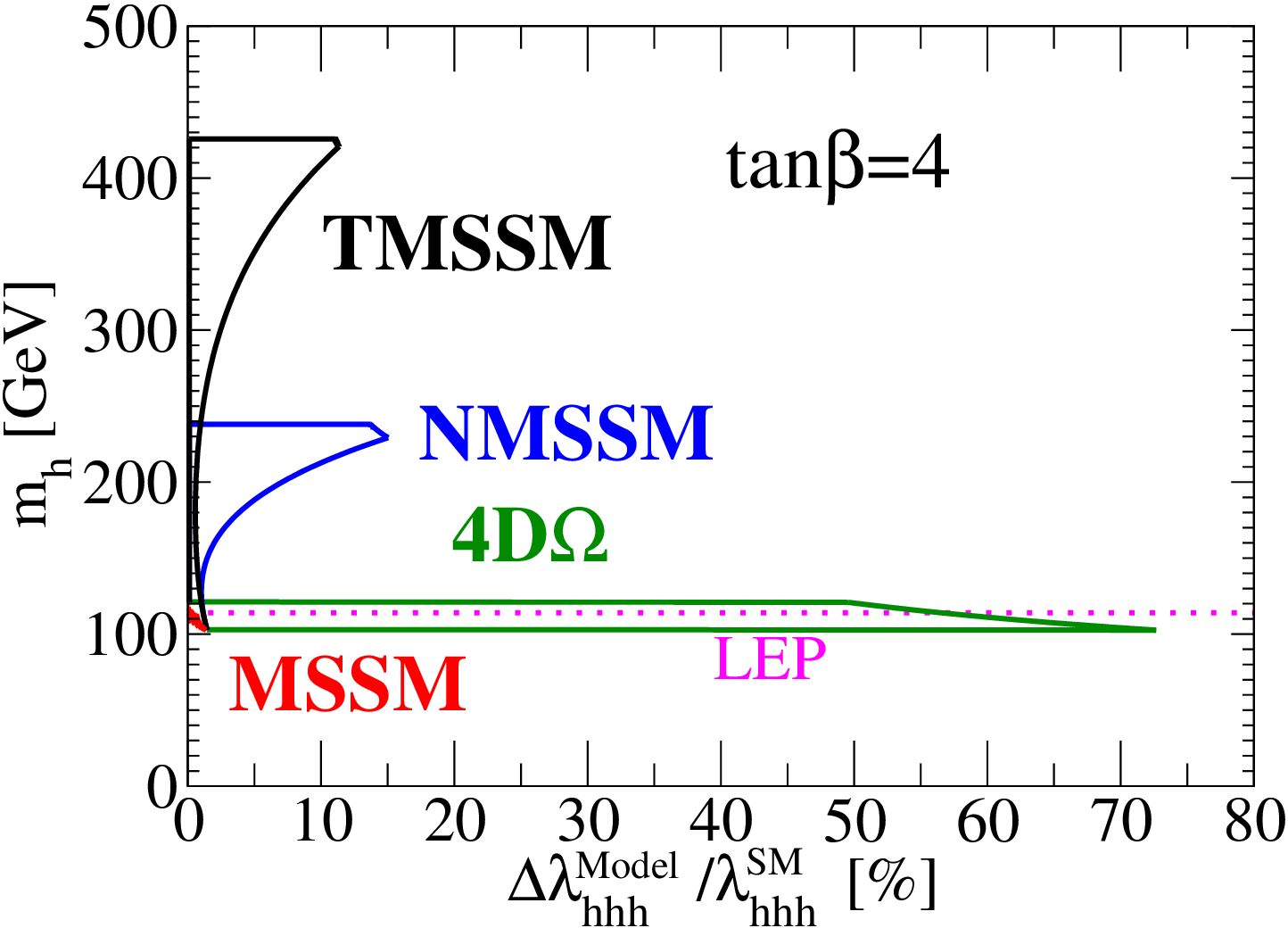}\\\vspace{7mm}
\includegraphics[width=65mm]{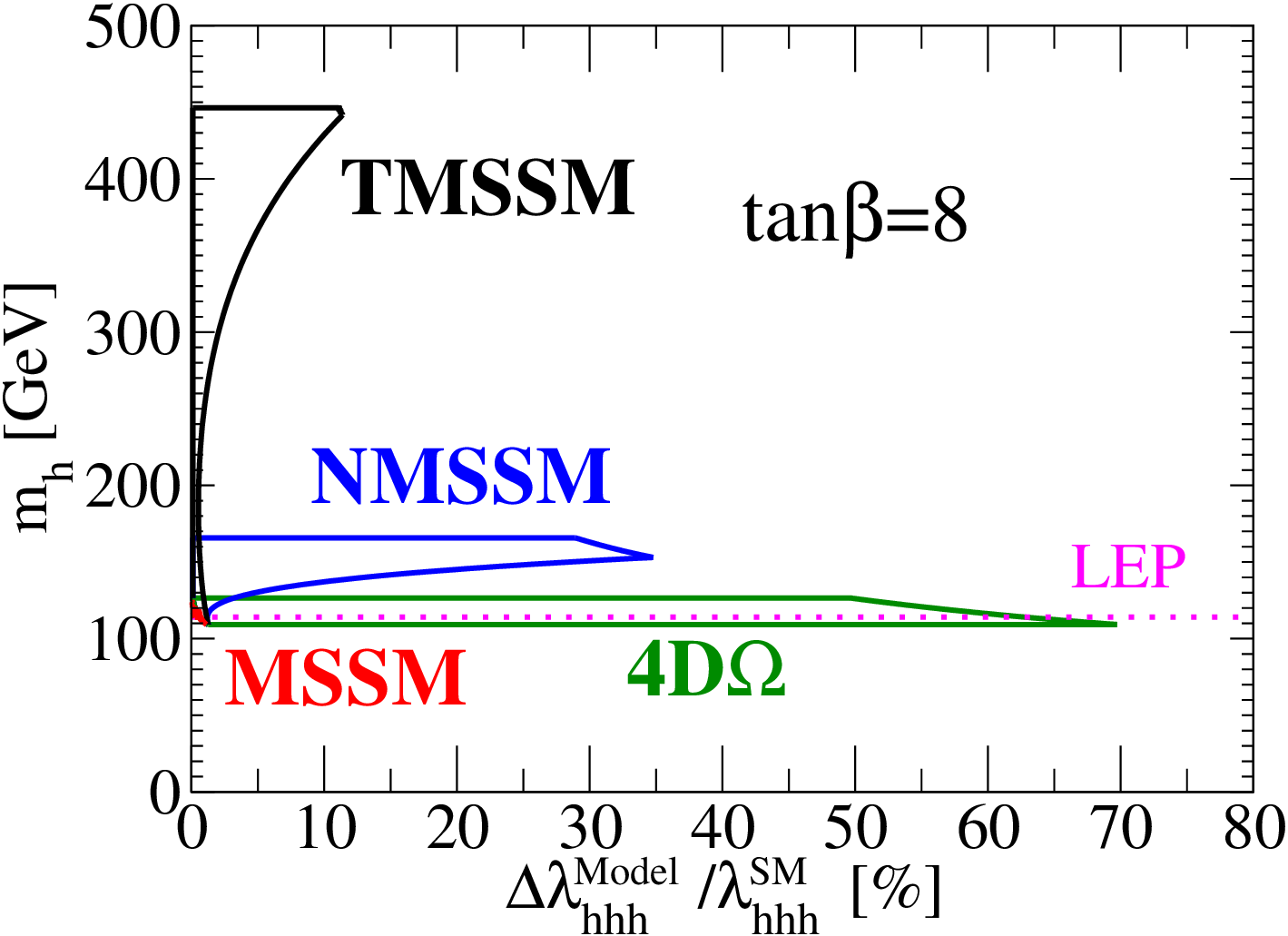}\hspace{3mm}
\includegraphics[width=65mm]{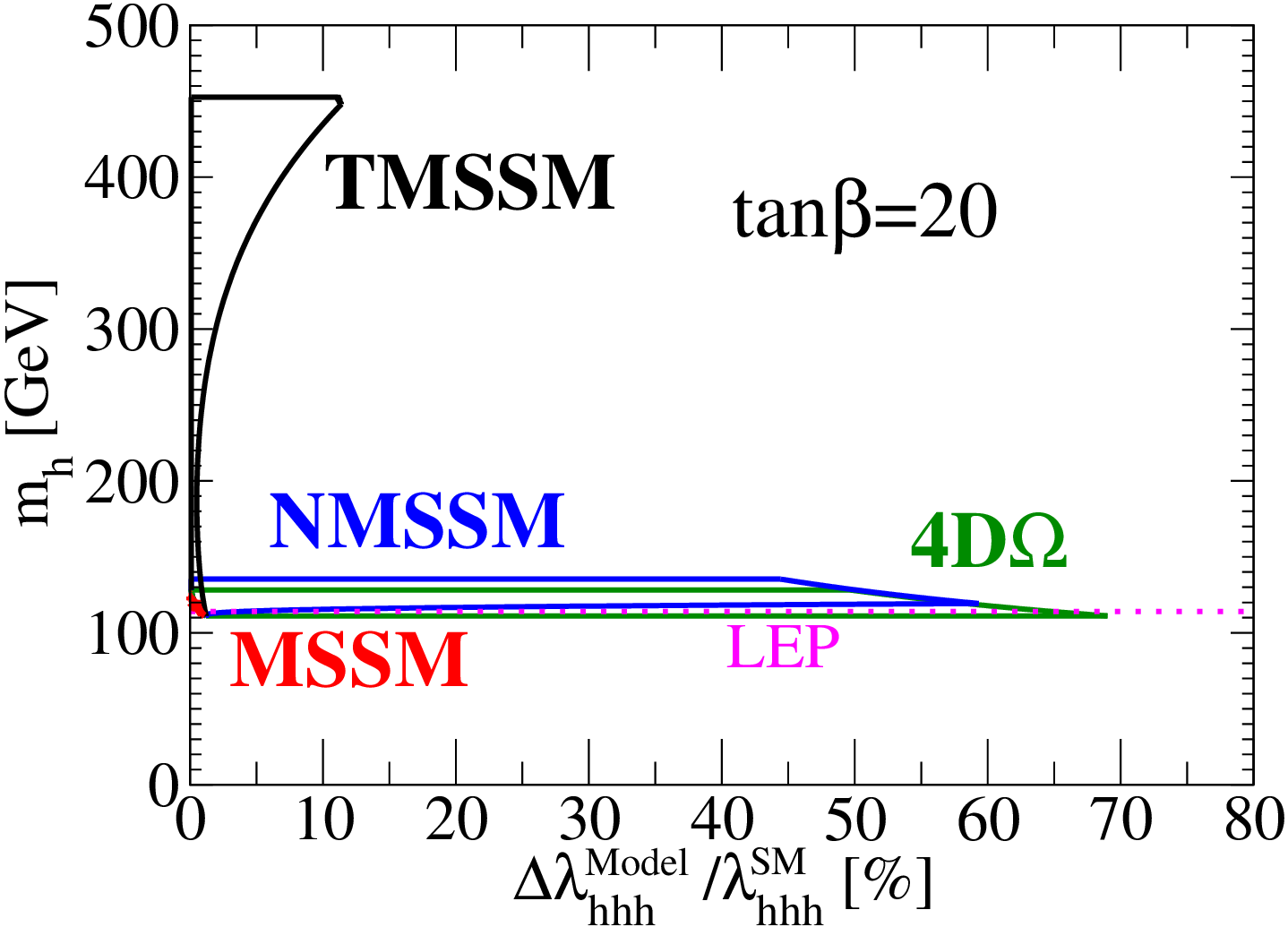}
\end{center}
\caption{
Possible allowed regions in the $m_h$-$(\Delta \lambda_{hhh}/\lambda_{hhh})$ plane  
in the MSSM, Model-1, Model-5 and Model-9 for each $\tan\beta$ value.
We scan the parameter space as
$0<\lambda_{HH\phi}<2.5$, 
$0.5\,\text{TeV}<m_{\tilde{t}_{1,2}}^{}<1.5\,\text{TeV}$,
and
$0.5\,\text{TeV}<m_{\phi}^{}$
for each model~\cite{ksy}, where $m_{\phi}^{}$ is the physical mass of the extra scalar bosons.
}
\label{ksy2}
\end{figure}

In  Fig.~\ref{ksy1}, the upper bounds on $m_h$ in the NMSSM and 
TMSSM are shown as a function of $\tan\beta$, and the possible allowed region in the 
MSSM is also indicated by the red-filled region. 
The coupling constants $\lambda_{HH \phi}$ 
($\phi=S$, $\Delta_L$ or $\Delta_R$) are taken as 
$0 < \lambda_{HH \phi} <2.5$. 
In the NMSSM with a fixed value of $\lambda_{HHS}^{}$, $m_h$ can be maximal for $\tan\beta = 1$, 
while in the TMSSM it becomes maximal for large values of $\tan\beta$ 
for a fixed value of $\lambda_{HH\Delta}$ ($=\lambda_{HH\Delta_L}=\lambda_{HH\Delta_R}$). 
The maximal value in the NMSSM becomes asymptotically the same 
as that in the MSSM in the large $\tan\beta$ limit 
up to the one-loop logarithmic contributions.


\clearpage

We scan the parameter space in each model to find allowed regions in the
$m_h$-$(\Delta \lambda_{hhh}^{\rm Model}/\lambda_{hhh}^{\rm SM})$ plane 
under the assumption of $\lambda_{HH\phi}<2.5$ at the EW scale,
where $\Delta \lambda_{hhh}^{\rm Model}=\lambda_{hhh}^{\rm Model}-\lambda_{hhh}^{\rm SM}$.
In Fig.~\ref{ksy2}, we show the possible allowed region 
for several value of $\tan\beta=1$, $4$, $8$ and $20$.
The coupling constants $\lambda_{HH\phi}$ 
($\phi=S$, $\chi_\pm$ and $\Omega_\pm$) are taken to be less than 2.5 as 
in Fig.~\ref{ksy1}.
The stop masses are scanned as $0.5\,\text{TeV}\leq m_{\tilde{t}_{1,2}}^{}\leq 1.5\,\text{TeV}$.
We also scan the physical masses of the extra scalar bosons as $0.5\,\text{TeV}\leq m_{\phi}^{}$.
The mass of fermion component is taken as same as the mass of the scalar component
for each extra field.
We note that the parameters are scanned such that the additional contributions to the 
rho parameter are negligible\footnote{For example, parameters in the stop-sbottom sector are 
taken to keep the rho parameter constraint satisfied.}.
The region in the MSSM is indicated as the red-filled one. 
The possible allowed region in the NMSSM depends largely on $\tan\beta$: 
for smaller (larger) $\tan\beta$, $m_h$ can be higher (lower) and  
$\Delta\lambda_{hhh}^{\rm Model\mbox{-}1}/\lambda_{hhh}^{\rm SM}$ is smaller (larger).
the TMSSM is relatively insensitive to the value 
of $\tan\beta$: $m_h$ can always be larger than 
about 300 GeV while $\Delta\lambda_{hhh}^{\rm TMSSM}/\lambda_{hhh}^{\rm SM}$ 
remains less than about 10 \%.
On the other hand, in the 4D$\Omega$, although the possible value of $m_h$ 
is similar to that in the MSSM, 
the deviation in the $hhh$ coupling can be very large: 
i.e.,  $\Delta\lambda_{hhh}^{{\rm 4D}\Omega}/\lambda_{hhh}^{\rm SM} \sim 30-60$ \%\footnote{
The definition of $\tan\beta$ in models with four Higgs doublets 
is that 
$\tan\beta = \sqrt{\langle H_u^0 \rangle^2+\langle H_u'^0 \rangle^2}/
\sqrt{\langle H_d^0 \rangle^2+\langle H_d'^0 \rangle^2}$.}. 
When we consider the higher value of $\Lambda$, which corresponds to the smaller upper bound on 
$\lambda_{HH\phi}^{}$, the possible allowed region becomes the smaller.

\begin{figure}[t]
\begin{center}
\includegraphics[width=80mm]{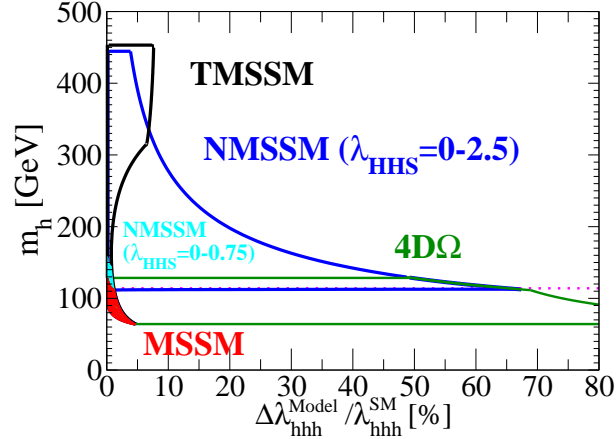}
\caption{Possible allowed regions in the $m_h$-$(\Delta \lambda_{hhh}/\lambda_{hhh})$ plane  
in the MSSM, the NMSSM, the TMSSM and the 4D$\Omega$ with scanned $\tan\beta$~\cite{ksy}.}
\label{ksy3}
\end{center}
\end{figure}

In Fig.~\ref{ksy3}, 
possible allowed regions with scanned $\tan\beta$ are shown in the 
$m_h$-$(\Delta \lambda_{hhh}^{\rm Model}/\lambda_{hhh}^{\rm SM})$ plane 
in the NMSSM, the TMSSM and the 4D$\Omega$ as well as the MSSM. 
The maximal values of $\lambda_{HH\phi}^{}$ in the NMSSM, the TMSSM
and the 4D$\Omega$ are taken to be the same as those in Fig.~\ref{ksy2}. 
The region in the MSSM (the NMSSM with $0\lesssim \lambda_{HHS} \lesssim 0.75$, which corresponds 
to $\Lambda \simeq 10^{16}$ GeV~\cite{mh-NMSSM2}) 
is indicated as the red-filled (cyan-filled) one. 
The possible allowed regions 
are different among the models so that 
the information of $m_h$ and $\Delta \lambda_{hhh}^{\rm Model}$ can be used to
classify the SUSY standard models.

\newpage
\section{SUSY Higgs sectors with quasi-nondecoupling effects}
In this section, we consider the 4HDM in order to examine the quasi-nondecoupling 
effect as discussed in the section 4.2.
Studies for the 4HDM have been done by several papers. 
The Higgs potential and the mass matrices in the 4HDM 
have been analyzed by Gupta and Wells~\cite{Gupta}. 
The collider phenomenology of the 4HDM has been analysed in Ref.~\cite{Marshall}. 
The dark matter physics in the framework of the 4HDM has been investigated in Ref.~\cite{Kawase}. 
In these papers,  
quasi-nondecoupling effects have not been studied. 

\subsection{Model}


We here discuss the 4HDM, in which two extra isospin-doublet chiral superfields 
$\hat{H}_d'$ ($Y=-1/2$) and $\hat{H}_u'$ ($Y=1/2$) are introduced to 
the MSSM in addition to the Higgs doublets $\hat{H}_d$ and $\hat{H}_u$. 
The general expression for the superpotential with the R parity
is given in terms of chiral superfields as 
\begin{align}
W=W_{\text{MSSM}}
&-(\hat{Y}_u^{\prime})_{ij}U_{Ri}^c\hat{H}_u'\cdot Q_{Lj}
+(\hat{Y}_d^{\prime})_{ij}D_{Ri}^c\hat{H}_d'\cdot Q_{Lj}
+(\hat{Y}_e^{\prime})_{ij}E_{Ri}^c\hat{H}_d'\cdot L_{Lj}
\nonumber\\
&-\mu_{14}\hat{H}_d\cdot \hat{H}_u'
-\mu_{32}\hat{H}_d'\cdot \hat{H}_u
-\mu_{34}\hat{H}_d'\cdot \hat{H}_u'\;, \label{eq:superpotential}
\end{align}
The most general holomorphic soft-SUSY-breaking terms with the R parity is 
\begin{align}
\mathcal{L}^{\text{soft}}&=\mathcal{L}^{\text{soft}}_{\text{MSSM}}-
(\bar{M}_-^2)_{33}^{}H_d^{\prime\dagger}H_d'
-(\bar{M}_+^2)_{44}^{}H_u^{\prime\dagger}H_u^\prime\notag\\
&-\left[
(\bar{M}_-^2)_{13}^{}H_d^{\dagger}H_d'
+(\bar{M}_+^2)_{24}^{}H_u^{\dagger}H_u'+\text{h.c.}
\right]\notag\\
&-\left[
-(A_{u}^{\prime})^{ij}\tilde{u}_{Ri}^* H_u'\cdot \tilde{Q}_{Lj}
+(A_{d}^{\prime})^{ij}\tilde{d}_{Ri}^* H_d'\cdot \tilde{Q}_{Lj}
+(A_{e}^{\prime})^{ij}\tilde{e}_{Ri}^* H_d'\cdot \tilde{L}_{Lj}
+\text{h.c.}\right]\notag\\
&-\left(
B_{34}\mu_{34} H_d'\cdot H_u'
+B_{14}\mu_{14} H_d^{}\cdot H_u'
+B_{32}\mu_{32} H_d'\cdot H_u^{}
+\text{h.c.}\right), \label{soft-term}
\end{align}
where $\mathcal{L}^{\text{soft}}_{\text{MSSM}}$ is given in Eq.~(\ref{Lsoft_mssm}). 
The Higgs potential can be obtained by Eqs.~(\ref{eq:superpotential}) and (\ref{soft-term}) as 
\begin{align}
V_{\text{H}}^{}=&
\begin{pmatrix}
H_d^{\dagger}&H_d^{\prime\dagger}
\end{pmatrix}
\begin{pmatrix}
(M_{-}^{2})_{11}^{}&
(M_{-}^{2})_{12}^{}\\
(M_{-}^{2})_{12}^{*}&
(M_{-}^{2})_{22}^{}\\
\end{pmatrix}
\begin{pmatrix}
H_d^{}\\ H_d^{\prime}
\end{pmatrix}
+
\begin{pmatrix}
H_u^{\dagger}&H_u^{\prime\dagger}
\end{pmatrix}
\begin{pmatrix}
(M_{+}^{2})_{11}^{}&
(M_{+}^{2})_{12}^{}\\
(M_{+}^{2})_{12}^{*}&
(M_{+}^{2})_{22}^{}\\
\end{pmatrix}
\begin{pmatrix}
H_u^{} \\ H_u^{\prime}
\end{pmatrix}
\displaybreak[0]
\nonumber\\
&
-\left(
\begin{pmatrix}
{H}_d^{}&
{H}_d^{\prime}
\end{pmatrix}
\begin{pmatrix}
B\mu&
B_{14}\mu_{14}\\
B_{32}\mu_{32}&
B_{34}\mu_{34}&
\end{pmatrix}
\cdot 
\begin{pmatrix}
{H}_u^{}\\
{H}_u^{\prime}
\end{pmatrix}
+\text{h.c.}\right)
\displaybreak[0]
\nonumber\\
&
+\frac{g^{\prime 2}+g^2}{8}\left(
H_u^{\dagger}H_u^{}
+H_u^{\prime\dagger}H_u^{\prime}
-H_d^{\dagger}H_d^{}
-H_d^{\prime\dagger}H_d^{\prime}
\right)^2
\displaybreak[0]
\nonumber\\
&
+\frac{g^2}{2}\left[
(H_d^{\dagger}H_u^{})(H_u^\dagger H_d^{})
+(H_d^{\dagger}H_u^{\prime})(H_u^{\prime\dagger}H_d^{})
+(H_d^{\prime\dagger}H_u^{})(H_u^\dagger H_d^{\prime})
+(H_d^{\prime\dagger}H_u^{\prime})(H_u^{\prime\dagger}H_d^{\prime})
\right.
\nonumber\\
&\left.
+(H_d^{\dagger}H_d^{\prime})(H_d^{\prime\dagger}H_d^{})
-(H_d^{\dagger}H_d^{})(H_d^{\prime\dagger}H_d^{\prime})
+(H_u^{\dagger}H_u^{\prime})(H_u^{\prime\dagger}H_u^{})
-(H_u^{\dagger}H_u^{})(H_u^{\prime\dagger}H_u^{\prime})
\right],  
\end{align}
where 
\begin{align}
(M_{-}^2)_{11}^{}=&(M_{-}^2)^{}+|\mu^{}|^2+|\mu_{14}^{}|^2\;,\nonumber\\
(M_{-}^2)_{22}^{}=&(\bar{M}_{-}^2)_{33}^{}+|\mu_{32}^{}|^2+|\mu_{34}^{}|^2\;,\nonumber\\
(M_{-}^2)_{12}^{}=&(\bar{M}_{-}^2)_{13}^{}+\mu^{*}\mu_{32}^{}+\mu_{14}^{*}\mu_{34}^{}\;,\nonumber\\
(M_{+}^2)_{11}^{}=&(M_{+}^2)^{}+|\mu^{}|^2+|\mu_{32}^{}|^2\;,\nonumber\\
(M_{+}^2)_{22}^{}=&(\bar{M}_{+}^2)_{44}^{}+|\mu_{14}^{}|^2+|\mu_{34}^{}|^2\;,\nonumber\\
(M_{+}^2)_{12}^{}=&(\bar{M}_{-}^2)_{24}^{}+\mu^{*}\mu_{14}^{}+\mu_{32}^{*}\mu_{34}^{}\;.
\end{align}

\begin{table}[t]
\begin{center}
{\renewcommand\arraystretch{1}
\begin{tabular}{|c||c|c|c|c|c|c|c|c|c||c|}\hline
      &$\hat{H}_d$&$\hat{H}_u$&$\hat{H}_d'$&$\hat{H}_u'$&$\hat{U}_R^c$&$\hat{D}_R^c$&$\hat{E}_R^c$&$\hat{Q}_L$&$\hat{L}_L$& $\hat{N}_R^c$ \\\hline\hline
Type A&$+$  &$+$  &$-$  &$-$  &$+$&$+$&$+$&$+$&$+$& $+$   \\\hline
Type B&$+$  &$+$  &$-$  &$-$  &$+$&$+$&$-$&$+$&$+$& $+$   \\\hline
Type C&$+$  &$+$  &$-$  &$-$  &$+$&$+$&$+$&$+$&$+$& $-$   \\\hline
Type D&$+$  &$+$  &$-$  &$-$  &$+$&$+$&$-$&$+$&$+$& $-$   \\\hline
\end{tabular}}
\caption{Classification for the charge assignment for
 the $Z_2$ symmetry in the 4HDM. Type C and Type D are introduced only when $N_{Ri}^c$
 are added to the model~\cite{aksy_4hdm}.}
\label{z2}
\end{center}
\end{table}

There are two Higgs doublets for each quantum number, so that they can mix with each other. 
The Yukawa sector then produces a dangerous FCNC via the scalar boson
exchange at the tree level.
There are several ways to eliminate such an excessive FCNC.
In non-SUSY extended Higgs sectors with multi-doublets,
a softly-broken discrete $Z_2$ symmetry is often imposed~\cite{GW}. 
In the general two Higgs doublet model, there are four 
types of Yukawa interactions under such a $Z_2$ symmetry 
depending on the assignment of the $Z_2$ charge~\cite{Barger,Grossman,typeX}.
The other possibility of eliminating the FCNC may be to
consider a certain of alignment in the Yukawa sector~\cite{pich}, but
we do not consider this possibility  in this paper. 
In the 4HDM, we also impose the $Z_2$ symmetry to eliminate the FCNC. 
There are two types of Yukawa interactions (Type A and Type B)
as shown in Table~1, assuming that all the Higgs doublet fields receive VEVs.
If we introduce additional chiral superfields $N_{Ri}^c$ for
right-handed neutrinos which are singlet under the SM gauge symmetries,
possible number of the type of Yukawa interaction becomes doubled
under the $Z_2$ symmetry, depending on the two possible 
assignment of the $Z_2$ charge for $N_{Ri}^c$.
We define additional two types in Table 1 (Type C and Type D) which
correspond to the $Z_2$ odd $N_{Ri}^c$.
%
Under the $Z_2$ symmetry,
some of the Yukawa coupling constants are forbidden for each type of Yukawa interaction.
For example, in the MSSM-like Yukawa interaction (Type A)
$\hat{Y}_u'=\hat{Y}_d'=\hat{Y}_e'=0$ is required,
while in the lepton specific one (Type B) we have $\hat{Y}_u'=\hat{Y}_d'=\hat{Y}_e=0$.
Marshall and Sher discussed phenomenology of the Type B Yukawa interaction
in the 4HDM~\cite{Marshall}. 
Notice that the dimensionful parameters are not forbidden as long
as the discrete symmetry is softly broken.
In this paper, we assume that the FCNC is sufficiently suppressed
by a softly-broken $Z_2$ symmetry. 
However, we do not specify the type of Yukawa interaction, because 
all the essential results in this paper 
do not depend on the types of Yukawa interaction.


From the Lagrangian in Eq.~(\ref{eq:lag}), we can extract the Higgs potential of the model,
in which neutral scalar components of $H_{u,d}$ and $H_{u,d}'$ receive
the VEV. However, because $H_d$ and $H_d'$ ($H_u$ and $H_u'$)
have the same quantum numbers under the $SU(2)\times U(1)$ gauge
symmetries, there are $U(2)$ symmetries in the D-terms in the
potential. By using the $U(2)$ symmetry, we may rotate the
fields $H_d$ and $H_d'$ as well as the fields $H_u$ and $H_u'$ and take the
basis in which only one of the doublets receives the VEV while
the other does not as
\begin{align}
\left(
\begin{array}{c}
H_1\\
H_3
\end{array}
\right)
=U_-
\left(
\begin{array}{c}
H_d\\
H_d'
\end{array}
\right),\quad
\left(
\begin{array}{c}
H_2\\
H_4
\end{array}
\right)
=U_+
\left(
\begin{array}{c}
H_u\\
H_u'
\end{array}
\right),
\end{align}
where $U_-$ and $U_+$ are the 2$\times$2 unitary matrices. 
Consequently, without loss of generality we can rewrite the
Higgs potential as 
\begin{align}
V_{\text{H}}^{}=&
\begin{pmatrix}
H_1^{\dagger}&H_1^{\prime\dagger}
\end{pmatrix}
\begin{pmatrix}
(M_{1}^{2})_{11}^{}&
(M_{1}^{2})_{12}^{}\\
(M_{1}^{2})_{12}^{*}&
(M_{1}^{2})_{22}^{}\\
\end{pmatrix}
\begin{pmatrix}
H_1^{}\\ H_1^{\prime}
\end{pmatrix}
+
\begin{pmatrix}
H_2^{\dagger}&H_2^{\prime\dagger}
\end{pmatrix}
\begin{pmatrix}
(M_{2}^{2})_{11}^{}&
(M_{2}^{2})_{12}^{}\\
(M_{2}^{2})_{12}^{*}&
(M_{2}^{2})_{22}^{}\\
\end{pmatrix}
\begin{pmatrix}
H_2^{} \\ H_2^{\prime}
\end{pmatrix}
\displaybreak[0]
\nonumber\\
&
-\left(
\begin{pmatrix}
{H}_1^{}&
{H}_1^{\prime}
\end{pmatrix}
\begin{pmatrix}
(M_{3}^2)_{11}&
(M_{3}^2)_{12}\\
(M_{3}^2)_{21}&
(M_{3}^2)_{22}&
\end{pmatrix}
\cdot 
\begin{pmatrix}
{H}_2^{}\\
{H}_2^{\prime}
\end{pmatrix}
+\text{h.c.}\right)
\displaybreak[0]
\nonumber\\
&
+\frac{g^{\prime 2}+g^2}{8}\left(
H_2^{\dagger}H_2^{}
+H_2^{\prime\dagger}H_2^{\prime}
-H_1^{\dagger}H_1^{}
-H_1^{\prime\dagger}H_1^{\prime}
\right)^2
\displaybreak[0]
\nonumber\\
&
+\frac{g^2}{2}\left[
(H_1^{\dagger}H_2^{})(H_2^\dagger H_1^{})
+(H_1^{\dagger}H_2^{\prime})(H_2^{\prime\dagger}H_1^{})
+(H_1^{\prime\dagger}H_2^{})(H_2^\dagger H_1^{\prime})
+(H_1^{\prime\dagger}H_2^{\prime})(H_2^{\prime\dagger}H_1^{\prime})
\right.
\nonumber\\
&\left.
+(H_1^{\dagger}H_1^{\prime})(H_1^{\prime\dagger}H_1^{})
-(H_1^{\dagger}H_1^{})(H_1^{\prime\dagger}H_1^{\prime})
+(H_2^{\dagger}H_2^{\prime})(H_2^{\prime\dagger}H_2^{})
-(H_2^{\dagger}H_2^{})(H_2^{\prime\dagger}H_2^{\prime})
\right],  \label{V_gi}
\end{align}
where $H_1$ ($Y=-1/2$) and $H_2$ ($Y=1/2$) have VEVs, while
$H_1'$ ($Y=-1/2$) and $H_2'$ ($Y=1/2$) do not. 
In Eq.~(\ref{V_gi}), we use following the reparametrization: 
\begin{align}
&
\begin{pmatrix}
(M_{1}^{2})_{11}^{}&
(M_{1}^{2})_{12}^{}\\
(M_{1}^{2})_{12}^{*}&
(M_{1}^{2})_{22}^{}\\
\end{pmatrix}
=U_-^{}
\begin{pmatrix}
(M_{-}^{2})_{11}^{}&
(M_{-}^{2})_{12}^{}\\
(M_{-}^{2})_{12}^{*}&
(M_{-}^{2})_{22}^{}\\
\end{pmatrix}
U_-^{\dagger}
\;,\nonumber\\
&
\begin{pmatrix}
(M_{2}^{2})_{11}^{}&
(M_{2}^{2})_{12}^{}\\
(M_{2}^{2})_{12}^{*}&
(M_{2}^{2})_{22}^{}\\
\end{pmatrix}
=
U_+^{}
\begin{pmatrix}
(M_{+}^{2})_{11}^{}&
(M_{+}^{2})_{12}^{}\\
(M_{+}^{2})_{12}^{*}&
(M_{+}^{2})_{22}^{}\\
\end{pmatrix}
U_+^{\dagger}
\;,\nonumber\\
&
\begin{pmatrix}
(M_{3}^{2})_{11}^{}&
(M_{3}^{2})_{12}^{}\\
(M_{3}^{2})_{21}^{}&
(M_{3}^{2})_{22}^{}\\
\end{pmatrix}
=
-U_-^*
\begin{pmatrix}
B^{}\mu^{}&
B_{14}^{}\mu_{14}^{}\\
B_{32}^{}\mu_{32}^{}&
B_{34}^{}\mu_{34}^{}\\
\end{pmatrix}
U_+^{\dagger}
\;.
\end{align}
Throughout this paper, we restrict ourselves in the CP invariant case. 
We thus  hereafter neglect all CP violating phases in the dimentionful parameters. 


The rotated Higgs doublet fields $H_1$,
$H_1'$, $H_2$ and $H_2'$
are expressed as 
\begin{align}
 H_1=\left[
 \begin{array}{c}
    \varphi_1^{0\ast} \\
    -\varphi_1^-\\
  \end{array}
 \right],
\quad 
 H_2=\left[
 \begin{array}{c}
    \varphi_2^+ \\
    \varphi_2^0\\
  \end{array}
 \right],
\quad
 H_1^\prime=\left[
 \begin{array}{c}
    \varphi_1^{\prime 0\ast} \\
    -\varphi_1^{\prime -}\\
  \end{array}
 \right],
\quad
 H_2^\prime=\left[
 \begin{array}{c}
    \varphi_2^{\prime +} \\
    \varphi_2^{\prime 0}\\
  \end{array}
 \right],
\end{align}
where the neutral scalar fields can be parameterized as 
\begin{align}
&\varphi_1^0 = \frac{1}{\sqrt{2}}\left(v_1^{}+\phi_1^{}+i\chi_1^{}\right)\;,\quad
\varphi_2^0 = \frac{1}{\sqrt{2}}\left(v_2^{}+\phi_2^{}+i\chi_2^{}\right)\;,\quad
\nonumber\\
&\varphi_1^{\prime 0} = \frac{1}{\sqrt{2}}\left(\phi_1^{\prime}+i\chi_1^{\prime}\right)\;,\quad
\varphi_2^{\prime 0} = \frac{1}{\sqrt{2}}\left(\phi_2^{\prime}+i\chi_2^{\prime}\right)\;,\quad
\end{align}
where the VEVs of these neutral components are given by 
$\langle \varphi_1^0\rangle = v_1^{}/\sqrt{2}$,
$\langle \varphi_2^0\rangle = v_2^{}/\sqrt{2}$, 
$\langle \varphi_1^{\prime 0}\rangle= 0$ and  
$\langle \varphi_2^{\prime 0}\rangle = 0$.  
Introducing
\begin{equation}
v=(\sqrt{2}G_F^{})^{-1/2}\simeq 246\;\text{GeV}\;, 
\end{equation}
and the mixing angle $\beta$, 
we express $v_1$ and $v_2$ as 
$v_1 \equiv v \cos\beta$ and
$v_2 \equiv v \sin\beta$.  
The vacuum conditions for the Higgs potential are given by 
\begin{align}
\frac{1}{v}\left.\frac{\partial V_H^{}}{\partial \phi_1^{}}\right|_{\phi_i^{}=0}=&
c_{\beta}\left((M_{1}^2)_{11}^{}+\frac{m_Z^2}{2}c_{2\beta}\right)
-s_{\beta}(M_{3}^2)_{11}^{}=0\;,
\displaybreak[0]\nonumber\\
\frac{1}{v}\left.\frac{\partial V_H^{}}{\partial \phi_2^{}}\right|_{\phi_i^{}=0}=&
s_{\beta}\left((M_{2}^2)_{11}^{}-\frac{m_Z^2}{2}c_{2\beta}\right)
-c_{\beta}(M_{3}^2)_{11}^{}=0\;,
\displaybreak[0]\nonumber\\
\frac{1}{v}\left.\frac{\partial V_H^{}}{\partial \phi_1^{\prime}}\right|_{\phi_i^{}=0}=&
c_{\beta}(M_{1}^2)_{12}^{}
-s_{\beta}(M_{3}^2)_{21}^{}=0\;,
\displaybreak[0]\nonumber\\
\frac{1}{v}\left.\frac{\partial V_H^{}}{\partial \phi_2^{\prime}}\right|_{\phi_i^{}=0}=&
s_{\beta}(M_{2}^2)_{12}^{}
-c_{\beta}(M_{3}^2)_{12}^{}=0\;.
\end{align}
Solving this set of conditions, one can eliminate
$(M_1^2)_{11}^{}$, $(M_2^2)_{11}^{}$, $(M_1^2)_{12}^{}$, and $(M_2^2)_{12}^{}$.


After imposing the vacuum conditions,
the mass matrices $M_A^2$, $M_{H^\pm}^2$ and
$M_H^2$ for the CP-odd, charged and
CP-even scalar component states are
respectively obtained in the basis of
$(\Phi_1, \Phi_2, \Phi^{\prime}_1, \Phi^{\prime}_2)$.    
It is however more useful to work
the mass matrices of the CP-odd scalar bosons
and the charged Higgs bosons in the 
gauge eigenstate basis (the so-called Georgi basis)  as~\cite{Georgi_Base}
\begin{align}
\bar{M}_A^2=&O_0^TM_A^2O_0
=
 \begin{pmatrix}
0&0&0&0\\
0&
\frac{2(M_3^2)_{11}}{s_{2\beta}}&
\frac{(M_3^2)_{21}}{c_{\beta}}&
\frac{(M_3^2)_{12}}{s_{\beta}}\\
0&
\frac{(M_3^2)_{21}}{c_{\beta}}&
(M_1^2)_{22}^{}+\frac{m_Z^2}{2}c_{2\beta}&
(M_3^2)_{22}^{}\\
0&
\frac{(M_3^2)_{12}}{s_{\beta}}&
(M_3^2)_{22}^{}&
(M_2^2)_{22}^{}-\frac{m_Z^2}{2}c_{2\beta}
  \end{pmatrix}\;, 
\end{align}
\begin{align}
\bar{M}_{H^{\pm}}^2=&O_0^TM_{H^{\pm}}^2O_0
=\bar{M}_A^2
+m_W^2\begin{pmatrix}
0&0&0&0\\
0&1&0&0\\
0&0&-c_{2\beta}&0\\
0&0&0&c_{2\beta}
\end{pmatrix}\;,
\end{align}
with the orthogonal matrix  
\begin{equation}
O_0=\begin{pmatrix}
c_{\beta}&s_{\beta}&0&0\\
-s_{\beta}&c_{\beta}&0&0\\
0&0&1&0\\
0&0&0&1
\end{pmatrix}\;,
\end{equation}
where we used the abbreviation such
as $\sin\theta=s_\theta$ and
$\cos\theta=c_\theta$. 
In this basis the massless modes, which are NG bosons to be
absorbed by the longitudinal modes of the weak gauge bosons,
are separated in the mass matrices. 
The basis taken here is essentially the same as that discussed in Ref.~\cite{Gupta}.
It is also useful to rotate the mass matrix for the CP-even scalar bosons as 
\begin{align}
\bar{M}_H^2=&O_0M_H^2O_0^T\nonumber\\
=&
\begin{pmatrix}
m_Z^2c_{2\beta}^2&-m_Z^2s_{2\beta}c_{2\beta}&0&0\\
-m_Z^2s_{2\beta}c_{2\beta}&
m_Z^2s_{2\beta}^2+\frac{2(M_3^2)_{11}}{s_{2\beta}}&
\frac{-(M_3^2)_{21}}{c_{\beta}}&
\frac{(M_3^2)_{12}}{s_{\beta}}\\
0&
\frac{-(M_3^2)_{21}}{c_{\beta}}&
(M_1^2)_{22}^{}+\frac{m_Z^2}{2}c_{2\beta}&
-(M_3^2)_{22}\\
0&
\frac{(M_3^2)_{12}}{s_{\beta}}&
-(M_3^2)_{22}&
(M_2^2)_{22}^{}-\frac{m_Z^2}{2}c_{2\beta}
\end{pmatrix}\;.
\end{align}


\subsection{Definition of the large mass limit}

 The soft-breaking mass parameters $(M_3^2)_{ij}$
 come from the B-terms in Eq.~(\ref{soft-term}).
When we consider the case with $(M_3^2)_{12}=(M_3^2)_{21}=0$, the
mass matrices $\bar{M}_A^2$, $\bar{M}_{H^\pm}^2$ and  $\bar{M}_H^2$
are block diagonal. The upper $2\times 2$ submatrix in each mass matrix
corresponds to that in the MSSM; i.e.,
$2(M_3^2)_{11}/s_{2\beta} \to m_A^2$, 
and the other $2\times 2$ submatrix corresponds to that for
the extra two scalar bosons.
They are separated completely in this case.
The model effectively becomes the MSSM
in the large mass limit of the extra scalar bosons.    
On the other hand, in the case with nonzero $(M_3^2)_{12}$ or $(M_3^2)_{21}$,
the masses of the light scalars $h$, $H$ and $H^\pm$  are modified 
from the MSSM predictions by the mixing via the B-terms
between $\Phi_1$ and $\Phi_1'$ or between $\Phi_2$ and $\Phi_2'$.
These effects are expected to be nonvanishing when
$(M_3^2)_{12}$ or $(M_3^2)_{21}$
grows with taking a similar value to
the 3-3 or 4-4 component in the mass matrices such as
$(M_1^2)_{22}$ or $(M_2^2)_{22}$. 
We here discuss these effects in
details in the following. 

We start from discussing the CP-odd scalar mass matrix. 
In order to examine the decoupling property of the mass matrices, we
further rotate $\bar{M}_A^2$ as  
\begin{equation}
\hat{M}_A^2=
\begin{pmatrix}
1&0&0&0\\
0&1&0&0\\
0&0&c_{\bar{\theta}}&s_{\bar{\theta}}\\
0&0&-s_{\bar{\theta}}&c_{\bar{\theta}}\\
\end{pmatrix}
\bar{M}_A^2
\begin{pmatrix}
1&0&0&0\\
0&1&0&0\\
0&0&c_{\bar{\theta}}&-s_{\bar{\theta}}\\
0&0&s_{\bar{\theta}}&c_{\bar{\theta}}\\
\end{pmatrix}
=
\begin{pmatrix}
0&0&0&0\\
0&\frac{2(M_3^2)_{11}^{}}{s_{2\beta}}&k^{\prime}M^2&kM^2\\
0&k^{\prime}M^2&M^2&0\\
0&kM^2&0&rM^2
\end{pmatrix}\;, 
\end{equation} 
where 
\begin{equation}
\tan2\bar{\theta}=\frac{2(M_3^2)_{22}}{(M_1^2)_{22}-(M_2^2)_{22}+m_Z^2c_{2\beta}}\;, 
\end{equation}
and $M^2$, $k$,
$k'$ and $r$ are defined
such that 
\begin{align}
&(M_3^2)_{21}=\kappa_{21}M^2\;,\quad
(M_3^2)_{12}=\kappa_{12}M^2\;,\quad\nonumber\\
&
(M_1^2)_{22}c_{\bar{\theta}}^2+
(M_2^2)_{22}s_{\bar{\theta}}^2
 +(M_3^2)_{22} s_{2\bar{\theta}}
+\frac{m_Z^2}{2}c_{2\beta}c_{2\bar{\theta}}=M^2\;,\quad\nonumber\\
&(M_1^2)_{22}s_{\bar{\theta}}^2+
(M_2^2)_{22}c_{\bar{\theta}}^2
-(M_3^2)_{22} s_{2\bar{\theta}}
-\frac{m_Z^2}{2}c_{2\beta}c_{2\bar{\theta}}=rM^2\;, 
\end{align}
and 
\begin{align}
k^{\prime}=\frac{c_{\bar{\theta}}}{c_{\beta}}\kappa_{21}+\frac{s_{\bar{\theta}}}{s_{\beta}}\kappa_{12}\;,\quad
k=-\frac{s_{\bar{\theta}}}{c_{\beta}}\kappa_{21}+\frac{c_{\bar{\theta}}}{s_{\beta}}\kappa_{12}\;.
\end{align}
These parameters are relevant to the extra doublets, then 
the decoupling limit is taken as  $M^2\to \infty$.
Here we assume that $m_A^2\ll M^2$ and we treat $m_A^2/M^2$ as an 
expansion parameter.
One of the eigenvalues of $\bar{M}_A^2$ should be $m_A^2$,
the mass of the lightest CP-odd
Higgs boson $A$, which should coincide
with $2(M_3^2)_{11}/s_{2\beta}$ in
the limit of $M \to \infty$ if $k=k'=0$. 
In generic cases, after diagonalizing the mass matrix,  
$(M_3^2)_{11}$ is expressed in terms of $m_A^2$ as 
\begin{equation}
\frac{2(M_3^2)_{11}}{s_{2\beta}}=
m_A^2\left\{ \frac{k^2+(1+k^{\prime 2})r^2}{r^2}+\mathcal{O}\left(\frac{m_A^2}{M^2}\right)\right\}
+M^2\frac{k^2+rk^{\prime 2}}{r}
\;.
\end{equation} 
The mass eigenvalues for heavier
states $A_1$ and $A_2$ are 
\begin{equation}
m_{A_1}^2\simeq a_1M^2\;,\quad
m_{A_2}^2\simeq a_2M^2\;,
\end{equation}
where $a_1$ and $a_2$ are given by 
\begin{align}
a_1=&\frac{k^2+rk^{\prime 2}+r(1+r)
-\sqrt{\left\{k^2+rk^{\prime 2}-r(r-1)\right\}^2+4k^2r(r-1)}}{2r}\;,\nonumber\\
a_2=&\frac{k^2+rk^{\prime 2}+r(1+r)
+\sqrt{\left\{k^2+rk^{\prime 2}-r(r-1)\right\}^2+4k^2r(r-1)}}{2r}\;.
\end{align}
We note that $a_1 \to 1$ and $a_2
\to r$ for $k=k' \to 0$.

\begin{figure}[t]
\begin{center}
\includegraphics[width=70mm]{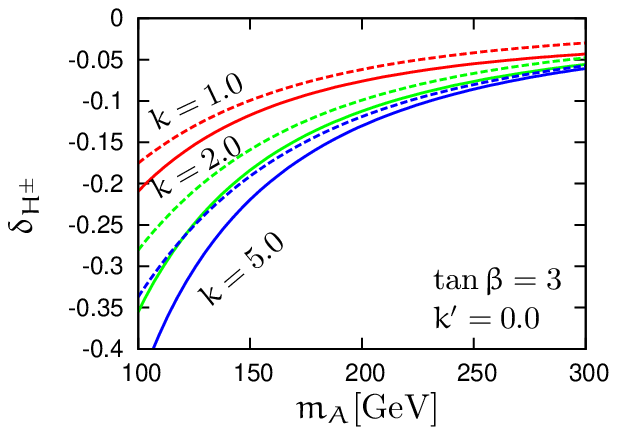}\hspace{3mm}
\includegraphics[width=70mm]{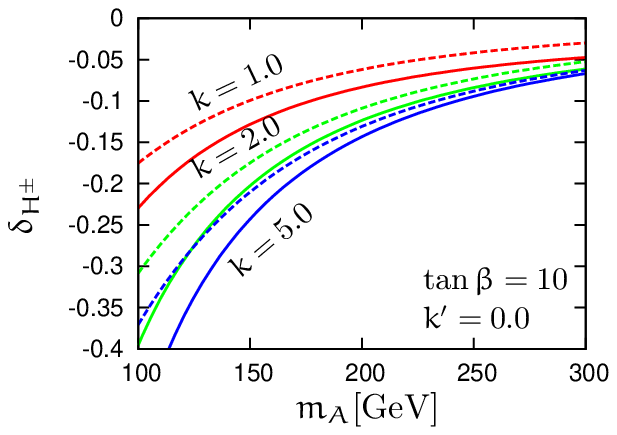}\\\vspace{6mm}
\includegraphics[width=70mm]{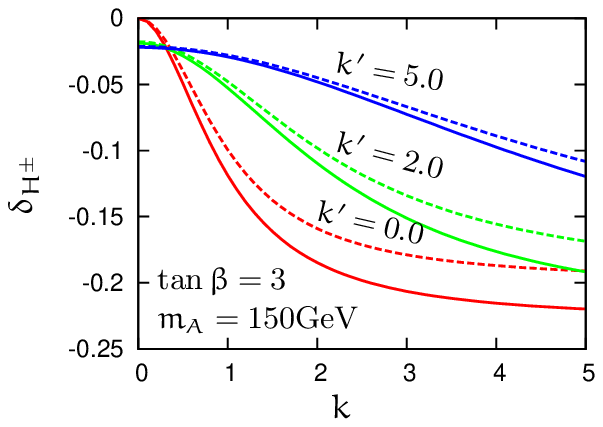}\hspace{3mm}
\includegraphics[width=70mm]{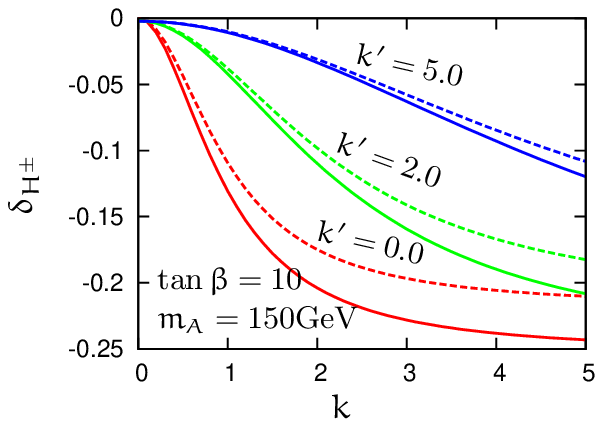}
\end{center}
\caption{The deviation $\delta_{H^\pm}$ defined in Eq.~(\ref{eq:delta})
 due to the quasi-nondecoupling effect of the B-term mixing
 parameterized by $k$ and $k'$ in the 4HDM.
 We here take $M=500$ GeV, $r=1$ and $\bar{\theta}=0$.
 The SUSY breaking scale for the MSSM particles is taken to be 1 TeV,
 and the trilinear soft-breaking parameters $A_t$ and $A_b$
 as well as the $\mu$ parameter are taken to be zero.
The upper figures: $\delta_{H^\pm}$ as a function of $m_A$ for $\tan\beta=3$
 (left) and $\tan\beta=10$ (right) for $k=1.0$, $2.0$ and $5.0$ with fixed $k'(=0.0)$. 
The lower figures: $\delta_{H^\pm}$ as a function of $k$ for $\tan\beta=3$
 (left) and $\tan\beta=10$ (right) for $k'=0.0$, $2.0$ and $5.0$
 with the fixed $m_A$ (= 150 GeV).
In all figures, the solid curves
 are the results from the full numerical calculation, while the dotted
 curves are those by using the approximated formula in Eq.~(\ref{delmch_approx})~\cite{aksy_4hdm}. 
 }
\label{fig:del_ch}
\end{figure}

\begin{figure}[t]
\begin{center}
\includegraphics[width=70mm]{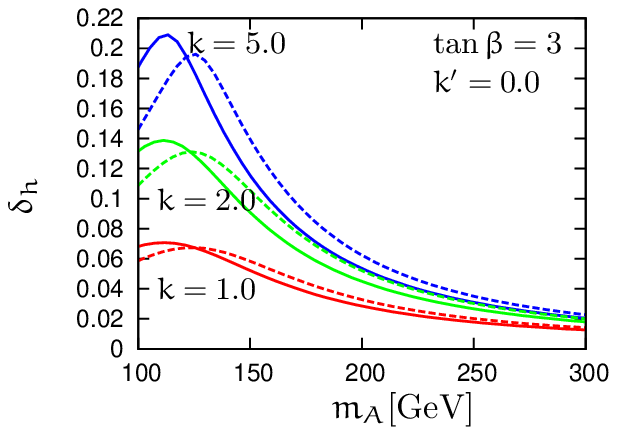}\hspace{3mm}
\includegraphics[width=70mm]{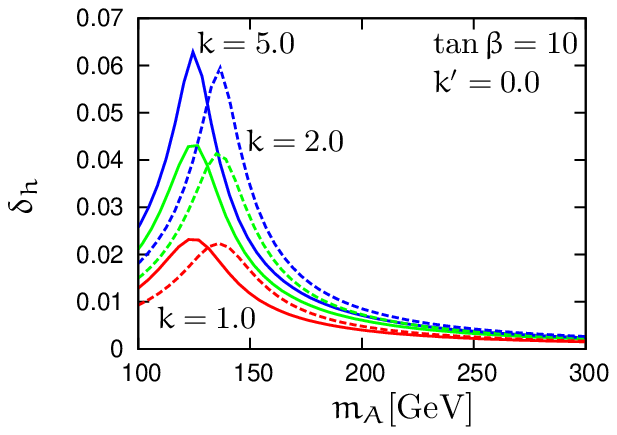}\\\vspace{6mm}
\includegraphics[width=70mm]{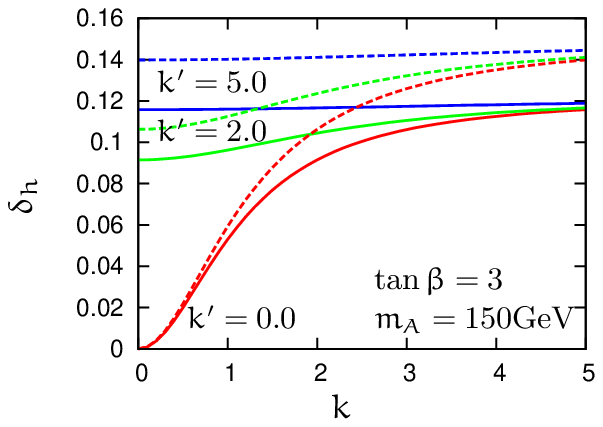}\hspace{3mm}
\includegraphics[width=70mm]{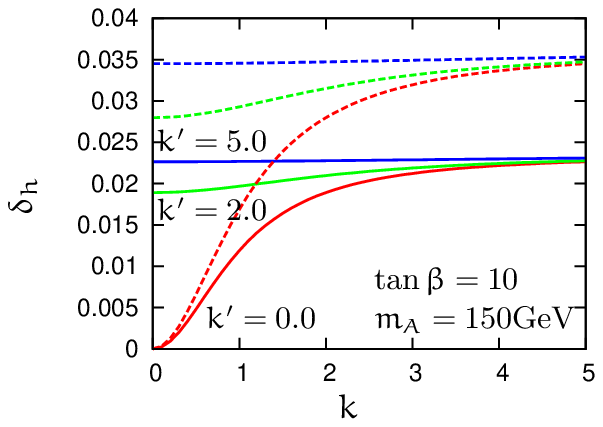}
\end{center}
\caption{The deviation $\delta_h$ in
 $m_h=m_h^{\text{MSSM}}(1+\delta_h)$ 
 due to the quasi-nondecoupling
 effect of the B-term mixing parameterized by $k$ and $k'$
  in the 4HDM, where  
 $m_h^{\text{MSSM}}$ is the renormalized mass of $h$.
  We here take $M=500$ GeV, $r=1$ and $\bar{\theta}=0$.
 The SUSY soft-breaking scale for the MSSM particles
 is taken to be 1TeV (solid curves) and 2 TeV (dotted curves),
 and the trilinear soft-breaking parameters $A_t$ and $A_b$ as well as 
 the $\mu$ parameter are taken to be zero.
The upper figures: $\delta_h$ as a function of $m_A$ for $\tan\beta=3$
 (left) and $\tan\beta=10$ (right) for $k=1.0$, $2.0$ and $5.0$ with fixed $k'(=0.0)$. 
The lower figures: $\delta_h$ as a function of $k$ for $\tan\beta=3$
 (left) and $\tan\beta=10$ (right) for $k'=0.0$, $2.0$ and $5.0$
 with the fixed $m_A$ (= 150 GeV)~\cite{aksy_4hdm}. }
\label{fig:del_h}
\end{figure}

For the charged Higgs mass matrix, via
the similar procedure to the case of
the CP-odd Higgs bosons, we obtain
the deviation in $m_{H^\pm}$,  
the mass eigenvalue for the lightest
charged scalar $H^\pm$, from
the MSSM prediction as 
\begin{equation}
m_{H^{\pm}}=\sqrt{(m_{H^{\pm}}^2)^{\rm MSSM}}(1+\delta_{H^\pm}), \label{eq:delta}
\end{equation}
where 
\begin{equation}
\delta_{H^\pm} = 
-\frac{1}{2}\frac{m_W^2}{m_A^2+m_W^2}
\frac{k^2+k^{\prime 2} r^2
-c_{2\beta}\left\{
	(k^2-k^{\prime 2}r^2)c_{2\bar{\theta}}
	+2kk^{\prime}rs_{2\bar{\theta}}
	\right\}}
{\left\{k^2+(1+k^{\prime 2})r^2\right\}}
+\mathcal{O}\left(\frac{m_A^2}{M^2}\right)\;, \label{delmch_approx}
\end{equation}
and $(m_{H^{\pm}}^2)^{\rm MSSM}$ is 
the prediction in the MSSM renormalized in the on-shell
scheme~\cite{dabelstein}, which is simply given by~\cite{Ref:KY,Ref:CKY}  
\begin{equation}
(m_{H^{\pm}}^2)^{\rm MSSM}=m_A^2+m_W^2
 - \Pi_{H^+H^-}^{\rm 1PI}(m_A^2+m_W^2)
 + \Pi_{AA}^{\rm 1PI}(m_A^2)
 + \Pi_{WW}^{\rm 1PI}(m_W^2), 
\end{equation}
where $\Pi_{\phi\phi}^{\rm 1PI}(p^2)$
represent the one particle irreducible diagram contributions
to the two point function of the field $\phi$ at the squared
momentum $p^2$.
Masses of the heavier charged scalar
bosons $H_1^\pm$ and $H_2^\pm$ are obtained as 
\begin{equation}
m_{H_1^{\pm}}^2\simeq a_1M^2\;,\quad
m_{H_2^{\pm}}^2\simeq a_2M^2\;.
\end{equation}

\begin{figure}\label{fig:del_H}
\begin{center}
\begin{tabular}{cc}
\includegraphics{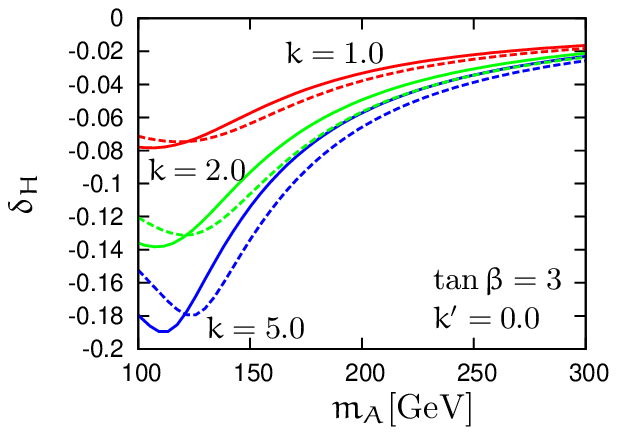}&
\includegraphics{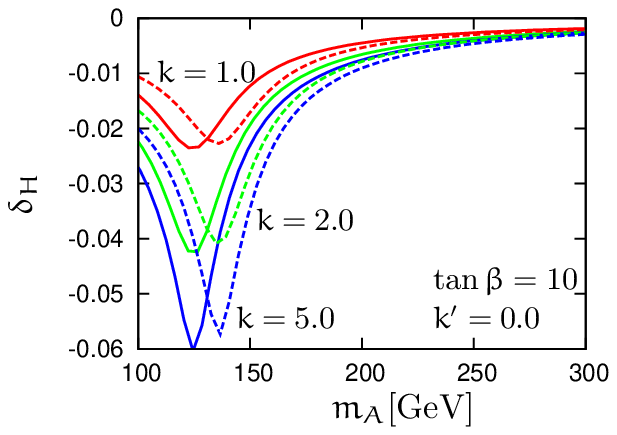}\\
\includegraphics{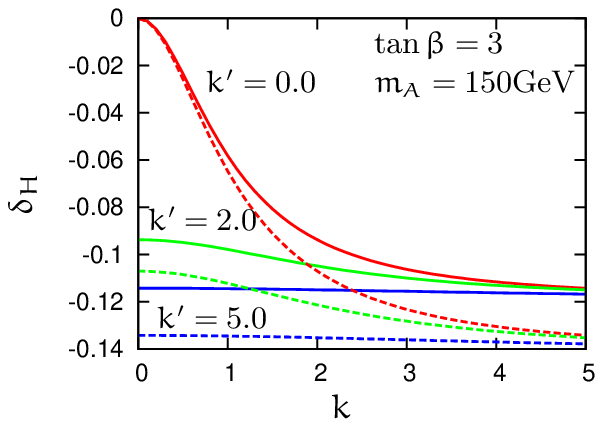}&
\includegraphics{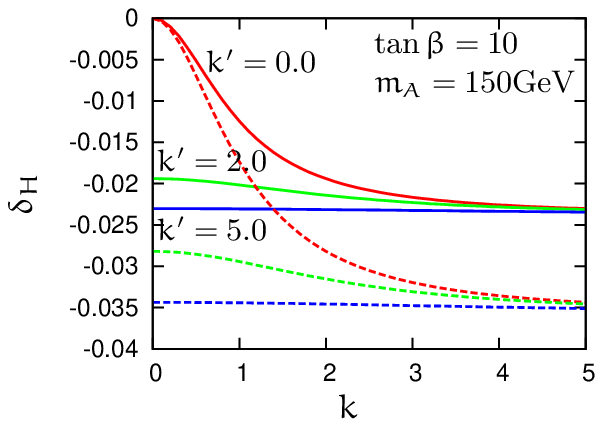}\\
\end{tabular}
\end{center}
 \caption{The deviation $\delta_H$ in $m_H=m_H^{\rm MSSM}(1+\delta_H)$ of
 the renormalized mass of the second lightest CP-even Higgs boson $H$
 due to the quasi-nondecoupling effect of the B-term mixing
 parameterized by $k$ and $k'$ in the 4HDM.
   We here take $M=500$ GeV, $r=1$ and $\bar{\theta}=0$.
The SUSY soft-breaking scale of the MSSM particles is taken to be 1 TeV
 (solid curves) and 2 TeV (dotted curves), and the trilinear
 soft-breaking parameters $A_t$ and $A_b$
 as well as the $\mu$ parameter are taken to be zero.
The upper figures: $\delta_H$ as a function of $m_A$ for $\tan\beta=3$
 (left) and $\tan\beta=10$ (right) for $k=1.0$, $2.0$ and $5.0$ with fixed $k'(=0.0)$. 
The lower figures: $\delta_H$ as a function of $k$ for $\tan\beta=3$
 (left) and $\tan\beta=10$ (right) for $k'=0.0$, $2.0$ and $5.0$
 with the fixed $m_A$ (= 150 GeV)~\cite{aksy_4hdm}. } 
\end{figure}

In Fig.~1, we show the numerical results for the deviation $\delta_{H^\pm}$ defined 
in Eq.~(\ref{eq:delta}) due to the quasi-nondecoupling
effect of the B-term mixing parameterized by $k$ and $k'$ in our model.
The solid curves in the figures 
represent the results from the full numerical calculation, while the dotted
 curves are those by using the approximated formula in
 Eq.~(\ref{delmch_approx}).
The deviation $\delta_{H^\pm}$ turns out to be negative, and amounts
to $-20$ \% for a relatively small value of $m_A$. The magnitude of
 the deviation is smaller for a larger value of $m_A$, but still a few times
$-1$ \% even for $m_A=300$ GeV.
On the other hand, the deviation is not very sensitive to $\tan\beta$.
We note that the results
are insensitive to the details of the MSSM parameters such as the soft-breaking mass
parameters, the $\mu$ parameter and the trilinear $A_{t,b}$ parameters.
In fact, when $\mu$ and $A_{t,b}$ are varied in the phenomenologically
acceptable regions, the radiative corrections vary at most
from $-2$ \% to $+2$ \%. We have confirmed that our results on 
the one-loop correction in the MSSM agree with those given in
Ref.~\cite{Ref:KY,Ref:CKY}. 
The mass of $H^\pm$ can be determined with the accuracy of
a few percent via the decays of $H^\pm \to \tau \nu $ and $H^\pm \to tb$ at
the LHC~\cite{Assamagan:2002in}, and with the statistical 
error of less than 1 \% via $e^+e^- \to H^+H^-$ at the ILC~\cite{Djouadi2}.
The mass of $A$ can also be determined with the resolution about 2\%
via the decays $A \to \mu^+\mu^-$ at the LHC, while at the ILC
it can be measured with the precision 0.2~\%
via $e^+e^- \to HA$~\cite{Djouadi2}.
Therefore, the quasi-nondecoupling effect on $m_{H^\pm}$ can be
extracted when both $m_{H^\pm}$ and $m_{A}$ are
measured at future collider experiments.
The prediction on $m_{H^\pm}$ (not on $\delta_{H^\pm}$) in the 4HDM is
 shown in Fig.~5 with the comparison of the result in the MSSM.

Next, the CP-even scalar mass matrix $\bar{M}_H^2$ can also be
diagonalized. We first define $\hat{M}_H^2$ by  
\begin{align}
\hat{M}_H^2=
\begin{pmatrix}
1&0&0&0\\
0&1&0&0\\
0&0&c_{\bar{\theta}}&-s_{\bar{\theta}}\\
0&0&s_{\bar{\theta}}&c_{\bar{\theta}}\\
\end{pmatrix}
\bar{M}_H^2
\begin{pmatrix}
1&0&0&0\\
0&1&0&0\\
0&0&c_{\bar{\theta}}&s_{\bar{\theta}}\\
0&0&-s_{\bar{\theta}}&c_{\bar{\theta}}\\
\end{pmatrix}\; , 
\end{align}
and  according to the usual mathematical procedure
$\hat{M}_H^2$ can be block-diagonalized 
 by rotating the basis with an appropriate orthogonal matrix $O_{MH}$ as    
\begin{align}
O_{MH}^T\hat{M}_H^2O_{MH}=&
\begin{pmatrix}
m_Z^2c_{2\beta}^2&-m_Z^2c_{2\beta}s_{2\beta}R&0&0\\
-m_Z^2c_{2\beta}s_{2\beta}R&m_A^2+m_Z^2s_{2\beta}^2R^2&
0&0\\
0&0&a_1M^2&0\\
0&0&0&a_2M^2
\end{pmatrix}
+\mathcal{O}\left(\frac{m_A^2}{M^2}\right)\;, \label{cpemm}
\end{align}
where $R$ is defined as 
\begin{equation}
R=\frac{1}{\sqrt{1+\frac{k^2}{r^2}+k^{\prime 2}}}\;.
\end{equation}
The upper $2 \times 2$ submatrix coincides to the mass matrix of
the two light scalar bosons $H$
and $h$ of the MSSM when $M\to \infty$ if $k=k'=0$. 
For the case with nonzero $k$ and $k'$, after diagonalizing the
$2\times 2$ submatrix by the mixing
angle $\alpha_{\rm eff}$ the mass eigenvalues of the CP-even
Higgs bosons are obtained as
\begin{align}
m_h^2=&
\frac{1}{2}\left[ m_A^2+m_Z^2\left(c_{2\beta}^2+R^2s_{2\beta}^2\right)
-\sqrt{\left\{(m_A^2-m_Z^2\left(1-(1-R^2)s_{2\beta}^2\right)\right\}^2
+4m_A^2m_Z^2s_{2\beta}^2R^2} \right. \nonumber\\
&\left. +m_A^2\mathcal{O}\left(\frac{m_A^2}{M^2}\right)+ \Delta_h^{\rm loop}
\right] 
\;,\nonumber\\
m_H^2=&
\frac{1}{2} \left[ m_A^2+m_Z^2\left(c_{2\beta}^2+R^2s_{2\beta}^2\right)
 +\sqrt{ \left\{m_A^2-m_Z^2\left(1-(1-R^2)s_{2\beta}^2\right)\right\}^2
 +4m_A^2m_Z^2s_{2\beta}^2R^2} \right. \nonumber\\
&\left. +m_A^2\mathcal{O}\left(\frac{m_A^2}{M^2}\right)
+ \Delta_H^{\rm loop}
\right] \;,  
\label{mh}
\end{align}
where $\Delta_h^{\rm loop}$ and $\Delta_H^{\rm loop}$ represent
the one-loop corrections in the MSSM.
The masses of heavier states
$H_1'$ and $H_2'$ are given by  
$m_{H_1'}^2 \simeq a_1M^2\left\{ 1+\mathcal{O}(m_A^2/M^2) \right\}$
and
$m_{H_2'}^2 \simeq a_2M^2\left\{ 1+\mathcal{O}(m_A^2/M^2) \right\}$.
The mixing angle $\alpha_{\rm eff}$
satisfies the relation  
\begin{equation}
\tan(\beta-\alpha_{\text{eff}})
=\frac{m_h^2-m_A^2-m_Z^2s_{2\beta}^2R^2}{m_Z^2c_{2\beta}s_{2\beta}R}
\left\{ 1+\mathcal{O}\left(\frac{m_A^2}{M^2}\right)
 + \Delta_{\tan(\beta-\alpha)}^{\rm loop}\right\}\;,  \label{tanbaeff} 
\end{equation}
where $\Delta_{\tan(\beta-\alpha)}^{\rm loop}$ is the one-loop
correction in the MSSM.
Notice that $m_h$ and $m_H$ given in Eq.~(\ref{mh}) and
$\tan(\beta-\alpha_{\rm eff})$ in Eq.~(\ref{tanbaeff}) do not depend on
the sign of $k$ and $k'$. 
\begin{figure}
\begin{center}
\begin{tabular}{cc}
\includegraphics{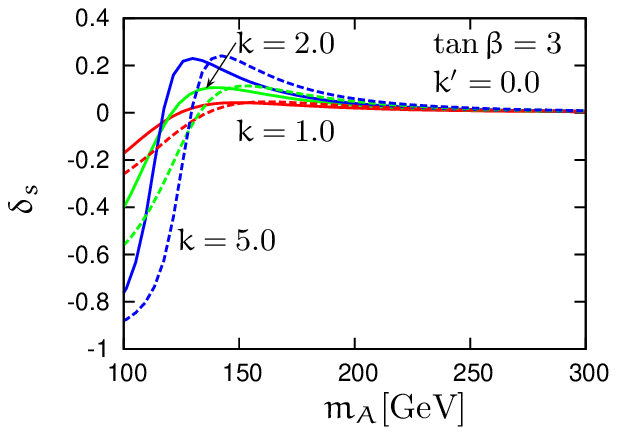}&
\includegraphics{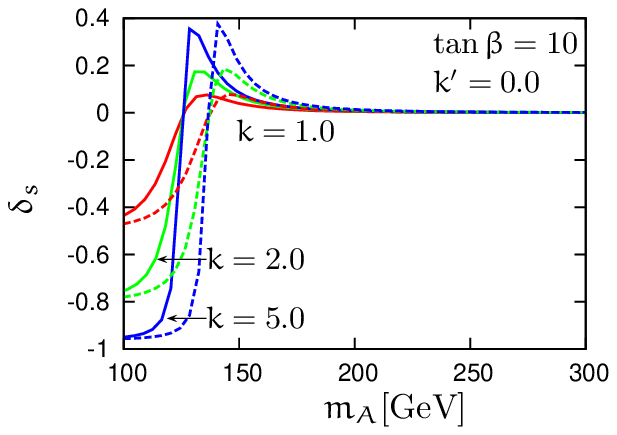}\\
\includegraphics{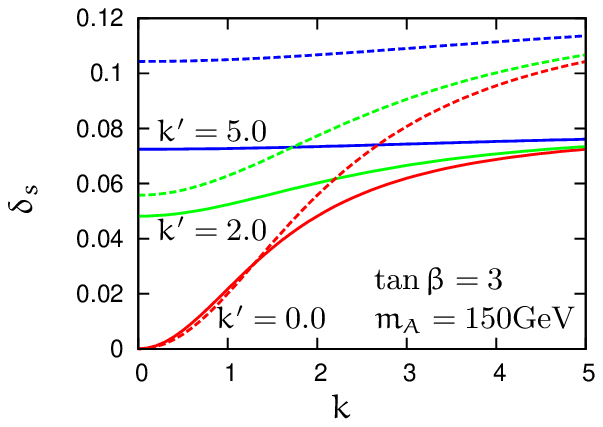}&
\includegraphics{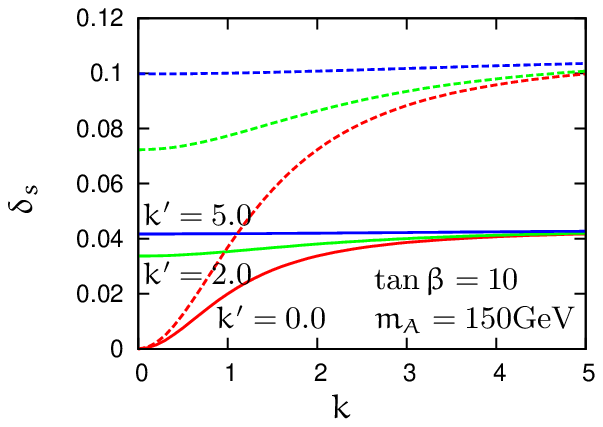}
\end{tabular}
\end{center}\label{fig:del_s}
 \caption{The deviation $\delta_s$ in Eq.~(\ref{eq:delta_s}). 
 due to the quasi-nondecoupling effect of extra doublet fields
 via the B-term mixing parameterized by $k$ and $k'$ in the 4HDM.
  We here take $M=500$ GeV, $r=1$ and $\bar{\theta}=0$.
 The SUSY soft-breaking scale of the MSSM particles is taken to be 1 TeV
 (solid curves) and 2 TeV (dotted curves), and the trilinear
 soft-breaking parameters $A_t$ and $A_b$ as well as
 the $\mu$ parameter are taken to be zero.
The upper figures: $\delta_s$ as a function of $m_A$ for $\tan\beta=3$
 (left) and $\tan\beta=10$ (right) for $k=1.0$, $2.0$ and $5.0$ with fixed $k'(=0.0)$. 
The lower figures: $\delta_s$ as a function of $k$ for $\tan\beta=3$
 (left) and $\tan\beta=10$ (right) for $k'=0.0$, $2.0$ and $5.0$
 with the fixed $m_A$ (= 150 GeV)~\cite{aksy_4hdm}. }
\end{figure}
We note that the effective mixing angle
$\alpha_{\rm eff}$ contains information of the B-term quasi-nondecoupling
effects between $\Phi_1$ and $\Phi_1'$ or
between $\Phi_2$ and $\Phi_2'$ by $k$ and $k'$, but for
$m_A^2 \ll M^2$ the tree level
formula with the angle $\alpha$ in
the MSSM can still hold by replacing
$\alpha$ by $\alpha_{\rm eff}$ in a good approximation. For
example, the coupling constants of
the two light CP-even Higgs bosons with the
weak gauge bosons $V$ ($V=W^\pm$ and $Z^0$) in the case 
with nonzero $k$ and $k'$ are given
by 
\begin{align}
\Gamma_{VVh}^{}=-\frac{m_V^2}{v} \left\{ c_{\beta}(O_H)_{12}
 +s_{\beta}(O_H)_{22}\right\}
&=\frac{m_V^2}{v}
 \sin(\beta-\alpha_{\text{eff}})
 \left( 1+  \Delta_{hVV}^{\rm loop}\right)\;, \\
\Gamma_{VVH}^{}=-\frac{m_V^2}{v}
 \left\{ c_{\beta}(O_H)_{11}+s_{\beta}(O_H)_{21} \right\}
&=\frac{m_V^2}{v} \cos(\beta-\alpha_{\text{eff}})
\left( 1+ \Delta_{HVV}^{\rm loop}\right)\;,
\end{align}
where 
$\Delta_{hVV}^{\rm loop}$ and $\Delta_{HVV}^{\rm loop}$ represent 
radiative corrections in the MSSM.
Finally, in general, magnitudes of $k$ and $k'$ are not
necessarily smaller than 1, still it
is helpful to deduce the approximate
formulae assuming that they are
small;  
\begin{align}
m_h^2=&
(m_h^2)^{\rm MSSM} 
\left(
1+
\frac{m_Z^2s_{2\beta}^2(\frac{k^2}{r^2}+k^{\prime 2})}{\sqrt{(m_A^2-m_Z^2)^2+4m_Z^2m_A^2s_{2\beta}^2}}
+\mathcal{O}(k^4,k^{\prime 4}, k^2k^{\prime 2})
+\mathcal{O}\left(\frac{m_A^2}{M^2}\right)
\right)
\;,\\
m_H^2=&
(m_H^2)^{\rm MSSM} 
\left(
1-
\frac{m_Z^2s_{2\beta}^2(\frac{k^2}{r^2}+k^{\prime 2})}{\sqrt{(m_A^2-m_Z^2)^2+4m_Z^2m_A^2s_{2\beta}^2}}
+\mathcal{O}(k^4,k^{\prime 4}, k^2k^{\prime 2})
+\mathcal{O}\left(\frac{m_A^2}{M^2}\right)
\right) 
\;, \\
\tan(\beta-\alpha_{\text{eff}})
=& [\tan(\beta-\alpha)]^{\rm MSSM} 
\left(
1+
\frac{(m_A^2-2m_h^2-m_Z^2s_{2\beta}^2)(\frac{k^2}{r^2}+k^{\prime 2})}
{2(m_A^2-m_h^2+m_Z^2s_{2\beta}^2)} 
+\mathcal{O}(k^4,k^{\prime 4}, k^2k^{\prime 2}) \right. \nonumber\\
&\left. \hspace{33mm}+\mathcal{O}\left(\frac{m_A^2}{M^2}\right)
\right)\;, 
\end{align}
where $(m_h^2)^{\rm MSSM}$, $(m_H^2)^{\rm MSSM}$ and
$[\tan(\beta-\alpha)]^{\rm MSSM}$ are the corresponding parameters
evaluated at the one-loop level assuming the MSSM.
In this paper, we have used the approximate one-loop formula given in
Ref.~\cite{dabelstein} in evaluating
 $(m_h^2)^{\rm MSSM}$, $(m_H^2)^{\rm MSSM}$ and
$[\tan(\beta-\alpha)]^{\rm MSSM}$.

In Fig.~2, we show the numerical results for
the deviation $\delta_h$ in $m_h=m_h^{\text{MSSM}}(1+\delta_h)$,
where $m_h^{\text{MSSM}}$ is the one-loop corrected mass
of $h$, due to the quasi-nondecoupling effect of the B-term mixing
parameterized by $k$ and $k'$ in the 4HDM.
The SUSY soft-breaking scale of the MSSM is taken to be 1 TeV and 2 TeV,
and the trilinear soft-breaking parameters $A_t$ and $A_b$ and the $\mu$
are taken to be zero.
It is found that $\delta_h$ is always positive.
This is understood from Eq.~(\ref{cpemm}).
The parameter $R$ is unity for $k=k'=0$, and is smaller
for larger values of $k$ and $k'$. A smaller value of $R$ ($R < 1$)  
reduces the value of the off-diagonal term in Eq.~(\ref{cpemm}), which
makes the mixing between the first two CP-even states weaker. Consequently,
the mass difference between $h$ and $H$ becomes smaller than the case
with the MSSM case with the same value of $m_A$ and $\tan\beta$.
The deviation takes its maximal 
values (6-20 \% for $\tan\beta =3$ and $2$-$5$ \% for $\tan\beta=10$)
around the crossing point ($m_A \sim 130$-$150$ GeV)
where the role of $h$ and $H$ are exchanged. 
For larger values of $m_A$ the magnitude of $\delta_h$ is smaller, but
it can be still 3-6 \% (about 1 \%) at $m_A=200$ GeV for $\tan\beta = 3$ $(10)$.
These values are substantial and can be tested by the precise
measurement of $m_h$ at the LHC (the ILC), where $m_h$ is expected to be
determined with about 0.1\% \cite{acc-mh-LHC} accuracy at the LHC, while 
at the ILC it is expected to be measured within 
less than 70 MeV \cite{acc-mh-ILC}) error.
The prediction on $m_{h}$ (not on $\delta_{h}$) in the 4HDM is 
shown in Fig.~5 with the comparison of the result in the MSSM.
We can see that in the 4HDM $m_{h}$ reaches its maximal value at a smaller $m_{A}$
than that in the MSSM, although the predicted upper bound on the $m_{h}$
is the same in both models.

In Fig.~3, we show 
the deviation $\delta_H$ in $m_H=m_H^{\text{MSSM}}(1+\delta_H)$,
where $m_H^{\text{MSSM}}$ is the one-loop corrected mass
of $H$, due to the quasi-nondecoupling effect of the B-term mixing
parameterized by $k$ and $k'$ in the 4HDM.
The SUSY parameters are taken as in the same way as Fig.~2.
As we discussed, the mixing of the light two CP-even states is
weakened by non-zero values of $k$ and $k'$, so that $m_H$
is smaller than the prediction in the MSSM. Therefore, 
$\delta_H$ is negative as we expect. The behavior of $\delta_H$
as a function of $m_A$ and $\tan\beta$ are similar to the case of
$\delta_h$ except for the sign. 
The magnitude is maximal around the crossing point ($m_A =130$-$150$
GeV), and amounts to $-18$ \% ($-5$ \%) for $\tan\beta=3$ (10).
At the LHC and the ILC, the mass of $H$ can be determined with the
similar precision to  that of $A$ mentioned in the previous paragraph.
The prediction on $m_{H}$ (not on $\delta_{H}$) in the 4HDM is 
shown in Fig.~5 with the comparison with the result in the MSSM.

In Fig.~4, we show the numerical results for
the deviation $\delta_s$ defined in Eq.~(\ref{eq:delta}),   
in which  $[\sin^2(\beta-\alpha)]^{\text{MSSM}}$ is the one-loop corrected
mixing factor $\sin^2(\beta-\alpha)$ evaluated in the MSSM. 
$\delta_s$ is the net deviation from the MSSM prediction 
due to the quasi-nondecoupling effect of the B-term mixing
parameterized by $k$ and $k'$ in the 4HDM.
The SUSY soft-breaking scale of the MSSM is taken to be 1 TeV and 2 TeV,
and the trilinear soft-breaking parameters $A_t$ and $A_b$ and the $\mu$
parameter are taken to be zero. 
In the figures, we can see that $\delta_s$ is negative when  
$m_A$ is smaller than the crossing point at $m_A \sim 130$-$150$ GeV,
while it is positive for larger $m_A$.  
The deviation can be as large as ${\mathcal O}(10)$ \% ($\tan\beta=3$)
and ${\mathcal O}(20)$ \% ($\tan\beta=10$) just above the crossing
point; i.e., at around 
$m_A \sim 140$-$150$ GeV.  
It is rapidly close to unity for larger values of $m_A$.
Notice that for larger soft-SUSY-breaking scale, a larger $\delta_s$
is possible. 
The prediction on $\sin^2(\beta-\alpha_{\rm eff})$ (not on
$\delta_{s}$)
in the 4HDM is shown in Fig.~5 with the comparison with the result in the MSSM.

\begin{figure}[t]
\begin{center}
\begin{tabular}{cc}
\includegraphics{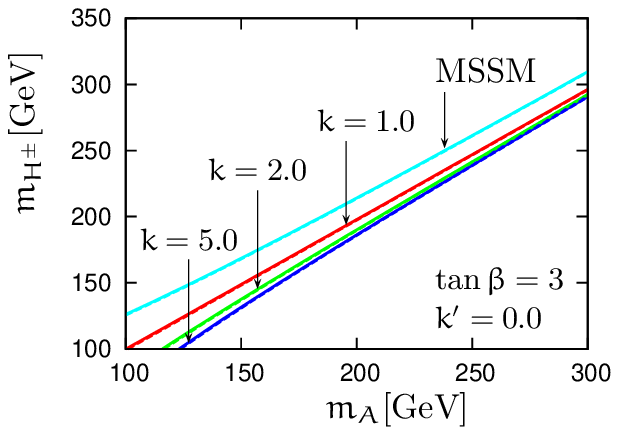}&
\includegraphics{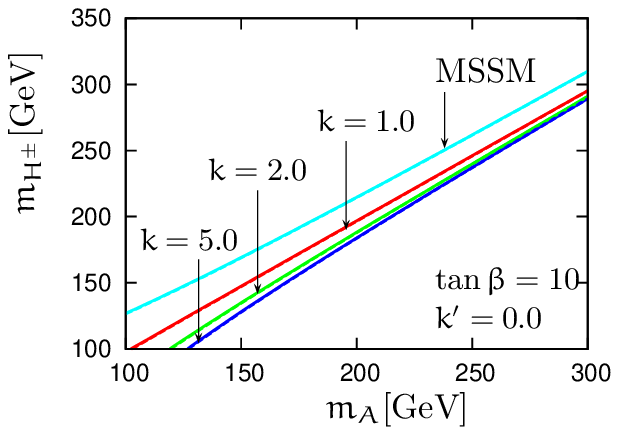}\\
\includegraphics{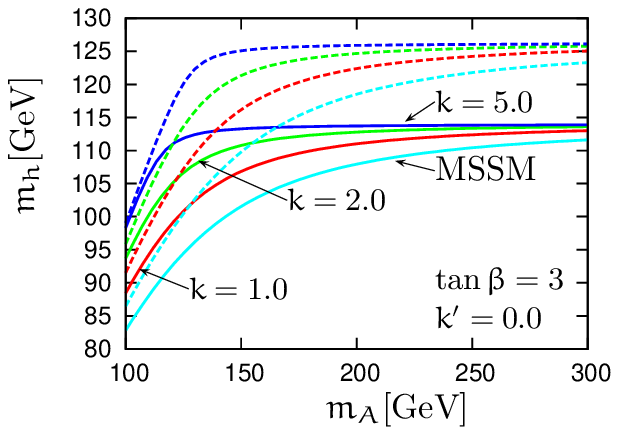}&
\includegraphics{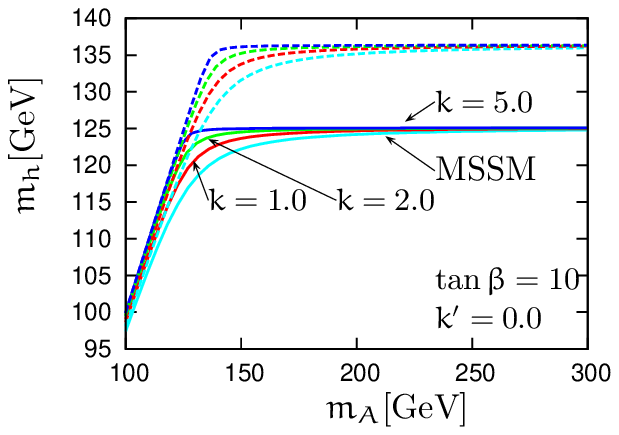}\\
\includegraphics{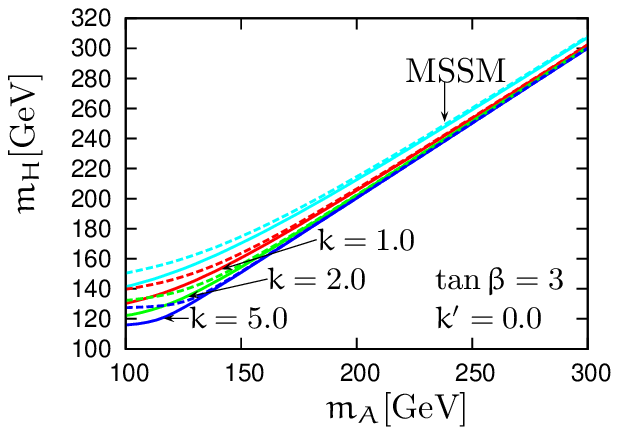}&
\includegraphics{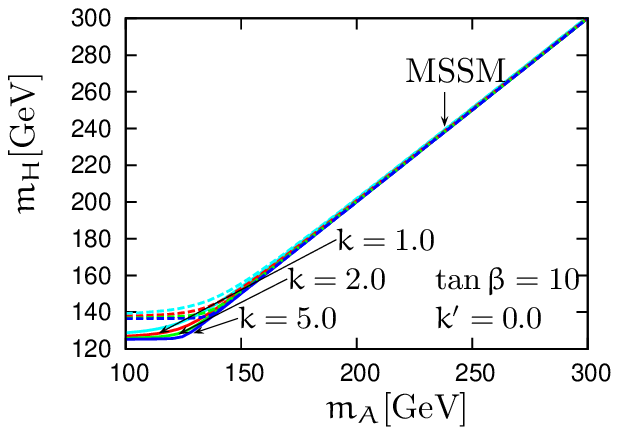}\\
\includegraphics{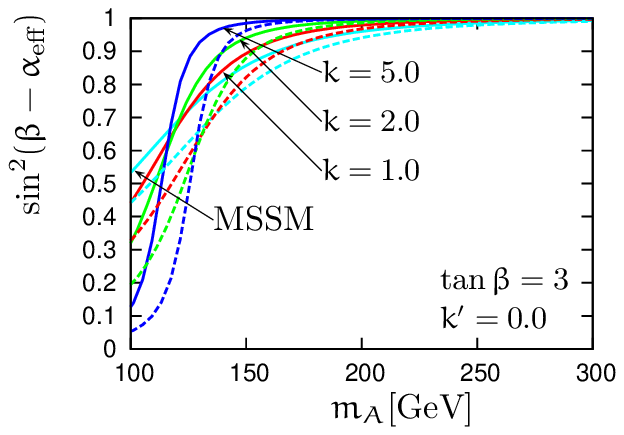}&
\includegraphics{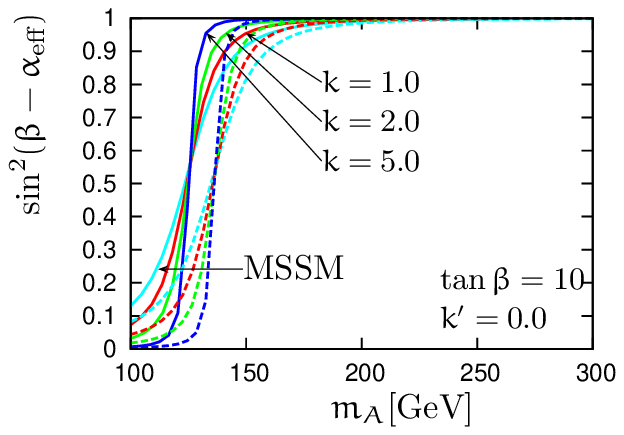}
\end{tabular}
\end{center}
\caption{The values of $m_{H^\pm}$,$m_{h}$,$m_{H}$ and
 $\sin^2(\beta-\alpha_{\rm eff})$ in the 4HDM and the MSSM 
 as a function of $m_A$ for $k=1.0$, $2.0$ and $5.0$.
 The soft-SUSY-breaking scale of the MSSM
 is set to be 1 TeV (solid curves) and 2 TeV (dotted curves).
 The trilinear soft-breaking parameters $A_t$ and $A_b$ as well as
 the $\mu$ parameter are taken to be zero.
 The other parameters are takes as 
 $M=500$ GeV, $r=1$, $\bar{\theta}=0$ and $k'=0$.
 Figures in the left column are for $\tan\beta=3$ and those in the right
 are for $\tan\beta=10$~\cite{aksy_4hdm}.}
\end{figure}
\clearpage

\newpage
\part{Models which can explain the phenomena beyond the SM at the TeV scale}
\chapter{Review of radiative seesaw models}
The neutrino oscillation has been established by experiments. 
This suggest that neutrinos have tiny masses as compared to the electroweak scale. 
This is clear evidence for physics beyond the SM. 
If neutrinos are the Majorana fermion, then 
the tiny masses of left-handed
neutrinos are generated from the dimension-five effective operators
\begin{align}
\mathcal{L}=\frac{c_{ij}}{2\Lambda}\overline{\nu_L^{ci}}\nu_L^j\phi^0\phi^0,\label{dim6}
\end{align}
where $\Lambda$ represents a mass scale for new physics, 
$c_{ij}$ are dimensionless coefficients, and $\phi^0$ is the Higgs
boson. 
After electroweak symmetry breaking due to the VEV of the Higgs boson $\langle \phi^0\rangle=v$,
the mass matrix $M_\nu^{ij}$
for left-handed neutrinos are generated as 
\begin{align}
M_\nu^{ij} = \frac{c_{ij}}{2\Lambda}v^2. 
\end{align}
Since we have already known $v\simeq 246$ GeV via the muon decay experiments, 
the coefficient $c_{ij}/\Lambda$ has to be of order $10^{-14}$ GeV$^{-1}$ to reproduce 
the observed tiny neutrino masses which are around $\mathcal{O}(0.1)$ eV. 
The seesaw mechanism is the simple scenario to obtain the operator in Eq.~(\ref{dim6}) 
in the low energy effective theory from the tree level diagram, 
where right-handed neutrinos~\cite{typeI_seesaw} are added to the SM. 
At the same time, the baryon asymmetry can be generated 
via the lepton number violation in the CP violating decays of 
right-handed Majorana neutrinos. 
In addition, in a SUSY extension of the model with such a heavy Majorana neutrinos, 
the lightest SUSY partner particle can be a candidate of the dark matter. 
In this model, if we assume the magnitude of the coefficient $c_{ij}$ 
to be $\mathcal{O}(1)$, then the mass of right-handed neutrinos 
has to be of $\mathcal{O}(10^{14})$ GeV to obtain the scale of left-handed neutrino masses. 
Although this scenario is simple, 
it requires another
hierarchy between the mass of right-handed neutrinos and the electroweak scale. 
In addition, physics at such a large mass scale is difficult to be tested at collider experiments.

As the other way to obtain the dimension six operator,  
radiative seesaw models~\cite{zee,zee-2loop,babu-2loop,Krauss:2002px,Ma:2006km,aks_prl} have been proposed, 
where neutrino masses are generated at the loop level. 
In this class of models, the coefficient $c_{ij}$ is naturally suppressed by the loop factor, so that 
masses of new particles in these models can 
be as low as the TeV scale. 
Therefore, they are expected to be directly testable at current and future 
collider experiments. 
One of the characteristic features of these models is an extended Higgs sector. 
Another feature is the Majorana nature, either introducing lepton 
number violating couplings or introducing right-handed neutrinos. 

The original model for radiative neutrino mass generation was first proposed  
by A.~Zee\cite{zee}, where neutrino masses are generated at the 
one-loop level by adding an extra  $\text{SU}(2)_{\text{L}}^{}$ doublet scalar field 
and a charged singlet scalar field with lepton number violating 
couplings to the SM particle entries. 
Phenomenology of this model has been studied in Ref.~\cite{zee-ph1,kkloy}. 
However, it turned out that it was difficult to reproduce the current 
data for neutrino oscillation in this original model\cite{zee-nu}. 
Some extensions have been discussed in Ref.~\cite{zee-variation}. 

The simplest successful model today may be the one proposed by 
A.~Zee\cite{zee-2loop} and K.~S.~Babu\cite{babu-2loop}, 
in which two kinds of $\text{SU}(2)_{\text{L}}^{}$ singlet scalar fields
are introduced; i.e., a singly charged scalar boson and a 
doubly charged one. 
These fields carry lepton number of two unit.  
In this model, which we refer to as the Zee-Babu model,  
the neutrino masses are generated at the two-loop level. 
Phenomenology of this model has been discussed in 
Refs.~\cite{macesanu,aristizabal,nebot,ohlsson,ak_majorana}. 
Although the Zee-Babu model can explain neutrino data, 
this model has not a dark matter candidate, because 
there are no new neutral particles in the particle contetent. 

Apart from the Zee-Babu model, there is also another type of radiative seesaw
models~\cite{Krauss:2002px,Ma:2006km,aks_prl}, where 
TeV-scale right-handed neutrinos are introduced with the odd charge
under the exact discrete $Z_2^{}$ symmetry. 
In these models, the $Z_2^{}$ symmetry protects the tree level Dirac Yukawa coupling among 
the lepton doublet, Higgs doublet and right-handed neutrino. 
At the same time, this $Z_2$ symmetry also protect 
the decay of the lightest
$Z_2^{}$ odd particle, which can be a candidate of dark matter. 
This is an advantage of this class of models\cite{knt_dm,ma_dm,aks_dm}.
In addition to the explanation of neutrino masses and dark matter, 
baryon asymmetry of the Universe may also be able to explain in the model proposed in Ref.~\cite{aks_prl}. 
In Ref.~\cite{aks-prd}, detailed phenomenological study of this model has been analyzed. 

These models give an explanation for tiny masses of neutrino, give a candidate for dark matter 
and/or explain baryon asymmetry of the Universe. 
However, the origin of masses of neutrinos and dark matter comes from different mass scales. 
In Refs.~\cite{Shimomura,Nabeshima1}, it has been proposed that 
both of these masses can be explained by the spontaneous breakdown of the gauged $U(1)_{B-L}$ symmetry.

In the above discussion, we consider the case where neutrinos are the Majorana fermion, while 
we can also consider the case where those are the Dirac fermion. 
A model which can generate the Dirac masses of neutrinos is invariant under the $U(1)$ lepton number such as the SM. 
Up to the present, phenomena suggesting lepton number violation have not been confirmed yet by experiments such as 
neutrino less double beta decay. 
Therefore, it is valuable to investigate a possibility that kind of models. 
In Refs.~\cite{Dirac_nu1,Dirac_nu2,Dirac_nu3,Dirac_nu4,Dirac_nu5,Dirac_nu6,Dirac_nu7,Nabeshima2}, 
radiative generation of masses for Dirac neutrinos has been proposed.

In Part~II, we first discuss the theoretical and experimental bounds of the model 
proposed in Ref.~\cite{aks_prl}. 
Second, we discuss the supersymmetric extension of the Zee-Babu model, and its phenomenology at the LHC. 
Finally, we discuss the model with an isospin doublet with $Y=3/2$ field $\Phi_{3/2}$. 
We study the phenomenology of the simple model which includes $\Phi_{3/2}$ at the LHC, 
and then we show the model with $\Phi_{3/2}$ can apply to the radiative seesaw model, where 
neutrino masses can generate at one-loop level as well as the dark matter candidate exists. 
\clearpage
\chapter{Three-loop neutrino mass model}

In this chapter, we discuss a TeV-scale model has been proposed by M.~Aoki, 
S.~Kanemura and O.~Seto~\cite{aks_prl},   
in which 1) tiny neutrino masses are generated without excessive fine tuning 
at the three-loop level by the dynamics of 
an extended Higgs sector and right-handed neutrinos under an unbroken $Z_2$ parity, 2)    
the $Z_2$ parity also guaranties the stability of a dark matter candidate which is a $Z_2$ odd 
scalar boson, and 3) the strongly first order phase transition for 
successful electroweak baryogenesis can be realized by the nondecoupling effect 
in the Higgs sector.  
Phenomenology of this model has been discussed in Ref.~\cite{aks-prd}, and the 
related collider physics~\cite{typeX,stefano,ak_majorana} and dark matter properties~\cite{aks_dm} 
have also been studied.  
In these papers, phenomenologically allowed parameter regions have been mainly discussed. 

We investigate the theoretical constraint on the parameter regions in this model~\cite{aks_prl}  
from the requirement of vacuum stability and perturbativity up to a given cutoff scale $\Lambda$ of the model~\cite{RGE_SM}. 
In the present model, there is no mechanism for cancellation of the quadratic divergences which appear in the 
renormalization calculation for the Higgs boson mass, so that a huge fine tuning is required 
if $\Lambda$ is much higher than the electroweak scale.  
To avoid such an unnatural situation, we need to consider $\Lambda$ to be at most ${\cal O}(10)$ TeV, 
above which the model would be replaced by a more fundamental theory~\cite{Murayama}. 
Hence, we have to study the theoretical consistency of the model up to such values for $\Lambda$. 
In particular, some of the neutrino Yukawa coupling constants are of order one in magnitude, as 
the scale of tiny neutrino masses is generated by loop dynamics so that we do not need fine tuning 
for the size of the coupling constants.  In addition, some of the coupling constants in the Higgs potential 
are of order one to realize the nondecoupling one-loop effect for strongly first order phase transition. 
Although the parameters discussed in the previous works satisfy the bound from tree-level unitarity~\cite{PU_thdm, PU_thdm2}, 
it is non-trivial that the model can be consistent at the quantum level 
with the theoretical requirements up to $\Lambda \sim 10$ TeV. 
Theoretical bounds from vacuum stability and perturbativity have been used to constrain parameters 
in extended Higgs sectors such as the THDM~\cite{VS_thdm,VS_thdm2} and the Zee model~\cite{Kasai}.
Here we apply the similar analysis to the model.
We prepare a full set of the renormalization group equations (RGEs) for dimensionless coupling constants  
in the model at the one-loop level, and analyze the behavior of running coupling constants. 
 
We also calculate the phenomenological constraint from lepton flavor violation (LFV) in the model. 
In the previous analysis~\cite{aks_prl,aks-prd} only 
the constraint from $\mu\to e\gamma$ data has been taken into account. 
Here, we also analyze the one-loop induced $\mu\to eee$ process, whose current experimental data~\cite{meee} 
turn out to give a stronger bound on the parameter 
space than those of $\mu \to e \gamma$~\cite{MEGA}.

\section{Model}
In this model, two Higgs doublets ($\Phi_1$ and $\Phi_2$) with hypercharge 
$Y=1/2$, charged scalar singlets ($S^\pm$), a real scalar singlet 
($\eta$) and right-handed neutrinos ($N_R^\alpha$ with $\alpha =1,2$) 
are introduced.
We impose two kinds of discrete symmetries; i.e., 
$Z_2$ and $\tilde{Z}_2$ to the model. 
The former, which is exact, is introduced in order to  
forbid the tree-level Dirac neutrino mass term 
and at the same time to guarantee the stability of dark matter. 
The latter one, which is softly broken, is introduced to avoid 
the tree-level flavor changing neutral current~\cite{GW}. 
Under the $\tilde{Z}_2$ symmetry there are four types of Yukawa 
interactions~\cite{Barger, Grossman}. In our model~\cite{aks_prl}, so-called the 
type-X Yukawa interaction~\cite{typeX,typeX2} is favored since the charged Higgs boson 
from the two doublets can be taken to be as light as around 100 GeV 
without contradicting the $b\to s \gamma$ data. 
Such a light charged Higgs boson is important to reproduce 
the correct magnitude of neutrino masses. 
The particle properties under the discrete symmetries are shown 
in Table~\ref{z2},
where $Q_L^i$, $u_R^i$, $d_R^i$, $L_L^i$ and $e_R^i$ are the $i$-th 
generation of the left-handed quark doublet, the 
right-handed up-type quark singlet, the 
left-handed lepton doublet and the right-handed charged lepton singlet, 
respectively.
\begin{table}[t]
\begin{center}
\begin{tabular}{|c||c|c|c|}\hline
&$Q_L^i\hspace{1mm}u_R^{i}\hspace{1mm}d_R^{i}\hspace{1mm}L_L^i \hspace{1mm}e_R^{i}$&$\Phi_1\hspace{1mm}\Phi_2$&$S^{\pm}\hspace{1mm}\eta \hspace{1mm}N_R^\alpha$\\\hline\hline
$Z_2$(exact)&$+\hspace{2mm}+\hspace{2mm}+\hspace{2mm}+\hspace{2mm}+$&$+\hspace{2mm}+$&$-\hspace{2mm}-\hspace{2mm}-$\\\hline
$\tilde{Z}_2$(softly broken)&$+\hspace{2mm}-\hspace{2mm}-\hspace{2mm}+\hspace{2mm}+$&$+\hspace{2mm}-$&$+\hspace{2mm}-\hspace{2mm}+$\\\hline
\end{tabular}
\caption{Particle properties under the discrete symmetries~\cite{aks_prl}.}
\label{z2}
\end{center}
\end{table}

The type-X Yukawa interaction is given by
\begin{align}
\mathcal{L}_{\text{yukawa}}^{\text{Type-X}}&=-\sum_{i,j}
\left[\left(\bar{Q}_L^i Y^{d}_{ij} \Phi_2 d_R^j \right) 
+ \left(\bar{Q}_L^i Y^{u}_{ij} \Phi^c_2u_R^j \right)
+ \left(\bar{L}_L^{i} Y^{e}_{ij} \Phi_1 e_R^j \right)\right]+{\rm h.c.},   \label{yukawa1}
\end{align}
where Yukawa coupling matrix for leptons is diagonal, 
$Y^e_{ij}={\rm diag}(y_{e^1}, y_{e^2}, y_{e^3})$.
The mass term and the Yukawa interaction for $N_R^\alpha$ are written as 
\begin{align}
\mathcal{L}_{N_R}&=\sum_{\alpha=1}^2
\frac{1}{2}m_{N_R^\alpha}
\overline{(N_R^{\alpha})^c}N_R^\alpha-\sum_{i=1}^{3}\sum_{\alpha=1}^2
\left[h_i^\alpha\overline{(e_R^i)^c}N_R^\alpha S^++{\rm h.c.}\right].   \label{yukawa2}
\end{align}
The scalar potential is given by
\begin{align}
V=&+\mu_1^2|\Phi_1|^2+\mu_2^2|\Phi_2|^2
-\Big[\mu_3^2 \Phi_1^\dagger\Phi_2+ {\rm h.c.} \Big]\notag\\
& +\frac{1}{2}\lambda_1|\Phi_1|^4+\frac{1}{2}
\lambda_2|\Phi_2|^4+\lambda_3|\Phi_1|^2|\Phi_2|^2
+\lambda_4|\Phi_1^\dagger\Phi_2|^2
+\frac{1}{2}\Big[\lambda_5 (\Phi_1^\dagger\Phi_2)^2 
           +  {\rm h.c.} \Big] \notag\\
&+\mu_S^2|S^-|^2+\rho_1|S^-|^2|\Phi_1|^2+\rho_2|S^-|^2|\Phi_2|^2+\frac{1}{4}\lambda_S|S^-|^4\notag\\
&+\frac{1}{2}\mu_\eta^2\eta^2+\frac{1}{2}\sigma_1\eta^2|\Phi_1|^2+\frac{1}{2}\sigma_2\eta^2|\Phi_2|^2+\frac{1}{4!}\lambda_\eta\eta^4\notag\\
&+\sum_{a,b=1}^2\Big[\kappa\epsilon_{ab}(\Phi_a^c)^\dagger\Phi_b S^-\eta+{\rm h.c.}\Big]+\frac{1}{2}\xi|S^-|^2\eta^2,  \label{pot}
\end{align}
where $\epsilon_{ab}$ are anti-symmetric matrices with $\epsilon_{12}=1$. 
The parameters $\mu_3^2$, $\lambda_5$, and $\kappa$ are complex numbers. 
Two of their phases can be absorbed by rephasing the fields, and the 
rest is a physical one. In this paper, we neglect this CP-violating 
phase for simplicity. 
The Higgs doublets are parameterized as
\begin{align}
\Phi_i=\left(\begin{array}{c}
w_i^+\\
\frac{1}{\sqrt{2}}(h_i+v_i+iz_i)
\end{array}\right),
\end{align}
where $v_i$ are vacuum expectation values (VEVs) of the Higgs fields,  
and these are constrained by $v (=\sqrt{v_1^2+v_2^2}) \simeq 246$ GeV. 
The ratio of the two VEVs is defined by $\tan\beta =v_2/v_1$. 
The physical scalar states $h$, $H$, $A$ and $H^\pm$ 
in the $Z_2$ even sector can be obtained mixing angles $\alpha$ and $\beta$,
\begin{align}
\left(\begin{array}{c}
w_1^\pm\\
w_2^\pm
\end{array}\right)
=R(\beta)\left(\begin{array}{c}
w^\pm\\
H^\pm
\end{array}\right),\hspace{3mm}\left(\begin{array}{c}
z_1\\
z_2
\end{array}\right)
=R(\beta)\left(\begin{array}{c}
z\\
A
\end{array}\right),\hspace{3mm}
\left(\begin{array}{c}
h_1\\
h_2
\end{array}\right)
=R(\alpha)\left(\begin{array}{c}
H\\
h
\end{array}\right),
\end{align}
where $w^\pm$ and $z$ are the NG bosons absorbed 
by the longitudinal weak gauge bosons, and 
the rotation matrix with the angle $\theta$ is given  by 
\begin{align}
R(\theta)=
\left(\begin{array}{cc}
\cos\theta & -\sin\theta\\
\sin\theta & \cos\theta
\end{array}\right).
\end{align}
The mass formulae of physical scalar states are given by
\begin{align}
m_A^2&=M^2-v^2\lambda_5, \label{mA}\\
m_{H^\pm}^2&=M^2-\frac{v^2}{2}(\lambda_4+\lambda_5),\label{mHpm}\\
m_H^2&=\cos^2(\alpha-\beta) M_{11}^2+2\sin(\alpha-\beta)\cos(\alpha-\beta) M_{12}^2+\sin^2(\alpha-\beta) M_{22}^2,\\
m_h^2&=\sin^2(\alpha-\beta) M_{11}^2-2\sin(\alpha-\beta)\cos(\alpha-\beta) M_{12}^2+\cos^2(\alpha-\beta) M_{22}^2,\\
m_{S}^2&=\mu_S^2+\frac{v^2}{2}\rho_1\cos^2\beta+\frac{v^2}{2}\rho_2\sin^2\beta, \label{mS}\\
m_\eta^2&=\mu_\eta^2+\frac{v^2}{2}\sigma_1\cos^2\beta+\frac{v^2}{2}\sigma_2\sin^2\beta,
\end{align}
where $M (=\mu_3/\sqrt{\sin\beta\cos\beta})$ is the soft breaking scale 
for the $\tilde{Z}_2$ symmetry, and 
\begin{align}
M_{11}^2&=v^2(\lambda_1\cos^4\beta+\lambda_2\sin^4\beta)+\frac{v^2}{2}\lambda\sin^22\beta,\\
M_{22}^2&=M^2+v^2\sin^2\beta\cos^2\beta(\lambda_1+\lambda_2-2\lambda),\\
M_{12}^2&=v^2\sin\beta\cos\beta(-\lambda_1\cos^2\beta+\lambda_2\sin^2\beta+\lambda\cos2\beta),
\end{align} 
with $\lambda=\lambda_3+\lambda_4+\lambda_5$. 
Notice that $M$, $\mu_S$ and $\mu_\eta$ are free mass parameters 
irrelevant to the electroweak symmetry breaking.

\section{Neutrino mass and mixing}

\begin{figure}[t]
\begin{center}
\includegraphics[width=120mm]{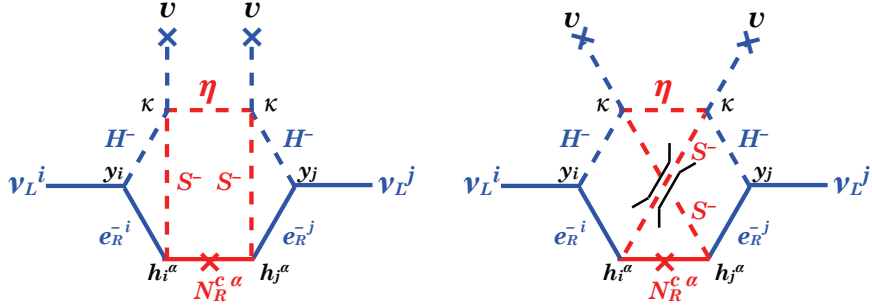}
\caption{The Feynman diagrams for generating tiny neutrino masses~\cite{aks_prl}. }
\end{center}
\label{diag-numass}
\end{figure}

The neutrino mass matrix $M^\nu_{ij}$ is generated by the three-loop diagrams in
FIG.~\ref{diag-numass}.
The absence of lower order loop contributions is guaranteed by the 
$Z_2$ symmetry.  
The resulting mass matrix is calculated as 
\begin{eqnarray}
M^\nu_{ij} &=& \sum_{\alpha=1}^2 
      4 \kappa^2 \tan^2\beta 
  (y_{\ell_i}^{\rm SM} h_i^\alpha) (y_{\ell_j}^{\rm SM} h_j^\alpha)
   F(m_{H^\pm}^{},m_{S}^{},m_{N_R^{\alpha}}, m_\eta), \label{eq:mij}
\end{eqnarray}
where the loop integral function $F$ is given by
\begin{eqnarray}
&&F(m_{H^\pm}^{},m_S^{},m_{N}, m_\eta) =
   \left(\frac{1}{16\pi^2}\right)^3 \frac{(-m_N^{})}{m_N^2-m_\eta^2}
 \frac{v^2}{m_{H^\pm}^4}\nonumber\\
 && \times \int_0^{\infty} x dx
  \left\{B_1(-x,m_{H^\pm}^{},m_S^{})-B_1(-x,0,m_S^{})\right\}^2
  \left(\frac{m_N^2}{x+m_N^2}-\frac{m_\eta^2}{x+m_\eta^2}\right),
\end{eqnarray}
with $y_{\ell_i}^{\rm SM}=\sqrt{2}m_{\ell_i}/v$, where 
$\ell_1$, $\ell_2$ and $\ell_3$ correspond to 
$e$, $\mu$ and $\tau$, respectively. 
The function $B_1$ is the tensor coefficient
in the formalism by Passarino-Veltman 
for one-loop integrals~\cite{Ref:PV}.
In the following discussion, we take $m_{N_R^1}=m_{N_R^2}\equiv m_{N_R}$, 
for simplicity. 
Numerically, the magnitude of the function $F$ is of order $10^{4}$ eV
in the wide range of parameter regions of our interest.  
Since $y_{\ell_i}^{\rm SM} < 10^{-2}$, the correct 
scale of neutrino masses can be naturally obtained from 
the three-loop diagrams.

The generated mass matrix 
$M^\nu_{ij}$ in Eq.~(\ref{eq:mij}) of neutrinos
can be related to the neutrino oscillation data by
\begin{eqnarray}
 M^\nu_{ij} = U_{is} (M^\nu_{\rm diag})_{st} (U^T)_{tj}, 
\end{eqnarray}
where $M^\nu_{\rm diag}$ $=$ ${\rm diag}(m_1, m_2, m_3)$.
For the case of the normal hierarchy we identify the mass eigenvalues
as  
$m_1=0$, $m_2=\sqrt{\Delta m_{\rm solar}^2}$, $m_3=\sqrt{\Delta m_{\rm
atm}^2}$, while  for inverted hierarchy $m_1=\sqrt{\Delta m_{\rm atm}^2}$,
$m_2=\sqrt{\Delta m_{\rm atm}^2 + \Delta m_{\rm solar}^2}$ and $m_3=0$
are taken.
The Maki-Nakagawa-Sakata matrix $U_{is}$~\cite{mns} is parameterized as 
\begin{eqnarray}
  U=\left[\begin{array}{ccc}
     1&0&0\\
     0&c_{23}^{}&s_{23}^{}\\
     0&-s_{23}^{}&c_{23}^{}\\
           \end{array}
     \right]
  \left[\begin{array}{ccc}
     c_{13}^{}&0&s_{13}^{}e^{i\delta}\\
     0& 1 &0\\
     -s_{13}^{}e^{-i\delta}&0&c_{13}^{}\\
           \end{array}
     \right]
  \left[\begin{array}{ccc}
     c_{12}^{}&s_{12}^{}&0\\
     -s_{12}^{}&c_{12}^{}&0\\
         0&0&1
          \end{array}
     \right]
    \left[\begin{array}{ccc}
     1&0&0\\
     0&e^{i\tilde{\alpha}}&0\\
         0&0&e^{i\tilde{\beta}}
          \end{array}
     \right],
\end{eqnarray}
where $s_{ij}$ and $c_{ij}$ represent $\sin\theta_{ij}$ and
$\cos\theta_{ij}$, respectively, with $\theta_{ij}$ to be the neutrino mixing angle 
between the $i$th and $j$th generations, and $\delta$ is the Dirac phase while
$\tilde{\alpha}$ and $\tilde{\beta}$ are Majorana phases.
For simplicity, we neglect the effects of these CP violating 
phases in the following 
analysis.
Current neutrino oscillation data give the following 
values~\cite{PDG};  
\begin{eqnarray}
\Delta m^2_{\rm solar} \simeq 7.59\times 10^{-5} \,{\rm eV}^2 \,,~~
|\Delta m^2_{\rm atm}| \simeq 2.43\times 10^{-3} \,{\rm eV}^2\,, \\
\sin^2\theta_{12} \simeq 0.32 \,,~~~~
\sin^2\theta_{23} \simeq 0.5 \,,~~~~
\sin^2\theta_{13} < 0.04\,.~~~~~~
\label{obs_para}
\end{eqnarray}

In the next section, we discuss parameter regions in which 
both neutrino data and the LFV data are satisfied.

\section{Lepton flavor violation}
  
\begin{figure}[t]
\begin{center}
\includegraphics[width=120mm]{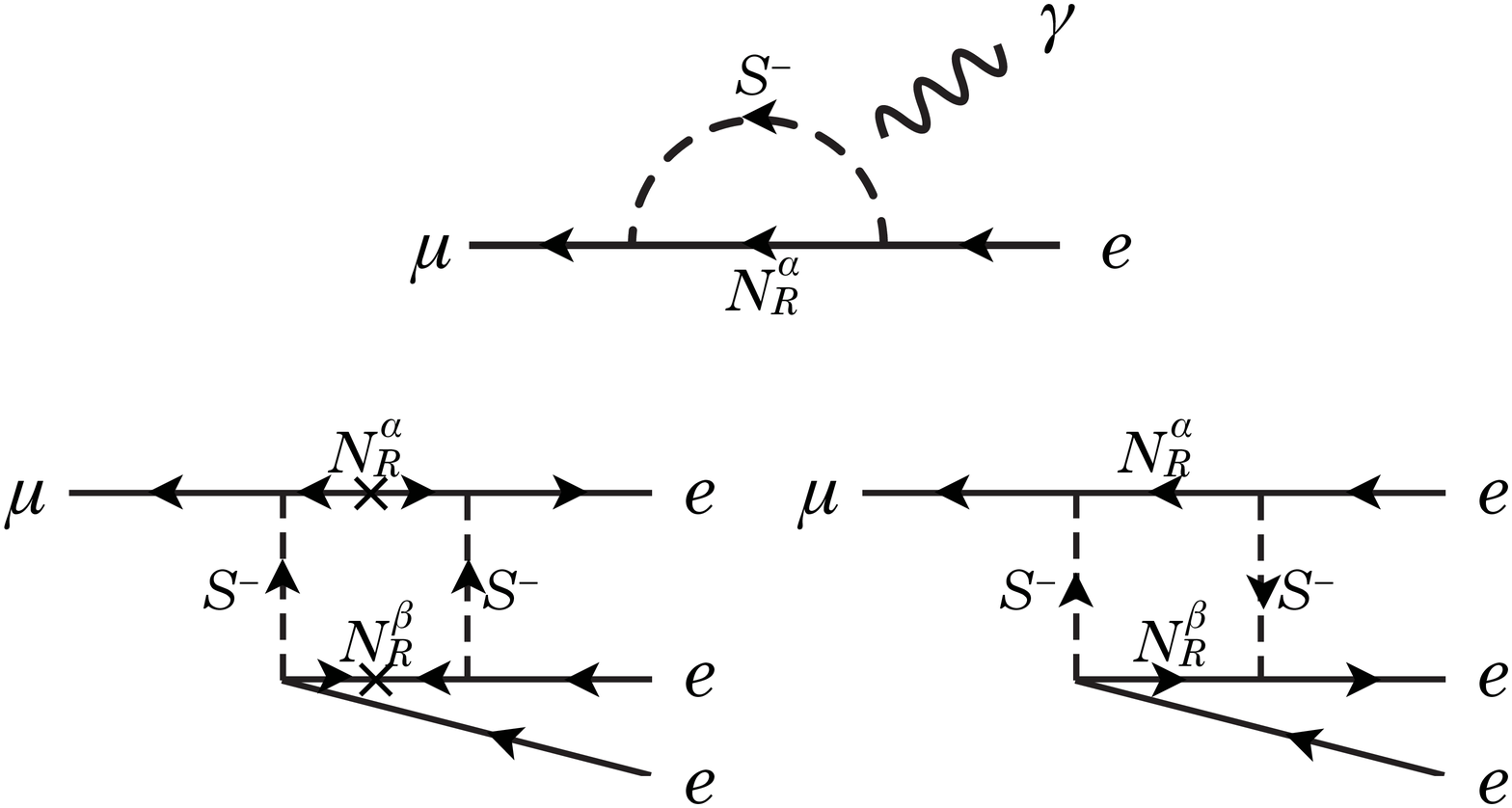}
\caption{The LFV processes~\cite{aky_rge}.}
\end{center}
\label{diagram}
\end{figure}

The model receives the severe constraints from the lepton-flavor violating processes of 
$\mu \to e \gamma$ and $\mu\to eee$: see Fig.~\ref{diagram}.
These processes are induced through one-loop diagrams by
$N_R^\alpha$  and $S^\pm$ with the Yukawa couplings $h_i^\alpha$
($i=e$ and $\mu$).
The branching ratio of $\mu \to e \gamma$ is given by
\footnote{The formula of Eq.(\ref{mer}) is different from Eq.(31) in Ref.~\cite{aks-prd}  which includes errors.
We have recalculated the values of $B(\mu\to e\gamma)$ 
by using the corrected formula and checked 
that the values of $B(\mu\to e\gamma)$ in the parameter sets in Table~II 
in Ref.~\cite{aks-prd} are still below the experimental bound.}
\begin{equation}
B(\mu\to e \gamma) \simeq  \frac{3\alpha_{\rm em} v^4}{32\pi m_{S}^4}\left|
\sum_{\alpha=1}^2 h_e^{\alpha\ast} h_\mu^\alpha F_2\left(\frac{m_{N_R^\alpha}^2}{m_{S}^2}\right)
\right|^2 ,
\label{mer}
\end{equation}
where $F_2(x)\equiv (1-6x+3x^2+2x^3-6x^2\ln x)/6(1-x)^4$.
For $\mu \to eee$, the branching ratio is calculated by 
\begin{align}
&B(\mu \to eee)=\frac{1}{64G_F^2}\left(\frac{1}{16\pi^2}\right)^2
\times\Bigg|\sum_{\alpha,\beta=1}^2\left(\frac{1}{m_{N_R^\alpha}^2-m_S^2}\right)\left(\frac{1}{m_{N_R^\beta}^2-m_S^2}\right)\notag\\
&\times\Bigg[h_\mu^{\alpha *} h_e^\alpha h_e^{\beta *}h_e^\beta 
\Big(\frac{m_{N_R^\alpha}^2m_{N_R^\beta}^2}{m_{N_R^\alpha}^2
-m_{N_R^\beta}^2}\log\frac{m_{N_R^\alpha}^2}{m_{N_R^\beta}^2}
-\frac{m_{N_R^\alpha}^2m_S^2}{m_{N_R^\alpha}^2-m_S^2}\log\frac{m_{N_R^\alpha}^2}{m_S^2}-\frac{m_{N_R^\beta}^2m_S^2}{m_{N_R^\beta}^2-m_S^2}\log\frac{m_{N_R^\beta}^2}{m_S^2}+m_S^2\Big)\notag\\
&+h_\mu^{\alpha *} h_e^{\alpha *} h_e^\beta h_e^\beta m_{N_R^\alpha}m_{N_R^\beta} 
\Big(
\frac{m_{N_R^\alpha}^2+m_S^2}{m_{N_R^\alpha}^2-m_S^2}\log\frac{m_{N_R^\alpha}^2}{m_S^2}
-\frac{m_{N_R^\alpha}^2+m_{N_R^\beta}^2}{m_{N_R^\alpha}^2
-m_{N_R^\beta}^2}\log\frac{m_{N_R^\alpha}^2}{m_{N_R^\beta}^2}
+\frac{m_{N_R^\beta}^2+m_S^2}{m_{N_R^\beta}^2-m_S^2}\log\frac{m_{N_R^\beta}^2}{m_S^2}-2\Big)\Bigg]\Bigg|^2. \label{meee1}
\end{align}
In particular, when $m_{N_R^1}=m_{N_R^2}$ $(=m_{N_R})$, the expression in Eq.~(\ref{meee1}) is reduced to  
\begin{align}
B(\mu \to eee)
&=\frac{1}{64G_F^2}\left(\frac{1}{16\pi^2}\right)^2\left(\frac{1}{m_N^2-m_S^2}\right)^4\notag\\
&\times\Bigg|\sum_{\alpha,\beta=1}^2h_\mu^{\alpha *} h_e^\alpha h_e^{\beta *}h_e^\beta 
\Big(m_N^2+m_S^2-\frac{2m_N^2m_S^2}{m_N^2-m_S^2}\log\frac{m_N^2}{m_S^2}\Big)\notag\\
&+2m_N^2\sum_{\alpha,\beta=1}^2h_\mu^{\alpha *} h_e^{\alpha *} h_e^\beta h_e^\beta \left(\frac{m_N^2+m_S^2}{m_N^2-m_S^2}\log\frac{m_N^2}{m_S^2}-2\right)\Bigg|^2. \label{meee2}
\end{align}
Notice that the contributions of the two diagrams to $\mu\to eee$ are constructive for all the parameter sets in Table II. 
We also note that there are additional contributions to $\mu\to eee$ from penguin diagrams, which are neglected because 
their contributions are much smaller. 

Assuming that $h_e^\alpha \sim {\cal O}(1)$, 
the masses of $N_R^\alpha$ and $S^\pm$ are strongly constrained from below. 
In particular, if we assume that $m_{S}\gtrsim 400$ GeV, $m_{N_R}\gtrsim {\cal O}(1)$ TeV 
is required to satisfy the current experimental bounds, $B(\mu\to e \gamma)< 1.2 \times 10^{-11} $~\cite{MEGA} 
and $B(\mu\to eee)< 1.0 \times 10^{-12} $~\cite{meee}.
Such a relatively heavier $S^\pm$ is favored from the discussion on 
the dark matter relic abundance and electroweak baryogenesis~\cite{aks_prl,aks-prd}.

\section{Typical scenarios}

In Table~\ref{h-numass}, we show four choices 
for the parameter sets, and resulting values for   
the neutrino Yukawa coupling constants $h_i^\alpha$ which satisfy 
the neutrino data and the LFV data. 
For all parameter sets, $m_S=400$ GeV and $m_{N_R}=5$ TeV are assumed.
Set A and Set B are taken as the normal hierarchy in the neutrino masses with
$\sin^2\theta_{13}=0$ and 0.03, respectively, while 
Set C and Set D are for the inverted hierarchy. 
The predictions on $B(\mu\to e\gamma)$ and $B(\mu \to eee)$ are also shown in the table\footnote{
In Table~\ref{h-numass}, 
we show the numbers of the $h_i^\alpha$ coupling constants 
with four digits for Set C,   
because the branching ratios of $\mu\to e\gamma$ 
and $\mu\to eee$ are sensitive to these numbers 
due to large cancellations.}.
The scenario with the inverted hierarchy requires the larger values for
$\kappa \tan\beta$, so that the normal hierarchy scenarios are more natural in our model.  


 \begin{table}[t]
\begin{center}
  \begin{tabular}{|c||c|c|c|c|c|c||c|c|}\hline
    & \multicolumn{6}{c||}{\mbox{Yukawa~couplings}} &
 \multicolumn{2}{c|}{\mbox{LFV}}  \\ \hline
 & $h_e^1$ &$h_e^2$&$h_\mu^1$&$h_\mu^2$&$h_\tau^1$&$h_\tau^2$&
   $B(\mu\!\!\to\!\! e\gamma)$&$B(\mu\!\!\to\!\! 3e)$
   \\\hline \hline
  A&1.2&1.3&0.024&-0.011&7.1$\times 10^{-4}$&-1.4$\times 10^{-3}$
   &$2.8\!\times \!10^{-14}$&$7.8\!\times \!10^{-13}$\\\hline 
  B&1.1&1.1&0.0028& 0.018&-5.5$\times 10^{-4}$&9.7$\times 10^{-4}$
   &$6.1\!\times \!10^{-14}$&$9.8\!\times \!10^{-13}$\\\hline\hline 
  C&3.500&3.474&0.01200&-0.01192&-7.136$\times 10^{-4}$&7.086$\times 10^{-4}$
&$4.4\!\times \!10^{-17}$&$8.2\!\times \!10^{-14}$\\\hline 
  D&2.1&2.2&6.4$\times 10^{-3}$  & -8.6$\times 10^{-3}$&-5.3$\times 10^{-4}$ & 3.5$\times 10^{-4}$
   &$3.5\!\times \!10^{-15}$&$9.3\!\times \!10^{-13}$\\\hline 
   \end{tabular}
\end{center}
\vspace{-2mm}  \caption{Values of $h_i^\alpha$ as well as those of 
branching ratios of $\mu\to e \gamma$ and $\mu\to 3e$ 
for $m_\eta=50$ GeV and $m_{H^\pm}^{}=100$ GeV for various scenarios 
which satisfy neutrino data:     
Set A and Set B are scenarios of the the normal hierarchy 
while Set C and Set D are those of the inverted hierarchy. 
For all sets, $m_S=400$ GeV and $m_{N_R}=5$ TeV are taken. 
Input parameters of ($\sin^2\theta_{13}$, $\kappa\tan\beta$) are taken to be 
(0, 54), (0.03, 76), (0, 80) and (0.03, 128) for Set A, Set B, Set C and Set D, respectively~\cite{aky_rge}. }
\label{h-numass}
\end{table}

\begin{figure}[t]
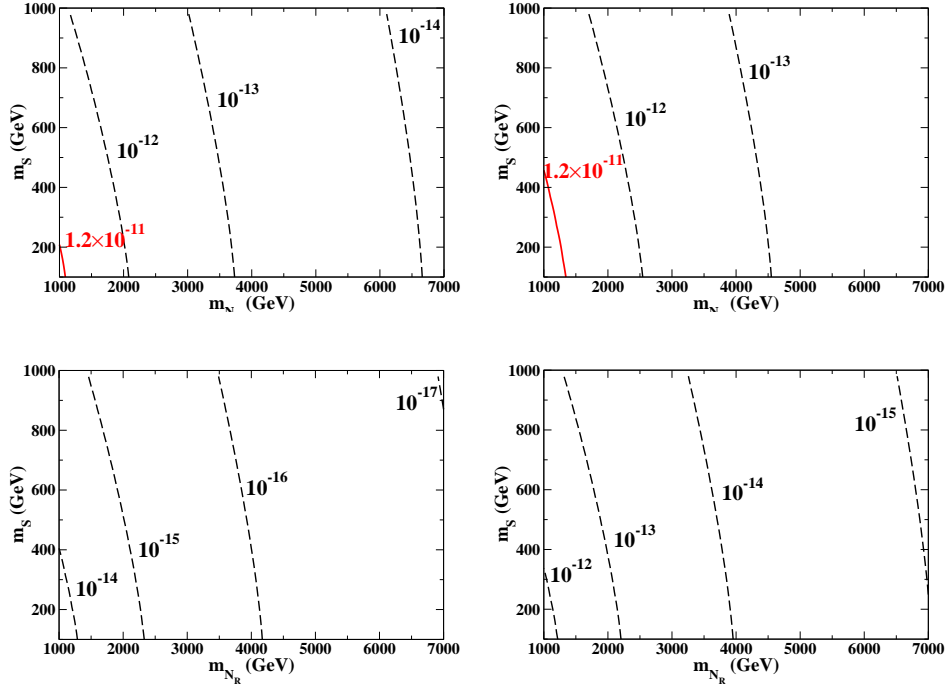

\begin{center}
\includegraphics[width=60mm]{meg_A.eps}\hspace{3mm}
\includegraphics[width=60mm]{meg_B.eps}\\ \vspace{5mm}
\includegraphics[width=60mm]{meg_C.eps}\hspace{3mm}
\includegraphics[width=60mm]{meg_D.eps}
\caption{Contour plots for the branching ratio of $\mu\to e\gamma$ for the neutrino Yukawa coupling constants in 
           Set A (top-left), Set B (top-right), Set C (bottom-left) and Set D (bottom-right). 
           The contour for the upper limit from the data is given as the (red) solid curve~\cite{aky_rge}.}  
\end{center}
\label{LFV_1}
\end{figure}

\begin{figure}[t]
\begin{center}
\includegraphics[width=60mm]{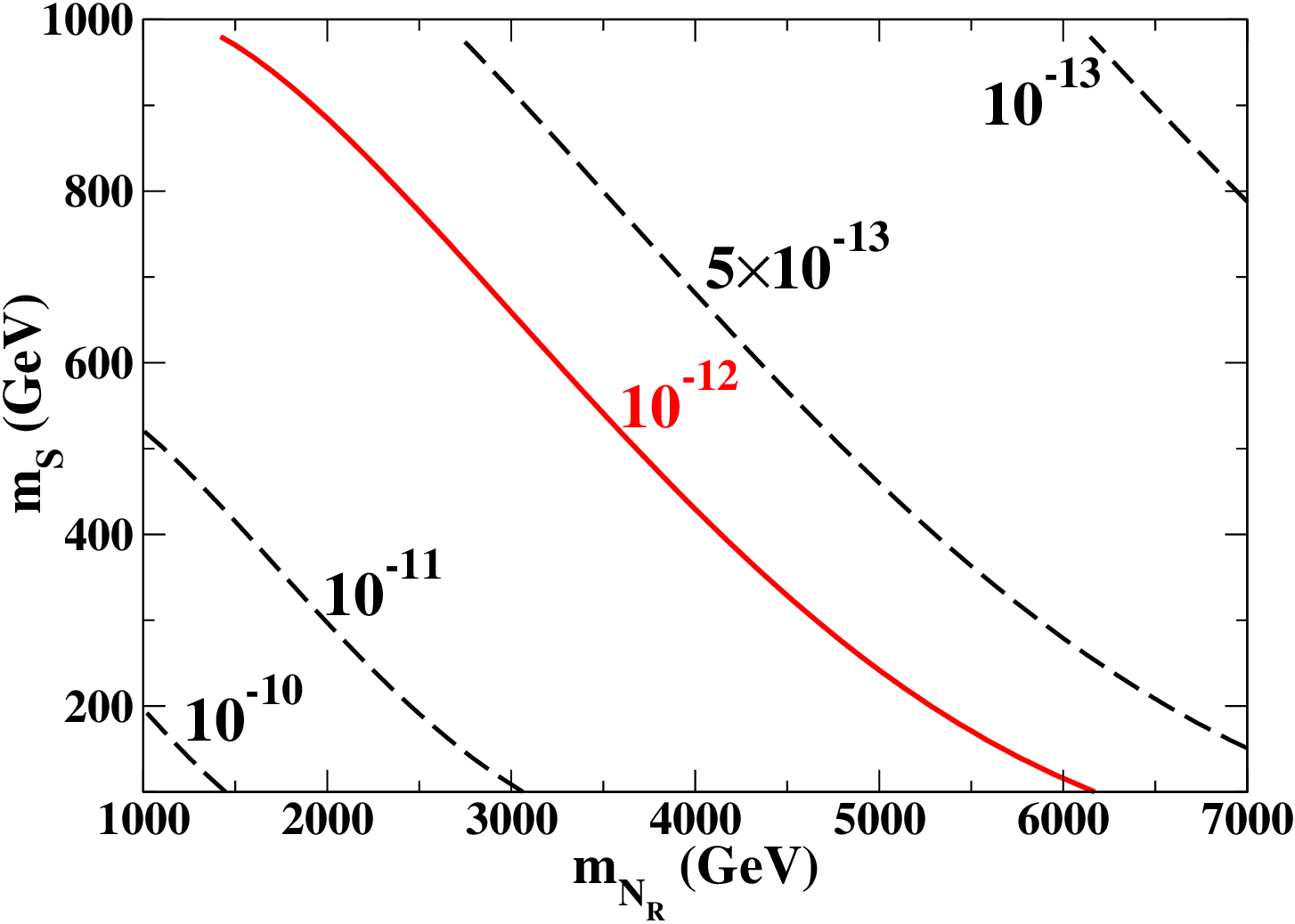}\hspace{3mm}
\includegraphics[width=60mm]{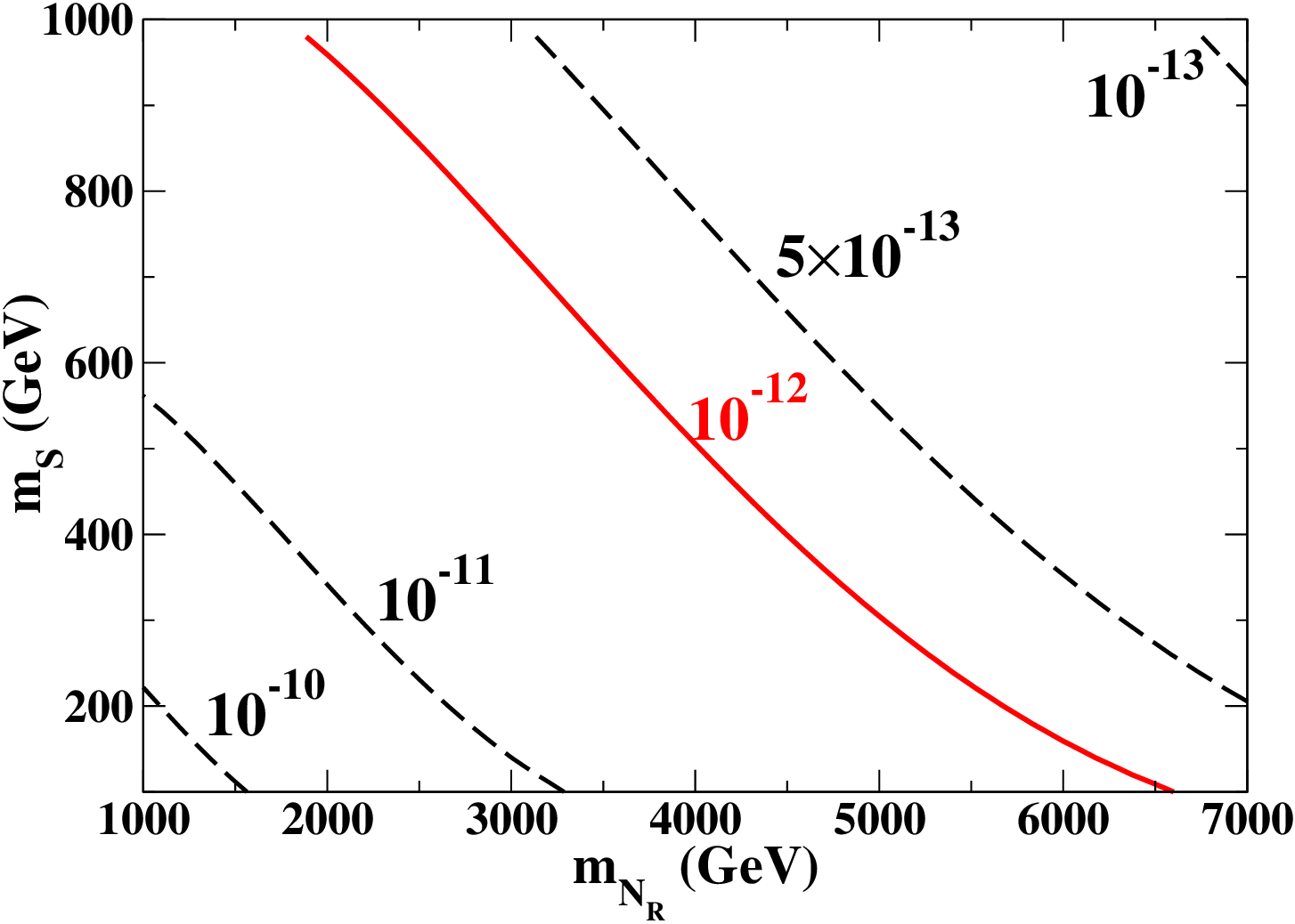}\\ \vspace{5mm}
\includegraphics[width=60mm]{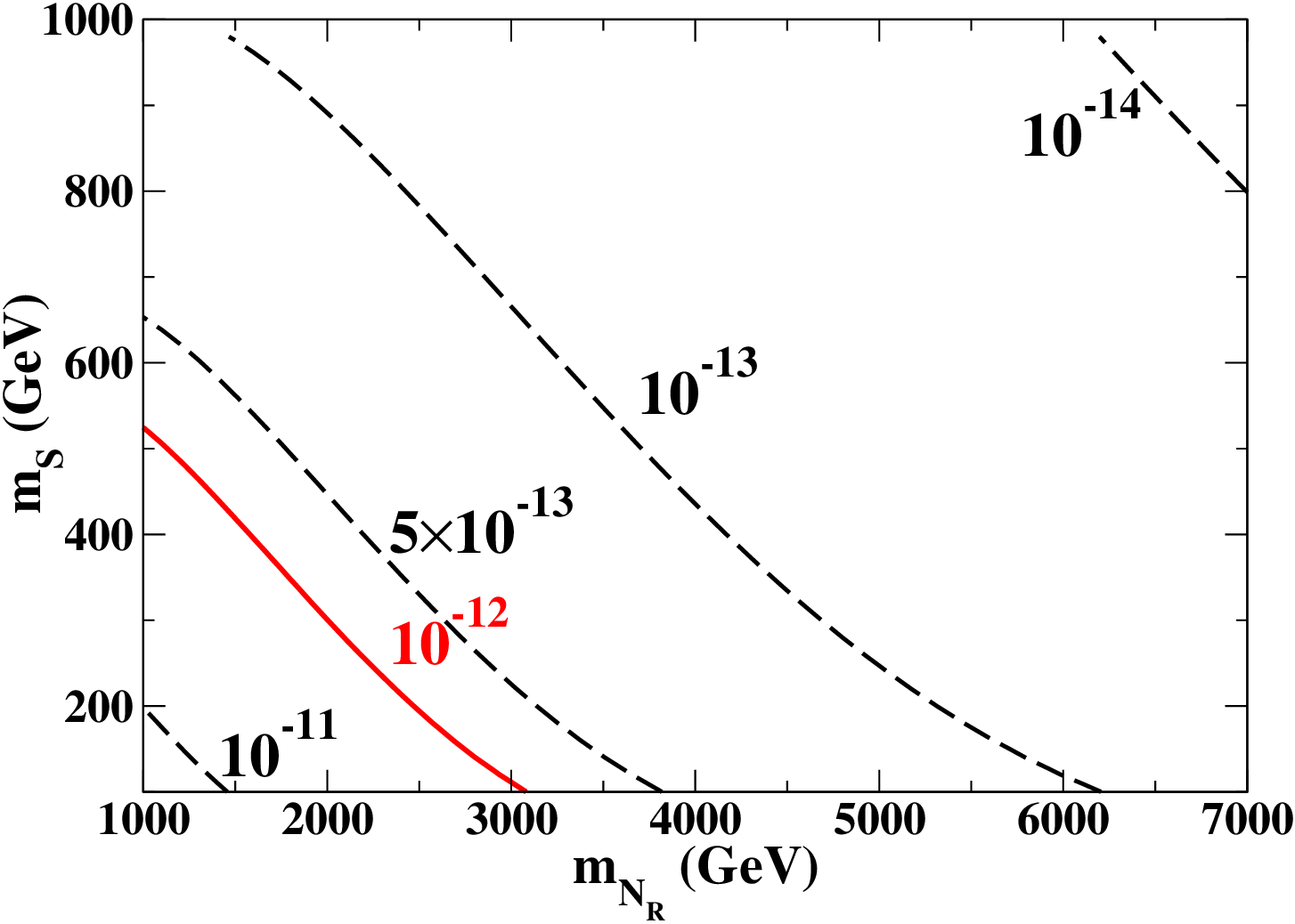}\hspace{3mm}
\includegraphics[width=60mm]{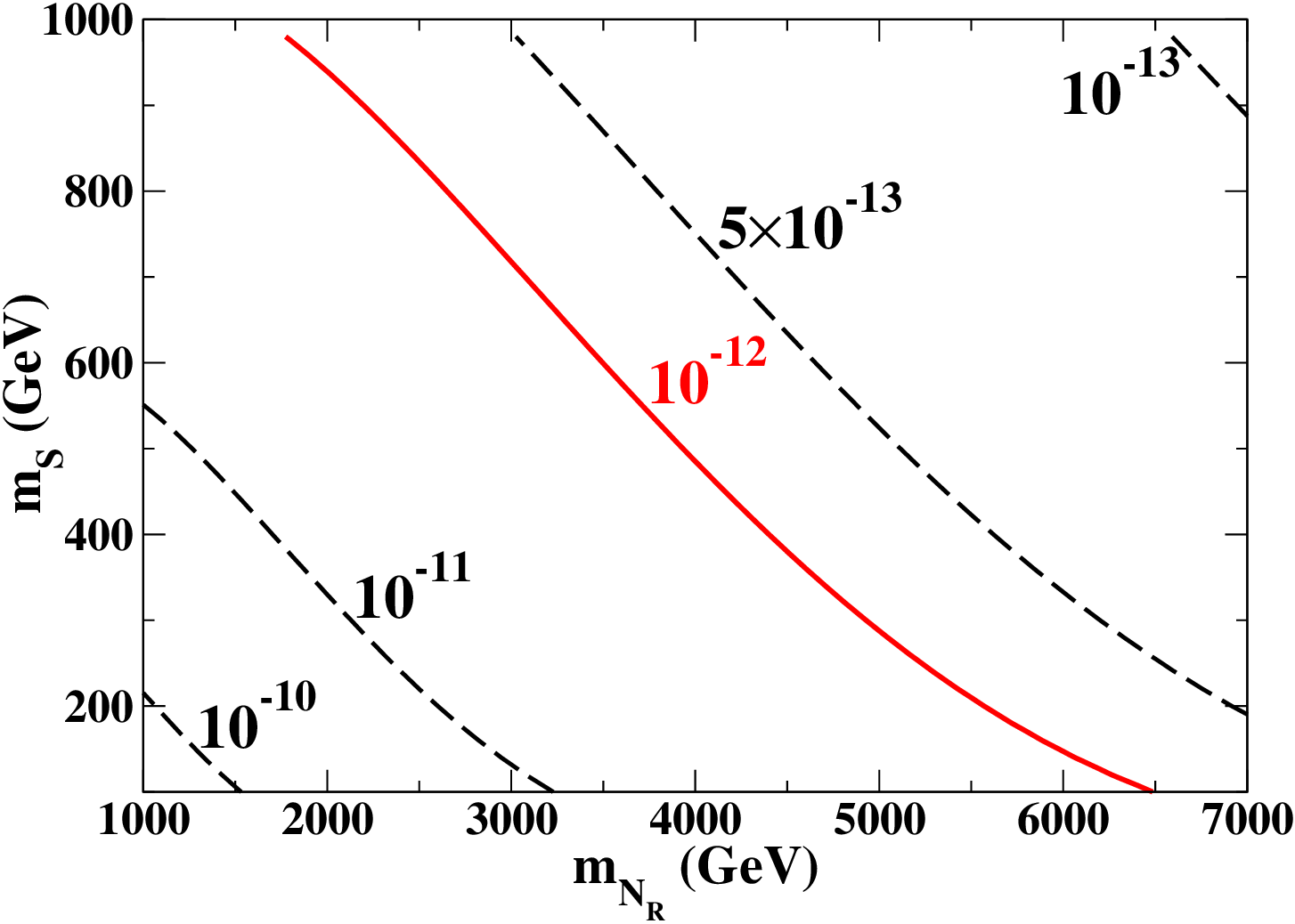}
  \caption{Contour plots for the branching ratio of $\mu\to eee$ for the neutrino Yukawa coupling constants in 
           Set A (top-left), Set B (top-right), Set C (bottom-left) and Set D (bottom-right). 
           The contour for the upper limit from the data is given as the (red) solid curve~\cite{aky_rge}.}
\end{center}
\label{LFV_2}
\end{figure}

In Fig.~\ref{LFV_1}, the contour plots of the branching ratio $B(\mu\to e\gamma)$ are shown in the $m_{S}^{}$-$m_{N_R}$ plane 
for the neutrino Yukawa coupling constants in Set A to Set D, 
while those of the branching ratio $B(\mu\to eee)$ are shown for these scenarios in Fig.~\ref{LFV_2}.  
The scale of the branching ratio of $\mu\to e\gamma$ is determined by $m_{N_R}$ and is insensitive to $m_{S}$, 
while that of $\mu\to eee$ largely depend on both $m_{N_R}$ and $m_{S}$ especially for Set A, Set B and Set D. 
It can easily be seen that a much stronger constraint comes from $\mu\to eee$ for all scenarios.

\section{Dark matter and electroweak phase transition}

From now on, we employ Set A in Table~\ref{h-numass} for further phenomenological analyses.
In this scenario, masses of $N_R^\alpha$ are at the multi-TeV scale, so that it may be natural that 
the rest $Z_2$-odd neutral field $\eta$ is the candidate of dark matter.  
Since $\eta$ is a singlet under the SM gauge group,
the interactions with $Z_2$-even particles are only through the Higgs coupling.
When $m_\eta < m_W$, the $\eta$ field predominantly annihilates into $b \bar{b}$ and $\tau^+\tau^-$ 
through $s$-channel Higgs boson ($h$ and $H$) mediations.  
Strong annihilation occurs at $m_\eta \simeq m_H^{}/2$
(and $m_\eta \simeq m_h/2$) due to the resonance of $H$ ($h$) mediation in the $s$-channel diagrams.
The pair annihilation into two photons through one-loop diagrams by $H^\pm$ and $S^\pm$ can also be important if $\kappa$ 
is of the order one.
The relic abundance becomes consistent with the data ($\Omega_{\rm DM} h^2 \sim 0.11$~\cite{DM}) for $m_\eta \sim 50-60$ GeV, when we take $m_H^{}=m_{H^\pm}^{}\simeq 100$ GeV, $m_h^{}\simeq 120$ GeV,
$m_{S}^{}\gtrsim 400$ GeV with $\kappa=1.5$, $\sigma_1=0.05$, $\sigma_2=0.03$, and $\tan\beta=36$. 
In such scenario, the typical spin-independent cross section for the scattering of dark matter with a proton 
is of order of $10^{-8}$ pb which is within the reach of the direct search experiments such as superCDMS and XMASS. 

When $m_\eta < m_h/2$, the (SM-like) Higgs boson $h$ can decay into a dark matter pair $\eta\eta$. 
The branching ratio of $h \to \eta\eta$ is evaluated as $34$ \% ($22$ \%) for $m_h=120$ GeV and $m_\eta=48$ ($55$) GeV  
when $\sigma_1=0.05$, $\sigma_2=0.03$ and $\tan\beta > 10$. 
The invisible decay of $h$ can be tested at the LHC when $B(h \to \eta\eta) > 25$ \%~\cite{LHCinv}. 
At the ILC, it is expected that the branching ratio for the invisible decay 
of a few \% can be detected~\cite{Schumacher:2003ss}. Therefore, the invisible decay in this model can be tested at the collider experiments.

Our model~\cite{aks_prl}  satisfies the conditions for baryogenesis~\cite{BAU}. 
The $B$ number violating interaction is the sphaleron interaction.  
The additional CP violating phases are in the Higgs sector and in the Yukawa interaction.
The condition of departure from thermal equilibrium can be realized by the strong first order electroweak phase transition,
which requires a large 
tri-linear coupling of the order parameter 
in the expression of the high temperature expansion~\cite{hte} where 
only the bosonic loop can contribute\footnote{We note that such a nondecoupling effect due to the bosonic loop 
can also affect the quantum correction to the triple Higgs boson coupling~\cite{ewbg-thdm2,KOSY}. Such 
a large correction to the Higgs self-coupling can be an important signature for successful 
electroweak baryogeneis at collider experiments.}.
In our model, there are many additional scalars running in the
loop so that the large coupling can be easily realized~\cite{ewbg-thdm2}.
The strong first order phase transition is possible for large $m_{S}^{}$ 
and/or $m_A^{}$ with the large nondecoupling effect:
{\it e.g.} 
$m_S^{} \sim 400$ GeV,
$m_A^{} \sim 100$ GeV, 
$M=100$ GeV and $\mu_S^{}=200$ GeV,
where $M$ and $\mu_S^{}$ are the 
invariant masses in Eq.(\ref{mA}) and Eq.(\ref{mS}), respectively. 
The result is not sensitive to $\tan\beta$.

\section{Bounds from triviality and vacuum stability}

There are scalar bosons in this model, so that quadratic divergences 
appear in the one-loop calculation for their masses. 
Because there is no mechanism by which such quadratic divergences 
are eliminated, enormous fine tuning is required to realize the 
renormalized Higgs boson mass being at the weak scale with a very high cutoff scale.   
Allowing the 1~\% fine tuning, the cutoff scale is at most $\Lambda \sim O(10)$ TeV, 
above which the theory would be replaced by a more fundamental one~\cite{Murayama}. 
Unless a mechanism of cancellation of the quadratic divergences such as supersymmetry is 
implemented, to avoid excessive fine tuning the model should be regarded as an effective 
theory, whose cutoff scale $\Lambda$ is between $m_{N_R^\alpha}$ and $O(10)$ TeV. 
We then need to confirm the theoretical consistency of the model up to $\Lambda$~\cite{RGE_SM}.
We here evaluate bounds on the parameter space from vacuum stability and perturbativity, 
and examine whether 
the theoretically allowed parameter region is consistent with that by the experimental 
data discussed in the previous sections. 

We have to consider these two bounds seriously because of the following reasons. 
First, this model includes many scalar fields, $e.g.$, $h$, $H$, $A$, $H^\pm$, $S^\pm$ and $\eta$, 
so that the scale dependent dimensionless coupling constants would be drastically changed  
by the loop corrections due to the scalar bosons. 
Second, some of the Yukawa coupling constants for right-handed neutrinos are necessarily 
of order one for a radiative generation of the tiny mass scale of the neutrinos at the three-loop level. 
Finally, to realize the first order electroweak phase transition, some of the scalar self-coupling 
constants has to be as large as of order one.  

In order to evaluate the vacuum stability bound and the triviality bound,  
we estimate the scale dependences of the dimensionless coupling constants 
by using the RGEs at the one-loop level. 
We have calculated the one-loop beta functions for all the coupling constants in this model.   
The full set of the beta functions is listed in Appendix.  
We take into account the threshold effects in the calculation of the scale dependent coupling constants. 
In the scale below the mass of $S^\pm$, we treat the theory without $N_R$ and $S^\pm$. 
In the scale between the masses of the $S^\pm$ and $N_R$, we treat the theory without $N_R$. 
In the scale higher than the mass of the $N_R$, we treat the theory with full particle contents.
 
\section{The conditions}

In this model, there are scalar fields $\Phi_i$ ($i=1,2$), $S^\pm$ and $\eta$, which contain 
eleven degrees of freedom which would share the order parameter. 
The four of them are eliminated because of the $SU(2)_L\times U(1)_Y $ gauge symmetry. 
In the remining seven dimensional parameter space, we require that for any direction 
the potential is bounded from below with keeping positiveness~\cite{VS_thdm}. 
In the SM, this requirement is satisfied when the Higgs self-coupling constant is positive. 
In this model, we put the following conditions on the the dimensionless coupling constants:   
\begin{align}
\lambda_1(\mu)>0,\quad \lambda_2(\mu)>0,\quad \lambda_S(\mu)>0,\quad \lambda_\eta(\mu) >0, \label{eq:vsc1}
\end{align}
\begin{align}
&\sqrt{\lambda_1(\mu)\lambda_2(\mu)}+\lambda_3(\mu)+\text{MIN}[0,\quad (\lambda_4(\mu)+\lambda_5(\mu)),\quad (\lambda_4(\mu)-\lambda_5(\mu))]>0,\notag\\
&\sqrt{\lambda_1(\mu)\lambda_S(\mu)/2}+\rho_1(\mu)>0,\quad \sqrt{\lambda_1(\mu)\lambda_\eta(\mu)/3}+\sigma_1(\mu)>0,\quad \sqrt{\lambda_2(\mu)\lambda_S(\mu)/2}+\rho_2(\mu)>0,\notag\\
&\sqrt{\lambda_2(\mu)\lambda_\eta(\mu)/3}+\sigma_2(\mu)>0,\quad \sqrt{\lambda_S(\mu)\lambda_\eta(\mu)/6}+\xi(\mu)>0, \label{eq:vsc2}
\end{align} 
\begin{align}
&2\lambda_1(\mu)+2\lambda_2(\mu)+4\lambda_3(\mu)+4\rho_1(\mu)+4\rho_2(\mu)+\lambda_S(\mu)+4\sigma_1(\mu)+4\sigma_2(\mu)\notag\\
&+\frac{2}{3}\lambda_\eta(\mu)+4\xi(\mu)-16\sqrt{2}|\kappa(\mu)| >0.\label{vs_kappa}  
\end{align}
The conditions in Eqs.~(\ref{eq:vsc1}) and (\ref{eq:vsc2}) are obtained by the similar way as in 
Ref.~\cite{Kasai}, while the last condition in Eq.~(\ref{vs_kappa}) is derived such that 
the term with the coupling constant $\kappa$ in the potential satisfies the positivity condition 
for the direction where the VEVs of the fields $\Phi_1$, $\Phi_2$, $S^\pm$ and $\eta$ are 
a common value.

We require that all the dimensionless running coupling constants do not blow up below $\Lambda$. 
Since we discuss the model within the scale where the perturbation calculation remains reliable,  
we here require that the running coupling constants do not exceed some critical value. 
In this paper, we impose the following criterion in the coupling constants in the Higgs potential Eq.~(\ref{pot})
and the Yukawa interaction in Eqs.~(\ref{yukawa1}) and (\ref{yukawa2}):
\begin{align}
&|\lambda_i(\mu)|, \hspace{2mm}|\sigma_i(\mu)|, \hspace{2mm}|\rho_i(\mu)|, \hspace{2mm} |\kappa(\mu)|, \hspace{2mm} |\xi(\mu)| < 8\pi,\notag\\
&y_t^2(\mu),\hspace{2mm} y_b^2(\mu),\hspace{2mm}  y_\tau^2(\mu),\hspace{2mm}  (h_i^\alpha)^2(\mu)< 4\pi.  \label{vs3}
\end{align}
The similar critical value has been adopted in the analyses in the two Higgs dobulet model~\cite{VS_thdm2,Kasai} 
and in the Zee model~\cite{kkloy}. 

\section{Allowed regions in the parameter space}

In this section, we evaluate allowed regions in parameter space,  
which satisfy the conditions of triviality and vacuum stability 
for each fixed cutoff scale $\Lambda$.
For the scenarios of the neutrino Yukawa coupling constants 
as well as the masses of right-handed neutrinos, we choose 
Set A in Table~\ref{h-numass}.  
We investigate the allowed regions in the $m_S$-$m_A$ plane, and 
the rest of the mass parameters in the scalar sector is fixed 
as
\begin{align}
&m_{H^+}=100\text{ GeV},\quad m_H=100\text{ GeV},\quad m_{h}=120\text{ GeV},\quad m_\eta=50\text{ GeV},\notag\\
&M=100\text{ GeV},\quad \mu_S=200\text{ GeV},\quad \mu_\eta=30\text{ GeV}. 
\end{align}
The initial values for the scalar coupling constants 
in the Higgs sector are taken to be 
\begin{align}
\lambda_1(m_Z)=0.24,\quad \lambda_2(m_Z)=0.24,\quad \lambda_3(m_Z)=0.24,\quad \kappa(m_S)\tan\beta=54,\notag\\
\rho_1(m_S)=0.1,\quad \lambda_S(m_S)=2,\quad \sigma_1(m_Z)=0.05,\quad \sigma_2(m_Z)=0.05,\quad \lambda_\eta=3,
\end{align}
and the mixing angle are set on $\sin(\beta-\alpha)=1$. 
We note that the initial value of $\lambda_4$, $\lambda_5$ and $\rho_2$ 
are determined by given values for the masses of $A$ and $S^\pm$ 
using Eqs.~(\ref{mA}), (\ref{mHpm}) and (\ref{mS}).
The rest parameter $\xi$ (the coupling constant for $|S^-|^2\eta^2$) 
is taken as $\xi = 3$ and $5$. 
The results in the case with $\xi=3$ is shown in Fig.~\ref{result1} for 
$\kappa=1.2$ (left figure) and $\kappa=1.5$ (right figure), while those with $\xi=5$ 
is in Fig.~\ref{result2} for the same values of $\kappa$.  

In Fig.~\ref{result1}, the shaded area in the figure is excluded due to the vacuum stability condition 
in Eq.~(\ref{vs_kappa}).  
In this area, the condition is not satisfied already at the electroweak scale, 
so that the excluded region is independent of $\Lambda$. 
The vacuum stability bound become stronger for a larger value of $\kappa$, 
although the area compatible with both theoretical conditions with $\Lambda=10$ TeV 
still exists for $\kappa=1.5$. 
On the other hand,  
the bound from perturbativity depends on $\Lambda$. 
In Fig.~\ref{result1}, the contour plots for $\Lambda=6$, $10$ and $15$ TeV, the scales 
where one of the coupling constants blows up and breaks the 
condition of perturbativity are shown for the case of $\xi=3$ 
in the $m_S$-$m_A$ plane.   
We find that there is the parameter region which satisfies both the conditions of 
vacuum stability and perturbativity with the blow-up scale to be above $\Lambda=10$ TeV.  
The area of the vicinity of $m_S \sim 400$ GeV and $m_A < 350$ GeV can also be 
consistent from the theoretical bounds.
We stress that this parameter region is favored for phenomenologically 
successful scenarios for neutrino masses, relic abundance 
for the dark matter, and the strongly first order phase transition. 

The similar figures but with $\xi=5$ are shown in Fig.~\ref{result2}. 
The contour plots are for $\Lambda=6$, $10$, $12$ and $14$ TeV in the $m_S$-$m_A$ plane.   
We find that there is the parameter region which satisfies both the conditions of 
vacuum stability and perturbativity with the blow-up scale to be above $\Lambda=10$ TeV.  
The vacuum stability bound is more relaxed as compared to that for $\xi=3$, while 
the bound from perturbatibity becomes rather strict.  
In the regions with $m_S < 400$ GeV, the running coupling constants blow up earlier 
than the case with $\xi=3$, because of the threshold effect at the scale $\mu=m_S$, above 
which the running of $\lambda_\eta$ becomes enhanced by the loop contribution of $S^\pm$. 
In the area of $300$ GeV $< m_S < 400$ GeV and $m_A < 350$ GeV, $\Lambda$  can be 
above $10$ TeV.



\begin{figure}[t]
\includegraphics[width=75mm]{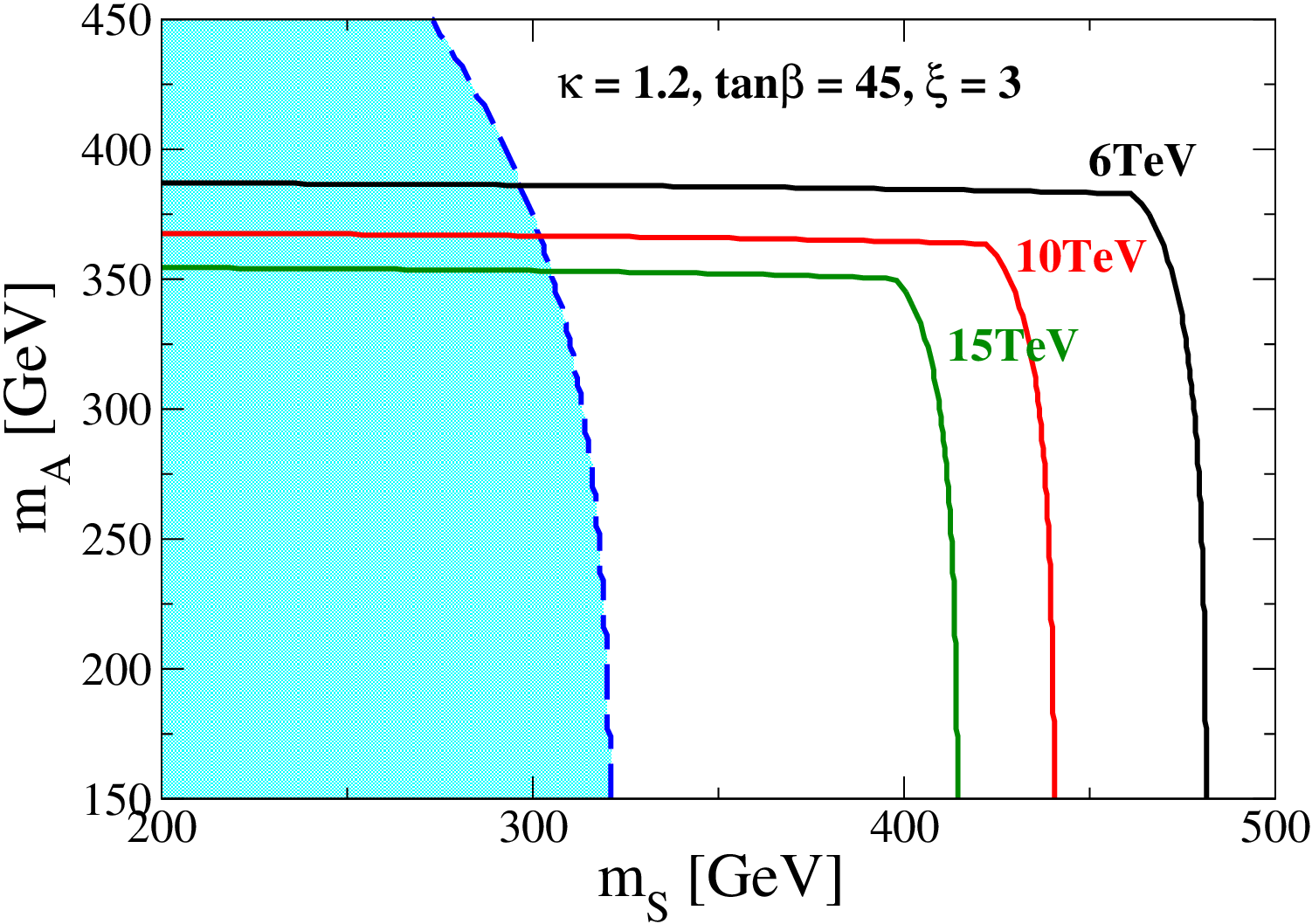}
\includegraphics[width=75mm]{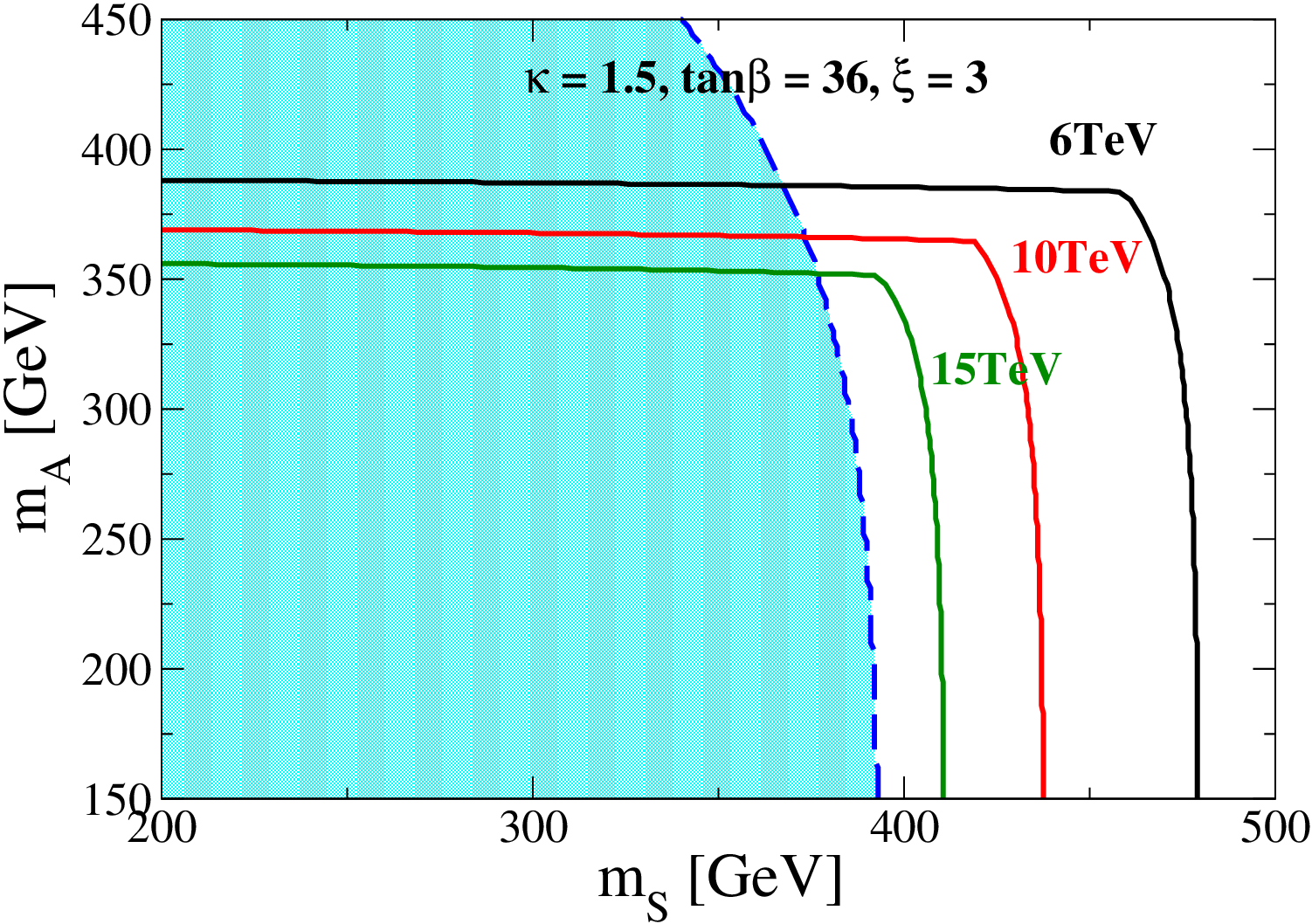}
\caption{Contour plots for the scale where condition of perturbativity is broken are shown 
         in the $m_S$-$m_A$ plane in the case of $(\kappa,\tan\beta)=(1.2, 45)$ (left figure), 
         and $(\kappa,\tan\beta)=(1.5, 36)$  (right figure).   
         The region excluded by the vacuum stability condition is also shown as the shaded area.
         The constant $\xi$ is taken to be 3 at the scale of $m_S$~\cite{aky_rge}.}
\label{result1}
\vspace{1cm}
\includegraphics[width=75mm]{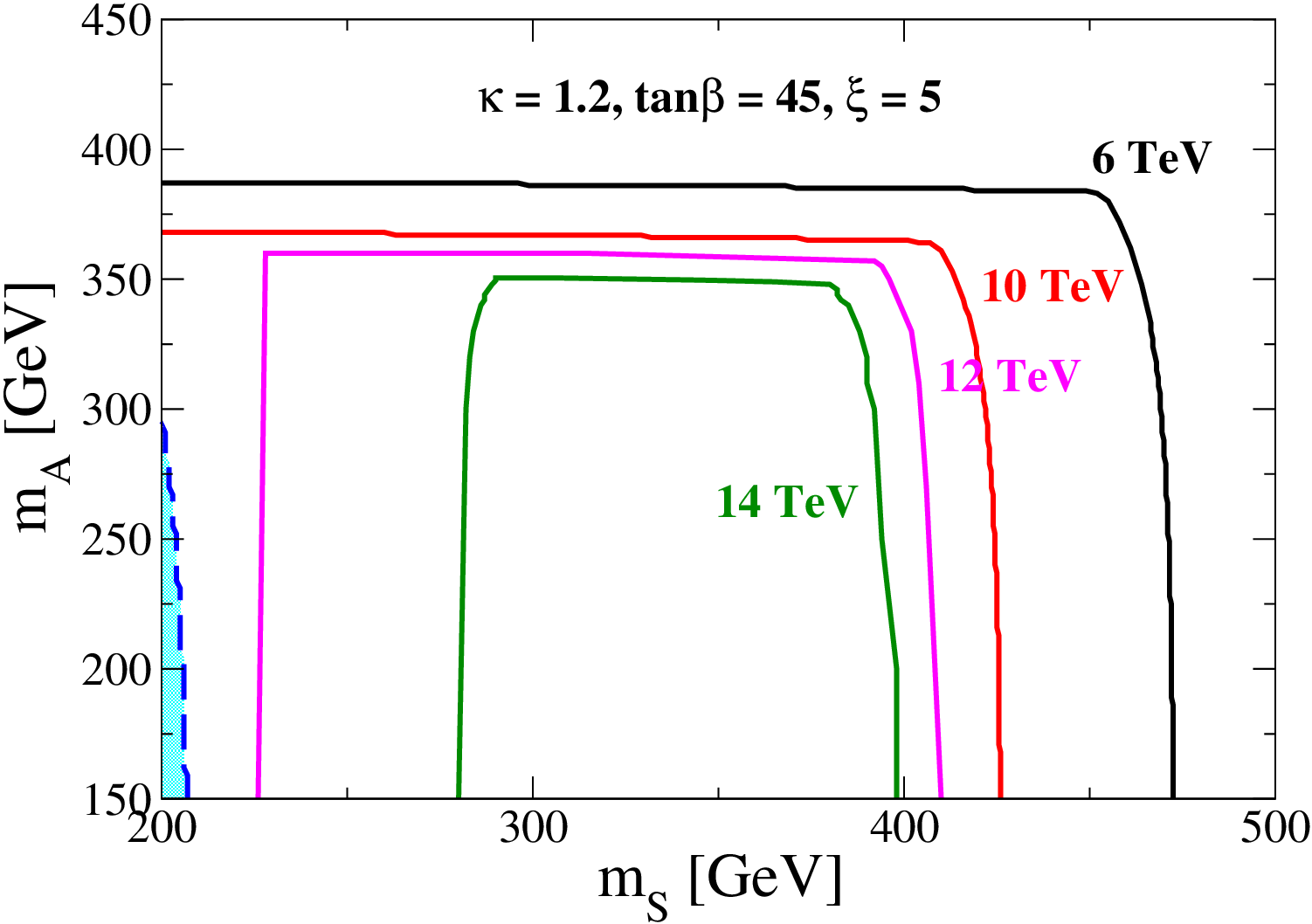}
\includegraphics[width=75mm]{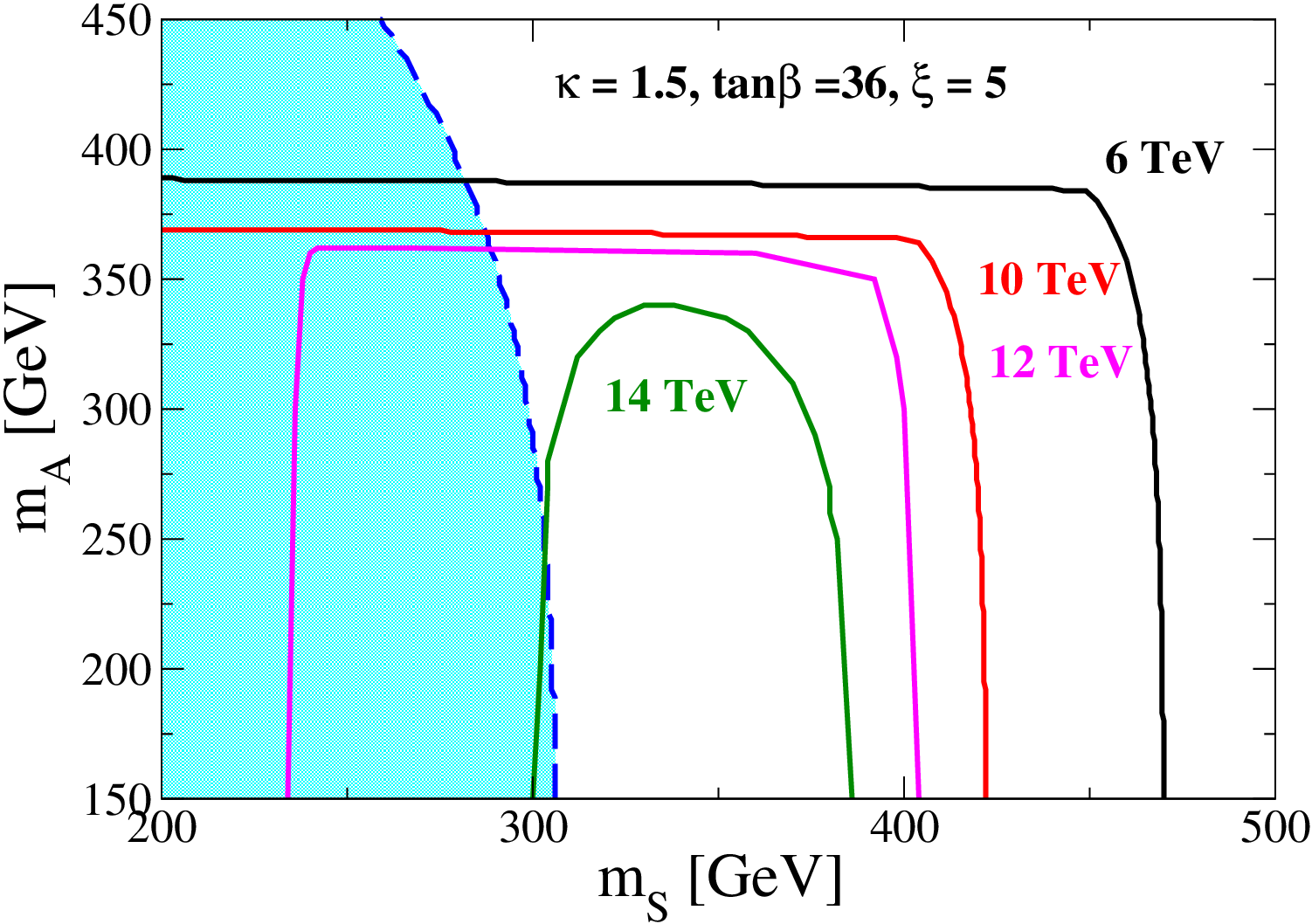}
\caption{Contour plots for the scale where condition 
 of perturbativity is broken are shown 
         in the $m_S$-$m_A$ plane in the case of $(\kappa,\tan\beta)=(1.2, 45)$ (left figure), 
         and $(\kappa,\tan\beta)=(1.5, 36)$  (right figure).   
         The region excluded by the vacuum stability condition is also shown as the shaded area.
         The constant $\xi$ is taken to be 5 at the scale of $m_S$~\cite{aky_rge}.}
\label{result2}
\end{figure}

\chapter{Supersymmetric extention of the Zee-Babu model}

The Zee-Babu model may be the simplest successful model 
which can be generated tiny neutrino masses at the two-loop level. 
In this model, isospin SU(2) singlet singly- and doubly-charged scalar bosons which carry lepton number of two unit 
are added to the SM. 
Although this model can explain neutrino oscillation data, 
this model cannot explain the hierarchy problem because there is no symmetry to 
forbid the quadratic divergence in radiative corrections to the Higgs boson mass. 
In addition to the hierarchy problem, this model does not have a dark matter candidate. 

In this section, we investigate a supersymmetric extension 
of the Zee-Babu model.  
By introducing SUSY, the quadratic divergence 
in the one-loop correction to the mass of the Higgs boson  
can be eliminated automatically.
In addition, a discrete symmetry, which is so called the R-parity, is
imposed in our model to forbid the term which causes the dangerous proton decay.
The R-parity also guarantees the stability of the lightest super partner
particle (LSP) such as the neutralino,
which may be identified as a candidate of DM. 

We find that there are allowed parameter regions in which 
the current neutrino oscillation data can be reproduced 
under the constraint from the lepton flavour
violation (LFV) data. 
In addition, this model provides quite interesting phenomenological signals
in the collider physics; i.e., the existence of singly as well as
doubly charged singlet scalar bosons and their SUSY partner fermions.
Such an allowed parameter region also appears even when new particles and their partners
are as light as the electroweak scale. 
We also discuss the outline of phenomenology for these particles at
the LHC.

\section{Model}
\begin{figure}[t]
\begin{center}
\includegraphics[width=100mm]{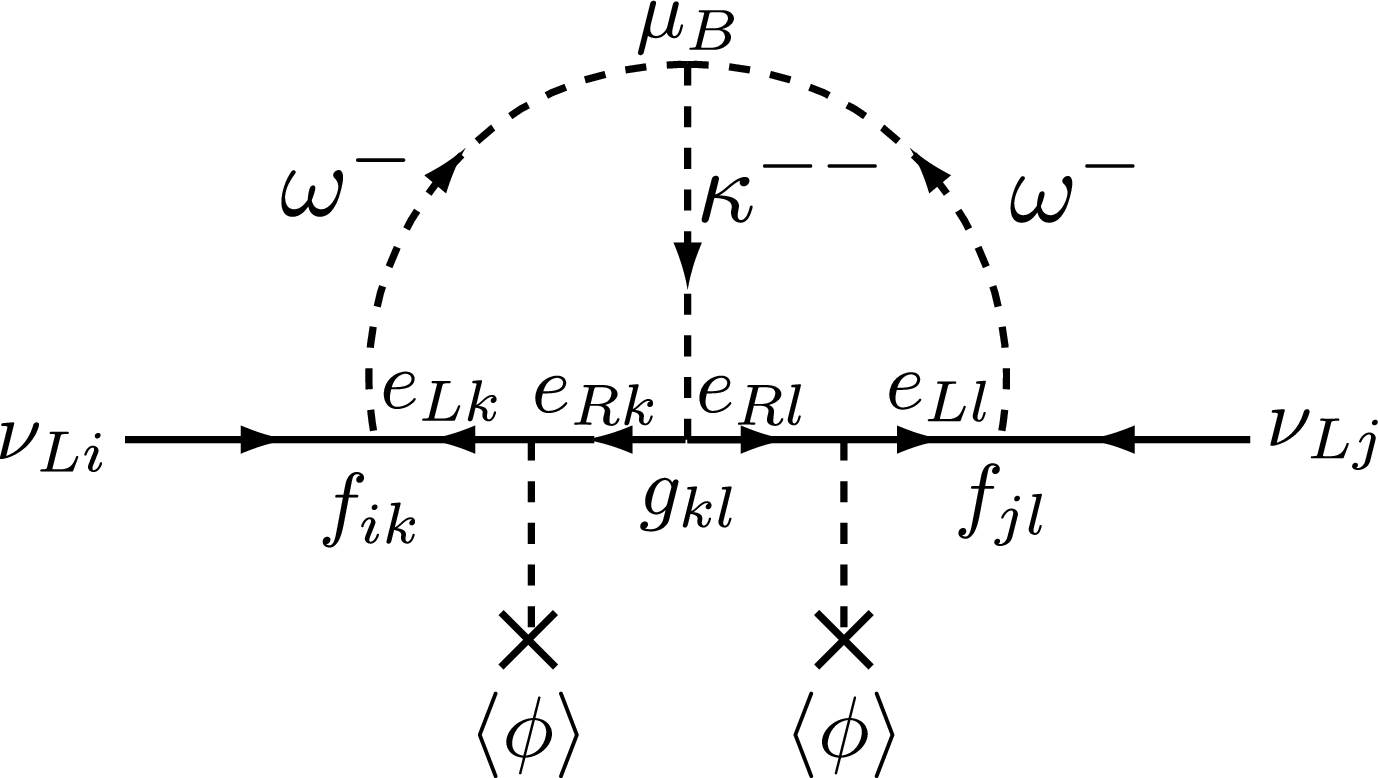}
\caption{The two-loop diagram relevant to the neutrino mass matrix~\cite{aksy_szb}.}
\end{center}
\label{fig:diagram0}
\end{figure}

In the original (non-SUSY) Zee-Babu model\cite{zee-2loop,babu-2loop},
two kinds of $\text{SU}(2)_{\text{L}}^{}$ singlet fields $\omega^{-}$ ($Y=-1$) and
$\kappa^{--}$ ($Y=-2$) are introduced.
The Yukawa interaction and the scalar potential are given by 
\begin{equation}
\mathcal{L}=
-\sum_{i,j=1}^3 f_{ij}^{}\bar{L}_L^{ic}\cdot L_{L}^{j}\omega^+
-\sum_{i,j=1}^3g_{ij}^{}\bar{e}_R^ie_R^{jc}\kappa^{--}
-\mu_B^{}\omega^{-}\omega^{-}\kappa^{++}+\text{h.c.}-V^{\prime}-V_{\text{SM}}^{}\;,
\end{equation}
where $V_{\text{SM}}^{}$ is the Higgs potential of the SM, 
the indices $i$, $j$ are the flavour indices 
and all the scalar couplings with 
respect to $\omega^-$ and $\kappa^{--}$ other than 
$\omega^-\omega^-\kappa^{++}$ are in $V^{\prime}$.
Notice that lepton number conservation is broken only by the term of $\mu_B ^{}$.
The neutrino mass matrix is generated via two-loop diagrams as shown in Fig.~\ref{fig:diagram0}.
The induced neutrino mass matrix is computed as\footnote{Our result for
the neutrino mass matrix is consistent with that in Ref.~\cite{nebot}
including the factor.} 
\begin{equation}
(m_{\nu}^{})_{ij}^{}=\sum_{k,l=1}^3
16\mu_B^{} f_{ik}^{}(m_{e}^{})_k^{} g_{kl}^{}(m_e^{})_{l}^{}f_{jl}^{}
 I(m_{\omega}^{},(m_e^{})_{k}^{}|m_{\omega}^{},(m_e^{})_l^{}|m_{\kappa}^{})\;,
\end{equation}
where $(m_e^{})_i$ are charged lepton masses,  and
 the induced mass matrix $(m_\nu^{})_{ij}^{}$ is defined in the effective
Lagrangian as
\begin{equation}
 \mathcal{L}_{\nu}^{} = - \sum_{i,j=1}^3 \frac{1}{2} (\overline{\nu}^c_L)_i^{} (m_\nu^{})_{ij} (\nu_L^{})_j^{} +
  {\rm h.c.}, 
\end{equation}
and $I(m_{11}^{},m_{12}^{}|m_{21}^{},m_{22}^{}|M)$ is the two-loop integral function
defined as
\begin{align}
&I(m_{11}^{},m_{12}^{}|m_{21}^{},m_{22}^{}|M)
\nonumber\\
&= 
\int \frac{d^4p}{(2\pi)^4}
\int \frac{d^4q}{(2\pi)^4}
\frac{1}{(p^2+m_{11}^2)}
\frac{1}{(p^2+m_{12}^2)}
\frac{1}{(q^2+m_{21}^2)}
\frac{1}{(q^2+m_{22}^2)}
\frac{1}{((p+q)^2+M^2)}\;.
\label{eq:I-func}
\end{align}
Following Refs.~\cite{vanderBij:1983bw},
one can evaluate the function $I(m_{11}^{},m_{12}^{}|m_{21}^{},m_{22}^{}|M)$ as
\begin{align}
&I(m_{11}^{},m_{12}^{}|m_{21}^{},m_{22}^{}|M)\nonumber\\
&=\frac{I(m_{12}^{}|m_{22}^{}|M)-I(m_{11}^{}|m_{22}^{}|M)
	-I(m_{12}^{}|m_{21}^{}|M)+I(m_{11}^{}|m_{21}^{}|M)}
{(m_{11}^2-m_{12}^2)(m_{21}^2-m_{22}^2)}\;,
\end{align}
where
\begin{equation}
I(m_1^{}|m_2^{}|M)=-
	m_1^2f\left(\frac{m_2^2}{m_1^2},\frac{M^2}{m_1^2}\right)
	-m_2^2f\left(\frac{m_1^2}{m_2^2},\frac{M^2}{m_2^2}\right)
	-M^2f\left(\frac{m_1^2}{M^2},\frac{m_2^2}{M^2}\right).
\end{equation}
The function $f(x,y)$ is given by
\begin{align}
f(x,y)=&-\frac{1}{2}\ln x\ln y -\frac{1}{2}\left(\frac{x+y-1}{D}\right)\nonumber\\
&\times\left\{
	\mathrm{Li}_2^{}\left(\frac{-\sigma_{-}}{\tau_{+}}\right)
	+\mathrm{Li}_2^{}\left(\frac{-\tau_{-}}{\sigma_{+}}\right)
	-\mathrm{Li}_2^{}\left(\frac{-\sigma_{+}}{\tau_{-}}\right)
	-\mathrm{Li}_2^{}\left(\frac{-\tau_{+}}{\sigma_{-}}\right)\right.\nonumber\\
	&\phantom{Space}\left.
	+\mathrm{Li}_2^{}\left(\frac{y-x}{\sigma_{-}}\right)
	+\mathrm{Li}_2^{}\left(\frac{x-y}{\tau_{-}}\right)
	-\mathrm{Li}_2^{}\left(\frac{y-x}{\sigma_{+}}\right)
	-\mathrm{Li}_2^{}\left(\frac{x-y}{\tau_{+}}\right)\right\}\;,
\end{align}
where $D$, $\sigma_{\pm}$ and $\tau_{\pm}$ are
\begin{align}
D=\sqrt{1-2(x+y)+(x-y)^2},~~~
\sigma_\pm=\frac{1}{2}(1-x+y\pm D),~~~
\tau_\pm=\frac{1}{2}(1+x-y\pm D),
\end{align}
and $\text{Li}_2(x)$ is the dilogarithm function defined as 
\begin{equation}
\text{Li}_2^{}(x)=-\int_0^x\frac{\ln(1-t)}{t}dt\;.
\end{equation}
We note that
in the limit of $m_{12}^{}=m_{22}^{}=0$ and $m_{11}^{}=m_{21}^{}=m_{\omega}^{}$ the
above function $I(m_{11}^{},m_{12}^{}|m_{21}^{},m_{22}^{}|M)$ 
has the same form as the function 
given in Refs.~\cite{macesanu,aristizabal},
\begin{equation}
I(m_{\omega}^{},0|m_{\omega}^{},0|m_{\kappa}^{})=
-\frac{1}{(16\pi^2)^2}\frac{1}{m_\kappa^2}\int_0^1dx\int_0^{1-x}dy\frac{r}{x+(r-1)y+y^2}\ln\frac{y(1-y)}{x+ry}\;,
\end{equation}
where $r=m_{\kappa}^2/m_{\omega}^2$.
Details of the Zee-Babu model have been studied 
in the literature\cite{macesanu,aristizabal,nebot,ohlsson}. 
It is known that the model can reproduce the present neutrino
data with satisfying constraints from the LFV.

We turn to the SUSY extension of the Zee-Babu model. 
The $\text{SU}(2)_{\text{L}}^{}$ singlet chiral superfields 
$\hat{\Omega}_R^c$,
$\hat{\Omega}_L^{}$,
$\hat{K}_L^{}$,
and $\hat{K}_R^c$
are added to the superfields in 
the MSSM, whose details are shown in 
Table.~\ref{mattercontent}.
Notice that although the non-SUSY Zee-Babu model includes only two  
$\text{SU}(2)_{\text{L}}^{}$ singlet scalars
these four chiral fields are required in the SUSY model.
If only $\hat{\Omega}_R^c$ and $\hat{K}_L^{}$ are introduced in the model, 
their fermion components are massless and the model is ruled out.
By introducing additional fields $\hat{\Omega}_L^{}$ and $\hat{K}_R^c$ such massless
fermions can be massive, and furthermore the model becomes anomaly free.
\begin{table}
\begin{center}
{\renewcommand\arraystretch{1.3}
\begin{tabular}{|c||c|c||c|c|c||c|c|}\hline
&Spin 0&Spin 1/2&$\text{SU(3)}_{\text{C}}^{}$&$\text{SU(2)}_{\text{L}}^{}$&
$\text{U(1)}_{\text{Y}}^{}$&Electric charge&Lepton number\\ \hline
$\hat{\Omega}_R^c$&$\omega_R^*$&$(\tilde{\omega}_R)^c$&1&1&1&1&$-2$\\ \hline
$\hat{\Omega}_L^{}$&$\omega_L^{}$&$\tilde{\omega}_L^{}$&1&1&$-1$&$-1$&2\\ \hline
$\hat{K}_L^{}$&$\kappa_L^{}$&$\tilde{\kappa}_L^{}$&1&1&$-2$&$-2$&2\\ \hline
$\hat{K}_R^c$&$\kappa_R^*$&$(\tilde{\kappa}_R)^c$&1&1&2&2&$-2$\\ \hline
\end{tabular}}
\end{center}
\caption{Particle properties of chiral superfields~\cite{aksy_szb}.}
\label{mattercontent}
\end{table}

The superpotential is given by\footnote{Hereafter we  omit
the summation symbol for simplicity.}
\begin{align}
W=&W_{\text{MSSM}}^{}+f_{ij}^{}\hat{L}_i^{}\cdot \hat{L}_j^{}\hat{\Omega}_R^c+g_{ij}^{}\hat{E}_i^c\hat{E}_j^c\hat{K}_L^{}
+\lambda_L^{}\hat{K}_L^{}\hat{\Omega}_R^c\hat{\Omega}_R^c
+\lambda_R^{}\hat{K}_R^c\hat{\Omega}_L^{}\hat{\Omega}_L^{}
\nonumber\\
&
+\mu_{\Omega}^{}\hat{\Omega}_R^c\hat{\Omega}_L^{}
+\mu_{K}^{}\hat{K}_L^{}\hat{K}_R^c\;,
\label{eq:W}
\end{align}
where $W_{\text{MSSM}}$ is the superpotential in the MSSM. 
The superfields in the superpotential are listed in Table.~\ref{mattercontent},
and the coupling matrices $f_{ij}$ and $g_{ij}$ are an
antisymmetric matrix $f_{ji}=-f_{ij}$  and a symmetric one $g_{ji}=g_{ij}$, respectively.
It is emphasized that 
we here impose the exact R-parity in order to protect the decay of the 
LSP,
so that the LSP is a candidate of the DM.
The soft SUSY breaking terms are given by 
\begin{equation}
\mathcal{L}_{\text{soft}}^{}=
\mathcal{L}_{\text{MSSM}}^{}
+\mathcal{L}_{\text{SZB}}^{}
+\mathcal{L}_{\text{C}}^{}\;,
\end{equation}
where $\mathcal{L}_{\text{MSSM}}$ represents the corresponding terms in the MSSM, 
\begin{align}
\mathcal{L}_{\text{SZB}}^{}=&
-M_+^2\omega_R^\ast\omega_R^{}
-M_-^2\omega_L^*\omega_L^{}
-M_{--}^2\kappa_L^*\kappa_L^{}
-M_{++}^2\kappa_R^*\kappa_R^{}
\nonumber\\
&
+\biggl(
-m_S^{}\tilde{f}_{ij}^{}\omega_R^*\tilde{L}_{L}^{i}\cdot \tilde{L}_{L}^{j}
-m_S^{}\tilde{g}_{ij}^{}\kappa_L\tilde{e}_{R}^{i*}\tilde{e}_{R}^{j*}
-m_S^{}\tilde{\lambda}_L^{}\kappa_L^{}\omega_R^*\omega_R^*
-m_S^{}\tilde{\lambda}_R^{}\kappa_R^*\omega_L^{}\omega_L^{}
\nonumber\\
&\phantom{Spac}
-B_{\omega}^{}\mu_{\Omega}^{}\omega_R^*\omega_L^{}
-B_{\kappa}^{}\mu_{K}^{}\kappa_L^{}\kappa_R^*
+\text{h.c.}\biggr)
\;,
\label{eq:Lsoft}
\end{align}
and
\begin{align}
\mathcal{L}_{\text{C}}^{}=&
-C_u^{}\omega_R^*H_u^{\dagger}H_d^{}
-C_d^{}\omega_L^{}H_d^{\dagger}H_u^{}
-(C_{\omega}^{})^{ij}\omega_L^*\tilde{L}_{L}^{i}\cdot \tilde{L}_{L}^{j}
+\text{h.c.}\;,
\label{eq:LsoftNH}
\end{align}
where $m_S^{}$ denotes a typical SUSY mass scale,
and $\tilde{f}_{ji}^{}=-\tilde{f}_{ij}^{}$ and $\tilde{g}_{ji}^{}=\tilde{g}_{ij}^{}$. 
$\mathcal{L}_{\text{SZB}}^{}$ is the standard soft-breaking terms
with respect to the new charged singlet fields, $\omega_{L,R}^{}$ and 
$\kappa_{L,R}^{}$.
$\mathcal{L}_{\text{C}}^{}$  contains the terms so-called the ``C-terms''\cite{Hall:1990ac}, where the scalar component and its
conjugation are mixed
\footnote{The singlet scalar C-terms 
break SUSY hard, while the
terms listed in the $\mathcal{L}_{\text{C}}^{}$ include non-singlet scalars
and the quadratic divergence does not occur.}.

\begin{figure}
\begin{center}
{\includegraphics[scale=0.8]{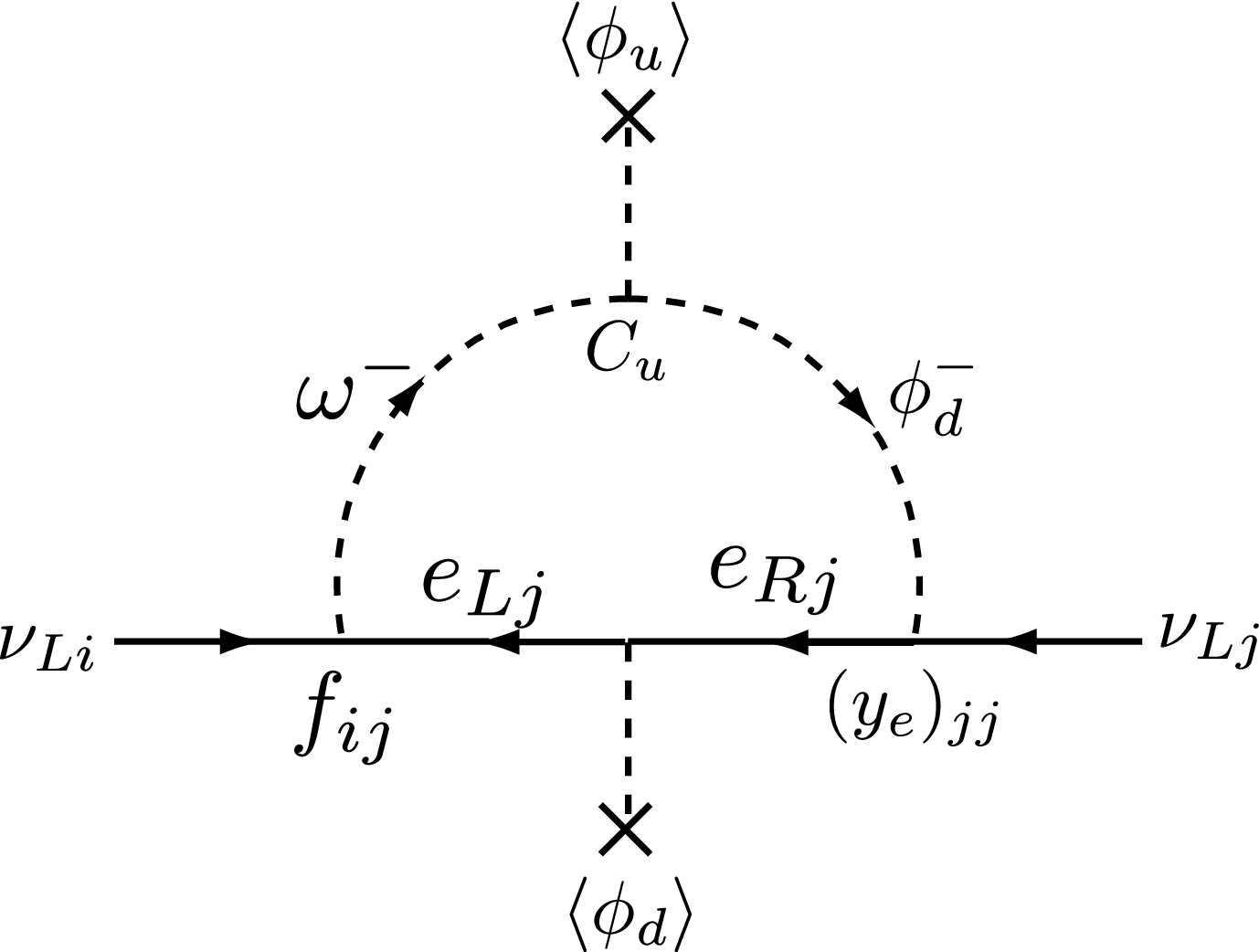}}
\end{center}
\caption{The one-loop diagram relevant to the neutrino mass matrix
 with the C-term $\omega^+H_u^*H_d^{}$. $(y_e^{})_{jj}^{}$ is 
 the charged lepton Yukawa coupling~\cite{aksy_szb}.}
\label{fig:diagram00}
\end{figure}

There are two possibilities in building a SUSY model with the charged
singlet fields, depending on whether or not the C-terms are 
switched on in a SUSY breaking scenario\footnote{
Many models derived by $N=1$ supergravity do not lead to the C-terms
and if they are absent at the cut off scale, they do not appear through the radiative corrections\cite{JackJonesKord}.
Thus the C-terms are usually ignored in the MSSM.
On the other hand, it is known that C-terms are induced in some models of SUSY breaking 
such as an intersecting brane model with a flux compactification\cite{Camara:2003ku}.}.
If we assume that $\mathcal{L}_{\text{C}}^{}$ is absent,
tiny neutrino masses are generated only by at least two loop diagrams as 
in the Zee-Babu model.
On the other hand, 
with the term $\omega_R^*H_u^{\dagger}H_d^{}$, tiny
neutrino masses are dominated by 
one loop diagrams in Fig.~\ref{fig:diagram00} just like in 
the original Zee model\cite{zee}.
Here, we focus on the case where the SUSY breaking mechanism does not lead to the soft SUSY breaking 
C-terms, so that all the neutrino masses are generated at the two loop
level.  

From the superfields $\hat{\Omega}_R^c$, $\hat{\Omega}_L^{}$, $\hat{K}_L^{}$ and 
$\hat{K}_R^c$, there appear singly charged ($Y=-1$)
and doubly charged ($Y=-2$) singlet scalar bosons,
$\omega_{R,L}^{}$ and $\kappa_{L,R}^{}$,  
as well as their superpartner fermions,
namely singly and doubly charged singlinos, 
$\tilde{\omega}$ and $\tilde{\kappa}$, respectively.
The superpotential and the soft SUSY breaking terms lead to 
the mass matrix for the singly charged scalars in the basis of
$(\omega_R^{},\omega_L^{})$ as\;,
\begin{equation}
M_{\omega}^2=
\begin{pmatrix}
M_+^2+|\mu_{\Omega}|^2-m_W^2\tan^2\theta_W^{}\cos2\beta& B_{\omega}^{}\mu_{\Omega}^{}\\
B_{\omega}^{}\mu_{\Omega}^{}&M_-^2+|\mu_{\Omega}|^2+m_W^2\tan^2\theta_W^{}\cos2\beta
\end{pmatrix}\;,
\end{equation}
and the mass matrix for the doubly charged singlet scalars in the basis of 
$(\kappa_L^{}, \kappa_R^{})$ as
\begin{equation}
M_{\kappa}^2=
\begin{pmatrix}
M_{--}^2+|\mu_{K}|^2+2m_W^2\tan^2\theta_W^{}\cos2\beta& B_{\kappa}^{}\mu_{K}^{}\\
B_{\kappa}^{}\mu_{K}^{}&M_{++}^2+|\mu_{K}|^2-2m_W^2\tan^2\theta_W^{}\cos2\beta
\end{pmatrix}\;,
\end{equation}
where $\tan\beta$ is 
a ratio of the two vacuum expectation values of the MSSM Higgs bosons as
$\tan\beta = \langle \phi_u^{}\rangle/\langle \phi_d^{}\rangle$.
As easily seen from the above expressions, 
$\omega_R^{}$  and 
$\omega_L^{}$ ($\kappa_L^{}$ and $\kappa_R^{}$) can mix with each other
by the soft-breaking ``B-term'', $(B_{\omega}^{}\mu_{\Omega}^{}) \omega_R^*\omega_L^{}$ 
($(B_{\kappa}^{}\mu_{K}^{})\kappa_L^{}\kappa_R^*$).
The mass eigenvalues of singly and doubly charged singlet scalar bosons
are obtained after diagonalizing their mass matrices $M_{\omega}^2$ 
and $M_{\kappa}^2$ by the unitary matrices $U_{\omega}^{}$ and 
$U_{\kappa}^{}$ as
\begin{equation}
U_{\omega}^{\dagger}M_{\omega}^2U_{\omega}^{}
=\begin{pmatrix}
(m_{\omega})_1^2&0\\
0&(m_{\omega})_2^2
\end{pmatrix}\;,\quad
U_{\kappa}^{\dagger}M_{\kappa}^2U_{\kappa}^{}
=\begin{pmatrix}
(m_{\kappa})_1^2&0\\
0&(m_{\kappa})_2^2
\end{pmatrix}\;.
\end{equation}
The mass eigenstates are then given by 
\begin{equation}
\omega_{a}^{}=(U_{\omega}^{\dagger})_{a1}^{}\omega_R^{}
+(U_{\omega}^{\dagger})_{a2}^{}\omega_L^{}\;,\quad
\kappa_{a}=(U_{\kappa}^{\dagger})_{a1}^{}\kappa_L^{}
+(U_{\kappa}^{\dagger})_{a2}^{}\kappa_R^{}\;,\quad (a=1,2)\;.
\end{equation}
The mass eigenstates of the singlinos are
\begin{equation}
\tilde{\omega}=\begin{pmatrix} \tilde{\omega}_L^{}\\ \tilde{\omega}_R^{}\end{pmatrix}\;,\quad
\tilde{\kappa}=\begin{pmatrix}\tilde{\kappa}_L^{}\\ \tilde{\kappa}_R^{}\end{pmatrix}\;,
\end{equation}
whose mass eigenvalues are given by 
the SUSY invariant parameters as 
$m_{\tilde{\omega}}^{}=\mu_{\Omega}^{}$ and $m_{\tilde{\kappa}}^{}=\mu_K^{}$,
respectively.

\begin{figure}
\begin{center}
\begin{tabular}{ccc}
\raisebox{20mm}{(a)}&\phantom{space}&\includegraphics[scale=0.7]{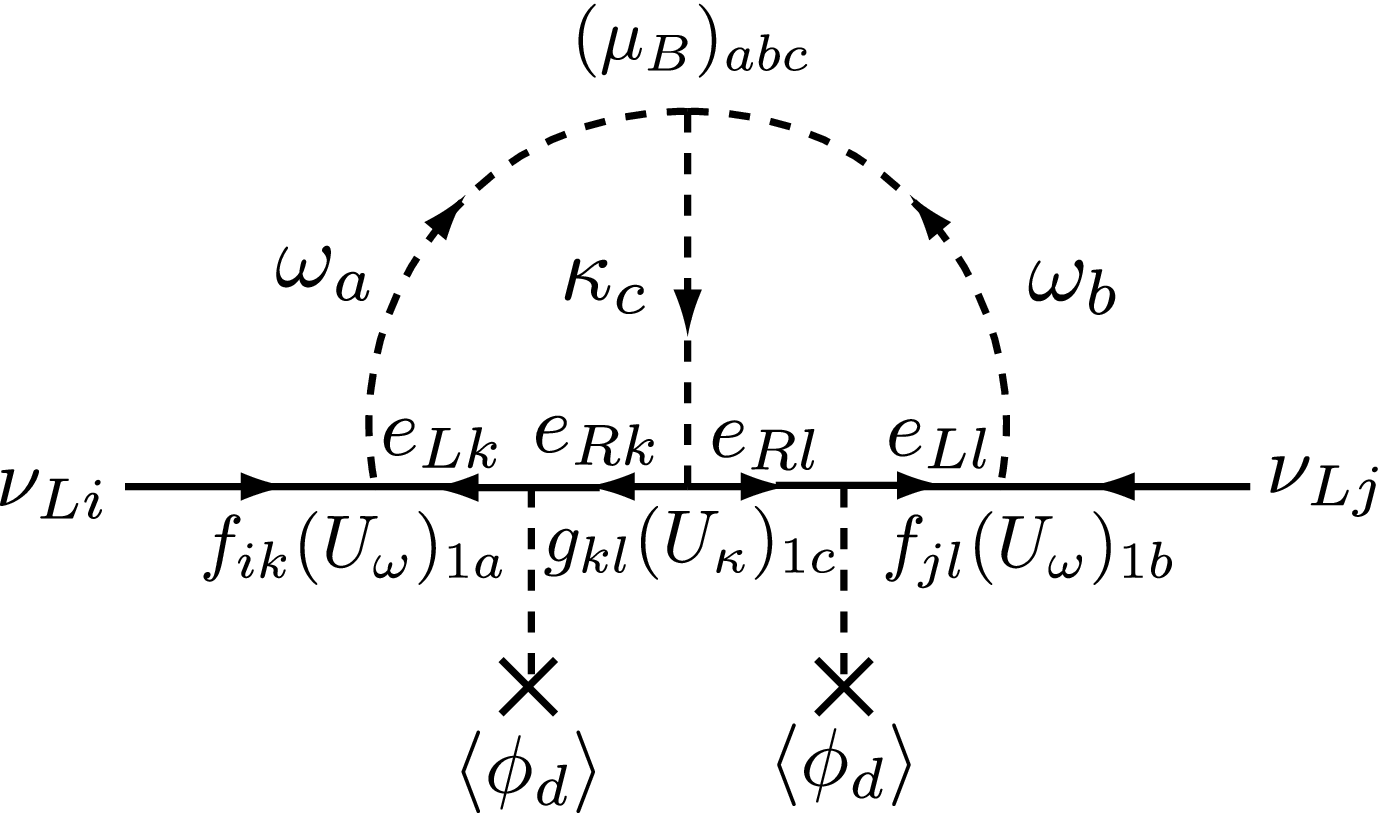}\\[2mm]
\raisebox{20mm}{(b)}&\phantom{space}&\includegraphics[scale=0.7]{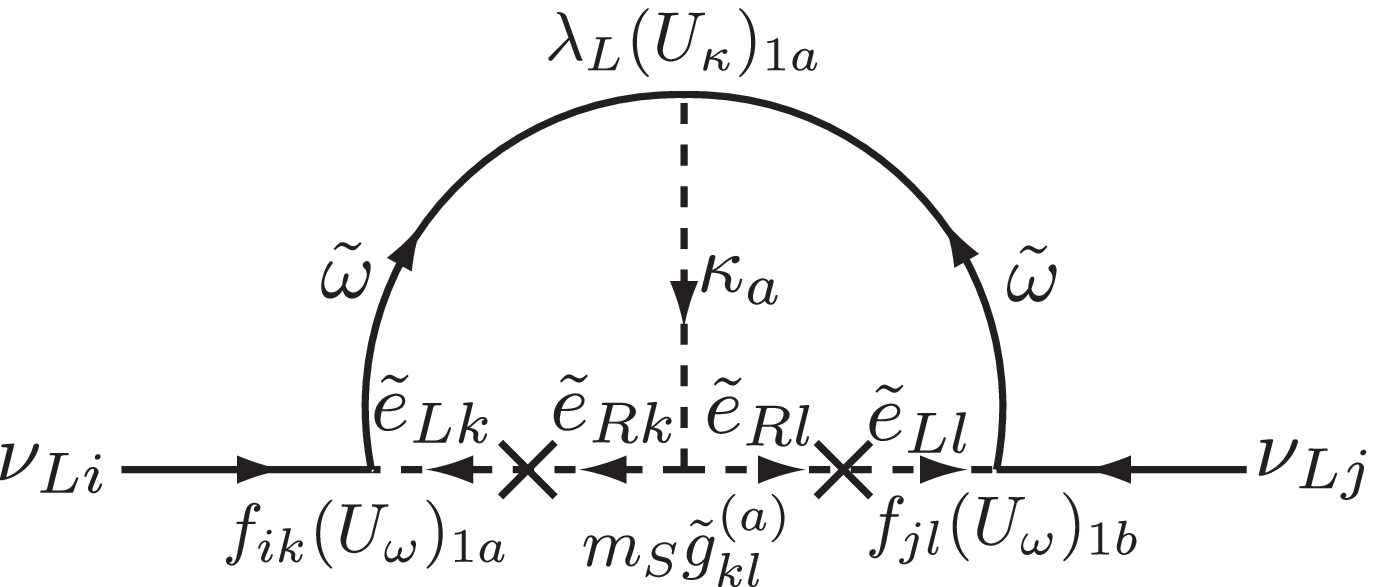}\\[2mm]
\raisebox{20mm}{(c)}&\phantom{space}&\includegraphics[scale=0.7]{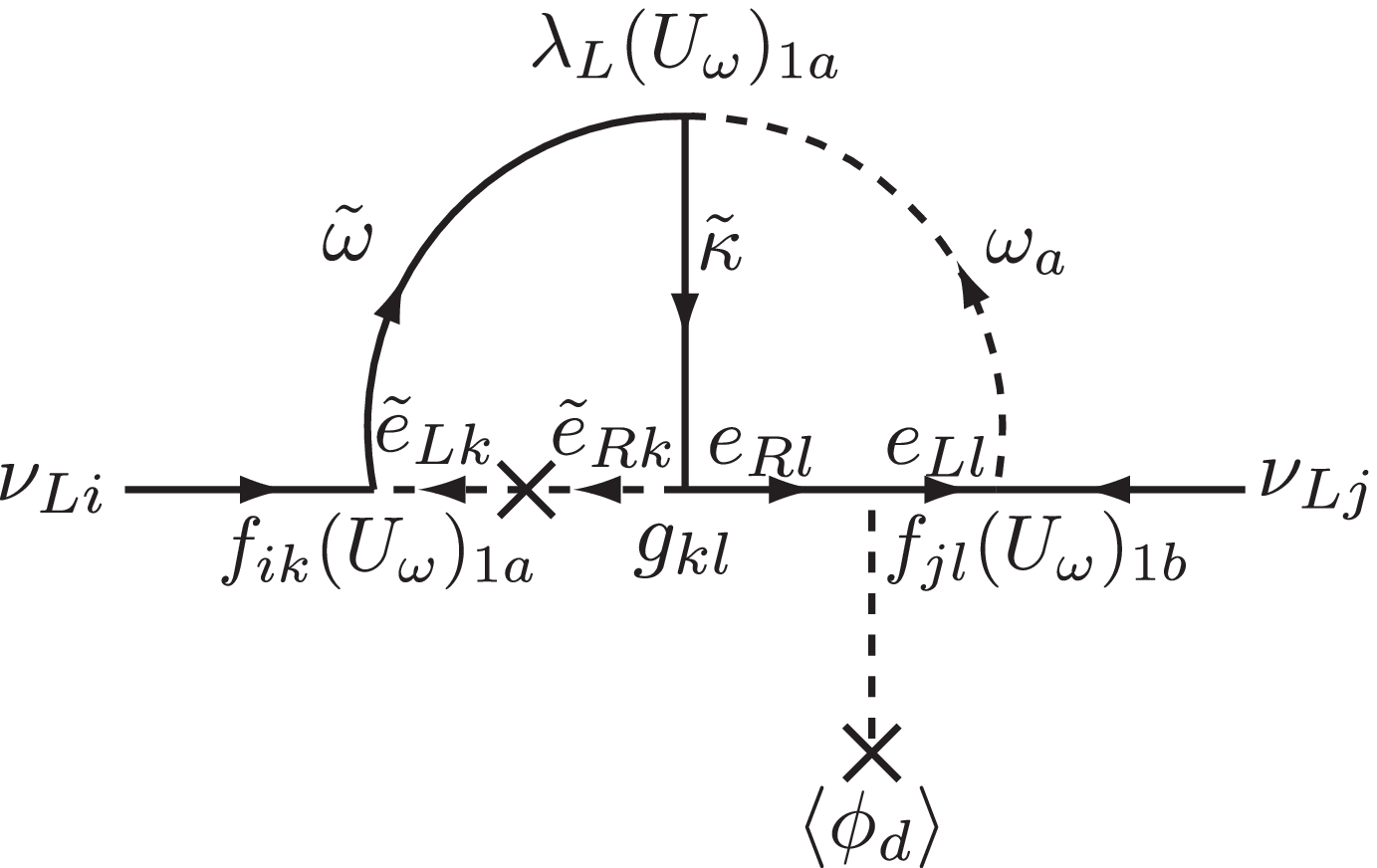}
\end{tabular}
\end{center}
\caption{The contributions to the neutrino mass generations. A type of a diagram (a) is 
the corresponding diagram to the non-SUSY Zee-Babu model. Diagrams (b) and (c) 
are new type of diagram in the SUSY model~\cite{aksy_szb}.}
\label{fig:diagram}
\end{figure}

The neutrino mass matrix is generated via 
the two-loop diagrams shown 
in Fig.~\ref{fig:diagram},
which can be written as
\begin{equation}
(m_{\nu}^{})_{ij}^{}=f_{ik}^{}(m_e)_{k}H_{kl}^{}(m_e)_{l}f_{jl}^{}\;,
\end{equation}
where the matrix $H_{kl}^{}$ is a symmetric matrix
\begin{align}
H_{kl}^{}=&
16(\mu_B^{})_{abc}^{}(U_{\omega}^{})^*_{1a}(U_{\omega}^{})^*_{1b} (U_{\kappa}^{})_{1c}^{}
g_{kl}^{} I((m_e^{})_{k}^{},(m_{\omega}^{})_a^{}|(m_e^{})_{l}^{},(m_{\omega}^{})_b^{}|(m_{\kappa})_c)
\nonumber\\
&
+16\frac{\lambda_{L}^*m_{\tilde{\omega}}^2}{m_S}(U_{\kappa}^{})^*_{1a}(U_{\kappa}^{})_{1a}^{}
\biggl\{
\frac{X_k^{}}{m_S^{}}
\tilde{g}_{kl}^{}
\frac{X_l^{}}{m_S^{}}
I((m_{\tilde{e}_R^{}}^{})_{k}^{},m_{\tilde{\omega}}^{}|(m_{\tilde{e}_{R}^{}})_l^{},m_{\tilde{\omega}}^{}|(m_{\kappa}^{})_a^{})
\nonumber\\
&\phantom{S}
+\frac{X_k^{}}{m_S^{}}
g_{kl}^{}
I((m_{\tilde{e}_{R}^{}})_k^{},m_{\tilde{\omega}}^{}|(m_{\tilde{e}_{L}^{}}^{})_l^{},m_{\tilde{\omega}}^{}|(m_{\kappa}^{})_a^{})
+g_{kl}
\frac{X_l^{}}{m_S^{}}
I((m_{\tilde{e}_{L}^{}}^{})_k^{},m_{\tilde{\omega}}^{}|(m_{\tilde{e}_{R}^{}}^{})_l^{},m_{\tilde{\omega}}^{}|(m_{\kappa}^{})_a^{})
\biggr\}
\nonumber\\
&
+\frac{8\lambda_{L}^{}m_{\tilde{\omega}}^{}m_{\tilde{\kappa}}^{}}{m_S^{}}(U_{\omega}^{})_{1a}^{}(U_{\omega}^{})_{1a}^*
\nonumber
\\
&\phantom{S}\times
\left\{
\frac{X_k^{}}{m_S^{}}
g_{kl}^{}
I((m_{\tilde{e}_{R}^{}}^{})_k^{},m_{\tilde{\omega}}^{}|(m_{e}^{})_l^{},(m_{\omega}^{})_a^{}|m_{\tilde{\kappa}}^{})
+
g_{kl}^{}
\frac{X_l^{}}{m_S^{}}
I((m_e^{})_{k}^{},(m_{\omega}^{})_a^{}|(m_{\tilde{e}_{R}^{}}^{})_l^{},m_{\tilde{\omega}}^{}|m_{\tilde{\kappa}}^{})
\right\}
\;,
\label{eq:Hkl}
\end{align}
where the indices $a,b,c$ run from 1 to 2, the mass eigenstates of the 
charged singlet scalars, 
$(m_{\tilde{e}_{R}^{}}^{})_i$ and $(m_{\tilde{e}_{L}^{}}^{})_i^{}$ are 
slepton masses,
the left-right mixing term in the slepton sector is parameterized as
$(m_{e}^{})_k^{}X_k^{}/m_S^{}$,
$I(m_{11}^{},m_{12}^{}|m_{21}^{},m_{22}^{}|M)$ is the loop function given in Eq.~(\ref{eq:I-func}), and
the other parameters are defined in the 
relevant Lagrangian as
\begin{align}
\mathcal{L}=&-2f_{ij}^{}(U_{\omega}^{})^*_{1a}\bar{\nu}^{ic}P_L^{}e^{j}\omega_a^*
-g_{ij}^{}(U_{\kappa}^{})_{1a}^{}\bar{e}_i^{}P_L^{}e^{jc}\kappa_a^{}
-2f_{ij}^{}\tilde{\nu}_{L}^{i*}\bar{\tilde{\omega}}P_L^{}e^{j}
-2f_{ij}^{}\bar{\nu}^{ic}P_L^{}\tilde{\omega}^c\tilde{e}_{L}^{j}
\nonumber\\
&
-2g_{ij}^{}\tilde{e}_{R}^{i*}\bar{e}^{j}P_L^{}\tilde{\kappa}
-\lambda_L^{}(U_{\kappa}^{})_{1a}^{}\bar{\tilde{\omega}}P_L^{}\tilde{\omega}^c\kappa_a^{}
-2 \lambda_L^{}(U_{\omega}^{})_{1a}^*\bar{\tilde{\omega}}P_L^{}\tilde{\kappa}\omega_a^*
-g_{ij}^{}(U_{\kappa}^{})_{1a}^{}(m_{e}^{})_j^{}\tilde{e}_{R}^{i*}\tilde{e}_{L}^{j*}\kappa_a^{}
\nonumber\\
&
-(\mu_B^{})_{abc}\omega_a^{}\omega_b^{}\kappa_c^*
-m_S^{}\tilde{g}_{ij}(U_{\kappa}^{})_{1a}\tilde{e}_{R}^{i*}\tilde{e}_{R}^{j*}\kappa_a^{}
+\text{h.c.}\;,
\end{align}
with
\begin{align}
(\mu_B)_{abc}\equiv&
A_L^*(U_{\omega}^{})_{1a}^{}(U_{\omega}^{})_{1b}^{}(U_{\kappa}^{})_{1c}^*
+A_R^{}(U_{\omega}^{})_{2a}^{}(U_{\omega}^{})_{2b}^{}(U_{\kappa}^{})_{2c}^*\;.
\end{align}
In the above expression, we assume that there is no flavour mixing in the
slepton sector.
In our model, there are two sources of the LFV processes.
One is the slepton mixing which also appear in the MSSM.
The other is the flavour mixing in the coupling with 
the charged singlet particles.
In order to concentrate on the latter contribution to the 
lepton flavour violating phenomena, the usual slepton mixing
effect is assumed to be zero.
The phenomenological constraints in our discussion strongly 
depend on this assumption.
If the assumption is relaxed, the phenomenological allowed 
parameters of the model can be changed to some extent.
Still we think our assumption is valuable to consider in order 
to obtain some definite physics consequences which are 
relevant to the new particles in our model.

\section{Allowed parameter region under the current constraint}

It is non-trivial whether there is an allowed parameter region in our
model except for the decoupling limit where masses of all the super partner
particles are set to be much larger than the electroweak scale.
Let us search for the parameter region where the neutrino mixing is consistent with the 
present oscillation data and the LFV constraints are satisfied.

Flavour violation in couplings between $\text{SU}(2)_{\text{L}}^{}$ singlet fields and leptons
should be large in order to generate large off-diagonal elements in the neutrino mass matrix.
These large flavour violation couplings enhance the LFV processes.
In particular doubly charged singlet scalar exchange tree level diagram contributes to the 
$e_i^+ \to e_j^+e_k^+e_l^- $ process. 
The predicted decay width of $e_i^+\to e_j^+e_k^+e_l^-$ in the model is calculated as\cite{macesanu,aristizabal}
\begin{equation}
\Gamma(e_i^+\to e_j^+e_k^+e_l^-)=
C_{jk}^{}\frac{1}{8}\frac{(m_e^{})_i^5}{192\pi^3}\left|
(U_{\kappa}^{})_{1a}^*(U_{\kappa}^{})_{1a}^{}\frac{g_{il}^{}g_{jk}^{*}}{(m_{\kappa}^{})_a^2}\right|^2\;,
\end{equation}
where $C_{jk}$ is a statistical factor as
\begin{equation}
C_{jk}=\begin{cases}
1&(j=k)\\
2&(j\neq k)
\end{cases}\;.
\end{equation}
There can be still large contributions to 
$\mu\to e\gamma$,
even if the constraint from $\mu^+\to e^+e^+e^-$ can be avoided.
The contribution is from one-loop diagrams. The decay width of $e_i\to e_j\gamma$ is evaluated as
\begin{equation}
\Gamma(e_i\to e_j \gamma)=\frac{\alpha_e}{4}(m_e^{})_i^5\left(|A_L^{ji}|^2+|A_R^{ji}
|^2\right)\;,
\end{equation}
with 
\begin{align}
A_L^{ji}=&
-\frac{1}{(4\pi)^2}
\left\{
(U_{\omega}^{})_{1a}^*(U_{\omega}^{})_{1a}^{}
\frac{4f_{kj}^*f_{ki}^{}}{(m_{\omega}^{})_a^2}F_2\left(\frac{(m_{\nu}^{})_k^2}{(m_{\omega}^{})_a^2}\right)
-\frac{4f_{kj}^*f_{ki}^{}}{(m_{\tilde{\nu}_{L}^{}})_k^2}F_1\left(\frac{m_{\tilde{\omega}}^2}{(m
_{\tilde{\nu}_{L}^{}}^{})_k^2}\right)
\right\}
\;,\\
A_R^{ji}=&
-\frac{1}{(4\pi)^2}
\left\{
(U_{\kappa})_{1a}^*(U_{\kappa})_{1a}
\frac{g_{kj}^{*}g_{ki}^{}}{(m_{\kappa}^{})_a^2}\left(2F_2\left(\frac{(m_{e}^{})_k^2}{(m_{\kappa}^{})_a^2}\right)
+F_1\left(\frac{(m_{e}^{})_k^2}{(m_{\kappa}^{})_a^2}\right)\right)
\right.
\nonumber\\
&\phantom{SpaceSpace}
\left.-\frac{g_{kj}^{*}g_{ki}^{}}{(m_{\tilde{e}_{R}^{}}^{})_k^2}
\left(
2F_1\left(\frac{m_{\tilde{\kappa}}^2}{(m_{\tilde{e}_{R}^{}}^{})_k^2}\right)
+F_2\left(\frac{m_{\tilde{\kappa}}^2}{(m_{\tilde{e}_{R}^{}}^{})_k^2}\right)\right)
\right\}\;,
\end{align}
where $(m_{\nu})_i$ are neutrino masses, and $(m_{\tilde{\nu}_L^{}})_i$ are sneutrino masses.
The loop functions $F_1(x)$ and $F_2(x)$ are\cite{inamilim} 
\begin{align}
F_1(x)=&\frac{x^2-5x-2}{12(x-1)^3}+\frac{x\ln x}{2(x-1)^4}\;,\\
F_2(x)=&\frac{2x^2+5x-1}{12(x-1)^3}-\frac{x^2\ln x}{2(x-1)^4}\;.
\end{align}
The coupling constants $f_{ij}$ only have nonzero values in flavour off-diagonal
 elements, and they tend to be large to reproduce the bi-large mixing.
Then the bound from  the data becomes severe.

Let us discuss how the LFV processes constrain the parameter space.
First of all, the tree level diagram contributing to the $\mu\to eee$ must be 
suppressed.
The present bound on the branching fraction  is
$B(\mu^+\to e^+e^+e^-)<1.0\times 10^{-12}$\cite{meee},
which gives very strong constraint on the model parameter space.
There are two possible cases to suppress the tree level contribution to the $\mu^+\to e^+e^+e^-$.
The first possibility is considering heavy doubly charged bosons $\kappa_1$ and $\kappa_2$.
If $g_{11}\sim g_{12}\sim 0.1$ is taken, the doubly charged bosons 
should be heavier than 15 TeV to avoid too large contribution.
The second option is suppressing a product of the couplings $|g_{12}g_{11}|$.
When the doubly charged bosons are $500$\;GeV, 
the upper bound on the product $|g_{12}g_{11}|$ is obtained as $|g_{12}g_{11}|<10^{-5}$.
The contributions to $\tau^+\to e^+e^+e^-$, 
$\tau^+\to e^+e^+\mu^-$, 
$\tau^+\to \mu^+\mu^+e^-$,
$\tau^+\to \mu^+\mu^+\mu^-$,
$\tau^+\to \mu^+ e^+e^-$, 
and
$\tau^+\to e^+\mu^+\mu^-$  can be computed in the same manner.
These flavour changing tau decays into three leptons are also enhanced in the model
with tree level contributions.
If future tau flavour experiments such as the high luminosity B factories\cite{superB} would discover a signal of such decays,
it could support the model.
In the phenomenological point of view, the scenario with a light doubly charged singlet scalar
is attractive because the scenario with such a light exotic particle is
testable at the LHC.
Therefore we have searched for a solution with a suppressed $|g_{12}g_{11}|$ and we have found that
the coupling $g_{11}$ can be taken to be so small that the tree level contribution 
to the $\mu^+ \to e^+e^+e^-$ process is negligible with reproducing the neutrino oscillation data.
In such a parameter space, the $B(\mu^+\to e^+e^+e^-)$ is suppressed by the electromagnetic coupling constant 
compared with $B(\mu\to e\gamma)$,
say $B(\mu^+\to e^+e^+e^-)\sim \alpha_eB(\mu\to e\gamma)$ where
the current upper limit is given by $B(\mu\to e\gamma) < 1.2\times
10^{-11}$ \cite{meg}.
The $B(\mu^+\to e^+e^+e^-)$ is below the experimental upper bound, if the constraint of $B(\mu\to e\gamma)$
is satisfied. 

In our analysis below, we work in the limit of 
$B_{\omega}^{}\mu_{\Omega}^{}\to 0$ and $B_{\kappa}^{}\mu_K^{}\to 0$ for simplicity.
If these terms are switched on, the mixings in the charged singlet 
scalar mass eigenstates take part in the neutrino mass generation.
However these mixings do not change our main results.
In this limit, the mixing matrices $U_{\omega}^{}$ and $U_{\kappa}^{}$ 
become  the unit matrix, and only $\omega_1^{}$ and $\kappa_1^{}$ 
contribute to the neutrino mass matrix and the LFV.
Below we simply write the relevant fields as 
$\omega\equiv\omega_1^{}$ and $\kappa\equiv\kappa_1^{}$,
and their masses are written as 
$m_{\omega}^{}\equiv (m_{\omega}^{})_1$ and 
$m_{\kappa}^{}\equiv (m_{\kappa}^{})_1$.

Following the above strategy, we search for an allowed 
parameter set.
An example of the allowed parameter sets is 
\begin{align}
&
f_{12}^{}=f_{13}^{}=\frac{f_{23}^{}}{2}=3.7\times 10^{-2}\;,\quad
\nonumber\\
&g_{11}^{}\simeq 0\;,\quad g_{12}^{}=4.8\times 10^{-7}\;,\quad
g_{13}^{}=2.1\times 10^{-7}\;,\quad
\nonumber\\
&
g_{22}^{}=-0.13\;,\quad
g_{23}^{}=6.1\times 10^{-3}\;,\quad
g_{33}^{}=-4.6\times 10^{-4}\;,\quad
\nonumber\\
&
\tilde{g}_{ij}^{}=g_{ij}^{}\;,\quad
\lambda_a^{}=1.0\;,\quad
\mu_B^{}=500\;\text{GeV}\;,\quad
\frac{X_k^{}}{m_S^{}}=1.0\;,\nonumber\\
&(m_{\tilde{e}_{L}^{}}^{})_k=(m_{\tilde{e}_{R}^{}}^{})_k=(m_{\tilde{\nu}_L^{}}^{})_k^{}=m_S^{}=1000\;\text{GeV}\;,\quad
\nonumber\\
&
m_{\omega}^{}=600\;\text{GeV}\;,\quad
m_{\tilde{\omega}}^{}=600\;\text{GeV}\;,\quad
m_{\kappa}^{}=300\;\text{GeV}\;,\quad
m_{\tilde{\kappa}}^{}=200\;\text{GeV}\;,\quad
\nonumber\\
&(m_{\omega}^{})_2^{}\gg m_{\omega}\;,\quad
(m_{\kappa}^{})_2^{}\gg m_{\kappa}\;.
\label{eq:benchmark}
\end{align}
On this benchmark point, the neutrino masses and mixing angles are given as
\begin{align}
&\sin^2\theta_{12}=0.33\;,\quad 
\sin^2\theta_{23}=0.5\;,\quad
\sin^2\theta_{13}=0.0\;,\nonumber\\
&\Delta m_{21}^2=7.6\times 10^{-5}\;\text{eV}^2\;,\quad
|\Delta m_{31}^2|=2.5\times 10^{-3}\;\text{eV}^2\;,
\end{align}
which are completely consistent with the present neutrino data:
the global data analysis\cite{neutrino-oscillation} of the neutrino oscillation experiments provide 
$\sin^2\theta_{12}=0.318^{+0.019}_{-0.016}$, 
$\sin^2\theta_{23}=0.50^{+0.07}_{-0.06}$, 
$\sin^2\theta_{13}=0.013^{+0.013}_{-0.009}$, 
$\Delta m_{21}^2=(7.59^{+0.23}_{-0.18})\times 10^{-5}\;\text{eV}^2$,
and
$|\Delta m_{31}^2|=(2.40^{+0.12}_{-0.11})\times 10^{-3}\;\text{eV}^2$.
Based on this benchmark point, our model predicts
$B(\mu\to e\gamma)=1.1\times 10^{-11}$ 
and $B(\tau^+\to  \mu^+ \mu^+ \mu^-)=1.3\times 10^{-8}$, 
both of which are just below the present experimental bounds.


\section{Phenomenology at the LHC}
We turn to discuss collider phenomenology in  the model assuming
the parameters of the benchmark scenario given in Eq.~(\ref{eq:benchmark}).
In our model, the new $\text{SU}(2)_{\text{L}}^{}$ charged singlet fields
are introduced, which can be accessible at collider experiments such
as the LHC unless they are too heavy.
In particular, the existence of the doubly charged singlet scalar
boson and its SUSY partner fermion (the doubly charged singlino)
provides discriminative phenomenological signals.
They are produced in pair ($\kappa^{++}\kappa^{--}$ or
$\tilde{\kappa}^{++}\tilde{\kappa}^{--}$) and 
each doubly charged boson (fermion) can be observed as 
a same-sign dilepton event, 
which would be a clear signature.
In this Letter, we focus on such events including doubly charged particles.
For the benchmark point given in Eq.~(\ref{eq:benchmark}),
almost all the $\kappa$ decays into the same-sign muon pair,
$\kappa^{\pm\pm}\to \mu^{\pm}\mu^{\pm}$.

\begin{figure}
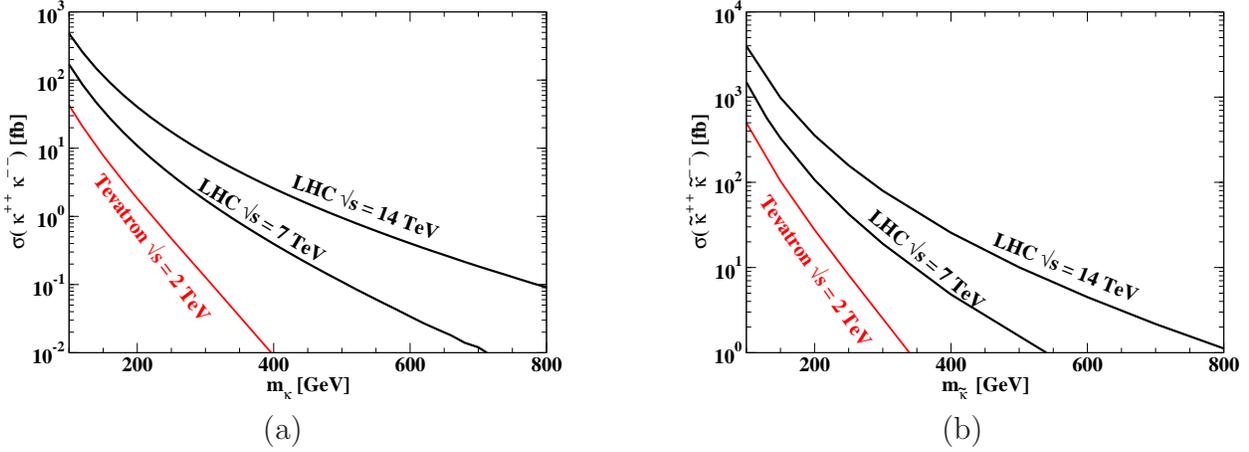

\begin{center}
\begin{tabular}{ccc}
\includegraphics[scale=0.3]{xs_k.eps}&\phantom{spac}
&\includegraphics[scale=0.3]{xs_kino.eps}\\
(a)&&(b) 
\end{tabular}
\end{center}
\caption{Production cross sections of 
 (a) $\kappa^{++}\kappa^{--}$ and   
 (b) $\tilde{\kappa}^{++}\tilde{\kappa}^{--}$,
via Drell-Yan processes at the LHC ($pp$)
 and the Tevatron ($p\overline{p}$).
 The production cross section at the LHC is evaluated 
for $\sqrt{s}=7\;\text{TeV}$ and $\sqrt{s}=14\;\text{TeV}$~\cite{aksy_szb}.
}
\label{fig:xs}
\end{figure}
At hadron colliders such as the LHC and the Tevatron,
the doubly charged singlet scalar $\kappa$ and the doubly charged singlino 
$\tilde{\kappa}$ are produced dominantly  in pair through 
the Drell-Yang processes. 
The production cross sections for 
$\kappa^{++}\kappa^{--}$ and 
$\tilde{\kappa}^{++}\tilde{\kappa}^{--}$ are shown as
in Fig.~\ref{fig:xs}(a) and Fig.~\ref{fig:xs}(b), respectively.
The first two plots from above correspond to the cross sections
at the LHC of $\sqrt{s}=14$ TeV and $\sqrt{s}=7$ TeV, and the
lowest one does to that at the Tevatron of $\sqrt{s}=2$ TeV.
We note that magnitudes of the production cross sections
for the pair of singly-charged singlet scalars  
$\omega^+\omega^-$ and that of singly-charged singlinos
$\tilde{\omega}^+\tilde{\omega}^-$ are (1/4) smaller than
those for $\kappa^{++}\kappa^{--}$ and 
$\tilde{\kappa}^{++}\tilde{\kappa}^{--}$ for the common mass
for produced particles.
%
%
At the LHC with $\sqrt{s}=7$ TeV with the integrated luminosity
${\cal L}$ of 1 fb$^{-1}$, about 100 of $\tilde{\kappa}^{++}\tilde{\kappa}^{--}$ pairs
can be produced when $m_{\tilde{\kappa}}^{} = 200$ GeV, while only
a couple of the $\kappa^{++}\kappa^{--}$ pair is expected for $m_\kappa^{} =300$ GeV.

\begin{figure}
\begin{center}
\includegraphics[scale=0.49]{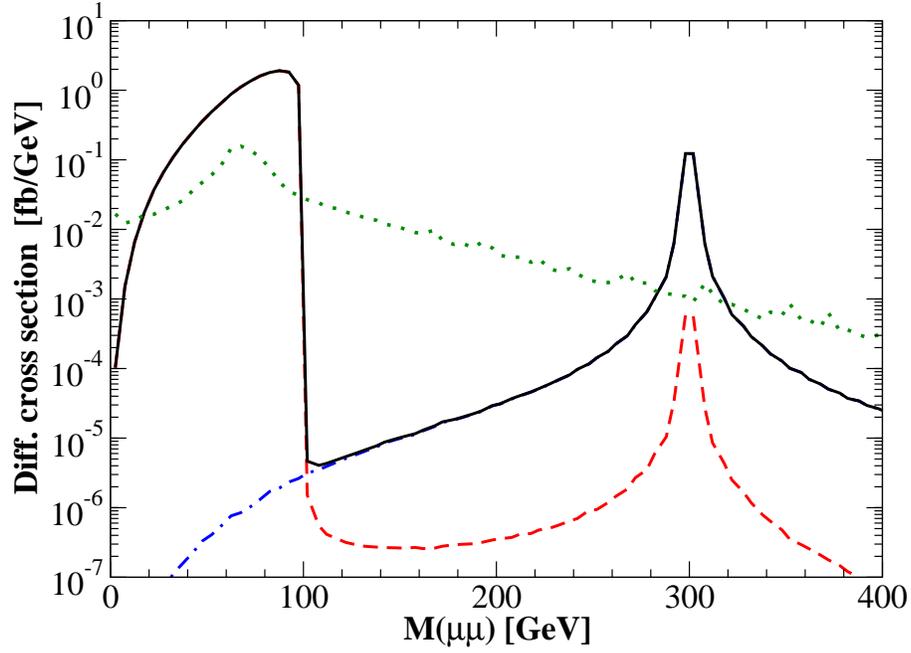}
\end{center}
\caption{
The invariant mass distribution of the same-sign dilepton event. 
The benchmark point in Eq.~(\ref{eq:benchmark}) is used
and the neutralino mass is taken as $m_{\tilde{\chi}^0}=100\;\text{GeV}$.
The dashed (red) curve corresponds to the events from 
$pp\to \tilde{\kappa}^{++} \tilde{\kappa}^{--}
   \to \tilde{\chi}^0 \kappa_1^{++} \tilde{\chi}^0 \kappa_1^{--} \to \tilde{\chi}^0
   \tilde{\chi}^0  \mu^+\mu^+\mu^-\mu^-$.
The dot-dashed (blue) curve shows the contributions from 
$pp\to \kappa^{++} \kappa^{--} \to \mu^+\mu^+\mu^-\mu^-$.
The solid (black) curve denotes total events from the both signal processes.
The dotted (green) curve shows the background events. 
For kinematical cut, see the text~\cite{aksy_szb}.
}
\label{fig:Mmm}
\end{figure}

In Fig.~\ref{fig:Mmm}, the distribution of the differential cross section for
four muon (plus a missing transverse momentum) final states as a function of 
the invariant mass $M(\mu^+\mu^+)$ of the same-sign muon pair is shown
assuming the bench mark scenario in Eq.~(\ref{eq:benchmark}) at the LHC with
$\sqrt{s}=7$ TeV.
In order to suppress background events, 
we select the muon events with 
the transverse momentum larger than 20\,GeV and 
the pseudo-rapidity less than 2.5.
The signal events come from both
$pp\to \kappa^{++} \kappa^{--} \to \mu^+\mu^+\mu^-\mu^-$ and
$pp\to \tilde{\kappa}^{++} \tilde{\kappa}^{--}
   \to \tilde{\chi}^0 \kappa_1^{++} \tilde{\chi}^0 \kappa_1^{--} \to \tilde{\chi}^0
   \tilde{\chi}^0  \mu^+\mu^+\mu^-\mu^-$.
The $M(\mu^+\mu^+)$ distribution can be a key to explore
the phenomena with the doubly charged particles. 
The doubly charged scalar mass and the mass difference between the
doubly charged singlino and the neutralino are simultaneously determined at the
LHC. A sharp peak is expected in the $M(\mu^+\mu^+)$ distribution at
$M(\mu^+\mu^+) = m_{\kappa}^{}$, because the same-sign muon pair 
 from the $\kappa$ decay is not associated with missing particles.
On the other hand, the doubly charged singlino decays as
$\tilde{\kappa}^{--}\to \tilde{\chi}^0\kappa^{--} \to \tilde{\chi}^0\mu^-\mu^-$ 
in the case that the lightest R-parity odd particle is a neutralino,
$\tilde{\chi}^0$, which is a DM candidate in the model.
In this Letter, we just assume that the LSP neutralino is Bino-like. 
In our analysis, we fix the neutralino mass as $m_{\tilde{\chi}^0}^{}=100\;\text{GeV}$.
The mass difference between $\tilde{\kappa}$ and $\tilde{\chi}^0$
can be measured by looking at a kink at
$M(\mu^+\mu^+)=m_{\tilde{\kappa}}^{}-m_{\tilde{\chi}^0}^{}$ in the $M(\mu^+\mu^+)$ distribution.
The main background comes from four muon events from the SM processes
where muons are produced via the $ZZ$, $\gamma\gamma$ and $\gamma Z$ production, 
or a pair production of muons with the $Z$ or $\gamma$ emission.
The expected background is also shown in Fig.~\ref{fig:Mmm}.
The events from signal dominate those from the background in the area
of $M(\mu^+\mu^+) < m_{\tilde{\kappa}}^{}-m_{\tilde{\chi}^0}^{}$ and
at around $M(\mu^+\mu^+) \sim m_{\kappa}^{}$.
The background events have been evaluated by using CalcHEP\cite{CalcHEP}. 
From this rough evaluation, one may expect that the event from the
signal can be identified even at the LHC with $\sqrt{s}=7$ TeV
and ${\cal L}=1$ fb$^{-1}$.
As for the case with $\sqrt{s}=14\,\text{TeV}$, 
the signal to background ratio becomes
larger and it will be more promising to explore our model.

There are other models in which the same-sign dilepton events are predicted.
The model with the complex triplet scalar fields is an example of 
such a class of models\cite{HTM_delm0, HTM_delm0_2, HTM_delm0_3, Kadastik:2007yd, Perez:2008ha}.
They can in principle be distinguished by looking at the decay products
from doubly charged fields.
In our scenario, $\kappa^{\pm\pm}$ can mainly decay into  
$\mu^{\pm}\mu^{\pm}$, 
while in the triplet models where the decay of doubly charged singlet 
scalars are
directly connected with the neutrino mass matrix, there is no solution
where only the $\mu^{\pm}\mu^{\pm}$ mode can be dominant decay mode.
The difference in such decay pattern can be used to discriminate our model
from the triplet models.

\newpage
\chapter{Models with the $Y=3/2$ doublet scalar field}

In this section, we consider the extended Higgs models, in which one of the isospin 
doublet scalar fields carries the hypercharge $Y=3/2$. 
Such a doublet field $\Phi_{3/2}$ is composed of a doubly charged scalar 
boson as well as a singly charged one.  

Contrary to the $Y=1$ triplet field $\Delta$ 
as well as the $Y=2$ singlet $S^{++}$, 
the Yukawa coupling between $\Phi_{3/2}$ 
and charged leptons is protected by the chirality.  
In addition, the component 
fields of $\Phi_{3/2}$ are both charged and do not receive a 
vacuum expectation value (VEV) as long as electric charge is conserved.  
Hence, the field decays via the mixing with the other scalar 
representations which can decay into the SM particles 
or via some higher order couplings.
This characteristic feature of $\Phi_{3/2}$ 
would give discriminative predictions at collider experiments. 
We therefore first study collider signatures of 
$\Phi_{3/2}$ at the LHC in the model (Model~I) of 
an extension from the SM with an extra $Y=1/2$ 
doublet and $\Phi_{3/2}$.  

We then consider a new model for radiatively generating neutrino 
masses with a dark matter candidate (Model~II), 
in which $\Phi_{3/2}$ and an extra $Y=1/2$ doublet 
as well as vector-like singlet fermions carry 
the odd quantum number for an unbroken discrete $Z_2$ symmetry. 
We also discuss the neutrino mass model (Model~III),  
in which  the exact $Z_2$ parity  in Model~II is softly broken. 

\section{Model I} 
The simplest model, 
where $\Phi_{3/2}$ is just added to the SM,  
can decay 
into SM particles only if lepton-number violating  
higher order operators are introduced~\cite{32_higer_order}. 
Thus, we here consider the model in which 
$\Phi_{3/2}$ is added to the model with two $Y=1/2$ Higgs 
doublet fields $\phi_1$ and $\phi_2$ (Model~I). 
The singly charged scalar state in $\Phi_{3/2}$
can decay into the SM particles via the mixing with 
the physical charged state from the $Y=1/2$ doublets. 
This model can be regarded as an effective theory of Model~III 
which we discuss later, or 
it may be that of the model with an additional heavier $\Delta$,    
in which the gauge coupling unification would be possible. 
In order to avoid flavor changing neutral current, 
a softly-broken $Z_2$ symmetry is imposed~\cite{GW}, under which 
the scalar fields are transformed as $\phi_1\to \phi_1$, 
$\phi_2 \to - \phi_2$, and $\Phi_{3/2} \to - \Phi_{3/2}$. 

The most general scalar potential is given by 
\begin{align}
&V=\sum_{i=1}^2 \mu_i^2|\phi_i|^2+(\mu_3^2\phi_1^\dagger \phi_2+\text{h.c.})
+\sum_{i=1}^2\frac{1}{2}\lambda_i|\phi_i|^4 
+\lambda_3|\phi_1|^2|\phi_2|^2+\lambda_4|\phi_1^\dagger\phi_2|^2
+\frac{1}{2}[\lambda_5(\phi_1^\dagger\phi_2)^2+\text{h.c.}]\notag\\
&+\mu_\Phi^2 |\Phi_{3/2}|^2+\frac{1}{2}\lambda_\Phi|\Phi_{3/2}|^4
+\sum_{i=1}^2 \rho_i|\phi_i|^2|\Phi_{3/2}|^2 
+\sum_{i=1}^2 \sigma_i |\phi_i^\dagger\Phi_{3/2}|^2 
+[\kappa(\Phi_{3/2}^\dagger\phi_1)(\phi_2\cdot \phi_1)+\text{h.c.}],
\label{potential}
\end{align}
where the $Z_2$ symmetry is softly broken at the $\mu_3^2$ term.  
We neglect the CP violating phase for simplicity.
The scalar doublets $\phi_1$, $\phi_2$ and $\Phi_{3/2}$ are 
parameterized as 
\begin{align}
\hspace{-0.2cm}\phi_i\!=\! \left[\begin{array}{c}
w_i^+ \\
\frac{1}{\sqrt{2}}(h_i+v_i+iz_i)
\end{array}\right]  (i=1,2), \; 
\Phi_{3/2} \!=\! \left[\begin{array}{c}
\Phi^{++} \\
\Phi^+
\end{array}\right], \nonumber
\end{align}
where the VEVs $v_i$ satisfy $\sqrt{v_1^2+v_2^2} = v \simeq 246$ GeV. 
Mass matrices for the neutral components are diagonalized 
as in the same way as 
those in the usual two Higgs doublet model (2HDM) with $\phi_1$ and $\phi_2$.
The mass eigenstates $h$ and $H$ for CP-even states are 
obtained by diagonalizing the mass matrix by the angle $\alpha$.  
By the angle $\beta$ ($\tan\beta \equiv v_2/v_1$), 
the mass eigenstates for the CP-odd states $z$ and $A$ 
are obtained, where $z$ is the NG boson 
and $A$ is the CP-odd Higgs boson. 
For simplicity $\sin(\beta-\alpha)=1$ is taken such that 
$h$ is the SM-like Higgs boson\cite{SMlike_thdm, KOSY}.
The existence of $\Phi_{3/2}$ affects the singly charged scalar sector. 
The mass eigenstates 
are obtained by mixing angles $\beta$ and $\chi$ as 
\begin{align}
\left[
\begin{array}{c}
w^\pm\\
H_1^\pm\\
H_2^\pm
\end{array}\right] 
=  
\left[
\begin{array}{ccc}
1 & 0 &0\\
0 & c_\chi & s_\chi \\
0&  -s_\chi &  c_\chi
\end{array}
\right] 
\left[
\begin{array}{ccc}
c_\beta & s_\beta & 0\\
-s_\beta & c_\beta & 0 \\
0&  0          & 1
\end{array}\right]
\left[
\begin{array}{c}
w_1^\pm\\
w_2^\pm\\
\Phi^\pm
\end{array}\right], \label{m+}
\end{align}
where $c_\theta=\cos\theta$ and $s_\theta=\sin\theta$, 
$w^\pm$ are the NG bosons   
absorbed by the longitudinal component of the $W^\pm$ bosons. 
$H_1^\pm$ and $H_2^\pm$  are physical mass eigenstates with 
the masses $m_{H_1^\pm}$ and $m_{H_2^\pm}$.

The Yukawa couplings for charged states are given by  
\begin{align}
&\mathcal{L}_Y = 
-\frac{\sqrt{2} V^{ij}_{\rm KM}}{v}
\bar{u}^i (m_{u^i} \xi_A^u d_L^j +m_{d^j}\xi_A^d d_R^j) \phi^+ 
\notag \\&
-\frac{\sqrt{2} m_{\ell^i}\xi_A^\ell}{v}\bar{\nu}^i \ell_R^i
\phi^+
  +\text{h.c.}, \;\; (\phi^+=H_1^+\cos\chi-H_2^+\sin\chi), \nonumber
\end{align}
where the coupling parameters $\xi^{u,d,\ell}_A$~\cite{typeX} 
depend on the $Z_2$ charges of quarks and leptons~\cite{Barger, Grossman}.
\begin{figure}[t]
\begin{center}
\includegraphics[width=80mm]{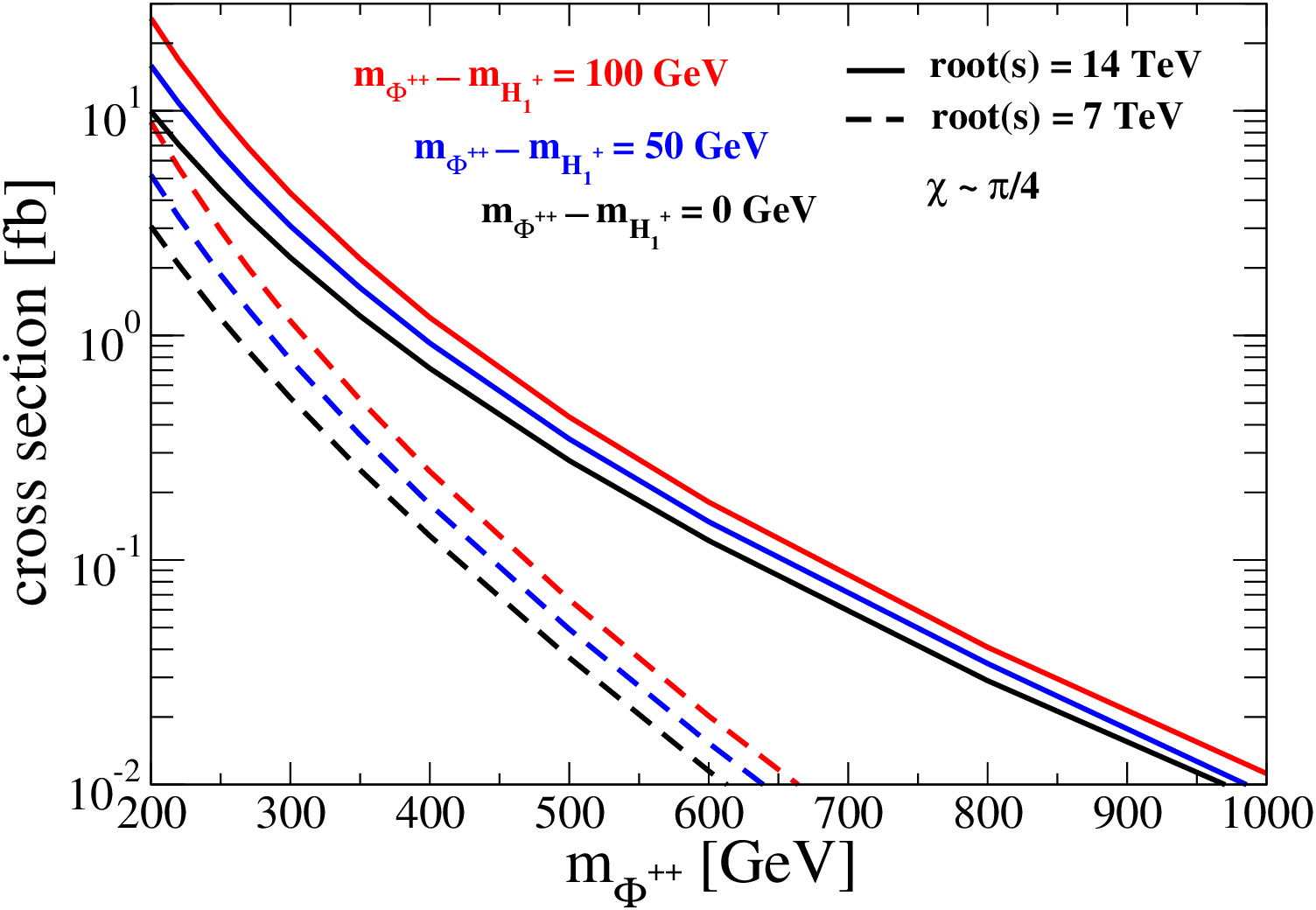}
\caption{Cross section of $pp \to X W^{+\ast} \to X \Phi^{++} H_1^-$~\cite{aky_32}. 
}\end{center}
\label{cs}
\end{figure} 
We are interested in the light charged scalar bosons 
such as $\mathcal{O}(100)$ GeV. 
To satisfy the $b\to s\gamma$ data~\cite{Ref:BELLE, Ref:BABAR, Ref:CLEO, Ref:HFAG}, 
we choose 
the Type-I Yukawa interaction with $\tan\beta\gtrsim 2$. 
Assuming $m_{H_{1,2}^\pm} < m_t+m_b$ and $m_{H^\pm_2}-m_{H^\pm_1} < m_Z$,  
the branching ratios for the main decay modes are evaluated as 
$B(H_{1,2}^\pm \to \tau^\pm \nu) \sim 0.7$ 
and $B(H_{1,2}^\pm \to cs) \sim 0.3$ when $\chi \neq  0$.  

\begin{figure}[t]
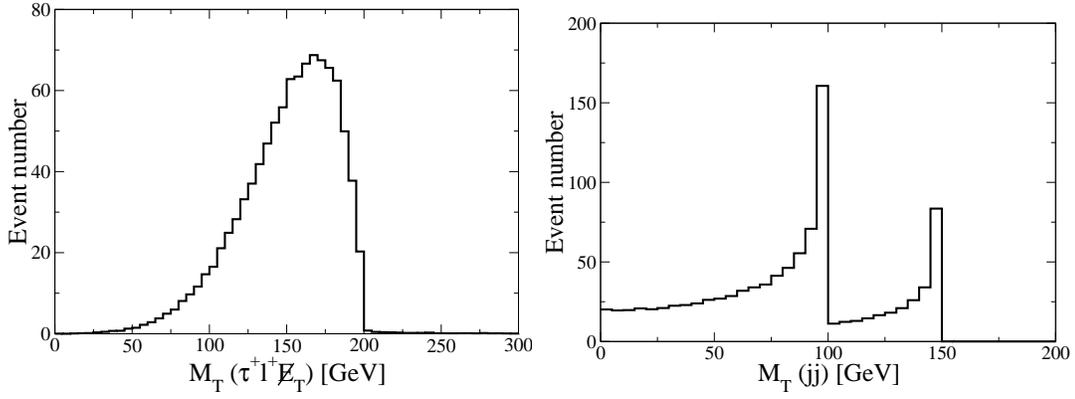

\begin{center}
\includegraphics[width=70mm]{FIG2a.eps} 
\includegraphics[width=70mm]{FIG2b.eps}
\caption{(Left) 
The transverse mass distribution for the 
$\tau^+\ell^+E_T {\!\!\!\!\!\!\!/} \hspace{3mm}$ system 
for the signal. 
(Right) That for the $jj$ system. 
The event number is taken to be 1000 for both figures~\cite{aky_32}. 
}\label{fig_jacobian-peak}
\end{center}
\end{figure}

At the LHC, $\Phi^{++}$ can be tested by using 
various processes such as the pair production and  
the associated production with $H_1^{-}$ or $H_2^{-}$. 
We here discuss an interesting signal 
via the process $u \overline{d} \to W^{+\ast} \to \Phi^{++} H_{1,2}^-$. 
The cross section is shown in Fig.~\ref{cs}.   
We may examine this process, for example, by the decay 
$\Phi^{++} \to H_{1,2}^+W^+ \to \tau^+\ell^+\nu\nu$ with
$H_{1,2}^- \to jj$, 
when $m_{\Phi^{\pm\pm}} > m_{H_1^{\pm}} + m_W$ with 
$m_{H_{1,2}^\pm} < m_t+m_b$ and $m_{H_2^{\pm}} - m_{H_1^{\pm}} < m_Z$, 
where $m_{\Phi^{\pm\pm}}$ is the mass of $\Phi^{\pm\pm}$.     
The signal is then $\tau^+\ell^+ jj$ plus a missing transverse 
momentum $E_T {\!\!\!\!\!\!\!/}\hspace{3mm}$ ($\ell^+=e^+$ or $\mu^+$). 
The signal cross section for 
$\tau^\pm\ell^\pm jj E_T {\!\!\!\!\!\!\!/}\hspace{3mm}$ 
is evaluated as 
4.0 fb (1.3 fb) for $\sqrt{s} 
=14$ TeV ($7$ TeV) 
for $m_{H_1^\pm}=100$ GeV, $m_{H_2^\pm}=150$ GeV, 
$m_{\Phi^{\pm\pm}}=200$ GeV and $\chi \simeq \pi/4$.

The mass for $\Phi^{++}$ can be determined from 
the Jacobian peak~\cite{Jacobian} in the distribution of the 
transverse mass, 
$M_T(\tau^+\ell^+E_T {\!\!\!\!\!\!\!/} \hspace{3mm}) 
= \sqrt{2 p_T^{\tau\ell} E_T {\!\!\!\!\!\!\!/} \hspace{3mm}
(1-\cos\varphi)}$, 
where $\varphi$ is the azimuthal angle between the transverse  
momentum $p_T^{\tau\ell}$ of the dilepton system and 
$E_T {\!\!\!\!\!\!\!/}\hspace{3mm}$. 
We show numerical results for the scenario with 
$m_{H_1^\pm}=100$ GeV, $m_{H_2^\pm}=150$ GeV, $m_{\Phi^{\pm\pm}}=200$ GeV, 
$\chi \simeq \pi/4$, $\tan\beta=3$, $\sin(\beta-\alpha)=1$ and 
$m_{H} = m_A = 127$ GeV, where $m_{H}$ and $m_A$ represent 
the masses of $H$ and $A$, respectively. 
The potential is then approximately custodial symmetric, 
so that the rho parameter constraint is satisfied with 
the mass of the SM-like Higgs boson $h$ to be 120 GeV. 
The end point in Fig.~\ref{fig_jacobian-peak}~(Left) 
indicates $m_{\Phi^{\pm\pm}}$, where the event number is 
taken to be 1000.   
One might think that the final decay products 
from the $\tau$ lepton should be discussed.  
We stress that the endpoint at $m_{\Phi^{++}}$ 
also appears in the distribution of 
$M_T(\ell^+\ell^+E_T {\!\!\!\!\!\!\!/} \hspace{3mm})$  
obtained from the leptonic decay of the $\tau^+$. 
The cross section for the 
$\ell^+\ell^+jjE_T {\!\!\!\!\!\!\!/} \hspace{3mm}$ signal 
is about 1.3 fb for $\sqrt{s}=14$ TeV (0.45 fb for $\sqrt{s}=7$ TeV).  
Furthermore, masses of singly charged Higgs bosons can also be 
measured by the distribution of $M_T(jj)$. 
In Fig.~\ref{fig_jacobian-peak}~(Right), 
the two Jacobian peaks at 100 and 150 GeV 
correspond to $m_{H_1^+}$ and $m_{H_2^+}$, respectively, where 
the event number is taken to be 1000. 
The SM background for $\ell^+\ell^+jjE_T {\!\!\!\!\!\!\!/} \hspace{3mm}$, 
which mainly comes from $u\bar{d}\to W^+W^+ jj$, 
is 3.95 fb for $\sqrt{s}= 14$ 
TeV (0.99 fb for $\sqrt{s}=7$ TeV). The cross section of the  
background is comparable to that for the signal before kinematic cuts. 
There is no specific kinematical structure in the $\ell^+\ell^+$ 
distribution in the background.   
All the charged scalar states can be measured simultaneously 
via this process unless their masses are too heavy 
if sufficient number of the signal event 
remains after kinematic cuts. 
While the detection at the LHC with 300 fb$^{-1}$ may be challenging, 
it could be much better at the upgraded version of the LHC with 3000 fb$^{-1}$.  

\section{Model II} 
We here present a new model   
in which $\Phi_{3/2}$ is introduced to naturally generate 
tiny neutrino masses at one-loop level. 
To this end, we again consider the scalar sector 
with $\phi_1$, $\phi_2$ and $\Phi_{3/2}$. 
In addition, we introduce two isospin singlet Dirac fermions $\psi^a$ 
($a=1,2$) with $Y=-1$. 
We impose the exact (unbroken) $Z_2$ parity, under which 
$\phi_2$, $\Phi_{3/2}$ and $\psi^a$ are odd while 
all the SM particles including $\phi_1$ are even.  
This $Z_2$ parity plays a role to forbid 
mixing terms of $\overline{\ell}_R \psi_L^a$ 
as well as couplings of $\overline{L}_L \phi_1 \psi_R^a$ and 
$\overline{L}_L \phi_2 \ell_R$, and to 
 guarantee the stability of a dark matter candidate; 
i.e., the lightest neutral $Z_2$ odd particle. 
Lepton numbers $L=-2$ and $+1$ are respectively assigned to 
$\Phi_{3/2}$ and $\psi^a$.

The scalar potential coincides that in Eq.~(\ref{potential})  
but $\mu_3^2=0$ due to the exact $Z_2$ parity. 
Without $\Phi_{3/2}$, 
the scalar sector is that of the  
inert doublet model~\cite{dark_higgs}, in which only $\phi_1$ receives 
the VEV yielding the SM-like Higgs boson $h$,  
while $\phi_2$ gives $Z_2$-odd scalar bosons $H$, $A$ and $H^\pm$. 
Including $\Phi_{3/2}$, 
$H^\pm$ can mix with $\Phi^{\pm}$ 
diagonalized by the angle $\chi$ 
in Eq.~(\ref{m+}) with $\beta=0$.
Masses and interactions for $\psi^a$ are given by 
\begin{align}
\mathcal{L}_Y= 
m_{\psi^a} \bar{\psi}_L^a \psi_R^a
+f_i^a\overline{(L_L^i)^c}\cdot\Phi_{3/2}\psi_L^a
+g_i^a\overline{L_L^i}\phi_2\psi_R^a + {\rm h.c.}. \label{yukawa_model2}
\end{align}

\begin{figure}[t]
\begin{center}
\includegraphics[width=100mm]{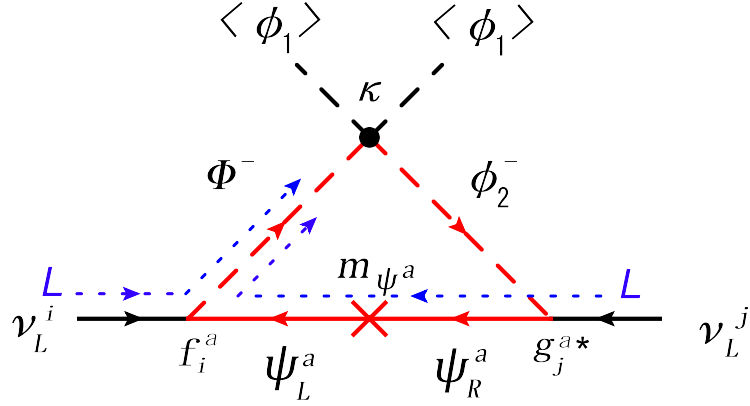}
\caption{Neutrino mass diagram~\cite{aky_32}.}\label{n_mass}
\end{center}
\end{figure}

The neutrino masses are generated via the one-loop diagram 
in Fig.~\ref{n_mass}. 
The flow of the lepton number is also indicated in the figure. 
The source of lepton number violation (LNV) is the coupling $\kappa$. 
This is similar to the model by Zee~\cite{zee}, although 
the diagram looks similar to the model by Ma~\cite{Ma:2006km} where 
Majorana masses of right-handed neutrinos $\nu_R$ 
is the origin of LNV.   
For $m_{\psi^a} \gg m_{H_1^\pm},m_{H_2^\pm}$, the mass matrix can be calculated as 
\begin{align}
(\mathcal{M}_\nu)_{ij}&\simeq \sum_{a=1}^2
\frac{1}{16\pi^2}\frac{1}{2m_{\psi^a}}(f_i^a g_j^{a*}+f_j^a g_i^{a*})
\frac{v^2\kappa}{m_{H_2^\pm}^2-m_{H_1^\pm}^2}\notag\\
& \times \left(m_{H_2^\pm}^2\log\frac{m_{\psi^a}^2}{m_{H_2^\pm}^2}
-m_{H_1^\pm}^2\log\frac{m_{\psi^a}^2}{m_{H_1^\pm}^2}\right). \nonumber
\end{align} 
For $m_\psi\sim 1$ TeV, 
$m_{H_1^\pm} \sim m_{H_2^\pm} \sim {\cal O}(100)$~GeV, and 
$f_i^a \sim g_i^a \sim \kappa \sim {\cal O}(10^{-3})$, 
the scale of neutrino masses ($\sim 0.1$ eV) can 
be generated.  The bound from LFV processes  
such as $\mu\to e\gamma$~\cite{meg} can easily be satisfied.
The neutrino data can be reproduced by introducing at least 
two fermions $\psi^1$ and $\psi^2$.
The lightest $Z_2$ odd neutral Higgs boson
(either $H$ or $A$) is a dark matter candidate~\cite{dark_higgs}. 
Assuming that $H$ is the lightest,  
its thermal relic abundance
can explain the WMAP data~\cite{DM} 
by the s-channel process 
$HH\to h \to b \overline{b}$ (or $\tau^+\tau^-$).  
The t-channel process 
$HH\to \overline{\ell}_L \ell_L$ with $\psi_R$ 
mediation 
is negligible. 
The direct search results can also be satisfied. 

Finally, we comment on the collider signature in Model~II. 
$\Phi_{3/2}$ is $Z_2$ odd,  
so that its decay product includes the dark matter $H$. 
For $m_H=50$ GeV, the mass of $h$  
would be about 115~GeV to satisfy the WMAP data~\cite{DM}.  
We then consider the parameter set; 
$m_{\Phi^{++}}=230$ GeV, $m_{H_2^+}=150$ GeV,  
$m_{H_1^+} \simeq m_A = 149$ GeV and $\chi=0.1$  
to satisfy the neutrino data and the LFV data. 
The signal at the LHC would be 
$W^+W^+W^-E_T {\!\!\!\!\!\!\!/} \hspace{3mm}$ via 
$u \overline{d} \to \Phi^{++} H_{i}^- 
\to (H_{i}^+ W^+)(W^- H) \to (H W^+W^+)(W^- H)$. 
The cross section of 
$W^\pm W^\pm W^\mp E_T {\!\!\!\!\!\!\!/} \hspace{3mm}$
is 23 fb for $\sqrt{s}=14$ TeV (7.3 fb for $\sqrt{s}=7$ TeV).
The main background comes from $W^\pm W^\pm W^\mp$, 
and the cross section 
is 135 fb for $\sqrt{s}=14$ TeV (76 fb for $\sqrt{s}=7$ TeV). 
The signal background ratio is not too small at all, 
and we can expect the signal would be detected after 
appropriate kinematic cuts.

\newpage
\chapter{Conclusion}
We have discussed the phenomenology of various Higgs sectors. 
Extended Higgs sectors often appear
in the new physics models, where problems which cannot explain within the SM such as
the hierarchy problem, neutrino masses, dark matter and baryon asymmetry of the Universe,
can be solved. Therefore, by studying extended Higgs sectors, we can determine the direction
of new physics models.

In Part~I, we have discussed the phenomenology of the THDM, the HTM and SUSY
Higgs sectors as an important examples of the extended Higgs sectors. 

In the THDM, we have discussed the discrimination among the types of Yukawa interaction
which appear under the softly-broken discrete $Z_2$ symmetry to avoid the FCNC at
the tree level. We have shown that the light charged Higgs boson of $\mathcal{O(100)}$ GeV is allowed
in the type-I and the type-X Yukawa interaction. We have discussed phenomenological discrimination
of the types of Yukawa interactions in the THDM at the LHC and the ILC. In
particular, we have mainly discussed the discrimination between the MSSM Higgs sector and
the type-X THDM in the relatively light charged Higgs boson scenario. At the LHC, the
type-X THDM can be discriminated from the MSSM by searching for the production and
decays of the extra Higgs bosons $A$, $H$ and $H^\pm$, such as $gg\to A/H\to \ell^+\ell^-$, where $\ell$
is $e$ or $\mu$ when $\sin(\beta-\alpha)\simeq 1$. 
We have also discussed the pair production processes $pp\to AH^\pm$, $HH^\pm$ and $AH$ 
to test the type-X THDM. These processes would provide distinctive four
lepton final states $\ell^+\ell^-\tau^\pm\nu$ and $\ell^+\ell^-\tau^+\tau^-$  
in the type-X THDM, while the MSSM Higgs
sector can be tested by $b\bar{b}\tau^\pm\nu$ and $b\bar{b}\tau^+\tau^-$. 
At the ILC, the type-X THDM is expected to 
be studied very well by the pair production $e^+e^-\to AH$. 
The signal should be four leptons ($\ell\ell\tau^+\tau^-$). 

In the HTM, a characteristic mass spectrum 
$m_{H^{++}}^2-m_{H^+}^2\simeq m_{H^+}^2-m_{\phi^0}^2$ $(\equiv \xi)$ is predicted
when $v_\Delta\ll v$. 
Therefore, by measuring this mass spectrum of the triplet-like scalar bosons,
the model can be tested at the LHC. We have investigated the collider signature in the
HTM with $\xi>0$ at the LHC. In this case, $H^{++}$
is the heaviest of all the triplet-like scalar
bosons. 
When $v_\Delta>10^{-4}$-$10^{-3}$ GeV, $H^{++}$ does not decay into the same sign dilepton so
that the limit of the mass of $H^{++}$ from the recent results at the LHC cannot be applied. 
We thus mainly have discussed the case of light triplet-like scalar bosons whose masses are of
$\mathcal{O}(100)$ GeV. In such a case, triplet-like scalar bosons mainly decay into 
$H^{++}\to H^+W^{+(*)}$, $H^+\to \phi^0W^{+(*)}$ and $\phi^0\to b\bar{b}$. 
We have found that all the masses of the triplet-like scalar
bosons may be able to be reconstructed by measuring the endpoint in the transverse mass
distribution and the invariant mass distribution of the systems which are produced via the
decay of the triplet-like scalar bosons.

We have investigated decoupling properties of SUSY Higgs sectors. The SUSY Higgs
sectors can be separated into the models with additional F-term contributions to the interaction
terms in the Higgs potential and those without such F-term contributions. The
former models have a nondecoupling property due to the F-term contribution. As an concrete
examples of the former model, we have discussed the NMSSM, the TMSSM and the
model with two additional Higgs doublets and the charged singlet fields (4D$\Omega$). While $m_h$
is at most 120-130 GeV in the MSSM, that in the NMSSM and the TMSSM can be much
larger. The deviation of hhh coupling from the prediction in the SM can be significant as
large as 30\%-60\% in the 4D$\Omega$. 
Therefore, even when only $h$ is observed in future, precision
measurements of $m_h$ and the $hhh$ coupling can help discriminate the SUSY Higgs sectors.
We have also discussed the latter SUSY models, where there are no additional F-term contributions in the Higgs potential at the tree level. We have considered the 4HDM as a simplest example. Even without interaction terms from the tree-level F-term contribution ,
significant quasi-nondecoupling effects of extra scalar fields can occur at the tree level due
to the B-term mixing among the Higgs bosons. We have deduced formulae for deviations
in the MSSM observables in the decoupling region for the extra heavy fields. The possible
modifications in the Higgs sector from the MSSM predictions have been studied numerically.
We have found that the quasi-nondecoupling effect from the B-term mixing can be significant
in the 4HDSSM, which can change the MSSM observables $m_{H^+}$, $m_h$, $m_H$ and $\sin^2(\beta-\alpha)$ to
a considerable extent. Detecting the deviations from the MSSM predictions on these MSSM
observables, the MSSM Higgs sector can be tested, and at the same time the possibility of
extended SUSY Higgs sectors including the 4HDM can be explored even when only the
MSSM particles are discovered in near future at the LHC and at the ILC.

In Part II, we discuss new physics models at the TeV scale, where neutrino masses, dark
matter and/or baryon asymmetry of the Universe can be explained.

First, we have discussed theoretical constraints on the parameter space under the conditions
from vacuum stability and triviality in the three-loop radiative seesaw model with
TeV-scale right-handed neutrinos which are odd under the $Z_2$ parity. It has been found that
the model can be consistent up to the scale above 10 TeV in the parameter region which
satisfies the neutrino data, the LFV data, the thermal relic abundance of dark matter as
well as the requirement from the strongly first order phase transition. We also reanalyzed
the constraint from the LFV data. The data from $\mu\to eee$ is found to be more severer than
that from $\mu\to e\gamma$. 

Second, we have discussed the SUSY extension of the Zee-Babu model under R-parity
conservation. In the model, it is not necessary to introduce very high energy scale as compared
to the TeV scale, and the model lies in the reach of the collider experiments and the
flavour measurements. We have found that the neutrino data can be reproduced with satisfying
the current bounds from the LFV even in the scenario where not all the superpartner
particles are heavy. The LSP can be a dark matter candidate. Phenomenology of doubly charged
singlet fields has also been discussed at the LHC.

Finally, we have studied various aspects of $\Phi_{3/2}$ including the signature at the LHC in
a few models. New TeV-scale models with $\Phi_{3/2}$ have been presented for generating tiny
neutrino masses, one of which also contains dark matter candidates. We have found that
$\Phi_{3/2}$ in these models shows discriminative and testable aspects at the LHC and its luminosity upgraded
version, so that models with $\Phi_{3/2}$ would be distinguishable from the other models
with doubly charged scalar states.
\appendix
\chapter{$W_LW_L\to W_LW_L$ scattering amplitude}
\begin{figure}[t]
\begin{center}
\includegraphics[width=120mm]{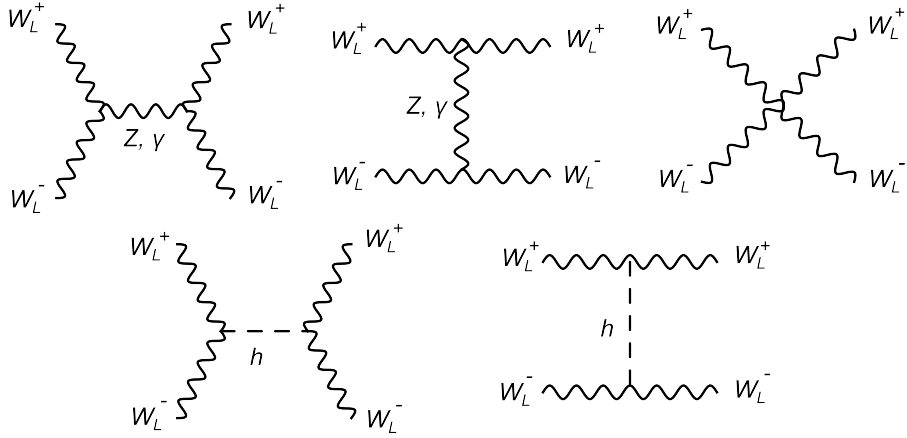}
\end{center}
\caption{Feynman diagrams for $W_L^+W_L^-\to W_L^+W_L^-$ scattering processes. }
\label{LQT_diag}
\end{figure}

In this appendix, we calculate the scattering amplitude of the process $W_LW_L\to W_LW_L$. 
Feynman diagrams of this process is shown in Fig.~\ref{LQT_diag}. 
In section 2.3, we have discussed the perturbative unitarity to obtain the upper bound of the Higgs boson, 
and we have calculated the $W_LW_L\to W_LW_L$ scattering amplitude in the high energy limit. 
Here, we calculate the amplitude up to the order $E^0$, where $E$ is the energy of each $W_L$. 
The amplitude of the s-channel photon exchange diagram is 
\begin{align}
\mathcal{M}(W_L^+W_L^- \to \gamma\to W_L^+W_L^-)_s
&=\frac{e^2}{m_W^4}\cos\theta\left(4E^4-3m_W^4\right)+\mathcal{O}(E^{-2}), 
\end{align}
where $\theta$ is the scattering angle. 
The amplitude of the s-channel $Z$ boson exchange diagram is 
\begin{align}
\mathcal{M}(W_L^+W_L^- \to Z\to W_L^+W_L^-)_s&=\frac{g^2\cos^2\theta_W}{m_W^4}\cos\theta\left(4E^4+m_Z^2E^2-3m_W^4\right)+\mathcal{O}(E^{-2}). 
\end{align}
The amplitudes of the t-channel contributions of photon and $Z$ boson exchange diagrams are
\begin{align}
&\mathcal{M}(W_L^+W_L^- \to \gamma\to W_L^+W_L^-)_t\notag\\
&=\frac{-e^2}{2m_W^4}\Big[(\cos2\theta-4\cos\theta-5)E^4
+16\cos\theta E^2m_W^2+\left(-\cos2\theta+11-\frac{8}{1+\cos\theta}\right)m_W^4\Big]+\mathcal{O}(E^{-2}),\\
&\mathcal{M}(W_L^+W_L^- \to Z\to W_L^+W_L^-)_t\notag\\
&=\frac{-g^2\cos^2\theta_W}{2m_W^4}\Big[(\cos2\theta-4\cos\theta-5)E^4
+(3m_Z^2+(16m_W^2-m_Z^2)\cos\theta)E^2\notag\\
&\hspace{20mm}+\left(-\cos2\theta+11-\frac{8}{1+\cos\theta}\right)m_W^4-\left(-\cos2\theta+20\cos\theta+5\right)\frac{m_W^2m_Z^2}{2(1+\cos\theta)}\Big].
\end{align}
The amplitude of the contact interaction diagram is 
\begin{align}
\mathcal{M}(W_L^+W_L^- \to W_L^+W_L^-)&=
\frac{g^2}{2m_W^4}[(\cos2\theta-12\cos\theta-5)E^4+4(1+3\cos\theta)E^2m_W^2].
\end{align}
Therefore, the amplitude of $W_L^+ W_L^-\to W_L^+ W_L^-$ without Higgs boson exchange diagram is 
\begin{align}
&i\mathcal{M}(W_L^+W_L^- \to W_L^+W_L^-)_{\text{w/o Higgs}}\notag\\
&=\frac{g^2}{m_W^4}\left[\frac{1}{2}(1-\cos\theta)E^2m_W^2+\left(\cos^2\theta-3\cos\theta-6+\frac{-\cos2\theta+20\cos\theta+21}{4(1+\cos\theta)}\right)m_W^4\right]+\mathcal{O}(E^{-2}).
\end{align}
This result shows that the $\mathcal{O}(E^4)$ dependence of the amplitude is cancelled but the $\mathcal{O}(E^2)$ dependence remains. 
The 0th partial wave amplitude can be calculated as 
\begin{align}
a_0(W_L^+ W_L^-\to W_L^+ W_L^-)_{\text{w/o Higgs}}&=\frac{1}{32\pi}\int_{-1}^1d\cos\theta \mathcal{M}(W_L^+W_L^- \to W_L^+W_L^-)_{\text{w/o Higgs}}\notag\\
&=\frac{\sqrt{2}G_FE_{\text{cm}}^2}{32\pi}+\mathcal{O}(E^0), 
\end{align}
where $E_{\text{cm}}=2E$ is the center of mass energy. 
By the condition in Eq.~(\ref{ll}), unitarity is violated around $E_{\text{cm}}\simeq$ 1.7 TeV. 

Next, we take into account the Higgs boson exchange diagrams. 
The amplitudes of s-channel and t-channel contributions are, respectively 
\begin{align}
i\mathcal{M}(W_L^+W_L^- \to h\to W_L^+W_L^-)_s
&=-i\frac{g^2}{m_W^2}\left[E^2-m_W^2+\frac{1}{4}m_h^2\right]+\mathcal{O}(E^{-2}), \\
i\mathcal{M}(W_L^+W_L^- \to h\to W_L^+W_L^-)_t
&=i\frac{g^2}{2m_W^2}\left[E^2(1+\cos\theta)-\frac{1}{2}m_h^2+m_W^2\cos\theta\right]+\mathcal{O}(E^{-2}).
\end{align}
The total amplitude of $W_L^+W_L^- \to W_L^+W_L^-$ is 
\begin{align}
i\mathcal{M}(W_L^+W_L^- \to W_L^+W_L^-)=
-i\frac{g^2m_h^2}{2m_W^2}+g^2\left[\cos^2\theta-\frac{5}{2}\cos\theta-6+\frac{-\cos2\theta+20\cos\theta+21}{4(1+\cos\theta)}\right]+\mathcal{O}(E^{-2}).
\end{align}
The $E^2$ dependence of the amplitude is cancelled by the Higgs boson exchange diagram. 

\chapter{Renormalization group equations}
In this appendix, we list one-loop level renormalization group equations (RGEs) 
in the SM, the THDM, the model by Aoki, Kanemura and Seto~\cite{aks_prl} and several SUSY Higgs models.
The beta function for the dimension less coupling constant $c$ is defined by 
\begin{align}
\beta(c)\equiv \frac{dc}{d\log Q}=Q\frac{dc}{dQ}, 
\end{align}
where $Q$ is an arbitrary scale. 
First, RGEs for the gauge couplings $g_i$ are written as  
\begin{align}
\beta(g_i)=\frac{g_i^3}{16\pi^2}\left[-\frac{11}{3}C(G)+\sum_{\text{fermions}}\frac{2}{3}C(F)+\sum_{\text{scalars}}\frac{1}{3}C(S)\right],
\end{align}
where $C(G)$ is defined by the structure constant $f^{abc}$ for the gauge group $G$, while 
$C(F)$ is defined by the generator $t^a$ which appears in fermion-fermion-gauge boson vertex as
\begin{align}
C(G)\delta^{ab}=f^{acd}f^{bcd}, \\
C(F)\delta^{ab}=\text{Tr}(t^at^b), 
\end{align}
and $C(S)$ is also defined in the same way as $C(F)$. 

\section{Beta functions in the SM}
Beta functions for the gauge couplings in the model with the number of the generation $N_g$
and that of the Higgs doublet $N_H$ are given by 
\begin{subequations}
\begin{align}
\beta(g_s)&=\frac{g_s^3}{16\pi^2}\left[-11+\frac{4}{3}N_g\right],\\
\beta(g)&=\frac{g^3}{16\pi^2}\left[-\frac{22}{3}+\frac{4}{3}N_g+\frac{1}{6}N_H\right],\\
\beta(g')&=\frac{g^{'3}}{16\pi^2}\left[\frac{20}{9}N_g+\frac{1}{6}N_H\right]. 
\end{align}
\label{beta_gg}
\end{subequations}
In the SM, ($N_g$, $N_H$) is (3, 1). 
We note that Eq.~(\ref{beta_gg}) is valid in the model 
whose Higgs sector is composed of only doublet fields with $Y=1/2$ and singlet fields with $Y=0$. 
The beta functions for the Higgs self-coupling $\lambda$ and the top Yukawa coupling $y_t$ are given by
\begin{align}
\beta(\lambda)&=\frac{1}{16\pi^2}\left[24\lambda^2+12y_t^2\lambda-6y_t^4+\frac{9}{8}g^4+\frac{3}{4}g^2g^{'2}+\frac{3}{8}g^{'4}-9\lambda g^2-3\lambda g^{'2}\right],\\
\beta(y_t)&=\frac{y_t}{16\pi^2}\Big[\frac{9}{2}y_t^2-8g_s^2-\frac{9}{4}g^2-\frac{17}{12}g^{'2}\Big].
\end{align}

\section{Beta functions in the THDM}
The beta functions for the Higgs self-couplings in the THDM and that for the top Yukawa coupling are given by
\begin{align}
\beta(\lambda_1)&=\frac{1}{16\pi^2}\Big[12\lambda_1^2+4\lambda_3^2+2\lambda_4^2+2\lambda_5^2+4\lambda_3\lambda_4
+\frac{9}{4}g^4+\frac{6}{4}g^2g^{'2}+\frac{3}{4}g^{'4}-9\lambda_1 g^2-3\lambda_1 g^{'2}\Big],\\
\beta(\lambda_2)&=\frac{1}{16\pi^2}\Big[12\lambda_2^2+4\lambda_3^2+2\lambda_4^2+2\lambda_5^2+4\lambda_3\lambda_4-12y_t^4+12y_t^2\lambda_2
\notag\\
&\quad\quad+\frac{9}{4}g^4+\frac{6}{4}g^2g^{'2}+\frac{3}{4}g^{'4}-9\lambda_2 g^2-3\lambda_2 g^{'2}\Big],\\
\beta(\lambda_3)&=\frac{1}{16\pi^2}\Big[6\lambda_1\lambda_3+2\lambda_1\lambda_4+6\lambda_2\lambda_3+2\lambda_2\lambda_4+4\lambda_3^2+2\lambda_4^2+2\lambda_5^2\notag\\
&\quad\quad+\frac{9}{4}g^4+\frac{3}{4}g^{'4}-\frac{6}{4}g^2g^{'2}-9\lambda_3g^2-3\lambda_3g^{'2}+6\lambda_3y_t^2\Big],
\end{align}
\begin{align}
\beta(\lambda_4)&=\frac{1}{16\pi^2}\Big[2\lambda_4(\lambda_1+\lambda_2+4\lambda_3+2\lambda_4)+8\lambda_5^2+3g^2g^{'2}
+\lambda_4\big(-9g^2-3g^{'2}\big)+6\lambda_4y_t^2\Big],\\
\beta(\lambda_5)&=\frac{1}{16\pi^2}\Big[2\lambda_5(\lambda_1+\lambda_2+4\lambda_3+6\lambda_4)+\lambda_5(-9g^2-3g^{'2})+6\lambda_5y_t^2\Big],\\
\beta(y_t)&=\frac{y_t}{16\pi^2}\Big[\frac{9}{2}y_t^2-8g_s^2-\frac{9}{4}g^2-\frac{17}{12}g^{'2}\Big].
\end{align}
The coupling constants $\lambda_1$-$\lambda_5$ are defined in Eq.~(\ref{pot_thdm2}). 
The beta functions for the gauge couplings are given in Eq.~(\ref{beta_gg}) with ($N_g$, $N_H$) to be (3, 2). 

\section{Beta functions in the three-loop neutrino mass model}
The beta functions for dimension less couplings in the model~\cite{aks_prl} are given as
\begin{align}
\beta(g_s)&=\frac{1}{16\pi^2}\left[-7g_s^3\right], \\
\beta(g)&=\frac{1}{16\pi^2}\left[-3g^3\right], \\
\beta(g')&=\frac{1}{16\pi^2}\Big[-\frac{22}{3}g^{'3}\Big],
\end{align}
%
%
\begin{align}
\beta(y_t)&=\frac{y_t}{16\pi^2}\Big[\frac{9}{2}y_t^2-8g_s^2-\frac{9}{4}g^2-\frac{17}{12}g^{'2}\Big],\\
\beta(h_i^\alpha)&=\frac{1}{16\pi^2}\Big[-5g^{'2}h_i^\alpha+\frac{1}{2}h_i^\alpha\sum_j(h_j^\alpha)^2+\frac{1}{2}h_i^\alpha\sum_\beta(h_i^\beta)^2+h_i^\alpha\sum_{j,\beta}(h_j^\beta)^2\Big],\\
\beta(\lambda_1)&=\frac{1}{16\pi^2}\Big[12\lambda_1^2+4\lambda_3^2+2\lambda_4^2+2\lambda_5^2+4\lambda_3\lambda_4+2\rho_1^2+\sigma_1^2+\frac{9}{4}g^4+\frac{6}{4}g^2g^{'2}+\frac{3}{4}g^{'4}\notag\\
&\hspace{15mm}-4y_\tau^4+(4y_\tau^2-9 g^2-3g^{'2})\lambda_1\Big], \\
\beta(\lambda_2)&=\frac{1}{16\pi^2}\Big[12\lambda_2^2+4\lambda_3^2+2\lambda_4^2+2\lambda_5^2+4\lambda_3\lambda_4+2\rho_2^2+\sigma_2^2+\frac{9}{4}g^4+\frac{6}{4}g^2g^{'2}+\frac{3}{4}g^{'4}\notag\\
&\hspace{15mm}-12y_t^4-12y_b^4+(12y_t^2+12y_b^2-9g^2-3 g^{'2})\lambda_2\Big], 
\end{align}
\begin{align}
\beta(\lambda_3)&=\frac{1}{16\pi^2}\Big[6\lambda_1\lambda_3+2\lambda_1\lambda_4+6\lambda_2\lambda_3+2\lambda_2\lambda_4+4\lambda_3^2+2\lambda_4^2+2\lambda_5^2+2\rho_1\rho_2+\sigma_1\sigma_2+4\kappa^2\notag\\
&\hspace{15mm}+\frac{9}{4}g^4+\frac{3}{4}g^{'4}-\frac{6}{4}g^2g^{'2}+(6y_t^2+6y_b^2+2y_\tau^2-9g^2-3g^{'2})\lambda_3\Big], \\
\beta(\lambda_4)&=\frac{1}{16\pi^2}\Big[2(\lambda_1+\lambda_2+4\lambda_3+2\lambda_4)\lambda_4 
 +8\lambda_5^2-8\kappa^2+3g^2g^{'2} \Big. \nonumber \\
& \hspace*{2cm}\Big. +(6y_t^2+6y_b^2+2y_\tau^2-9g^2-3g^{'2})\lambda_4\Big], \\
\beta(\lambda_5)&=\frac{1}{16\pi^2}\Big[2(\lambda_1+\lambda_2+4\lambda_3+6\lambda_4)\lambda_5+(6y_t^2+6y_b^2+2y_\tau^2-9g^2-3g^{'2})\lambda_5\Big],
\end{align}
\begin{align}
\beta(\rho_1)&=\frac{1}{16\pi^2}\Big[6\lambda_1\rho_1+4\lambda_3\rho_2+2\lambda_4\rho_2+2\rho_1\lambda_S+4\rho_1^2+\sigma_1\xi+8\kappa^2+3g^{'4}\notag\\
&\hspace{15mm}+(-\frac{15}{2}g^{'2}-\frac{9}{2}g^2+2\sum_{i,\alpha}(h_i^\alpha)^2+2y_\tau^2)\rho_1\Big], \\
\beta(\rho_2)&=\frac{1}{16\pi^2}\Big[6\lambda_2\rho_2+4\lambda_3\rho_1+2\lambda_4\rho_1+2\rho_2\lambda_S+4\rho_2^2+\sigma_2\xi+8\kappa^2+3g^{'4}\notag\\
&\hspace{15mm}+(-\frac{15}{2}g^{'2}-\frac{9}{2}g^2+2\sum_{i,\alpha}(h_i^\alpha)^2+6y_t^2+6y_b^2)\rho_2\Big], \\
\beta(\lambda_S)&=\frac{1}{16\pi^2}\Big[8\rho_1^2+8\rho_2^2+5\lambda_S^2+2\xi^2+24g^{'4}-12g^{'2}\lambda_S \Big. \notag \\ 
&\Big. \hspace*{2cm}+4\sum_{i,\alpha}(h_i^\alpha)^2\lambda_S-8\sum_{i,j}\sum_{\alpha,\beta}h_i^\alpha h_i^\beta h_j^\beta h_j^\alpha\Big], 
\end{align}
\begin{align}
\beta(\sigma_1)&=\frac{1}{16\pi^2}\Big[6\lambda_1\sigma_1+(4\lambda_3+2\lambda_4)\sigma_2+\sigma_1\lambda_\eta+2\rho_1\xi+16\kappa^2+(-\frac{9}{2}g^2-\frac{3}{2}g^{'2}+2y_\tau^2)\sigma_1\Big], \\
\beta(\sigma_2)&=\frac{1}{16\pi^2}\Big[6\lambda_2\sigma_2+(4\lambda_3+2\lambda_4)\sigma_1+\sigma_2\lambda_\eta+2\rho_2\xi+16\kappa^2\Big. \notag\\ 
& \hspace*{2cm} \Big. +(-\frac{9}{2}g^2-\frac{3}{2}g^{'2}+6y_t^2+6y_b^2)\sigma_2\Big], \\
\beta(\lambda_\eta)&=\frac{1}{16\pi^2}\Big[12(\sigma_1^2+\sigma_2^2)+3\lambda_\eta^2+6\xi^2\Big], 
\end{align}
\begin{align}
\beta(\kappa)&=\frac{1}{16\pi^2}\kappa\Big[2\lambda_3-2\lambda_4+2\xi+2\sigma_1+2\sigma_2+2\rho_1+2\rho_2+\sum_{\alpha,i}(h_i^\alpha)^2\Big. \notag\\
& \hspace*{2cm} \Big.-\frac{9}{2}g^2-\frac{9}{2}g^{'2}+3y_t^2+3y_b^2+y_\tau^2\Big], \\
\beta(\xi)&=\frac{1}{16\pi^2}\Big[4\rho_1\sigma_1+4\rho_2\sigma_2+2\lambda_S\xi+\lambda_\eta\xi+4\xi^2-6g^{'2}\xi+2\sum_{\alpha,i}(h_i^\alpha)^2\xi\Big], 
\end{align}
where definitions of these coupling constants are given in Eq.~(\ref{pot}). 

\section{Beta functions in SUSY models}
In the SUSY standard model, beta functions for the gauge couplings given in Eq.~(\ref{beta_gg}) are replaced by 
\begin{subequations}
\begin{align}
\beta(g_s)&=\frac{g_s^3}{16\pi^2}\left[-9+2N_g\right],\\
\beta(g)&=\frac{g^3}{16\pi^2}\left[-6+2N_g+\frac{1}{2}N_H\right],\\
\beta(g')&=\frac{g^{'3}}{16\pi^2}\left[\frac{10}{3}N_g+\frac{1}{2}N_H\right]. 
\end{align}
\label{beta_ggg}
\end{subequations}
We note that Eq.~(\ref{beta_ggg}) is valid in the model 
whose superpotential is composed of the MSSM chiral superfields with or without 
extra doublet chiral superfields with $Y=1/2$ and extra neutral chiral superfields.  
\subsection{MSSM}
In the MSSM, beta functions for the gauge couplings are obtained by Eq.~(\ref{beta_ggg}) with ($N_g$, $N_H$) is (3, 2), 
and that for the top Yukawa coupling is expressed as 
\begin{align}
\beta(y_t)&=\frac{y_t}{16\pi^2}\Big[6y_t^2-\frac{16}{3}g_s^2-3g^2-\frac{13}{9}g^{'2}\Big].
\end{align}
\subsection{NMSSM}
In the NMSSM, beta functions for the gauge couplings are the same as those in the MSSM. 
The other beta functions are given as 
\begin{align}
\beta(y_t)&=\frac{y_t}{16\pi^2}\Big[6y_t^2-\frac{16}{3}g_s^2-3g^2-\frac{13}{9}g^{'2}+\lambda_{HHS}^2\Big],\\
\beta(\lambda_{HHS})&=\frac{\lambda_{HHS}}{16\pi^2}\Big[4\lambda_{HHS}^2+2\kappa^2+3y_t^2-3g^2-g^{'2}\Big],\\
\beta(\kappa)&=\frac{\kappa}{16\pi^2}\Big[6\lambda_{HHS}^2+6\kappa^2\Big].
\end{align}
\subsection{TMSSM}
In the TMSSM, beta functions for the gauge couplings are given as
\begin{align}
\beta(g_s)&=-\frac{3g_s^3}{16\pi^2},\\
\beta(g)&=\frac{4g^3}{16\pi^2},\\
\beta(g')&=\frac{17g^{'3}}{16\pi^2}. 
\end{align}
The other beta functions are given by 
\begin{align}
\beta(y_t)&=\Big[6y_t^2-\frac{16}{3}g_s^2-3g^2-\frac{13}{9}g^{'2}+6\lambda_{HH\Delta_R}^2\Big],\\
\beta(\lambda_{HH\Delta_L})&=\frac{\lambda_{HH\Delta_L}}{16\pi^2}\Big[14\lambda_{HH\Delta_L}^2-7g^2-3g^{'2}\Big],\\
\beta(\lambda_{HH\Delta_R})&=\frac{\lambda_{HH\Delta_R}}{16\pi^2}\Big[14\lambda_{HH\Delta_R}^2-7g^2-3g^{'2}\Big].
\end{align}
\subsection{4D$\Omega$}
In the 4D$\Omega$, beta functions for the gauge couplings are given as
\begin{align}
\beta(g_s)&=-\frac{3g_s^3}{16\pi^2},\\
\beta(g)&=\frac{g^3}{16\pi^2},\\
\beta(g')&=\frac{13g^{'3}}{16\pi^2}. 
\end{align}
The other beta functions are given by 
\begin{align}
\beta(y_t)&=\Big[6y_t^2-\frac{16}{3}g_s^2-3g^2-\frac{13}{9}g^{'2}+\lambda_{HH\Omega_L}^2\Big],\\
\beta(\lambda_{HH\Omega_R})&=\frac{\lambda_{HH\Omega_R}}{16\pi^2}\Big[4\lambda_{HH\Omega_R}^2-3g^2-3g^{'2}\Big],\\
\beta(\lambda_{HH\Omega_L})&=\frac{\lambda_{HH\Omega_L}}{16\pi^2}\Big[4\lambda_{HH\Omega_L}^2-3g^2-3g^{'2}\Big].
\end{align}

\chapter{Decay rates}
In this appendix, we list the analytic fomulae for decay rates of Higgs bosons in the SM, the THDM and the HTM.  
\section{Decay rates of the Higgs boson in the SM}
\label{sm_decay}
The decay rates of the Higgs boson decaying into the fermion pair are given by
\begin{align}
\Gamma(h\to f\bar{f})&=\sqrt{2}G_F\frac{m_hm_f^2}{8\pi}N_c^f\beta\left(\frac{m_f^2}{m_h^2}\right)^3,
\end{align}
where $N_c^f$ is the color factor with $N_c^q=3$, $N_c^\ell=1$ and 
\begin{align}
\beta(x)=\sqrt{1-4x}. \label{decay_beta}
\end{align}
The decay rates of the Higgs boson decaying into the gauge boson $V$ pair  ($V=W$ or $Z$) are given by 
\begin{align}
\Gamma(h\to VV)&=\sqrt{2}G_F\frac{m_h^3}{32\pi}\delta_V\left[1-\frac{4m_V^2}{m_h^2}+\frac{12m_V^4}{m_h^4}\right]\beta\left(\frac{m_V^2}{m_h^2}\right), 
\end{align}
where $\delta_W=2$ and $\delta_Z=1$. 
The decay rates of the three body decay modes can be calculated as 
\begin{align}
\Gamma(h\to W^+W^{-*}\to W^+f\bar{f}') &=
\frac{G_F^2m_W^4m_h}{96\pi^3 } F\left(\frac{m_W^2}{m_h^2}\right), 
\end{align}
where the function $F(x)$ is given as 
\begin{align}
F(x)&=-|1-x|\left(\frac{47}{2}x-\frac{13}{2}+\frac{1}{x}\right)+3(1-6x+4x^2)|\log \sqrt{x}|\notag\\
&+\frac{3(1-8x+20x^2)}{\sqrt{4x-1}}\cos^{-1}\left(\frac{3x-1}{2x^{3/2}}\right).\label{decay_F}
\end{align}
In this decay mode, fermion pair $f\bar{f}'$ in the final state can be 
$(e^-\bar{\nu}_e)$, $(\mu^-\bar{\nu}_\mu)$, $(\tau^-\bar{\nu}_\tau)$, $(d\bar{u})$ and $(s\bar{c})$. 
By summimg all these final states, we obtain 
\begin{align}
\Gamma(h\to W^+W^{-*}) &\equiv \sum_{f}\Gamma(h\to W^+W^{-*}\to W^+ f\bar{f}')
=\frac{3G_F^2m_W^4m_h}{32\pi^3 } F\left(\frac{m_W^2}{m_h^2}\right). 
\end{align}
The decay rate of $h\to ZZ^*\to Zf\bar{f}$ mode can be calculated as 
\begin{align}
\Gamma(h\to ZZ^*\to Zf\bar{f}) &=\frac{G_F^2m_Z^4m_h}{48\pi^3}F\left(\frac{m_Z^2}{m_h^2}\right)
(I_f^2+2\sin^4\theta_WQ_f^2-2I_f\sin^2\theta_WQ_f), 
\end{align}
where $I_f$ and $Q_f$ are the third component of the isospin and the electromagnetic charge of the final state fermion $f$.  
In this decay mode, fermion pair $f\bar{f}$ in the final state can be 
$(\ell^+\ell^-)$, $(\nu_\ell\bar{\nu}_\ell)$, $(u\bar{u})$, $(d\bar{d})$, $(c\bar{c})$, $(s\bar{s})$ and $(b\bar{b})$ with $\ell^\pm =e^\pm$, $\mu^\pm$ or $\tau^\pm$. 
By summimg all these final states, we obtain
\begin{align}
\Gamma(h\to ZZ^*)\equiv \sum_f\Gamma(h\to ZZ^*\to Zf\bar{f}) &=\frac{G_F^2m_Z^4m_h}{64\pi^3}F\left(\frac{m_Z^2}{m_h^2}\right)\notag\\
&\times\left(7-\frac{40}{3}\sin^2\theta_W+\frac{160}{9}\sin^4\theta_W\right).
\end{align} 
There are one-loop induced decay processes, e.g., $h\to \gamma\gamma$, $h\to gg$ and $h\to \gamma Z$. 
These decay rates can be expressed by
\begin{align}
\Gamma(h\to \gamma\gamma)&=\frac{\sqrt{2}G_F\alpha_{\text{em}}^2m_h^3}{64\pi^3 }|\sum_fQ_f^2N_c^fI_f(m_h)+I_W(m_h)|^2 ,\\
\Gamma(h\to gg)&=\frac{\sqrt{2}G_F\alpha_s^2m_h^3}{128\pi^3}|\sum_{f=q}I_f|^2,\\
\Gamma(h\to Z\gamma)&=\frac{\sqrt{2}G_F\alpha_{\text{em}}^2m_h^3}{128\pi^3}\left(1-\frac{m_Z^2}{m_h^2}\right)^3
|\sum_f Q_fJ_f(m_h)+J_W(m_h)|^2,
\end{align}
where the loop functions are
\begin{subequations}
\begin{align}
I_f(m_h)&= -\frac{4m_f^2}{m_h^2}\left[1-\frac{m_h^2}{2}\left(1-\frac{4m_f^2}{m_h^2}\right)C_0(0,0,m_h^2,m_f,m_f,m_f)\right],\\
I_W(m_h)&= 1+\frac{6m_W^2}{m_h^2}-6m_W^2\left(1-\frac{2m_W^2}{m_h^2}\right)C_0(0,0,m_h^2,m_f,m_f,m_f),\\
J_f(m_h)&=-\frac{2N_c^f}{\sin\theta_W\cos\theta_W}(I_f(m_h)-2Q_f\sin^2\theta_W)[J_1(m_f)-J_2(m_f)],\\
J_W(m_h)&=-\cot\theta_W\notag\\
&\hspace{-6mm}\times\left\{4\left(3-\tan^2\theta_W\right)J_2(m_W)+\left[\left(1+\frac{m_h^2}{2m_W^2}\right)\tan^2\theta_W-\left(5+\frac{m_h^2}{2m_W^2}\right)\right]J_1(m_W)\right\},\\
J_1(m)&=\frac{2m^2}{m_h^2-m_Z^2}\Big[1+2m^2C_0(0,m_Z^2,m_h^2,m,m,m)\notag\\
&\hspace{20mm}+\frac{m_Z^2}{m_h^2-m_Z^2}\{B_0(m_h^2,m,m)-B_0(m_Z^2,m,m)\}\Big],\\
J_2(m)&=m^2C_0(0,m_Z^2,m_h^2,m,m,m). 
\end{align}
\label{xxxx}
\end{subequations}
In Eq.~(\ref{xxxx}), 
$C_0$ and $B_0$ functions are Passarino-Veltman function~\cite{Ref:PV}. 
The analytic formula for the $B_0$ function is given in Appendix~D. 
Here, we give expressions for the special case of the $C_0$ function which is used in the above decay rates: 
\begin{align}
C_0(0,0,m_h^2,m,m,m)&=\frac{-2}{m_h^2}f\left(\frac{4m^2}{m_h^2}\right),\\
C_0(0,m_Z^2,m_h^2,m,m,m)&=\frac{-2}{m_h^2-m_Z^2}\left[f\left(\frac{4m^2}{m_h^2}\right)-f\left(\frac{4m^2}{m_Z^2}\right)\right],
\end{align}
with 
\begin{align}
f(x)=\left\{
\begin{array}{c}
[\arcsin(1/\sqrt{x})]^2, \quad \text{if }x\geq 1,\\
-\frac{1}{4}[\ln \frac{1+\sqrt{1-x}}{1-\sqrt{1-x}}-i\pi]^2, \quad \text{if }x< 1
\end{array}\right.. 
\end{align}

\section{Decay rates of the Higgs bosons in the THDM}
In the THDM, there are five physical scalar bosons, i.e., the CP-even scalar bosons $h$ and $H$, 
the CP-odd scalar boson $A$ and the charged scalar bosons $H^\pm$. 
When we take $\sin(\beta-\alpha)=1$, then $h$ behaves the SM-like Higgs boson. 

First, decay rates for neutral scalar bosons decaying into fermions are given as
\begin{align}
\Gamma(\mathcal{H}\to f\bar{f})&=
\sqrt{2}G_F(\xi_{\mathcal{H}}^f)^2\frac{m_f^2m_\mathcal{H}}{8\pi}N_c^f\beta\left(\frac{m_f^2}{m_{\mathcal{H}}^2}\right)^3,\\
\Gamma(A \to f\bar{f})
&=\sqrt{2}G_F(\xi_A^f)^2\frac{m_f^2m_\mathcal{H}}{8\pi}N_c^f\beta\left(\frac{m_f^2}{m_A^2}\right),
\end{align}
where $\mathcal{H}$ is $h$ or $H$ and $\xi_{\mathcal{H},A}^f$ are listed in Table~{\ref{yukawa_tab}}. 

Decay rates for neutral scalar bosons decaying into gauge bosons are given as
\begin{align}
&\Gamma(\mathcal{H}\to VV)=\eta_{\mathcal{H}}^2\sqrt{2}G_F\frac{m_h^3}{32\pi}\delta_V\left[1-\frac{4m_V^2}{m_h^2}+\frac{12m_V^4}{m_h^4}\right]\beta\left(\frac{m_V^2}{m_h^2}\right), \\
&\text{with } \eta_{\mathcal{H}}=
\left\{
\begin{aligned}
\sin(\beta-\alpha)\hspace{3mm}\text{for}\hspace{3mm}\mathcal{H}=h,\\
\cos(\beta-\alpha)\hspace{3mm}\text{for}\hspace{3mm}\mathcal{H}=H
\end{aligned}\right..
\end{align}
In the CP conserving THDM, 
the decay rate for $A$ decaying into gauge bosons are zero at the tree level, since $AV_\mu V^\mu$ vertex is absent.  

If the mass of $H^\pm$ is larger than $m_t+m_b$, $H^+$ can decay into $t\bar{b}$ according to the decay rate: 
\begin{align}
\Gamma(H^+\to t\bar{b})&=\sqrt{2}G_F\frac{m_{H^+}}{8\pi}
\lambda\left(\frac{m_t^2}{m_{H^+}^2},\frac{m_b^2}{m_{H^+}^2}\right)^{1/2}\notag\\
&\times\left\{[m_b^2(\xi_A^{d})^2+m_t^2(\xi_A^{u})^2]\left(1-\frac{m_t^2+m_b^2}{m_{H^+}^2}\right)-\frac{4m_b^2m_t^2\xi_A^d\xi_A^u}{m_{H^+}^2}\right\},
\end{align}
where the function $\lambda(x,y)$ is given by
\begin{align}
\lambda(x,y)&=1+x^2+y^2-2xy-2x-2y,\label{decay_lam}
\end{align}
and $\xi_A^{u,d}$ are listed in Table~\ref{yukawa_tab}. 

if kinematically allowed, 
a scalar boson ($S_0$) can decay into the other scalar boson ($S_1$) plus a vector boson $V$. 
When $m_{S_0}>m_{S_1}+m_V$ where $m_{S_0}$, $m_{S_1}$ and $m_V$ are the masses of $S_0$, $S_1$ and $V$, respectively, 
this type of decay rates is evaluated as 
\begin{align}
\Gamma(\mathcal{H}\to H^\pm W^\mp)&=\sqrt{2}G_F\frac{m_\mathcal{H}^3\bar{\eta}_{\mathcal{H}}^2}{8\pi}\lambda^{3/2}
\left(\frac{m_{H^+}^2}{m_\mathcal{H}^2},\frac{m_W^2}{m_\mathcal{H}^2}\right),\\
\Gamma(A\to H^\pm W^\mp)&=\sqrt{2}G_F\frac{m_A^3}{8\pi}\lambda^{3/2}\left(\frac{m_{H^+}^2}{m_A^2},\frac{m_W^2}{m_A^2}\right),\\
\Gamma(H^\pm \to \mathcal{H}W^\pm )&=\sqrt{2}G_F\frac{m_{H^+}^3\bar{\eta}_{\mathcal{H}}^2}{8\pi}\lambda^{3/2}
\left(\frac{m_\mathcal{H}^2}{m_{H^+}^2},\frac{m_W^2}{m_{H^+}^2}\right),\\
\Gamma(H^\pm\to AW^\pm)&=\sqrt{2}G_F\frac{m_{H^+}^3}{8\pi}\lambda^{3/2}\left(\frac{m_A^2}{m_{H^+}^2},\frac{m_W^2}{m_{H^\pm}^2}\right),
\end{align}
\begin{align}
\Gamma(\mathcal{H}\to AZ)&
=\sqrt{2}G_F\frac{m_\mathcal{H}^3\bar{\eta}_{\mathcal{H}}^2}{8\pi}\lambda^{3/2}
\left(\frac{m_A^2}{m_\mathcal{H}^2},\frac{m_Z^2}{m_\mathcal{H}^2}\right),\\
\Gamma(A\to \mathcal{H}Z)&=\sqrt{2}G_F\frac{m_A^3\bar{\eta}_{\mathcal{H}}^2}{8\pi}\lambda^{3/2}
\left(\frac{m_\mathcal{H}^2}{m_A^2},\frac{m_Z^2}{m_A^2}\right),\\
&\hspace{-15mm}\text{with } \bar{\eta}_{\mathcal{H}}=
\left\{
\begin{aligned}
\sin(\beta-\alpha)\hspace{3mm}\text{for}\hspace{3mm}\mathcal{H}=H,\\
\cos(\beta-\alpha)\hspace{3mm}\text{for}\hspace{3mm}\mathcal{H}=h
\end{aligned}\right..
\end{align}
When $m_{S_1}+m_V>m_{S_0}>m_{S_1}$, $S_0$ can also decay into $S_1$ and the off-shell $V$. 
These decay rates are given by
\begin{align}
\Gamma(\mathcal{H}\to H^\pm W^{\pm *})&=\frac{9G_F^2m_W^4}{16\pi^3}\bar{\eta}_{\mathcal{H}}^2m_{\mathcal{H}}
G\left(\frac{m_{H^+}^2}{m_{\mathcal{H}}^2},\frac{m_W^2}{m_{\mathcal{H}}^2}\right),\\
\Gamma(A\to H^\pm W^{\pm *} )&=\frac{9G_F^2m_W^4}{16\pi^3}m_{A}
G\left(\frac{m_{H^+}^2}{m_{A}^2},\frac{m_W^2}{m_{A}^2}\right),\\
\Gamma(H^\pm \to \mathcal{H}W^{\pm *})&=\frac{9G_F^2m_W^4}{16\pi^3}\bar{\eta}_{\mathcal{H}}^2m_{H^+}
G\left(\frac{m_\mathcal{H}^2}{m_{H^+}^2},\frac{m_W^2}{m_{H^+}^2}\right),\\
\Gamma(H^\pm \to AW^{\pm *} )&=\frac{9G_F^2m_W^4}{16\pi^3}m_{H^+}
G\left(\frac{m_A^2}{m_{H^+}^2},\frac{m_W^2}{m_{H^+}^2}\right),\\
\Gamma(\mathcal{H}\to AZ^*)&=\frac{3G_F^2m_Z^4}{32\pi^3}\bar{\eta}_{\mathcal{H}}^2m_\mathcal{H}
\left(7-\frac{40}{3}\sin{\theta_W}^2+\frac{160}{9}\sin{\theta_W}^4\right)G\left(\frac{m_A^2}{m_\mathcal{H}^2},\frac{m_Z^2}{m_{\mathcal{H}}^2}\right),\\
\Gamma(A\to \mathcal{H}Z^*)&=\frac{3G_F^2m_Z^4}{32\pi^3}\bar{\eta}_{\mathcal{H}}^2m_A
\left(7-\frac{40}{3}\sin{\theta_W}^2+\frac{160}{9}\sin{\theta_W}^4\right)G\left(\frac{m_\mathcal{H}^2}{m_A^2},\frac{m_Z^2}{m_A^2}\right),
\end{align}
where the function $G(x,y)$ is given as
\begin{align}
G(x,y)&=\frac{1}{12y}\Bigg\{2\left(-1+x\right)^3-9\left(-1+x^2\right)y+6\left(-1+x\right)y^2\notag\\
&+6\left(1+x-y\right)y\sqrt{-\lambda(x,y)}\left[\tan^{-1}\left(\frac{-1+x-y}{\sqrt{-\lambda(x,y)}}\right)+\tan^{-1}\left(\frac{-1+x+y}{\sqrt{-\lambda(x,y)}}\right)\right]\notag\\
&-3\left[1+\left(x-y\right)^2-2y\right]y\log x\Bigg\}.  \label{g_func}
\end{align}

If kinematically allowed, there also are (Scalar $\to$ Scalar$'$ + Scalar$''$) type decay modes whose decay rates can be expressed as
\begin{align}
\Gamma(\phi_i\to \phi_j\phi_k)=(1+\delta_{jk})\frac{\lambda_{\phi_i\phi_j\phi_k}^2}{16\pi m_{\phi_1}}\lambda^{1/2}\left(\frac{m_{\phi_j}^2}{m_{\phi_i}^2},\frac{m_{\phi_k}^2}{m_{\phi_i}^2}\right), 
\end{align}
where $\lambda_{\phi_i\phi_j\phi_k}$ is the coefficient of the scalar three-point vertices which are defined by  
\begin{align}
\mathcal{L} = -\lambda_{\phi_i\phi_j\phi_k}\phi_i\phi_j\phi_k+\cdots. 
\end{align}
In Ref.~\cite{KOSY}, $\lambda_{\phi_i\phi_j\phi_k}$ are listed. 

The decay rates for the loop-induced decay modes can be calculated as 
\begin{align}
\Gamma(\mathcal{H}\to \gamma\gamma)&=\frac{\sqrt{2}G_F\alpha_{\text{em}}^2m_\mathcal{H}^3}{64\pi^3 }
|I_{H^\pm}(m_\mathcal{H})+\sum_fQ_f^2N_c^f\xi_\mathcal{H}^fI_f(m_\mathcal{H})+\eta_\mathcal{H}I_W(m_\mathcal{H})|^2 ,\\
\Gamma(A\to \gamma\gamma)&=\frac{\sqrt{2}G_F\alpha_{\text{em}}^2m_A^3}{64\pi^3 }
|\sum_fQ_f^2N_c^f\xi_A^fI_f^A(m_A)|^2 ,\\
\Gamma(\mathcal{H}\to gg)&=\frac{\sqrt{2}G_F\alpha_s^2m_\mathcal{H}^3}{128\pi^3}|\sum_{f=q}\xi_\mathcal{H}^fI_f(m_\mathcal{H})|^2,\\
\Gamma(A\to gg)&=\frac{\sqrt{2}G_F\alpha_s^2m_A^3}{128\pi^3}|\sum_{f=q}\xi_A^fI_f^A(m_A)|^2,
\end{align}
where
\begin{align}
I_{H^\pm}(m_\mathcal{H})&=\frac{v\lambda_{\varphi H^+ H^-}}{m_\mathcal{H}^2}[1+2m_{H^+}^2C_0(0,0,m_\mathcal{H}^2;m_{H^+}^2,m_{H^+}^2,m_{H^+}^2)],\\
I_{f}^A(m_A)&=2m_f^2C_0(0,0,m_h^2;m_f^2,m_f^2,m_f^2),
\end{align}

In the THDM, the $W^\pm H^\mp Z $ vertex does not appear at the tree-level, but it appears at the one-loop level 
as we have discussed in section~3.3. 
The effective $W^\pm H^\mp Z $ vertex can be expressed in Eq.~(\ref{HWV}) and in Fig.~\ref{hwvf}. 
We can caluclate the decay rate of $H^\pm\to ZH^\pm$ process in terms of $F$, $G$ and $H$ as
\begin{align}
\Gamma(H^\pm\to W^\pm Z)
&=\frac{g^2m_{H^\pm}}{16\pi }\lambda\left(\frac{m_W^2}{m_{H^+}^2},\frac{m_Z^2}{m_{H^+}^2}\right)^{1/2}\notag\\
&\times\Bigg[\frac{m_{H^+}^2}{4m_Z^2}
\Big|F^*\left(1-\frac{m_W^2}{m_{H^+}^2}-\frac{m_Z^2}{m_{H^+}^2}\right)
+\frac{G^*m_{H^+}^2}{2m_W^2}\lambda\left(\frac{m_W^2}{m_{H^+}^2},\frac{m_W^2}{m_{H^+}^2}\right)\Big|^2
\notag\\
&+\frac{2m_W^2}{m_{H^+}^2}|F|^2+\frac{m_{H^+}^2}{2m_W^2}|H|^2\lambda\left(\frac{m_W^2}{m_{H^+}^2},\frac{m_Z^2}{m_{H^+}^2}\right)\Bigg].
\end{align} 
\section{Decay rates of the Higgs bosons in the HTM}
In the HTM, in addition to the SM-like Higgs bososn $h$, 
there are the doubly-charged scalar bosons $H^\pm$, the singly-charged scalar bosons $H^\pm$, the 
neutral CP-even (odd) scalar boson $H$ ($A$). 
Here we list the formulae of decay rates for $H^{\pm\pm}$, $H^\pm$, $H$ and $A$ in order. \\
\subsection{Decay rates of $H^{\pm\pm}$}
The decay rates for $H^{\pm\pm}$ can be evaluated as 
\begin{align}
\Gamma(H^{\pm\pm} \to \ell_i^\pm \ell_j^\pm)
&=S_{ij}|h_{ij}|^2\frac{m_{H^{++}}}{4\pi}\left(1-\frac{m_i^2}{m_{H^{++}}^2}-\frac{m_j^2}{m_{H^{++}}^2}\right)\left[\lambda\left(\frac{m_i^2}{m_{H^{++}}^2},\frac{m_j^2}{m_{H^{++}}^2}\right)\right]^{1/2},\label{a1}\\
\Gamma(H^{\pm\pm} \to W^\pm W^\pm)&=\frac{g^4v_\Delta^2m_{H^{++}}^3}{64\pi m_W^4}
\left(1-\frac{4m_W^2}{m_{H^{++}}^2}+\frac{12m_W^4}{m_{H^{++}}^4}\right)\beta\left(\frac{m_W^2}{m_{H^{++}}^2}\right),\\
\Gamma(H^{\pm\pm} \to H^\pm W^\pm )&=\frac{g^2m_{H^{++}}^3}{16\pi m_W^2}\cos^2\beta_\pm\left[\lambda\left(\frac{m_W^2}{m_{H^{++}}^2},\frac{m_{H^+}^2}{m_{H^{++}}^2}\right)\right]^{3/2},\\
\Gamma(H^{\pm\pm} \to W^\pm W^{\pm *}) 
&=\frac{3g^6m_{H^{++}}}{512\pi^3}\frac{v_{\Delta}^2}{m_W^2}F\left(\frac{m_W^2}{m_{H^{++}}^2}\right),\\
\Gamma(H^{\pm\pm} \to H^\pm W^{\pm *})&=
\frac{9g^4m_{H^{++}}}{128\pi^3}\cos^2\beta_\pm G\left(\frac{m_{H^+}^2}{m_{H^{++}}^2},\frac{m_W^2}{m_{H^{++}}^2}\right),
\end{align}
where $m_i$ is the lepton mass ($i=e,\mu$ or $\tau$) and $S_{ij}=1$, $(1/2)$ for $i\neq j$, $(i=j)$. 
The functions $\beta(x)$, $\lambda(x,y)$, $F(x)$ and $G(x,y)$ are given in Eqs.~(\ref{decay_beta}), (\ref{decay_lam}), 
(\ref{decay_F}) and (\ref{g_func}), respectively. 
Although the expression in Eq.~(\ref{g_func}) is different from that in Ref.~\cite{Djouadi:1995gv}, 
we have confirmed that the numerical value by using Eq.~(\ref{g_func}) coincides with 
that by using CalcHEP. 

\subsection{Decay rates of $H^{\pm}$}
The decay rates for $H^{\pm}$ can be evaluated as 
\begin{align}
\Gamma(H^\pm \to q\bar{q}')&=\frac{3m_{H^+}^3}{8\pi v^2}\sin^2\beta_\pm 
\left[\left(\frac{m_q^2}{m_{H^+}^2}+\frac{m_{q'}^2}{m_{H^+}^2}\right)\left(1-\frac{m_q^2}{m_{H^+}^2}-\frac{m_{q'}^2}{m_{H^+}^2}\right)-4\frac{m_q^2}{m_{H^+}^2}\frac{m_{q'}^2}{m_{H^+}^2}\right]\notag\\
&\hspace{23mm}\times\left[\lambda\left(\frac{m_q^2}{m_{H^+}^2},\frac{m_{q'}^2}{m_{H^+}^2}\right)\right]^{1/2},\\
\Gamma(H^\pm \to \ell_i^\pm\nu_j)&=\delta_{ij}\frac{m_i^2m_{H^+}}{8\pi v^2}\sin^2\beta_\pm \left(1-\frac{m_i^2}{m_{H^+}^2}\right)^2+|h_{ij}|^2\frac{m_{H^+}}{8\pi}\cos^2\beta_\pm\left(1-\frac{m_i^2}{m_{H^+}^2}\right)^2,\\
\Gamma(H^\pm \to W^\pm Z)
& =\frac{g^2g_Z^2}{32\pi m_{H^+}}v_\Delta^2\cos^2\beta_\pm\left[\lambda\left(\frac{m_W^2}{m_{H^+}^2},\frac{m_Z^2}{m_{H^+}^2}\right)\right]^{1/2}
\notag\\
&\times\left[2+\frac{m_{H^+}^4}{4m_W^2m_Z^2}\left(1-\frac{m_W^2}{m_{H^+}^2}-\frac{m_Z^2}{m_{H^+}^2}\right)^2\right],\\
\Gamma(H^\pm \to W^\pm Z^* )&=\frac{3g^2g_Z^4}{1024\pi^3m_{H^+}}v_\Delta^2\cos^2\beta_\pm H\left(\frac{m_W^2}{m_{H^+}^2},\frac{m_Z^2}{m_{H^+}^2}\right)\notag\\
&\times\left(7-\frac{40}{3}\sin^2\theta_W+\frac{160}{9}\sin^4\theta_W\right),\\
\Gamma(H^\pm \to W^{\pm *} Z)&=\frac{9g^4g_Z^2}{512\pi^3 m_{H^+}}v_\Delta^2\cos^2\beta_\pm H\left(\frac{m_Z^2}{m_{H^+}^2},\frac{m_W^2}{m_{H^+}^2}\right),\\
\Gamma(H^\pm \to \hat{\varphi} W^\pm ) & =\frac{g^2m_{H^+}^3}{64\pi^2m_W^2}\xi_{H^+ W^- \hat{\varphi}}^2\left[\lambda\left(\frac{m_W^2}{m_{H^+}^2},\frac{m_{\hat{\varphi}}^2}{m_{H^+}^2}\right)\right]^{3/2},\label{a2}\\
\Gamma(H^\pm \to \hat{\varphi} W^{\pm *}  ) & =\frac{9g^4m_{H^+}}{512\pi^3}\xi_{H^+ W^- \hat{\varphi}}^2G\left(\frac{m_{\hat{\varphi}}^2}{m_{H^+}^2},\frac{m_W^2}{m_{H^+}^2}\right),\label{a3}
\end{align}
where $g_Z=g/\cos\theta_W$. 
The function $H(x,y)$ is
\begin{align}
&H(x,y)=\frac{\tan^{-1}\left[\frac{1-x+y}{\sqrt{-\lambda(x,y)}}\right]
+\tan^{-1}\left[\frac{1-x-y}{\sqrt{-\lambda(x,y)}}\right]}{4x \sqrt{-\lambda(x,y)}}
\Big[-3x^3+(9y+7)x^2-5(1-y)^2x+(1-y)^3\Big] \notag\\
&+\frac{1}{24xy}\Bigg\{(-1+x)[6y^2+y(39x-9)+2(1-x)^2]
-y[y^2+2y(3x-1)-x(3x+4)+1]\log x\Bigg\}. \label{hfunc}
\end{align}
We have confirmed that the numerical value by using Eq.~(\ref{hfunc}) coincides with 
that by using CalcHEP. 
In Eq.~(\ref{a2}) and Eq.~(\ref{a3}), 
$\hat{\varphi}$ denotes $h$, $H$ or $A$ and $\xi_{H^+W^- \hat{\varphi}}$ is expressed as 
\begin{align}
\xi_{H^+W^- h}=\cos\alpha\sin\beta_\pm-\sqrt{2}\sin\alpha\cos\beta_\pm,\notag\\
\xi_{H^+W^- H}=\sin\alpha\sin\beta_\pm+\sqrt{2}\cos\alpha\cos\beta_\pm,\notag\\
\xi_{H^+W^- A}=\sin\beta_0\sin\beta_\pm+\sqrt{2}\cos\beta_0\cos\beta_\pm.
\end{align}

\subsection{ Decay rates of $H$}
The decay rates for $H$ can be evaluated as 
\begin{align}
\Gamma(H\to f\bar{f})&=\frac{N_c^fm_f^2m_H}{8\pi v^2}\sin^2\alpha\left[\beta\left(\frac{m_f^2}{m_H^2}\right)\right]^3,\\
\Gamma(H \to \nu \nu)&=\Gamma(H \to \nu^c \bar{\nu})+\Gamma(H \to \bar{\nu}^c \nu)
=\sum_{i,j=1}^3S_{ij}|h_{ij}|^2\frac{m_H}{4\pi}\cos^2\alpha,\\
\Gamma(H\to W^+W^-)&=\frac{g^4 m_H^3}{16\pi m_W^4}\left(\frac{v}{2}\sin\alpha -v_\Delta \cos\alpha\right)^2\left(\frac{1}{4}-\frac{m_W^2}{m_H^2}+\frac{3m_W^4}{m_H^4}\right)\beta\left(\frac{m_W^2}{m_H^2}\right),\\
\Gamma(H\to ZZ)&=\frac{g_Z^4 m_H^3}{32\pi m_Z^4}\left(\frac{v}{2}\sin\alpha -2v_\Delta \cos\alpha\right)^2\left(\frac{1}{4}-\frac{m_Z^2}{m_H^2}+\frac{3m_Z^4}{m_H^4}\right)\beta\left(\frac{m_Z^2}{m_H^2}\right),\\
\Gamma(H\to WW^*)&=\frac{3 g^6 m_H}{512\pi^3}\frac{(\frac{v}{2}\sin\alpha-v_\Delta \cos\alpha)^2}{m_W^2}F\left(\frac{m_W^2}{m_H^2}\right),\\
\Gamma(H\to ZZ^*)&=\frac{g_Z^6 m_H}{2048\pi^3}\frac{(\frac{v}{2}\sin\alpha-2v_\Delta \cos\alpha)^2}{m_Z^2}\notag\\
&\times\left(7-\frac{40}{3}\sin^2\theta_W+\frac{160}{9}\sin^4\theta_W\right)F\left(\frac{m_Z^2}{m_H^2}\right),\\
\Gamma(H\to hh)&=\frac{\lambda_{Hhh}^2}{8\pi m_H}\beta\left(\frac{m_h^2}{m_H^2}\right),
\end{align}
where 
\begin{align}
\lambda_{Hhh}&=\frac{1}{4v^2}\Big\{2v_\Delta\left[-2M_\Delta^2+v^2(\lambda_4+\lambda_5)\right]\cos^3\alpha+v^3\left[-12\lambda_1+4(\lambda_4+\lambda_5)\right]\cos^2\alpha\sin\alpha \notag\\
&+4v_\Delta\left[2M_\Delta^2+v^2(3\lambda_2+3\lambda_3-\lambda_4-\lambda_5)\right]\cos\alpha\sin^2\alpha-2v^3(\lambda_4+\lambda_5)\sin^3\alpha\Big\}\notag\\
&\simeq \frac{1}{4v^2}\Big\{2v_\Delta\left[-2M_\Delta^2+v^2(\lambda_4+\lambda_5)\right]\cos^3\alpha+v^3\left[-12\lambda_1+4(\lambda_4+\lambda_5)\right]\cos^2\alpha\sin\alpha\Big\}.
\end{align}

\subsection{Decay rates of $A$}
The decay rates for $H$ can be evaluated as 
\begin{align}
\Gamma(A\to f\bar{f})&=\sin^2\beta_0\frac{N_c^f m_f^2 m_A}{8\pi v^2}\beta\left(\frac{m_f^2}{m_A^2}\right),\\
\Gamma(A\to \nu\nu)&=\Gamma(A\to \nu^c\bar{\nu})+\Gamma(A\to \bar{\nu}^c\nu)=\sum_{i,j=1}^3S_{ij}|h_{ij}|^2\frac{m_A}{4\pi}\cos^2\beta_0,\\
\Gamma(A\to hZ)&=\frac{g_Z^2m_A^3}{64\pi m_Z^2}(\cos\alpha\sin\beta_0-2\sin\alpha\cos\beta_0)^2\left[\lambda\left(\frac{m_h^2}{m_A^2},\frac{m_Z^2}{m_A^2}\right)\right]^{3/2},\\
\Gamma(A\to hZ^*)&=\frac{3g_Z^4m_A}{1024\pi^3}(\cos\alpha\sin\beta_0-2\sin\alpha\cos\beta_0)^2\notag\\
&\times G\left(\frac{m_h^2}{m_A^2}, \frac{m_Z^2}{m_A^2}\right)\left(7-\frac{40}{3}\sin^2\theta_W+\frac{160}{9}\sin^4\theta_W\right).
\end{align}

\chapter{One-loop functions}
In this appendix, we introduce the one-loop functions according to Passarino and Veltman~\cite{Ref:PV} and Ref.~{\cite{hhkm}}. 
First, we define A and B functions: 
\begin{align}
A(m_1)&=\int\frac{\overline{d^Dk}}{i\pi^2}\frac{1}{D_1},\\
[B_0,B^\mu,B^{\mu\nu}](p^2,m_1,m_2)&=\int\frac{\overline{d^Dk}}{i\pi^2}\frac{[1,k^\mu,k^\mu k^\nu]}{D_1D_2},  
\end{align}
where $D=4-2\epsilon$, 
\begin{align}
\overline{d^Dk}=\Gamma(1-\epsilon)(\pi\mu^2)^\epsilon d^Dk, 
\end{align}
is the $\overline{\text{MS}}$ regularization~\cite{MSbar1,MSbar2}, and the propagator factors are
\begin{align}
D_1=k^2-m_1^2+i\varepsilon,\quad D_2=(k+p)^2-m_2^2+i\varepsilon. 
\end{align}
The $A$ function is given by 
\begin{align}
A(m)&=m^2(\Delta+1-\ln m^2).
\end{align}
The vector and the tensor functions are reduced to the scalar functions as  
\begin{align}
B^\mu(p^2,m_1,m_2)&=p^\mu B_1(p^2,m_1,m_2),\label{b1}\\
B^{\mu\nu}(p^2,m_1,m_2)&=p^\mu p^\nu B_{21}(p^2,m_1,m_2)+g^{\mu\nu}B_{22}(p^2,m_1,m_2).\label{b2}
\end{align}
The coefficients of the vector and tensor functions ($B_1$, $B_{21}$ and $B_{22}$) 
can be expressed in terms of the functions $B_0$ and $A_0$ as  
\begin{subequations}
\begin{align}
B_1&=\frac{1}{2p^2}[A(m_1)-A(m_2)+f_1B_0(p^2,m_1,m_2)],\\
B_{21}&=\frac{1}{p^2(D-1)}\left[\frac{D}{2}f_1B_1(p^2,m_1,m_2)+\left(\frac{D}{2}-1\right)A(m_2)-m_1^2B_0(p^2,m_1,m_2)\right],\\
B_{22}&=\frac{1}{D-1}\left[\frac{1}{2}A(m_2)+m_1^2B_0(p^2,m_1,m_2)-\frac{1}{2}f_1B_1(p^2,m_1,m_2)\right], 
\end{align}
\end{subequations}
where $f_1=m_2^2-m_1^2-p^2$. 
By plugging $D=4-2\epsilon$ to the above two expression, we obtain 
\begin{subequations}
\begin{align}
B_{21}&=\frac{1}{3p^2}\left[2f_1B_1(p^2,m_1,m_2)+A(m_2)-m_1^2B_0(p^2,m_1,m_2)+\frac{f_1}{2}-m_2^2+\frac{2}{3}p^2\right]+\mathcal{O}(\epsilon),\\
B_{22}&=\frac{1}{3}\left[\frac{1}{2}A(m_2)+m_1^2B_0(p^2,m_1,m_2)-\frac{1}{2}f_1B_1(p^2,m_1,m_2)\right]+\frac{1}{6}(m_1^2+m_2^2)-\frac{p^2}{18}+\mathcal{O}(\epsilon).
\end{align}
\end{subequations}
It is convenient to introduce the following four $B$ functions in addition to $B_0$ and $B_1$ as
\begin{subequations}
\begin{align}
B_2(p^2,m_1,m_2)&=B_{21}(p^2,m_1,m_2),\\
B_3(p^2,m_1,m_2)&=-B_1(p^2,m_1,m_2)-B_2(p^2,m_1,m_2),\\
B_4(p^2,m_1,m_2)&=-m_1^2B_1(p^2,m_2,m_1)-m_2^2B_1(p^2,m_1,m_2),\\
B_5(p^2,m_1,m_2)&=A(m_1)+A(m_2)-4B_{22}(p^2,m_1,m_2).
\end{align}
\end{subequations}
The $B_n$ ($n = 0,1,\cdots,5$) function can be decomposed into the infinite part $\Delta$ and the finite part as
\begin{subequations}
\begin{align}
B_0(p^2,m_1,m_2)&=\Delta-F_0(p^2,m_1,m_2),\\
B_1(p^2,m_1,m_2)&=-\frac{1}{2}\Delta+F_1(p^2,m_1,m_2),\\
B_2(p^2,m_1,m_2)&=\frac{1}{3}\Delta-F_2(p^2,m_1,m_2),\\
B_3(p^2,m_1,m_2)&=\frac{1}{6}\Delta-F_3(p^2,m_1,m_2),\\
B_4(p^2,m_1,m_2)&=\frac{m_1^2+m_2^2}{2}\Delta-F_4(p^2,m_1,m_2),\\
B_5(p^2,m_1,m_2)&=\frac{p^2}{3}\Delta-F_5(p^2,m_1,m_2).
\end{align}
\end{subequations}
In the $\overline{\text{MS}}$ 
renormalization scheme the singular piece $\Delta$ i
n these function is simply replaced by a logarithm of the unit of mass $\mu$:
\begin{align}
\Delta\to\ln \mu^2. 
\end{align}
The $F_0$, $F_3$ and $F_A$ functions are given by
\begin{align}
F_0(p^2,m_1,m_2)&=\ln(m_1m_2)-\delta\ln\frac{m_2}{m_1}-2+\beta L,\\
F_3(p^2,m_1,m_2)&=\frac{1}{6}\ln(m_1m_2)-\frac{3\sigma-2\delta^2}{6}\delta\ln\frac{m_2}{m_1}-\frac{5}{18}
-\frac{\sigma-\delta^2}{3}+\frac{1+\sigma-2\delta^2}{6}\beta L,\\
F_A(p^2,m_1,m_2)&=-(\sigma-\delta^2)\ln\frac{m_2}{m_1}+\delta(1-\beta L), 
\end{align}
where 
\begin{align}
&\sigma=\frac{m_1^2+m_2^2}{p^2},\quad \delta=\frac{m_1^2-m_2^2}{p^2},\\
&\beta=\left\{
\begin{array}{l}
\sqrt{1-2\sigma+\delta^2}\text{ for }p^2<(m_1-m_2)^2\text{ or }p^2>(m_1+m_2)^2,\\
i\sqrt{2\sigma-\delta^2-1}\text{ for }(m_1-m_2)^2<p^2<(m_1+m_2)^2,
\end{array}\right.\\
L&=\left\{
\begin{array}{l}
\frac{1}{2}\ln\frac{1+\beta-\sigma}{1-\beta-\sigma}-i\pi\text{ for }p^2>(m_1+m_2)^2,\\
\frac{1}{2}\ln\frac{1+\beta-\sigma}{1-\beta-\sigma}\text{ for }p^2<(m_1-m_2)^2,\\
-i\left(\arctan\frac{1-\delta}{|\beta|}+\arctan\frac{1+\delta}{|\beta|}\right)\text{ for }(m_1-m_2)^2<p^2<(m_1+m_2)^2.
\end{array}\right.
\end{align}
The other $F_1$, $F_2$, $F_4$ and $F_5$ functions can be written in terms of the $F_0$, $F_3$ and $F_A$ functions as:
\begin{subequations}
\begin{align}
F_1(p^2,m_1,m_2)&=\frac{1}{2}[F_0-F_A](p^2,m_1,m_2)\label{f1},\\
F_2(p^2,m_1,m_2)&=\Big[\frac{1}{2}(F_0-F_A)-F_3\Big](p^2,m_1,m_2)\label{f2},\\
F_4(p^2,m_1,m_2)&=\Big[\frac{m_1^2+m_2^2}{2}F_0+\frac{m_1^2-m_2^2}{2}F_A\Big](p^2,m_1,m_2)\label{f4},\\
F_5(p^2,m_1,m_2)&=\Big[p^2(F_0-4F_3)+(m_1^2-m_2^2)F_A\Big](p^2,m_1,m_2)\label{f5}.
\end{align}
\end{subequations}
In the case with $m_1^2=m^2>p^2$ and $m_2^2\simeq 0$, we obtain 
\begin{subequations}
\begin{align}
F_0(p^2,m,0)&=\ln p^2 +R\ln R-2-(R-1)\ln(R-1),\\
F_3(p^2,m,0)&=\frac{1}{6}\ln p^2+\frac{R^2}{6}(3-2R)\ln R -\frac{5}{18}-R\frac{1-R}{3}+\frac{1}{6}(1-3R^2+2R^3)\ln (R-1),\\
F_A(p^2,m,0)&=R(1-R)\ln R+R +R(R-1)\ln(R-1),
\end{align}
\end{subequations}
where $R\equiv m^2/p^2$.
In the case with $m_1=m_2=m$, these expressions are reduced to 
\begin{subequations}
\begin{align}
F_0(p^2,m,m)&=2\ln m-2+\beta L,\\
F_1(p^2,m,m)&=\ln m-1+\frac{1}{2}\beta L,\\
F_2(p^2,m,m)&=\frac{2}{3}\ln m-\frac{13}{18}+\frac{2m^2}{3p^2}+\frac{p^2-m^2}{p^2}\beta L,\\
F_3(p^2,m,m)&=\frac{1}{3}\ln m-\frac{5}{18}
-\frac{2m^2}{3p^2}+\frac{p^2+2m^2}{6p^2}\beta L,\\
F_4(p^2,m,m)&=2m^2\ln m-2m^2+m^2\beta L,\\
F_5(p^2,m,m)&=\frac{2}{3}p^2\ln m -\frac{8}{9}p^2+\frac{8}{3}m^2+\frac{p^2-4m^2}{3}\beta L,\\
F_A(p^2,m,m)&=0,
\end{align}
\end{subequations}
with
\begin{align}
\beta=
\sqrt{1-\frac{4m^2}{p^2}},\quad
L=\left\{
\begin{array}{l}
\frac{1}{2}\ln\frac{1+\beta-\sigma}{1-\beta-\sigma}-i\pi\text{ for }p^2>4m^2,\\
-2i\arctan\frac{1}{|\beta|}\text{ for }p^2<4m^2.
\end{array}\right.
\end{align}
In the case with $p^2=0$, $F_0$, $F_3$ and $F_A$ functions are given by 
\begin{subequations}
\begin{align}
F_0(0,m_1,m_2)&=\ln(m_1m_2)-\frac{m_1^2+m_2^2}{m_1^2-m_2^2}\ln\frac{m_2}{m_1}-1\label{f0},\\
F_1(0,m_1,m_2)&=\frac{1}{2}\ln (m_1m_2)-\frac{1}{2}-\frac{1}{4}\frac{m_1^2+m_2^2}{m_1^2-m_2^2}+\frac{-m_1^4+m_2^4-2m_1^2m_2^2}{2(m_1^2-m_2^2)^2}\ln\frac{m_2}{m_1},\\
F_2(0,m_1,m_2)&=\frac{1}{3}\ln (m_1m_2)-\frac{13}{36}-\frac{3m_1^4-3m_2^4+4m_1^2m_2^2}{12(m_1^2-m_2^2)^2}
+\frac{m_1^6+3m_1^4m_2^2-3m_1^2m_2^4+m_2^6}{3(m_1^2-m_2^2)^3},\\
F_3(0,m_1,m_2)&=\frac{1}{6}\ln(m_1m_2)-\frac{5}{36}+\frac{1}{3}\frac{m_1^2m_2^2}{(m_1^2-m_2^2)^2}\notag\\
&-\frac{1}{6}\frac{(m_1^2+m_2^2)(m_1^4+m_2^4-4m_1^2m_2^2)}{(m_1^2-m_2^2)^3}\ln\frac{m_2}{m_1}\label{f3},\\
F_4(0,m_1,m_2)&=-\frac{1}{4}(m_1^2+m_2^2)-\frac{1}{2}\frac{m_1^4+m_2^4}{m_1^2-m_2^2}\ln\frac{m_2}{m_1}+\frac{m_1^2+m_2^2}{2}\ln(m_1m_2),\\
F_5(0,m_1,m_2)&=\frac{1}{2}(m_1^2+m_2^2)+\frac{2m_1^2m_2^2}{(m_1^2-m_2^2)}\ln\frac{m_2}{m_1}\label{b5tof},\\
F_A(0,m_1,m_2)&=\frac{1}{2}\frac{m_1^2+m_2^2}{m_1^2-m_2^2}+\frac{2m_1^2m_2^2}{(m_1^2-m_2^2)^2}\ln\frac{m_2}{m_1}.\label{fa}
\end{align}
\end{subequations}
In the limit of $p^2\to 0$ and $m_1=m_2=m$, we obtain 
\begin{subequations}
\begin{align}
F_0(0,m,m)&=\ln m^2,\\
F_1(0,m,m)&=\frac{1}{2}\ln m^2,\\
F_2(0,m,m)&=\frac{1}{3}\ln m^2,\\
F_3(0,m,m)&=\frac{1}{6}\ln m^2,\\
F_4(0,m,m)&=m^2\ln m^2,\\
F_5(0,m,m)&=F_A(0,m,m)=0.
\end{align} 
\end{subequations}

\chapter{Gauge boson self-energies and vertex corrections}
In this appendix, 
analytic expressions for the gauge boson self-energies at the one-loop level are listed 
in terms of the Passarino-Veltman functions~\cite{Ref:PV}. 
The gauge boson propagators $D_{\mu\nu}^V(p)$ ($V=\gamma,Z$ and $W$) and the photon-$Z$ boson mixing 
can be expressed as~\cite{hollik_sm}
\begin{align}
D_{\mu\nu}^V(p^2)&=-ig_{\mu\nu}\left(\frac{1}{p^2-m_V^2}-\frac{1}{p^2-m_V^2}\Pi_{\mu\nu}^{VV}(p^2)\frac{1}{p^2-m_V^2}\right),\\
D_{\mu\nu}^{\gamma Z}(p^2)&=+ig_{\mu\nu}\frac{1}{p^2-m_Z^2}\Pi_{\mu\nu}^{\gamma Z}\frac{1}{p^2}, 
\end{align}
where $\Pi^{VV}_{\mu\nu}(p^2)$ is the amplitude of the gauge boson two-point functions at the one-loop level. 
The functions $\Pi^{VV}_{\mu\nu}(p^2)$ can be decomposed into the transverse part 
and the longitudinal part as
\begin{align}
(\Pi^{VV})^{\mu\nu}(p^2)=\left(-g^{\mu\nu}+\frac{p^\mu p^\nu}{p^2}\right)\Pi_T^{VV}(p^2)+\frac{p^\mu p^\nu}{p^2}\Pi_L^{VV}(p^2). 
\end{align} 
The transverse part of the gauge boson two point functions $\Pi_T^{VV}$ 
are composed of the fermionic-loop contributions and the bosonic-loop contributions as
\begin{align}
\Pi_T^{VV}(p^2)=\Pi_{T,F}^{VV}(p^2)+\Pi_{T,B}^{VV}(p^2). 
\end{align} 
\section{Fermionic-loop contributions}
The fermionic-loop contributions to the transverse part of the gauge boson two point functions are 
calculated as 
\begin{align}
\Pi_{T,F}^{WW}(p^2)&=\frac{g^2}{16\pi^2}\sum_{f}N_c^f[-B_4+2p^2B_3](p^2,m_f,m_{f'}),\\
\Pi_{T,F}^{\gamma\gamma}(p^2)&=\frac{e^2}{16\pi^2}\sum_{f}N_c^fQ_f^2[8p^2B_3](p^2,m_f,m_f),\\
\Pi_{T,F}^{\gamma Z}(p^2)&=-\frac{g^2}{16\pi^2}\frac{\hat{s}_W}{\hat{c}_W}\sum_{f}N_c^f2p^2[-4\hat{s}_W^2Q_f^2+2I_fQ_f]B_3(p^2,m_f,m_f),\\
\Pi_{T,F}^{ZZ}(p^2)
&=\frac{g_Z^2}{16\pi^2}\sum_f N_c^f\Big[2p^2(4\hat{s}_W^4Q_f^2-4\hat{s}_W^2Q_fI_f+2I_f^2)B_3-2I_f^2m_f^2B_0\Big](p^2,m_f,m_f).
\end{align}
\section{Bosonic-loop contributions}
In the HTM, the bosonic-loop contributions to the transverse part of the gauge boson two point functions are listed below.  
The W boson two-point function is calculated as
\begin{align}
&\left(\frac{1}{16\pi^2}\right)^{-1}\Pi_{T,B}^{WW}(p^2)
=\left(\frac{1}{16\pi^2}\right)^{-1}\Pi_{T,B}^{WW}(p^2)_\text{SM}\notag\\
&+g^4\left(\frac{v_\Phi}{2}c_\alpha+v_\Delta s_\alpha\right)^2B_0(p^2,m_h,m_W)
+g^4\left(-\frac{v_\Phi}{2}s_\alpha+v_\Delta c_\alpha\right)^2B_0(p^2,m_H,m_W)\notag\\
&+2g^4v_\Delta^2B_0(p^2,m_{H^{++}},m_W)\notag\\
&+\frac{g^4}{2\hat{c}_W^2}v_\Delta^2c_{\beta_\pm}^2B_0(p^2,m_{H^+},m_W)
+\frac{g^4}{\hat{c}_W^2}\left[\frac{v_\Phi}{2}\hat{s}_W^2c_{\beta_\pm}
+\frac{v_\Delta}{\sqrt{2}}(1+\hat{s}_W^2)s_{\beta_\pm}\right]^2B_0(p^2,m_Z,m_W)\notag\\
&+\frac{g^2e^2}{4}(v_\Phi^2+2v_\Delta^2)B_0(p^2,0,m_W)\notag\\
&+g^2c_{\beta_\pm}^2B_5(p^2,m_{H^{++}},m_{H^+})+g^2s_{\beta_\pm}^2B_5(p^2,m_{H^{++}},m_W)\notag\\
&+\frac{g^2}{4}(c_\alpha s_{\beta_\pm}-\sqrt{2}s_\alpha c_{\beta_\pm})^2B_5(p^2,m_{H^+},m_h)
+\frac{g^2}{4}(c_\alpha c_{\beta_\pm}+\sqrt{2}s_\alpha s_{\beta_\pm})^2B_5(p^2,m_W,m_h)\notag\\
&+\frac{g^2}{4}(s_\alpha s_{\beta_\pm}+\sqrt{2}c_\alpha c_{\beta_\pm})^2B_5(p^2,m_{H^+},m_H)
+\frac{g^2}{4}(s_\alpha c_{\beta_\pm}-\sqrt{2}c_\alpha s_{\beta_\pm})^2B_5(p^2,m_W,m_H)\notag\\
&+\frac{g^2}{4}(s_{\beta_0} s_{\beta_\pm}+\sqrt{2}c_{\beta_0} c_{\beta_\pm})^2B_5(p^2,m_{H^+},m_A)
+\frac{g^2}{4}(s_{\beta_0} c_{\beta_\pm}-\sqrt{2}c_{\beta_0} s_{\beta_\pm})^2B_5(p^2,m_W,m_A)\notag\\
&+\frac{g^2}{4}(-c_{\beta_0} s_{\beta_\pm}+\sqrt{2}s_{\beta_0} c_{\beta_\pm})^2B_5(p^2,m_{H^+},m_Z)
+\frac{g^2}{4}(c_{\beta_0} c_{\beta_\pm}+\sqrt{2}s_{\beta_0} s_{\beta_\pm})^2B_5(p^2,m_W,m_Z). 
\end{align}
The photon two-point function is calculated as
\begin{align}
\left(\frac{1}{16\pi^2}\right)^{-1}\Pi_{T,B}^{\gamma\gamma}(p^2)&=
\left(\frac{1}{16\pi^2}\right)^{-1}\Pi_{T,B}^{\gamma\gamma}(p^2)_\text{SM}\notag\\
&+\frac{e^2g^2}{2}(v_\Phi^2+2v_\Delta^2)B_0(p^2,m_W,m_W)\notag\\
&+4e^2B_5(p^2,m_{H^{++}},m_{H^{++}})+e^2B_5(p^2,m_{H^+},m_{H^+})+e^2B_5(p^2,m_W,m_W). 
\end{align}
The photon-Z boson mixing is calculated as
\begin{align}
\left(\frac{1}{16\pi^2}\right)^{-1}\Pi_{T,B}^{\gamma Z}(p^2)
&=\left(\frac{1}{16\pi^2}\right)^{-1}\Pi_{T,B}^{\gamma Z}(p^2)_\text{SM}\notag\\
&+g^4\frac{\hat{s}_W}{\hat{c}_W}\sqrt{v_\Phi^2+2v_\Delta^2}\left[\frac{v_\Phi}{2}\hat{s}_W^2c_{\beta_\pm}+\frac{v_\Delta}{\sqrt{2}}(1+\hat{s}_W^2)s_{\beta_\pm}\right]
B_0(p^2,m_W,m_W)\notag\\
&-2g^2\frac{\hat{s}_W(\hat{c}_W^2-\hat{s}_W^2)}{\hat{c}_W}B_5(p^2,m_{H^{++}},m_{H^{++}})\notag\\
&-\frac{g^2}{2}\frac{\hat{s}_W}{\hat{c}_W}(\hat{c}_W^2-\hat{s}_W^2-c_{\beta_\pm}^2)B_5(p^2,m_{H^+},m_{H^+})\notag\\
&-\frac{g^2}{2}\frac{\hat{s}_W}{\hat{c}_W}(\hat{c}_W^2-\hat{s}_W^2-s_{\beta_\pm}^2)B_5(p^2,m_W,m_W),
\end{align}
where $\Pi_{T,B}^{VV}(p^2)_\text{SM}$ are the SM gauge boson loop contributions. 
These are calculated as 
\begin{align}
\left(\frac{1}{16\pi^2}\right)^{-1}\Pi_{T,B}^{WW}(p^2)_\text{SM}&=
-g^2\hat{s}_W^2[6(D-1)B_{22}+p^2(2B_{21}+2B_1+5B_0)](p^2,0,m_W)\notag\\
&-g^2\hat{c}_W^2[6(D-1)B_{22}+p^2(2B_{21}+2B_1+5B_0)](p^2,m_Z,m_W)\notag\\
&+g^2(D-1)\left[\hat{c}_W^2A(m_Z)+A(m_W)\right]\notag\\
&+2e^2\left[B_{22}(p^2,0,m_W)+\frac{\hat{c}_W^2}{\hat{s}_W^2}B_{22}(p^2,m_Z,m_W)\right]\notag\\
&-4g^2(p^2-m_W^2)[\hat{c}_W^2B_0(p^2,m_W,m_Z)+\hat{s}_W^2B_0(p^2,m_W,0)]
,\\
\left(\frac{1}{16\pi^2}\right)^{-1}\Pi_{T,B}^{\gamma \gamma}(p^2)_\text{SM}&=
-e^2[6(D-1)B_{22}+p^2(2B_{21}+2B_1+5B_0)](p^2,m_W,m_W)\notag\\
&+2e^2(D-1)A(m_W)+2e^2B_{22}(p^2,m_W,m_W)\notag\\
&-4e^2p^2B_0(p^2,m_W,m_W),\\
\left(\frac{1}{16\pi^2}\right)^{-1}\Pi_{T,B}^{\gamma Z}(p^2)_\text{SM}&=
+e^2\frac{\hat{c}_W}{\hat{s}_W}[6(D-1)B_{22}+p^2(2B_{21}+2B_1+5B_0)](p^2,m_W,m_W)\notag\\
&-2e^2\frac{\hat{c}_W}{\hat{s}_W}(D-1)A(m_W),\notag\\
&-2e^2\frac{\hat{c}_W}{\hat{s}_W}B_{22}(p^2,m_W,m_W)\notag\\
&+4g^2\frac{\hat{s}_W}{\hat{c}_W}\left(p^2-\frac{m_W^2}{2}\right)B_0(p^2,m_W,m_W). 
\end{align}

\section{Vertex correction}
The form factors $\Gamma_{V,A}^{Z\bar{f}f}(p^2)$ of the $Z\bar{f}f$ vertex can be calculated as \cite{hhkm}
\begin{align}
\Gamma_V^{Z\bar{f}f}(p^2)=\Gamma_L^{Z\bar{f}f}(p^2)+\Gamma_R^{Z\bar{f}f}(p^2),\quad
\Gamma_A^{Z\bar{f}f}(p^2)=\Gamma_L^{Z\bar{f}f}(p^2)-\Gamma_R^{Z\bar{f}f}(p^2),
\end{align}
where $\Gamma_{L}^{Z\bar{f}f}(p^2)$ ($\Gamma_{R}^{Z\bar{f}f}(p^2)$) is $Z\bar{f}_Lf_L$ ($Z\bar{f}_Rf_R$) vertex form factor. 
These are given as
\begin{align}
\Gamma_{L}^{Z\bar{f}f}(p^2)&=-\frac{1}{16\pi^2}\Bigg\{(I_f-Q_f\hat{s}_W^2)\left[\frac{g^2}{\hat{c}_W^2}(I_f-Q_f\hat{s}_W^2)^2\Gamma_{1Z}^f(p^2)+\frac{g^2}{2}\Gamma_{1W}^{f'}(p^2)\right]
\notag\\
&+I_f\left[-\hat{c}_W^2g^2\bar{\Gamma}_{2W}^{f'}(p^2)+\frac{g^2}{2}\Gamma_{mW}^{f'}(p^2)\right]
\Bigg\},\\
\Gamma_{R}^{Z\bar{f}f}(p^2)&=\frac{1}{16\pi^2}\frac{g^2}{\hat{c}_W^2}Q_f^3\hat{s}_W^6\Gamma_{1Z}^f(p^2),
\end{align}
where 
\begin{align}
\Gamma_{1Z}^f(p^2)&=\Gamma_1(p^2,m_f,m_Z,m_f)-\Sigma'(m_f^2,m_f,m_Z),\\
\Gamma_{1W}^{f'}(p^2)&=[\Gamma_1+\Gamma_{1m}](p^2,m_{f'},m_W,m_{f'})-\Sigma'(m_f^2,m_{f'},m_W),\\
\bar{\Gamma}_{2W}^{f'}(p^2)&=[\Gamma_1+\Gamma_{1m}](p^2,m_{f'},m_W,m_{f'})-\Gamma_2(p^2,m_W,m_{f'},m_W)+2\text{Re}B_0(p^2,m_W,m_W),\\
\Gamma_{mW}^{f'}(p^2)&=\Gamma_{1m}(p^2,m_{f'},m_W,m_{f'})+\Gamma_{2m}(p^2,m_W,m_{f'},m_W),
\end{align}
with
\begin{align}
\Sigma'(p^2,m,M)&=-\left(2+\frac{m^2}{M^2}\right)B_1(p^2,m,M)-1,\\
\Gamma_1(p^2,m,M,m)&=\left[2p^2(C_{11}+C_{23})+4C_{24}-\frac{m^4}{M^2}C_0\right](0,0,p^2,m,M,m)-2,\\
\Gamma_2(p^2,M,m,M)&=2\left[p^2(C_{11}+C_{23})+\left(6+\frac{m^2}{M^2}\right)C_{24}+(p^2-m^2)C_0\right](0,0,p^2,M,m,M)-2,\\
\Gamma_{1m}(p^2,m,M,m)&=\frac{m^2}{M^2}\Big\{[p^2(C_{12}+C_{23})+2C_{24}-2M^2C_0](0,0,p^2,m,M,m)-\frac{1}{2}\Big\},\\
\Gamma_{2m}(p^2,M,m,M)&=\frac{2m^2}{M^2}[2M^2C_0-C_{24}](0,0,p^2,M,m,M).
\end{align}


\end{document}